\documentclass[final,1p,11pt]{elsarticle}

\usepackage{epubtk}    % calls hyperref, fancyvrb, and LRR macros
\usepackage{booktabs}  % possibly also \usepackage{longtable}
\usepackage{graphicx}  % must come after epubtk
\usepackage{xcolor,color,graphics}
\usepackage{makeidx}
\usepackage{epsfig,subfigure}
\usepackage{epstopdf}
\usepackage{bm}
\usepackage{amsmath}
\usepackage{amssymb}
\usepackage{amsthm}
\usepackage{color}
\usepackage{ucs}
\usepackage[utf8]{inputenc}
\usepackage{fontenc}
\usepackage[numbers]{natbib}
\usepackage{bbm}
\usepackage{calligra}
\numberwithin{equation}{section}
\usepackage{amssymb}
\journal{Physics Reports}

\newcommand{\be}{\begin{equation}}
\newcommand{\ee}{\end{equation}}
\newcommand{\bea}{\begin{eqnarray}}
\newcommand{\eea}{\end{eqnarray}}

\newcommand{\refeq}[1]{(\ref{#1})}
\newcommand{\etal}{{\it et al.}}

\newcommand{\lsim}   {\mathrel{\mathop{\kern 0pt \rlap
  {\raise.2ex\hbox{$<$}}}
  \lower.9ex\hbox{\kern-.190em $\sim$}}}
\newcommand{\gsim}   {\mathrel{\mathop{\kern 0pt \rlap
  {\raise.2ex\hbox{$>$}}}
  \lower.9ex\hbox{\kern-.190em $\sim$}}}

\newcommand{\n}{\~n}
\newcommand{\af}{u}

\newcommand{\mS}{{\mathcal S}}
\newcommand{\mP}{{\mathcal P}}
\newcommand{\mR}{{\mathcal R}}
\newcommand{\mQ}{{\mathcal Q}}
\newcommand{\mG}{{\mathcal G}}
\newcommand{\mT}{{\mathcal T}}
\newcommand{\mU}{{\mathcal U}}
\newcommand{\mM}{{\mathcal M}}
\newcommand{\mX}{{\mathcal X}}

\newcommand{\Ss}[1]{{\mathcal S_{\rm #1}}}
\newcommand{\lag}{{\mathcal L}}

\newcommand{\qinv}{\big(\hat{q}^{-1}\big)}

\newcommand{\m}[1]{{\hat{#1}}}
\newcommand{\intd}{\int{\rm d}^4x}
\newcommand{\intmu}[1]{\int{\rm d}^4x\sqrt{-#1}}

\newcommand{\Tr}{{\rm Tr}}

\newcommand{\Id}{\mathbbm 1}
\newcommand{\Od}{\mathcal{O}}

\def\d{{\rm d}}

\newcommand{\mpl}{M_{\rm Pl}}
\newcommand{\mbi}{M_{\rm BI}}
\newcommand{\Mbi}{\bar{M}_{\rm BI}}

\newcommand{\tGam}{\tilde{\Gamma}}

\newcommand{\pb}{\bar{p}}
\newcommand{\rhob}{\bar{\rho}}

\begin{document}

\begin{frontmatter}



\title{\huge{Born-Infeld inspired modifications of gravity}}


\author{Jose Beltr\'an Jim\'enez}
\address{Aix Marseille Univ, Universit\'e de Toulon, CNRS, CPT, Marseille, France.}
\ead{jose.beltran@cpt.univ-mrs.fr}

\author{Lavinia Heisenberg}
\address{Institute for Theoretical Studies, ETH Zurich, Clausiusstrasse 47, 8092 Zurich, Switzerland.}
\ead{lavinia.heisenberg@eth-its.ethz.ch}

\author{Gonzalo J. Olmo}
\address{Depto. de F\'isica Te\'orica and IFIC, Centro Mixto Universidad de Valencia-CSIC,  \\Burjassot-46100, Valencia, Spain.\\Departamento de F\'isica, Universidade Federal da Para\'iba, 58051-900 Jo\~ao Pessoa, Para\'iba, Brazil.}
\ead{gonzalo.olmo@uv.es}

\author{Diego Rubiera-Garcia}
\address{Instituto de Astrof\'isica e Ciencias do Espa\c{c}o, Universidade de Lisboa, \\ Faculdade de Ciencias, Campo Grande, PT1749-016 Lisboa, Portugal.}
\ead{drgarcia@fc.ul.pt}

\begin{abstract}
General Relativity has shown an outstanding observational success in the scales where it has been directly tested. However, modifications have been intensively explored in the regimes where it seems either incomplete or signals its own limit of validity. In particular, the breakdown of unitarity near the Planck scale strongly suggests that General Relativity needs to be modified at high energies and quantum gravity effects are expected to be important. This is related to the existence of spacetime singularities when the solutions of General Relativity are extrapolated to regimes where curvatures are large. In this sense, Born-Infeld inspired modifications of gravity have shown an extraordinary ability to regularise the gravitational dynamics, leading to non-singular cosmologies and regular black hole spacetimes in a very robust manner and without resorting to quantum gravity effects. This has boosted the interest in these theories in applications to stellar structure, compact objects, inflationary scenarios, cosmological singularities, and black hole and wormhole physics, among others. We review the motivations, various formulations, and main results achieved within these theories, including their observational viability, and provide an overview of current open problems and future research opportunities.
\end{abstract}

\vspace{-3cm}

\end{frontmatter}



\newpage

\tableofcontents

\newpage

\section{Preamble} \label{sec:introduction}

\subsection{Motivations and introduction}

General Relativity (GR) is nowadays firmly established as the standard theory to describe the gravitational interaction with the same mathematical framework and physical principles as those used by Einstein more than one hundred years ago. After all this time, it still stands out as the most successful theory able to explain all the gravitational phenomena in a wide range of scales. Direct tests comprise from sub-milimeter %\cite{Hoyle:2000cv}
to Solar System scales, where the Parameterised Post-Newtonian formalism has allowed to constrain deviations from GR in the weak field limit at the level of $\sim 10^{-5}$ \cite{Will:2014kxa}. Moreover, the amazing direct observation of gravitational waves by the LIGO collaboration is also compatible with the prediction of GR for the merging of two black holes, where strong field effects are relevant \cite{Abbott:2016blz,TheLIGOScientific:2016pea}. On the other hand, we have witnessed how the accurate measurements of the CMB anisotropies and galaxy surveys have established $\Lambda$CDM as the standard model of cosmology, which is based on a homogeneous and isotropic Universe governed by GR as the theoretical framework for gravity. This picture requires an unobserved cold dark matter source plus a tiny cosmological constant to account for the current accelerated expansion of the Universe. Furthermore, the $\Lambda$CDM model needs to be supplemented with the inflationary paradigm so that the primordial perturbations are generated during a short period of accelerated expansion at very early times. For a review on the current status of the $\Lambda$CDM model, its challenges and possible alternatives, see Bull et al. \cite{Bull:2015stt}. Further observational tests, for instance via the Euclid satellite \cite{Amendola:2012ys}, will hopefully shed light on all the additional elements above and their contributions to fundamental physics.

Despite its observational success, there are strong arguments supporting and/or motivating to seek for theories beyond GR. These arguments are of two kinds. On the theoretical side, GR itself predicts the unavoidable existence of spacetime singularities, i.e., events where our ability to make predictions comes to an end \cite{Senovilla:2014gza}. Such singularities are unavoidably developed during the gravitational collapse of a fuel-exhausted star to form a black hole \cite{Joshibook}, as well as during the cosmological evolution in the early Universe. In this sense, the requirement that ``nothing should  cease to exist suddenly" and that ``nothing should emerge out of nowhere" should be seen as basic consistency conditions for any physical theory, including GR. The existence of singularities in GR unavoidably leads to the breakdown of these conditions, and gives clear indications that we have pushed the theory beyond its regime of validity. According to the standard lore \cite{Burgess:2003jk}, GR is a good effective field theory up to a scale somewhere near the Planck mass and, therefore, those singular behaviours are regarded as manifestations that the higher order operators should be included. For this reason, quantum gravity is usually expected to regularise such singularities, although it is possible that high energy modifications of GR might allow to classically regularise some of those singularities
before reaching the cut-off of the theory without invoking any quantum gravity effects.

On the phenomenological side, the unprecedented experimental precision reached by observational cosmology requires the aforementioned {\it ad hoc} extra ingredients in order to account for the observations. While the cosmological constant is fundamental part of the theory and its difficulty resides in its {\it aesthetic} value that  poses naturalness problem, dark matter and inflation require the introduction of new physics and, as a consequence, a large degeneracy among all the proposed models. This degeneracy is more prominent owing to the lack of experimental signatures from laboratory experiments and particle accelerators, despite the existence of different ongoing galactic \cite{Strigari:2013iaa,Ackermann:2015zua}, cosmic rays \cite{Aramaki:2015pii}, CMB \cite{Ade:2015rim}, collider \cite{Carpenter:2012rg} and underground laboratory \cite{Akerib:2016vxi} searches.

In view of the above situation, one may wonder if the difficulties and lack of naturalness faced in GR indicates that a new framework to describe gravity is needed, which would yield different astrophysical and cosmological observational signatures from the $\Lambda$CDM model \cite{Joyce:2014kja}. From a conservative perspective, one may stick to the point of view that gravitation is a manifestation of the curvature of spacetime, but one that is not sufficiently well described by GR. As a matter of fact, the common factor to all the issues discussed above is the extrapolation of GR to regions where it has not been directly well tested and this may introduce significant bias in the interpretation of astrophysical and cosmological observations. The consideration of additional curvature contributions to the Einstein-Hilbert action, usually under the form of curvature invariants, has been used in the literature as a way to enlarge the phenomenology of gravity. This typically involves a number of problems such as higher-order field equations, which usually entail the presence of ghost-like instabilities \cite{Stelle:1977ry,Stelle:1976gc,Nunez:2004ts,Chiba:2005nz}, or the difficulty to make these models compatible with solar system tests due to the existence of new degrees of freedom \cite{Olmo:2006eh,Chiba:2006jp}\footnote{Those models avoiding these shortcomings and, at the same time, being able to provide a consistent cosmological expansion which is coherent with the GR limit are usually termed as \emph{viable}, see e.g. \cite{Amendola:2006kh,Cognola:2007zu,delaCruz-Dombriz:2015tye}.}. The arbitrariness in the choice of curvature invariants also implies a strong lack of naturalness in these models. The main references regarding such models and their applications are provided by de Felice and Tsujikawa \cite{DeFelice:2010aj}, Capozziello and de Laurentis \cite{Capozziello:2011et}, and Nojiri and Odintsov \cite{Nojiri:2010wj} (see also Faraoni and Sotiriou \cite{Sotiriou:2008rp}).

The difficulties with ghost-like instabilities in higher curvature modifications of gravity can be avoided by formulating those theories in the so-called Palatini or metric-affine formalism \cite{Olmo:2011uz}. Though this approach is sometimes viewed as a shortcut to obtain the field equations of GR (and rightly so for some specific Lagrangians), it actually represents an inequivalent formulation of gravity in which metric and affine structures are regarded as independent geometrical entities. The fact that, when formulated \`{a} la Palatini \cite{Ferraris1982}, metric and connection are compatible in the case of GR has spread the view that such condition should always hold regardless of the form of the gravity Lagrangian. However, this is not true in general. In the metric-affine approach, the specific relation between metric and connection is determined by the field equations, not imposed a priori by mathematical conventions. In fact, whether the affine connection is determined by the metric degrees of freedom or not is as fundamental a question as the number of spacetime dimensions or the existence of supersymmetry.

The metric-affine or Palatini approach, therefore, avoids the problems with ghosts that affect extensions of GR in the usual metric formulation. In vacuum configurations, the field equations of these theories boil down to Einstein's equations with an effective cosmological constant \cite{Ferraris:1992dx} which, apparently, supports their compatibility with orbital motion tests (see \cite{Olmo:2005zr,Olmo:2008ye} for a discussion). Though this mathematical framework cannot solve on its own the arbitrariness in the choice of gravity Lagrangian, a novel class of extensions of GR with a solid motivation for a high-energy completion of gravity has been proposed and explored with much interest in the last few years. These models are motivated by the Born-Infeld approach to electrodynamics, where a modification of Maxwell's Lagrangian is introduced to set an upper bound on the electromagnetic field intensity \cite{BI1934}, with the result that the divergence of the self-energy of a point-like charge is regularised. This type of high-energy modification is analogous to the transformation that leads from a free particle in Newtonian mechanics to a free relativistic particle, whose maximum speed is bounded by the speed of light.
%and leads to the causal structure generated by the Lorentz group \cite{Ferraro_PRD?}.
The same Lagrangian structure describes the electromagnetic fields of $p$-branes in string theories \cite{Gibbons:1997xz,Brecher:1998su,Callan:1997kz}. It is natural to wonder whether such an approach, now fully defined in terms of geometrical objects, could play a similar role in order to avoid divergences and spacetime singularities in the high-energy/curvature regime and, accordingly, different proposals have been considered in the literature. Indeed, a major reason for the investigation of such models is the fact that, using standard matter sources satisfying the energy conditions, they naturally lead to non-singular cosmologies, inflationary scenarios without the need for scalar fields, and black hole spacetimes without singularities, among other appealing results. Moreover, the physics of these gravity theories has been studied in numerous astrophysical, black hole and cosmological scenarios where high-energy physics is relevant.

In this work we shall refer to this kind of models, which are close to the original spirit of Born-Infeld electrodynamics, as \emph{Born-Infeld inspired modifications of gravity}. They are defined by the following basic principles:

\begin{itemize}
  \item \emph{Square-root form}: Some geometric object(s) appears under a square-root with a determinantal structure in the action which defines the gravitational theory, alongside with some new mass/length scale.
  \item \emph{Consistency}: No obvious pathologies are present, among which the absence of ghost-like instabilities is of utmost importance. In turn, this almost unavoidably enforces the use of a metric-affine formulation.
  \item \emph{High-energy modification}: The modifications of GR mostly occur in the ultraviolet regime, i.e., in regions of large mass/curvature or short scales. This implies that GR is recovered in the low-energy limit.
\end{itemize}

Nonetheless, as there are available proposals in the literature for these theories that run away from one (or both) of the two last requirements, for completeness of this work we shall also discuss such proposals. A more precise description and classification of such theories will be presented in section \ref{Sec:BITheories}, alongside a criticism of each of them.

%Original paragraph:
% It should be stressed that a proposal for a theory of gravity where the affine connection plays a fundamental role was considered as early as in the twenties by Eddington \cite{Eddington24}. In such a proposal, an action defined in terms of the square-root of the determinant of the Ricci tensor and formulated in a purely affine approach is introduced, but where the coupling to matter is made though a (auxiliary) metric tensor that can be integrated out of the action. Built upon these principles, the influential paper by BaÃ±ados and Ferreira \cite{Banados:2010ix} modified Eddington's proposal with the addition of a contribution of the metric tensor under the square-root and a matter contribution in a standard way (i.e., as a separate piece of the action), and putting on equal footing the metric and affine structures (Palatini approach). This proposal, dubbed here as \emph{Eddington-inspired Born-Infeld gravity}, has had a great success in the literature and, indeed, most of the results on astrophysics, black holes and cosmology included in this review will revolve around it. Investigations on this model, its extensions, and other Born-Infeld inspired modifications of gravity might provide new insights on long-standing problems of gravitational physics, specially in the regime of high energy densities, while at the same time being consistent with the recovery of GR and its predictions in the low-energy density region.

This review is intended to fill a gap in the recent literature of Born-Infeld inspired modifications of gravity by providing a comprehensible account of the many different scenarios on which these classes of theories have been considered, including the astrophysics and internal structure of compact objects, solar physics constraints, modifications on black hole structure, non-singular black holes and wormholes, early universe and bouncing solutions, inflation, and dark energy, among others. Its aim is to summarise, classify and unify the different theoretical approaches, to clarify the assumptions on which the different approaches to build the theory are formulated, to discuss the numerous theoretical and phenomenological results, to highlight the experimental constraints these theories are subjected to, to clarify some existing misunderstandings, and to provide an overview of the future research opportunities. It is designed to be useful both for pure theorists and for astrophysicists/cosmologists working on alternatives to the $\Lambda$CDM (plus inflation) model.

% We hope the reader will enjoy and profit from his/her time spent reading it.

For a review on modified gravity in cosmology mainly focused on infrarred modifications of gravity in connection with late-time solutions (but with little contact with Born-Infeld-inspired theories or the Palatini formalism), see instead Clifton et al. \cite{Clifton:2011jh}. For additional astrophysical and cosmological observational constraints over different modified theories of gravity deviating from GR predictions, see Berti et al. \cite{Berti:2015itd}.

\subsection{Outline}

The main content of this review is split in four sections, according to the context on which Born-Infeld-inspired theories of gravity have been investigated.

In section \ref{Sec:BITheories} we will briefly review the original Born-Infeld electrodynamics theory from which the motivation for analogue constructions within gravity emerges. After explaining the early attempts that resulted in pathological theories, we will introduce what represents the most extensively studied theory of gravity with the Born-Infeld structure. The slightly different formulations of such a theory will be discussed as well as the main equations. Along the way, we will spend some time discussing the two frames existing in these theories and clarify the physical meaning of the different geometrical objects arising in them. We will end this section with a survey on the different Born-Infeld inspired theories of gravity existing in the literature and we will provide a general mathematical framework for these theories. The general developments introduced in this section will serve as starting points for the practical applications discussed in the subsequent sections.

In section \ref{sec:astrophysics}  some attempts to place observational constraints on the Born-Infeld theory using stellar models are reviewed. We will make special emphasis on the central role played by the energy density in the modified dynamics of this theory, which affects in a nontrivial way the mass-radius relation and maximum mass limit of compact objects, the energy transport mechanisms and oscillation frequencies of stars, the intensity of neutrino fluxes from the Sun, \ldots providing numerous tests to confront the theory with observations. The need for a careful description of the outermost layers of compact objects is also discussed in detail, considering for this purpose some relevant examples in which the peculiarities of metric-affine theories demand additional modeling beyond the canonical approaches of GR.

In section \ref{sec:Blackholes} we will review the counterparts of the Schwarzschild and Reissner-Nordstr\"om black hole solutions of GR, %mainly focused on the proposal of Eddington-inspired Born-Infeld gravity by Ba\~nados and Ferreira \cite{Banados:2010ix},
where a coupling to a Maxwell field is considered. We will spend some time explaining the procedure for derivation of the corresponding solutions, so as to highlight some important subtleties. Then we will explain the main differences of such solutions as compared to the GR ones, in particular, regarding the modifications on the horizon structure, which bear some resemblance to that of black holes supported by Born-Infeld electrodynamics in GR. On the other hand, we will study how these black holes may affect the description of strong gravitational lensing as well as the physics regarding mass inflation. An important issue will be the existence of non-singular geometries in these theories, whose nature and properties is tested using different well-established criteria. We also review some wormhole solutions constructed out of anisotropic fluids. Finally, different extensions to higher and lower dimensions, as well as to magnetically charged solutions will be discussed.

% Preferiría esperar a la siguiente sección para introducir la terminología EiBI y las referencias asociadas.

The section \ref{Sec:Cosmology} will be devoted to the effects of Born-Infeld inspired theories of gravity in cosmological scenarios. %As in the precedent sections, most of the cosmological applications have been considered for the EiBI model.
We will discuss the existence of homogeneous and isotropic solutions free from Big Bang singularities with standard matter sources as well as couplings of these theories to other types of fields. Anisotropic models and inhomogeneous perturbations will also be discussed. Since the Born-Infeld inspired theories are designed to modify gravity in the high curvatures regime, their natural domain of applicability is the early universe. However, there have also been studies where Born-Infeld theories are considered for late time cosmology and we will revisit them. %Finally, we will review some cosmological applications of other Born-Infeld inspired theories of gravity.

We will end in section \ref{Sec:Conclusions} by giving a summary of all the material presented in the core of this review. We will discuss the most outstanding achievements and will make special emphasis on the open questions that remain as well as the prospects for future research within the field.

\subsection{Preliminaries}
%\addcontentsline{toc}{section}{Preliminaries}
In this section we will review some basic ingredients of differential geometry that we will use throughout the different parts of this review. We will assume that the reader is familiarized with the concepts presented here and the main purpose of this section will be to fix the notation and the conventions for the different choices of signs and numerical factors in the definitions of relevant geometrical objects. It does not intend to be an exhaustive and rigorous exposition, but rather it should be regarded as a brief compendium of useful concepts and formulae. For a more detailed treatment we urge the reader to consult her/his favourite book on differential geometry or General Relativity or, in the lack thereof, see e.g. \cite{schouten1954ricci,misner1973gravitation,Waldbook}. One reference particularly useful and with numerous applications in gravitation and gauge theories is \cite{EGUCHI1980213}.

\subsection*{Connection, curvature and torsion conventions}
The theories that will be considered throughout the present review will be formulated either in (pseudo-)Riemannian or non-Riemannian geometries. In order to construct the necessary geometrical framework, we first introduce a 4-dimensional manifold $\mM$ that will eventually constitute our spacetime. In that spacetime we introduce a general connection $\Gamma$ that defines the covariant derivative of a 1-form $A_\mu$ as
\be
 \nabla_\mu A_\nu=\partial_\mu A_\nu-\Gamma^\lambda_{\mu\nu}A_\lambda.
\ee
This definition results in the following covariant derivative for a vector field $A^\mu$:
\be
 \nabla_\mu A^\nu=\partial_\mu A^\nu+\Gamma^\nu_{\mu\lambda}A^\lambda.
\ee
These expressions can then be easily generalised to arbitrary tensors $T^{\mu_1\cdots\mu_p}{}_{\nu_1\cdots\nu_q}$ so that
\begin{align}
\nabla_\alpha T^{\mu_1\cdots\mu_p}{}_{\nu_1\cdots\nu_q}=\partial_\alpha T^{\mu_1\cdots\mu_p}{}_{\nu_1\cdots\nu_q}&-\Gamma^\lambda_{\alpha\nu_1}T^{\mu_1\cdots\mu_p}{}_{\lambda\nu_2\cdots\nu_q}-\cdots-\Gamma^\lambda_{\alpha\nu_q}T^{\mu_1\cdots\mu_p}{}_{\nu_1\cdots\nu_{q-1}\lambda}\nonumber\\
&+\Gamma^{\mu_1}_{\alpha\lambda}T^{\lambda\mu_2\cdots\mu_p}{}_{\nu_1\cdots\nu_q}+\cdots+\Gamma^{\mu_p}_{\alpha\lambda}T^{\mu_1\cdots\mu_{p-1}\lambda}{}_{\nu_1\cdots\nu_q}\,.
\label{Eq:CDtensor}
\end{align}

In addition to objects with tensorial transformation properties under changes of coordinates, we will also find objects with other transformation properties throughout this review. In particular, we will encounter vector densities, which pick up some power of the Jacobian under a change of coordinates. If ${\mathcal {A}}^\mu$ is a vector density of weight $w$, it transforms as\footnote{This is true for true tensorial densities. For pseudo-tensorial densities the transformation also picks up a sign for parity odd transformations.}
\be
\tilde{\mathcal{A}}^\mu=\left(\det \frac{\partial \tilde{x}^\alpha}{\partial x^\beta}\right)^w \frac{\partial \tilde{x}^\mu}{\partial x^\nu}\mathcal{A}^\nu.
\ee
This modified transformation property makes necessary to add a piece to the definition of the covariant derivative to maintain its tensorial character, that reads
\be
\nabla_\mu \mathcal{A}^\nu=\partial_\mu \mathcal{A}^\nu+\Gamma^\nu_{\mu\lambda}\mathcal{A}^\lambda+w\Gamma^\lambda_{\mu\lambda}\mathcal{A}^\nu.
\label{Eq:covmA}
\ee
Again, this formula can be generalized  for an arbitrary tensorial density $\mathcal{T}^{\mu_1\cdots\mu_p}{}_{\nu_1\cdots\nu_q}$ by adding a term $w\Gamma^\lambda_{\alpha\lambda}\mathcal{T}^{\mu_1\cdots\mu_p}{}_{\nu_1\cdots\nu_q}$ in \refeq{Eq:CDtensor}.

After introducing the connection, we can start computing geometrical objects from the commutator of covariant derivatives acting on different tensorial fields. The first commutator we can compute is that of two covariant derivatives acting on a scalar field, which reads
\be
\big[\nabla_\mu,\nabla_\nu\big]\phi=-\mT^\lambda_{\mu\nu}\partial_\lambda\phi
\ee
with
\be
\mT^\lambda_{\mu\nu}\equiv\Gamma^\lambda_{\mu\nu}-\Gamma^\lambda_{\nu\mu}
\ee
the torsion tensor. Let us notice that it has tensorial transformation properties because it can be seen as the difference of two connections. The next geometrical important object is obtained by computing the commutator of two covariant derivatives acting on a vector field, which can be written as
\be
\big[\nabla_\mu,\nabla_\nu\big]A^\alpha=\mR^\alpha{}_{\beta\mu\nu}A^\beta-\mT^\lambda_{\mu\nu}\nabla_\lambda A^\alpha
\ee
where we have introduced the curvature Riemann tensor, defined as
\be
\mR^\alpha{}_{\beta\mu\nu}\equiv \partial_\mu\Gamma^\alpha_{\nu\beta}-\partial_\nu\Gamma^\alpha_{\mu\beta}+\Gamma^\alpha_{\mu\lambda} \Gamma^\lambda_{\nu\beta}-\Gamma^\alpha_{\nu\lambda} \Gamma^\lambda_{\mu\beta}
\ee
Out of this general Riemann tensor, we can build two independent traces, namely the Ricci tensor defined as usual $\mR_{\mu\nu}=\mR^\alpha{}_{\mu\alpha\nu}$ and the homothetic tensor given by $\mQ_{\mu\nu}=\mR^\alpha{}_{\alpha\mu\nu}$. While the Ricci tensor does not have any symmetry (even for a torsion-free connection), the homothetic tensor is antisymmetric. A quantity that we will need to compute field equations is the variation of the Ricci tensor under an infinitesimal displacement of the connection $\Gamma\rightarrow\bar{\Gamma}+\delta\Gamma$, which reads
\be
\delta \mR_{\mu\nu}=\bar{\nabla}_\lambda\delta\Gamma^\lambda_{\nu\mu}-\bar\nabla_\nu\delta\Gamma^\lambda_{\lambda\mu}+\bar{\mT}^\lambda_{\rho\nu}\delta\Gamma^\rho_{\lambda\mu}
\ee
where the bars denote quantities corresponding to the background connection $\bar{\Gamma}$. This relation reduces to the usual Ricci identity for torsion-free connections.

\subsection*{Metric convention}
After setting-up the notation and convention for the objects directly related to the connection, we will turn to the conventions for the metric tensor $g_{\mu\nu}$. This object is assumed to be non-degenerate and its inverse is denoted with upper indices $g^{\mu\nu}$ so that $g^{\mu\alpha}g_{\alpha\nu}=\delta^\mu_\nu$ and so on. Furthermore, this object is used to raise and lower indices of arbitrary tensors (i.e. it establishes an isomorphism between the tangent and the co-tangent spaces). We will use the mostly plus signature for the metric so that the Minkowski metric is $\eta_{\mu\nu}={\rm diag}(-,+,+,+)$. The covariant derivative of the metric  defines the non-metricity tensor $Q_{\alpha\mu\nu}$ as
\be
\nabla_\alpha g_{\mu\nu}=Q_{\alpha\mu\nu}.
\ee
Notice that the non-metricity is symmetric in the last two indices. This expression can be solved in the usual way to write the connection as
\be
\Gamma^\alpha_{\mu\nu}=\frac12 g^{\alpha\lambda}\Big(\partial_\nu g_{\mu\lambda}+\partial_\mu g_{\lambda\nu}-\partial_\lambda g_{\mu\nu}\Big)+L^\alpha_{\mu\nu}(Q)+K^\alpha_{\mu\nu}(\mT)
\ee
where the first term is the standard Levi-Civita piece, the second term depends on the non-metricity and the last term (usually called contorsion) is determined by the torsion. If the non-metricity vanishes and the connection is symmetric (i.e. vanishing torsion), the connection reduces to the Levi-Civita connection given by the Christoffel symbols. With a metric at hand, there is yet a third rank-2 tensor we can construct from the Riemann tensor of the full connection, known as co-Ricci tensor and defined as $\mP^\alpha{}_\mu\equiv g^{\beta\nu}\mR^\alpha{}_{\beta\mu\nu}$. Of course, for the Levi-Civita connection all three objects coincide up to a sign so the only independent trace of the Riemann is the Ricci tensor $R_{\mu\nu}$. Throughout this review we will denote with calligraphic letters $\mR_{\mu\nu},...$ the objects corresponding to an arbitrary connection, while the curvature objects associated to the Levi-Civita connection will be denoted with normal characters $R_{\mu\nu},...$ 

The determinant of the metric $\det g_{\mu\nu}\equiv g$ is a tensorial density of weight $-2$ so that $\sqrt{-g}$ is a tensorial density of weight $-1$  whose covariant derivative is given by
\be
\nabla_\mu\sqrt{-g}=\partial_\mu\sqrt{-g}-\Gamma^\lambda_{\mu\lambda}\sqrt{-g}.
\ee
We can thus use $\sqrt{-g}$ to {\it tensorialize} tensorial densities. For instance, if $\mathcal{A}^\mu$ is a tensorial density of weight $w$, then $A^\mu\equiv (\sqrt{-g})^w\mathcal{A}^\mu$ has weight zero. Another important use of this object is to construct invariant volume elements. Since $\d V$ generates a Jacobian under a change of coordinates, we can compensate for that by adding a factor of $\sqrt{-g}$ so that $\sqrt{-g}\d V$ will be invariant. Let us notice that this is a choice and actually we could use $\varphi \d V$ with $\varphi$ being whatever scalar density of weight $-1$. For instance, $\sqrt{\det a_{\mu\nu}}$ with $a_{\mu\nu}$ being an arbitrary rank 2 tensor will do the job.

The totally antisymmetric tensor is defined as
\be
\varepsilon_{\mu\nu\rho\sigma}=\sqrt{-g}\big[\mu\nu\rho\sigma\big]
\ee
with $\big[\mu\nu\rho\sigma\big]$ the totally antisymmetric Levi-Civita symbol with $[0 1 2 3]=1$. The contravariant version of it is
\be
\varepsilon^{\mu_1 \mu_2 \mu_3 \mu_4}=g^{\mu_1\nu_1}g^{\mu_2\nu_2}g^{\mu_3\nu_3}g^{\mu_4\nu_4}\varepsilon_{\nu_1 \nu_2 \nu_3 \nu_4}=-\frac{1}{\sqrt{-g}}\big[\mu_1 \mu_2 \mu_3 \mu_4\big].
\ee

The Levi-Civita tensor allows to introduce the Hodge dual that establishes an isomorphism between\footnote{Let us remember that a form is nothing but a completely antisymmetric tensor.} $p$-forms and $(D-p)$-forms. If $F_{\mu_1\cdots\mu_p}$ is a $p$-form, its dual is defined as
\be
\tilde{F}^{\mu_1\cdots\mu_{D-p}}=\frac{1}{p!}\epsilon^{\mu_1\cdots\mu_{D-p}\nu_1\cdots\nu_p}F_{\nu_1\cdots\nu_p}.
\ee
As a specific example that we will use throughout the review, the dual of a 2-form $F_{\mu\nu}$ in four dimensions is given by
\be
\tilde{F}^{\mu\nu}=\frac12 \epsilon^{\mu\nu\alpha\beta} F_{\alpha\beta}.
\ee
For an antisymmetric rank 2 tensor we can introduce the so-called electric $E_\mu$ and magnetic $B_\mu$ components relative to an observer with 4-velocity $u^\mu$ as
\be
E_\mu=F_{\mu\nu}u^\nu\quad\quad{\rm and}\quad\quad B_\mu=\tilde{F}_{\mu\nu}u^\nu.
\ee
For an observer with $u^\mu=(1,\vec{0})$ these definitions reduce to the usual expressions $F_{0i}=E_i$ and $F_{ij}=\frac12 \epsilon_{ijk} B^k$.

\subsubsection*{Tetrads formulation}
An alternative language to describe the geometrical framework of gravity theories is provided by the formalism of frames. We start by introducing a set of vectors defined on the tangent space $e_a=e_a{}^\mu\partial_\mu$ with a Lorentz index $a$ so that they satisfy the following orthonormality condition
\be
e_a{}^\mu e_b{}^\nu g_{\mu\nu}=\eta_{ab}.
\label{normalization}
\ee
with respect to the Minkowski metric $\eta_{ab}$. These objects receive several aliases in the literature: {\it tetrads}, {\it vierbein} or {\it frames}. The corresponding dual objects $e^a=e^a{}_\mu \d x^\mu$ belonging to the cotangent space are defined in the usual way $e^a_{\mu} e_b{}^\mu=\delta^a_b$. This relation in turns also implies $e^a_{\mu} e_a{}^\nu=\delta^\nu_\mu$. They are sometimes interpreted as the {\it square root } of the metric because $g_{\mu\nu}$ can be expressed as
\be
e^a{}_\mu e^b{}_\nu \eta_{ab}=g_{\mu\nu}.
\label{normalization2}
\ee
The vierbein can be used to {\it transform} tangent space indices into spacetime indices for arbitrary tensors. All the geometrical objects introduced above thus have their corresponding object in the tetrads formulation. If we introduce the so-called spin connection given by the set of 1-forms $\omega^{a}_{\mu\,b}$, the associated curvature 2-form is given by
\be
\mR^a{}_b=\d\omega^a{}_b+\omega^a{}_m\wedge\omega^m{}_b
\ee
where $\d$ is the exterior derivative and $\wedge$ stands for the exterior product. The existence of the tetrad allows to define the torsion 2-form as
\be
\mT^a=\d e^a+\omega^a{}_b\wedge e^b.
\ee
Applying the exterior derivative on this expression we obtain a consistency condition
\be
\d\mT^a+\omega^a{}_b\wedge T^b=\mR^a{}_b\wedge e^b
\ee
that relates all the relevant objects, namely, the tetrads, the spin connection, the torsion and the curvature. Taking a second exterior derivative of this expression will yield the usual Bianchi identities, which we do not need to display here. Instead, let us focus on two special connections that will be of relevance for this review. The first one is defined by the condition of being torsion-free, so it is defined by $\d e^a+\omega^a{}_b\wedge e^b=\mR^a{}_b\wedge e^b=0$ and it is the relevant one for the usual formulation of General Relativity. The second connection is curvature-free so we have $\d\omega^a{}_b+\omega^a{}_m\wedge\omega^m{}_b=d\mT^a+\omega^a{}_b\wedge T^b=0$ and defines the so-called Weitzenb\"ock space. This is the natural place for the Teleparallel formulation of GR.

\subsubsection*{Energy conditions}

A perfect fluid can be defined as one in which the energy-momentum tensor is locally seen as isotropic and it is fully determined by its density $\rho$ and its pressure $p$. According to this definition, the energy-momentum tensor of a perfect fluid as seen by an observer with 4-velocity $u^\mu$ ($u^2=-1$) is given by.
\be
T_{\mu\nu}=(\rho+p)u_\mu u_\nu+pg_{\mu\nu}
\ee
where it is immediate to see that $\rho=T_{\mu\nu}u^\mu u^\nu$ and $p=\frac13 (g^{\mu\nu}+u^\mu u^\nu)T_{\mu\nu}$. For a comoving observer with $u^\mu\propto\partial_t$ we have that $T^0{}_0=-\rho$ and $T^i{}_j=p\delta^i{}_j$.

For a general energy-momentum tensor, there is a set of conditions known as energy conditions that play an important role in theories of gravity in relation with singularity theorems, instabilities, superluminal propagation or entropy bounds. In the following we list them for future reference:

\begin{itemize}

\item Weak Energy Condition (WEC). This condition states that $T_{\mu\nu}v^\mu v^\nu\geq0$ for every time-like vector $v^\mu$ ($v^2<0$). For a perfect fluid, it implies the positivity of the energy density $\rho\geq0$ as measured by any observer and $\rho+p\geq0$.

\item Dominant Energy Condition (DEC). This condition is satisfied if $T_{\mu\nu}w^\mu w^\nu\geq0$ for every causal vector $w^\mu$ ($w^2\leq0$) and $-T^\mu{}_\nu w^\nu$ is a future-oriented causal vector. For a perfect fluid, this condition translates into $\rho\geq\vert p\vert$.

\item Strong Energy Condition (SEC). The SEC is satisfied if $T_{\mu\nu} v^\mu v^\nu\geq-\frac12 T$ for every time-like vector $v^\mu$ ($v^2<0$). A perfect fluid satisfies this conditions if $\rho+p\geq0$ and $\rho+3p\geq0$.

\item Null Energy Condition (NEC). The NEC is satisfied if for any null vector $n^\mu$ ($n^2=0$) the condition $T_{\mu\nu}n^\mu n^\nu\geq0$ holds. For a perfect fluid this implies $\rho+p\geq0$. This condition is satisfied for all known types of matter and it is saturated by a cosmological constant.

\end{itemize}

\subsubsection*{Matrix notation}
Given a rank-2 tensor, we will often use a hat to denote the corresponding matrix. Thus, the metric tensor $g_{\mu\nu}$ will also appear as $\m{g}$ and its inverse $g^{\mu\nu}$ will be denoted by $\m{g}^{-1}$ and similarly for other objects. The determinant of a matrix $\m{M}$ will be explicitly spelled out as $\det (\m{M})$ or will be alternatively denoted as $\vert\m{M}\vert$ where no confusion with absolute value should occur. In the special case of a metric $g_{\mu\nu}$, we will alternatively use the broadly used notation $g$ for its determinant. Analogously, for the trace of a matrix we will use either the explicit notation $\Tr(\m{M})$ or the more compact notation $[\m{M}]$ where, again, the context should clarify when the square brackets  stand for the trace or simply play the role of actual brackets.

A recurrent matrix formula that we will use throughout this review is the expansion valid for an arbitrary $n\times n$ matrix $\m{M}$ given by
\be
\det\Big(\Id+\m{M}\Big)=\sum_{i=0}^ne_i(\m{M})
\label{Eq:expansionen}
\ee
where $\Id$ is the $n\times n$ identity and $e_i$ the elementary symmetric polynomials which, for the case of interest here of $n=4$, read:
\begin{eqnarray}
e_0(\m{M}) &=& 1,\nonumber\\
e_1(\m{M}) &=& [\m{M}] ,\nonumber\\
e_2(\m{M}) &=& \frac{1}{2!}\Big([\m{M}]^2- [\m{M}^2]\Big),  \nonumber\\
e_3(\m{M}) &=&\frac{1}{3!}\Big( [\m{M}]^3- 3[\m{M}][\m{M}^2]+2[\m{M}^3] \Big),\nonumber\\
e_4(\m{M}) &=&\frac{1}{4!}\Big([\m{M}]^4-6[\m{M}]^2[\m{M}^2]+8[\m{M}][\m{M}^3]+3[\m{M}^2]^2-6[\m{M}^4]   \Big).
\label{Eq:defen}
\end{eqnarray}
It is useful to notice that the last elementary symmetric polynomial coincides with the determinant of $\m{M}$. Moreover, if $\m{M}$ is antisymmetric its trace is identically zero and, thus, $e_1$ and $e_3$ vanish.

\subsubsection*{Units and constants}
Unless otherwise stated, we will use units with $\hbar=c=1$. We will mostly use the reduced Planck mass, related to Newton's constant as $\mpl^{-2}=8\pi G_{\rm N}$. We will also make use of the Einstein's constant $\kappa^2=8\pi G_{\rm N}$.

\newpage
\section{Born-Infeld theories}
\label{Sec:BITheories}
The class of theories that generally go under the name of Born-Infeld all share the same basic feature of being defined in terms of some square root structure aimed at regularising the presence of divergences. The inception of these theories originated from the pioneering works by Born and Infeld in the 1930's \cite{Born:1933qff,Born410,1933Natur.132.1004B,Born:1934gh} where they assumed a principle of finiteness, according to which physical quantities are always bounded and can never become infinite. The self-energy of the electron, or a general point-like charged particle, is infinite in the classical Maxwell's theory so they searched for a non-linear modification capable of regularising this divergence as to comply with the principle of finiteness, i.e., a non-linear theory where point-like charges had finite self-energy\footnote{We should perhaps remark here that, at the time when Born and Infeld developed their theory for electromagnetism, the full machinery of quantum electrodynamics and the renormalization techniques were not available. Today we know that quantum electrodynamics is a renormalizable quantum field theory where physical quantities are finite and, in particular, the charge of a particle acquires radiative corrections at high energies owed to virtual processes.}. Motivated by the existence of an upper bound for the velocities of particles in relativistic mechanics, in the summer of 1933 Born proposed to introduce the same square root structure for electromagnetism in order to have an upper bound for the electric fields \cite{Born:1933qff,Born410}.  A few months later Infeld joined Born and together worked on a better version of this construction because they wanted a theoretically better motivated argument for such a theory and, then, they argued that the square root structure should come in from symmetry arguments. In analogy with mechanics where going from Newtonian to relativistic mechanics means upgrading Galilean transformations to the fully relativistic Lorentz group, Born and Infeld assumed that the Lorentz symmetry of Maxwell's theory should be enlarged in the new theory. They considered the new symmetry to be the full group of coordinate transformations which, after imposing the recovery of Maxwell's theory in the appropriate limit, led to the non-linear theory now known as Born-Infeld electromagnetism, expressed as the square root of a certain determinant \cite{1933Natur.132.1004B,Born:1934gh}. It is no surprise that the use of symmetries as a guiding principle gave rise to a remarkable theory of non-linear electromagnetism which, not only classically regularises the self-energy of point like charges, but it also shares some interesting features with Maxwell's theory and found a natural arena in the realm of other theoretically appealing theories, like e.g. string theory \cite{polchinski1998stringI,polchinski1998stringII,zwiebach2009first}.

Given the success of Born-Infeld theory to classically regularise divergences in electromagnetism, it is perhaps surprising that the same ideas were applied to resolve the singularities of General Relativity (GR) only in the late 1990's\footnote{A possible reason for this was the relative lack of interest in these topics until the seminal works by Hawking and Penrose \cite{Penrose:1964wq,Penrose:1969pc,Hawking:1966vg} concerning the singularity theorems in GR. On the other hand, the extraordinary success of quantum field theory perhaps motivated to invoke quantum gravity effects as the most likely mechanism that should regularise gravity in the high curvatures regime.}. The first attempt in this direction came about in a work by Deser and Gibbons \cite{Deser:1998rj}, where they finally took over the idea and tried to apply it to the case of gravity. However, as usual with gravity, things can very quickly go wrong when one tries to modify the Einstein-Hilbert action. The most straightforward application of the Born-Infeld philosophy by introducing a square root structure of a determinant involving the Ricci tensor gives rise to the presence of ghosts owing to the Ostrogradski instability associated to higher order equations of motion \cite{Woodard:2006nt,Woodard:2015zca}. In order to resolve the ghost problem, they proposed to add an additional term to remove the ghost order by order so that, when expanding the full action in the curvature, only the corresponding Lovelock term remains. This avoids the problem of the ghost, but the large freedom remaining in the choice of the additional piece and the lack of any guiding principle, makes the construction less appealing than the case of Born-Infeld electromagnetism. An obvious way to get around the ghost problem is to only use the Ricci scalar and apply the Born-Infeld construction to this quantity. This would lie within the class of $f(R)$ theories that contain one extra degree of freedom with respect to GR and, thus, it would deviate from the original Born-Infeld spirit where the theory is modified in some high energy regime by changing the structure of the theory in that regime instead of adding additional modes.

Some years later, Vollick re-considered Born-Infeld type of actions for gravity from a different perspective \cite{Vollick:2003qp}. Similarly to Deser and Gibbons, Vollick also resorted to a straightforward translation of the Born-Infed action to the case of gravity. However, instead of adopting the metric formalism, he considered the action within a metric-affine approach so that the connection is left arbitrary and promoted to an independent field. Within that formalism, the problem of the ghosts encountered in the metric formalism are avoided and, thus, no additional terms to remove undesired interactions are needed. This approach can actually be seen as a combination of the Born-Infeld ideas together with the original purely affine theory of gravity proposed by Eddington. Later on, Ba{\n}ados and Ferreira took on Vollick's approach with a slight modification of the original action, that now goes under the name of Eddington-inspired Born-Infeld gravity (EiBI), and showed the existence of non-singular cosmological and black hole solutions. This particular realisation of Born-Infeld gravity theories has since then received a considerable attention and has been extensively explored in different contexts with promising results.

The proposal by Vollick and its relative by Ba{\n}ados and Ferreira finally succeeded to implement the ideas of Born-Infeld electrodynamics to the case of gravity. However, it is fair to say that this initial proposal merely consisted in obtaining a gravitational action \`a la Born-Infeld, but it lacked any underlying guiding principle, based on symmetries like in Born-Infeld electrodynamics or any other equally valid motivation. In fact, it is very simple to come up with more general actions that could also be catalogued as Born-Infeld theories and could be considered on the same footing as EiBI. It does not come as a surprise then that very soon, modifications, extensions or alternative implementations of the Born-Infeld ideas to gravity appeared in the literature.

In this section we will review in detail the developments discussed above that led to the formulation of Born-Infeld gravity theories. We will start by reviewing Born-Infeld electrodynamics as a good starting point to motivate the search for analogous theories within gravitational contexts. We will show how the first attempts formulated in the metric formalism did not succeed due to the presence of ghosts. After that, we will turn to the formulation of Born-Infeld actions for gravity within a metric-affine approach and explain how the ghost issue is avoided. The general properties of these theories will be discussed in detail and, in particular, we will explain the existence of two frames. We will end this section by performing a classification of the different Born-Infeld inspired theories of gravities considered in the literature so far and briefly discuss them.

\subsection{Born-Infeld electromagnetism in a nutshell}
\label{Sec:BIEM}

\begin{figure}[h!]
\begin{center}
\includegraphics[width=15cm]{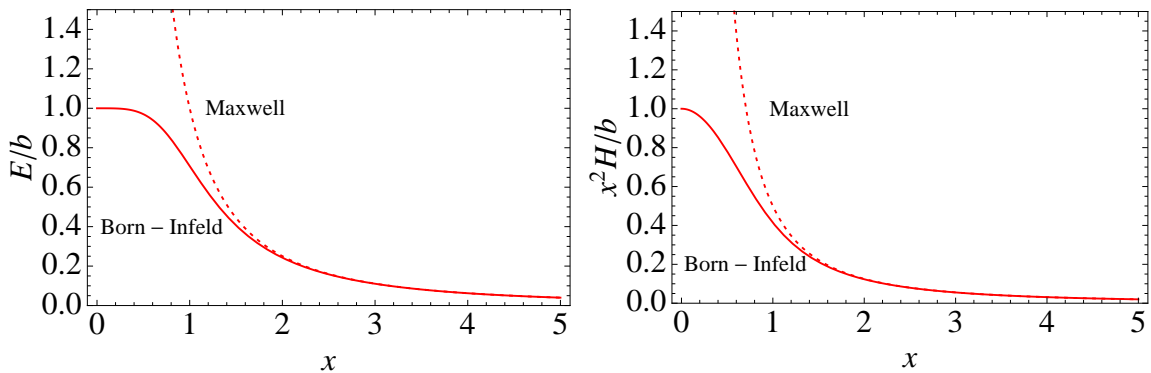}
\caption{In this plot we show the regularisation occurring in Born-Infeld electromagnetism (solid lines) as compared to the case of Maxwell's theory (dotted lines). In the left panel we show the profile (as a function of $x\equiv\sqrt{4\pi b/Q}r$) for the electric field generated by a point-like charge. We can clearly see the change from the usual $1/r^2$ behaviour at large distances to the saturation for the electric field due to the Born-Infeld corrections on small scales. In the right panel we show how this modified behaviour at small scales also regularises the energy density of the particle.}
\end{center}
\label{Fig:BIregularization}
\end{figure}

The underlying idea used by Born and Infeld to develop a modification of the Maxwell action as a potential mechanism to regularise some divergences associated to point-like charges was motivated by  the appearance of an upper bound for the speed of particles when upgrading Newtonian mechanisms to relativistic mechanics. In that case, the Newtonian Lagrangian for a massive particle of mass $m$ is simply $L=\frac12 m^2 v^2$, where $v$ is its velocity and can take any value. When including the principles of relativistic mechanics, the Lagrangian for the massive particle becomes $L=-m^2c^2\sqrt{1-(v/c)^2}$, where the speed of light $c$ makes its appearance as an upper bound for the velocities due to the square root. Taking inspiration from this, Born came up with the idea of modifying Maxwell's Lagrangian in such a way that the divergences of the Coulomb potential are automatically regularised due to the existence of a natural upper bound in the theory. In \cite{Born:1933qff,Born410}, he followed the most  straightforward application of this idea and proposed  the following replacement of Maxwell's Lagrangian:
\be
\lag=-\frac14 F_{\mu\nu}F^{\mu\nu}\to \lag=b^2\left[\sqrt{1-\frac{1}{2b^2} F_{\mu\nu}F^{\mu\nu}}-1\right]\,,
\label{Eq:Born1}
\ee
with $b$ representing the desired upper limit of possible electric fields. Although this simple replacement could do the job of regularising the infinities associated to point-like charges, it is not completely satisfactory from a theoretical point of view since there is no guiding principle for it other than the principle of finiteness. That is the reason that motivated Born, this time in collaboration with  Infeld, to look for a more theoretically appealing modification of Maxwell electromagnetism. They noted that, when going from classical mechanics to relativistic mechanics, the symmetry group is enlarged from the Galileo to the Lorentz group and it is precisely this group structure that nicely introduces the desired square root. Born and Infeld embraced this line of reasoning and looked for a non-linear theory of electromagnetism enlarging the group of special relativity as the relevant one. The idea is then that, very much like Newtonian mechanics is the limit of special relativity for small velocities and the Lorentz group reduces to the Galilean transformations, Maxwell electromagnetism should be the limit of some theory with a larger group of symmetries which, in some suitable limit, should reduce to the usual relativistic Lorentz transformations. Motivated by recent developments in gravity where the relevant group was shown by Einstein to be general coordinate transformations, they opted by enlarging the symmetry group of electromagnetism from the Lorentz group to the full group of general coordinate transformations\footnote{Incidentally, they were aware and noticed similarities with earlier attempts by Einstein, Weyl and Eddington, among others, in the same direction as a way to unify gravity and electromagnetism in a geometrical theory. However, Born and Infeld motivation was completely different and, as themselves claimed, their approach had nothing to do with those theories, except for some formal analogies, specially with Eddington's developments in \cite{eddington1924mathematical}. Remarkably, Eddington's theory eventually served as guidance to develop gravity theories \`a la Born-Infeld, as we will see in the section \ref{Sec:EiBI}.}.  Then, to have general covariance, the action should be constructed as $\mS=\intd \sqrt{\det a_{\mu\nu}}$, with $a_{\mu\nu}$ some rank-2 covariant tensor whose symmetric part can be identified with the metric tensor and its antisymmetric part is identified with the electromagnetic field strength $F_{\mu\nu}$. After imposing that Maxwell's theory should be recovered for small electromagnetic fields and neglecting some boundary terms, they arrived at the celebrated Born-Infeld action
\be
\Ss{BI}=-b^2\left[\intd\sqrt{-\det\Big(\eta_{\mu\nu}+\frac{1}{b}F_{\mu\nu}\Big)}-1\right].
\label{Eq:BIelectro}
\ee
This action has the properties they were after, namely, it introduces the square root structure by means of enlarging the symmetry group of Maxwell's theory. The constant $b$ is the only free parameter of the theory and it precisely gives the maximum allowed value for electric fields. Born and Infeld assumed the value of $b$ to be such that the corrections arise at the electron radius, although that value is now experimentally ruled out (see \cite{PhysRevD.93.093020} for a recent review on experimental bounds for non-linear electromagnetism). In order to see the appearance of a maximum value for the electric field,  let us notice that the action can be written in several useful ways by expanding the determinant in \refeq{Eq:BIelectro} to obtain
\begin{align}
\Ss{BI}=&-b^2\int \d^4x \left[\sqrt{1+\frac{1}{2b^2}F_{\mu\nu}F^{\mu\nu}-\frac{1}{16b^4}(F_{\mu\nu}\tilde{F}^{\mu\nu})^2}-1\right]\\\nonumber
=&-b^2\int \d^4x \left[\sqrt{1-\frac{\vec{E}^2-\vec{B}^2}{b^2}-\frac{(\vec{E}\cdot\vec{B})^2}{b^4}}-1\right]\,,
\end{align}
with $\tilde{F}^{\mu\nu}\equiv\frac12\varepsilon^{\mu\nu\alpha\beta} F_{\alpha\beta}$ the dual of the field strength, $\vec{E}$ and $\vec{B}$ the corresponding electric and magnetic parts and we have used the matrix identity
\begin{align}
\det\left(\delta^\mu{}_\nu+\frac{1}{b}F^\mu{}_\nu\right)=1+\frac{1}{2b^2} F_{\mu\nu} F^{\mu\nu}-\frac{1}{16b^4}\Big(F_{\mu\nu}\tilde{F}^{\mu\nu}\Big)^2.
\end{align}
Notice that this implies a $\mathbb{Z}_2$ symmetry $F_{\mu\nu}\rightarrow -F_{\mu\nu}$ owed to the property of the determinant $\det(\Id+\m{M})=\det(\Id-\m{M})$ for an arbitrary matrix $\m{M}$. From the above expression it is straightforward to see that Maxwell's electromagnetism is recovered for electromagnetic fields much smaller than $b$ and that, for configurations without magnetic field, we also recover the first Lagrangian \refeq{Eq:Born1} considered by Born. Furthermore, written in this way, we can easily understand why the self-energy of point-like charged particles is regularised. Since a particle at rest (or in its own rest frame) only generates electric field, the Lagrangian reduces to
\be
\lag_{\rm BI}=-b^2\sqrt{1-\frac{\vec{E}^2}{b^2}}
\ee
and we clearly see that the electric field is bounded by $b$. Given the gauge character of the theory, we still have the constraint equation generating the gauge symmetry (or the equivalent of Gauss' law) given by
\be
\vec{\nabla}\cdot \vec{\Pi}=\rho\quad\quad{\rm with}\quad\quad\vec{\Pi}=\frac{\partial\lag_{\rm BI}}{\partial \vec{E}}=\frac{\vec{E}}{\sqrt{1-\frac{\vec{E}^2}{b^2}}}
\label{Eq:PiToE}
\ee
and $\rho$ is the density of electric charge. As usual, for a point-like particle of charge $Q$ we can integrate the equation over a sphere to obtain
\be
\oint\vec{\nabla}\cdot \vec{\Pi}\;\d^3x=Q\quad\Rightarrow\quad \vert\vec{\Pi}\vert=\frac{Q}{4\pi r^2},
\label{Eq:solPi}
\ee
 where $Q$ is the total charge enclosed by the sphere $Q=\oint\rho\d^3x$. By inverting the relation \refeq{Eq:PiToE} between $\vec{\Pi}$ and $\vec{E}$ we can obtain the solution for the electric field generated by the particle
\be
\vec{E}=\frac{\vec{\Pi}}{\sqrt{1+\frac{\vec{\Pi}^2}{b^2}}}.
\label{Eq:solEBI}
\ee
As promised, for $\vert \vec{\Pi}\vert\ll b$ we have $\vert\vec{\Pi}\vert\simeq \vert\vec{E}\vert\propto 1/r^2$ which is the usual result in Maxwell's electromagnetism, while in the opposite regime with $\vec{\Pi}\gg b$ the electric field saturates to the value $\vert\vec{E}\vert=b$. This saturation is in turn the responsible for the regularization of the self-energy of the particle, that is given by
\be
\mathcal{U}=\int\d^3x\mathcal{H}=b^2\int\d^3x\left(\sqrt{1+\frac{\vec{\Pi}^2}{b^2}}-1\right)=4\pi b^2\int_0^\infty r^2\d r \left(\sqrt{1+\left(\frac{Q}{4\pi br^2}\right)^2}-1\right)\,,
\label{Eq:UBI1}
\ee
where we have used the expression for the Hamiltonian density\footnote{For the more careful reader, let us clarify that the Hamiltonian density including the interaction between the electric potential and the charge is $\mathcal{H}=\vec{\Pi}\cdot\dot{\vec{A}}-\lag_{\rm BI}+A_0\rho$. However, we can use the definition of the electric field $\vec{E}=\dot{\vec{A}}-\vec{\nabla} A_0$ to express the Hamiltonian density, up to total derivatives giving rise to boundary terms, as $\mathcal{H}=\vec{\Pi}\cdot\vec{E}-\lag_{\rm BI}+A_0(\rho-\vec{\nabla}\cdot\vec{\Pi})$. The term depending on $A_0$ will then be responsible for the gauge constraint giving Gauss' law that vanishes on-shell, so that it will not modify the self-energy of the particle.} $\mathcal{H}=\vec{\Pi}\cdot\vec{E}-\lag$ and the corresponding solution \refeq{Eq:solPi}. The integral diverges in the case of Maxwell electromagnetism due to the unbounded contributions from the small scales where one has ${\mathcal H}_{\rm Maxwell}\sim \vec{E}^2\propto r^{-4}$. In the Born-Infeld case however, the small scales region is modified and we have ${\mathcal H}_{\rm BI}\sim\vec{\Pi}\propto r^{-2}$ which makes the integral convergent (see Fig. \ref{Fig:BIregularization}). The integral can be exactly computed in terms of the gamma function $\Gamma(x)$ and the total finite result is
\be
\mathcal{U}=\frac{\Gamma^2(1/4)}{12\pi}\sqrt{b Q^3}.
\label{Eq:UBI2}
\ee

Let us return to the solution for the electric field given in \refeq{Eq:solEBI} and express it directly in terms of the generating charge as\footnote{For the amusement of the reader familiarised with screening mechanisms in modified gravity, let us notice that this solution realises a screening mechanism for the electromagnetic interaction resembling the so called K-mouflage or Kinetic screening of scalar fields.}
\be
\vert\vec{E}\vert=\frac{1}{ \sqrt{1+\left(\frac{Q}{4\pi br^2}\right)^2}}\frac{Q}{4\pi r^2}.
\label{Eq:solEBI}
\ee
This expression allows for an alternative interpretation of Born-Infeld electromagnetism. Instead of having modified Maxwell equations in the sector of the electromagnetic field, we can equivalently interpret Born-Infeld electromagnetism as a modification in the source term, i.e., the way in which charges generate electric fields is modified on small scales. In other words, we can interpret it as an effective scale-dependent charge, showing a certain formal resemblance with the renormalisation of the charge when radiative corrections are included in standard QED, but here from a purely classical standpoint without any quantum effect. This re-interpretation of Born-Infeld electromagnetism will be useful for the case of gravity where the Born-Infeld inspired modified gravity theories will also admit an interpretation as a modification of the way in which matter gravitates at high energies.

We will conclude by stressing that the resulting theory turned out to have a series of remarkable features that make the Born-Infeld action be very special among all possible non-linear extensions of electrodynamics. Such properties are related to its special structure, giving additional motivation and support to the idea of implementing the principle of finiteness by enlarging the symmetry group of Maxwell theory. This is nothing but another example of the power of using symmetries as guiding principles to formulate physical theories. In order to avoid further delays in entering into the main topic of this review, namely Born-Infeld inspired theories of gravity, we will abstain our desire of going through all the fascinating features of Born-Infeld electromagnetism and we will content ourselves with briefly enumerating some of its more remarkable properties. For more detailed information we refer to \cite{Plebanski:106680,Gibbons:2001gy,Ketov:2001dq} or standard textbooks on string theory where the Born-Infeld Lagrangian naturally appears, as e.g. \cite{polchinski1998stringI,polchinski1998stringII,zwiebach2009first}:

\begin{itemize}

\item The Born-Infeld action arises in string theory from $T$-duality invariance when describing an open string in an electromagnetic field, i.e., the Born-Infeld action is the appropriate one to couple strings to electromagnetic (or more general gauge) fields.

\item Born-Infeld electromagnetism shares with its Maxwellian relative (and other non-linear theories of electromagnetism) the so-called electric-magnetic self-duality \cite{BialynickiBirula:1984tx,Gibbons:1995cv}. This is a highly non-trivial invariance of the theory corresponding to a dual transformation of the electric and magnetic fields. See \cite{Aschieri:2008ns} for a review on many interesting aspects of duality rotations and theories with duality symmetry.

\item Despite the highly non-linear character of the Born-Infeld action, the corresponding equations of motion give rise to causal propagation and avoid the presence of shock waves and birrefringence phenomena.

\item The equations of Born-Infeld electromagnetism admit solitonic solutions with finite energy, known as BIons \cite{Callan:1997kz, Gibbons:1997xz}.

\end{itemize}

As we can see, the Born-Infeld theory for electromagnetism not only conforms to the task it was devised for, namely the regularisation of divergences associated to point-like charges, but it is kind enough as to also provide a number of additional gifts that were not required {\it a priori}. In the remaining of this section we will overview the attempts to apply similar ideas to the case of gravity. In general, we could say that, by the time of writing, there is not a gravitational analogue of Born-Infeld electromagnetism exhibiting all the successes and remarkable properties discussed above, but the search for it has nevertheless yielded very interesting gravitational theories \`a la Born-Infeld, both from a theoretical and a phenomenological points of view. We will start our tour however by reviewing the first attempts in this direction that led to pathological theories.

\subsection{The Deser-Gibbons proposal: The ghost problem of the metric formalism }
\label{Sec:D&G}

The original idea by Born and Infeld to regularise divergences in electromagnetism was taken over by Deser and Gibbons \cite{Deser:1998rj} as a potential mechanism to regularise the singularities that commonly appear in General Relativity, like e.g. the divergences at the center of black holes or the original Big Bang singularity. Following the same scheme, they considered an action for the gravitational interaction including the same determinantal and square root structures that appear in Born-Infeld electromagnetism. A straightforward translation of the Born-Infeld action for electromagnetism to the case of gravity would be the naive replacement of field strength $F_{\mu\nu}$ by the Ricci tensor $R_{\mu\nu}$ so that the first naive tentative action for a gravitational version of Born-Infeld electromagnetism would be
\be
\Ss{}=\intd\sqrt{-\det\Big(ag_{\mu\nu}+bR_{\mu\nu}\Big)}\,,
\ee
where $a$ and $b$ some parameters, $g_{\mu\nu}$ the spacetime metric and $R_{\mu\nu}$ the Ricci tensor of the corresponding Levi-Civita connection. However, this naive procedure leads to a theory plagued by ghost-like instabilities. The reason is clear from the well-known fact that an arbitrary action containing a non-linear function of the Ricci tensor will give rise to higher order gravitational field equations and, thus, it will be prone to the Ostrogradski instability \cite{Woodard:2006nt}. In order to avoid the presence of ghosts in the theory, Deser and Gibbons considered instead the action
\be
\Ss{DG}=\intd\sqrt{-\det\Big(ag_{\mu\nu}+bR_{\mu\nu}+cX_{\mu\nu}\Big)}\,,
\label{Eq:actionDG}
\ee
where the {\it fudge tensor} $X_{\mu\nu}$ must be tuned in order to get rid of the ghost. The form  $X_{\mu\nu}$ can be obtained perturbatively to remove the ghost at a given order and its effects are then pushed to higher orders. We can use the identity
\begin{eqnarray}
\sqrt{\det\Big(\Id+\m{M}\Big)}&=&\sqrt{1+[\m{M}]+\frac12\Big([\m{M}]^2-[\m{M}^2]\Big)+\Od(\m{M}^3)} \nonumber\\
&=&1+\frac12[\m{M}]+\frac18[\m{M}]^2-\frac14[\m{M}^2]+\Od(\m{M}^3)\, ,
\end{eqnarray}
valid for an arbitrary matrix $\m{M}$, to expand the action in powers of the curvature as
\be
\Ss{DG}=\intd\sqrt{-\det (ag_{\mu\nu})}\left[1+\frac{\Big(bR+cX\Big)}{2a}+\frac{\Big(bR+cX\Big)^2}{8a^2}-\frac{\Big(bR_{\mu\nu}+cX_{\mu\nu}\Big)^2}{4a^2}+\cdots\right]
\ee
where $R=g^{\mu\nu}R_{\mu\nu}$ is the Ricci scalar and $X=X^\alpha{}_\alpha$. In this expression we can see that, omitting $X_{\mu\nu}$ for a moment, we have a cosmological constant at zeroth order, while at first order we recover the usual Einstein-Hilbert term. At higher orders however the appearance of the quadratic terms $R_{\mu\nu}R^{\mu\nu}$ will lead to higher order equations of motion, thus rendering the theory unstable due to the presence of ghosts. Since we know that, at quadratic order, only the Gauss-Bonnet prevents the appearance of such ghosts, we must use the leading order contribution from $X_{\mu\nu}$ in order to remove the undesired terms. We can then assume an expansion in curvatures starting at quadratic order\footnote{We could also add lower order terms for the fudge tensor as, e.g. $X^{(1)}_{\mu\nu}=Rg_{\mu\nu}$, but that will not introduce the discussion other than adding some more terms in the equations.} for the fudge tensor of the form $X_{\mu\nu}=X_{\mu\nu}^{(2)}+\cdots$ and choose $X_{\mu\nu}^{(2)}$ to satisfy
\be
cX^{(2)\mu}{}_\mu+\frac{b^2}{4a}\Big(R^2-2R_{\mu\nu}R^{\mu\nu}\Big)=\alpha \Big(R_{\mu\nu\rho\sigma}R^{\mu\nu\rho\sigma}-4R_{\mu\nu}R^{\mu\nu}+R^2\Big),
\ee
with $\alpha$ some constant. The above choice thus only leaves the Gauss-Bonnet contribution at second order. By iterating this procedure one could remove the ghosts at any desired order. However, we already see at quadratic order that only the trace of $X_{\mu\nu}$ is determined and, therefore, a large variety of fudge tensors can do the job (see \cite{Gullu:2015cha,Gullu:2014gza} for explicit constructions). In fact, except for some singular actions, one can presumably write almost any gravitational action in the form of \refeq{Eq:actionDG} by means of an appropriate choice of $X_{\mu\nu}$. We can exemplify this by taking the Born-Infeld gravity theory developed by Nieto in \cite{Nieto:2004qj}. Motivated by the MacDowell-Mansouri formalism, Nieto considered a spacetime manifold endowed with a connection giving rise to a total curvature $\mR^a{}_\mu$ that can be split as $\mR^a{}_\mu=R^a{}_\mu+\lambda e^a{}_\mu$, where $R^a{}_\mu$ is the usual curvature of the Levi-Civita connection, $e^a{}_\mu$ is the vielbein field and $\lambda$ a constant parameter. For this connection, he then considers a Lagrangian in $D$ dimensions given by
\be
\lag=\det \mR^a{}_\mu.
\ee
If we use the previous splitting, we can write the Lagrangian as
\be
\lag=\lambda^D \,e \det\left(\delta^\mu{}_{\nu}+\frac{1}{\lambda}R^\mu{}_{\nu}\right) =\lambda^D e\sum_{n=0}^D L^{(n)}(R)\,,
\ee
where $e=\det e^a{}_\mu$ and we have used the matrix identity $\det (\Id+\m{M})=\sum_{n=0}^D e_n(\m{M})$, with $e_n(\m{M})$ denoting the $n$-th elementary symmetric polynomial of the matrix $\m{M}$ (see \refeq{Eq:expansionen}). In the present case, the matrix is the Ricci tensor and its elementary symmetric polynomials are precisely the Lovelock invariants, that we denote by $L^{(n)}(R)$, so that the considered action is nothing but a combination of all the Lovelock terms and, thus, the theory is ghost-free. One can then rewrite this Lagrangian in the Deser-Gibbons form by simply defining a matrix $\m{G}$ given by $\m{G}=-\m{R}^2$ so that the Lagrangian can be alternatively written as
\be
\lag\propto \sqrt{-\det\left(g_{\mu\nu}+\frac{2}{\lambda} R_{\mu\nu}+\frac{1}{\lambda^2} R_{\mu\alpha}R_\mu{}^\alpha\right)}\,,
\ee
where we have used the commutativity of the determinant and the square root (whenever it exists). This is the form found by Nieto and which he then related to Born-Infeld gravity. However, as we have seen, it is nothing but Lovelock gravity written in an obscure way. Furthermore, no additional work is necessary to know that the theory does not contain any ghosts. This example perfectly illustrates the necessity of a better defined strategy to construct theories of gravity \`a la Born-Infeld in order not to be deluded with well-known healthy theories in mysterious disguises.

\subsection{Other proposals in the metric formalism}
\label{Sec:Othermetric}
In the procedure presented in the precedent section, we have been careful to impose that only the Lovelock invariants should remain at a given order in the expansion. This is a crucial requirement for the consistency of the theory, as the presence of ghosts invalidates any background classical solution. The approach followed by Deser and Gibbons can be seen as a way to make sense of the theory by pushing the scale at which the ghost becomes relevant at higher scales, but the lack of any other guiding principle obstructs the construction of an appealing and well-defined full theory.

One might however take a less demanding approach and impose instead a weaker condition without compromising the stability of the theory due to the presence of ghosts, but at the expense of partially abandoning the original Born-Infeld spirit.  For instance, instead of using the fudge tensor to only leave Lovelock invariants at each order, one could allow for some arbitrary functions of them. Thus, at quadratic order we could have allowed for terms involving some linear combination of the squares of the Ricci scalar and the Gauss-Bonnet term. This would find motivation in the fact that arbitrary functions of these two scalar quantities are known to be particular cases where the Ostrogradski instability is bypassed. In the end, this would be nothing but a complicated way of rewriting the class of theories described by an arbitrary function $f(R,\mG)$, with $R$ and $\mG$ the Ricci scalar and the Gauss-Bonnet term respectively. Although perfectly legitimate, these theories introduce additional scalar degrees of freedom and, thus, they can hardly be considered as genuine Born-Infeld modifications of gravity. Of course, this does not mean that those alternatives are uninteresting, but rather they should be regarded as belonging to another class of theories.

In case one is interested in obtaining gravitational theories with an upper bound for the curvature, then one can simply write a specific model of an $f(R)$ theory where the function $f$ presents a branch cut at some high but finite curvature $R_0$. The square root function typical from Born-Infeld would achieve this, but other functions involving e.g. logarithms could serve as well. Feigenbaum \etal \cite{Feigenbaum:1997pf} explored this route in two dimensions where the curvature is fully determined by the Ricci scalar and they studied some black holes solutions. In a subsequent work \cite{Feigenbaum:1998wy}, Feigenbaum extended the analysis to four dimensions where he considered an action of the following type:
 \be
 \lag=R+\beta \sqrt{1-k_1 R_{\mu\nu\rho\sigma}R^{\mu\nu\rho\sigma}-k_2 R_{\mu\nu} R^{\mu\nu}-k_3 R^2}\,,
 \ee
 with $k_i$ and $\beta$ some constants. Again he studied black hole solutions that we will briefly review in section \ref{sec:ssswm}. However, the problem of ghosts arising from the explicit dependence on the full Riemann and the Ricci tensors is not discussed. In fact, from the own equations of motion given in \cite{Feigenbaum:1998wy}, one can see that they are fourth order and, thus, it would be expected to have ghosts. This pathology  renders the black hole solutions of limited physical interest, as the perturbations around them are likely to be unstable. The same problem applies to the theories considered by Comelli and Dolgov in \cite{Comelli:2004qr} constructed in terms of the Lagrangian
 \be
\lag=\det\sqrt{A(R) g_{\mu\nu}+B(R) R_{\mu\nu}}\,,
\ee
with $A$ and $B$ some given functions of the Ricci scalar. This Lagrangian combines the Deser  and Gibbons proposal with $f(R)$-type of theories, but without taming the presence of ghosts so that the obtained cosmological solutions are again of limited realistic applicability.

A more interesting proposal that is also closer to the Born-Infeld spirit was given by Wohlfarth in \cite{Wohlfarth:2003ss}. The theory is based on a symmetric object defined as
\be
R_{AB}\equiv R_{[a_1a_2][b_1b_2]}\,,
\ee
where the indices $A\equiv[a_1a_2]$, $B\equiv[b_1b_2]$ should be regarded as ordered pairs of indices. He then introduces the new metric and Kronecker delta
\bea
g_{AB}&\equiv& g_{a_1b_1}g_{a_2b_2}-g_{a_2b_1}g_{a_1b_2}\\
\delta^A_B&\equiv& \delta^{a_1}_{b_1} \delta^{a_2}_{b_2}-\delta^{a_2}_{b_1} \delta^{a_1}_{b_2}
\eea
that are then used in the usual way to manipulate capital indices. Moreover, one has the identity $\det g_{AB}=(\det g_{ab})^{d-1}$ valid in $d$ dimensions. The proposed Lagrangian within this formalism is
\be
\lag=\sqrt{-g}\left[\det\Big(\delta^A{}_B+\lambda R^A{}_B\Big)\right]^\zeta\,,
\ee
with $\lambda$ some constant and $\zeta$ a parameter with the only restriction to be a fractional number in order to allow for a regularization of curvature divergences. This represents an extension of Deser and Gibbons construction since more general curvature invariants appear in the Lagrangian. However, it shares the same problematic of containing ghosts (typically appearing at the scale determined by $\lambda$) which is then resolved in a similar fashion, i.e., the Lagrangian is corrected as
\be
\lag=\sqrt{-g}\left[\det\Big(\delta^A{}_B+\lambda M^A{}_B+\lambda^2 N^A{}_B\Big)\right]^\zeta\,,
\ee
where $M^A{}_B$ and $N^A{}_B$ are general expressions containing linear and quadratic curvature terms, respectively. The relative parameters among all the terms must be tuned to remove the ghosts at quadratic order, although one would expect to find again the ghost at higher orders. Thus, similarly to the Deser and Gibbons construction, additional requirements are necessary to find a satisfactory Born-Infeld theory of gravity within this formalism.

Another approach that has been taken in the literature consists in choosing the fudge tensor $X_{\mu\nu}$ such that some specific gravity theories are recovered in the low curvatures regime. In \cite{Gullu:2010pc}, the authors followed this path to construct a Born-Infeld extension of the so-called New Massive Gravity theory \cite{Bergshoeff:2009hq}, whose action is given by
\be
\Ss{NMG}=\frac{1}{\kappa^2}\int\d^3x\sqrt{-g}\left[-R+\frac{1}{m^2}\Big(R_{\mu\nu}R^{\mu\nu}-\frac38 R^2\Big)\right]
\ee
and describes a massive graviton in 3 dimensions\footnote{Since the graviton propagator trivialises in 3 dimensions, the problem of the potential ghosts discussed above are less virulent.}. One can then see that this action is recovered at quadratic order from \refeq{Eq:actionDG} in 3 dimensions by choosing $X_{\mu\nu}$ proportional to $R g_{\mu\nu}$ and appropriately tuning the parameters (with the possible addition of a cosmological constant). Interestingly, the resulting action that they consider recovers at cubic order the extension of New Massive Gravity found in \cite{Sinha:2010ai} by imposing the existence of a $c-$theorem. The same authors pursued a similar approach in \cite{Gullu:2010wb} to construct theories that recover Horava's gravity \cite{Horava:2009uw,Horava:2009if} in 3 dimensions at quadratic order.

\subsection{Eddington-Born-Infeld gravity}
\label{Sec:EiBI}
In the previous sections we have seen that a straightforward implementation of the Born-Infeld idea to the case of gravity is not an obvious task. It is not difficult to convince oneself that the main difficulty is the avoidance of ghosts and this is hardwired in the use of the metric formalism in the action. One can however seek for Born-Infeld inspired modifications of gravity within the realm of affine theories of gravity where the connection is regarded as an independent object. Within this framework, it is very natural to remember the purely affine theory of gravity introduced by Eddington and described by the following action\footnote{Deser and Gibbons already made reference to this approach in \cite{Deser:1998rj}, but they did not consider it any further in favour of a metric formalism.} \cite{eddington1924mathematical}:
\be
\Ss{E}=\intd\sqrt{\vert\det \mR_{(\mu\nu)}(\Gamma)\vert}\,,
\ee
where $\mR_{(\mu\nu)}(\Gamma)$ is the symmetric part of the Ricci tensor of an arbitrary connection $\Gamma^\alpha_{\mu\nu}$. In vacuum, this theory is equivalent to GR\footnote{The recovery of the GR equations in vacuum is not specific of Eddington's theory and, in fact, it is a general result for any  theory of gravity. The generality of this result actually boils down to the covariance of the field equations which imposes that, in vacuum, the Ricci tensor must be proportional to the metric. In theories of gravity with additional degrees of freedom, the extra fields should be regarded as matter fields and the recovery of GR in vacuum also applies. Another complementary way of understanding this general result is provided by the fact that GR is the only Lorentz invariant and unitary theory for a self-interacting massless spin-2 field in 4 dimensions, usually called graviton. Thus, if by gravity we understand a theory for such a particle, we will inevitably find GR in vacuum. Differences can however show up when including matter fields, as we will discuss later.}. This is easy to understand, since this theory can be seen as GR after integrating out the metric tensor. If we start with GR in the presence of a cosmological constant and in the Palatini formalism, we have
 \be
 \mS=\frac12 \mpl^2\intd \sqrt{-g}\Big( \mR(\Gamma)+2\Lambda\Big)
\ee
that gives the Einstein equations
\be
\mR_{(\mu\nu)}-\frac12\mR g_{\mu\nu}=\Lambda g_{\mu\nu}.
\ee
We can now take the trace to obtain $\mR=-4\Lambda$, which allows to rewrite the equations as $\mR_{(\mu\nu)}=-\Lambda g_{\mu\nu}$. This relation can be used in the action to remove the dependence on the metric tensor and we then recover the Eddington action. This procedure of integrating out the metric tensor is also valid when including matter fields as long as they couple minimally, i.e., the metric tensor will only enter algebraically. In that case, the resulting action will be more involved, but it allows to write a fully affine theory of gravity, as it was Eddington's original idea.

An important consequence of using the connection as a fundamental geometrical object in Eddington's theory is the avoidance of introducing ghosts associated to higher order equations of motion for the metric tensor. This is not  a specific feature of Eddington's theory, but it is a general result for theories formulated \`a la Palatini. In view of these results, Eddington's action seems to be a better suited starting point to implement the Born-Infeld construction for theories of gravity. This approach was taken by Vollick \cite{Vollick:2003qp}, who considered the action\footnote{Here we use the dimension 1 parameter $\mbi$ as the Born-Infeld scale instead of the constant $b$ used in \cite{Vollick:2003qp}. The relation between both is $b=\mbi^{-2}$.}
\be
\Ss{EBI}=\mbi^2\mpl^2\intd\left[\sqrt{-\det\Big( g_{\mu\nu}+\frac{1}{\mbi^2} \mR_{\mu\nu}(\Gamma)\Big)}-\sqrt{-\det g_{\mu\nu}}\right]\,,
\label{Eq:actionEBI}
\ee
where $\mbi$ is a mass scale determining when high curvature corrections are important. The second term is introduced to remove a cosmological constant, thus allowing for Minkowski solutions in vacuum. The above action for a theory of gravity combines the ideas of Eddington's theory with the Born-Infeld construction, resulting in a theory of gravity formulated in a metric-affine approach and incorporating the square root and determinantal structures characteristic of Born-Infeld electrodynamics.

Before entering into further developments, let us check that GR is indeed recovered when the curvature is much smaller than the scale $\mbi^2$. When taking that limit, the leading order correction is
\be
\Ss{EBI}(\vert \mR_{\mu\nu}\vert\ll \mbi^2)\simeq \frac12\mpl^2\intd\sqrt{-g} g^{\mu\nu} \mR_{\mu\nu}(\Gamma)
\label{Eq:GRlimitEBI}
\ee
thus reproducing the Einstein-Hilbert action in the first order formalism, which is known to coincide with GR on-shell and provided the matter fields couple minimally\footnote{The equivalence between the metric and the Palatini approaches has also been considered for more general actions in, e.g. \cite{Exirifard:2007da,Borunda:2008kf,Dadhich:2010dg}. A particularly interesting result is that the equivalence of both formulations extends to the whole series of Lovelock invariants, among which the Einstein-Hilbert action represents nothing but the lowest order term.} (see for instance \cite{Hehl1978,ortin2007gravity}). Let us pause here for a moment and seize the opportunity to clarify some subtleties concerning this point which are well-known in the community but are still source of a little confusion in some works (see for instance the discussion at this respect in section 2.3.1 of \cite{Clifton:2011jh}). When considering the Einstein-Hilbert action in the Palatini formalism in the presence of minimally coupled fields, the field equations of the connection can be recast as a metric compatibility condition for the metric tensor\footnote{See for instance \cite{ortin2007gravity} for details. We will also show more details on how this is achieved in section \ref{Sec:EBIextensionsGeneral} within the context of more general theories.} and, thus, a solution of the equations is the Levi-Civita connection of the spacetime metric. An important point to note however is that such a solution represents {\it a} solution, but the most general solution for the connection field equations involves an arbitrary 1-form, which can be taken to be the trace of the non-metricity or the trace of the torsion tensor. This is of course nothing but a reflection of the fact that the metric compatibility condition obtained from the connection field equations does not fully determine the connection and the Levi-Civita connection is only obtained after imposing a symmetric condition. It is sometimes stated that such a condition must be supplemented for the Einstein-Hilbert action to give GR in the Palatini formalism. However, one must also notice that the Einstein-Hilbert action has a projective invariance\footnote{In section \ref{Sec:BIfieldequations} we will show that this symmetry is shared by all theories defined in terms of the symmetric part of the Ricci tensor and we will compute the associated conserved current.} $\Gamma^\lambda_{\mu\nu}\rightarrow \Gamma^\lambda_{\mu\nu}+\xi_\mu\delta^\lambda_\nu$ which also involves an arbitrary 1-form $\xi_\mu$, and this is precisely the undetermined mode obtained when solving the connection equation. The gauge character of the projective invariance is discussed in great detail in \cite{Julia:1998ys,Dadhich:2010xa}.

In the case of the action (\ref{Eq:actionEBI}), the projective invariance is only obtained as a low curvature accidental symmetry, but it is generally broken by higher order interactions, unless the initial theory is defined only in terms of the symmetric part of the Ricci tensor, in which case the projective invariance is a symmetry of the full theory. Considering only the symmetric part of the Ricci tensor is a widely adopted (and very convenient) option in the literature and, in addition, it would be closer to Eddington's original theory. This is the option adopted by Ba{\n}ados and Ferreira in \cite{Banados:2010ix}\footnote{Here we prefer to restore all the dimensionful constants as opposed to \cite{Banados:2010ix}, where the authors set $8\pi G=1$. Furthermore, we correct a typo in form of a factor of 2 appearing there, which has propagated in the literature.}, where they considered the action
\be
\Ss{EiBI}=\mbi^2\mpl^2\intd\left[\sqrt{-\det\Big( g_{\mu\nu}+\frac{1}{\mbi^2} \mR_{(\mu\nu)}(\Gamma)\Big)}-\lambda\sqrt{-\det g_{\mu\nu}}\right]
\label{Eq:actionEiBI}
\ee
that has now become the standard version of the so-called Eddington-inspired-Born-Infeld gravity (EiBI). In this version, it is customary to let a cosmological constant term be encoded in the parameter $\lambda$ as $\Lambda=(\lambda-1)\mbi^2$. An important notational convention that might lead to some misinterpretations but is very common in the community is to use $\mR_{\mu\nu}$ to denote the symmetric part of the Ricci tensor without the explicit symmetrisation. To avoid any confusion, we will always make explicit the corresponding symmetrisation.

\subsection{Field equations}
\label{Sec:BIfieldequations}
In the literature there is a number of subtle points in the derivation of the field equations that are sometimes overlooked or omitted, so we will provide a detailed derivation here. The main differences that one can encounter eventually boil down to whether only the symmetric part or the full Ricci tensor is considered and if the connection is assumed to be symmetric a priori or not. The former condition is related to the presence of a projective invariance, while the latter has to do with the presence of torsion. In many practical applications, these differences do not  make a huge impact in the results, but one should nevertheless be careful to obtain the correct field equations. Let us then consider the action
\be
\Ss{}=\mbi^2\mpl^2\intd\left[\sqrt{-\det\Big( g_{\mu\nu}+\frac{1}{\mbi^2} \mR_{\mu\nu}(\Gamma)\Big)}-\lambda\sqrt{-\det g_{\mu\nu}}\right]+\Ss{M}[\Psi,g_{\mu\nu},\Gamma]
\label{Eq:actiongeneral}
\ee
where no assumptions are made a priori on the connection and the full Ricci tensor $\mR_{\mu\nu}$ with both its symmetric and its antisymmetric parts. Let us stress again that Vollick \cite{Vollick:2003qp} used the full Ricci tensor but constrained the connection to be symmetric, while Ba{\n}ados and Ferreira left the connection fully undetermined but considered only the symmetric part of the Ricci. We have also added general matter fields $\Psi$ that can, in principle, couple to both the metric and the connection. Then, we will detail where the differences arise when making the different assumptions. For later convenience and to comply with standard notation in the literature, let us introduce the notation
\be
q_{\mu\nu}\equiv g_{\mu\nu}+\frac{1}{\mbi^2} \mR_{\mu\nu}.
\ee
Then, the variation of the action \refeq{Eq:actiongeneral} can be expressed as\footnote{In the literature of Born-Infeld theories it is customary to denote the inverse of the matrix $q_{\mu\nu}$ simply as $q^{\mu\nu}$, in accordance with the usual convention of denoting the inverse of a metric with upper indices. Since we will have two metrics, we prefer to explicitly keep the inverse for the moment in order to avoid any confusion to the unfamiliar reader in these first steps into the formalism of Born-Infeld theories, since $q^{\mu\nu}$ could very well be confused with $g^{\mu\alpha} g^{\mu\beta} q_{\alpha\beta}$. We will eventually drop the explicit mention for the inverse of $\m{q}$ to alleviate the notation and whenever there is no risk of confusion.}
\be
\delta\Ss{}=\frac{\mbi^2\mpl^2}{2}\intd\left[\sqrt{-q}\big(\m{q}^{-1}\big)^{\nu\mu}\left(\delta g_{\mu\nu}+\frac{1}{\mbi^2} \delta \mR_{\mu\nu}\right)-\lambda\sqrt{-g}g^{\mu\nu}\delta g_{\mu\nu}\right]
+\delta\Ss{M}[\Psi,g_{\mu\nu},\Gamma]
\label{Eq:var1}
\ee
where $q=\det \m{q}$ and we have used the formula
\be
\delta\sqrt{-\det \m{M}}=\frac12\sqrt{-\det \m{M}}\,\Tr\Big[\m{M}^{-1}\delta\m{M}\Big]
\ee
valid for an arbitrary matrix $\m{M}$. The field equations for the metric tensor are then immediately seen to be
\be
\sqrt{-q} \big(\m{q}^{-1}\big)^{(\mu\nu)}=\sqrt{-g} \left(\lambda g^{\mu\nu}-\frac{1}{\mpl^2\mbi^2}T^{\mu\nu}\right)
\label{Eq:metricequations}
\ee
with the energy-momentum tensor of the matter fields defined as
\be
T^{\mu\nu}\equiv\frac{2}{\sqrt{-g}}\frac{\delta\Ss{M}}{\delta g_{\mu\nu}}\Big\vert_\Gamma.
\ee
Notice that this energy-momentum tensor is defined {\it at constant connection}. For minimally-coupled bosonic fields this is not relevant and the energy-momentum tensor will have the standard form. However, when considering fermionic and non-minimally coupled bosonic fields, the expression for the energy-momentum tensor will be in general different from the one that would be obtained in a purely metric formalism. It is important to note the symmetrisation of the object $\m{q}^{-1}$ in the field equations as a consequence of the symmetry of the metric tensor. Had we considered only the symmetric part of the Ricci tensor in the starting action, this symmetrisation would be innocuous. Furthermore, as said before, in most practical applications in cosmological contexts or spherically symmetric solutions, the matrix $\m{q}$ is symmetric and then one could omit the symmetrisation, but in the general case it is important to properly include it. We will come back to this point in section \ref{Sec:EBIextensionsGeneral} for more general Lagrangians.

The derivation of the connection field equations requires a bit more of work. In order to obtain them, we need the variation of the Ricci tensor:
\be
\delta \mR_{\mu\nu}=\nabla_\lambda\delta\Gamma^\lambda_{\nu\mu}-\nabla_\nu\delta\Gamma^\lambda_{\lambda\mu}+\mT^\lambda_{\rho\nu}\delta\Gamma^\rho_{\lambda\mu},
\label{Eq:deltaR}
\ee
where $\mT^\lambda_{\rho\nu}=\Gamma^\lambda_{\rho\nu}-\Gamma^\lambda_{\nu\rho}$ is the torsion tensor. Equipped with this relation, we can now proceed to compute the variation with respect to the connection. Leaving aside the variation of the matter sector for a moment, we have
\begin{align}
\delta\Ss{\Gamma}=&\frac{\mpl^2}{2}\intd\sqrt{-q}\big(\m{q}^{-1}\big)^{\nu\mu}\delta \mR_{\mu\nu}\nonumber\\
=&\frac{\mpl^2}{2}\intd\sqrt{-q}\big(\m{q}^{-1}\big)^{\nu\mu}\Big(\nabla_\lambda\delta\Gamma^\lambda_{\nu\mu}-\nabla_\nu\delta\Gamma^\lambda_{\lambda\mu}+\mT^\lambda_{\rho\nu}\delta\Gamma^\rho_{\lambda\mu}\Big)\nonumber\\
=&-\frac{\mpl^2}{2}\intd\Big\{\nabla_\lambda\Big[\sqrt{-q}\big(\m{q}^{-1}\big)^{\nu\mu}\Big]\delta\Gamma^\lambda_{\nu\mu}-\nabla_\nu\Big[\sqrt{-q}\big(\m{q}^{-1}\big)^{\nu\mu}\Big]\delta\Gamma^\lambda_{\lambda\mu}-\sqrt{-q}\big(\m{q}^{-1}\big)^{\nu\mu}\mT^\lambda_{\rho\nu}\delta\Gamma^\rho_{\lambda\mu}\Big\}\nonumber\\
&+\frac{\mpl^2}{2}\intd\left\{\nabla_\lambda\Big[\sqrt{-q}\big(\m{q}^{-1}\big)^{\nu\mu}\delta\Gamma^\lambda_{\nu\mu}\Big]-\nabla_\nu\Big[\sqrt{-q}\big(\m{q}^{-1}\big)^{\nu\mu}\delta\Gamma^\lambda_{\lambda\mu}\Big]\right\}.
\label{Eq:var2}
\end{align}
Let us take a moment here to elaborate on the terms in the last line. Usually in (pseudo-)Riemannian geometries without torsion, these terms correspond to total derivatives that can be simply dropped and do not contribute to the equations of motion. However, the divergence of a vector density $\mathcal{A}^\mu$ of weight $w=-1$ for a general connection is given by (see equation \refeq{Eq:covmA})
\be
\nabla_\mu \mathcal{A}^\mu=\partial_\mu \mathcal{A}^\mu+\mT^\lambda_{\lambda\mu}\mathcal{A}^\mu.
\ee
Since $\sqrt{-q}$ is indeed a scalar density of weight $w=-1$, we then see that the usual boundary terms generated when integrating by parts, actually contribute non-trivially to the field equations whenever torsion is present. Let us stress that the crucial element here is the torsion, i.e., even if there is non-metricity, the boundary terms would not contribute to the equations in the absence of torsion. This is in fact one of the important differences arising from considering a torsion-free connection from the beginning. After taking into account these considerations in the variation \refeq{Eq:var2} we obtain
\begin{align}
\delta_\Gamma\Ss{}=&-\frac{\mpl^2}{2}\intd\Big\{\nabla_\lambda\Big[\sqrt{-q}\big(\m{q}^{-1}\big)^{\nu\mu}\Big]\delta\Gamma^\lambda_{\nu\mu}-\nabla_\nu\Big[\sqrt{-q}\big(\m{q}^{-1}\big)^{\nu\mu}\Big]\delta\Gamma^\lambda_{\lambda\mu}-\sqrt{-q}\big(\m{q}^{-1}\big)^{\nu\mu}\mT^\lambda_{\rho\nu}\delta\Gamma^\rho_{\lambda\mu}\Big\}\nonumber\\
&+\frac{\mpl^2}{2}\intd\left\{\sqrt{-q}\big(\m{q}^{-1}\big)^{\nu\mu}\delta\Gamma^\beta_{\nu\mu}-\sqrt{-q}\big(\m{q}^{-1}\big)^{\beta\mu}\delta\Gamma^\lambda_{\lambda\mu}\right\}\mT^\alpha_{\alpha\beta}.
\label{Eq:var3}
\end{align}
After an appropriate re-shuffling of the indices, the connection field equations  can finally be expressed as
\begin{eqnarray}
&&\nabla_\lambda\Big[\sqrt{-q}\big(\m{q}^{-1}\big)^{\mu\nu}\Big]-\delta_\lambda^\mu\nabla_\rho\Big[\sqrt{-q}\big(\m{q}^{-1}\big)^{\rho\nu}\Big]\nonumber\\
&&=\Delta^{\mu\nu}_\lambda+\sqrt{-q}\Big[\mT^\mu_{\lambda\alpha}(\m{q}^{-1}\big)^{\alpha\nu}+\mT^\alpha_{\alpha\lambda}(\m{q}^{-1}\big)^{\mu\nu}-\delta^\mu_\lambda \mT^\alpha_{\alpha\beta}(\m{q}^{-1}\big)^{\beta\nu} \Big]
\label{Eq:connectioneq}
\end{eqnarray}
where, for completeness, we have added the {\it hypermomentum} of the matter fields
\be\label{Eq:defhyper}
\Delta^{\mu\nu}_\lambda\equiv\frac{2}{\mpl^2}\frac{\delta \mS_m}{\delta \Gamma^\lambda_{\mu\nu}}\Big\vert_{g_{\mu\nu}}.
\ee
Analogously to the energy-momentum tensor, the hypermomentum must be computed at constant metric. In most of the cases, we deal with minimally coupled bosonic fields in which case we have $\Delta_{\lambda}^{\mu\nu}=0$. However, the standard way of coupling fermionic fields to gravity is by resorting to the vierbeins formalism that allows to generalise the definition of the gamma matrices to curved spacetime. In that formalism, the fermions couple directly to the spin connection and, thus, contributions to the hypermomentum typically arise. We will leave this case aside and will assume vanishing hypermomentum. For this simplified case, we have the full set of equations for the Born-Infeld gravity that we display grouped together here for future reference
\begin{align}
&\sqrt{-q} \big(\m{q}^{-1}\big)^{(\mu\nu)}=\sqrt{-g} \left(\lambda g^{\mu\nu}-\frac{1}{\mpl^2\mbi^2}T^{\mu\nu}\right)\,,\\
\;\nonumber\\
&\nabla_\lambda\Big[\sqrt{-q}\big(\m{q}^{-1}\big)^{\mu\nu}\Big]-\delta_\lambda^\mu\nabla_\rho\Big[\sqrt{-q}\big(\m{q}^{-1}\big)^{\rho\nu}\Big]=\sqrt{-q}\Big[\mT^\mu_{\lambda\alpha}(\m{q}^{-1}\big)^{\alpha\nu}+\mT^\alpha_{\alpha\lambda}(\m{q}^{-1}\big)^{\mu\nu}-\delta^\mu_\lambda \mT^\alpha_{\alpha\beta}(\m{q}^{-1}\big)^{\beta\nu} \Big]
\end{align}
where
\be
q_{\mu\nu}\equiv g_{\mu\nu}+\frac{1}{\mbi^2} \mR_{\mu\nu}.
\ee
These will be the fundamental set of equations that need to be solved in Born-Infeld gravity. In most practical situations, the equations are greatly simplified and the general case is rarely required. Thus, instead of tackling the full set of equations directly, let us first first consider a simplified case where most of the results will be sufficient for the astrophysical and cosmological applications discussed in the subsequent sections.

\subsubsection{Simplified case: Vanishing torsion and projectively invariant case}
\label{Sec:SimplifiedEqEiBI}
We will start by considering the simplest possible case with vanishing torsion {\it a posteriori} and where the action is constructed out of the symmetric part of the Ricci tensor solely, and we will postpone the general case for later. The busy reader rushing to explore the different applications of the theory will be able to grasp the essential details in this section, since this simplified scenario is the most extensively considered case in spherically symmetric and cosmological solutions. The thorough reader will hopefully be satisfied with the more detailed discussion provided in section \ref{Sec:EBIextensionsGeneral} for more general theories (where in fact we will see that getting rid of the torsion does not represent any limitation for a class of theories among which we find EiBI). Let us notice that the assumption on $\mR_{\mu\nu}$ refers to the own definition of the theory while the torsion-free condition restricts the considered class of solutions within the theory.

The fact that we only consider the symmetric part of the Ricci tensor in the action has two important consequences. On one hand, the object $q_{\mu\nu}$ will inherit the symmetry of the Ricci tensor (along with that of the metric $g_{\mu\nu}$). On the other hand, we are enlarging the symmetries of the theory by introducing a projective invariance and, thus, this condition can be naturally introduced by imposing such a symmetry in the gravitational sector. The projective invariance corresponds to a shift in the connection of the form
\be
\Gamma^\lambda_{\mu\nu}\rightarrow \Gamma^\lambda_{\mu\nu}+\xi_\mu\delta^\lambda_\nu
\ee
for an arbitrary 1-form $\xi_\mu$. That this is in fact a symmetry of the theory containing only the symmetric part of the Ricci tensor can be easily seen from \refeq{Eq:deltaR} by taking $\delta_\xi\Gamma^\lambda_{\mu\nu}=\xi_\mu\delta^\lambda_\nu$ to obtain that, under a projective transformation, the full Ricci tensor transforms as
\be
\delta_\xi \mR_{\mu\nu}=\nabla_\mu\xi_\nu-\nabla_\nu\xi_\mu+\mT^\lambda_{\mu\nu}\xi_\lambda.
\ee
We can clearly see from here that the variation of the Ricci tensor under a projective transformation of the connection is antisymmetric and, thus, its symmetric part is invariant $\delta_\xi \mR_{(\mu\nu)}=0$. A consequence of this symmetry is that one of the traces of the connection field equations vanishes identically, i.e., the constraint associated to the projective symmetry is
\be
\delta^\lambda{}_\nu\frac{\delta \mS}{\delta \Gamma^\lambda_{\mu\nu}}=0.
\ee
Let us stress here that the projective symmetry will not be broken by the presence of minimally coupled fields. Bosonic fields with minimal couplings will only couple to the metric, so the projective invariance is obvious. On the other hand, minimally coupled fermions do couple to the connection, but such a coupling still respects the projective symmetry (see for instance \cite{Hehl:1994ue}). Finally, it is also interesting to note that the projective invariance is so-called because it is in fact a symmetry of the geodesics equations, since its effect can be re-absorbed into a re-definition of the affine parameter. For minimally coupled fields this is irrelevant because they are only sensitive to the Levi-Civita part of the full connection.

The field equations under the conditions at hand now reduce to
\begin{align}
&\sqrt{-q} \big(\m{q}^{-1}\big)^{\mu\nu}=\sqrt{-g} \left(\lambda g^{\mu\nu}-\frac{1}{\mpl^2\mbi^2}T^{\mu\nu}\right)\label{Eq:metricsimplified}\\
\;\nonumber\\
&\nabla_\lambda\Big[\sqrt{-q}\big(\m{q}^{-1}\big)^{\mu\nu}\Big]-\delta_\lambda^\mu\nabla_\rho\Big[\sqrt{-q}\big(\m{q}^{-1}\big)^{\rho\nu}\Big]=0\label{Eq:connectionsimplified}
\end{align}
where we have set $\mT^\lambda_{\mu\nu}=0$ and dropped the explicit symmetrization for $\qinv^{\mu\nu}$ since it is automatically symmetric. We can check that the trace of the connection equations with respect to $\lambda$ and $\nu$ vanishes identically, as a consequence of the projective symmetry, while the trace with respect to $\lambda$ and $\mu$ gives
\be
\nabla_\lambda\Big[\sqrt{-q}\big(\m{q}^{-1}\big)^{\lambda\nu}\Big]=0.
\ee
This constraint can then be plugged back into the connection equations to finally obtain
\be
\nabla_\lambda\Big[\sqrt{-q}\big(\m{q}^{-1}\big)^{\mu\nu}\Big]=0.
\label{Eq:Metriceqsimplified}
\ee
Since the action only depends on the symmetric part of the Ricci, the object $q_{\mu\nu}$ is symmetric and the above equation tells us that the connection must be compatible with the {\it auxiliary metric} $q_{\mu\nu}$, i.e., the connection is given by the Levi-Civita connection of the metric $q_{\mu\nu}$. It is important to notice that the metric compatibility condition only determines the symmetric part of the connection and, in general, leaves a vector component of the antisymmetric part undetermined. However, the assumption of a symmetric condition fixes this undetermined part. At this point, one could fairly object that we have not solved the connection yet, as the auxiliary metric $q_{\mu\nu}$ is defined in terms of the Ricci tensor, which depends on the connection itself. The resolution to this comes about by going back to the metric field equations \refeq{Eq:metricsimplified}. From there, we can see that the auxiliary metric can be fully expressed in terms of the spacetime metric $g_{\mu\nu}$ and the matter content through its energy-momentum tensor\footnote{Again, remember that we are considering minimally coupled fields, so the energy-momentum tensor does not depend on the connection, but only on the matter fields and, perhaps, the spacetime metric $g_{\mu\nu}$.} $T^{\mu\nu}$, so that the solution for the connection has actually been achieved. An important feature of this procedure that should not go unnoticed is that the connection has been obtained by solving algebraic equations and, therefore, no degrees of freedom are actually associated to it. In other words, there are no additional boundary conditions that we need to provide to solve for the connection, which means that it is nothing but an auxiliary field. This is the reason why the Born-Infeld theory modifies gravity without introducing new degrees of freedom. We will come back to this point later for more general cases.

Now that we have the solution for the connection, we can proceed to complete the resolution of the problem. This is not a very difficult task, since the field equations determining the auxiliary metric (that then gives the connection) are simply
\be
R_{\mu\nu}(\m{q})=\mbi^2\Big(q_{\mu\nu}-g_{\mu\nu}\Big)
\label{Eq:equationsforq}
\ee
where we need to remember that $\m{q}=\m{q}(\m{g},\Psi)$ is obtained from the metric field equations. However, instead of using these equations directly in this form, it is convenient to work them out a little bit to recast them into a more suitable form for direct applications. Let us begin by introducing some additional notation that is commonly used in the literature and which will allow to make contact with more general theories. We will denote by $\m{\Omega}$ the {\it deformation matrix} relating the auxiliary and the spacetime metrics as
\be
q_{\mu\nu}=g_{\mu\alpha}\Omega^\alpha{}_\nu
\label{Eq:defOmega}
\ee
or, in matrix notation, $\m{q}=\m{g}\m{\Omega}$. In the present case, this matrix is simply $\m{\Omega}=\Id+\frac{1}{\mbi^2}\m{g}^{-1}\m{\mR}$, obtained from the definition of $\m{q}$. However, the advantage of introducing this notation is that we can very easily solve the metric field equations \refeq{Eq:metricsimplified} for $\m{\Omega}$. When plugging \refeq{Eq:defOmega} into \refeq{Eq:metricsimplified}, we obtain the relation
\be
\m{\Omega}^{-1}=\frac{1}{\sqrt{\det\m{\Omega}}}\left(\lambda\Id-\frac{1}{\mpl^2\mbi^2}\m{T}\m{g}\right).
\label{Eq:equationsOmega}
\ee
Now, we can multiply \refeq{Eq:equationsforq} by $\m{q}^{-1}$ and use \refeq{Eq:defOmega} to obtain
\be
\m{q}^{-1}\m{R}(q)=\frac{1}{\mpl^2\sqrt{\det\m{\Omega}}}\left[\mpl^2\mbi^2\left(\sqrt{\det\m{\Omega}}-\lambda\right)\Id+\m{T}\m{g}\right].
\label{Eq:qequationscanonical}
\ee
This will be the starting point for many of the discussions in the subsequent sections devoted to astrophysical, black holes and cosmological applications. Let us stress that the components of $\m{\Omega}$ will be obtained as solutions of the metric field equations now expressed as \refeq{Eq:equationsOmega}. Thus, the solution of the problems is achieved in two steps. One first solves the set of algebraic equations \refeq{Eq:equationsOmega} to obtain $\m{\Omega}=\m{\Omega}(\m{T}\m{g})$, i.e., in terms of the metric and the matter content through the combination $\m{T}\m{g}$. For some important material contents, this combination does not depend on the metric but only on the energy density $\rho$ and the pressure $p$. This is the case for instance for a perfect fluid or an electromagnetic field (see sections \ref{sec:wormholes}, \ref{sec:Geons} and \ref{sec:general_framework_cosmology}). In that case, solving \refeq{Eq:equationsOmega} will yield $\m{\Omega}=\m{\Omega}(\rho,p)$. After obtaining these expressions, one can then complete the resolution of the problem by solving the differential equations \refeq{Eq:qequationscanonical}.

To end this section, let us notice that the equations \refeq{Eq:qequationscanonical} admit yet another formulation in terms of the Born-Infeld gravitational Lagrangian defined by means of $\mS_{BI}=\intd\sqrt{-g}\lag_{BI}$. If we restore the components notation, we have
\be
R^\mu{}_\nu(q)=\frac{1}{\mpl^2\sqrt{\det\m{\Omega}}}\Big(\lag_{BI}\delta^\mu{}_\nu+T^\mu{}_\nu\Big),
\label{Eq:qequationscanonical2}
\ee
where we have used the metric $q_{\mu\nu}$ to raise the first index of the Ricci tensor, i.e., $R^\mu{}_\nu(q)\equiv q^{\mu\alpha}R_{\alpha\nu}$. The interest of writing the equations in this form is twofold. Firstly, as we will see in section \ref{Sec:EBIextensionsGeneral}, this form of the field equations is valid not only for the Born-Infeld gravity theory considered here, but also for a large variety of theories formulated in the Palatini formalism. Thus, given a certain specific theory, we can immediately obtain the corresponding field equations by using \refeq{Eq:qequationscanonical2} directly. Secondly, this will be the starting point for many of the developments for practical applications that will be discussed in the subsequent sections of this review.

Another important feature of \refeq{Eq:qequationscanonical2} is that we can directly compare it with the usual Einstein equations of GR written as
\be
R^\mu{}_\nu=\frac{1}{\mpl^2}\left(T^\mu{}_\nu-\frac12 T \delta^\mu{}_\nu\right).
\ee
We can then see that the equations for Born-Infeld gravity written as \refeq{Eq:qequationscanonical2} can be interpreted as the usual Einstein equations for the auxiliary metric but with a modified source term, i.e., matter fields {\it gravitate} in a non-standard way. This closely resembles the analogous interpretation for Born-Infeld electromagnetism given from \refeq{Eq:solEBI}. A perhaps more apparent way of showing this is by re-writing \refeq{Eq:qequationscanonical2} in a more familiar form. If we take the trace, we obtain the relation
\be
R(q)=\frac{1}{\mpl^2\sqrt{\det\m{\Omega}}}\Big(4\lag_{BI}+T\Big)
\ee
so the Einstein tensor associated to the auxiliary metric $G^{\mu}{}_{\nu}(q)=R^{\mu}{}_{\nu}(q)-\frac12 R(q)\delta^\mu{}_\nu$ can be expressed as
\be
G^{\mu}{}_{\nu}(q)=\frac{1}{\mpl^2\sqrt{\det\m{\Omega}}}\left[T^\mu{}_\nu-\frac12\Big(T+2\lag_{BI}\Big)\delta^\mu{}_\nu \right].
\label{Eq:qEinsteineq}
\ee
These equations show even more clearly how the Born-Infeld theory can be seen as usual gravity {\it for the auxiliary metric} with a modified source term (let us remember once again that $\Omega$ is algebraically related to the matter content through \refeq{Eq:equationsOmega}). Furthermore, from here we can also easily understand a very distinguishing property of the theory. If we now use the relation between the two metrics $\m{q}=\m{g}\m{\Omega}$ to expand the Einstein tensor in \refeq{Eq:qEinsteineq} in terms of $\m{g}$-related objects we can immediately see that, since the Einstein tensor contains up to second derivatives of the metric, we will end up with up to second derivatives of the deformation matrix $\m{\Omega}$. This deformation matrix depends on the energy momentum tensor through \refeq{Eq:equationsOmega}) so that the evolution equations for the spacetime metric $g_{\mu\nu}$ will contain derivatives of the energy-momentum tensor components\footnote{One could object that second derivatives of $T_{\mu\nu}$ will give rise to higher than second order derivatives of the matter fields because the energy-momentum tensor typically contains first derivatives and, thus, the system might be prone to the very same Ostrogradski instabilities we claimed to be avoided. However, one should keep in mind that the matter fields will have their own second order field equations so they will in any case propagate the correct number of degrees of freedom.}. This is a very distinctive feature of these theories  that gives rise to new effects and, among others, a dependence of the gravitational potential on the local density and not only on an integrated density as in the usual case (see \refeq{eq:potential}). In fact, this effect has been claimed to lead to very serious drawbacks. We will give a careful discussion about this issue in section \ref{sec:Newtonianlimit}. Finally, this feature will also be the responsible for a dependence of the background cosmology evolution on the sound speed and not only on the equations of state parameter as in ordinary gravitational theories. We will see in section \ref{Sec:EBIextensions} that these properties are in fact shared by a large class of theories.
This non-standard interplay between the gravitational sector and the matter fields has been noticed and extensively used in the literature. See for instance \cite{Delsate:2012ky} for a devoted discussion on this point and \cite{Pani:2013qfa} where it is shown that gravity theories with generic auxiliary fields exhibit these properties.

 In the next subsection we will re-obtain this result in a slightly different and complementary way that will allow to clarify the role played by both metrics. Already here we can sense that the auxiliary metric carries physical relevance and it is not simply a mathematical object. We will postpone a thorough discussion about this point for the next subsection. Let us notice now that, very much like for the electromagnetic case, when we take curvatures  much smaller than $\mbi^2$ (or, equivalently, densities much smaller than $\mbi^2\mpl^2$), the deformation matrix is approximately the identity $\m{\Omega}=\Id+\Od(R/\mbi^2)$ so that $q_{\mu\nu}$ and $g_{\mu\nu}$ coincide up to corrections $\Od(R/\mbi^2)$. In that case, we also have $\lag_{BI}\simeq\frac12\mpl^2R$ and \refeq{Eq:qEinsteineq} reduces to the usual Einstein equations, confirming that the modifications only appear when the curvatures become order one as compared to the Born-Infeld scale $\mbi$. Equivalently, the Born-Infeld modifications will appear when $\vert T^\mu{}_\nu\vert\sim \mpl^2\mbi^2$.

To end this section, we will give some good news that will appease the less thorough reader. Despite  having neglected the torsion, all the results obtained here are completely valid for the general case with torsion provided the projective symmetry is imposed. We will show this explicitly in section \ref{Sec:EBIextensionsGeneral}

\subsection{The two frames of Born-Infeld gravity and the physical relevance of the auxiliary metric}
\label{Sec:Frames}
We have seen that Born-Infeld gravity naturally leads to the appearance of two metric tensors, namely the spacetime metric $g_{\mu\nu}$ and the auxiliary metric $q_{\mu\nu}$. The former plays the role of the metric to which matter fields couple, while the latter has been introduced as an auxiliary object to solve the equations so that the connection is the one compatible with it. So far, we have not provided this object with any physical meaning and it simply appeared as a mathematical trick to facilitate the resolution of the field equations or, equivalently, it appears as a result of integrating out the connection. The aim of this section will be to clarify the role of this object and unveil its physical significance.

The bi-metric character of the theory can be better appreciated by rewriting the EiBI action in the equivalent form\footnote{Here we consider the $\lambda-$term as part of the matter sector, where it will contribute as a cosmological constant.}
\be
\Ss{EiBI}=\frac12\mpl^2\mbi^2\intd\sqrt{-q}\left[\qinv^{\mu\nu}\left(g_{\mu\nu}+\frac{1}{\mbi^2}\mR_{(\mu\nu)}(\Gamma)\right)-2\right] +\Ss{M}[\Psi,g_{\mu\nu}]
\label{Eq:bimetricEiBI}
\ee
where we have introduced an auxiliary field that we have suspiciously called $q_{\mu\nu}$. To see that this is in fact equivalent to the EiBI action we can compute the field equations for this auxiliary field
\be
-\frac12\left[\qinv^{\mu\nu}\left(g_{\mu\nu}+\frac{1}{\mbi^2}\mR_{(\mu\nu)}\right)-2\right]q_{\alpha\beta}+g_{\alpha\beta}+\frac{1}{\mbi^2}\mR_{(\alpha\beta)}=0.
\ee
If we contract this equation with $\qinv^{\alpha\beta}$ we obtain the relation
\be
\qinv^{\mu\nu}\left(g_{\mu\nu}+\frac{1}{\mbi^2}\mR_{(\mu\nu)}\right)=4
\ee
 which can be plugged into the field equation to obtain the solution $q_{\mu\nu}=g_{\mu\nu}+\frac{1}{\mbi^2}\mR_{\mu\nu}$, that justifies our original name for this auxiliary field, since it turns out to be nothing but the auxiliary metric defined above. If we insert the solution into \refeq{Eq:bimetricEiBI} we see that we recover the original determinantal form of the EiBI action after integrating out the auxiliary field $q_{\mu\nu}$, proving the equivalence of both representations. The bimetric representation however provides a more orderly arrangement of the two metrics that allows to unveil their role in the theory. The role of the metric tensor is already clear from the beginning as the metric seen by matter fields and, therefore, determining their causal structure. In particular, point-like particles will follow the geodesics of the Levi-Civita connection corresponding to $g_{\mu\nu}$. There is nothing special here as this is a consequence of considering minimally coupled fields, the only difference with respect to the usual case being that  the solution for the metric tensor will be different. In order to properly identify the physical role of the auxiliary metric, let us notice two important features in \refeq{Eq:bimetricEiBI}. The first one is that the spacetime metric $g_{\mu\nu}$ only enters the action algebraically, i.e., without any derivatives. This means that $g_{\mu\nu}$ is an auxiliary field that can be integrated out. In fact, its equation of motion is given by
 \be
 \sqrt{-q}\qinv^{\mu\nu}=-\frac{1}{\mpl^2\mbi^2}\sqrt{-g}T^{\mu\nu}
 \label{Eq:equationg}
\ee
which allows to obtain $g_{\mu\nu}$ algebraically in terms of the matter fields and the auxiliary metric $q_{\mu\nu}$. For some types of matter fields this step might not be possible and, thus, the following discussion would not apply. Barring these singular cases, we can integrate out the spacetime metric and we will end up with an action of the form
\be
\Ss{EiBI}=\frac12\mpl^2\intd\sqrt{-q}\qinv^{\mu\nu}\mR_{(\mu\nu)}(\Gamma) +\tilde{\mS}_{\rm M}[\Psi,q_{\mu\nu}]
\label{Eq:bimetricEiBI2}
\ee
where $\tilde{\mS}_{\rm M}$ represents the new form of the matter sector after replacing the solution for $g_{\mu\nu}$ obtained from \refeq{Eq:equationg}. We thus arrive at an equivalent action with the Einstein-Hilbert term in the Palatini formalism to describe the dynamics of the auxiliary metric $q_{\mu\nu}$, but now the coupling of this auxiliary metric to the matter fields will have a complicated form. As discussed above, the Einstein-Hilbert sector will state that the connection $\Gamma$ must correspond to the Levi-Civita connection of the auxiliary metric $q_{\mu\nu}$, which is again the result obtained when working with the determinantal form of the action. This version of the action reveals a more profound role for the auxiliary metric since now we can see that the gravitational waves can be straightforwardly interpreted as the tensor part of the perturbations of the auxiliary metric. To further clarify this point, let us assume that we have a background configuration for both metrics given by $\bar{g}_{\mu\nu}$ and $\bar{q}_{\mu\nu}$. In this background geometry, the matter fields will propagate in the metric $\bar{g}_{\mu\nu}$ that will determine the corresponding causal structure. In particular, the light cone for photons will be determined by this metric. Furthermore, massive objects will be coupled in the standard way to the gravitational potentials and will follow the geodesics of $\bar{g}_{\mu\nu}$. On the other hand, gravitational waves will propagate on the background metric $\bar{q}_{\mu\nu}$ and it is this auxiliary metric that determines the causal structure for them so that gravitons will follow the geodesics of the auxiliary metric $\bar{q}_{\mu\nu}$. Since matter fields couple in a non-standard way to this metric, the interaction of the gravitational potentials encoded in the perturbations of $q_{\mu\nu}$ with the matter fields will differ from the usual case. We can then summarise this discussion by saying that the spacetime metric determines the propagation of matter fields and the auxiliary metric determines the propagation of gravitons.

The result obtained here and that boils down to the equivalent action \refeq{Eq:bimetricEiBI2} for EiBI gravity is equivalent to the finding presented at the end of \ref{Sec:SimplifiedEqEiBI} where the field equations were eventually written as \refeq{Eq:qEinsteineq} in the form of Einstein equations for the metric $q_{\mu\nu}$ with a modified source term. This is exactly what the action \refeq{Eq:bimetricEiBI2} is telling us, since the corresponding field equations will consist of the Einstein tensor obtained from varying the gravitational sector which will then be sourced by the energy-momentum tensor of the matter sector {\it as computed with respect to the metric} $q_{\mu\nu}$. In other words, the field equations are\footnote{Of course, we need to integrate out the connection. Since for the Einstein-Hilbert term at hand we know that the connection is given by the Levit-Civita connection of the metric $q_{\mu\nu}$, we omit this step here and assume that this operation has already been carried out.}
\be
G_{\mu\nu}=\frac{1}{\mpl^2}\tilde{T}_{\mu\nu}
\ee
where $\tilde{T}_{\mu\nu}$ is the effective energy-momentum tensor defined as
\be
\tilde{T}^{\mu\nu}\equiv\frac{2}{\sqrt{-q}}\frac{\delta\tilde{\mS}_{\rm M}}{\delta q_{\mu\nu}}\Big\vert_\Gamma.
\ee
We thus recover the field equations for Born-Infeld gravity written in an Einsteinian form as in \refeq{Eq:qEinsteineq} where we need to identify the non-standard source term in the right hand side with the effective energy-momentum tensor $\tilde{T}_{\mu\nu}$, which is non-trivially related to $T_{\mu\nu}$. It is important to notice that both energy-momentum tensors will satisfy their corresponding conservation equations, namely:  $\tilde{\nabla}_\mu\tilde{T}^{\mu\nu}=\nabla_\mu T^{\mu\nu}=0$ with $\tilde{\nabla}$ and $\nabla$ the covariant derivatives associated to $q_{\mu\nu}$ and $g_{\mu\nu}$ respectively. The result found here will help explaining why singular solutions like the Big Bang and/or black holes can be regularised without violating the null energy condition, because the object that will need to violate an {\it effective} null energy condition is not the standard energy-momentum tensor of the matter fields (see also sections \ref{sec:wormholes}, \ref{sec:Geons} and \ref{sec:general_framework_cosmology}).

A certain familiarity with modified gravity allows to appreciate a close analogy between the above discussion and the existence of two frames in scalar-tensor theories. In the Jordan frame matter fields are minimally coupled to the metric, but gravity is described by  a scalar-tensor theory. In the Einstein frame however gravity is described by the Einstein-Hilbert term, but matter fields couple to a conformal metric whose conformal factor depends on the scalar field. In the case of Born-Infeld, the situation is alike, but with the crucial difference that there are no additional propagating degrees of freedom. In the original description of the theory, that we will call the {\it Born-Infeld frame}, matter fields couple in the standard way to the metric but the gravitational action is non-standard. In this frame, we have that gravity reacts differently to the presence of matter when the densities are very high and particles follow the geodesics of the metric just as in standard gravity. In the alternative description exposed in this section, that we will call {\it Einstein frame} for obvious reasons, gravity has the standard Einstein-Hilbert action, but now the couplings of the matter fields to gravity are not the usual ones, i.e., we cannot simply follow the usual minimal coupling rule from flat spacetime and replace the Minkowski metric by the curved one appearing in the Einstein-Hilbert term.

The existence of the Einstein frame also helps understanding the Born-Infeld inspired gravity theories from a particle physics perspective. The common wisdom says that GR is the only consistent\footnote{By consistent one usually means unitary and Lorentz invariant, although locality is a frequent implicit condition. See for instance \cite{Biswas:2011ar} for consistent theories including non-localities. See also \cite{Hertzberg:2016djj,Hertzberg:2017abn} for other constructions based on higher but finite derivatives.} theory for a massless spin 2 field in 4 dimensions and this is usually used to state that modifications of gravity either introduce additional degrees of freedom or they reduce to GR. As we have seen, the Born-Infeld theories modify gravity without introducing additional degrees of freedom so we seem to face an apparent paradox. However, the more precise statement about GR being the unique theory for a massless spin 2 field concerns the IR regime and, thus, it is modifying gravity in the IR what requires the introduction of additional degrees of freedom. This is what usually happens in models of dark energy based on IR modifications of gravity. On the other hand, the high energy regime is not locked by the consistent requirements and, as we understand now from the Einstein frame, the Born-Infeld theories precisely modify this regime of gravity.

The Einstein frame also permits a more clear interpretation of the different regimes that we encounter in Born-Infeld inspired theories of gravity. As we have seen, these theories are characterised by two different scales, namely the Planck mass $\mpl$ and the Born-Infeld scale $\mbi$. These two scales are assumed to satisfy $\mbi\ll\mpl$ and this hierarchy introduces yet another relevant scale in the problem given by their geometrical mean $\Mbi=\sqrt{\mpl\mbi}$. The introduced hierarchy has the purpose of having Born-Infeld corrections before hitting the quantum gravity regime, that takes place at some scale near $\mpl$, so that we can have a range of scales between $\Mbi$ and $\mpl$ where gravity behaves differently but the quantum gravity effects can still be safely neglected. From the action in the Einstein frame expressed as \ref{Eq:bimetricEiBI2}, we see that the Born-Infeld scale $\mbi$ can be completely moved to the matter sector and, in combination with $\mpl$ through $\Mbi$, it controls the scale at which the generated non-linear interactions of the matter fields become relevant\footnote{Since the source of gravity in most situations is the energy density $\rho$, the transition between the usual GR and the Born-Infeld regimes in the gravitational sector is expected to occur when $\rho\sim \Mbi^4$, as we will confirm in the numerous applications studied in the subsequent sections. However, we should point out that this is only true in the simplest scenarios, but, in general, the Born-Infeld corrections will become relevant whenever some interactions reach $\Mbi$. To give an example, one could imagine a situation where the densities are small as compared to $\Mbi^4$, but some anisotropic stresses or heat fluxes are of order 1 as compared to the scale $\Mbi$.}. Interestingly, even fields that do not interact directly in the Born-Infeld frame will couple in the Einstein frame and the coupling will again be controlled by $\Mbi$. The fact that all fields will be generically coupled in the Einstein frame and the coupling constant $\Mbi$ is universal can be nicely interpreted as a consequence of dealing with a gravitational theory, i.e., as a sort of additional {\it Born-Infeld equivalence principle}. In other words, the Born-Infeld inspired theories have the usual equivalence principle, according to which all matter fields couple to gravity with a universal coupling constant $\mpl$ (fully valid on scales below $\Mbi$), and what we have called Born-Infeld equivalence principle, according to which all the generated couplings in the matter sector come in with another universal coupling constant $\Mbi$. Since we have not observed any anomalous interactions beyond those of the standard model at LHC, we can straightforwardly impose the very conservative constraint $\Mbi\gtrsim 10$ TeV, which translates into $\mbi\gtrsim 10^{-1}$eV so that the Born-Infeld corrections can only have effects in regions of spacetime where the curvature is larger than $10^{-2}$ eV$^2$.  

The couplings generated in the matter sector bring about one important point that is usually overlooked in the literature and has not been properly addressed yet, namely whether, or under which conditions, the quantum corrections can remain under control in the Born-Infeld regime. This is not obvious {\it a priori} because the couplings generated in the matter sector controlled by $\Mbi$ will usually contain non-renormalisable operators and, in fact, one would naively expect $\Mbi$ to play the role of a strong coupling scale and, thus, the effects at that scale will require non-perturbative analysis for large background field configurations. This however does not necessarily mean that the Born-Infeld regime will inevitably face strong coupling problems. We will give here a taste on a possible situation where the Born-Infeld regime can be safe, but a more careful analysis should definitely be performed. If we consider a massless scalar field in the Born-Infeld frame, in the Einstein frame we will have a $K$-essence type of theory where the interactions will be controlled by $\Mbi$. The strong coupling scale in these theories around a trivial background is $\Mbi$ and one can apply the standard perturbative analysis because the background value of the field is smaller than $\Mbi$. The worry comes when the background field takes values near $\Mbi$ and non-perturbative effects would be expected to become relevant. However, around these non-trivial backgrounds the vacuum value of the scalar field re-dresses the strong coupling scale so that it can be pushed to values higher than $\Mbi$. This mechanism is at work for instance in theories featuring a K-mouflage/Kinetic or Vainshtein screening (see for instance \cite{deRham:2013qqa,deRham:2012ew,Burrage:2011cr,deRham:2014naa,Heisenberg:2014rka,deRham:2014wfa,Brax:2016jjt}). In these situations the non-linear classical solutions can be trusted in the Born-Infeld regime. In this case the scalar will also couple to other matter fields through $\Mbi$, but again the coupling scale will be re-dressed by the background value of the scalar field, so that these interactions can also remain small. As we have emphasised, this is only a potential resolution of the strong coupling problems that one would expect in these theories, but one should carefully check whether this is the actual situation.

Let us end this section by noting that the discussion presented here is not particular of the Born-Infeld gravity, but it is a feature of a general class of gravity theories formulated \`a la Palatini. We will show this explicitly in section \ref{Sec:EBIextensionsGeneral}

\subsection{Classes of Born-Infeld inspired gravity.}
\label{Sec:EBIextensions}
The Born-Infeld theory of gravity discussed in the previous sections are naturally formulated on a spacetime manifold endowed with a general affine connection. Thus, given the richness offered by this geometrical framework, it is of no surprise that the Eddington-Born-Infeld theory described so far has found extensions in different directions. Unlike the case of Born-Infeld electrodynamics, the Eddington-Born-Infeld theory has not been singled out by resorting to symmetries principles or any other guiding criteria, but rather it originates from a straightforward transcription of the Born-Infeld Lagrangian for electromagnetism to gravity and taking inspiration from the Eddington affine theory. Thus, Born-Infeld inspired gravity is more prone to modifications and extensions than its electromagnetic relative. However, before proceeding to review the existing models and in view of the zoology of Born-Infeld inspired gravity theories found in the literature, we find it convenient to introduce some taxonomic system. We will classify the theories according to their proximity to the original Born-Infeld spirit, consisting in modifying the high curvature regime of gravity without introducing additional fields or pathologies. Furthermore, we will take the EiBI theory as the baseline because it is the most extensively studied model. After these considerations, we have decided to make the following classification:

\begin{itemize}
\item {\bf Class 0}. We start our classification with a class comprising all those early attempts of building gravity theories \`a la Born-Infeld which did not succeed due to the presence of pathologies. Subsequent proposals sharing these pathologies will also be considered to belong to this class.

\item {\bf Class I}. Here we will include the EiBI theory and the modifications that are the closest to the Born-Infeld spirit and do not introduce additional ingredients, be it new degrees of freedom or additional geometrical objects.

\item {\bf Class II}. A next step with respect to the Class I is to allow for more general geometrical objects, but respecting the Born-Infeld philosophy, i.e., only the high curvature regime is modified and no additional degrees of freedom are present.

\item {\bf Class III}. Under this category we will classify those models where the Born-Infeld structure remains but additional degrees of freedom are included.

\item {\bf Class IV}. Finally, in this class we will include theories that, although resemble Born-Infeld theories in some aspect, they could be very well classified within a different class of theories.

\end{itemize}

The above classification does not intend to be exhaustive nor having sharp edges. For instance, sometimes the presence of additional degrees of freedom might depend on some subtle assumptions on the theory or its solutions so that the same theory can have slightly different versions belonging to different classes. In those cases, we have opted by classifying it according to the most extensively used version in practical applications.

A substantial part of the formal developments and equations for many of the Born-Infeld inspired theories share numerous similarities among them and with the theory discussed so far. For that reason, prior to discussing specific theories we will present a general framework applicable to most of them.

\subsubsection{General mathematical framework}
\label{Sec:EBIextensionsGeneral}
In this section we will discuss some features that are common to a large class of theories, that include many of the proposed extensions and which are shared with EiBI gravity. Let us consider a general theory of the form
\be
\mS=\frac12\mpl^2\mbi^2\intd\sqrt{-g}F\big(g^{\mu\nu}, \mR_{\mu\nu}(\Gamma)\big)
\ee
where $F$ is a function of the inverse of the metric and the Ricci tensor. Notice that we have included a factor $\sqrt{-g}$ in the measure so that $F$ behaves as a true scalar. For simplicity, we will assume that the function will only depend on the combination $P^\mu{}_\nu=g^{\mu\alpha} \mR_{\alpha\nu}/\mbi^2$, where we have introduced the scale $\mbi$ for dimensional reasons. This is also the usual case in the literature so it will suffice for us\footnote{A treatment of more general theories can be found for instance in \cite{Jimenez:2015caa}.}. An important consequence of the function being a scalar is that $F(\m{A}^{-1}\m{P}\m{A})=F(\m{P})$ for any non-degenerate transformation $\m{A}$. Furthermore, the independent scalars built out of $\m{P}$ can be expressed as traces of powers of $[\m{P}^n]$. By using the Cayley-Hamilton theorem, we can express any power higher than 4 in terms of lower powers so that the action could in principle be written as $F(X_1,X_2,X_3,X_4)$ with $X_n=[\m{P}^n]$. This is useful to show some general properties of this general class of theories. We will not make extensive use of the advantages introduced by writing the action in this form and we will instead consider the action written as
\be
\mS=\frac12\mpl^2\mbi^2\intd\sqrt{-g}F\big(\m{P}\big).
\ee
In order to recover GR in the limit $\vert P^\mu{}_\nu\vert\ll 1$ we need to impose
\be
\frac{\partial F}{\partial P^\mu{}_\nu}\Big\vert_{\m{P}=0}=\delta_\mu{}^\nu.
\label{Eq:LimitlowR}
\ee
Notice that this is not really a constraint and any analytic function will satisfy it up to a constant factor that can be absorbed into $\mbi^2$. The Einstein-Hilbert action is recovered for $F(\m{P})=P^\alpha{}_\alpha$, in which case the above relation is exactly fulfilled for all values of $\m{P}$ and not only at $\m{P}=0$. To be completely precise we should say that the above condition will guarantee the existence of one branch of solutions that will be continuously connected with GR at low curvatures. Nevertheless, the non-linearity of the equations can, in general, present several branches and some of them will give a different behaviour for the low curvatures regime. We will encounter specific examples where this situation occurs when studying explicit solutions.

For the general action considered, we can obtain the corresponding field equations by taking with respect to both the metric and the connection, yielding
\be
\delta\mS=\frac12\mpl^2\mbi^2\intmu{g}\left[-\frac12Fg_{\mu\nu}\delta g^{\mu\nu}+\frac{1}{\mbi^2}\frac{\partial F}{\partial P^\mu{}_\alpha}\Big(\delta g^{\mu\nu}R_{\nu\alpha}+g^{\mu\nu}\delta R_{\nu\alpha} \Big)\right].
\label{Eq:dSgeneral}
\ee
For the subsequent developments, it is convenient to write the above variation in matrix notation
\be
\delta\mS=\frac12\mpl^2\mbi^2\intmu{g}\Tr\left[-\frac12F\m{g}\delta \m{g}^{-1}+
\frac{1}{\mbi^2}\left(\frac{\partial F}{\partial \m{P}}\m{R}^T\delta\m{g}^{-1}+\m{g}^{-1}\frac{\partial F}{\partial\m{P}}\delta\m{R}^T\right)\right]
\label{Eq:dSgeneralmatrix}
\ee
Now it will be useful to introduce some definitions before proceeding any further. First, let us define
\be
\sqrt{-q}q^{\alpha\nu}\equiv\sqrt{-g}g^{\nu\mu}\frac{\partial F}{\partial P^\mu{}_\alpha}
\label{Eq:defqgeneral1}
\ee
or in matrix notation
\be
\sqrt{-q}\m{q}^{-1}\equiv\sqrt{-g}\left(\m{g}^{-1}\frac{\partial F}{\partial \m{P}}\right)^T.
\label{Eq:defqgeneral2}
\ee
This definition is not an innocent choice and we will see later that $q_{\mu\nu}$ will actually play the role of the auxiliary metric determining the connection, as in the EiBI case. We can take determinants in both sides of \refeq{Eq:defqgeneral2} to obtain the relation
\be
\frac{g}{q}=\frac{1}{\det \m{F}_{\m{P}}}
\ee
where we have introduced the notation $\m{F}_{\m{P}}\equiv\partial F/\partial \m{P}$. Then, we can re-write the definition \refeq{Eq:defqgeneral2} as
\be
\m{q}^{-1}=\frac{1}{\sqrt{\det \m{F}_{\m{P}}}}\left(\m{g}^{-1}\frac{\partial F}{\partial \m{P}}\right)^T,
\label{Eq:defqgeneral3}
\ee
or, if we invert both sides, we finally obtain an expression for $q_{\mu\nu}$ as follows:
\be
\m{q}=\sqrt{\det \m{F}_{\m{P}}}\left[\left(\frac{\partial F}{\partial \m{P}}\right)^{-1}\m{g}\right]^T\,.
\label{Eq:defqgeneral4}
\ee
For the Einstein-Hilbert term, the derivative of $F$ gives the identity and $\m{q}$ exactly coincides with the spacetime metric, as expected.  If we consider $f(\mR)$ types of theories for which $F_{f(\mR)}=F(P^\alpha{}_\alpha)$, the derivative gives
\be
\frac{\partial F_{f(\mR)}}{\partial \m{P}}=F'_{f(\mR)}\Id
\ee
so we have that $q_{\mu\nu}=F'_{f(\mR)}g_{\mu\nu}$, recovering the known result that in these theories the two metrics are conformally related. Finally, in the case of the EiBI action \refeq{Eq:actiongeneral} we have $F_{\rm EBI}=2\sqrt{\det\big(\Id+\m{P}\big)}$ so its derivative is
\be
\left(\frac{\partial F_{\rm EBI}}{\partial \m{P}}\right)^T=\sqrt{\det\big(\Id+\m{P}\big)}\Big(\Id+\m{P}\Big)^{-1}.
\ee
When inserting this expression into the definition \refeq{Eq:defqgeneral2} we obtain
\be
\sqrt{-q}\m{q}^{-1}=\sqrt{-g}\sqrt{\det\left(\Id+\m{P}\right)}\Big(\Id+\m{P}\Big)^{-1}\m{g}^{-1}=
\sqrt{-\det\left(\m{g}+\frac{1}{\mbi^2}\m{\mR}\right)}\left(\m{g}+\frac{1}{\mbi^2}\m{\mR}\right)^{-1}
\ee
and we recover that $q_{\mu\nu}=g_{\mu\nu}+\frac{1}{\mbi^2}\mR_{\mu\nu}$ as it should. After this little satisfaction, we can continue with the computation of the field equations. Another useful relation for the variation of the action that we can obtain from the definition of $q_{\mu\nu}$ is the following:
\be
\frac{\sqrt{-g}}{\mbi^2}\frac{\partial F}{\partial \m{P}}\m{R}^T=\sqrt{-q}\left(\m{g}\m{P}\m{q}^{-1}\m{g}\right)^T.
\ee
With the new jargon, we can re-write the variation \refeq{Eq:dSgeneralmatrix} as
\be
\delta\mS=\frac12\int\d^4x\Tr\left[\left(\sqrt{-g}\lag_{G}\m{g}^{-1}\delta \m{g}-
\mpl^2\mbi^2\sqrt{-q}\qinv^T\m{P}^T\right)\delta \m{g}+\mpl^2\sqrt{-q}\m{q}^{-1}\delta \m{R}
\right]
\label{Eq:dSgeneralmatrix2}
\ee
where we have used the ciclic property of the trace and the identity $\m{g}\delta\m{g}^{-1}=-\delta\m{g}\m{g}^{-1}$. Furthermore we have re-introduced the Lagrangian $\lag_{G}=\frac12\mpl^2\mbi^2F$. From the last term we can already sense that $q_{\mu\nu}$ will be related to the metric generating the connection, since that piece resembles the variation one would obtain from the Einstein-Hilbert action in the Palatini formalism with a metric $q_{\mu\nu}$. A word of caution is necessary though, since $q_{\mu\nu}$ does not need to be symmetric at this point. Again, if we assume a projective symmetry so only the symmetric Ricci enters, only the symmetric part of $q_{\mu\nu}$ will contribute and, thus, it will exactly be the auxiliary metric. Prior to the discussion of the connection field equations, let us first write the metric field equations:
\be
\frac12\mpl^2\mbi^2\sqrt{-q}\left[\m{P}\m{q}^{-1}+\Big(\m{P}\m{q}^{-1}\Big)^T\right]-\sqrt{-g}\lag_{G}\m{g}^{-1}=\sqrt{-g}\m{T}
\label{Eq:metricgeneralmatrix}
\ee
where the symmetrization follows from the symmetry of $g_{\mu\nu}$ and we have also added the energy-momentum tensor of the matter sector. For the sake of completeness, we will also give the expression of this equation in components
\be
\mpl^2\mbi^2\sqrt{-q}q^{\alpha(\mu}P^{\nu)}{}_\alpha-\sqrt{-g}\lag_{G}g^{\mu\nu}=\sqrt{-g}T^{\mu\nu}\; .
\label{Eq:metricgeneralcomp}
\ee
As one of our favourite exercises, let us check that we recover the expected results when the above equation is particularised to known cases. For the Einstein-Hilbert action, we have already seen that $q_{\mu\nu}=g_{\mu\nu}$ and it is immediate to see that \refeq{Eq:metricgeneralmatrix} reduces to $\mR_{(\mu\nu)}-\frac12 \mR g_{\mu\nu}=\frac{1}{\mpl^2} T_{\mu\nu}$. For the Born-Infeld inspired theory with $F_{\rm EBI}=2\sqrt{\det\big(\Id+\m{P}\big)}$ we have also shown that $\m{q}$ reduces to the expected result. In that case, it is easy to see from $\m{g}^{-1}\m{q}=\Id+\m{P}$ that $\m{P}\m{q}^{-1}=\m{g}^{-1}-\m{q}^{-1}$. If we insert this relation into \refeq{Eq:metricgeneralmatrix} and use that $\sqrt{-g}\lag_{G}=\mpl^2\mbi^2\sqrt{-q}$, we can see that the equations reduce to $-\mpl^2\mbi^2\sqrt{-q}q^{(\mu\nu)}=\sqrt{-g} T^{\mu\nu}$, in agreement with \refeq{Eq:metricequations} (taking $\lambda=0$).

Let us pause a bit before moving on to the connection field equations to discuss the structure of the metric field equations. In general, the symmetry of the metric results in a set of ten independent equations. The general treatment of theories \`a la Palatini requires the use of these equations to solve for the Ricci tensor (or connection-dependent objects for more general theories) in terms of the metric and the matter fields. This step is algebraic and it is crucial for the subsequent resolution of the connection as the Levi-Civita connection of some auxiliary metric. However, while the Ricci tensor has in general sixteen components, the metric field equations are limited to ten and, therefore, the full Ricci cannot be obtained from them. This means that the method to solve the connection as the Levi-Civita connection will fail. This motivates considering theories with the projective symmetry for simplicity reasons.

Let us know turn to the computation of the connection field equations. By looking at the last piece of \refeq{Eq:dSgeneralmatrix2} we can see that it reads exactly the same as the corresponding variation for the Born-Infeld case in \refeq{Eq:var2}. Hence, the derivation will follow analogously and we can simply use the equations already obtained in \refeq{Eq:connectioneq}
\begin{eqnarray}
&&\nabla_\lambda\Big[\sqrt{-q}\big(\m{q}^{-1}\big)^{\mu\nu}\Big]-\delta_\lambda^\mu\nabla_\rho\Big[\sqrt{-q}\big(\m{q}^{-1}\big)^{\rho\nu}\Big]\nonumber\\
&&=\Delta^{\mu\nu}_\lambda+\sqrt{-q}\Big[\mT^\mu_{\lambda\alpha}(\m{q}^{-1}\big)^{\alpha\nu}+\mT^\alpha_{\alpha\lambda}(\m{q}^{-1}\big)^{\mu\nu}-\delta^\mu_\lambda \mT^\alpha_{\alpha\beta}(\m{q}^{-1}\big)^{\beta\nu} \Big]
\end{eqnarray}
where we only need to remember that now $\m{q}$ is defined in \refeq{Eq:defqgeneral1}
and we have added the {\it hypermomentum} of the matter fields defined in \refeq{Eq:defhyper}. If only the symmetric part of the Ricci tensor enters in the action, so that we have a projective symmetry $\Gamma^\alpha_{\mu\nu}\rightarrow\Gamma^\alpha_{\mu\nu}+\xi_\mu\delta^\alpha_\nu$, we can see from \refeq{Eq:dSgeneral} that only the symmetric part of $q^{\mu\nu}$ will contribute to the connection field equations. In that case we can easily see that the trace with respect to $\rho$ and $\nu$ vanishes identically, as a consequence of the projective invariance. Sometimes, this is regarded as a flaw of these theories because, in case the symmetry is not present in the matter sector, there is no reason to expect to have $\Delta_\rho^{\mu\rho}=0$ and this would be the source of an inconsistency in the equations. However, there is an obvious way to evade this apparent problem by assuming that matter fields do not couple to the connection directly so that we actually have that the full hypermomentum vanishes. Again, this is the case for minimally coupled bosonic fields, but complications might arise due to fermions. In any case, even if we need to have $\Delta_\rho^{\mu\rho}=0$, this should be regarded as a constraint in the matter sector and there is no reason a priori to assume that solutions satisfying that constraint cannot be found\footnote{In order to illustrate this point, let us remember the case of a Proca field coupled to conserved currents whose equations read $\partial_\nu F^{\mu\nu}+m^2A^\mu=J^\mu$. The gauge invariance of the charged sector implies the conservation of the current $\partial_\mu J^{\mu}=0$, while the mass term for the vector field breaks the gauge invariance in the vector field sector. However, this does not introduce any inconsistency in the equations as, by taking their divergence one obtains the constraint $\partial_\mu A^\mu=0$ which, not only it does not represent an inconsistency, but it plays in fact a crucial role to remove additional polarizations for the massive vector.}. For simplicity and to comply with most of the literature we will take $\Delta_\rho^{\mu\nu}=0$
in the following.

On the other hand, if there is no projective symmetry in the action, the object $q^{\mu\nu}$ will not have, in general, any defined symmetry. In that case, the equations have a formal resemblance with non-symmetric gravity theories \cite{Moffat79, Damour:1992bt,ValkenburgThesis,BeltranJimenez:2012sz} so one could try to apply the same techniques to solve the equations. However, the similarities are purely formal and, in fact, there are profound conceptual differences between the non-symmetric gravity theories and the ones under study here, mainly the absence of an actual non-symmetric metric.

We will manipulate the equations to recast them in more useful forms. We can first take the trace of the  equations with respect to $\mu$ and $\lambda$ to obtain that
\be
\nabla_\lambda\Big(\sqrt{-q}q^{\lambda\nu}\Big)=\frac12\mT^\lambda_{\lambda\alpha}\sqrt{-q} q^{\alpha\nu}
\ee
If we plug this relation back into the equations we obtain
\be
\nabla_\lambda\Big(\sqrt{-q}q^{\mu\nu}\Big)+\sqrt{-q}\left(\frac13\delta^\mu_\lambda \mT^\alpha_{\alpha\beta}q^{\beta\nu}-\mT^\mu_{\lambda\alpha}q^{\alpha\nu}-\mT^\alpha_{\alpha\lambda}q^{\mu\nu}\right)=0.
\label{Eq:coneq2}
\ee
Now, it is convenient to introduce the shifted connection
\be
\tGam^\alpha_{\mu\nu}=\Gamma^\alpha_{\mu\nu}-\frac13 \mT^\lambda_{\mu\lambda}\delta^\alpha_\nu
\ee
that satisfies $\tGam^\alpha_{\alpha\mu}=\tGam^\alpha_{\mu\alpha}$ and it is invariant under a projective transformation of the original connection $\Gamma^\alpha_{\mu\nu}$, i.e., we have that $\tGam^\alpha_{\mu\nu}\rightarrow\tGam^\alpha_{\mu\nu}$ when $\Gamma^\alpha_{\mu\nu}\rightarrow\Gamma^\alpha_{\mu\nu}+\xi_\mu\delta^\alpha_\nu$ for an arbitrary $\xi_\mu$. This will play a crucial role in the following because it means that the connection $\tGam$ will only determine $\Gamma$ up to a projective transformation and we will see that it is $\tGam$ what is determined by the equations. In terms of the shifted connection, the equations \refeq{Eq:coneq2} read
\be
\frac{1}{\sqrt{-q}}\partial_\lambda\Big(\sqrt{-q} q^{\mu\nu}\Big)+\tGam^\mu_{\alpha\lambda}q^{\alpha\nu}+\tGam^\nu_{\lambda\alpha}q^{\mu\alpha}-\tGam^\alpha_{\lambda\alpha}q^{\mu\nu}=0.
\label{Eq:coneq3}
\ee
If we take the two possible traces of these equations and subtract them we find
\be
\partial_\lambda\Big(\sqrt{-q}q^{[\mu\lambda]}\Big)=0
\ee
and, thus, the antisymmetric part of $q^{\mu\nu}$ satisfies a Maxwell-like equation. Another useful relation is obtained by multiplying \refeq{Eq:coneq3} by $\sqrt{-q} q_{\mu\nu}$ to obtain
\be
\partial_\lambda\log\sqrt{-q}=\tGam^\alpha_{\alpha\lambda},
\ee
which can then be used in \refeq{Eq:coneq3} to finally write the equations as
\be
\partial_\lambda q^{\mu\nu}+\tGam^\mu_{\rho\lambda}q^{\rho\nu}+\tGam^\nu_{\lambda\rho}q^{\mu\rho}=0
\label{Eq:coneq4}
\ee
or, if we multiply by $q_{\alpha\mu} q_{\nu\beta}$, in the equivalent way
\be
\partial_\lambda q_{\alpha\beta}-\tGam^\mu_{\beta\lambda}q_{\alpha\mu}-\tGam^\mu_{\lambda\alpha}q_{\mu\beta}=0\;.
\label{Eq:coneq5}
\ee
These equations will determine the connection $\tGam$ in terms of $q_{\mu\nu}$ and, thus, the original connection $\Gamma$ up to the aforementioned projective mode. We can do a bit better by following the usual procedure to compute the connection in terms of the metric, i.e., we subtract appropriate permutations of indices from \refeq{Eq:coneq5} to write it in the following form:
\be
q_{(\mu\lambda)}\tGam^\mu_{\alpha\beta}=\frac12\Big(\partial_\alpha q_{\beta\lambda}+\partial_\beta q_{\lambda\alpha}-\partial_\lambda q_{\alpha\beta}\Big)+q_{[\alpha\mu]}\tGam^\mu_{\beta\lambda}+q_{[\mu\beta]}\tGam^\mu_{\lambda\alpha}\;.
\label{Eq:soltGam}
\ee
This expression is crucial to understand many features of the theories under consideration that will in turn determine many of their properties. Let us stress that we have not considered any simplifying assumption, so our result is completely general. This pays our debt to the meticulous reader, who was promised a more thorough analysis in section \ref{Sec:SimplifiedEqEiBI}. The first thing to notice is that, for a symmetric $q_{\mu\nu}$, the solution for the connection $\tGam$ is nothing but the usual Levi-Civita connection of $q_{\mu\nu}$. Of course, the matrix $q_{\mu\nu}$ as defined in \refeq{Eq:defqgeneral1} depends on the Ricci and, thus, on the curvature. As usual, the resolution to this is that $q_{\mu\nu}$ can be algebraically solved from the metric field equations \refeq{Eq:metricgeneralcomp}. The connection $\tGam$ is thus solved as the Christoffel symbols of $q_ {\mu\nu}$ and this is how $\m{q}$ earns its denomination of auxiliary metric in the general case.  All this reasoning is however based on the assumption that $q_{\mu\nu}$ is symmetric, but this is not an outrageous wish to ask and, in fact, it will be nicely granted by the projective invariance. To see this, we can express the definition of $q^{\mu\nu}$ in terms of derivatives with respect to the Ricci tensor as follows
\be
\sqrt{-q}q^{\alpha\nu}\equiv\sqrt{-g}g^{\nu\mu}\frac{\partial F}{\partial P^\mu{}_\alpha}
=\sqrt{-g}g^{\nu\mu}\frac{\partial F}{\partial \mR_{\rho\sigma}}\frac{\partial \mR_{\rho\sigma}}{\partial P^\mu{}_\alpha}=\mbi^2\sqrt{-g}\frac{\partial F}{\partial \mR_{\nu\alpha}}
\ee
where we have used the definition of $\m{P}$ to compute its derivative with respect to $\m{\mR}$. From here, we see that the matrix $q^{\mu\nu}$ will inherit the symmetries of the Ricci tensor. In particular, if only the symmetric part of the Ricci enters the action, then $q^{\mu\nu}$ will automatically be symmetric and its Levi-Civita connection will be the solution for $\tGam$. Equivalently, if only the symmetric part of the Ricci appears in the action, only the symmetric part of $q^{\mu\nu}$ will contribute to the connection field equations. On the other hand, it is also very easy to see that, in that case, the metric field equations permit to obtain $q^{\mu\nu}$ (the number of equations will coincide with the number of components of $\m{q}$) and, thus, the usual procedure giving the connection as the Levi-Civita of the auxiliary metric $q_{\mu\nu}$ is fully consistent.

This is an appropriate place to make some remarks on these results. The first one is that the connection has only been obtained up to a projective mode. However, this does not represent a flaw and, in fact, rather the opposite for theories based on a symmetric Ricci. For those theories, there is a projective gauge symmetry that will necessary be responsible for the presence of undetermined modes in the solutions. In other words, the apparent undetermined projective mode will be innocuous and can be removed by a simple gauge fixing. This also applies to the case of the Einstein-Hilbert action and it is precisely the discussion we exposed below \refeq{Eq:GRlimitEBI}. Hence, for theories with the projective symmetry, the whole resolution of the field equations is consistent and, at least formally, achievable.

So far we have discussed the case when $q_{\mu\nu}$ is symmetric by definition. Things can be quite different when this condition is abandoned. In that case, we find problems in the two sets of equations, namely the metric and the connection equations. For the metric equations, we find the trouble already discussed above that, while the metric field equations provide ten independent equations, $q_{\mu\nu}$ will in general have sixteen independent components and, therefore, it cannot be fully expressed in terms of the spacetime metric and the matter content. Concerning the connection field equations and its polished expression in \refeq{Eq:soltGam}, simply obtaining $\tGam$ in terms of the non-symmetric $q_{\mu\nu}$ is an arduous task. In fact, in theories with non-symmetric metrics, the solution is usually obtained only perturbatively with respect to the antisymmetric part of the metric \cite{Moffat79, Damour:1992bt,ValkenburgThesis,BeltranJimenez:2012sz}. This gives further motivation to consider only theories with the projective symmetry, but theories without it will definitely present a much richer structure. In particular, they will likely contain additional degrees of freedom, among which there could be propagating torsion. Additionally, the results obtained here give support to the simplifying assumption of vanishing torsion upon which the results of section \ref{Sec:SimplifiedEqEiBI} were obtained.

For the projectively invariant theories, we can also make contact with the previous formalism developed in the case of EiBI theories and the definition of the deformation matrix relating the spacetime and the auxiliary metrics. If we remember the relation between both metrics defined in \refeq{Eq:defOmega} as $\m{q}=\m{g}\m{\Omega}$ we see that we can re-write \refeq{Eq:defqgeneral3} in a similar form by defining
\be
\m{\Omega}^{-1}=\frac{1}{\sqrt{\det \m{F}_{\m{P}}}}\left(\frac{\partial F}{\partial \m{P}}\right)^T.
\label{Eq:defOmegageneral}
\ee
As one would require, the condition \refeq{Eq:LimitlowR} imposed to recover GR in the low curvatures regime implies that $\m{\Omega}\simeq\Id$ in that limit, so that both metrics coincide when $\vert\m{P}\vert\ll1$. An important derived relation is that, in four dimensions, we have $\det\m{\Omega}=\det \m{F}_{\m{P}}$. Let us also notice that the Lorentzian signature for the auxiliary metric will be guaranteed as long as the derivative $\m{F}_{\m{P}}$ is positive definite or, equivalently, if the deformation matrix $\m{\Omega}$ is positive definitive. In Born-Infeld inspired theories of gravity, this is usually related to the existence of the square root of a matrix characteristic of those theories, which is then imposed as a condition on physical solutions. In the general case, we will need to impose the deformation matrix be positive definite for physical solutions. This will in turn guarantee that $\sqrt{\det \m{F}_{\m{P}} }$ is a real quantity. We can now follow the same procedure as we did with the EiBI theory and obtain an algebraic equation for the deformation matrix by introducing $\m{q}^{-1}=\m{\Omega}^{-1}\m{g}^{-1}$ into \refeq{Eq:metricgeneralmatrix} and multiplying by $\m{g}$ to obtain
\be
\m{\Omega}^{-1}\m{P}=\frac{1}{\mbi^2\mpl^2\sqrt{\det\m{\Omega}}}\Big(\lag_G\Id+\m{T}\m{g}\Big),
\label{Eq:Omegaeqgen}
\ee
where we have used that $\m{P}$ and $\m{\Omega}$ commute and the property\footnote{Both properties can be easily proven by assuming that $F$ is an analytic function so that $\m{F}_{\m{P}}$ and, as a consequence, $\m{\Omega}$ are analytic matrix functions of $\m{P}$. If we have an arbitrary analytic function $\m{\mathcal{F}}$ of $\m{P}$ we can expand it as $\m{\mathcal{F}}=\sum_n c_n \m{P}^n$ from where it is trivial to see that it commutes with $\m{P}$. Furthermore, we can also show that $\m{g}\m{\mathcal{F}}(\m{P})\m{g}^{-1}=\m{\mathcal{F}}(\m{g}\m{P}\m{g}^{-1})=\m{\mathcal{F}}(\m{P}^T)=\m{\mathcal{F}}^T(\m{P})$,
where we have used that $\m{g}\m{P}\m{g}^{-1}=\mbi^{-2}\m{R}\m{g}^{-1}=\m{P}^T$ which is valid whenever the Ricci tensor $\m{R}$ is symmetric or, as in our case, when only its symmetric part is considered. From this relation we can obtained the desired property by simply taking $\m{\mathcal{F}}=\m{\Omega}^{-1}\m{P}$.
} $g^{-1}(\m{\Omega}^{-1}\m{P})^T\m{g}=\m{\Omega}^{-1}\m{P}$. This is an algebraic equation for the deformation matrix provided $\m{P}$ can be expressed in terms of $\m{\Omega}$ by inverting \refeq{Eq:defOmegageneral}. Now, if we use that $\m{q}^{-1} \m{R}=\mbi^2\m{q}^{-1}\m{g} \m{P}=\mbi^2\m{\Omega}^{-1}\m{P}$ we finally obtain the differential equations satisfied by the auxiliary metric
\be
R^\mu{}_\nu(q)=\frac{1}{\mpl^2\sqrt{\det\m{\Omega}}}\Big(\lag_G\delta^\mu{}_\nu+T^\mu{}_\nu\Big)
\label{Eq:qeqgeneralF}
\ee
in complete analogy with the equations \refeq{Eq:qequationscanonical2} obtained for the EiBI case. This proves our claim that those equations are valid for general theories. Furthermore, the same conclusions drawn there are automatically valid for this more general case. In particular, in the low curvatures regime the deformation matrix is the identity and the Lagrangian is $\lag_G\simeq\frac12 \mpl^2R$ by construction and, thus, we recover the usual Einstein equations.

To end this section, let us extend the discussion on the existence of two frames shown for the Born-Infeld case to the more general theories considered here. In view of the discussions so far about the structure of the theories, it should be clear by now that assuming a projective symmetry would be a wise decision on the grounds of simplicity. Very much like we did for EiBI, let us go to a bi-metric representation of the theory by introducing an auxiliary field $\Sigma_{\mu\nu}$ as follows:
\be
\mS=\frac 12\mpl^2\mbi^2\intd\sqrt{-g}\left[F(g^{\mu\nu},\Sigma_{\mu\nu})+\frac{\partial F}{\partial \Sigma_{\mu\nu}}\left(\frac{1}{\mbi^2}\mR_{(\mu\nu)}-\Sigma_{\mu\nu}\right)\right]+\Ss{M}[\Psi,g_{\mu\nu}]
\ee
We can see that $\Sigma_{\mu\nu}$ can be integrated out by solving its own equation of motion and we recover the original action. We have considered the case with projective invariance to simplify the analysis which in turn implies that $\Sigma_{\mu\nu}$ is symmetric. Now we can introduce a field re-definition as
\be
\sqrt{-q}q^{\mu\nu}=\sqrt{-g} \frac{\partial F}{\partial \Sigma_{\mu\nu}}
\label{Eq:relqSigma}
\ee
that can be used to obtain $\Sigma_{\mu\nu}=\Sigma_{\mu\nu}(\m{g},\m{q})$ so that the action can be expressed as
\be
\mS=\frac12\mpl^2\intd\Big[\sqrt{-q}q^{\mu\nu}\mR_{\mu\nu}+\sqrt{-g}\mbi^2\mU(\m{g},\m{q})\Big]+\Ss{M}[\Psi,g_{\mu\nu}]
\label{Eq:biF2}
\ee
with
\be
\mU(\m{g},\m{q})=F-\frac{\partial F}{\partial \Sigma_{\mu\nu}}\Sigma_{\mu\nu}.
\label{Eq:defU}
\ee
In this action, the spacetime metric $g_{\mu\nu}$ appears as an auxiliary field (provided the matter fields are minimally coupled) so it can be integrated out. Its equation is simply
\be
\frac{\partial \mU}{\partial g^{\mu\nu}}-\frac12 \mU g_{\mu\nu}=\frac{1}{\mpl^2\mbi^2}T_{\mu\nu}
\label{Eq:geq}
\ee
which allows to solve algebraically for $g_{\mu\nu}$ in terms of $q_{\mu\nu}$ and the matter fields, similarly to the case of Born-Infeld. Thus, the original action can be alternatively expressed as
\be
\mS=\frac12\mpl^2\intd\sqrt{-q}q^{\mu\nu}\mR_{\mu\nu}(\Gamma)+\tilde{\mS}_{\rm M}[\Psi,q_{\mu\nu}]
\label{Eq:FToGR}
\ee
and we see again that the theory is equivalent to GR but with modified couplings to the matter fields. Hence, the same discussion presented in section \ref{Sec:Frames} applies to the more general class of theories considered here. The Born-Infeld frame introduced in that section naturally extends to a more general {\it affine frame} within the framework of the general class of theories discussed here. This naturally motivates an extension of the Born-Infeld equivalence principle to a more general {\it affine equivalence principle} with the same theoretical and phenomenological consequences, in particular the constraint $\mbi\gtrsim10^{-1}$ eV obtained by imposing the absence of anomalous interactions at LHC also applies here. Notice however that some exceptions exist where this argument fails, since we are assuming that \refeq{Eq:defU} can be inverted to express $\Sigma_{\mu\nu}$ in terms of $g_{\mu\nu}$ and $q_{\mu\nu}$ and similarly for \refeq{Eq:geq} that allows to integrate out $g_{\mu\nu}$. One important example of theories where this argument is not applicable is the case of $f(\mR)$ theories. In that case, only the trace of $\Sigma$ enters \refeq{Eq:relqSigma} so it is not possible to invert it and obtain $\Sigma_{\mu\nu}=\Sigma_{\mu\nu}(\m{g},\m{q})$. As it is well-known, in that case it is a better idea to add a scalar field in the Legendre transformation instead of $\Sigma_{\mu\nu}$. As we showed above, for these theories $q_{\mu\nu}$ and $g_{\mu\nu}$ are conformally related.

After the general considerations discussed in this section, let us turn to considering specific examples of extensions of Born-Infeld gravity corresponding to the classification introduced above.

\subsubsection{Class 0}
We will classify under this category those theories aiming at modifying GR in the high curvature regime with a Born-Infeld type of modification, but which fail in fulfilling some crucial consistency requirement, like the presence of unavoidable ghost-like instabilities. Into this class will go the first attempts towards Born-Infeld gravity explained in the sections \ref{Sec:D&G} and \ref{Sec:Othermetric}  that were based on the metric formalism. As extensively discussed there, the higher order field equations for the metric arising in those theories compromise their stability due to the presence of ghosts.

As another example of Born-Infeld inspired gravity theories that would belong to this class we can mention  theories consisting of a Born-Infeld sector formulated in the affine approach (similar to the EiBI Lagrangian) supplemented with another sector formulated in the metric formalism. This type of action was already considered by Ba{\n}ados in \cite{Banados:2008rm} and some phenomenological consequences were explored in \cite{Rodrigues:2008kv,Banados:2008fi,Banados:2008fj}. In view of the analysis performed in section \ref{Sec:EBIextensionsGeneral}, it is clear that these theories are generally plagued by ghost-like instabilities, similarly to the original attempts made in the pure metric formalism. The problem with these theories is precisely the presence of the metric sector. We can repeat the same construction leading to \refeq{Eq:biF2}, but now with the additional sector formulated in the metric formalism. If we take such a sector to consist of an Einstein-Hilbert term for the metric\footnote{The same will however apply if we consider more general metric sectors like, e.g., $f(R)$ terms.} $g_{\mu\nu}$, as it was the case considered in \cite{Banados:2008rm}, we end up with the equivalent action
\be
\mS=\frac12\mpl^2\intd\Big[\sqrt{-g}g^{\mu\nu}R_{\mu\nu}(g)+\sqrt{-q}q^{\mu\nu}\mR_{\mu\nu}+\sqrt{-g}\mbi^2\mU(\m{g},\m{q})\Big]+\Ss{M}[\Psi,g_{\mu\nu}]\,,
\label{Eq:EHEiBI}
\ee
so we have a bi-metric theory where both metrics are coupled through $\mU(\m{g},\m{q})$. If there was no metric sector explicitly making $g_{\mu\nu}$ a propagating field, the spacetime metric could be integrated out and we would be left with only one propagating metric, as we obtained in \refeq{Eq:FToGR}. However, having an independent Einstein-Hilbert term for the spacetime metric makes it a propagating field and, thus, the action \refeq{Eq:EHEiBI} shows that we can no longer integrate the metric $g_{\mu\nu}$ out. The result of this is that we have a bi-metric theory where the two metrics are dynamical and interact through the potential $\mU(\m{g},\m{q})$. Unless the interactions encoded in that potential belong to the class of ghost-free bi-gravity type \cite{deRham:2010kj,Hassan:2011zd}, the theory will contain the so-called Boulware-Deser ghost \cite{Boulware:1973my} and, therefore, the theory will be unstable. In general, the absence of ghosts will then be guaranteed if the following condition holds
\be
\mU(\m{g},\m{q})=F-\frac{\partial F}{\partial \Sigma_{\mu\nu}}\Sigma_{\mu\nu}=\sum_{n=0}^4\beta_ne_n\big(\sqrt{\m{g}^{-1}\m{q}}\big)\,,
\label{Eq:FTobiG}
\ee
where the terms in the last sum are the massive gravity and bi-gravity potentials written in terms of the  the elementary symmetric polynomials $e_n$ defined in \refeq{Eq:defen}. One can easily check that the EiBI action does not fulfill this condition and, thus, the theory will contain the undesired ghostly mode. A construction with auxiliary fields that somehow connect the EiBI Lagrangian with bi-gravity theories as different branches of the same underlying theory was presented in \cite{Schmidt-May:2014tpa}, but our discussion here differs from the one given there.

We should notice that this is in fact a general result for theories mixing sectors formulated in the metric and in the affine formalism that go under the name of hybrid theories \cite{Capozziello:2015lza}. We see that the hybrid theories containing the Ricci tensor will either be unstable or equivalent to massive bi-gravity if \refeq{Eq:FTobiG} holds. As with simple affine theories, actions built out of the Ricci scalar alone do not fall within this general result since, as pointed out below \refeq{Eq:FToGR}, in those cases the construction fails. The way to go for those theories is introducing a scalar auxiliary field that makes the two metrics be conformally related. That explains why we will not eventually obtain two propagating metrics so that those particular hybrid theories will avoid the ghost. This was obtained for the perturbative degrees of freedom around relevant backgrounds in \cite{Koivisto:2013kwa} (see also \cite{Capozziello:2015lza}), but we can see from our analysis here that this also extends fully non-linearly.

We can conclude that the presence of ghost-like instabilities is a generic pathology of the theories belonging to the class 0 and represents a serious drawback for their phenomenological consequences. In fact, the very existence of such pathologies could be taken as the defining property of this class.

\subsubsection{Class I}
\label{Sec:ClassI}
We identify this class as the one containing the most extensively studied case in the literature, i.e., the EiBI reviewed in the precedent subsections, as well as its extensions. The most immediate class of extensions of the EiBI gravity is to consider some sort of functional extension. As we have extensively seen above, the fundamental object in EiBI gravity is the determinant
\be
\det\left[g_{\mu\nu}+\frac{1}{\mbi^2} \mR_{\mu\nu}(\Gamma)\right]
\ee
in terms of which the action is written. One of the reasons to introduce the determinant is to guarantee the diffeomorphisms invariance of the volume element because the determinant of a rank-2 covariant tensor transforms as a scalar density of weight $w=-2$. Thus, in order to introduce functional extensions of the EiBI theory, it is more convenient to rewrite the action in the following form:
\be
\Ss{EiBI}=\mbi^2\mpl^2\intd\sqrt{-g}\sqrt{\det\left(\Id+\m{P}\right)}
\label{Eq:EiBIOmega}
\ee
with
\be
 P^\mu{}_\nu\equiv\frac{1}{\mbi^2} g^{\mu\alpha}\mR_{\alpha\nu}.
\ee
From here one can straightforwardly perform functional extensions in different directions that are discussed in the following sections. There can be slightly different versions of the theory depending on whether the connection is assumed symmetric a priori or if only the symmetric part of the Ricci tensor is considered, as we have seen above, and this could also be the origin of differences in the formulation of the theories.

\vspace{0.3cm}
$\bullet$ {\bf  Arbitrary function of the determinant}
\vspace{0.15cm}

It is a common practice in modified gravity to generalise theories by introducing arbitrary functions of the defining quantities as it is done for instance in $f(R)$ or $f(R,G)$ theories where arbitrary functions of the Ricci scalar and/or the Gauss-Bonnet term are considered. Thus, probably the first extension one could think of for the EiBI theory is taking an arbitrary function of the defining determinant. This was done in \cite{Odintsov:2014yaa} where the following extension of Born-Infeld was considered\footnote{We adapt the notation of that reference to be consistent with the notation of this review, so we reserve $\m{q}$ and $\m{\Omega}$ for the auxiliary metric and the deformation matrix respectively.}
\be
\Ss{}=\mbi^2\mpl^2\intd\sqrt{-g}f\big(\mX\big)
\ee
with $\mX=\det(\Id+\m{P})$.  The EiBI theory is recovered for $f(\mX)=\mX^{1/2}$. As usual, the function $f$ should be chosen so that we recover GR at small curvatures. The action in that limit can be obtained by expanding the function around $\mX=1$ so we have
\be
\Ss{}\simeq\mbi^2\mpl^2\intd\sqrt{-g}f'(1)\Big(\mX-1\Big),
\ee
which only differs from the original EiBI theory by the factor $f'(1)$ so we need to impose $f'(1)=1$ to have the correct limit at low curvatures. The generality introduced by considering an arbitrary function of the determinant can be handled in the usual way by introducing a Legendre transformation with an auxiliary field $\phi$ as
\be
\Ss{}=\mbi^2\mpl^2\intd\sqrt{-g}\Big[f(\phi)+f_{\phi}\Big(\mX-\phi\Big)\Big],
\ee
followed by a field redefinition $\varphi=f_\phi$ so that the action can be written in the equivalent way
\be
\Ss{}=\mbi^2\mpl^2\intd\sqrt{-g}\Big[\varphi\mX-V(\varphi)\Big].
\ee
The field equation of the scalar field imposes the constraint
\be
\mX=V_{,\varphi}
\ee
which can be eventually incorporated in the final form of the equations. The procedure presented above for general theories can be straightforwardly applied to this case and one finds that the auxiliary metric reads
\be
q_{\mu\nu}=\varphi\sqrt{\mX}\left(g_{\mu\nu}+\frac{1}{\mbi^2}\mR_{(\mu\nu)}\right)
\ee
while the deformation matrix $\m{\Omega}$ relating the two metrics is given by
\be
\m{\Omega}=\varphi\sqrt{\mX}\Big(\Id+\m{P}\Big).
\ee
After some more manipulations along the lines of the general case depicted in the previous section, one can finally write the equations as
\be
R^\mu{}_\nu(q)=\frac{1}{2\mpl^2\varphi^2\mX^{3/2}}\Big(\mpl^2\mbi^2f(\mX)\delta^\mu{}_\nu+T^\mu{}_\nu\Big),
\ee
which, as shown above in \refeq{Eq:qeqgeneralF}, is the standard form for these theories and will permit the direct applications that will be discussed in detail in the next sections.

\vspace{0.3cm}
$\bullet$ {\bf  Extension to all the elementary symmetric polynomials}
\vspace{0.15cm}

A slightly different way of writing the EiBI action leads to another class of extensions. By commuting the square root and the determinant in \refeq{Eq:EiBIOmega}, we can alternatively write the action as
\be
\Ss{EiBI}=\mbi^2\mpl^2\intd\sqrt{-g}\det\m{M}
\label{Eq:EiBIdetM}
\ee
where the matrix $\m{M}$ has been defined as $\m{M}\equiv\sqrt{\m{\Omega}}=\sqrt{\Id+\m{P}}$.
Now, since the determinant of a matrix is nothing but the invariant elementary symmetric polynomial of highest degree, the EiBI action rewritten as \refeq{Eq:EiBIdetM} calls for a natural extension including the full series of elementary symmetric polynomials
of the fundamental matrix $\m{M}$. This is the path taken in \cite{Jimenez:2014fla} that led to the family of Born-Infeld inspired theories described by the following actions
\begin{equation}\label{gen_Born_Infeld}
\Ss{GBI}=\mbi^2\mpl^2\int \d^4x\sqrt{-g}\sum_{n=0}^4 \beta_ne_n(\hat{M}).
\end{equation}
with $\beta_n$ some dimensionless constants and $e_n(\m{M})$ the elementary symmetric polynomials defined as
\begin{eqnarray}\label{eq:polynomials}
e_0(\m{M}) &=& 1,\nonumber\\
e_1(\m{M}) &=& [\m{M}] ,\nonumber\\
e_2(\m{M}) &=& \frac{1}{2!}\Big([\m{M}]^2- [\m{M}^2]\Big),  \nonumber\\
e_3(\m{M}) &=&\frac{1}{3!}\Big( [\m{M}]^3- 3[\m{M}][\m{M}^2]+2[\m{M}^3] \Big),\nonumber\\
e_4(\m{M}) &=&\frac{1}{4!}\Big([\m{M}]^4-6[\m{M}]^2[\m{M}^2]+8[\m{M}][\m{M}^3]+3[\m{M}^2]^2-6[\m{M}^4]   \Big).
\end{eqnarray}
As commented above, the fourth symmetric polynomial coincides with the determinant i.e. $e_4(\hat{M})=\det\hat {M}$ so that the $\beta_4$ term contributes the usual EiBI Lagrangian.
The low curvature limit $\vert g^{\mu\alpha}R_{\alpha\nu}\vert\ll\mbi^2$ gives
\begin{eqnarray}\label{genPal}
\mathcal S\simeq\mpl^2\mbi^2\int \d^4x\sqrt{-g}\Big[ \Big(\beta_0+4\beta_1+6\beta_2+4\beta_3+\beta_4  \Big) \nonumber\\
+\frac{1}{2\mbi^2}\Big(\beta_1+3\beta_2+3\beta_3+\beta_4  \Big) g^{\mu\nu}\mR_{\mu\nu}(\Gamma)  \Big]
\end{eqnarray}
which coincides with the Einstein-Hilbert action in the Palatini formalism supplemented with a cosmological constant term (which can be cancelled by tuning the parameter $\beta_0$), provided we impose $\beta_1+3\beta_2+3\beta_3+\beta_4=1$. The projective symmetry always appears here as an accidental symmetry of the low curvature action and it will only be a symmetry of the full theory if the elementary symmetric polynomials are constructed in terms of the symmetric Ricci tensor.
Let us now consider the high curvature limit where $\vert g^{\mu\alpha}\mR_{\alpha\nu}\vert\gg\mbi^2$. This means that $\hat{M}\simeq\sqrt{\m{P}}$ and, therefore, the action in this regime turns into a combination of the elementary symmetric polynomials of $\sqrt{\m{P}}$. In the presence of all the polynomials, this regime will be dominated by the fourth one and we will recover an Eddington-like action
\be
\mS\simeq\beta_4\frac{\mpl^2}{\mbi^2}\int\d^4x\sqrt{\det R_{\mu\nu}(\Gamma)}.
\ee
In the general case, the Born-Infeld regime will be determined by the highest degree polynomial present in the action. The case of $e_2$ admits an amusing interpretation since its Born-Infeld regime gives
\be
\mS=\tilde{m}^2\int \d^4x\sqrt{-g} \left( \Big[\hat{g}^{-1}\hat{R}\Big]-\Big[\sqrt{\hat{g}^{-1}\hat{R}}\Big]^2 \right)
\ee
with $\tilde{m}$ some scale. This theory could even be treated in the metric formalism. Now if we interpret the operation of tracing as a type of averaging, the above action can be interpreted as being the variance of $\sqrt{\hat{g}^{-1}\hat{R}}$. Despite its amusing interpretation, its physical viability is dubious since it likely gives rise to observational conflicts and a lack of hyperbolicity in the field equations might. However, these  issues should be explored before reaching a definite conclusion.

Again, we can apply the machinery developed above for the general case to this particular family of theories with the identification
\be
F(\m{P})=2\sum_{n=0}^4\beta_ne_n(\m{M})
\ee
where $\m{M}=\sqrt{\Id+\m{P}}$. The derivative of this function can be computed as
\be
\frac{\partial F}{\partial\m{P}}=2\sum_{n=1}^4\beta_n\sum_{k=1}^4\frac{\partial e_n}{\partial [\m{M}^k]}\frac{\partial[\m{M}^k]}{\partial\m{M}}\frac{\partial \m{M}}{\partial\m{P}}
\ee
where we have made extensive used of the chain rule and dropped the term with $n=0$ because that is just a cosmological constant term. Now, we will introduce the notation $E_n^k=\partial e_n/\partial[M^k]$, whose explicit form is given by
\be
E_n^k=\left( \begin{array}{cccc}
                      e_0 & 0 & 0 & 0 \\
                      e_1 & -\frac{e_0}{2} & 0 & 0 \\
e_2 & -\frac{e_1}{2} & \frac{e_0}{3} & 0 \\
e_3 & -\frac{e_2}{2} & \frac{e_1}{3} &  -\frac{e_0}{4} \end{array}\right) \ ,
\ee
 and use that $\partial[\m{M}^k]/\partial\m{M}=k\big(\m{M}^{k-1}\big)^T$ and $\partial\m{M}/\partial\m{P}=\frac12\big(\m{M}^{-1}\big)^T$ to finally obtain the auxiliary metric as given in \refeq{Eq:defqgeneral2} for the present case:
\be
\sqrt{-q}\m{q}^{-1}=\sqrt{-g}\left(\m{g}^{-1}\frac{\partial F}{\partial\m{P}}\right)^T=\sqrt{-g}\left(\sum_{n=1}^4\beta_n\sum_{k=1}^4E_n^k\m{M}^{k-2}\right)\m{g}^{-1}.
\label{eq:qelementary}
\ee
This is precisely the result found in \cite{Jimenez:2014fla}, where the sums in the brackets corresponds to the matrix $\m{W}$ defined in that reference. Since the sum over $k$ runs from 1 to 4, the right hand side of \refeq{eq:qelementary} will contain powers of $\m{M}$ from $-1$ to $2$. This allows to re-write \refeq{eq:qelementary} in the more useful form
\be
\sqrt{-q}\m{q}^{-1}=\sqrt{-g}\left(f_1\m{M}^{-1}+f_2\Id+f_3\m{M}+f_4\m{M}^2 \right)\m{g}^{-1}
\label{eq:qelementary2}
\ee
with
\begin{eqnarray}
f_1&=& \beta_1 e_0+\beta_2 e_1+\beta_3 e_2+\beta_4 e_3\\
f_2&=& -(\beta_2 e_0+\beta_3 e_1+\beta_4 e_2)\\
f_3&=& \beta_3 e_0+\beta_4 e_1\\
f_4&=& -\beta_4 e_0.
\end{eqnarray}
From these expressions one can now straightforward adapt the general formalism for this family of theories and obtain all the relevant equations, which, of course, coincide with those in \cite{Jimenez:2014fla}. As a particularly simple case, we can take a theory containing only $e_1$ so that the action reads
\begin{equation}
\Ss{Min}=\mbi^2\mpl^2\int \d^4x\sqrt{-g} \Tr\left[\sqrt{\mathbbm 1+\mbi^{-2}\hat{g}^{-1}\hat{\mR}}-\mathbbm 1\right].
\label{Eq:gminimal}
\end{equation}
This is the model that was studied in more detail in \cite{Jimenez:2014fla} and subsequently used in \cite{Jimenez:2015jqa} to develop an inflationary scenario. For that case, we have that $f_2=f_3=f_4=0$ and $f_1=\beta_1$ is a constant, which is set to 1 in order to recover GR at low curvatures. In this very simple case, we can easily compute the deformation matrix from \refeq{Eq:defOmegageneral}, which yields
\be
\m{\Omega}=\frac{1}{\sqrt{\det\m{M}}}\m{M}.
\label{Eq:OmegaToMminimal}
\ee
If we use this relation together with $\m{P}=\m{M}^2-\Id$ obtained from the definition of $\m{M}$, the equation for the deformation matrix given in \refeq{Eq:Omegaeqgen} can be written in terms of $\m{M}$ as
\be
\m{M}^{-1}-\m{M}-\Big[\Tr\Big(\m{M}-\Id\Big)\Big]\Id=\frac{1}{\mbi^2\mpl^2}\m{T}\m{g}
\label{Eq:MinimalMatrix}
\ee
which exactly coincides with the equation found in \cite{Jimenez:2014fla}. This equation will give the matrix $\m{M}$ in terms of the matter sector and then one can follow the common procedure to solve the equations. This will be explicitly done in section \ref{subsection_Minext_BI}, where the cosmology of this model will be studied.

\subsubsection{Class II}\label{subsubsectionClassII}
There is a second class of extensions of the EiBI theories that makes use of additional geometrical objects. Let us remind that the original EiBI theory only utilizes the Ricci tensor and the metric and its natural arena is a non-Riemannian geometry. The metric affine formulation of the theory implies the presence of a completely independent connection and its associated curvature encoded in the Riemann tensor $\mR^\alpha{}_{\beta\mu\nu}$. For this general Riemann, we can take three independent traces, namely: the Ricci $\mR_{\beta\nu}=\mR^\alpha{}_{\beta\alpha\nu}$, the homothetic tensor $\mQ_{\mu\nu}=\mR^\alpha{}_{\alpha\mu\nu}$ and the co-Ricci $\mP^\alpha{}_\mu=g^{\beta\nu}\mR^\alpha{}_{\beta\mu\nu}$. The traces of these three objects are all the same and give the Ricci scalar $\mR=g^{\mu\nu}\mR_{\mu\nu}$. We can see that the EiBI theory only makes use of the Ricci tensor, but a much larger variety is possible thanks to the rich geometrical structure at our disposal. For instance, the determinantal form of EiBI can be extended to include an arbitrary combination of the three different traces of the Riemann tensor so we could consider actions of the type
\be
\mS=\mpl^2\mbi^2\intd\sqrt{-\det\left[a_1g_{\mu\nu}+\frac{1}{\mbi^2}\Big(a_2\mR_{\mu\nu}+a_3\mQ_{\mu\nu}+a_4\mP_{\mu\nu}\Big)\right]}
\label{Eq:GenClassII}
\ee
where $a_i$ can be arbitrary scalar functions of curvature invariants. In the simplest case we could take $a_i=a_i(\mR)$, but other scalars like Gauss-Bonnet combinations or $\mR_{\mu\nu}\mR^{\mu\nu}$ could also be envisaged. Obviously, here we encounter once again a similar obstacle as in the Deser and Gibbons construction discussed in section \ref{Sec:D&G} (although this time avoiding the ghost problem), namely the lack of a guiding principle. Thus, very much like in that case, one can foreshow that any gravitational theory (except for some singular cases) can be recast in the above form by appropriately tuning the free functions $a_i$. The EiBI theory corresponds to possibly the simplest among the possible theories described by the action \refeq{Eq:GenClassII}. Let us stress that we always remain within the Born-Infeld spirit, so we leave out here well-known theories, like those written in terms of Lovelock invariants, rewritten in a way that resemble the characteristic square root structure of Born-Infeld theories. Let us notice that we can consider even more general actions by including generalised determinants for the Riemann tensor itself.

Faced with the obstruction of lacking some motivation to select extensions of EiBI within the Class II, people have resorted to the always welcomed principle of simplicity. In this case, it means that extensions along the lines depicted here have predominantly resorted to adding new terms only containing the Ricci scalar. This has been considered in two fashions, either by writing new $\mR$-dependent terms outside the EiBI action, as in the Born-Infeld plus $f(\mR)$ models introduced in \cite{Makarenko:2014lxa}, or by modifying the determinantal structure with only $\mR$-dependent terms, as was considered in the appendix of that same work \cite{Makarenko:2014lxa} and in \cite{Chen:2015eha} for a specific case.

Another possibility that differs more profoundly from the one sketched so far is to consider other geometrical frameworks. At this respect,  an interesting class of theories formulated on a Weitzenb\"ock space was introduced in \cite{Ferraro:2009zk,Fiorini:2013kba}. It is well-known that GR admits a formulation in a Weitzenb\"ock space, where the connection is constrained to have vanishing curvature and all the gravitational effects are encoded in the torsion tensor $\mT^\lambda_{\mu\nu}$. This construction goes under the name of Teleparallel Equivalent of General Relativity (TEGR) and it has been used as the starting point of some modifications of gravity, among them some Born-Infeld inspired gravity theories that are of interest for us (see section \ref{sec:hdim3d} for applications of this theory on black holes). An extensive and comprehensive review on TEGR can be found in \cite{TelparallelAldrovandi}. Even though TEGR is GR in disguise, this mask shows an interesting face for GR as a gauge theory of the inhomogeneous part of the Poincar\' e group where the vierbeins are precisely the gauge fields of translations. Thus, TEGR provides a very appealing starting point for Born-Infeld modifications of gravity that deserves to be explored.

After discussing some of the different approaches that can be taken to obtain Born-Infeld inspired theories of gravity within the Class II, let us briefly review some specific examples.

\vspace{0.3cm}
$\bullet$ {\bf  Born-Infeld plus $f(\mR)$}
\vspace{0.15cm}

A simple extension within this class is to combine the Born-Infeld action with the well-known $f(\mR)$ theories in the metric-affine formalism, as considered in \cite{Makarenko:2014lxa,Makarenko:2014fla,Makarenko:2014nca,Elizalde:2016vsd}. The resulting action adapted to our notation is given by
\begin{align}
\mS=&\frac12\mpl^2\mbi^2\intd\left[2 \sqrt{-\det\left(g_{\mu\nu}+\frac{\alpha}{\mbi^2}\mR_{(\mu\nu)}\right)}+\sqrt{-g}f(\mR)\right]\nonumber\\
=&\frac12\mpl^2\mbi^2\int\d^4x\sqrt{-g}\left[2 \sqrt{\det\left(\Id+\alpha\m{P}\right)}+f([\m{P}])\right]
\label{Eq:BIf(R)}
\end{align}
with $\alpha$ some dimensionless constant. Since the small curvature limit of the EiBI sector in the above action already gives the Einstein-Hilbert term, we need to impose $\alpha+f_{[P]}(0)=0$ to recover GR at low curvatures. The above action is simply a combination of the EiBI and the $f(\mR)$ and, as such, the corresponding solutions are expected to interpolate between these two cases. The general formulae obtained in \ref{Sec:EBIextensionsGeneral} can be straightforwardly applied to this case. For instance, the definition of the auxiliary metric given in \refeq{Eq:defqgeneral1} yields
\be
\sqrt{-q}\m{q}^{-1}=\sqrt{-\det\left(\m{g}+\frac{\alpha}{\mbi^2}\m{\mR}\right)}\left(\m{g}+\frac{\alpha}{\mbi^2}\m{\mR}\right)^{-1}+\sqrt{-g}f_{[P]}\m{g}^{-1},
\ee
which coincides with the result found in the literature.

A second possibility to extend EiBI gravity by including the Ricci scalar is to include it in the determinantal structure. This was considered in an appendix in \cite{Makarenko:2014lxa} where the authors considered an action of the form
\be
\mS=\mpl^2\mbi^2\intd\sqrt{-\det\left[\Big(1+f(\mR)\Big)g_{\mu\nu}+\frac{\alpha}{\mbi^2}\mR_{(\mu\nu)}\right]}
\label{Eq:BRI1}
\ee
as another example of the addition of an $f(\mR)$ piece to the EiBI action. In this case, recovering GR at low curvatures requires to have $4\mbi^2f_\mR(0)+\alpha=0$. In  \cite{Chen:2015eha} this path was considered in more detail and the authors explored the cosmology of the following specific case:
\be
\mS=\mpl^2\mbi^2\intd\sqrt{-\det\left[\Big(1+\frac{\beta}{\mbi^2}\mR\Big)g_{\mu\nu}+\frac{\alpha}{\mbi^2}\mR_{(\mu\nu)}\right]}
\label{Eq:BRI}
\ee
with $\alpha$ and $\beta$ some constants satisfying $\alpha+4\beta=0$ in order to recover GR in the limit of small curvatures.

\vspace{0.3cm}
$\bullet$ {\bf  Born-Infeld actions in Weitzenb\"ock spaces.}
\vspace{0.15cm}

Another example of extensions that we classify within the Class II, but which take a different direction, are those based on the teleparallel equivalent of GR. In this description of GR, one makes a fundamental use of the vierbein language and Weitzenb\"ock spaces, characterised by having a curvature-free connection so that the torsion is the only relevant object. In terms of the vierbein $e^a{}_\mu$ and its inverse $e^\mu{}_a$, the connection is given by $\Gamma^\lambda_{\mu\nu}=e^\lambda{}_a\partial_\nu e^a{}_\mu$ so that the torsion tensor reads $\mT^\lambda_{\mu\nu}=e^\lambda{}_a(\partial_\nu e^a{}_\mu-\partial_\nu e^a{}_\mu)$. From the torsion we can built a useful quantity called the super-potential and that is given by
\be
S_\rho{}^{\mu\nu}=\frac14\Big(\mT^{\nu\mu}{}_\rho+\mT{}^{\nu\mu}{}_\rho-\mT_\rho{}^{\mu\nu}\Big)+\frac12\Big(\delta^\mu_\rho \mT^{\alpha\nu}{}_\alpha-\delta^\nu_\rho \mT^{\alpha\mu}{}_\alpha\Big).
\label{Eq:DefSuperS}
\ee
With this object, we can construct the Weitzenb\"ock invariant defined as
\be
\mT=S_{\rho}{}^{\mu\nu}\mT^\rho{}_{\mu\nu}.
\label{Eq:DefTinv}
\ee
Then, the so-called Teleparallel Equivalent of General Relativity is described by the action
\be
\Ss{TEGR}=\frac12\mpl^2\int\d^4x e \mT
\label{Eq:TEGR}
\ee
where $e=\det e^a{}_\mu$. That this action is equivalent to GR can be seen from the fact that $\mT$ for the Weitzenb\"ock connection differs from the Ricci scalar of GR by a total divergence, so that both theories give rise to the same equations of motion. The TEGR, however, serves as an alternative starting point to develop modifications of gravity \`a la Born-Infeld. This was pursued in \cite{Ferraro:2009zk,Fiorini:2013kba,Fiorini:2016zrt}, where the authors considered a general expression of the form
\be
\Ss{BITG}=\mpl^2\mbi^2\intd\left[\sqrt{\left\vert g_{\mu\nu}+
\frac{1}{\mbi^2}\Big(\alpha_1 S_\mu{}^{\lambda\rho}\mT_{\nu\lambda\rho}+\alpha_2 S_{\lambda\mu}{}^\rho T^\lambda{}_{\nu\rho}+\alpha_3 g_{\mu\nu}\mT \Big)\right\vert}-\lambda\sqrt{-g}\right],
\label{Eq:SBITG}
\ee
with $\alpha_i$ some parameters that must satisfy $\alpha_1+\alpha_2+4\alpha_3=1$ in order to recover \refeq{Eq:TEGR} in the limit of $\mT\ll\mbi^2$. Similarly to the case of EiBI gravity, the parameter $\lambda$ controls the presence of a cosmological constant. Unlike the proposals discussed so far, this class of theories must be formulated with the vierbein being the fundamental fields and the general framework presented in section \ref{Sec:EBIextensionsGeneral} cannot be applied to this case. The Born-Infeld extensions along these lines are substantially less explored than those based on the EiBI formulation. However, the Born-Infeld theories based on the TEGR also show interesting features and, furthermore, could be seen to be closer to Born-Infeld electromagnetism, since TEGR can be seen as a gauge theory where the vierbeins play the role of gauge fields associated to the translations group. One might be concerned however with the fact that, similarly to what happens in the  models belonging to the class 0 formulated in the metric formalism, generic theories described by the action \refeq{Eq:SBITG} will introduce instabilities. In particular, the loss of local Lorentz symmetry when going from TEGR to the action \refeq{Eq:SBITG} will likely introduce additional degrees of freedom. At the time of writing this review, it lacks a full analysis of the fields content and their stability around relevant backgrounds of those theories.

\subsubsection{Class III}
In this category we will include theories based on the Born-Infed structure but which make use of additional fields. The most natural example of these theories would be to combine EiBI gravity with its electromagnetic predecessor or with a Dirac-Born-Infeld scalar field $\phi$, resulting in actions of the form
\be
\mS=\mpl^2\mbi^2\intd\sqrt{-\det\left[g_{\mu\nu}+\frac{1}{\mbi^2}\Big(b_1\mR_{\mu\nu}+b_2F_{\mu\nu}+b_3\partial_\mu\phi\partial_\mu\phi\Big)\right]}\,.
\label{Eq:GenClassIII1}
\ee
This type of actions are perhaps the most natural combination of Born-Infeld actions for gravity, electromagnetism and/or scalar fields. Already Vollick considered a combination of this type in \cite{Vollick:2005gc}.  A different approach is to simply add the corresponding Lagrangians and consider actions of the form
\begin{align}
\mS=\mpl^2\mbi^2\intd&\left[\sqrt{-\det\left(g_{\mu\nu}+\frac{1}{\mbi^2}\mR_{\mu\nu}\right)}+c_1\sqrt{-\det\left(g_{\mu\nu}+\frac{1}{\mbi^2}F_{\mu\nu}\right)}\right.\nonumber\\
&\left.+c_2\sqrt{-\det\left(g_{\mu\nu}+\frac{1}{\mbi^4}\partial_\mu\phi\partial_\mu\phi\right)}\;\right]\,.
\label{Eq:GenClassIII2}
\end{align}
This was considered for instance in \cite{Jana:2015cha,Jana:2016uvq}. More general actions that belong to this class can be obtained from the EiBI action formulated in higher dimensions after a dimensional reduction, for instance by compactifying one extra dimension as done in \cite{Fernandes:2014bka}.

\subsubsection{Class IV}
Besides the extensions or variations around the EiBI theory discussed so far, there are other alternatives that make use of some of the ideas characteristic of Born-Infeld theories, but they could be classified as belonging to other classes of theories. Within this category we could mention some of the early attempts to build a Born-Infeld inspired gravity theory in the metric formalism already discussed in section \ref{Sec:Othermetric}. Among them, we could cite here specific $f(R)$ models with a square root or some other bounded function (see also \cite{Kruglov:2012ja,Kruglov:2014gva}). Although the square root structure introduces some resemblance with Born-Infeld theories, those models could be classified as belonging to the $f(R)$ class of theories. The same would apply to theories involving not only the Ricci scalar, but also the higher order Lovelock invariants, in particular, the Gauss-Bonnet term $\mG$ which is the only relevant one in four dimensions besides the Ricci scalar. It is possible to construct theories of the type $f(R,\mG)$ that incorporate some square root structure, as it is considered in\footnote{Actually, in \cite{Comelli:2005tn}, the author considers a family of theories that would generically belong to the Class 0, but a particular model is eventually selected that would belong to the Class IV and is the one we refer to here.} \cite{Comelli:2005tn}. These theories would then belong to the general class of $f(R,\mG)$ theories. The same would apply to theories based on the teleparallel equivalent of GR as for instance in \cite{Ferraro:2006jd} that can be classified as belonging to the $f(T)$ extensions of teleparallel theories \cite{Cai:2015emx}. We will also include in this class theories making use of the determinantal structure characteristic of Born-Infeld theories, but which reduce to other types of theories, either completely or in its regime of validity. This is for instance the case of \cite{Nieto:2004qj} that secretly describes Lovelock gravity, as discussed in section \ref{Sec:D&G}.

\subsection{Final remarks}

In this section we have provided the reader with a general framework for the study of Born-Infeld theories, as well as an overview of the different classes of these theories existing in the literature. We have started by briefly reviewing Born-Infeld electromagnetism and surveyed the attempts to adapt the same ideas to the case of gravity as potential mechanisms to regularise the divergences appearing in GR. The first attempts formulated in a metric formalism faced serious shortcomings due to the presence of ghosts. In order to bypass these pathologies, one can introduce higher order corrections to remove the ghosts at a given order, but the large freedom existing in the choice of the counter-terms renders the procedure unappealing. It is fair to say that, to date, there is no compelling theory free from ghosts that comply with the Born-Infeld philosophy in the metric formalism. A step forward was given when considering Born-Infeld types of actions in the affine approach. In that case, the ghost is not present from the beginning and the theory can really be regarded as a proper Born-Infeld theory of gravity, meaning that it modifies the gravitational interaction at high curvatures where a natural bound appears. We would like to remark once again that other attempts that resemble Born-Infeld theories actually contain additional degrees of freedom so that they deviate from what we consider should be the spirit of Born-Infeld theories. At this respect, we have introduced a classification of the different Born-Infeld inspired theories of gravity attending to their closeness to the Born-Infeld realm.

Since the most extensively explored theory within the framework of Born-Infeld extensions of gravity is the EiBI model, we have devoted a substantial effort to showing in detail its main properties, although we have later shown that the same features are shared by a much larger class of theories. We have provided a detailed derivation of the field equations and highlighted the importance of the projective symmetry in the construction of the theories. In particular, we have seen that theories with that invariance can be fully solved in terms of an auxiliary metric and the torsion only enters as an irrelevant projective mode (under some assumptions on the matter sector). Even though this auxiliary metric makes its first appearance as an object allowing to solve the connection, we have seen that it carries physical relevance. This was apparent when we discussed the two frames existing in these theories. From there, we clearly saw that, while the spacetime metric determines the causal structure of matter fields, the auxiliary metric determines the causal structure of the gravitational waves. This in turn implies that while photons travel along null geodesics of the spacetime metric, gravitons move along null geodesics of the auxiliary metric and, thus, even if both particles are massless, their motion will differ in regions where the curvature is large as compared to the Born-Infeld scale.

An issue that remains within the affine formulation of Born-Infeld gravity is the lack of clear guiding principles to select a unique family of theories. Born and Infeld followed a symmetry principle that allowed them to single out their non-linear electrodynamics, which was later shown to have a number of remarkable features and it was even related to string theory. The same is currently lacking for Born-Infeld inspired theories of gravity. In fact, modifications and extensions of Born-Infeld gravity have flourished in several directions. By studying some of the proposed extensions, one can convince oneself that some families of theories seem to lead to much simpler equations than others. While this simplicity principle can be useful to explore the physical consequences of these families of theories, a more profound and appealing principle would be desired.

\section{Astrophysics} \label{sec:astrophysics}

%\subsection{Introduction}\label{sec:intro}

A generic feature of extended theories of gravity in which the connection is regarded as independent of the metric (Palatini approach) is the emergence of a dependence of the metric on the local stress-energy densities. This property was soon noticed in the case of $f(R)$ theories \cite{Olmo:2011uz} and its extensions with Ricci-squared $R_{\mu\nu}R^{\mu\nu}$ terms \cite{Olmo:2009xy,Olmo:2012nx}, and is also present in Born-Infield inspired theories of gravity and its known generalisations, see section \ref{Sec:BIfieldequations}. This local dependence on the matter fields may at first appear as something exotic but is such a basic and fundamental issue in metric-affine theories of gravity that it must be properly understood in order to handle these theories correctly and properly define strategies to test their viability.

% The point is that there exist situations of physical interest in which the averaging in the matter sector is not accompanied by a corresponding smooth averaging of the metric. As a result, the theory may lead to absurd predictions, which requires a careful reconsideration of the steps taken in the description of certain scenarios in order to obtain a valid physical description.  \\

In this section we will explore situations of astrophysical interest in which the dependence of the metric on the local densities of energy and momentum manifests itself very clearly. In fact, numerous observables of stellar objects are very sensitive to the physical processes taking place in their interiors, whose properties strongly depend  on the local density. This is the case, for instance, of the mass-radius relation, the mechanisms of energy transport,  the seismic properties of stars, the type and intensity of neutrino fluxes, the speed of sound profile of acoustic waves in the sun, the potential existence of phase transitions in terms of ordered (crystalline) and superfluid phases inside neutron stars, the deconfinement of quarks or the mechanisms of generation of very large magnetic fields. For some reviews on these topics see e.g. \cite{Heiselberg:1999mq,Glendenning:2001pe}. This dependence on the local density can thus be used to efficiently test some aspects of this type of modified theories of gravity but it may also lead to unexpected subtleties. In particular, we will see that the fluid approximation and some models regularly used in the context of GR must be handled with care or conveniently adapted in order to avoid fictitious forces induced by the averaging procedure employed in the transition from the microscopic description to the continuous limit. This will be particularly relevant in the discussion of the outer boundaries of some stelar models both in the relativistic and in the non-relativistic limit.

We will begin this section by considering the weak-field, slow-motion limit of the Eddington-inspired Born-Infield (EiBI) theory first introduced in section \ref{Sec:EiBI}\footnote{For the purpose of this section, we shall consider just the case of both symmetric connection and Ricci tensor for this theory, a case discussed in detail in section \ref{Sec:SimplifiedEqEiBI}.}, and its implications for non-relativistic stars. This will allow us to visualise in a very simple way where the  subtleties of the fluid approximation may arise, which will help us better understand the peculiarities of these theories and identify situations in which an improved description of the matter sector may be necessary in order to construct realistic models. We will then move to consider relativistic stars, their structure, and their observational properties.

\subsubsection*{A word on the notation of this section}

For operational convenience and to make contact with existing literature, both in this section and in the black holes section \ref{sec:Blackholes}, we shall redefine part of the notation employed in section \ref{Sec:EiBI} and redefine Born-Infield mass as $\mbi=1/\epsilon$ and reintroduce Einstein's constant in the action via $\mpl=1/(8\pi G)=1/\kappa^2$. This way, by dimensional consistency $\epsilon$ has dimensions of length squared, while the Einstein-Hilbert action of GR reads $\Ss{GR}=\frac{1}{2\kappa^2}\int d^4x \sqrt{-g} R$.

%The extreme densities reached in the interior of neutron stars also offers an excellent opportunity to test the predictions of the Born-Infield dynamics in the relativistic regime. A glance at the modeling of the external regions of compact objects will make it apparent that the nonlinearities associated to the matter density may have a nontrivial impact on the  averaging procedure when going from the microscopic to the macroscopic description of some systems. A clear understanding of these aspects is essential in order to build realistic models which can be used to confront the theory with observations.

\subsection{Newtonian limit and fluid approximation} \label{sec:Newtonianlimit}

\subsubsection{The modified Poisson equation}

%\subsection{Newtonian limit} \label{sec:Newtonianlimit}

To better visualise the local dependence of the spacetime metric on the stress-energy densities, it is useful to study the weak field, non-relativistic limit of EiBI theory given by Eq.(\ref{Eq:actionEiBI}). For this theory one finds that, to leading order in the EiBI parameter $\epsilon$, the right-hand side of the Ricci tensor on the field equations (\ref{Eq:qequationscanonical}) takes the form

\begin{equation}\label{eq:Rmn-qw}
R_{\mu\nu}(q)\approx \kappa^2 \left(T_{\mu\nu}-\frac{1}{2}g_{\mu\nu}T \right)+\epsilon \kappa^4\left(S_{\mu\nu}-\frac{S}{4}g_{\mu\nu}\right) \ ,
\end{equation}
where $S_{\mu\nu}={T^\alpha}_\mu T_{\alpha\nu}-\frac{1}{2}T T_{\mu\nu}$, while $T$ and $S$ are the trace of $T_{\mu\nu}$ and $S_{\mu\nu}$, respectively. This equation indicates that the deviation of the auxiliary metric $q_{\mu\nu}$ from the Minkowski metric will be determined by the total amount of energy-momentum appearing on the right-hand side of this equation. For weak sources, therefore, $q_{\mu\nu}$ will be given by the Minkowski metric plus corrections which depend on integrals of the elements on the right-hand side. Now, since $g_{\mu\nu}$ is related to $q_{\mu\nu}$ via the deformation matrix $\hat\Omega$ as defined in Eq.(\ref{Eq:defOmega}), which in the low density limit is given by

\begin{equation}
{\Omega^\mu}_\nu\approx {\delta^\mu}_\nu+\epsilon \kappa^2 \left({T^\mu}_\nu-\frac{T}{2}{\delta^\mu}_\nu \right)
\end{equation}
the relation between the perturbations in  $q_{\mu\nu}\approx \eta_{\mu\nu} +t_{\mu\nu}$ and $g_{\mu\nu}\approx \eta_{\mu\nu} +h_{\mu\nu}$ turns out to be

\begin{equation}
t_{\mu\nu}=h_{\mu\nu}+\epsilon\kappa^2 \left(T_{\mu\nu}-\frac{1}{2}\eta_{\mu\nu}T \right) \ .
\end{equation}
The left-hand side of (\ref{eq:Rmn-qw}), once the standard gauge choice $\partial_\lambda (h^\lambda_\mu-\frac{h}{2}\delta^\lambda_\mu)=0$ is made, leads to $R_{\mu\nu}(\eta+t)\approx -\frac{1}{2}\Box t_{\mu\nu}$, where $\Box$ is the flat d'Alembertian. For weak sources, therefore,
the above equations lead to

\begin{equation}\label{eq:weak-field}
-\frac{1}{2}\Box t_{\mu\nu}=\kappa^2 \left(T_{\mu\nu}-\frac{1}{2}\eta_{\mu\nu}T \right) \ ,
\end{equation}
where only the leading order contributions on the right-hand side have been kept. For the (weak field and slow-motion) Newtonian limit we just focus on the $t_{00}$-component assuming, as usual, a pressureless fluid with $T_{\mu\nu}\approx \rho u_\mu u_\nu$, where $\rho$ is the energy density of the fluid. Defining $t_{00}=-2\bar\phi_N$ and $h_{00}=-2\phi_N$, such that $\bar\phi_N=\phi_N-\frac{\epsilon \kappa^2}{4}\rho$, the above equation in the non-relativistic limit can be written as

\begin{equation}\label{eq:Phi-N}
\nabla^2 \phi_N=\frac{\kappa^2}{2}\rho+\frac{\epsilon \kappa^2}{4}\nabla^2\rho \ ,
\end{equation}
which admits a general solution of the form

\begin{equation}\label{eq:potential}
\phi_N(t,\vec{x})=\frac{\kappa^2}{8\pi}\int d^3\vec{x}'\frac{\rho(t,\vec{x}')}{|\vec{x}-\vec{x}'|}+\frac{\epsilon \kappa^2}{4}\rho(t,\vec{x}) \ .
\end{equation}
The first term in (\ref{eq:Phi-N}) represents the standard Newtonian source, while the second one corresponds to a new source of gravity that involves derivatives of the matter density. Whenever those gradients become important, significant deviations from Newtonian gravity will arise. To estimate the scale at which such deviations occur and the kind of effects one may find, it is illuminating to take the Fourier transform of (\ref{eq:Phi-N}) \cite{Avelino:2012qe}, which leads to

\begin{equation}\label{eq:Fourier_Phi-N}
k^2 \tilde{\phi}_N(\vec{k})=\frac{\kappa^2}{2}\left(\frac{\epsilon k^2}{2}-1\right)\tilde{\rho}(\vec{k}) \ ,
\end{equation}
where $\tilde{\phi}_N(\vec{k})$ and $\tilde{\rho}(\vec{k})$ are the momentum space counterparts of $\phi_N$ and $\rho$. It is clear from this expression that in the GR limit, $\epsilon\to 0$, the right-hand side of (\ref{eq:Fourier_Phi-N}) is always negative. For any finite (but positive) $\epsilon$, however, one finds a scale $k_J=\sqrt{2/\epsilon}$ beyond which the right-hand side of (\ref{eq:Fourier_Phi-N}) flips its sign, thus leading to repulsive rather than attractive gravity. This allows us to interpret the effective {\it Jeans length} $\lambda_J=2\pi/k_J$ as the critical scale below which the collapse of pressureless dust is not possible due to the dominance of repulsive interactions. One obvious consequence of this is that for $\epsilon<0$ nothing seems to prevent the possibility of complete gravitational collapse for pressureless fluids (within this approximation), which is a significant change of behavior as compared to the case $\epsilon>0$. Another important consequence that can be derived from the $\epsilon<0$ case is that the growth of the intensity of the gravitational field at small scales may lead to equality between electric and gravitational forces at length scales $\sim 10^{-15}-10^{-14}$ m unless $8\pi G \epsilon< 10^{-3}$ m$^5$s$^{-2}$kg$^{-1}$ \cite{Avelino:2012qe}. Our current understanding of nuclear and particle physics, therefore, requires $\epsilon<6\times 10^5$ m$^2$ or, equivalently, $\sqrt{\epsilon}\lesssim 800$ m.

%is that the mere existence of atomic nuclei requires that the effective Jeans length be smaller than typical atomic radii, $\lambda_J<10^{-15}-10^{-14}$ m, which implies the bound\footnote{For nuclear matter of density $\rho_N=2.3\times 10^{17}$ kg/m$^3$, the associated gravitational length scale squared is $l_N^2=1/\kappa^2\rho c^2\approx 2.3 \times 10^8$ m$^2$, which should be compared to $\epsilon<10^6$ m$^2$ in this example.} $8\pi G \epsilon< 10^{-3}$ m$^5$s$^{-2}$kg$^{-1}$ (or $\epsilon<10^6$ m$^2$), since otherwise gravitational repulsion would overcome the attractive nuclear forces.

\subsubsection{Non-relativistic fluids}

The application of the modified Poisson equation (\ref{eq:Phi-N}) to the study of non-relativistic self-gravitating fluids  was first carried out in \cite{Pani:2011mg} (see also \cite{Pani:2012qb} for more details and some clarifications). %An important first result was the observation of the existence of stable, pressureless, self-gravitating objects in the case $\epsilon>0$, where analytical solutions were found.
For fluids in hydrostatic equilibrium, one must supplement the modified Poisson equation (\ref{eq:Phi-N}) with the fluid conservation equation $\nabla_\mu {T^\mu}_\nu=0$ in the appropriate limit. For spherically symmetric systems, the conservation equation boils down to $dp/dr=-(\rho+p)\Gamma_{tr}^r$ (where $p$ is the pressure of the fluid), and for weak sources $\Gamma_{tr}^r\approx -\frac{1}{2}\partial_r h_{tt}$ leads to

\begin{equation}\label{eq:TOV-Newton}
p_r=-\frac{\kappa^2}{8\pi}\frac{m(r)\rho}{r^2}-\frac{\epsilon \kappa^2}{4}\rho \rho_r \ ,
\end{equation}
where $p_r\equiv dp/dr$, $\rho_r\equiv d\rho/dr$, $m(r)=4\pi\int^r dx\rho(x)x^2$, and an equation of state $p=p(\rho)$ must be specified.

\begin{figure}[h]
\centering
\includegraphics[width=0.60\textwidth]{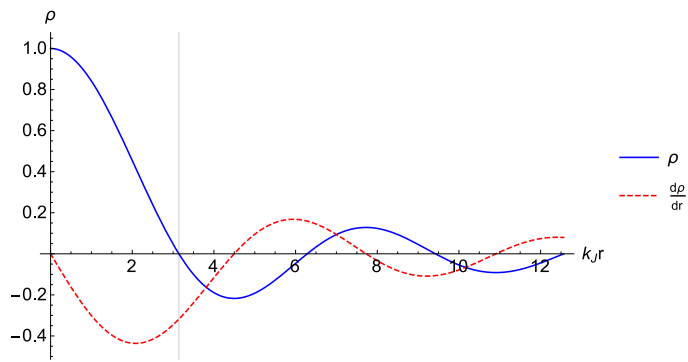}
\caption{Density profile (\ref{eq:P-less}) of the pressureless configurations. Note that the density is not positive definite beyond $k_J r>\pi$. % and that $d\rho/dr$ is not zero at that point, which causes a discontinuity in this function if the density is set to zero as of that point.
\label{fig:Pressureless}}
\end{figure}

An immediate solution of this equation corresponds to the case in which $p(r)=0$. Unlike Newtonian gravity, where pressureless solutions cannot be in hydrostatic equilibrium, the above equation yields a nontrivial solution when $\epsilon>0$.  This case simply requires solving the equation $m(r)=-2\pi \epsilon r^2\rho_r$. Applying on this equation a radial derivative, it can be cast as the Lane-Emden equation of a polytrope with index $n=1$ (recall that polytropes have equation of state $p(\rho)=K\rho^{1+\frac{1}{n}}$, where $K$ and $n$ are real positive constants, and $n$ is the so-called polytropic index). If the regularity condition $\rho(0)=\rho_c$ is imposed at the centre to get rid of the $1/r$ term, the solution to this equation takes the form

\begin{equation}\label{eq:P-less}
\rho(r)=\rho_c \frac{\sin(k_J r)}{k_Jr} \ .
\end{equation}
As is standard in the study of polytropes, the authors in  \cite{Pani:2011mg,Pani:2012qb}  restricted the range of validity of this solution to the interval $r\in [0,\pi/k_J]$ to avoid the presence of a negative energy density beyond the first zero at $r= \pi/k_J$ (see Fig.\ref{fig:Pressureless}). Though this restriction is natural and harmless in the usual Newtonian theory, the fact is that it forces a discontinuity in $\rho_r$ at $r= \pi/k_J$, thus causing a divergence on the right-hand side of (\ref{eq:Phi-N}).

\subsubsection{The issue with the matter profiles at a star surface}

The example above illustrates an important property of this type of theories of gravity, namely, that the matter fields must satisfy certain differentiability conditions that are not necessary in the context of GR. The matter/energy profiles must be continuous and differentiable up to some degree. This requirement may certainly be {\it inconvenient}, because it forces us to pay more attention to the modeling of our energy sources in certain applications, but is {\it not a fundamental problem} because matter and radiation are ultimately described in terms of quantum fields, which are sufficiently smooth to comply with the differentiability requirements of these theories. Therefore, the solution (\ref{eq:P-less}) admits two possible interpretations: 1) that we are dealing with an unconventional fluid or 2) that an improved description of the matter fields (with a different fluid or even beyond the fluid approximation) is necessary near the surface at $k=\pi/k_J$ to avoid undesired or fictitious unphysical effects.

\begin{itemize}

\item Regarding option 1), note that in the transition from the (relativistic) weak-field approximation (\ref{eq:weak-field}) to (\ref{eq:Phi-N}) we (implicitly) assumed that the stress-energy tensor $T_{\mu\nu}$ of the matter fields could be averaged to yield that of a perfect fluid without causing any harm to the theory. In this process, the fluid we had in mind was some distribution of localised particles (or wavepackets) such that when averaged over a certain scale should yield a continuum distribution characterisable by the $T_{\mu\nu}$ of a perfect fluid.  In particular we expected a positive definite density function $\rho(x)$, which turns out to be in conflict with our solution (\ref{eq:P-less}) beyond $r=\pi/k_J$. Our microscopic interpretation of the fluid, therefore, does not fit well with the predictions of this theory, which indicates that we are dealing with an unconventional matter source. Note in this sense that the authors in \cite{Pani:2011mg,Pani:2012qb} argued in favor of this solution representing some kind of dark matter, which might give plausibility to this result. The effects on the galactic metric of a dark matter density profile of this kind has been studied in detail in \cite{Harko:2013xma}.

\item Regarding option 2), if the fluid is interpreted as made out of standard particles, an improved microscopic description of those particles should be considered near the outer boundary (where the density is close to zero) to get a smooth transition to the exterior region in the neighborhood of $r=\pi/k_J$. Thus, a refinement of the physics near the surface is necessary to build a complete solution. As mentioned above, this might be inconvenient but is not a fundamental problem. In fact, as shown in \cite{Avelino:2012qe}, different averaging procedures in the transit to the continuum fluid approximation may lead to different (fictitious) acceleration fields associated to the specific weight functions employed in the averaging. The emergence of negative densities in the outer regions of these solutions can thus be interpreted as a manifestation of fictitious effects which should be regarded as unphysical and avoidable by an improved description.

\end{itemize}

The view that one should go beyond the fluid approximation or consider a suitable transition thick shell in the description of the surface region is further reinforced by the analysis presented in \cite{Pani:2012qb} regarding the process of dust collapse. Starting with generic static profiles, it was found that the fate of the system is to reach a universal configuration which oscillates around the pressureless solution (\ref{eq:P-less}) with a period that coincides with the fundamental mode of proper oscillations of the pressureless case. This means that the configurations provided by (\ref{eq:P-less}) are not a fine-tuned solution of an exotic matter field but, rather, they are a universal, regular final state for the collapse of reasonable matter sources. The role of the EiBI dynamics is, clearly, to stabilise the object against collapse by generating repulsive gravitational forces at short scales. The problems arising on the surface can be regarded as artifacts of the particular fluid approximation considered.
%Thick or thin???

\subsubsection{Limitations and improvements of the polytropic description}

Similar problems affecting the exterior boundary of some polytropic stellar models were also found in \cite{Pani:2012qd}. Due to the divergence of derivatives of the energy density with respect to the pressure as $p\to 0$ near the surface, quantities such as the Ricci curvature scalar diverge (this also happens in the Newtonian model above if one imposes a discontinuity in $\rho_r$ at the surface). This occurs, in particular, for polytropic indices $\gamma=1+\frac{1}{n}>3/2$, which include the case of a gas of degenerate non-relativistic electrons ($\gamma=5/3$) or the case $\gamma=10/3$ used to model the atmosphere of white dwarfs. This result led to claim that these divergences could not even be cured by abandoning the fluid approximation, because a microscopic description of the matter sources would increase the differential order of the field equations in the matter sector, thus making the curvature even more sensitive to sharp variations in the matter fields \cite{Pani:2012qd}.
%10/3????

A number of objections can be presented to the pessimistic view of \cite{Pani:2012qd}. Firstly, the claim that a microscopic description cannot cure these problems was just a conjecture which has never been explicitly proven. In fact, the accumulated evidence so far goes in the opposite direction. For instance, self-gravitating solutions of isolated charged particles in the EiBI theory do not show any pathologies neither at high nor at low curvatures, see black holes section \ref{sec:Regular}. If individual particles are well behaved, it is difficult to conceive how a collection of them (a fluid) could develop pathologies in the low density regime, where the interparticle separations are large and the isolated particle description is better justified. Similar results are found in the case of self-gravitating scalar fields, which possess a solitonic structure compatible with the idea of isolated neutral particles \cite{Afonso}. Since the microscopic constituents are individually well behaved, the curvature divergences on the surface of polytropes are likely to be a manifestation of fictitious accelerations induced by the continuum approximation \cite{Avelino:2012qe}. Secondly, a careful analysis of the validity of the polytropic equation of state near regions of divergent curvature was carried out in \cite{Kim:2013nna}. The idea was not to estimate the corrections due to finite temperature or electromagnetic repulsion between charged particles, as is necessary in realistic models to properly account for the opacities in stellar atmospheres, but to explore how the microscopic definition of pressure could be affected near curvature divergences. By analysing the geodesic deviation equation (\ref{eq:geoeqdev}), the frequency of the interactions between a particle and a nearby (fictitious) wall\footnote{Note that a particle and a wall are necessary to define the pressure microscopically.} was found to increase with respect to the corresponding statistical estimate in flat space-time. This represents an additional pressure which changes the effective equation of state for the case $3/2<\gamma<2$ and avoids the original curvature divergence. For the case $2<\gamma<3$, which is also problematic, it is found that the fluid is repelled from the surface. It is then argued that such fluids would not be appropriate to describe the surface and that some other type of matter should be necessary. The conclusion is that the fluid reacts as the curvature grows on the surface and that an improved description of the matter there is necessary.

We thus see that the fluid approximation and/or the modeling of certain objects in the EiBI theory of gravity may require some refinements to avoid unphysical effects that arise at the outer boundaries of some solutions. This occurs when the derivatives of the matter density diverge too rapidly as the pressure goes to zero or when the matter profile and its derivatives are abruptly set to zero at some point in order to match with the external (idealised) Schwarzschild solution. By smoothing the behavior of the matter profiles, these problems can, in principle, be overcome. Though this is certainly an inconvenience, it is not that far from what realistic models require. In fact, in order to qualitatively and quantitatively understand numerous observational features of neutron stars, such as their electromagnetic spectra, envelope composition, X-ray bursts, surface temperature profiles, etc, it is not only necessary, it is essential, to carefully describe the microphysics of the outer layers. Some of these layers are very thin as compared to the radius of the star, with a height of $\sim 0.1-10$ cm and  density, $\rho\sim 10^{-2}-10^{3}$ g/cm$^3$ \cite{Ho:2006uk} in the photosphere, and densities always below $10^{10} $g/cm$^2$ on the outer $10^4$ cm of the envelope. The composition of this region is dominated by a gas of (partially) ionized atoms and electrons plus radiation, with the electron equation of state transitioning from an ideal to a degenerate gas as one goes deeper into the star \cite{Chang:2002wy,1983ApJ267315P}, which has a crucial effect on the efficiency of the different energy transport mechanisms and, thus, dramatically affects the observable features of the star  \cite{Potekhin:2004jr}.
We thus see that in these layers, finite temperature, radiation fields, chemical composition, electromagnetic repulsion, magnetic fields, etc, induce significant deviations from the basic polytropic equations of state \cite{Koester:2008sh}, which are nonetheless very useful to estimate the gross properties of these objects. Though models with this level of refinement have not been yet constructed in the EiBI gravity scenario, as we will see below, the evidence so far indicates that there is no fundamental reason to believe they are not possible.

% aqui citamos estos papers: Potekhin_2004_ApJ_612_1034 y White dwarf spectra and atmosphere models - 2010MmSAI__81__921K

%We thus see that an apparently innocent equation is forcing us to reconsider our set up if we are to use this solution in physical applications. \\

%The first thing we need to note is that in the transition from the (relativistic) weak-field approximation (\ref{eq:weak-field}) to (\ref{eq:Phi-N}) we (implicitly) assumed that the stress-energy tensor $T_{\mu\nu}$ of the matter fields could be averaged to yield that of a perfect fluid without causing any harm to the theory. In this process, the fluid we had in mind was some distribution of localized particles (or wavepackets) such that when averaged over a certain scale should yield a continuum distribution characterizable by the $T_{\mu\nu}$ of a perfect fluid.  In particular we expected a positive definite density function $\rho(x)$, which turns out to be in conflict with our solution (\ref{eq:P-less}) beyond $r=\pi/k_J$. Our microscopic interpretation of the fluid, therefore, does not fit well with the predictions of this theory. If one insists on retaining the $r< \pi/k_J$ region as physically meaningful, then a more careful description of the matter sources in the vicinity of  $r= \pi/k_J$ should be taken into account to avoid artificial effects involving negative energy densities.

\subsection{Non-relativistic stars}

From the discussion in the previous section, it is now clear that the external boundary of stars should be modeled in such a way that the matter and pressure as well as its first and second-order derivatives should smoothly vanish to guarantee a correct matching with the exterior empty solution. This refinement should be done if one is really interested in understanding observational features of the models such as electromagnetic spectra, but can be overlooked in situations in which only structural aspects are important. In this sense, the standard approach in which the stellar surface is identified as the region where the pressure is {\it sufficiently} low can be retained as valid, as long as one accepts that a thin transition shell should be added to correctly complete the model.

\begin{figure}[h]
\centering
\includegraphics[width=0.60\textwidth]{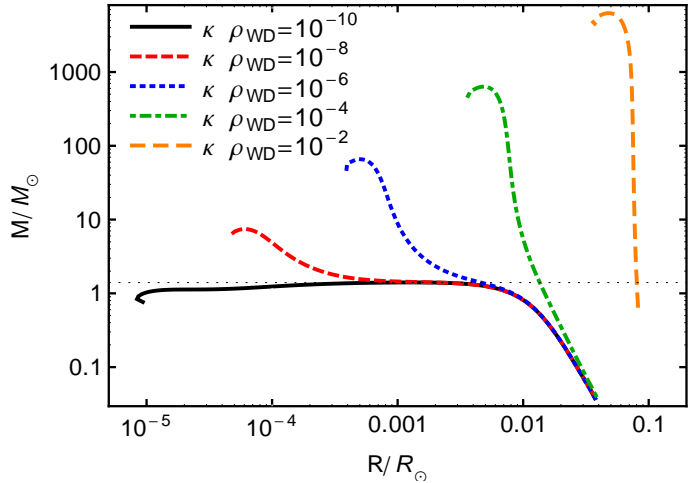}
\caption{Zero-temperature relativistic white dwarfs in EiBI theory (in this plot, $\epsilon \rightarrow \kappa$) in units of the typical density $\rho_{WD}=10^9 kg/m^3$. The horizonal line denotes the Chandrasekhar limit, $M=1.4 M_\odot$. Note that an upper limit on the mass of these stars arises but can be much larger than in GR. Figure taken from Ref.\cite{Pani:2012qb}.
\label{fig:Pani2012qb}}
\end{figure}

Having understood the peculiarities that the matter profiles should satisfy on the outer boundaries of stars, we now focus on the information that stellar models can provide to test the viability of EiBI gravity and constrain its parameters. The results of \cite{Pani:2012qb} establish a limitation for the existence of polytropic solutions with regular boundary condition at the centre, $\rho\approx \rho_c+\rho_2 r^2$, which requires

\begin{equation}\label{eq:boundPol}
\epsilon \kappa^2>-4K\left(1+\frac{1}{n}\right)\rho_c^{-1+\frac{1}{n}} \ .
\end{equation}
The reason for this bound in static configurations is related to the monotonicity of $\rho(r)$, which requires $\rho_2<0$. An expansion of (\ref{eq:TOV-Newton}) around the centre puts forward that if the bound (\ref{eq:boundPol}) is not satisfied, then $\rho_2>0$. Going beyond polytropic
models, a non-rotating, zero temperature white dwarf model with parametric equation of state

\begin{eqnarray}
p(x)&=& \frac{\pi m_e^4c^5\mu_e m_P}{3h^3}\left[x(2x^2-3)\sqrt{1+x^2}+3\sinh^{-1} x\right] \ , \\
\rho(x)&=& \frac{8\pi m_e^3c^3\mu_e m_P}{3h^3}x^3 \ ,
\end{eqnarray}
was studied for different values of $\epsilon$ \cite{Pani:2012qb}. It was found that for $\epsilon>0$, the mass of these objects is not limited by the Chandrasekhar bound $M\approx 1.4 M_{\odot}$. It turns out that the mass can be arbitrarily large while the radius tends to a minimum value which scales as $\sqrt{\epsilon}$. In the relativistic version of these objects, however, an upper bound for the mass does appear, though it can be much larger than in GR (see Fig.\ref{fig:Pani2012qb}).

\subsubsection{Solar physics constraints} \label{Sec:ssc}

A closer confrontation with observations is certainly possible by considering the effects of the modified Poisson equation on the properties of the Sun \cite{Casanellas:2011kf}. Since the hydrostatic equilibrium and energy transport ultimately depend on this equation, any correction would have an impact on the thermal balance and temperature profile inside the star, which can leave observable traces. In fact, neutrino fluxes are very sensitive to the temperature profile inside the Sun \cite{Bahcall:1996vj,TurckChieze:2010gc}. An increase or decrease of the innermost conditions due to a modified Poisson equation will necessarily leave a trace on the amounts of emitted neutrinos, which are relatively well understood observationally. Something similar happens with helioseismic data, which provide very accurate information on the solar acoustic modes, the sound speed profile, and the depth of the convective envelope, see e.g. \cite{ChristensenDalsgaard:2002ur} for a review.  In order to extract the necessary information to use solar neutrinos and helioseismic data to test EiBI gravity, the hydrostatic equilibrium equation (\ref{eq:TOV-Newton}) and the continuity equation, $dm/dr=4\pi r^2\rho(r)$, must be supplemented with the conservation of thermal energy equation

\begin{equation}
\frac{dL}{dm}=q(r)-r\frac{ds}{dt} \ ,
\end{equation}
where $q(r)$ represents the rate of heating from nuclear reactions and $s$ is the entropy per unit mass \cite{Clayton1968}, plus the corresponding equation for the convective energy transport, which takes the form

\begin{equation}
\frac{dT}{dm}=-\left[\frac{G m(r)}{4\pi r^4}+\frac{\epsilon\kappa^2\rho}{4}\frac{d\rho}{dm}\right]\frac{T}{p}\tau \ .
\end{equation}
Here $\tau\equiv d\log T/d\log p$ is the temperature gradient, which for adiabatic changes becomes $\tau=(\Gamma_2-1)/\Gamma_2$, where $\Gamma_2$ is one of the adiabatic exponents \cite{2004cgpsbook}. In the radiative zone, the transport energy equation is unmodified

\begin{equation}
\frac{dT}{dm}=-\frac{3\lambda_R}{16\sigma T^3}\frac{L}{16\pi^2 r^4} \ ,
\end{equation}
where $\lambda_R$ is the Rosseland mean opacity and $\sigma$ the Boltzmann constant.

Implementing the above equations in CESAM \cite{Morel1997}, a numerical code for stellar structure and evolution, the authors of \cite{Casanellas:2011kf} constructed a number of models able to fit the solar properties with an accuracy of $10^{-5}$ in the interval $-0.032G R_{\odot}^2< \epsilon \kappa^2<0.02 GR_{\odot}^2$. For smaller values of $\epsilon \kappa^2$, no equilibrium stars were found, in agreement with the constraint (\ref{eq:boundPol}) for polytropic models. For $\epsilon \kappa^2>0.02 GR_{\odot}^2$, solutions do exist but are unable to match simultaneously the observed values for age, radius, mass, luminosity and metallicity of the Sun.

Qualitatively, models with $\epsilon>0$ show lower central density and temperature than in GR ($\epsilon=0$), whereas for $\epsilon<0$ those magnitudes grow due to the larger attractiveness of the modified potential. An increment in the central density and temperature imply a raise in the thermonuclear reactions, which must be followed by a modification in the  neutrino emission. In the inner $10\%$ radius, the pp-III chain produces high-energy neutrinos associated to the generation of ${^8}B$ with an intensity that scales as $\phi_{{^8}B}\propto T_c^{18}$. This flux is currently measured with high precision by neutrino telescopes: $(5.046\pm0.16)\times 10^6$ cm$^{-2}$ s$^{-1}$. From the numerics one observes a decay in the neutrino flux for $\epsilon>0$ and a growth for $\epsilon<0$, such that with the current data it is possible to conclude that $\epsilon\kappa^2\lesssim -0.024G R_{\odot}^2$ is ruled out.

The precision with which acoustic modes are currently measured by helioseismic missions allows to probe the solar interior revealing the sound speed and density profiles down to $10\%$ of the solar radius \cite{Dziembowski:1998nb}. The separation between the frequencies of modes with different degree $l$ and radial order $n$, $\delta\nu_{n,l}=\nu_{n,l}-\nu_{n-1,l+2}$, is a quantity very sensitive to the temperature gradient. The case of $l=0$ is particularly important because it corresponds to waves that traveled through the entire solar radius, carrying valuable information about the density profile. The small uncertainties associated with these quantities allow to rule out the regions $\epsilon \kappa^2>0.016GR_\odot^2$ and $\epsilon \kappa^2<-0.01GR_\odot^2$. On the other hand, the agreement between the sound speed profile and the solar standard model reaches an accuracy better than $1\%$ in most of the solar interior, with larger uncertainties right below the convective envelope. Comparison with this model and the data, one can safely rule out the region $\epsilon \kappa^2>0.012GR_\odot^2$. Constraints on the depth of the convective envelope and the helium surface abundance, which also follow from helioseismic data, imply that $-0.016GR_\odot^2<\epsilon \kappa^2<0.013GR_\odot^2$  and $\epsilon \kappa^2>-0.018GR_\odot^2$, respectively. These examples clearly illustrate the capabilities of solar observations to constrain modifications of gravity with new couplings in the matter sector.

\subsection{Relativistic stars} \label{Sec:RelStars}

White dwarfs and neutron stars are by far the natural scenarios where the highest pressures can be achieved, which offers an excellent opportunity to test our theories about nuclear matter and also the modified dynamics of alternative gravity theories. It is well known that the masses and radii of neutron stars depend critically on the equation of state of dense matter \cite{GlendenningCS,Lattimer:2006xb}. For a given equation of state, a mass-radius relation and a maximum mass can be derived. The so-called {\it stiffness} of the equation of state depends on how many bosons are present. Since bosons do not contribute to the Fermi pressure, they tend to soften the equation of state, which leads to low maximum neutron star masses ($\sim 1.5 M_\odot$). GR sets a maximum mass $\sim 3.2 M_\odot$, and it is expected that the maximum achievable mass in nature is of order $\sim 2.5 M_\odot$, but this depends on the stiffness of the equation of state \cite{2006Msngr12627K} and is thus open to observational scrutiny. The density-dependent modifications induced by the EiBI dynamics can be seen as the effect of a modified source \cite{Delsate:2012ky} and, for this reason, must also leave an imprint in the mass/radius relation of these compact objects. In this section we consider the efforts carried out so far to understand the impact of the EiBI modified dynamics on the structure and stability of neutron stars as well as some strategies to distinguish its predictions from those of GR.

\subsubsection{Stellar structure}

In the EiBI gravity scenario, the equations of stellar structure in the full relativistic case have been studied in numerous works \cite{Pani:2012qb,Sham:2012qi,Harko:2013wka,Sotani:2014goa}, and for a line element of the form

\begin{equation}\label{eq:ds2}
\text{d}s^2=-e^{\phi(r)}dt^2+e^{\lambda(r)}dr^2+f(r)d\Omega^2 \ ,
\end{equation}
can be written as
\begin{eqnarray}
\frac{dm}{dr}&=&\frac{r^2}{4\epsilon }\left(2-\frac{3}{ab}+\frac{a}{b^3}\right) \\
\frac{dp}{dr}&=&-\frac{\left[\frac{2m}{r^2}+\frac{r}{2\epsilon}\left(\frac{1}{ab}+\frac{a}{b^3}-2\right)\right] }{\left[1-\frac{2m}{r}\right]\left[\frac{2}{\rho+p}+\frac{\epsilon}{2}\left(\frac{3}{b^2}+\frac{1}{a^2c_s^2}\right)\right]}\label{eq:hydroRel}
\end{eqnarray}
with
\begin{eqnarray}
f(r)&=& \frac{r^2}{ab}\\
a&\equiv& \sqrt{1+\epsilon \kappa^2 \rho}\\
b&\equiv& \sqrt{1-\epsilon \kappa^2 p}\\
c_s^2&\equiv & \frac{dp}{d\rho}  \ .
\end{eqnarray}
Given a barotropic equation of state $p=p(\rho)$ and appropriate boundary conditions, concrete models can be studied.  The boundary conditions at the centre  typically involve the specification of the central density $\rho_c$ (or the central pressure $p_c$), and the condition $m(0)=0$. For rough structural considerations, the radius of the star is defined by the condition $p(R)\leq p_0$, where $p_0$ is ideally zero but in practice is represented by a small positive number. At that point one assumes that the Schwarzschild solution should take over, provided a sufficiently smooth transition profile is used.  This last step is usually assumed to hold and can be omitted (more details on this will come later).

From the definitions of the functions $a$ and $b$ above, the constraints
\begin{eqnarray}
\epsilon \kappa^2 p_c <1  \ , \ \text{for } \epsilon>0 \\
|\epsilon| \kappa^2 \rho_c <1  \ , \ \text{for } \epsilon<0
\end{eqnarray}
appear naturally for stellar models. Assuming that $\rho_c$ in neutron stars is of the order $\rho_c\sim 10^{18}$ kg/m$^3$ and $p_c\sim 10^{34}$ N/m$^2$, we get that $|\epsilon \kappa^2|\lesssim 1$ m$^5$ kg$^{-1}$ s$^{-2}$. Numerically one verifies that compact objects only exist if $p(r)\approx p_c+p_2r^2$ near the centre has $p_2<0$. This leads to a condition compatible with (\ref{eq:boundPol}).

The case of pressureless relativistic fluids was considered in \cite{Pani:2012qb} as an extension of the Newtonian case, finding that solutions always exist if $\epsilon>0$. These objects have a maximum compactness of $GM/R\sim 0.3$ and a maximum mass and radius that linearly grow with $\epsilon$. The fact that the current cosmological model requires a significant component of cold dark matter particles with $p=0$ makes these solutions particularly interesting, since they indicate that such particles could aggregate in structures with the typical compactness and mass of most neutron stars.

\begin{figure}[h]
\centering
\includegraphics[width=0.50\textwidth]{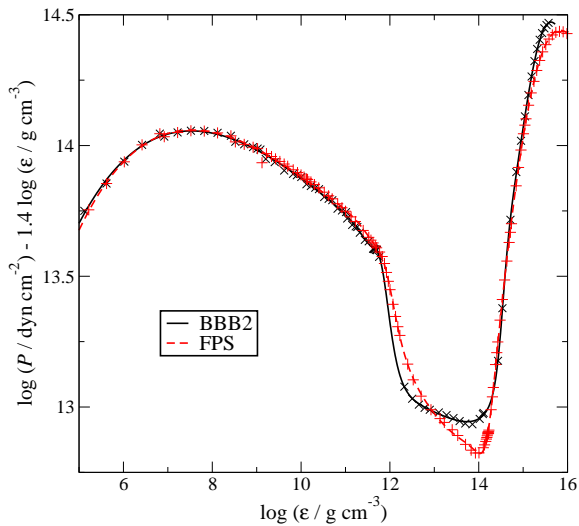}
\caption{Analytic fit of the BBB2 and FPS models, taken from Ref.\cite{Sham:2012qi}. The crosses and pluses represent the data points in the EOS
tables and the lines are the analytic fit functions.
\label{fig:fig1_1208_1314}}
\end{figure}

Models in EiBI gravity with realistic equations of state based on nuclear physics have been studied in detail in \cite{Sham:2012qi} (see also \cite{Sotani:2014goa} and \cite{Pani:2012qb}). These equations of state are usually presented in tabulated form and require numerical interpolation for their implementation in the codes. Though this is not a problem in the case of GR, the interpolation method may introduce numerical noise and artificial effects which should be avoided. In \cite{Sham:2012qi} this was solved by using smooth analytic functions to model the tabulated equations of state (see Fig.\ref{fig:fig1_1208_1314}), while in  \cite{Pani:2012qb} a piecewise polytropic interpolation was implemented. As a general feature, it is observed that the mass of the solutions increases with the central density until a certain maximum value. This maximum mass is larger than the GR prediction if $\epsilon>0$ and smaller if $\epsilon<0$. The maximum appears at a lower central density than in GR if $\epsilon>0$. As pointed out in  \cite{Pani:2012qb}, the larger mass predicted by models with $\epsilon>0$ could serve to prevent ruling out some softer equations of state for which the observation of a neutron star with mass $M\approx 1.97 M_\odot$ was a critical test (see also \cite{Qauli:2016vza} for updated data on massive pulsars). Different examples were studied in \cite{Harko:2013wka}, including a model with a causal stiff fluid for which the speed of sound equals the speed of light, a radiation-type equation of state, a polytrope of index $n=3$ (relativistic neutrons), and the quark matter equation of state. An exact (and exotic) analytical solution of the relativistic equations was also found there.  More recently, the influence of hyperons in the equation of state has been explicitly considered in \cite{Qauli:2016vza} to illustrate that the ``hyperon problem" found in neutron star models within GR may be avoided in the EiBI theory. Their conclusions are in agreement with the previous literature on this topic. Slowly rotating relativistic stars were also considered in \cite{Pani:2012qb} using the perturbative approach introduced by Hartle \cite{Hartle:1967he}.

\subsubsection{Stability}

A detailed analysis of the stability under radial perturbations of relativistic stars was carried out in \cite{Sham:2012qi} and \cite{Sotani:2014xoa}, both focusing on the fluid modes and neglecting the space-time modes. In \cite{Sham:2012qi} a fixed, static physical background was assumed but it was noted that the auxiliary metric could develop a non-zero contribution in the $t-r$ sector due to the perturbations in the fluid. The approach of \cite{Sotani:2014xoa} is different, as the author adopts a crude Cowling approximation forcing both the physical metric perturbation, $\delta g_{\mu\nu}$, and the auxiliary metric perturbation, $\delta q_{\mu\nu}$, to vanish.

\begin{figure}[h]
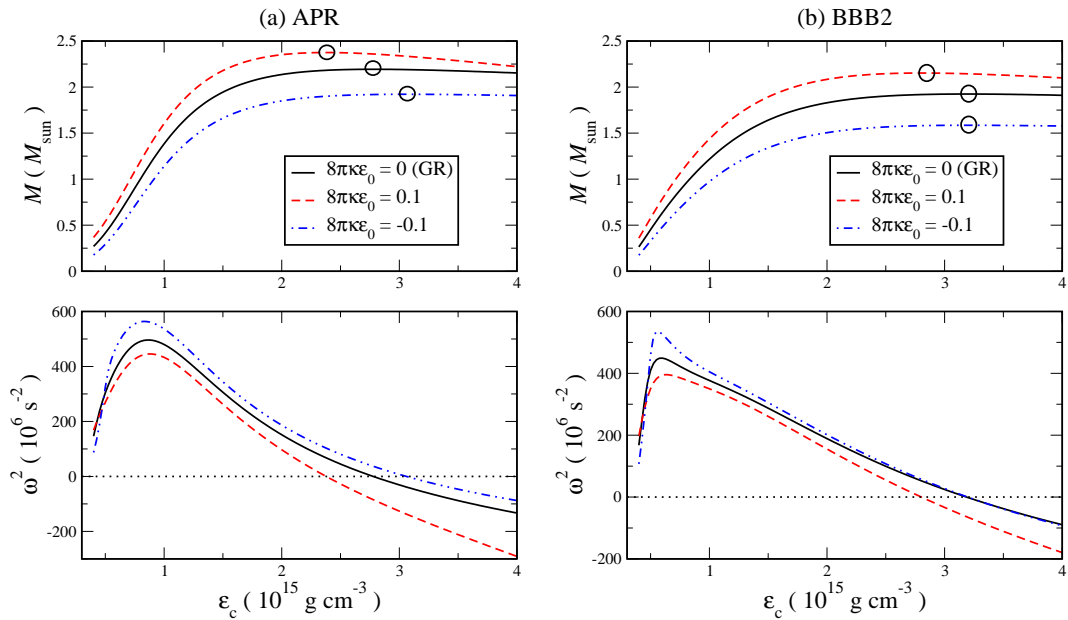

\centering
\begin{tabular}{cc}\includegraphics[width=0.45\textwidth]{fig2a_1208_1314.eps} & \includegraphics[width=0.45\textwidth]{fig2b_1208_1314.eps} \end{tabular}
\caption{Gravitational mass $M$ and fundamental mode frequency square $\omega^2$ plotted against the central density $\rho_c$ (denoted as $\varepsilon_c$ in the plots) for two EOS models: (a) APR and (b) BBB2. Three different values of the Born-Infield parameter $\epsilon$, denoted as $\kappa$ in the plots (not to be confused with the Einstein constant, as follows from the notation employed in this section) are considered. The circle on each $M$-density curve corresponds to the maximum-mass configuration. Plots taken from \cite{Sham:2012qi}. \label{fig:fig1_1208_1314bc}}
\end{figure}

Denoting by $\xi$ the radial Lagrangian displacement and $\dot\xi$ its time derivative, the four-velocity of the fluid is given by $u^\mu=(-e^{\phi/2},e^{-\phi/2}\dot\xi,0,0)$, which to linear order induces an off-diagonal perturbation in ${T^\mu}_\nu$ given by ${T^r}_t=-(\bar{\rho}+\bar{p})\dot\xi$, with the bar denoting background values. Assuming a time dependence $e^{i\omega t}$ for all the perturbed quantities, the relevant eigenvalue equation for the radial oscillation modes can be written as

\begin{equation}
\chi''=-W_1 \chi-W_2 \chi' \ , \
\end{equation}
where $\chi\equiv r^2 Q_1(\bar{\rho}+\bar{p})\xi$, and the functions $W_1, W_2,$ and $Q_1$ depend on the background fields and the frequency $\omega^2$ (see Appendix A of \cite{Sham:2012qi} for the explicit expressions of these functions). The analogous equation for linear radial perturbations in the non-relativistic limit was studied in  \cite{Pani:2012qb} in the context of polytropic fluids finding a more tractable expression:

\begin{equation}
-\frac{1}{\bar{\rho}}\left[\frac{\gamma \bar{p}}{r^2}(r^2\xi)'\right]'+\frac{4}{\bar{\rho} r}\xi \bar{p}'+\frac{\epsilon \kappa^2}{4}\left[\frac{2\xi \bar{\rho}'}{r}-\xi' \bar{\rho}'-\left(\frac{\bar{\rho}}{r^2}(r^2\xi)'\right)'\right]=\omega^2\xi \ ,
\end{equation}
where $\gamma$ defines the adiabatic index of the perturbations. In both cases, the resulting eigenvalue equation must be solved subject to standard regularity condition at the centre, $\xi(0)=0=\xi'(0)$, being $\xi(R)$ finite at the surface. An instability corresponds to an eigenmode with $\omega^2<0$.

In the relativistic case, the stability of compact stars is investigated using four different equations of state (APR \cite{Akmal:1998cf}, BBB2 \cite{1997AA328274B}, FPS \cite{Lorenz:1992zz}, and SLy4 \cite{Douchin:2000kx}). The results confirm in a robust manner that $\omega^2$ remains positive up to the value of the central density  at which the stellar mass reaches its maximum (see Fig.\ref{fig:fig1_1208_1314bc}). This critical density sets the onset of a dynamical instability against radial perturbations. At lower densities, the solutions are linearly stable. This behavior is qualitatively identical to that already observed in GR.  The numerical results in Fig.\ref{fig:fig1_1208_1314bc} show that in the EiBI gravity theory the mass of the solutions may attain a local minimum at larger central densities. The location of this extremum coincides with a zero in the frequency square of the first radial harmonic. While the frequency of the first and higher harmonics depend on the specific value of $\epsilon$ chosen, for a given mass, the fundamental mode is quite insensitive to $\epsilon$.

For non-relativistic stars,  in the presureless case one finds that, for a given $\epsilon$, there is one fundamental mode, which is numerically determined as $\omega=\alpha \rho_c^{1/2}$, where $\alpha\approx 2.1866$ is a factor independent of $\epsilon$ \cite{Pani:2012qb}. These solutions do not have unstable modes, confirming that they are linearly stable. For polytropic models, the stability is improved as compared to the case of GR. In GR, models with adiabatic index $\gamma=4/3$ are marginally stable for any polytropic index $n$, whereas in the Born-Infeld theory these models are always stable if $\epsilon>0$ and unstable for $\epsilon<0$.

%It is important to note that in the derivation of the effective Poisson equation (\ref{eq:Phi-N}) the possibility of describing the matter fields in terms of a continuous fluid was implicitly assumed, as is common in this type of discussions. This assumption, however, may not always be justified for this type of theories, as first pointed out in  \cite{Avelino:2012qe}. The reason is that different microscopic distributions which in the continuum average to the same fluid description need not yield the same acceleration profile for test particles because the averaging procedure in the matter sector may generate fictitious accelerations.

%What are the different EoS good for?

\subsubsection{Observational discriminations from GR}

The exceptional conditions of matter inside neutron stars, with densities that may be several times above nuclear densities, turn these objects into natural laboratories from which to extract valuable information on the properties and behavior of nuclear matter. The study of their structure, rotation, and magnetic fields can thus offer a powerful window on the microscopic properties of matter, complementing in this way the knowledge obtained from laboratory experiments. In the context of GR, this could help constrain the form of the equation of state of nuclear matter, which has led to the study of empirical relations between the structural properties of neutron stars (mass, radius, moment of inertia, ...) and their equations of state. Alternatively, the potential existence of relations weakly dependent on the equation of state could also be used to constrain the gravitational dynamics.

The possibility of discriminating between the predictions of GR and those of EiBI gravity using a special kind of neutron stars was raised in \cite{Sotani:2014goa}. As mentioned above, the mass-radius relation of neutron stars is intimately linked to the details of the equations of state and the internal gravitational dynamics. Extracting useful information about these two inputs of the theory requires not only measuring the radii of neutron stars for a broad range of masses but also breaking the potential degeneracies that may arise between the matter and the gravity sectors. In this respect, a normal neutron star with $M\sim 1.4 M_\odot$ is more sensitive to the high density properties of the equation of state than a low mass star with $M\sim 0.5 M_\odot$, and the uncertainties in the corresponding equations of state of each model are very different.
The importance of the analysis of \cite{Sotani:2014goa} lies on the observation that the radii of neutron stars with $M\sim 0.5 M_\odot$ are strongly correlated with the neutron skin thickness\footnote{In the determination of the relation between pressure and energy density in nuclear matter equations of state, the so-called {\it symmetry energy} quantifies the change in nuclear energy associated with modifying the neutron-proton asymmetry. Accurate determination of the thickness of the neutron skin of neutron rich heavy nuclei would provide crucial experimental constraints on the symmetry energy and, as a consequence, on the structural properties of neutron stars.} of $^{208}Pb$ nuclei in a way which is independent of the particular equation of state \cite{Carriere:2002bx}.  This suggests that laboratory measurements of the neutron skin thickness of $^{208}Pb$ combined with the precise observational determination of the radii of neutron stars with $0.5 M_\odot$ could provide a direct estimate of $\epsilon$.

\begin{figure}[h]
\centering
\includegraphics[width=0.6\textwidth]{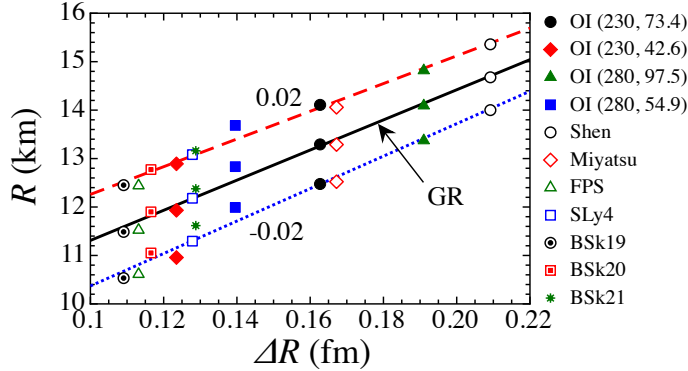}
\caption{Radii of neutron stars with $0.5M_\odot$, $R_{05}$ in Eq.(\ref{eq:RdeltaR}), as a function of the neutron skin thickness of $^{208}$Pb for
$\epsilon\kappa^2\rho_0$ = $-0.02, 0$, and $0.02$, using various EOS as shown in \cite{Sotani:2014goa}. The curves are insensitive to the EOS but do depend on the Born-Infield parameter $\epsilon$.  \label{fig:skin-R_1404_5369}}
\end{figure}

Quantitatively,  \cite{Sotani:2014goa} finds that the radius of $0.5M_\odot$ neutron stars, $R_{05}$ (measured in km), for different equations of state admits a linear parametrisation of the form (see Fig.\ref{fig:skin-R_1404_5369})
\begin{equation}\label{eq:RdeltaR}
R_{05}=c_0+c_1 \Delta R \ ,
\end{equation}
where $\Delta R$ (measured in fm) represents the neutron skin thickness of $^{208}Pb$. It should be noted that the value of $\Delta R$ depends on the particular equation of state considered but is independent of $\epsilon$. The constants $c_0$ and $c_1$ depend on the value of $\epsilon \kappa^2 \rho_0$, where $\kappa^2\rho_0/8\pi=1.99 \times 10^{-4}$ km$^{-2}$ follows from taking $\rho_0=2.68 \times 10^{14}$ g/cm$^3$ (nuclear saturation density). The dependence of these parameters on  $\epsilon \kappa^2 \rho_0$ is explored in the range $-0.02\leq \epsilon \kappa^2 \rho_0\leq 0.04$, finding that

\begin{eqnarray}
\frac{c_0}{\text{km}}&=& 8.21 + 60.3\times  (\epsilon \kappa^2 \rho_0) \\
\frac{c_1}{\text{km/fm}}&=& 31.0 -125.8\times  (\epsilon \kappa^2 \rho_0) \ . \end{eqnarray}
Combining these relations with Eq.(\ref{eq:RdeltaR}), one finds that

\begin{equation}
\epsilon \kappa^2 \rho_0=\frac{R_{05}-8.21-31.0 \Delta R}{60.3-125.8 \Delta R} \ ,
\end{equation}
where, recall, $R_{05}$ is measured in km while $\Delta R$ in fm. Using this expression, with observational values of $R_{05}$ and $\Delta R$ one can, in principle, determine the value of $\epsilon$. In this regard, assuming that $R_{05}$ and $\Delta R$ had $\pm 10 \%$ variances, the uncertainties in the determination of $\epsilon \kappa^2 \rho_0$ would reach $\pm 0.04$ for $R_{05}=12$ km and $\pm 0.06$ for $R_{05}=14$ km even in the case of GR. According to the best current data, the situation is even worse because $\Delta R=0.33^{+0.16}_{-0.18} $fm does not allow to constrain $\epsilon$ even if $R_{05}$ were exactly known. Given that the measurement of $R_{05}$ is expected to be more difficult than that of $\Delta R$, because $0.5 M_\odot$ is exceptionally small for a neutron star, the use of this approach to constrain the theory is really challenging. Nonetheless, there is still hope that the observation of neutron stars with $\sim 0.7 M_\odot$, for which this qualitative analysis is still valid, could be used in the future (the lowest mass of neutron stars observed so far is $(0.87\pm 0.07)M_\odot$ \cite{Rawls:2011jw}).

\subsubsection{Phase transitions}

Let us now focus our attention on phenomena taking place in the high density family of neutron stars. The potential existence of phase transitions in the nuclear matter of massive neutron stars could have more dramatic effects in Born-Infield theories of gravity than in GR due to the role that matter gradients play in this theory. The relativistic hydrostatic equilibrium equation (\ref{eq:hydroRel}) and the study of stellar pulsations put forward the appearance in the equations of terms associated with the sound speed and its first derivative, which are dependent on the first and second-order derivatives of the pressure with respect to the matter density. This, in part, motivated the use of interpolating functions to approximate the tabulated equations of state in order to avoid artificial numerical effects. However, should first-order (or second-order) phase transitions take place in the interior of neutron stars,  discontinuities in the matter density (sound speed) would occur. The potential effects of first-order phase transitions have been investigated in \cite{Sham:2013sya}.

The first thing to note in the case of first-order phase transitions, is that in the limit in which $c_s^2\to 0$, Eq.(\ref{eq:hydroRel}) behaves as $dp/dr\propto -c_s^2/\epsilon$ thus implying that $d\rho/dr=c_s^{-2}dp/dr$ is finite even when $dp/dr$ vanishes. If $\epsilon>0$,  one finds that in the region of constant pressure $d\rho/dr$ is continuous, constant, and negative, generating in that way a discontinuity in the function $p=p(\rho)$. This region of constant pressure is self-supported due to the repulsive gravity generated by the strong density gradient, in much the same way as  pressureless solutions are stabilized. The continuity of $\rho(r)$ in this scenario contrasts with the case of GR, where a discontinuity in $\rho$ is unavoidable. If $\epsilon<0$, both $dp/dr$ and $d\rho/dr$ are positive at the $c_s^2\approx 0$ region, yielding a completely different qualitative behavior. The relevant term in the hydrostatic equilibrium equation is now of the form

\begin{equation}
\frac{dp}{dr}\propto -\left[\frac{2}{\rho+p}+\frac{\kappa^2\epsilon}{16\pi}\left(\frac{3}{b^2}+\frac{1}{a^2c_s^2}\right)\right]^{-1} \ .
\end{equation}
Given that the $\rho+p$ term is positive and that the other one is negative and grows as $c_s^2\to 0$, a divergence in $dp/dr$ is unavoidable, which indicates the impossibility of having equilibrium static solutions when $\epsilon<0$.

For the $\epsilon>0$ case, one can verify that the metric functions are smooth and finite but the Ricci scalar develops a delta-type divergence due to its dependence on the radial derivative of $c_s^2$, which is discontinuous at the phase transition. The physical implications of this divergence have not been studied in detail (see the section \ref{sec:curvaEibIint} on black holes for closely related discussions), though as suggested in \cite{Kim:2013nna} they could induce a backreaction able to avoid them.
In fact, as acknowledged in \cite{Sham:2013sya}, there is no evidence whatsoever that compact stars in nature exhibit phase transitions in their interiors. As a final comment, we just note that curvature divergences of this type are common in many physical problems involving thin shells, in which a certain thick boundary is idealised in the form of a hypersurface that separates two regions \cite{Musgrave:1995ka,Dias:2010uh,Garcia:2011aa}. The delta-like divergences are expected to disappear on physical grounds once small perturbations are allowed in the density/pressure profiles.

\subsubsection{Universality relations: $f$-mode and $I$-Love-$Q$.}

Aside from the mass-radius relation in the low range band of neutron stars, other empirical relations connecting parameters of neutron stars have been proposed. In particular, a correlation between the scaled moment of inertia $I/MR^2$ and the compactness $M/R$ has been observed \cite{Bejger,Lattimer:2004nj}. Also the frequency and damping rate of the quadrupolar $f-$mode, associated to internal fluid oscillations, can be related to global properties such as $M$ and $I$ in a way that depends very weakly on the equation of state \cite{1996PhRvL..77.4134A,Andersson:1997rn,Benhar:1998au,BFG2004,Tsui:2004qd,Lau:2009bu}.  Also, the values of $M, R$, and the moment of inertia $I$ can be accurately inferred from the $f-$mode gravitational wave signals \cite{Lau:2009bu}.  More recently, a universal relation involving $Q M/J^2$ and $M/R$, where $Q$ is the spin-induced quadrupole moment and $J$ the angular momentum, has been found. Other universal relations between $I, Q$ and the tidal and rotational Love numbers $\lambda_{tid}$ and $\lambda_{rot}$ have also been discovered \cite{Yagi:2013awa,Yagi:2013bca}. These so-called $I$-Love-$Q$ relations are relevant for the understanding of gravitational wave signals in neutron stars binary mergers.

The $f-$mode universality relations and the $I$-Love-$Q$ relations of \cite{Yagi:2013awa,Yagi:2013bca} have been investigated in the context of Born-Infield gravity in \cite{Sham:2013cya}. For this purpose, the authors wrote the field equations in GR-like form following the approach of \cite{Delsate:2012ky} and computed the oscillation frequencies also in that representation. The consistency of this approach to the problem of stellar perturbations with that provided in \cite{Sham:2012qi} was also confirmed, within numerical accuracy, putting forward the usefulness of this representation of the field equations.

Let us first focus on the properties of the $f$-mode. Neutron star oscillations are damped by the emission of gravitational waves, which implies the existence of a complex part in the oscillation eigenfrequencies (quasi-normal modes), $\omega=\omega_r+i\omega_i$, with $\omega_i$ representing the damping rate of the oscillation mode. Within the frame of GR, it turns out that $\omega_r$ and $\omega_i$ of the $f$-modes (fluid oscillations) can be related to global parameters of the star according to

\begin{eqnarray}
M \omega_r&=& -0.0047+0.133\eta+0.575 \eta^2 \label{eq:omer}\\
\frac{I^2}{M^5}\omega_i &=& 0.00694-0.0256\eta^2 \label{eq:omei}\ ,
\end{eqnarray}
where the factor $\eta\equiv \sqrt{M^3/I}$ is dimensionless (in the appropriate units). These relations are more insensitive to the equation of state than previous relations where the radius $R$ was chosen as a parameter. The motivation for this choice comes from the fact that $R$ is sensitive to the low-density part of the equation of state, while the moment of inertia $I$ measures the mass distribution globally, which is more closely related to the $f$-mode oscillations of the star.  The approach of  \cite{Sham:2013cya} consisted on writing the Born-Infield field equations in GR-like form \cite{Delsate:2012ky}, solving the stellar structure equations for several nuclear matter equations of state, and then computing perturbations around the different backgrounds obtained to identify the $f$-mode frequency using well-established methods developed in GR \cite{Kokkotas:1999bd}. Considering the cases $\epsilon \kappa^2\rho_0= -0.1, 0.0, 0.1$, with $\rho_0\equiv 10^{15} g/cm^3$, for which $M$ can change up to $30\%$, it was found that the relations $M\omega_r(\eta)$ and $I^2 \omega_i/M^5(\eta)$ are essentially independent of the chosen equation of state and $\epsilon$, being in excellent agreement with  Eqs. (\ref{eq:omer}) and (\ref{eq:omei}), respectively.

\begin{figure}[h]
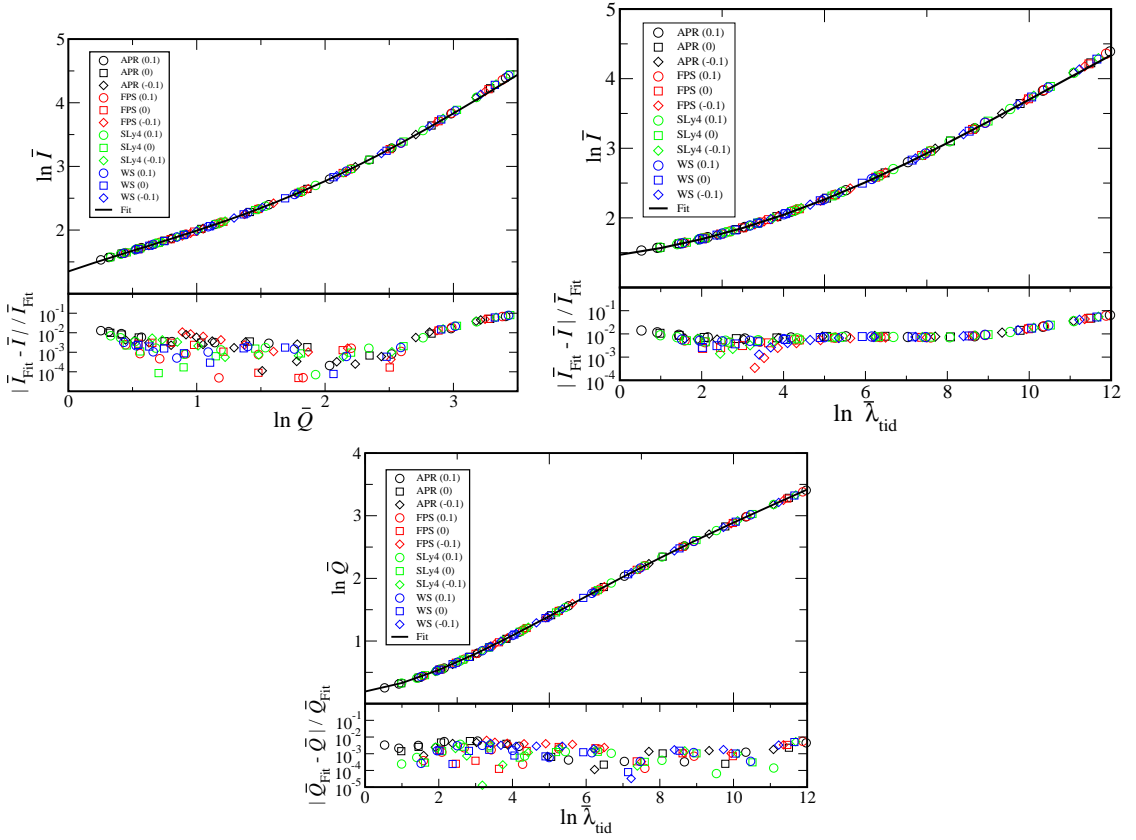

\centering
\begin{tabular}{cc}\includegraphics[width=0.45\textwidth]{I_Q_1312_1011.eps} & \includegraphics[width=0.50\textwidth]{I_Love_1312_1011.eps} \end{tabular}
\includegraphics[width=0.45\textwidth]{Q_Love_1312_1011.eps}
\caption{Universality relations between the moment of inertia, the spin-induced quadrupole moment, and the tidal Love number as shown in \cite{Sham:2013cya}. The curves are insensitive to both the EOS and the Born-Infield parameter $\epsilon$.  \label{fig:1312_1011}}
\end{figure}

The moment of inertia of a star is defined by $I\equiv J/\Omega$, where $J$ and $\Omega$ are the angular momentum and the angular velocity of the star, respectively. For a given $J$, $I$ determines how fast a star can spin and, for this reason, it is expected to be correlated with the spin-induced quadrupole moment $Q$ of the star. Interestingly, in \cite{Yagi:2013awa,Yagi:2013bca} it was found a relation between $I$ and $Q$ which is independent of the equation of state. Related to this, the (traceless) quadrupole moment induced on a neutron star by a nearby companion is determined by $Q_{ij}=-\lambda_{tid} \mathcal{E}_{ij}$, where $\mathcal{E}_{ij}$ is the tidal tensor and $\lambda_{tid}$ is the so-called Love tidal number. Though, in principle, there is no reason to expect an equation-of-state-independent relation between the variables $\bar I\equiv I/M^3$, $\bar Q\equiv -Q/(M^3\chi^2)$, and $\bar \lambda_{tid}\equiv \lambda_{tid}/M^5$, where $\chi\equiv J/M^2$, it turns out that they are related by an expression of the form

\begin{equation}
\ln y_i=a_i+b_i \ln x_i+c_i (\ln x_i)^2+d_i(\ln x_i)^3+e_i(\ln x_i)^4 \ ,
\end{equation}
where the pairs $(x_i,y_i)$ represent $(\bar \lambda_{tid},\bar I)$, $(\bar \lambda_{tid},\bar Q)$, and $(\bar Q, \bar I)$ (see Figs.\ref{fig:1312_1011}) and the coefficients  $a_i, b_i,c_i,d_i,$ and $e_i$ are constant. The numerical analysis puts forward that the $I$-Love-$Q$ relations for the EiBI theory of gravity are the same as the GR ones for the range of parameters explored, $|\epsilon \kappa^2\rho_0|\leq 0.1$ and, therefore, they cannot be used to observationally discriminate between these two theories.

%The detection of gravitational waves from neutron star binaries could break the degeneracy between the neutron stars quadrupole moment and the individual spins thanks to the $I$-Love-$Q$ relations.

For the sake of completeness, we briefly comment now on the approach of \cite{Delsate:2012ky} used to study the above universality relations.  Following our notation and manipulations, the field equations of the theory can be written in Einstein-like form as (recall Eq.(\ref{Eq:qEinsteineq}))

\begin{equation}\label{eq:GmnL}
{G^\mu}_\nu(q)=\frac{\kappa^2}{|\hat\Omega|^{1/2}}\left[{T^\mu}_\nu-\left(\mathcal{L}_G+\frac{T}{2}\right){\delta^\mu}_\nu\right] \ ,
\end{equation}
where $\mathcal{L}_G$ represents the gravity Lagrangian,
being $\mathcal{L}_G=(|\hat\Omega|^{1/2}-\lambda)/(\kappa^2\epsilon)$ in the Born-Infield theory, and  $|\hat\Omega|$ is the determinant of the deformation matrix that relates $q_{\mu\nu}$ and $g_{\mu\nu}$, as defined by Eq.(\ref{Eq:defOmega}). In the case of GR one gets $\mathcal{L}_G=R/(2\kappa^2)=-T/2$ leaving the ${T^\mu}_\nu$ term alone on the right-hand side of the equations. Note that while the Einstein tensor on the left-hand side is defined in terms of the auxiliary metric $q_{\alpha\beta}$, the matter terms  on the right-hand side (including $\mathcal{L}_G$ and $|\hat\Omega|^{1/2}$) depend on the physical metric $g_{\mu\nu}$. Since the (algebraic) relation between these two metrics depends on ${T^\mu}_\nu$, which can have some dependence on $g_{\mu\nu}$ (typically through kinetic terms), the resulting field equations may become highly nonlinear in the matter variables. The case of a perfect fluid, with ${T^\mu}_{\nu}=(\rho+p)u^\mu u_\nu +p{\delta^\mu}_\nu$, is particularly simple because the metric dependence of ${T^\mu}_{\nu}$ only appears through the covariant vector $u_\nu\equiv g_{\nu\alpha}u^\alpha$. For this matter source Eq.(\ref{eq:GmnL}) takes the explicit form

\begin{equation}\label{eq:GmnLexpl}
{G^\mu}_\nu(q)={\kappa^2}\left[\frac{(\rho+p)}{|\hat\Omega|^{1/2}}{u^\mu}u_\nu+\frac{1}{|\hat\Omega|^{1/2}}\left(\frac{\rho-p}{2}+\mathcal{L}_G\right){\delta^\mu}_\nu\right] \ ,
\end{equation}
with $|\hat\Omega|^{1/2}=(\lambda+\epsilon\kappa^2\rho)^{1/2}(\lambda-\epsilon\kappa^2p)^{3/2}$. This representation suggests a redefinition of variables such that the right-hand side of (\ref{eq:GmnLexpl}) can be interpreted as an effective perfect fluid coupled to the geometry defined by $q_{\mu\nu}$. The proposal of \cite{Delsate:2012ky} thus follows naturally, defining

\begin{eqnarray}
p_q&=& \frac{1}{|\hat\Omega|^{1/2}}\left(\frac{\rho-p}{2}+\mathcal{L}_G\right) \\
\rho_q+p_q&=& \frac{(\rho+p)}{|\hat\Omega|^{1/2}} \\
v^\mu v_\nu &= & u^\mu u_\nu \label{eq:vv=uu}\  ,
\end{eqnarray}
where $v^\mu$ is normalised using the auxiliary metric, $q_{\mu\nu} v^\mu v^\nu=-1$, and $v_\nu\equiv q_{\nu\alpha}v^\alpha$. Using Eq.(\ref{eq:vv=uu}) and the fact that in this model
$q_{\alpha\beta}=\Omega_0 g_{\alpha\beta} +\Omega_1 u_\alpha u_\beta$, with

\begin{eqnarray}
\Omega_0 &=& \sqrt{(\lambda+\epsilon\kappa^2\rho)(\lambda-\epsilon\kappa^2p)} \\
\Omega_1 &=& \epsilon \kappa^2(\rho+p)\sqrt{\frac{\lambda-\epsilon\kappa^2p}{\lambda+\epsilon\kappa^2\rho}}\ ,
\end{eqnarray}
the relation between $v^\mu$ and $u^\nu$ can be readily established.
In fact, contracting (\ref{eq:vv=uu}) with $v^\nu$, one finds $v^\mu=-(v\cdot u)u^\mu$, where $v\cdot u\equiv v^\mu u^\nu g_{\mu\nu}$ is the usual scalar product between vectors. If the contraction is done with  $u^\nu$, one finds instead $v_\nu=-u_\nu/(v\cdot u)$. Using the definition of $q_{\mu\nu}$ above to compute $u^\alpha v_\alpha=(v\cdot u)(\Omega_0-\Omega_1)$ and the relation $v_\nu=-u_\nu/(v\cdot u)$ to get the alternative expression $u^\alpha v_\alpha=1/(v\cdot u)$, one finds that $(v\cdot u)^2=1/(\Omega_0-\Omega_1)$, which completely specifies the relation between $v^\mu, v_\nu$ and $u^\mu, u_\nu$. These new variables have mapped the EiBI gravity theory into the usual Einstein equations, which can now be manipulated and solved using standard methods. The spacetime metric follows from the relation $g_{\mu\nu}=(q_{\mu\nu}-\Omega_1 u_\mu u_\nu)/\Omega_0$. This approach should, in principle, be applicable to other matter sources as well.

%The observational features of neutron stars are intimately related to their magnetic fields. The non-thermal component of the X-ray spectrum observed from many radio pulsars is believed to originate from the pulsar's magnetosphere,

% From ASTRO-PH/0206025.

\subsubsection{Magnetic fields}

Magnetic fields are thought to play an important role in supernova explosions \cite{Woosley:2006ie}, gamma-ray bursts \cite{Piran:2005qu}, soft gamma repeaters and quasi-periodic oscillations, anomalous X-ray pulsars \cite{Thompson:2001ig,Kouveliotou:1998ze}, etc, and are also fundamental to understand basic observational features of neutron stars. In particular, it is well known that the spectrum of radiation emergent from a neutron star atmosphere can significantly differ from a blackbody spectrum, and its angular distribution be far from isotropic due to the presence of magnetic fields \cite{Zavlin:2002ed}. In this sense, it is important to note that the radiation properties of neutron stars are strongly conditioned by their superficial layers \cite{Ozel:2012wu}, which can be in a gaseous state (atmosphere) or condensed state (liquid or solid) depending on surface temperature, magnetic field, and chemical composition. A condensed surface, for instance, may arise at low temperatures and very strong magnetic fields ($T\lesssim 10^5 $K and $B=10^{13}$G or $T\lesssim 10^6 $K and $B=10^{14}$G). On the other hand, the strong gravitational field on the surface layers, which is usually regarded as constant and of order $g\sim 10^{14-15}cm/s^2$, rapidly sinks the heaviest elements, leaving the lightest available ones at the surface, which will then be responsible for the radiative properties of the atmosphere, critically affecting its spectrum. A thin layer of Hydrogen of just $10^{-20}M_\odot$, for instance, is sufficient to condition the whole spectrum. This is so because magnetic fields are able to shift the ionization energy of Hydrogen up to $160$ eV if $B=10^{12}$G (or $310$ eV if $B=10^{13}$G). The intensity of magnetic fields on the neutron star outer layers is thus essential to understand the features of their radiation spectra, polarization, and thermal conductivity. The presence of magnetic fields above $B\sim 10^9-10^{10}$G, therefore, may affect the opacity of the outer layers resulting in a nonuniform surface temperature distribution, which may lead to pulsations of the thermal radiation due to rotation. At lower intensities, however, its impact on the opacity is negligible and can be safely neglected.

The effects of the Born-Infield gravitational dynamics on the magnetic fields of neutron stars have been investigated in \cite{Sotani:2015tya} focusing on the axisymmetric dipole configurations, which are expected to dominate in old neutron stars, and assuming spherically symmetric configurations. This assumption implies that the magnetic energy in the star is much smaller than the gravitational binding energy, which allows to neglect any deformation induced by the magnetic pressure. The stellar structure is thus determined by the fluid, while the magnetic field is just computed  on top of the resulting geometry. The equations governing the magnetic field follow from Maxwell's equations

\begin{eqnarray}
F_{[\mu\nu;\alpha]}&=&0 \\
\nabla_\mu F^{\mu\nu}&=& 4\pi J^\mu \ ,
\end{eqnarray}
while the coupling between the fluid and the magnetic field result from the conservation equation $\nabla_\mu T^{\mu\nu}=0$, which in the ideal magneto-hydrodynamic approximation takes the form

\begin{equation}
(\rho+p)u^\nu \nabla_\nu u_\mu+(\delta^\nu_\mu+u^\nu u_\mu)\partial_\nu p=F_{\mu\nu}J^\mu \ .
\end{equation}
With the appropriate gauge condition, $A_\mu$ can be written as $A_\mu=(0,A_r,0,A_\varphi)$, and expanding $A_\varphi$ as $A_\varphi=a_l(r)\sin\theta \partial_\theta P_l(\cos\theta)$, where $P_l(\cos\theta)$ is the Legendre polynomial of order $l$, the equation describing the dipole magnetic field ($l=1$) becomes

\begin{equation}\label{eq:a1}
a_1''+\frac{(\phi'-\lambda')}{2}a_1'+\left(\zeta^2e^{-\phi}-\frac{2}{f}\right)e^\lambda a_1=-4\pi e^\lambda j_1 \ ,
\end{equation}
where the line element (\ref{eq:ds2}) has been used, prime denotes radial derivative,  $j_1\equiv c_0 f(r)(\rho+p)$, with $c_0$ a constant, and the constant $\zeta$ is related to $A_r=\zeta e^{(\lambda-\phi)/2}a_l P_l$. The components of the magnetic field, $B_\mu=\epsilon_{\mu\nu\alpha\beta} u^\nu F^{\alpha\beta}/2$ can thus be written as

\begin{eqnarray}
B_r&=& \frac{2a_1 e^{\lambda/2}}{f}\cos\theta \\
B_\theta&=& -{a_1' e^{-\lambda/2}}\sin\theta \\
B_\varphi &=& -\zeta {a_1 e^{-\phi/2}}\sin^2\theta \ ,
\end{eqnarray}
from which it is apparent that $\zeta$ controls the strength of the toroidal magnetic field. Assuming that the exterior geometry is described by the Schwarzschild solution, the external poloidal magnetic field $(\zeta=0)$ is determined by

\begin{equation}
a_1^{(ex)}=-\frac{3\mu_b r^2}{8M^3}\left[\ln\left(1-\frac{2M}{r}\right)+\frac{2M}{r}+\frac{2M^2}{r^2}\right] \ ,
\end{equation}
where $\mu_b$ is the magnetic dipole moment at infinity. This solution sets the external boundary condition for $a_1$ and $a_1'$.
From (\ref{eq:a1}), one finds that at the centre $a_1(r)\approx \alpha_0 r^2+O(r^4)$, with $\alpha_0$ a constant. The constants $\alpha_0$ and $c_0$ (which appear in $j_1$) should be chosen so as to guarantee the continuity of $a_1$ and $a_1'$ at the surface.  The magnetic field strength can thus be written as

\begin{equation}
B\equiv (B_\mu B_\nu g^{\mu\nu})^{1/2}=f^{-1}[4 a_1^2 \cos^2\theta+a_1'^2 f e^{-\lambda}\sin^2\theta+\zeta^2a_1^2f e^{-\phi}\sin^2\theta]^{1/2} \ .
\end{equation}
At the stellar centre, one finds that $B_0=2\alpha_0 \sqrt{1+\epsilon \kappa^2\rho_c}\sqrt{1-\epsilon \kappa^2p_c}$.

The analysis of \cite{Sotani:2015tya} considered stellar models with $M=1.4 M_\odot$, a range of parameters $|\epsilon \kappa^2 \rho_s|<0.05$, with $\rho_s=2.68 \times 10^{14} g/cm^3$ representing the nuclear saturation density,  and two different realistic equations of state for nuclear matter, FPS \cite{Lorenz:1992zz} and SLy4 \cite{Douchin:2000kx}. This choice was necessary in order to compare the effects of the modified dynamics with those of different equations of state in different regions of the star. The magnetic distributions observed in the pure poloidal case, $\zeta=0$, are qualitatively the same as in GR, with deviations smaller than $10\%$ in some regions and reaching departures of less than $0.5\%$  in the crust. The mixed case, $\zeta\neq 0$, manifests some peculiar features depending on the value of $\zeta$, but roughly are also very similar to those of GR. Thus, the differences with respect to GR in the internal regions are comparable to the uncertainties due to the equation of state. The magnetic fields on the crust, however, depend very weakly on the coupling constant $\epsilon$, while properties of this region such as its thickness are very sensitive to the equation of state. It was suggested in \cite{Sotani:2015tya} that this could be used to extract information on the equation of state by exploring physical processes associated to the crust, such as stellar oscillations. However, given that stellar oscillations are also very insensitive to the Born-Infield parameter due to the universality relations discussed in  \cite{Sham:2013cya}, it seems that the magnetic field is a poor probe for this type of theories.

\subsection{Final remarks}

The use of astrophysical objects to constrain the magnitude of the non-linearity parameter in the EiBI theory of gravity has shown that with current data reasonable bounds can be placed on the theory. However, several important degeneracies arise which make it difficult to distinguish the theory from GR or discriminate its effects from those coming from the matter sector. The exploration of other Born-Infeld inspired theories in these scenarios could help better understand whether these degeneracies are proper of the EiBI or are common to a larger family of gravity theories. For all such theories, a realistic and satisfactory modeling of the transient from the top layers of the star to the external (idealized) vacuum solution is still missing.

\section{Black Holes} \label{sec:Blackholes}

According to General Relativity (GR), a fuel-exhausted star with a mass exceeding the refined Tolman-Oppenheimer-Volkoff limit, which may raise up to $\sim 2.5 M_\odot$, depending on the equation of state for dense matter (see section \ref{Sec:RelStars}) may end up its lifetime collapsing to form a region of spacetime causally disconnected from asymptotic observers, and which is called a black hole \cite{ShapTeuk}. The three-dimensional null hypersurface marking the boundary of this region, which acts as a one-way membrane, is the \emph{event horizon}. According to the unicity theorems formulated by Israel \cite{Israel:1967za,Israel:1967wq}, Carter \cite{Carter:1971zc} and Hawking \cite{Hawking:1971vc} and others \cite{Robinson:1975bv} (together with the no-hair conjecture, see \cite{Misner:1957mt}), starting from any initial (non-necessarily symmetric) configuration the final state of the gravitational collapse corresponds to a stationary and axisymmetric object solely described in terms of three parameters: mass, charge and angular momentum, leading to the Kerr-Newman family of solutions \cite{Kerr:1963ud,Newman:1965my} (see \cite{Joshi} for a review on gravitational collapse). With the recent detection of gravitational waves ascribed to black hole merger processes by the LIGO collaboration \cite{Abbott:2016blz}\footnote{Indeed, the existence of gravitational waves was already indirectly hinted by the observations of the Hulse-Taylor binary pulsar \cite{Hulse:1974eb} and others.}, which is added to the classical observations from compact X-ray sources (with Cygnus X-1 as the first historical and most influential example \cite{Orosz:2011np}), the astrophysics of compact objects has entered into a golden era, where GR can be tested with an unprecedent precision in new regimes \cite{TheLIGOScientific:2016src}.

Black holes have been and still are a very active area of research as they pose a number of challenges to our comprehension of gravitational interaction. These problems are of different kinds. First, it has been convincingly established in the literature that, if one assumes the validity of the Einstein's equations all the way down to the innermost region of a black hole, a spacetime singularity unavoidably develops \cite{Penrose:1964wq}. Moreover, this result is not due to an artifact of an excessively simplified modelling, but instead grounded on some physically reasonable restrictions \cite{Senovilla:2014gza}  (spacetime singularities and non-singular black holes in the context of Born-Infeld inspired modifications of gravity will be extensively discussed in section \ref{sec:Regular}). To avoid the breakdown of predictability and determinism, Penrose introduced the cosmic censorship conjecture \cite{Penrose:1969pc}, by which it is assumed that an event horizon covering the singularity is always developed during the gravitational collapse process, and thus a naked singularity cannot be seen from external observers.  Second, there is a tension between the classical description of gravitational phenomena provided by GR and the fundamental tenet of quantum mechanics, namely, unitarity, as given by the apparent disappearance of information inside a black hole, known as the black hole \emph{information loss problem} \cite{Hawking:1976ra,Marolf:2017jkr}. On the other hand, the very connection between Hawking's radiation \cite{Hawking:1974sw} and standard thermodynamic systems still calls for an understanding in terms of hypothetical black hole microscopic degrees of freedom, and the controversy about the potential existence of firewalls at the event horizon still goes on \cite{Almheiri:2012rt}. Finally there are apparent counterexamples of solutions with hair when adding the new ingredient of superradiance \cite{Herdeiro:2014goa} (see \cite{Herdeiro:2015waa} for a recent review), with related intensive searches for observational discriminations from the Kerr solution \cite{Cardoso:2016ryw}.

As black holes allow to test the strong field limit of GR, determining the deviations of black hole solutions from the Kerr one of GR and comparing them with astrophysical observations has become a major test on the viability of any modification of gravity\footnote{See Berti et.al. \cite{Berti:2015itd} for an overview on experimental constraints on the many gravitational modifications of GR proposed in the literature.}. Their study could shed new light on the understanding of all the open questions discussed above. In the context of Born-Infeld inspired theories of gravity we have already seen that the vacuum solutions, in Palatini approach, yield the same dynamics of GR with a cosmological constant term. Thus, the class of static, spherically symmetric vacuum black hole solutions of such theories is represented by the Schwarzschild one, characterized by mass $M$. In order to excite the dynamics contained in the new couplings of this theory one needs to couple it to some matter source. The available literature so far amounts to two such sources, namely, electrovacuum fields\footnote{It should be stressed that, though in astrophysically realistic situations the amount of net electric charge is negligible, its consideration for black holes may yield relevant lessons regarding gravitational physics beyond GR, in particular, on the spacetime singularities issue.} and anisotropic fluids. In this section we shall review in detail the corresponding deviations from the GR solutions and their contributions to fundamental and observational issues of black hole physics.

\subsection{Spherically symmetric solutions with matter in Born-Infeld gravities} \label{sec:ssswm}

Along the years, a number of Born-Infeld inspired actions have been considered in the literature regarding the search for spherically symmetric solutions. A quick review on some of the first proposals will prepare us to deal with the Eddington-inpired Born-Infeld gravity introduced by Ba\~nados and Ferreira \cite{Banados:2010ix}, for which most of the research on black hole physics in the literature has been carried out. Note that in the original proposal of Deser and Gibbons \cite{Deser:1998rj} the field equations of Born-Infeld gravity were derived using a purely metric variation, which results in fourth order equations of motion and presence of ghosts (see section \ref{Sec:D&G}), rendering the problem of finding exact solutions to such equations almost intractable. Nonetheless, Feigenbaum \cite{Feigenbaum:1998wy} considered the metric formulation of the four-dimensional, Class-0 action (recall the classification of theories of section \ref{Sec:EBIextensions}):

\begin{equation} \label{eq:actionFein}
\Ss{}=\int d^4x \sqrt{-g} \left[R+\beta \left( \sqrt{1-k_1 R^2 -k_2 R_{\mu\nu}R^{\mu\nu} -k_3 {R^\alpha}_{\beta\mu\nu}{R_\alpha}^{\beta\mu\nu}}-1 \right) \right]
\end{equation}
where $\beta,k_1,k_2,k_3$ are some constants. Despite the unavoidable trouble with ghosts, Feigenbaum investigated spherically symmetric solutions in the approximation $R_{\mu\nu} \simeq 0$ and with the additional simplification of taking $k_1=k_2=0$, which imposes the constraint of ${R^\alpha}_{\beta\mu\nu}{R_\alpha}^{\beta\mu\nu} \leq \frac{1}{k}$ upon the Kretschman scalar as long as $\beta \neq 0$. Given the limited physical interest of this scenario due to the ghost problem, let us just mention that Feigenbaum obtained analytical solutions (perturbatively to lowest order in $\epsilon$) under the form

\begin{eqnarray}
\text{d}s^2&=&-f^2(r)dt^2+\frac{dr^2}{h^2(r)} +r^2 d\Omega^2 \\
f(r)&=&\sqrt{1-\frac{2M}{r}}\left[1-\frac{8k^2 M^3 \beta}{r^9} \left(\frac{8r-11M}{r-2M} \right)  + \mathcal{O}\left(k^3 \beta^2 M^3 \right)  \right] \label{eq:phiFein1} \\
h(r)&=&1-\frac{8k^2 M^3\beta}{r^9} \left(\frac{36r-67M}{r-2M} \right) + \mathcal{O}\left(k^3 \beta^2 M^3 \right) \ . \label{eq:phiFein2}
\end{eqnarray}
In this expression $M$ is the total mass of spacetime, as seen from a far away observer. In the limit of negligible $\beta$ these solutions reduce to the Schwarzschild black hole of GR. When increasing the constant $\beta$ the event horizon disappears (the transition value follows from a non-trivial relation between $k$, $\beta$ and $M$ that can only be numerically determined) and a kind of ``bare mass" objects free of curvature divergences arise. We will see later that the existence of solutions without curvature divergences turns out to be a feature of other Born-Infeld inspired theories of gravity as well.

The explicit addition of matter, and the corresponding search for spherically symmetric black hole solutions, was first explored with some detail by Vollick. In \cite{Vollick:2005gc} he considered the following action:

\begin{equation} \label{eq:actionVoll1}
\Ss{}=\frac{1}{\kappa \beta} \int d^4x \left( \sqrt{ \vert g_{\mu\nu} + \beta \mathcal{R}_{\mu\nu} +\kappa \beta M_{\mu\nu} \vert } - \sqrt{ \vert g_{\mu\nu} \vert} \right)
\end{equation}
where $\beta$ is a constant (whose interpretation shall be clear later), $M_{\mu\nu}$ contains the matter contribution and the connection is taken to be symmetric.

When $M_{\mu\nu}=0$, the purely metric variation of this action (Class-0) has been considered by Feigenbaum, Freund and Pigli \cite{Feigenbaum:1997pf}, and Feigenbaum \cite{Feigenbaum:1998wy}. Working in the Palatini approach (Class-III theories), Vollick finds electrostatic, spherically symmetric solutions. In this case, one takes the matter contribution $M_{\mu\nu}=\alpha F_{\mu\nu}$, where $\alpha$ is a constant and $F_{\mu\nu}= \partial_{\mu}A_{\nu} -\partial_{\nu}A_{\mu}$ is the field strength tensor of the vector potential $A_{\mu}$. In order to obtain a system of equations that can be solved exactly Vollick assumes sufficiently weak fields and compute the field equations up to quadratic terms in the fields as

\begin{equation} \label{eq:fieldeqVol1}
G_{\mu\nu}(g)=\frac{\beta}{8} \left[g_{\mu\nu}\mathcal{R}^2-4\mathcal{R}\mathcal{R}_{\mu\nu}-2g_{\mu\nu}\mathcal{R}_{\alpha\beta}\mathcal{R}^{\alpha\beta}+8\mathcal{R}_{\mu\alpha}{\mathcal{R}^\alpha}_{\nu} \right] - \alpha^2 \kappa^2 \beta \left[ {F_\mu}^{\alpha}F_{\nu\alpha}-\frac{1}{4}g_{\mu\nu}F^{\alpha\beta}F_{\alpha\beta} \right] \ .
\end{equation}
Since the last term in brackets in this expression corresponds to the energy-momentum tensor of a Maxwell field, in order to obtain Einstein's equations to lowest order in $\beta$ one must take $\alpha^2=1/(\kappa \beta)$, which implies the positivity of $\beta$. Now let us consider (electrostatic) spherically symmetric solutions using the gauge $A_{\mu}=(\phi(r),0,0,0)$, which implies that the only non-vanishing component of the field strength tensor is $F_{tr}(r) \neq 0$. After a bit of algebra, this restriction allows to cast the field equations (\ref{eq:fieldeqVol1}) as

\begin{equation} \label{eq:GmunuVol1}
G_{\mu\nu}(g)=\kappa \left(\frac{{F_\mu}^{\alpha}F_{\nu\alpha}}{\sqrt{1-\frac{F^2}{2b^2} }}- b^2g_{\mu\nu} \left[1-\sqrt{1-\frac{F^2}{2 b^2}} \right] \right)
\end{equation}
where by convenience we have introduced a new constant as $b^2=1/(\kappa \beta)$ and $F^2=F^{\alpha\beta}F_{\alpha\beta}$ denotes the electromagnetic field invariant. Note that the contribution on the right-hand-side of these equations is formally similar to that of the energy-momentum tensor of Born-Infeld theory of electrodynamics \cite{Born:1934gh} with a reversed sign in front of second term and inside the square root. The Einstein equations (\ref{eq:GmunuVol1}) have to be compatible with the equations for the matter, which follow from variation of the action (\ref{eq:actionVoll1}) with respect to $A_{\mu}$ as $\nabla_{\mu} \left[\vert \hat{q} \vert \left( \hat{q}^{-1} \right)^{[\mu\nu]} \right]=0$, where the object $q_{\mu\nu} \equiv g_{\mu\nu}+\beta \mathcal{R}_{\mu\nu}+\sqrt{\kappa \beta} F_{\mu\nu}$. To quadratic order, and in the notation above, these equations become

\begin{equation} \label{eq:emqadVol1}
\nabla_{\mu} \left[ \frac{F^{\mu\nu}}{\sqrt{1-\frac{F^2}{2b^2} }} \right]=0
\end{equation}
which is nothing but the field equations of Born-Infeld electrodynamics with a reversed sign inside the square root. In Vollick's solutions, the mass function is given by (by convenience, we shall absorb here the factor $4\pi$ from the integration of the electromagnetic field equations as $Q \rightarrow 4\pi Q$)

\begin{equation} \label{eq:dmdrvol}
\frac{dm(r)}{dr}= b^2\left[\sqrt{r^4-Q^2/b^2}-r^2 \right]
\end{equation}
again with the reversed sign inside the square-root.

The main novelty of Vollick's reversed sign solution is that it is only defined beyond a radius $r=r_c$, where $r_c^2=\sqrt{ \vert Q \vert/b}$. The replacement of the point-like singularity of GR by a finite-size structure is a feature that will re-appear later when discussing electromagnetic geons in section \ref{sec:Geons}. In the present case, at the radius $r=r_c$ one finds that the curvature scalar behaves as

\begin{equation}
R=-8b^2 \left[1-\frac{r^4-r_c^2}{r^2\sqrt{r^4-r_c^4}} \right]
\end{equation}
and thus there is a curvature singularity, displaced here from $r=0$ to a finite radius. To find the horizons of these solutions one considers the zeros of the metric component $g_{tt}$, which can be found by solving the equation  $h(r)=r-2M+2 b^2( \vert Q \vert/b)^{3/2} \int_{r/r_c}^{\infty} \left[u^2-\sqrt{u^4-1} \right] du=0$.  A careful analysis of this equation reveals the presence of charged black holes with either two horizons, a single (degenerate) one or none (and a time-like singularity at $r=r_c$), or black holes with a single horizon and a time-like, space-like or null singularity, depending on the parameters of the solutions.

Black holes with a cosmological constant $\lambda$ can also be implemented within this framework via the Class-III action \cite{Vollick:2006qd}

\begin{equation} \label{eq:actionVoll2}
\Ss{}=\frac{1}{\kappa \beta} \int d^4x \left( \sqrt{ \vert g_{\mu\nu} + \beta \mathcal{R}_{\mu\nu} + \sqrt{\kappa \beta } F_{\mu\nu} + \beta \lambda g_{\mu\nu} \vert } -\sqrt{ \vert g_{\mu\nu} \vert } \right) \ .
\end{equation}
Now Vollick consider both electrostatic, $E(r) \equiv F_{tr}$, and magnetostic fields, $B(r) \equiv F_{\theta\varphi}$, via the two field invariants, $F^2=F_{\mu\nu}F^{\mu\nu}$ and $G^2=F_{\mu\nu}\tilde{F}^{\mu\nu}$. In analogy with the solutions above, now the Lagrangian, $\lag=\lag(F,G)$, corresponding to the energy-momentum on the right-hand-side of the gravitational field equations, is obtained as

\begin{equation} \label{eq:BINED}
\lag_{BI}=b^2\left(1-\sqrt{1-\frac{F^2}{2b^2} -\frac{G^2}{16b^4}} \right)
\end{equation}
with the same redefinitions as in the asymptotically flat case above. One can still assume a spherically symmetric line element given by Eq.(\ref{eq:ssssym}) and follow a similar procedure to solve the field equations, which yields the expression for the mass function

\begin{equation}
\frac{dm(r)}{dr}=\frac{\Lambda}{2}r^2+b^2(1+\tilde{\lambda})^{-1} \left[\sqrt{r^4-\tilde{Z}^2/b^2}-r^2 \right]
\end{equation}
where $\Lambda=\lambda \left(\frac{2+\beta \lambda}{1+\beta \lambda}\right)$ plays the role of the cosmological constant term, while we have defined $\tilde{\lambda}=\lambda/(\kappa b^2)$, $\tilde{Z}^2=\tilde{Q}^2+\tilde{p}^2$, with $\tilde{Q}=(1+\tilde{\lambda})Q$ and $\tilde{p}=(1+\tilde{\lambda})p$ being the re-scaled electric and magnetic charges, respectively. From the computation of the Ricci scalar constructed out of the spacetime metric

\begin{equation}
\mathcal{R}=-4\Lambda-8b^2 (1+\tilde{\lambda})^{-1} \left[1-\frac{r^4-r_c^4/2}{r^2\sqrt{r^4-r_c^4}} \right]
\end{equation}
it follows that a curvature singularity is still present at the finite radius $r=r_c=\sqrt{ \vert \tilde{Z} \vert/b}$. To close this part, let us briefly mention that spherically symmetric solutions were investigated in the context of $f(R)$ models with a square root (Class-IV), see \cite{Kruglov:2012ja}, but only mundane (Anti-)de Sitter solutions were found.

\subsubsection{Born-Infeld black holes in General Relativity} \label{sec:BIBHGRs}

There is a remarkable parallelism between the modifications on the structure of horizons for some of the solutions above and those of Born-Infeld electrodynamics coupled to GR. As this parallelism will re-appear later in the literature, it is instructive to consider the spherically symmetric solutions of Born-Infeld electrodynamics. In this sense, the framework of Einstein's gravity coupled to non-linear electrostatic fields has been developed to a great detail in the literature, particularly for Born-Infeld electrodynamics \cite{Salazar:1987ap, Demianski:1986wx, deOliveira:1994in, Breton:2003tk, Fernando:2003tz}. The action is written as

\begin{equation}
\Ss{}=\int d^4x \sqrt{-g} \left[\frac{R}{2\kappa^2} - \lag(F^2) \right]
\end{equation}
where the case of Born-Infeld electrodynamics is given by Eq.(\ref{eq:BINED}) with $G=0$. Due to the symmetry of the energy-momentum tensor for electrostatic solutions, ${T^t}_t={T^r}_r$, one can write a line element

\begin{equation} \label{eq:ssssym}
\text{d}s^2 =-\left(1-\frac{2m(r)}{r} \right)dt^2+\left(1-\frac{2m(r)}{r} \right)^{-1}dr^2 +r^2 d\Omega^2
\end{equation}
where $d\Omega^2=d\theta^2 + \sin^2(\theta) d\varphi^2$ is the solid angle element, and the mass function $m(r)$ is determined through the resolution of the Einstein's equations as

\begin{equation}
\frac{dm(r)}{dr}= r^2 {T^t}_t(r)= b^2\left[\sqrt{r^4+Q^2/b^2}-r^2 \right]
\end{equation}
(compare this equation with (\ref{eq:dmdrvol})). This can be explicitly integrated (with the constraint of recovering Schwarzschild black hole as $r \rightarrow \infty$) as

\begin{equation} \label{eq:massfuncBI}
m(r)=M-\frac{4\pi b^2 r}{3}\left[r^2 -\sqrt{r^4+Q^2/b^2}+\frac{2Q^2}{b^2r^2} {_{2}{F_1}} \left(\frac{1}{4},\frac{1}{2},\frac{5}{4},-\frac{Q^2}{b^2 r^4} \right) \right]
\end{equation}
where $M$ is the Schwarzschild mass. Due to the finiteness of the self-energy associated to a point-like charge in Born-Infeld electrodynamics, see Eq.(\ref{Eq:UBI2}), the behaviour of the metric component $g_{tt}$ at $r=0$, with the expressions (\ref{eq:ssssym}) and (\ref{eq:massfuncBI}), becomes there:

\begin{equation} \label{eq:metriccenterBIGR}
g_{tt}=g_{rr}^{-1}\underset{r\rightarrow 0}{\simeq}  -\left(1-8\pi bQ - \frac{2(M-\tilde{\mathcal{U}})}{r} +\mathcal{O}(r^2) \right)
\end{equation}
where $\tilde{\mathcal{U}}=4\pi^{3/2}\mathcal{U}$, with $\mathcal{U}$ defined in Eq.(\ref{Eq:UBI2}) and the factor $4\pi^{3/2}$ comes from the redefinition $Q \rightarrow 4\pi Q$ above. The zeros of $g_{tt}$ in (\ref{eq:metriccenterBIGR}) set the location of the horizons. In the (asymptotically flat) Reissner-Nordstr\"om solution of GR, such horizons are obtained as $r_{\pm}=M \pm \sqrt{M^2-Q^2}$, where the signs $\pm$ refer to the outer (event) and inner (Cauchy) horizons, respectively. For these horizons to exist, the inequality $M^2>Q^2$ has to be fulfilled (when this bound is saturated, $M^2=Q^2$, one has an extreme black hole with a degenerated horizon), otherwise one ends up into a naked singularity. In the Born-Infeld electrodynamics case, due to the finite character of $\tilde{\mathcal{U}}$\footnote{Note in passing by that finiteness of the self-energy can be achieved in other non-linear theories of electrodynamics, which indeed share most of the features regarding the structure of horizons and behaviour of curvature scalars when coupled to GR \cite{DiazAlonso:2009ak}.}, it turns out that the behaviour of the metric at the center determines the existence of three classes of configurations depending on the hierarchy between $M$ and $\mathcal{U}$. In this sense, if $M< \tilde{\mathcal{U}}$ the solutions resemble the Reissner-Nordstr\"om configurations of GR, in that two, a single (degenerate) horizon or none can be found, while for $M> \tilde{\mathcal{U}}$ a single horizon is always found, with similar features to those of the Schwarzschild black hole of GR. Finally, when $M= \tilde{\mathcal{U}}$ the metric at the center is finite and equal to $-(1-8\pi bQ)$, which consequently yields either a single horizon or none. This description is depicted in Fig.\ref{fig:effectivepotential} for a particular choice of $b=Q=1/2$. In all these cases a curvature divergence is always present at $r=0$ and this way Born-Infeld electrodynamics fails to solve the singularity problem within GR. We will see that quite a similar structure of horizons arises when considering Eddington-inspired Born-Infeld gravity in section \ref{sec:Geons}, while the issue with singularities will be reviewed in section \ref{sec:Regular}.

\begin{figure}[h]
\centering
\includegraphics[width=0.60\textwidth]{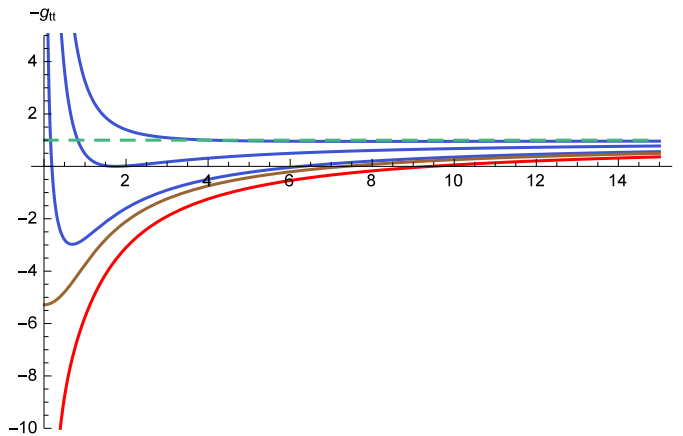}
\caption{Born-Infeld black holes in GR for $b=Q=1/2$ ($\tilde{\mathcal{U}} \simeq 3.882$). From top to bottom we find: naked singularities ($M=\tilde{\mathcal{U}}-3.5$), extreme black holes ($M\simeq \tilde{\mathcal{U}} -2.1152$), two-horizon black holes ($M=\tilde{\mathcal{U}}-0.5$), finite-metric solutions with a single horizon ($M=\tilde{\mathcal{U}}$), and black holes with a single horizon ($M=\tilde{\mathcal{U}}+1$). Note the transition between Reissner-Nordstr\"om-like configurations ($M<\tilde{\mathcal{U}}$, blue curves) to Schwarzschild-like black holes ($M>\tilde{\mathcal{U}}$, red) via the critical case $M=\tilde{\mathcal{U}}$ (brown). Solutions with $M=\tilde{\mathcal{U}}$ and no horizons are also possible. All solutions are asymptotically flat (horizontal dashed green line). \label{fig:effectivepotential}}
\end{figure}

\subsection{Eddington-inspired Born-Infeld black hole solutions}

We now turn our attention upon to the most widely employed proposal in the literature for Born-Infeld inspired modifications of gravity, and where the influential electrovacuum black hole solutions of Ba\~nados and Ferreira were found \cite{Banados:2010ix}. This proposal is defined via the action (\ref{Eq:actionEiBI}), and nowadays is usually known as Eddington-inspired Born-Infeld gravity (EiBI), which is a Class-I action (see section \ref{Sec:EBIextensions} for details on this classification). By convenience, let us write this action in the notation employed in this section as

\be
\Ss{EiBI}=\frac{1}{\kappa^2 \epsilon} \intd\left[\sqrt{-\det\Big( g_{\mu\nu}+ \epsilon \mR_{(\mu\nu)}(\Gamma)\Big)}-\lambda\sqrt{-\det g_{\mu\nu}}\right] + \Ss{M}(g_{\mu\nu},\psi_m)
\label{Eq:actionEiBIblackholes}
\ee
where $\psi_m$ denote the matter fields. A few remarks are in order: for the purpose of this section we shall assume hereafter that the (symmetric) connection $\Gamma$ is not coupled to the matter sector in the action (\ref{Eq:actionEiBIblackholes}), in agreement with Einstein's equivalent principle, that dictates that free-falling particles should follow geodesics of the background geometry $g_{\mu\nu}$ (see \ref{eq:geodesicbehaviour} for specific details). On the other hand, in vacuum, $\Ss{}_M=0$, the equation of motion for $g_{\mu\nu}$ implies $g_{\mu\nu}= \frac{\epsilon}{\lambda-1} \mathcal{R}_{(\mu\nu)}$ so that an effective cosmological constant term emerges as $\Lambda= \frac{\lambda-1}{\epsilon}$ (thus asymptotically flat solutions correspond to $\lambda=1$). This is consistent with the non-relativistic limit described in section \ref{sec:Newtonianlimit}, where post-newtonian corrections only emerge under variations on the energy density of the matter fields.

\subsubsection{Geometry and properties} \label{sec:geoandprop}

\begin{figure}[h]
\centering
\includegraphics[width=0.45\textwidth]{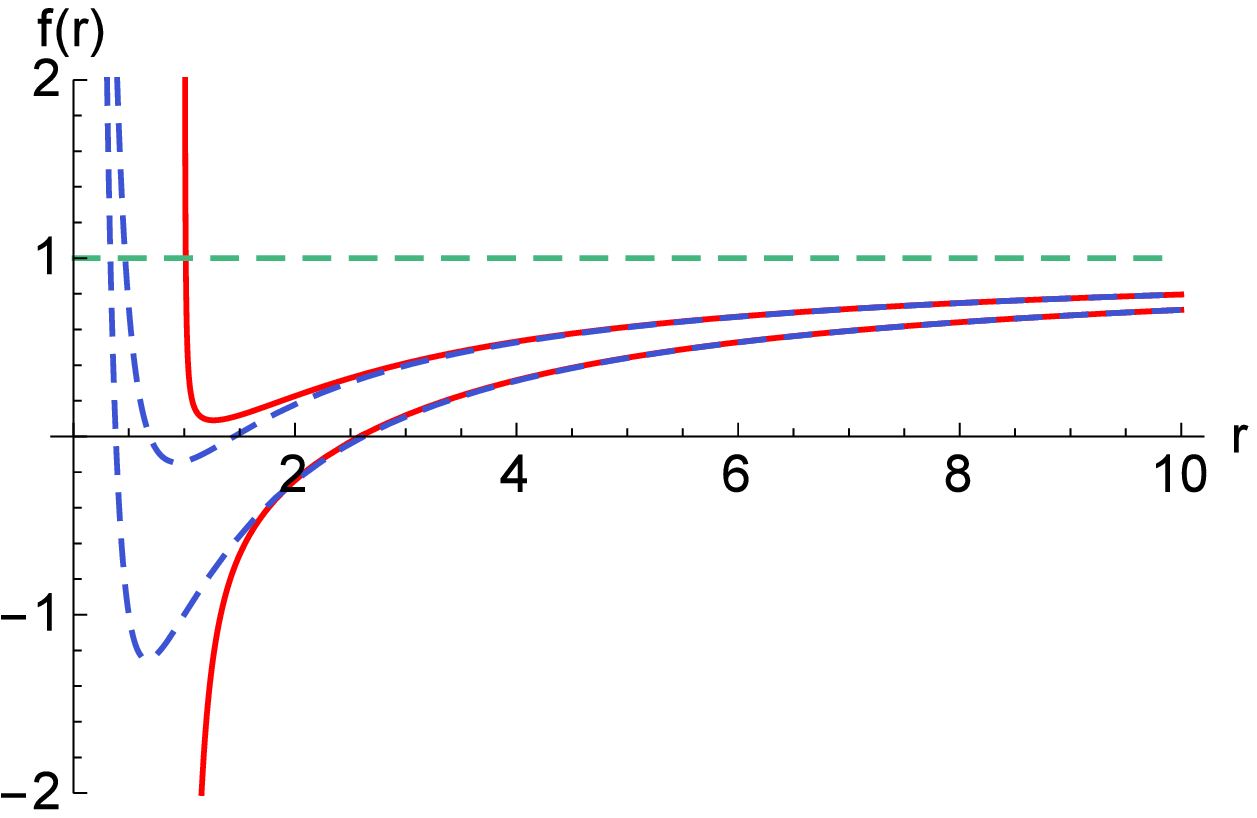}
\includegraphics[width=0.45\textwidth]{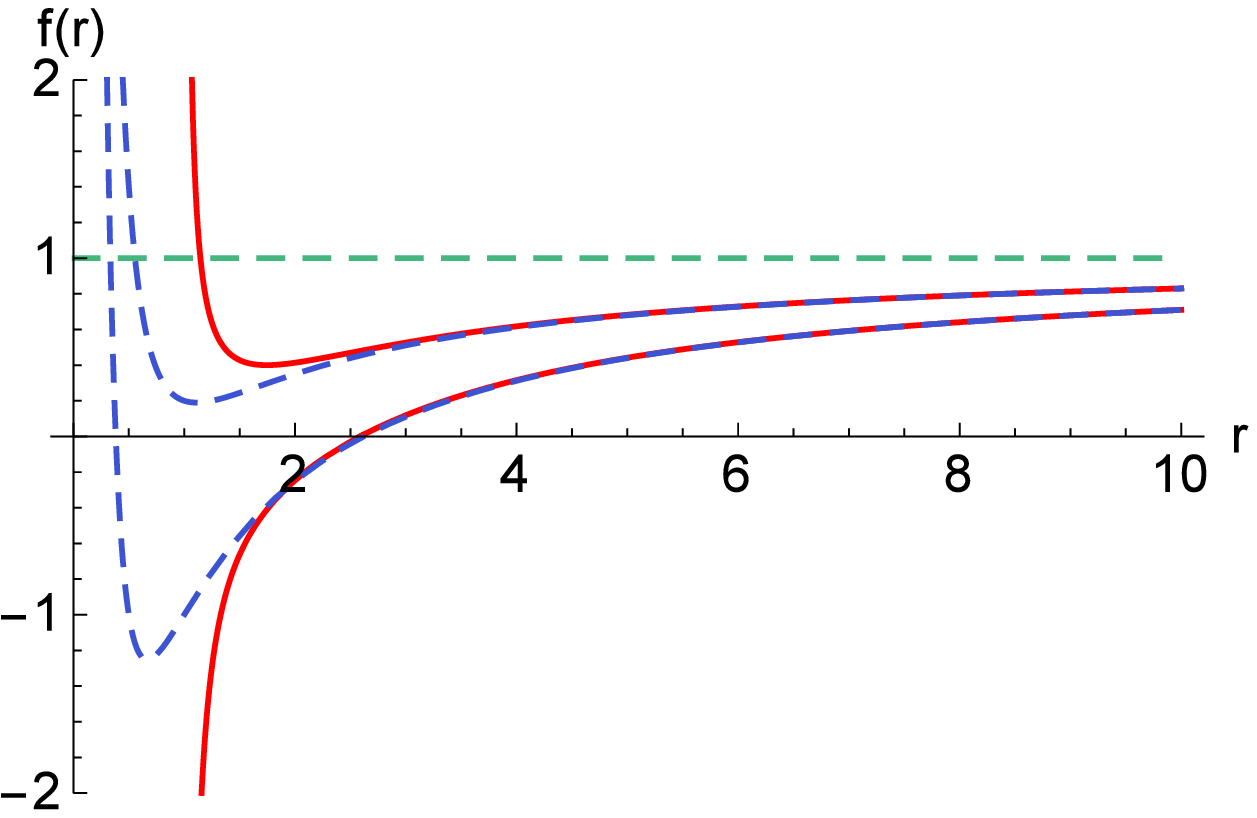}
\caption{Behaviour of the function $f(r)$ in Eq.(\ref{eq:fEiBI}) for EiBI gravity with $\epsilon>0$ (solid red curves), compared to the Reissner-Nordstr\"om black holes of GR (dashed blue curves). In these plots we fix $\epsilon=Q=1$ and vary the mass. Left plot: a Reissner-Nordstr\"om black hole with two horizons  may transform either into a naked singularity ($M=1.07$) or in a Schwarzschild-like black hole with a single horizon ($M=1.15$) in EiBI gravity. Right plot: a naked singularity in Reissner-Nordstr\"om black hole ($M=0.95$) always remains a naked singularity in EiBI gravity. Note that EiBI solutions are only defined beyond a certain radius $r>r_c$ with $r_c^2=\sqrt{\epsilon} Q$. All solutions are asymptotically flat (horizontal dashed green line). \label{fig:EiBIbh1}}
\end{figure}

Let us now study spherically symmetric configurations in EiBI gravity sourced by electrovacuum (Maxwell) fields, as given by the action

\begin{equation} \label{eq:actionMaxwell}
\Ss{}_{M} = -\frac{1}{16 \pi} \int d^4x \sqrt{-g} F_{\mu\nu}F^{\mu\nu} \ .
\end{equation}
The energy-momentum tensor for this source is written as

\begin{equation} \label{eq:Tmunuem}
T_{\mu\nu}=\frac{1}{4\pi} \left(F_{\mu\sigma}{F^\sigma}_{\nu}-\frac{1}{4} g_{\mu\nu}F_{\sigma \rho}F^{\sigma \rho} \right) \ .
\end{equation}
Bañados and Ferreira considered the spherically symmetric line element for the metric $g_{\mu\nu}$ as

\begin{equation} \label{eq:lineBFgEiBI}
\text{d}s_g^2=-\psi(r)^2 f(r)dt^2 + \frac{dr^2}{f(r)} +r^2 d \Omega^2
\end{equation}
and solved the EiBI equations for an asymptotically flat geometry, $\lambda=1$. It is instructive to consider in detail the obtention of the field equations, which is not provided in \cite{Banados:2010ix}, but derived in detail by Wei et. al. in Ref. \cite{Wei:2014dka} for arbitrary $\lambda$. This will be useful to understand the different results obtained in similar but slightly different scenarios in EiBI gravity. In many of such scenarios it is much simpler to solve the field equations for the auxiliary metric $q_{\mu\nu}$, and then transform the solution back to the spacetime metric $g_{\mu\nu}$ using Eq. (\ref{Eq:defOmega}). In the present case one proposes a line element for $q_{\mu\nu}$ as

\begin{equation} \label{eq:lineBFqEiBI}
ds_q^{2}=-G^{2}(r)F(r)dt^{2}+\frac{1}{F(r)}dr^{2}+H^{2}(r)d\Omega^2 \ .
\end{equation}
The five metric functions $\{\psi(r),f(r), G(r), F(r), H(r)\}$ are to be determined via the field equations (\ref{Eq:qequationscanonical2}) and the transformations (\ref{Eq:defOmega}). The gravitational field equations form a compatible set with the electromagnetic ones, $\partial_r(\psi^{-1}r^2E)=0$, which gives the result $E(r)=\frac{Q}{r^2} \psi(r)$, where $Q$ arises as an integration constant associated to the electric charge. Note that there is one redundant equation between the former and the latter, and consequently there are several ways to proceed. Wei et al. \cite{Wei:2014dka} choose to replace the expression for the electromagnetic field into the energy-momentum tensor (\ref{eq:Tmunuem}), and insert the result into the field equations for the auxiliary metric (\ref{Eq:qequationscanonical2}), which yields
\begin{eqnarray}
 4\frac{G'}{G}\frac{H'}{H}+2\frac{F'}{F}\frac{H'}{H}+3\frac{G'}{G}\frac{F'}{F}
   +2\frac{G''}{G}+\frac{F''}{F}
     &=&\frac{1}{\epsilon F}\bigg(\frac{1}{\lambda-\frac{\epsilon Q^{2}}{r^{4}}}-1\bigg),\label{Eieq1}\\
 4\frac{H''}{H}+2\frac{F'}{F}\frac{H'}{H}+3\frac{G'}{G}\frac{F'}{F}
   +2\frac{G''}{G}+\frac{F''}{F}
     &=&\frac{2}{\epsilon F}\bigg(\frac{1}{\lambda-\frac{\epsilon Q^{2}}{r^{4}}}-1\bigg),\label{Eieq2}\\
 -\frac{1}{H^{2}F}+\frac{F'}{F}\frac{H'}{H}+\frac{G'}{G}\frac{H'}{H}
                 +\frac{H'^{2}}{H^{2}}+\frac{H''}{H}&=&\frac{1}{\epsilon F}
                 \bigg(\frac{1}{\lambda+\frac{\epsilon Q^{2}}{r^{4}}}-1\bigg) \ , \label{Eieq3}
\end{eqnarray}
where primes stand for derivatives with respect to the radial coordinate $r$. On the other hand, the transformations (\ref{Eq:defOmega}) lead to the relations between the metric functions in each line element

\begin{equation}
G=\psi\bigg(\lambda-\frac{\epsilon Q^{2}}{r^{4}}\bigg)\hspace{0.1cm};\hspace{0.1cm} F= f\bigg(\lambda-\frac{\epsilon Q^{2}}{r^{4}}\bigg)^{-1} \hspace{0.1cm};\hspace{0.1cm} H=r\sqrt{\lambda+\frac{\epsilon Q^{2}}{r^{4}}} \ . \label{feqG}
\end{equation}
Now one just needs to solve the field equations (\ref{Eieq1}), (\ref{Eieq2}) and (\ref{Eieq3}) with the relations (\ref{feqG}), imposing the asymptotic GR limit:

\begin{equation} \label{eq:asymlimitEiBI}
\psi(r \rightarrow \infty) \rightarrow 1 \hspace{0.1cm};\hspace{0.1cm} f(r \rightarrow \infty) \rightarrow 1-\frac{2M}{r} + \frac{Q^2}{r^2} - \Lambda \frac{r^2}{3}
\end{equation}
which is nothing but the Reissner-Nordstr\"om-Anti-de Sitter solution, corresponding to the spacetime geometry outside a spherical distribution characterized by mass $M$, charge $Q$, and cosmological constant $\Lambda$. Now a bit of algebra yields the following expressions for the metric components and the electromagnetic field in the EiBI case

\begin{eqnarray}
\psi(r)&=&\frac{r^2}{\sqrt{r^4+ (\epsilon/\lambda) Q^2}} \label{eq:psiEiBI} \\
f(r) &=& \frac{r \sqrt{\epsilon  Q^2+\lambda  r^4}} {\lambda  r^4-\epsilon  Q^2}\bigg[\frac{\left(3 r^2-Q^2-(\lambda-1)r^4/\epsilon \right) \sqrt{\epsilon   Q^2+\lambda  r^4}}{3r^3}+\frac{1}{3}\sqrt{\frac{Q^{3}}{\pi\sqrt{\epsilon\lambda}}}\Gamma^{2}(1/4)\nonumber\\
  &+&\frac{4}{3} \sqrt{\frac{i Q^3}{\sqrt{\epsilon  \lambda }}} F\left(i\text{arcsinh}\left(\sqrt{\frac{i}{Q}\sqrt{\frac{\lambda}{\epsilon }}}r\right),-1\right)-2 \sqrt{\lambda} M \bigg] \label{eq:fEiBI} \\
E(r)&=&\frac{Q}{\sqrt{r^4 + (\epsilon/\lambda)  Q^2}} \label{eq:EBF}
\end{eqnarray}
where $F(\Phi,m)=\int_0^{\Phi} (1-m\sin^2 \theta)^{-1/2} d \theta$ (with $-\pi/2<\Phi<\pi/2$) is the elliptic integral of first kind. These explicit expressions were given in Ref.\cite{Wei:2014dka}, and refine that of $f(r)$ appearing under the form of an integral in Ba\~nados and Ferreira paper \cite{Banados:2010ix}, besides correcting a factor $2$ under the square root of the function $\psi(r)$ of the latter.  Regarding the horizon structure, one finds the remarkable result that, for any value of the EiBI parameter $\epsilon$, its mere presence induces a change in the causal structure of these black holes (see Fig.\ref{fig:EiBIbh1}), moving from the two-horizons description of the Reissner-Nordstr\"om solution of GR to a configuration with a single horizon (resembling the Schwarzschild solution of GR) or none, depending on the combination of parameters \cite{Wei:2014dka}.

Exploring further EiBI black holes, the expression for the electric field (\ref{eq:EBF}) bears a remarkable similarity with that obtained in Born-Infeld electromagnetism  \cite{Born:1934gh}. In the present case, despite the finiteness of the electric field everywhere, the metric functions are singular at the finite radius $r=r_c$, where $r_c^2=\sqrt{\epsilon} Q$, which may be hidden or not behind an event horizon. In Ref.\cite{Wei:2014dka} Wei et al. compute, for asymptotically flat solutions, $\lambda=1$, the following curvature scalars:

\begin{eqnarray}
R(g) &\equiv& g^{\mu\nu}R_{(\mu\nu)}(g) \propto \frac{1}{(r^4-r_c^4)^3} \\
R(g,q) &\equiv& q^{\mu\nu} g^{\mu\nu}R_{(\mu\nu)}(q)=g^{\mu\nu}(q_{\mu\nu}-g_{\mu\nu})/\epsilon=8\epsilon  \\
R(q) &\equiv& q^{\mu\nu} R_{(\mu\nu)}(q)=\frac{8(r^4+r_c^4/\sqrt{2})(r^4-r_c^4/\sqrt{2}) }{(r^4+r_c^4)(r^4-r_c^4)} \ .
\end{eqnarray}
It is thus immediately seen that the curvature scalar constructed either out of the metric $g_{\mu\nu}$ or of $q_{\mu\nu}$ blows up as the surface of radius $r=r_c$ is approached. However, no interpretation on the nature of such a surface is given, and the presence of divergences on curvature scalars could be interpreted as signal of the breakdown of the geometry and thus of the presence of a physical singularity. To overcome this point, Bañados and Ferreira argue that the geometry (\ref{eq:lineBFgEiBI}) describes just the exterior of a charged object, so a realistic model should consider the process of gravitational collapse to explore such a question in detail. Nonetheless, we shall see later when discussing non-singular solutions in section (\ref{sec:Regular}) that EiBI gravity hides some surprises regarding the singularity issue.

\subsubsection{Geodesic motion} \label{eq:geodesicbehaviour}

The new non-trivial gravitational dynamics introduced by EiBI gravity, that modifies the shape of the geometry, necessarily has its impact upon the geodesic behaviour of both null (associated to light rays) and timelike (associated to massive particles) geodesics. As already mentioned, the fact that in EiBI action (\ref{Eq:actionEiBI}) the connection does not couple directly to the matter sector, implies that Einstein's equivalence principle holds (see section \ref{Sec:Frames}). This way, the equations of motion for a geodesic curve $\gamma^{\mu}=x^{\mu}(\af)$, where $\af$ is some affine parameter, can be derived from the action

\begin{equation} \label{eq:actionparticle}
\Ss{}= \int d\af \lag   = \frac{1}{2} \int d\af \sqrt{g_{\mu\nu}\frac{dx^{\mu}}{d \af}\frac{dx^{\nu}}{d\af}}
\end{equation}
from which the geodesic equation, in a coordinate system, follows as \cite{Waldbook,Chandrabook}

\begin{equation} \label{eq:geodef}
\frac{d^2 x^\mu}{d \af^2}+\Gamma^\mu_{\alpha\beta}(g) \frac{d x^\alpha}{d \af}\frac{d x^\beta}{d \af}=0 \ ,
\end{equation}
where $\Gamma^\mu_{\alpha\beta}(g)$ is the affine connection constructed as the Christoffel symbols of the spacetime metric $g_{\mu\nu}$. Eq.(\ref{eq:geodef}) represents a set of second-order differential equations that provide a unique solution once initial conditions, $\{x^{\mu}(0),dx^{\mu}/d\af\vert_0\}$, are given. Now, replacing the line element (\ref{eq:lineBFgEiBI}) for $g_{\mu\nu}$ into the Lagrangian density of Eq.(\ref{eq:actionparticle}) one gets the result\footnote{In general, imposing a symmetry and obtaining the equations of motion do not commute. The conditions under which these two operations do commute are established by the Palais criticality theorem \cite{Palais:1979rca}.}

\begin{equation} \label{eq:geoEiBIout}
\lag=-\psi^2(r)f(r)\dot{t}^2+f(r)^{-1}\dot{r}^2 + r^2(\dot{\theta}^2 + \sin^2 (\theta) \dot{\varphi}^2)
\end{equation}
where dots denote derivatives with respect to the affine parameter $u$. From the Hamiltonian description of the system it follows that there are two conserved quantities, namely, $E=\psi(r)f(r) \dot{t}^2$ and $L=r^2 \sin(\theta) \dot{\varphi}$.  For timelike observers these quantities can be interpreted as the energy per unit mass and angular momentum per unit mass, respectively, while for null geodesics we can identify $b=L/M$ as an apparent impact parameter from asymptotic infinity. In addition, due to spherical symmetry one can assume the motion to be confined to a plane, that can be chosen to be $\theta=\pi/2$ without loss of generality. Now, the equation of the radial motion of a particle in the background geometry (\ref{eq:lineBFgEiBI}), can be deduced from Eq.(\ref{eq:geodef}) as

\begin{equation} \label{eq:radmodEibIint}
\psi^2 \left(\frac{dr}{d\af} \right)^2=E^2-V^2 \ .
\end{equation}
This is just the equation of motion of a one-dimensional particle moving in an effective potential of the form

\begin{equation} \label{eq:potEiBIint}
V(r)=\sqrt{f \psi^2\left(k+\frac{L^2}{r^2} \right)} \ ,
\end{equation}
where $\psi$ and $f$ are defined in Eqs.(\ref{eq:psiEiBI}) and (\ref{eq:fEiBI}), respectively, while the causal vector $u^{\mu}=dx^{\mu}/d\af$ satisfies $u_{\mu}u^{\mu}=-k$, with $k=0(+1)$ for null (time-like) geodesics.  Now, if one considers the circular motion of a test massive particle ($k=+1$) around an electrically charged EiBI black hole, this implies the constraint $dr/d\af=0$ which, via Eq.(\ref{eq:radmodEibIint}), yields $E=V(r)$. This orbit is realised, indeed, at the minimum of the effective potential $V(r)$. In \cite{Sotani:2014lua} Sotami and Miyamoto perform a numerical analysis of such a motion, using fixed values of $Q/M$ and $\epsilon/M^2$ and varying the ratio $L/M$, depicted in Fig.\ref{fig:effectivepotential} (left). The main result is that as the ratio $L/M$ decreases, the maximum of the effective potential decreases as well, while its minimum gets closer to the centre of the EiBI black hole, in such a way that there is a minimum bound for $L/M$ (depending on $\epsilon$), below which no minimum of the potential occurs. This bound determines the innermost stable circular orbit (ISCO), which is the minimum radius below which no stable circular orbit of a test massive particle can exist around an EiBI black hole. These results are qualitatively similar to those of GR, though the specific quantitative details depend on the particular value of the EiBI constant $\epsilon$.

\begin{figure}[h]
\centering
\includegraphics[width=0.45\textwidth]{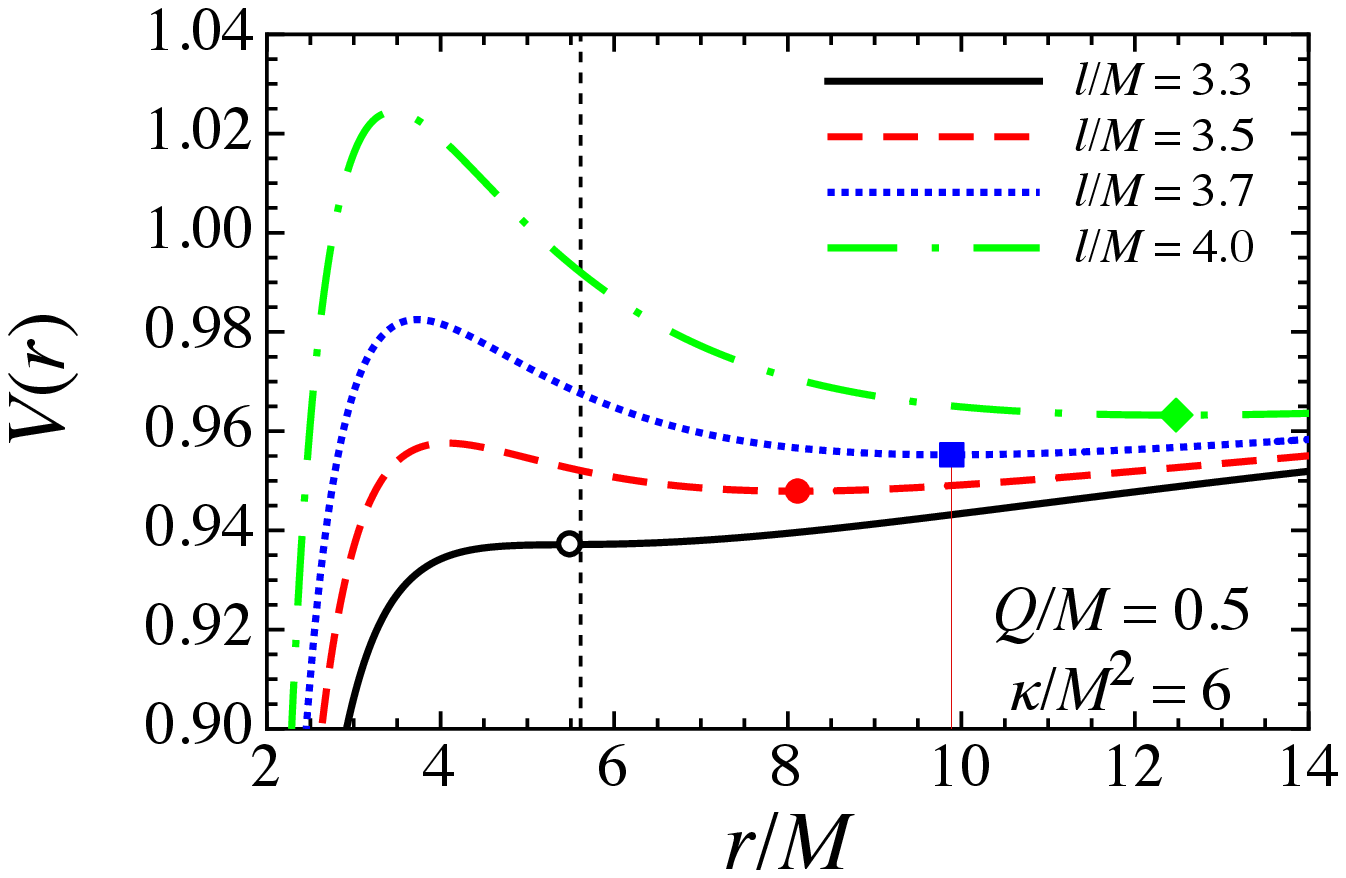}
\includegraphics[width=0.45\textwidth]{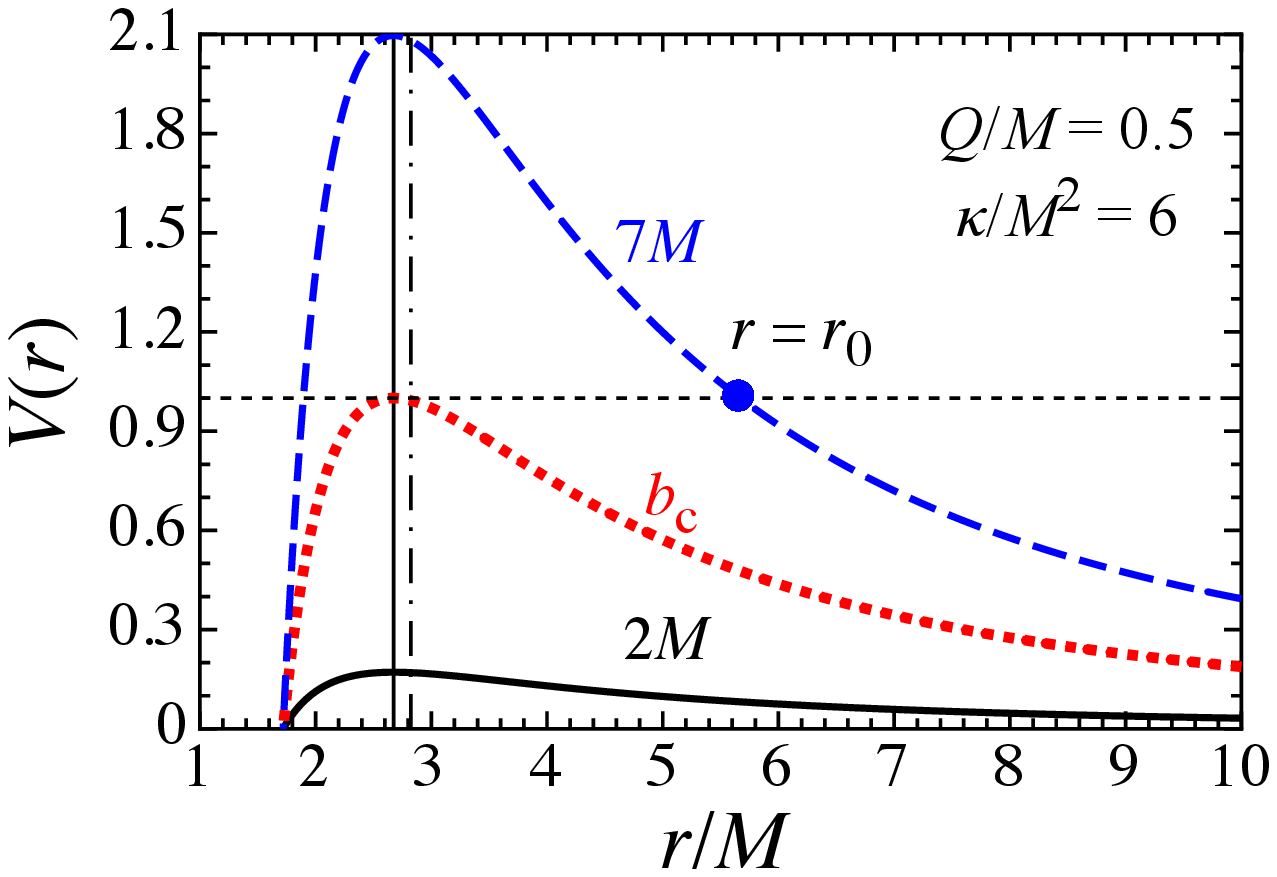}
\caption{Effective potential $V(r)$ for time-like geodesics, $k=+1$ (left) and null geodesics, $k=0$ (right) in Eq.(\ref{eq:potEiBIint}) for the choice $Q/M=0.5$ and $\epsilon/M^2=6$ (in the notation of this plot, $\epsilon \rightarrow \kappa$). Left figure: four values of the impact parameter $b \equiv L/M=3.3,3.5,3.7,4$ (in the notation of this plot, $L \rightarrow l$) have been depicted. On each of such curves the open circle corresponds to the radius of the innermost circular orbit (ISCO). Right figure: three values of the impact parameter $b=2M, b_c,7M$. On the $b=7M$ curve the photon is scattered by the black hole at $r=r_0$. Figures taken from Ref.\cite{Sotani:2014lua} and \cite{Sotani:2015ewa}, respectively. \label{fig:effectivepotential}}
\end{figure}

\subsubsection{Strong gravitational lensing} \label{sec:gravlen}

Two works \cite{Wei:2014dka,Sotani:2015ewa} have been carried out in the literature to determine the effect of the parameter $\epsilon$ of charged EiBI black holes regarding the lensing in a strong gravitational field. Gravitational lensing is indeed a powerful test to determine the nature of a compact object, which may allow to find deviations from GR predictions in the strong field regime \cite{Psaltis:2008bb}. For a massless particle, $k=0$, Eq.(\ref{eq:radmodEibIint}) can be conveniently rewritten as

\begin{equation} \label{eq:potEiBInull}
\psi^2 \left(\frac{dr}{d\af} \right)^2=1-\frac{b^2f\psi^2}{r^2}
\end{equation}
where we have redefined $\af \rightarrow \af/E$. To characterise the orbits of photons in the effective potential (\ref{eq:potEiBIint}) one first establishes the existence of the photon sphere, namely, the innermost region for a photon in orbit around a black hole, which for static, spherically symmetric spacetimes coincides with the unstable circular orbit (UCO) radius. According to the analysis carried out by Virbhadra and Ellis \cite{Virbhadra:1999nm,Claudel:2000yi,Virbhadra:2002ju}, for a line element of the form (\ref{eq:lineBFgEiBI}) this radius is simply defined by the equation $(\psi f^2)'r=2\psi^2 f$. Explicitly, for the EiBI black hole metric defined by the functions (\ref{eq:psiEiBI}) and (\ref{eq:fEiBI}), the UCO radius, $r_{UCO}$, corresponds to the solution of the equation (in units $2M=1$, which is equivalent to making dimensionless the black hole parameters as $r \rightarrow r/(2M)$, $Q \rightarrow Q/(2M)$ and $\kappa \rightarrow \kappa/(2M)^2$):

\begin{eqnarray} \label{eq:photonsphereEiBI}
  &&8 \epsilon ^{5/4} Q^2 r^2 \left(r^2-Q^2\right) \sqrt{\epsilon
   Q^2+r^4} =
   \left(3 \epsilon ^2 Q^4+2 \epsilon  Q^2 r^4+3 r^8\right)\\
   &\times&
   \bigg[-4 \sqrt{i} Q^{3/2}rF(i\text{arcsinh}(\sqrt{\frac{i}{\sqrt{\kappa}Q}}r),-1)
   -\frac{Q^{3/2} r \Gamma
   \left(\frac{1}{4}\right)^2}{\sqrt{\pi }}+\sqrt[4]{\epsilon}
   \left(3r-2 \sqrt{\epsilon  Q^2+r^4}\right)\bigg] \nonumber ,
\end{eqnarray}
which is consistent with the fact that, when the charge $Q=0$, the above equation yields the result $r_{UCO}=3/2$ (restoring units, this is the well known result $r_{UCO}=3M$), which corresponds to that of the Schwarzschild black hole. For non-vanishing $Q$, however, finding analytic solutions to (\ref{eq:photonsphereEiBI}) is highly non-trivial. One may note instead that the integration of the photon sphere equation above, and comparison with the effective potential (\ref{eq:potEiBInull}), tells us that the UCO radius $r_{UCO}$ corresponds to the solution of the equation $dV/dr=0$ with $d^2 V/dr^2<0$. This way photons will be swallowed by the black hole if $V(r_{UCO})<1$, be scattered by it at some radius $r=r_0$ if $V(r_{UCO})>1$, and move indefinitely around it in absence of perturbation if $V(r_{UCO})=1$. This can be translated into the condition $b^2 \lesseqgtr b_c^2$, where $b_c$ is a critical number that depends non-trivially on the black hole parameters and the EiBI constant. As a comparison, in the Reissner-Nordstr\"om case of GR, $\epsilon=0$, one has $r_{UCO}=3M(1+\sqrt{1-8Q^2/(9M^2)})/2$ and $b_c^2=r_{UCO}^4/[(r_{UCO}-r_{+})(r_{UCO}-r_-)]$. In Fig.\ref{fig:effectivepotential} (right) the effective potential for the choice $\epsilon/M^2=0.6$ for EiBI black holes is depicted, with the presence of the ISCO radius (marked by solid and dashed vertical lines, for EiBI and GR, respectively) and scattering radius $r_0$.  The dependence of the UCO radius $r_{UCO}$ at fixed charge with the EiBI constant can also be studied numerically, with the result that it monotonically decreases with increasing $\epsilon$ \cite{Sotani:2015ewa,Wei:2014dka}, meaning that it is harder to capture a photon by the EiBI black hole than in the Reissner-Nordstr\"om black hole of GR.

\begin{figure}[h]
\centering
\includegraphics[width=0.45\textwidth]{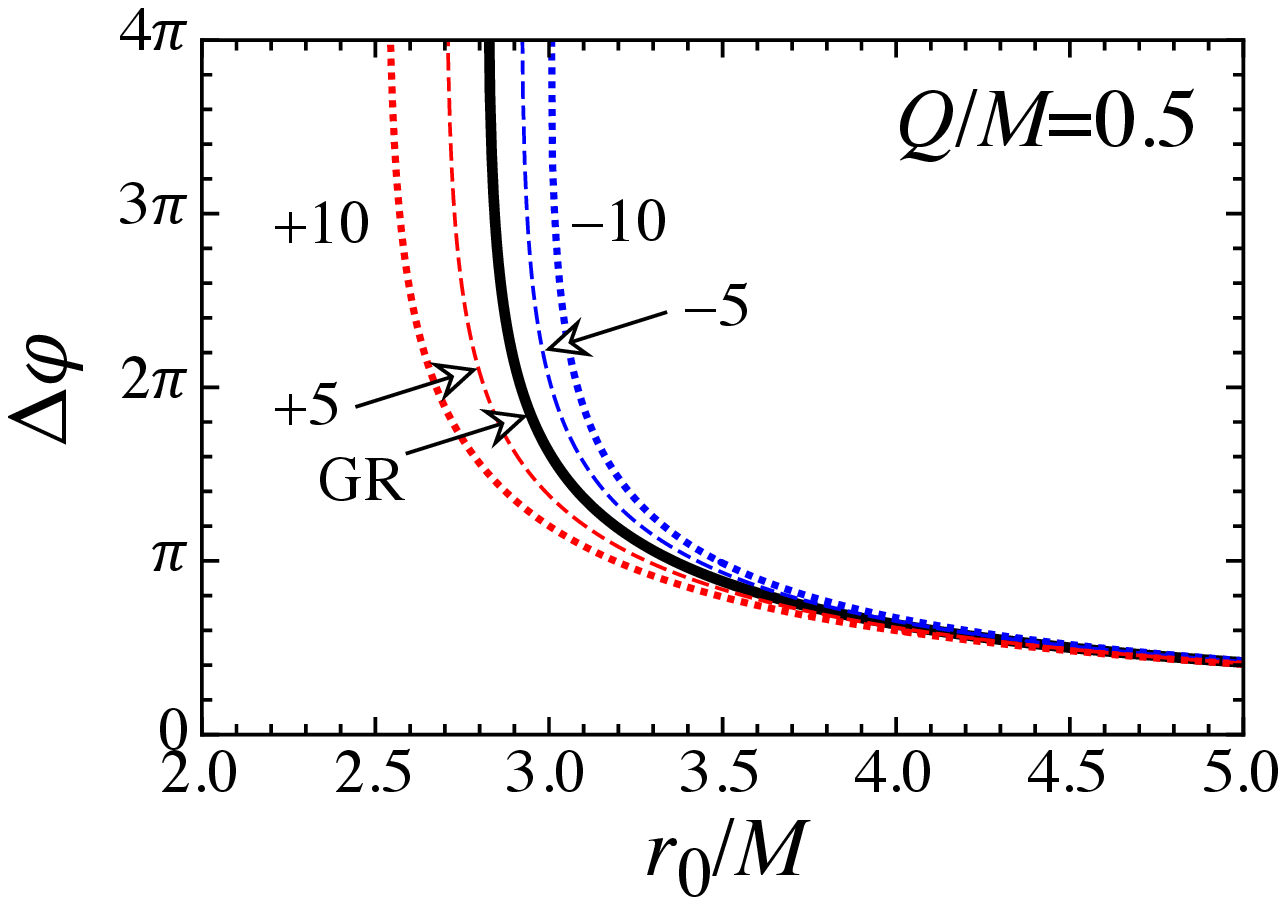}
\includegraphics[width=0.45\textwidth]{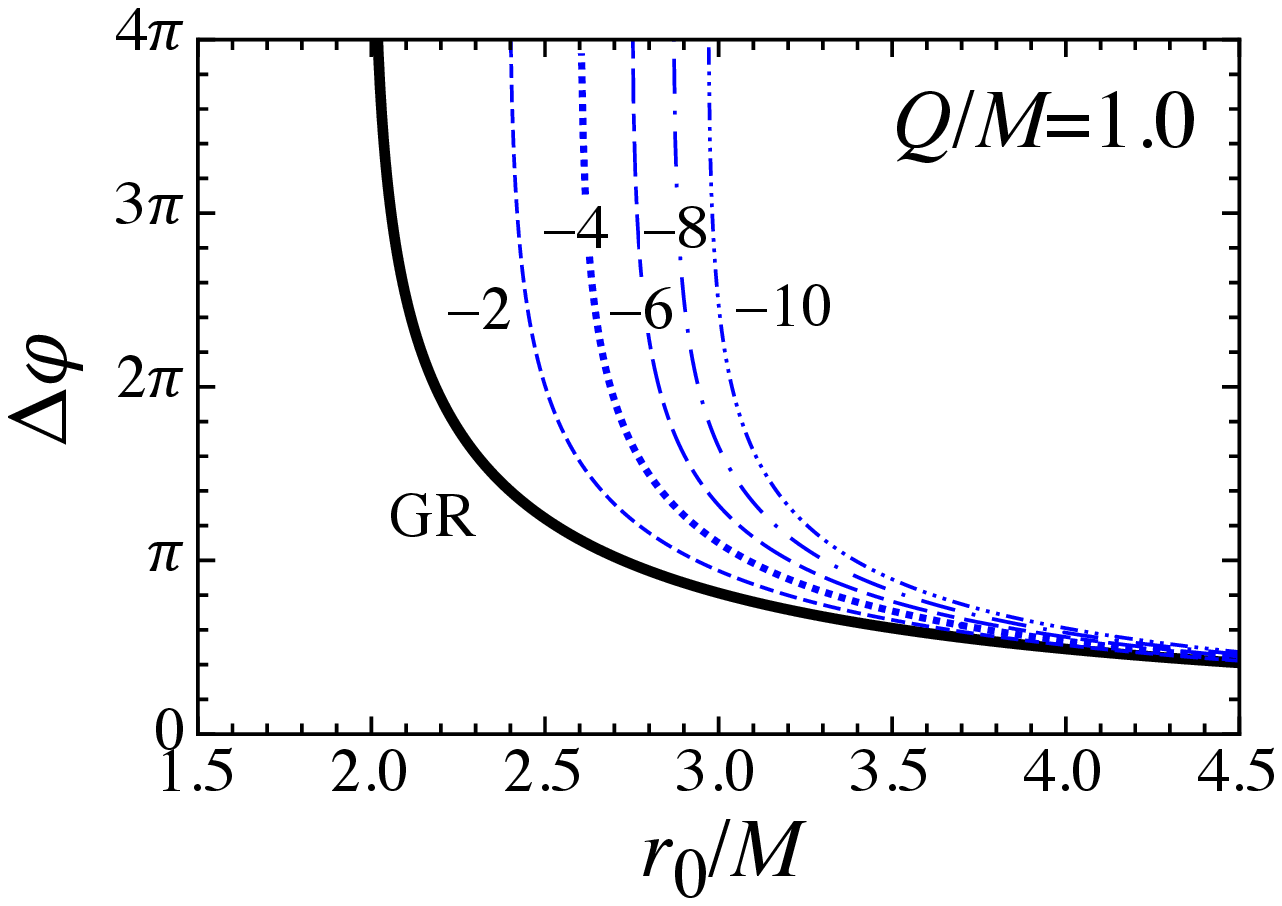}
\caption{Deflection angle $\Delta\varphi$ in Eq.\ref{eq:deflecang} for fixed charge $Q/M=0.5$ (left) and $Q/M=1.0$ (right) for the EiBI black hole as a function of $r_0/M$ for different values of $\epsilon/M^2$ (dashed lines) as compared to the GR case (solid), corresponding to $\epsilon=0$. Figures taken from Ref.\cite{Sotani:2015ewa}. \label{fig:deflecang}}
\end{figure}

Let us now consider the scattering process of a photon by the electrically charged EiBI black hole, which can only take place for $b>b_c$. First, from Eqs.(\ref{eq:radmodEibIint}) and (\ref{eq:potEiBIint}) we obtain the equation

\begin{equation} \label{eq:nullgeoEiBIBF}
\frac{d\varphi}{dr}=\frac{b \psi}{r\sqrt{r^2-f\psi^2b^2}} \ .
\end{equation}
We assume a photon that travels from infinity, is scattered at $r=r_0$ and $\varphi=0$ (see Fig.\ref{fig:effectivepotential}, right), and returns to infinity. By construction, this turn-around point satisfies $dr/d\varphi=0$, which implies $b^2=r_0^2/(f(r_0)\psi^2(r_0))$, where the subindex $0$ means that functions are being evaluated at $r_0$. This way, the integration of (\ref{eq:nullgeoEiBIBF}) yields the result

\begin{equation}
\phi(r)-\phi(r_0)=\int_{r_0}^r \frac{b\psi}{r\sqrt{r^2-b^2 \psi^2 f}}dr \ .
\end{equation}
With this expression, the deflection angle $\alpha(r_0)$ of the photon, which is defined as $\Delta(\varphi)(r_0)=2\phi(\infty)-\pi$ \cite{Virbhadra:1998dy}, can be written for the EiBI metric as

\begin{equation} \label{eq:deflecang}
\Delta(\varphi)(r_0)= 2b\int_{r_0}^{\infty} \frac{\psi}{r\sqrt{r^2 -b^2 \psi^2 f}}dr-\pi \ .
\end{equation}
Despite the presence of a pole in the integrand of (\ref{eq:deflecang}) at $r=r_0$, this can be isolated and properly handled using the variable $z=1-r_0/r$, which finally yields a finite result (see \cite{Sotani:2015ewa} for details). This way, for the EiBI black hole the deflection angle can be numerically computed and compared to the GR solution, and the result is plotted in Fig.\ref{fig:deflecang}. There it is seen that the deflection angle increases as $r_0$ decreases. As the ratio $r_0/M$ decreases the deflection angle increases until it reaches the value $2\pi$ corresponding to the point where the massless particle completes a loop around the black hole before reaching the asymptotic observer. By decreasing further the ratio $r_0/M$ one gets subsequent values $2\pi n$ ($n$ an integer number) of the deflection angle, which means that the photon performs $n$ loops around the black hole before escaping from it. Indeed, should $r_0$ be able to reach the UCO radius $r_{UCO}$, then the deflection angle would diverge, meaning that the photon would turn indefinitely around the EiBI black hole, again, in absence of any perturbation. These light rays passing close to the UCO radius give rise to multiple images on both sides of the optical axis, called \emph{relativistic images}. The position of such images in this case depends strongly on the value of the EiBI parameter $\epsilon/M$, i.e., on the gravitational theory. Thus, this \emph{strong gravitational lensing} represents a promising scenario to experimentally test EiBI gravity in the strong field limit.

As already mentioned, when $r_0=r_{UCO}$ the integrand in (\ref{eq:deflecang}) diverges, and it has to be handled with care via the new variable $z=1-r_0/r$. In both Refs.\cite{Wei:2014dka,Sotani:2015ewa} this allows to perform the integration of (\ref{eq:deflecang}) around the region $r_0 \simeq r_{UCO}$, with the (finite) result

\begin{equation}
\Delta\varphi(b)=-a_1 \log\left(\frac{b}{b_c}-1 \right) + a_2 + O\left(b-b_c \right)^{1/2}
\end{equation}
(alternatively one can write this expression in terms of $r_{UCO}$, as it is done by Wei et al. \cite{Wei:2014dka}) where the strong deflection coefficients $a_1$ and $a_2$ depend on the EiBI parameter $\epsilon$ in a non-trivial way (see \cite{Sotani:2015ewa} for details). For fixed charge, it turns out that increasing (and positive) $\epsilon$ implies an increasing of the deflection angle as compared to the Schwarzschild black hole (see Fig.5 of Ref.\cite{Wei:2014dka}), that could be used to obtain information on $\epsilon$  using strong gravitational lensing.

\begin{figure}[h]
\centering
\includegraphics[width=0.45\textwidth]{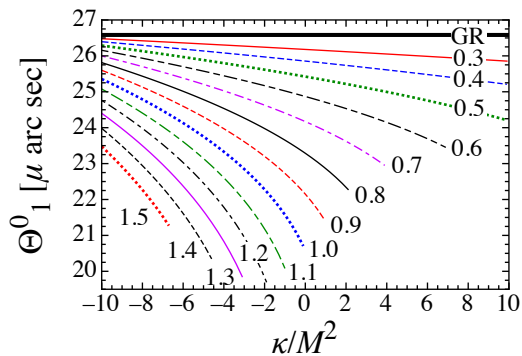}
\caption{First relativistic image in Eq.(\ref{eq:firstrelimagEiBI}) as function of the EiBI parameter $\epsilon/M^2$ (in this figure, $\epsilon \rightarrow \kappa$) for different values of the electric charge (set of curves), as compared to GR (solid thick line). Figure taken from Ref.\cite{Sotani:2015ewa}.   \label{fig:firstrelimageEiBI}}
\end{figure}

Next, to investigate the position and magnification of the relativistic images in strong gravitational lensing, one considers the lens geometry, where it is assumed that the black hole lens is located between the source and the observer, which are both required to be far away from the black hole so that the gravitational fields there are weak enough to be described by a flat metric. Under such constraints the form of the lens equation was found by Virbhadra and Ellis  \cite{Virbhadra:1999nm} as

\begin{equation}
\tan \omega =\tan \theta -\frac{D_{LS}}{D_{OS}} \left[\tan(\Delta (\varphi)-\theta)+\tan(\Theta) \right]
\end{equation}
where $\omega$ and $\Theta$ correspond to the lens/source and the lens/observer angular separation between, respectively, while $D_{LS}$ and $D_{OS}$ stand from the distance between lens and source, and observer and source, respectively. In the strong deflection limit source, lens and observer can be assumed to be highly aligned, i.e., $\omega \ll 1$ and $\Theta \ll 1$ (and $(\Delta \varphi_n - \Theta) \ll 1$, where $\Delta \varphi_n  \equiv \Delta -2\pi n$ is the deflection angle when all the loops of photons around the EiBI black hole are removed \cite{Bozza:2002zj}), and using also that in the lens geometry $b \simeq D_{OL} \Theta $ one gets the deflection angle

\begin{equation}
\Delta\varphi(\Theta)=-a_1 \log\left(\frac{D_{OL} \Theta}{b_c}-1 \right)+a_2 \ .
\end{equation}
The relativistic images correspond to $\Delta\varphi(\Theta)=2\pi n$, which yields

\begin{equation} \label{eq:firstrelimagEiBI}
\Theta_n^0=\frac{b_c}{D_{OL}}\left[1+\exp\left(\frac{a_2-2n\pi}{a_1} \right) \right]
\end{equation}
where $\Theta_n^0$ is the angle of the $n$th relativistic image. Due to the exponential contribution the first relativistic image, $\Theta_1^0$, is the brightest one, while the other are greatly demagnified. In Fig.\ref{fig:firstrelimageEiBI} the position of such an image is depicted as a function of the EiBI parameter $\epsilon/M^2$ for several values of the electric charge (set of curves) with assumed values of $D_{OL}=8.5$ kpc and $M=4.4 \times 10^6 M_{\odot}$, corresponding to the supermassive black hole at the centre of the Milky Way \cite{Genzel:2010zy}. From this figure it is clear that the deviation from the GR prediction increases with stronger EiBI coupling $\epsilon/M^2$ (in the range $\sim 3\%-5\%$ for $Q/M=0.5$ and $\vert \epsilon/M^2 \vert =10 $), which is consistent with the fact that the location of the scattering radius $r_0$ decreases as the EiBI parameter increases.

There are, in addition, other  quantities that can be constructed to be compared with astronomical observations. In order to take the simplest situation for observation, one can assume that the first relativistic image $\Theta_1^0$ can be resolved from the others, that are collectively packed at $\Theta_{\infty}^0$ \cite{Bozza:2002zj}. This way, one finds three observables: the position of the relativistic images except the first one, $\Theta_{\infty}^0$, and the two quantities

\begin{eqnarray}
s &\equiv& \Theta_1^0-\Theta_{\infty}^0= \Theta_{\infty}^0 \exp \left(\frac{a_2-2\pi}{a_1} \right) \\
R &=& \exp\left(2\pi/a_1 \right)
\end{eqnarray}
corresponding to the angular separation between the first image and all the others, and to the ratio between the flux of the first image and all the others, respectively. The latter defines a more convenient observable, $R_m=2.5 \log_{10}R$, which is the relative magnification of the images. This way, given an EiBI parameter $\epsilon$ one can numerically compute the strong deflection coefficients $a_1$ and $a_2$ and thus the three observables above. By comparing them with astronomical observations one can test the nature of black holes via gravitational lensing and, in particular, put experimental constraints on the value of the EiBI constant $\epsilon$. This has been explored, for $\epsilon>0$, by Wei et al.\cite{Wei:2014dka} by assuming that the EiBI black hole describes the supermassive black hole at the centre of our galaxy, and compare it to the description provided by Schwarzschild black hole \cite{Virbhadra:1999nm}. In table 1 of that paper, an explicit computation of these three observables for different values of $\epsilon$ has been done. The main result is that these observables fulfill the inequalities $\Theta_{\infty}^{EiBI}<\Theta_{\infty}^{RN}<\Theta_{\infty}^{Sch}$, $R^{EiBI}<R^{RN}<R^{Sch}$ and $s^{Sch}<s^{RN}<s^{EiBI}$ (for $\epsilon<0$ the inequality on $s$ do not necessarily hold for all values of $\epsilon$ \cite{Sotani:2015ewa}). For instance, the difference in the observable $\Theta_{\infty}$ between the charged EiBI black hole and the Schwarzschild black hole is of order $\sim 4$ $\mu$arcsecs, which seems to be far from the reach of current astronomical instruments \cite{Bozza:2012by}. On the other hand, the relative magnification $R_m$ may significantly deviate from the GR prediction, for instance, with the choice $Q/M=0.5$ one obtains a $5.5\%-12.7\%$ deviation with respect to GR for the EiBI parameter choice $ \epsilon/M^2=\mp 10$. This way, strong gravitational lensing can complement other techniques for testing deviations from the Kerr solution such as the measurement of the iron $K \alpha$ line observed in the X-ray fluorescence spectrum produced by the illumination of a cold accretion disk by a hot corona of (stellar-mass or supermassive) black hole candidates \cite{Jiang:2015dla, Jiang:2014loa, Bambi:2012at}.

\subsubsection{Mass inflation}

The innermost structure of black holes in the presence of accretion has been studied for decades, with the striking result first found by Israel and Poisson \cite{Poisson:1989zz,Poisson:1990eh}, and further extended by Ori \cite{Ori:1991zz} and others, that over the inner (Cauchy) horizon of a rotating black hole there occurs an exponential growth of the local Misner-Sharp mass, which in turns induces un unbounded growth of the curvature, a phenomenon known as \emph{mass inflation} (see \cite{Hamilton:2008zz} for a review on the topic). It is triggered by the relativistic counter-streaming effects between ingoing and outgoing streams, which occurs not only in the context of GR, but also in black hole solutions of other theories of gravity. In the case of EiBI gravity, this question has been investigated by Avelino \cite{Avelino:2016kkj} using electric charge instead of rotation in order to simplify the problem. The reason for this choice lies on the fact that the interior structure of a charged black hole closely resembles that of rotating black hole, where the negative pressure generated by the electric field yields a gravitational repulsion analog to that produced by the centrifugal force in a rotating black hole.

Since the inner structure of charged EiBI black holes can be drastically affected by the accretion of mass, one has to employ some simplifying assumptions in order to obtain analytic solutions. In particular, the homogeneous approximation assumes the ingoing and outgoing streams to be equal. This implies that all relevant quantities can be written as a function of a radial (timelike) coordinate, which has been shown to be useful for studying some of the most important aspects of mass inflation \cite{Hansen:2005am,Avelino:2011ee,Avelino:2009vv}. This allows to write two spherically symmetric line elements as

\begin{eqnarray}
ds_q^2&=&A(r)dt^2+B(r)dr^2 +H^2(r) d\Omega^2 \label{eq:dsqmiEiBI} \\
ds_g^2&=&g_{tt}dt^2+g_{rr}dr^2 +r^2 d\Omega^2 \label{eq:dsgmiEiBI}
\end{eqnarray}
where $A(r)$, $B(r)$, $H(r)$, $g_{tt}(r)$ and $g_{rr}(r)$ are functions of the radial coordinate $r$ alone. The total energy-momentum tensor is split into two pieces

\begin{equation} \label{eq:TmunufluidEiBI}
{T^\mu}_{\nu}= {^eT^\mu}_\nu + {^fT^\mu}_\nu
\end{equation}
where ${^eT^\mu}_\nu$ and ${^fT^\mu}_\nu$ are the electromagnetic and fluid contributions, respectively. The components of such an energy-momentum tensor can be written as

\begin{equation}
{T^r}_r=-\rho=-\rho_e-\rho_f \hspace{0.1cm};\hspace{0.1cm}
{T^t}_t=p_{\parallel}=-\rho_e + w_\parallel \rho_f \hspace{0.1cm};\hspace{0.1cm}
{T^\theta}_\theta={T^\phi}_\phi=p_{\perp}=\rho_e + w_\perp \rho_f
\end{equation}
where $\rho_e=\frac{Q^2}{8\pi r^4}$ is the electromagnetic energy density, $\rho_{f}$ the fluid energy density and the factors $\{w_\parallel,w_\perp\}$ are the fluid equations of state for the radial and tangential pressures, respectively. Since the electromagnetic and fluid contributions are assumed to be conserved independently, the conservation equation of the energy-momentum tensor of the latter can be explicitly integrated as $\rho_{f,f} =  \rho_{f,i} \left(\frac{g_{tt,i}}{g_{tt}}\right)^{(1+w_{\parallel})/2} \left(\frac{r_i}{r}\right)^{2(1+w_{\perp})}$, where the subscripts $\{i,f\}$ mean that physical quantities are evaluated at some initial and final radius, respectively. The above setup describes a charged EiBI black hole that accretes mass, the latter being described by a fluid, from an initial state which is the Reissner-Nordstr\"om solution of GR. Now, using Eqs.(\ref{Eq:defOmega}) the following relations between the metric functions in the line elements (\ref{eq:dsqmiEiBI}) and (\ref{eq:dsgmiEiBI}) are obtained

\begin{eqnarray}
A&=&g_{tt} \frac{(1+{\bar \epsilon} \rho)^{1/2} (1-{\bar \epsilon} p_{\perp})}{(1-{\bar \epsilon} p_{\parallel})^{1/2}}\,, \\
B&=&g_{rr} \frac{(1-{\bar \epsilon} p_{\parallel})^{1/2} (1-{\bar \epsilon} p_{\perp})}{(1+{\bar \epsilon} \rho)^{1/2}}\,,\\
H&=&r(1+{\bar \epsilon}\rho)^{1/4}(1+{\bar \epsilon} p_{\parallel})^{1/4},
\end{eqnarray}
where $\bar \epsilon \equiv 8\pi \epsilon$. When the fluid energy density vanishes, $\rho_f=0$, the solution reduces to the Reissner-Nordstr\"om one of GR.

\begin{figure}[h]
\centering
\includegraphics[width=0.45\textwidth]{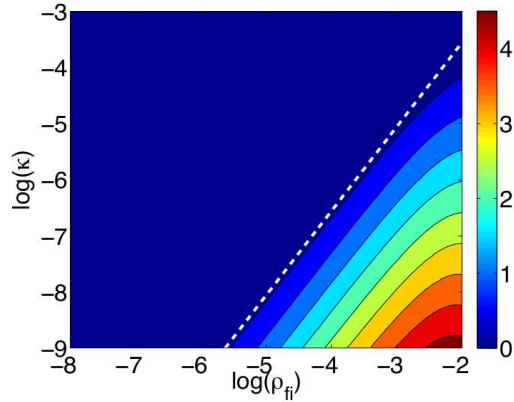}
\caption{The maximum value of Misner-Sharp mass $M_{\rm MS}$ as a function of the logarithm of the final density $\rho_{f,i}$ and the EiBI parameter $\epsilon$, assuming $\omega_{\perp}=1$ (left),
%and $\omega_{\parallel}=0$ (right),
resulting from a numerical simulation (taking $r_i \sim 0.95r_{-}$ and a Reissner-Nordstr\"om solution of GR as initial conditions) of the field equations to obtain the metric component $g_{rr \rm [end]}$. Figure taken from Ref.\cite{Avelino:2015fve}.  \label{fig:massinflationEiBI}}
\end{figure}

To obtain analytical solutions one must introduce additional constraints. In particular, Avelino \cite{Avelino:2016kkj} studies the mass inflation regime in which $w_\parallel \sim 1$ and $|{\bar \epsilon}| \rho \ll 1$, which simplifies the relations between metrics as $A=g_{tt}$, $B=g_{rr}$ and $H=r$. In addition, it is assumed that mass inflation takes place near the inner horizon, $r \sim r_{-}$, and since during this regime the energy density becomes much larger than that of the electromagnetic field (so $\rho \sim \rho_f$) one can approximate $H \sim r (1+{\bar \epsilon} \rho)^{1/4} (1- {\bar \epsilon} \rho)^{1/4} \sim r_- [1- ({\bar \epsilon} \rho/2)^2 ]$. Under these conditions, the $tt$ and $rr$ components of the field equations read

\begin{eqnarray}
-\frac{H'}{H}\frac{B'}{B}-\frac{B}{H^2}-\left(\frac{H'}{H}\right)^2+2\frac{H''}{H}&=&8\pi B {T^t}_t\,, \label{eq:MI1EiBI}\\
-\frac{H'}{H}\frac{A'}{A}+\frac{B}{H^2}-\left(\frac{H'}{H}\right)^2&=&8\pi B{T^r}_r \label{eq:MI2EiBI}\,.
\end{eqnarray}
Mass inflation takes place for $r_- {\bar \epsilon}^2 \rho |\rho'| \ll 1$, where one obtains the additional simplifications $H' \sim 1$, $A \sim  g_{tt}$, and $B \sim  g_{rr}$. This way, Eqs.(\ref{eq:MI1EiBI}) and (\ref{eq:MI2EiBI}) become approximately

\begin{equation}
\frac{g_{rr}'}{g_{rr}}\sim-8\pi r_- \rho g_{rr} \label{eq:grrMIEiBI} \hspace{0.1cm};\hspace{0.1cm} \frac{g_{tt}'}{g_{tt}}\sim-8\pi r_- \rho g_{rr} \ .
\end{equation}
Combining the last two equations and integrating the results one gets $\frac{g_{rr}}{g_{tt}}|_{\rm [MI]} \sim {\rm constant}$ where MI stands for quantities evaluated during mass inflation. This equation means that $\frac{g_{rr}}{g_{tt}}|_{\rm [start]} \sim \frac{g_{rr}}{g_{tt}}|_{\rm [end]}$ and we recall that $g_{tt[\rm start]} \sim - g_{rr[\rm start]}^{-1}$ (for the Reissner-Nordstr\"om solution of GR). Now, since mass inflation starts when the energy density of the fluid begins to dominate over the electromagnetic contribution, for the sake of finding analytical solutions one can assume $\rho_f=\alpha \rho_e$ where $\alpha$ is some constant of order unity. The combination of the above equations implies that the ratio between metric components during mass inflation satisfies

\begin{equation}
\left.\frac{g_{rr}}{g_{tt}}\right|_{\rm [MI]}\sim-\frac{\alpha^2 Q^4}{64 \pi^2 \rho_{f,i}^2 \, g_{tt,i}^2 \, r_-^{4(1-w_{\perp})} r_i^{4(1+w_{\perp})}} \ . \label{eq:ratioMIEiBI}
\end{equation}
Finally, assuming that mass inflation ends at $r_-\frac{{\bar \epsilon}^2}{2} \rho |\rho'| = \beta$, where $\beta$ is another constant of order unity, one gets the maximum energy density attained at the end of mass inflation:

\begin{equation}
\rho_{\rm [end]}\sim\frac{\beta^{1/2}}{2 \pi^{1/2}\alpha} \frac{g_{tt,i}^{1/2} r_-^{2-w_{\perp}}r_i^{1+w_{\perp}}  }{ Q^2 } \frac{\rho_{f,i}^{1/2}}{|\epsilon|}\ , \label{eq:rhoMIEiBI}
\end{equation}
which implies the presence of a threshold of energy density for mass inflation not to be triggered in EiBI gravity, i.e.,

\begin{equation}
\frac{\rho_{f,i}^{1/2}}{|\epsilon|} < \frac{\alpha^2}{4 \pi^{1/2}\beta^{1/2}}   \frac{Q^4}{g_{tt,i}^{1/2}r_-^{6-w_{\perp}} r_i^{1+w_{\perp}} }\ . \label{eq:threshold}
\end{equation}
This threshold depends on the solution's mass and charge, the accretion rate, and the EiBI parameter $\epsilon$. Note that mass inflation can always occur if the accretion rate is large enough, independently of the value of $\epsilon$. To see the effect of this threshold in the behaviour of the local mass inside a sphere of radius $r$ in the innermost region of these solutions, one considers the Misner-Sharp mass (MS), defined as  $M_{\rm MS}=\frac{r}{2}\left(1+\frac{Q^2}{r^2}-\frac{1}{g_{rr}}\right)$ \cite{Misner:1964je}, whose maximum is attained at the end of mass inflation, $g_{rr\rm [end]}$. The calculation of this mass in the present case yields the result

\begin{equation}
M_{\rm MS[end]} \sim  -\frac{r_-}{2 g_{rr[{\rm end}]}} \sim \frac{16 \pi^{3/2}\beta^{1/2}}{\alpha^3}
 \frac{g_{tt,i}^{3/2} r_-^{9-3w_{\perp}} r_i^{3+3w_{\perp}}}{Q^6} \frac{\rho_{f,i}^{3/2}}{|\epsilon|} \ . \label{eq:MSmassEiBI}
\end{equation}
These analytical calculations complement and are in agreement with the numerical analysis presented also by Avelino in \cite{Avelino:2015fve}. As depicted in Fig.\ref{fig:massinflationEiBI}, for small values of $\rho_{f,i}$ the slope of the contours indicates that the Misner-Sharp mass is a function of $\rho_{f,i}/\epsilon^{3/2}$ and that no significant mass inflation occurs below a threshold on the fluid energy density, which is fully consistent with the analytic result obtained in Eq.(\ref{eq:MSmassEiBI}). The conclusion of this analysis is that, under the restricted conditions considered in these works, in EiBI gravity there is a minimum accretion rate below which no mass inflation occurs, no matter how close the theory is to GR (which is obtained in the limit $\epsilon \rightarrow 0$). The underlying physical reason for this result still remains to be clarified.

\subsection{Wormholes} \label{sec:wormholes}

Lorentzian wormholes are geometric structures representing a shortcut or tunnel between two asymptotically flat regions of spacetime. Such a geometry, for a static spherically symmetric and \emph{traversable} (i.e. without horizons) solutions can be written as \cite{Visserbook}

\begin{equation} \label{eq:WHgeometries}
ds^2=-e^{2\Phi(r)}dt^2 + \frac{1}{1-\frac{b(r)}{r}}dr^2 + r^2 d\Omega^2
\end{equation}
where the (gravitational) redshift function $\Phi(r)$ and the (wormhole) shape function $b(r)$ characterise the geometry. In order to describe a wormhole, two charts for the two asymptotically flat regions are needed, $r \in [r_{0}, + \infty)$, where $r_{0}$ is the radius of the minimum area surface at which the two regions are joined. This defines the throat of the wormhole, for which $b(r_0)=r_0$ is fulfilled. In addition, from embedding calculations of the wormhole geometry, it follows that for the throat to be a minimum the \emph{flare-out} condition

\begin{equation} \label{eq:flareoutWH}
\frac{b(r)-b'(r)r}{b^{2}(r)} > 0 \ ,
\end{equation}
must be satisfied there by any wormhole geometry \cite{Morris:1988cz}. In GR, the flare-out condition at the wormhole throat (\ref{eq:flareoutWH}) implies the violation of the null convergence condition via Raychaudhuri equation which, for a congruence of light rays with vanishing shear and rotation, is given by (for further details see \cite{Waldbook}, chapter 9)

\begin{equation} \label{eq:RaychaudhuriGR}
\frac{d\hat{\theta}}{du}+\frac{1}{2}\hat{\theta}^2+\mathcal{R}_{\alpha\beta}\hat{u}^\alpha \hat{u}^\beta=0,
\nonumber
\end{equation}
where $\hat{u}^{\mu}$ is the four-velocity of a light ray and $\hat{\theta}$ the expansion of the congruence. In turn, via the Einstein equations, the Raychaudhuri equation entails the violation of the null energy condition \cite{Visserbook}, implying that in the context of GR wormholes are unavoidable sustained by exotic matter. However, such a restriction does not necessarily apply to extensions of GR and thus one could, in principle, obtain wormhole geometries without violations of the energy conditions. To investigate this issue in the context of EiBI gravity it is useful to write the field equations as

\begin{equation}
{G^\mu}_{\nu}={R^\mu}_{\nu}-\frac{1}{2} {\delta^\mu}_{\nu} R= \kappa^2 {S^\mu}_{\nu}
\end{equation}
where ${R^\mu}_{\nu} \equiv {R^\mu}_{\nu}(q)$ and $R={R^\mu}_{\mu}$, and the effective energy-momentum tensor ${S^\mu}_{\nu}$ is given by

\begin{equation} \label{eq:tauWHEiBI}
{S^\mu}_{\nu}=\tau {T^\mu}_{\nu} - \left( \frac{1-\tau}{\kappa^2 \epsilon} + \frac{\tau}{2} T \right) {\delta^\mu}_{\nu}
\end{equation}
with $\tau \equiv \sqrt{g/q}= \vert {\delta^\mu}_{\nu}-\kappa^2 \epsilon {T^\mu}_{\nu} \vert^{-1/2}$ and $T=g_{\mu\nu}T^{\mu\nu}$ is the trace of the energy-momentum tensor. This representation of the field equations makes clear that the effective energy-momentum tensor ${S^\mu}_{\nu}$, assumed to be exotic, could be able to sustain wormhole geometries without violations of the null energy condition on the physical energy-momentum tensor ${T^\mu}_{\nu}$.

In this section we shall consider the construction of such wormhole geometries in EiBI gravity. Consider a static spherically symmetric geometry, described by the line elements of the physical and auxiliary metrics as

\begin{eqnarray}
\text{d}s_g^2&=&-e^{\nu(r)}dt^2 + e^{\xi(r)}dr^2 + f(r)d\Omega^2 \label{eq:sgwhEiBI} \\
\text{d}s_q^2&=&-e^{\beta(r)}dt^2 + e^{\alpha(r)}dr^2 + r^2 d\Omega^2 \label{eq:sqwhEiBI}
\end{eqnarray}
where $\{\nu(r),\xi(r),f(r),\beta(r),\alpha(r)\}$ are some functions of the radial coordinate $r$. Observe that the gauge freedom has been imposed in this setup upon the line element for $q_{\mu\nu}$ in order to obtain two free functions  there, which contrast with the Ba\~nados-Ferreira geometry, where this restriction is made instead upon $g_{\mu\nu}$ (see Eqs.(\ref{eq:lineBFgEiBI}) and (\ref{eq:lineBFqEiBI}) in section \ref{sec:geoandprop}). As a matter source, let us consider an anisotropic fluid given by the energy-momentum tensor

\begin{equation} \label{eq:anisotropicfluid}
T_{\mu\nu}=(\rho+p_t)u_{\mu}u_{\nu} + p_t g_{\mu\nu} + (p_r-p_t)\chi_{\mu}\chi_{\nu}
\end{equation}
where $u^{\mu}$ is the four velocity in the metric $g_{\mu\nu}$, normalized as $u^{\mu}u^{\nu}g_{\mu\nu}=-1$,  $\chi^{\mu}$ is the unit vector in the radial direction, i.e. $\chi^{\mu}=e^{\xi/2} {\delta^\mu}_{r}$, while $\{\rho(r),p_{t}(r),p_{r}(r)\}$ are the energy density, tangential pressure (measured in the direction of $\chi^{\mu}$) and radial pressure (measured in the orthogonal direction to $\chi^{\mu}$) of the fluid, respectively. With the line element (\ref{eq:sqwhEiBI}), and assuming asymptotic flatness, $\lambda=1$, the gravitational field equations for the auxiliary metric $q_{\mu\nu}$ read \cite{Harko:2013aya}

\begin{eqnarray}
\frac{1}{ r^2} - \frac{e^{-\alpha}}{r^2} +\frac{\alpha' e^{-\alpha}}{r} &=& \frac{1}{2 \epsilon }\left( \frac{a}{hc^2} - \frac{h}{ac^2}-\frac{2}{ah} + 2  \right)  ,
     \label{eq:fieldWHEiBILobo1}  \\
- \frac{1}{ r^2}  + \frac{e^{-\alpha}}{r^2} +\frac{\beta' e^{-\alpha}}{r} &=& \frac{1}{2 \epsilon }\left( \frac{a}{hc^2} - \frac{h}{ac^2}+\frac{2}{ah} - 2  \right) \label{eq:fieldWHEiBILobo2}\\
\frac{e^{-\alpha}}{r} \left[2\beta'' r - \left(\alpha'-\beta' \right)\left(  2+\beta' r  \right)   \right]
&=& \frac{2}{\epsilon }\left( \frac{a}{hc^2} + \frac{h}{ac^2}-2  \right) , \label{eq:fieldWHEiBILobo3}
\end{eqnarray}
with the functions $a=\sqrt{1+\kappa^2 \epsilon \rho }$, $h=\sqrt{1-\kappa^2 \epsilon p_r}$, and $c=\sqrt{1-\kappa^2 \epsilon p_t}$, respectively, while we have $\tau=(ahc^2)^{-1/2}$. Like in the Ba\~nados-Ferreira solutions, two of the metric functions can be removed using the relations (\ref{Eq:defOmega}), which imply $e^{\beta } = \frac{h c^{2}}{a}e^{\nu}$, $e^{\alpha }= \frac{ac^2}{h}e^{\xi}$, and $f=\frac{r^{2}}{ah}$. In addition, from the assumption of minimal coupling of the matter to the spacetime metric, the energy-momentum tensor of the fluid satisfies the conservation equation $\nabla_{\mu}T^{\mu\nu}=0$, computed with the  covariant derivative constructed with the spacetime metric $g_{\mu\nu}$. This equation reads explicitly

\begin{equation}
\frac{d\nu }{dr}=\frac{4}{r} \frac{p_t -p_r}{\rho + p_r}  -\frac{2}{p_r+\rho }\frac{dp_r}{dr}= \frac{4}{r} \frac{h^2-c^2}{a^2-h^2}+ \frac{4d}{a^{2}-h^{2}}\frac{dh}{dr}.  \label{eq:conserWHEiBILobo}
\end{equation}
Now, the flare-out condition (\ref{eq:flareoutWH}), which can be written in this case as $\xi' e^{-\xi}<0$, together with the field equations (\ref{eq:fieldWHEiBILobo1}) and (\ref{eq:fieldWHEiBILobo2}), and the relations above between $q_{\mu\nu}$ and $g_{\mu\nu}$, imply that, for the energy conditions to be satisfied in these geometries, the inequality

\begin{equation} \label{eq:enconWHLobo}
\kappa^2 \epsilon (\rho +p_r) < \frac{\epsilon h^2}{r} \, \frac{(c^2)'}{c^2}\, \left( 1- \frac{b}{r} \right)
\end{equation}
(where we have redefined $e^{-\xi(r)}=1-b(r)/r$ to convert (\ref{eq:sgwhEiBI}) into the standard form of the wormhole geometry (\ref{eq:WHgeometries})) must be satisfied. Evaluation of this condition at the throat $b(r)=r_0$, implies that if the factor $(c^2)'/c^2$ is finite, then (\ref{eq:enconWHLobo}) is violated, which means that exotic matter is needed in order to thread these geometries, like in GR. However, if $(c^2)'/c^2$ diverges or, alternatively, $(c^2)'/c^2 e^{-\xi} \rightarrow K$ (with $K$ some constant) as $r \rightarrow r_0$, then the condition (\ref{eq:enconWHLobo}) is satisfied if $0<\rho+p_r < K$.

It is important to note that the set of field equations and relations provided so far do not constitute a closed system, since there are more independent functions than equations. Thus some restrictions have to be made. Harko et al \cite{Harko:2013aya} provide a particular wormhole geometry in this framework by introducing the equation of state

\begin{equation}
p_r(r)=\frac{\rho(r)}{1+\kappa^2 \epsilon \rho(r)} \ ,  \label{eq:radialpWHLobo}
\end{equation}
which is equivalent to choosing the restriction $a(r)h(r)=1$ on the matter components, and in turn implies $f(r)=r^2$ via the transformations between the metric $q_{\mu\nu}$ and $g_{\mu\nu}$ above. To close the system of solutions one introduces the additional constraint $\beta=0$, and upon solving of the field equations one obtains the result

\begin{equation} \label{eq:WHHarkoEiBI}
ds^2=-dt^2+ \left(\frac{1+2\epsilon r_0^2/r^4}{1-r_0^2/r^2} \right)dr^2 + r^2 d\Omega^2
\end{equation}
where $\epsilon>0$ has been assumed. This geometry describes two asymptotically flat spacetimes connected with a wormhole throat located at $r_0$, so that $r_0<r<+\infty$. Alternatively one can describe both sides of the wormhole using the radial coordinate $l$ defined as $r^2=l^2+r_0^2$, so now $-\infty < l < + \infty$ and the throat is located at $l=0$. The wormhole geometry (\ref{eq:WHHarkoEiBI}) reduces, in the GR limit $\epsilon \rightarrow 0$, to the Ellis and Bronnikov (EB) wormhole sustained by an exotic (phantom) scalar field \cite{Bronnikov:2004ax}. In the present case, the energy density, $\rho(r)=\frac{1}{\kappa^2 \epsilon} \left(\frac{1}{1+2\epsilon r_0^2/r^4}-1\right)$ is negative throughout all space, so the NEC is violated everywhere no matter the value of the EiBI constant $\epsilon$, which is an outrageous result. On other hand, the flare-out condition at the throat, $\xi' e^{-\xi}=-\frac{2r_0}{2\epsilon +r_0^2}<0$, is satisfied. It should be stressed that if $\epsilon<0$ then the flare-out condition can only be satisfied if $r_0^2>2 \vert \epsilon \vert$, which suggests a lower bound of $r_0=\sqrt{2\vert \epsilon \vert }$ for the wormhole throat in this case.

On each side of the wormhole throat $l=0$ the masses seen by an observer can be computed as \cite{Visser:2003yf}: $M^{\pm}= \pm 4\pi  \int_0^{\pm \infty} \rho r^2 \frac{dr}{dl} dl$, which in the case under consideration yields the result \cite{Tamang:2015tmd}: $M^{\pm}=\pm \frac{r_0}{2} \pm \frac{2\kappa}{20r_0} \mp \frac{5\kappa^2}{36r_0^2} + \ldots$. Thus, despite the fact that on each side of the throat an observer orbiting the wormhole would measure a mass $M^{\pm}$, the total mass $M=M^+ + M^-$ adds exactly to zero, which is a manifestation of the \emph{mass-without-mass} mechanism proposed by Wheeler \cite{Wheeler:1955zz} long ago (see section \ref{sec:Geons} for a more complete discussion of this issue).

Regarding the effects on physical observers crossing the wormhole throat, Tamang et al. \cite{Tamang:2015tmd} analyse the effect of $\epsilon$ on the tidal forces experienced by a free falling observer by considering the relative tidal acceleration, $\Delta a_j$, between two nearby parts of the observer falling into the wormhole. In an orthonormal basis $\{e_{\hat{0}},e_{\hat{1}},e_{\hat{2}},e_{\hat{3}} \}$ of the observer radially moving towards the wormhole, this acceleration is given by \cite{Morris:1988cz}

\begin{equation} \label{eq:tidalacceleration}
\Delta a_j=-\mathcal{R}_{\hat{0}\hat{j}\hat{0}\hat{p}} \xi^p
\end{equation}
where $\xi^p$ is the deviation vector between these two parts and $\mathcal{R}_{\hat{i}\hat{j}\hat{k}\hat{l}}$ are the components of the Riemann tensor. For the wormhole geometry (\ref{eq:WHHarkoEiBI}) one has \cite{Tamang:2015tmd} $\mathcal{R}_{\hat{0}\hat{j}\hat{0}\hat{p}} = \gamma/(2\epsilon_+ r_0^2) $ (where $\gamma=(1-v^2/c^2)^{-1/2}$ and $v=\pm \sqrt{\vert g_{rr}/g_{tt} \vert} dr/dt $), which is finite for any non-vanishing $\epsilon$, and thus the presence of a throat at $r_0$ may avoid the infinitely large tidal forces found in the EB black hole.

\begin{figure}[h]
\centering
\includegraphics[width=0.45\textwidth]{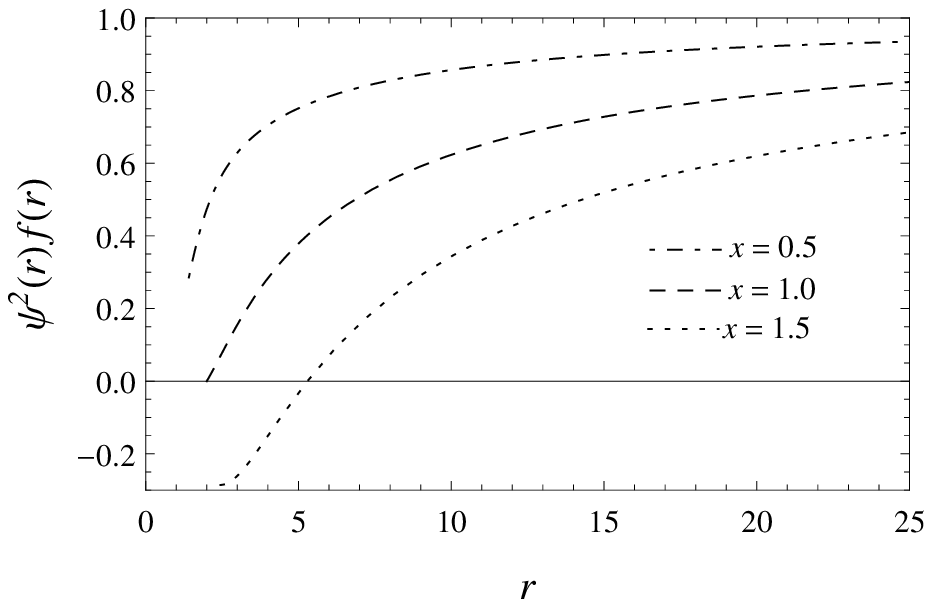}
\includegraphics[width=0.45\textwidth]{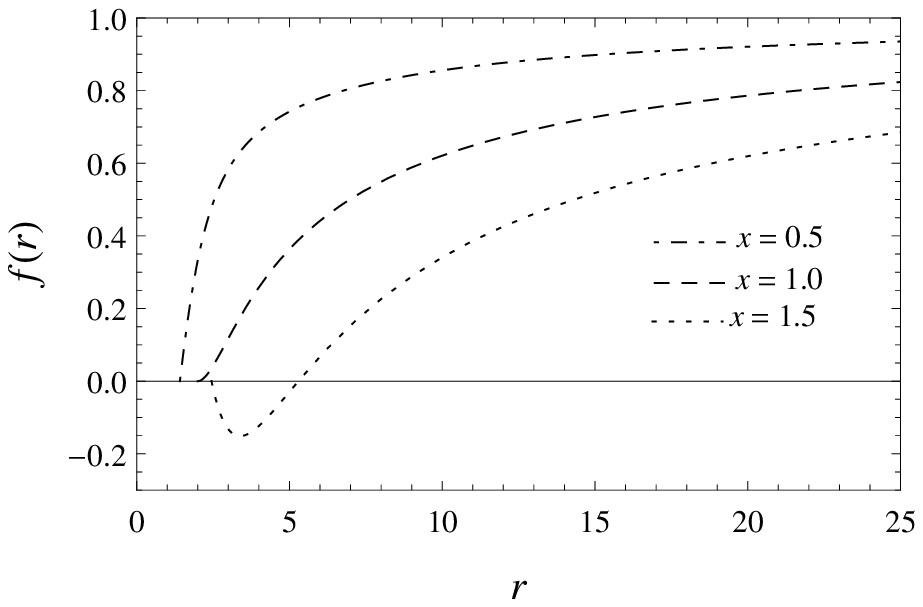}
\caption{The metric functions $g_{tt}=\psi^2(r)f(r)$ and $g_{rr}^{-1}=f(r)$ for the case $\alpha=3/4$ taking an EiBI parameter $\epsilon=-4$. In these plots, $x=r_0^2/ \vert \epsilon \vert$, in such a way that $x=1$ sets the appearance/dissappeareance of a horizon. When $x>1$, no horizon is found and the minimum value of the radial coordinate corresponds to $r_0$, where the wormhole throat is located. Figure taken from Ref.\cite{Shaikh:2015oha}. \label{fig:gttgrrShai}}
\end{figure}

Shaikh \cite{Shaikh:2015oha} also uses an anisotropic fluid (\ref{eq:anisotropicfluid}) to investigate wormhole structures within EiBI gravity, taking the equations of state $p_r=-\rho$ and $p_t=\alpha \rho$ (where $0\leq \alpha \leq 1$ in order for the energy conditions to be satisfied). This approach differs from the one of Harko et al. \cite{Harko:2013aya} in the gauge used for the line elements:

\begin{eqnarray}
\text{d}s_g^2&=&-\psi^2(r)f(r)dt^2+\frac{dr^2}{f(r)}+r^2(d\theta ^2+\sin ^2\theta d\phi ^2) \label{eq:physmetricShai} \\
\text{d}s_q^2&=&-G^2(r)F(r)dt^2+\frac{dr^2}{F(r)}+H^2(r)(d\theta ^2+\sin ^2\theta d\phi ^2).
\label{eq:auxmetricShai}
\end{eqnarray}
Integration of the conservation equation $\nabla_{\mu}T^{\mu\nu}=0$ yields the result $\rho=\frac{C_0}{r^{2(\alpha+1)}}$, where $C_0$ is a constant (of dimension $2(1-\alpha)$) whose explicit form will be determined from the asymptotic behaviour of the metric. Following the same strategy as in the previous spherically symmetric spacetimes considered in this section, the field equations (with $\lambda=1$) provide the relations between the metric functions in the line elements (\ref{eq:physmetricShai}) and (\ref{eq:auxmetricShai}) as

\begin{equation}
f(r)=F(r)(1-\bar{\epsilon}\alpha\rho), \hspace{0.2cm};\hspace{0.2cm} \psi(r)=G(r)(1-\bar{\epsilon}\alpha\rho)^{-1} \hspace{0.2cm};\hspace{0.2cm} H(r)=r\sqrt{1+\bar{\epsilon}\rho} \ ,
\end{equation}
where $\bar{\epsilon} \equiv \kappa^2 \epsilon$. With these relations, Eq.(\ref{Eq:defOmega}) can be explicitly written for this case, giving a set of three independent differential equations. Together with the fluid conservation equation, one can obtain the following solutions for the components of the line element (for $\epsilon<0$) as \cite{Shaikh:2015oha}

\begin{eqnarray}
\psi(r)&=&\left[1+\frac{r_0^{2(\alpha+1)}}{r^{2(\alpha+1)}}\right]^{-\frac{1}{2}} \label{eq:psianishai} \\
f(r)&=&\frac{1-\frac{r_0^{2(\alpha+1)}}{r^{2(\alpha+1)}}}{1+\alpha\frac{r_0^{2(\alpha+1)}}{r^{2(\alpha+1)}}}
\left[1-\frac{r_0^{2(\alpha+1)}}{3|\epsilon|r^{2\alpha}}-\frac{2M}{r\sqrt{1-\frac{r_0^{2(\alpha+1)}}{r^{2(\alpha+1)}}}}-
\frac{2(\alpha+1)r_0^{2(\alpha+1)}}{3|\epsilon|r\sqrt{1-\frac{r_0^{2(\alpha+1)}}{r^{2(\alpha+1)}}}}I(r)\right], \label{eq:fanishai}
\end{eqnarray}
where $r_0^{2(\alpha+1)} \equiv \vert \bar{\epsilon} \vert  C_0 $, the constant $-2M$ arises from imposing the asymptotically flat behavior of the metric component $f(r)$, and the function $I(r)$ satisfies

\begin{equation}
\frac{dI}{dr}= \frac{1}{r^{2\alpha}\sqrt{1-\frac{r_0^{2(\alpha+1)}}{r^{2(\alpha+1)}}}} \ .
\end{equation}
The interpretation of the radius $r_0$ is that of the minimum value of the radial coordinate, at which $\psi(r_0) \rightarrow \infty$. The reason to choose $\epsilon<0$ follows from the analysis of the Raychaudhuri equation (\ref{eq:RaychaudhuriGR}) for a radial null ray travelling towards the wormhole in the equatorial plane $\theta=\pi/2$. For the line element (\ref{eq:physmetricShai}) one has $\hat{u}^t=1/(\psi^2f)$ and $\hat{u}^r=\pm 1/\psi$ and thus the different contributions to (\ref{eq:RaychaudhuriGR}) read

\begin{equation}
\hat{\theta}=\pm \frac{2}{r}\sqrt{1+\frac{\bar{\epsilon} C_0}{r^{2(\alpha+1)}}} \hspace{0.1cm};\hspace{0.1cm} \frac{d\hat{\theta}}{d\lambda}=-\frac{2}{r^2}\left[1+\frac{(\alpha+2)\bar{\epsilon} C_0}{r^{2(\alpha+1)}}\right] \hspace{0.1cm};\hspace{0.1cm}
R_{(\alpha\beta)}\hat{u}^\alpha \hat{u}^\beta=\frac{2(\alpha+1)\bar{\epsilon} C_0}{r^{2(\alpha+2)}},
\end{equation}
with $\pm$ for ingoing (outgoing) rays. From these expressions it is clear that, since one needs to have $C_0>0$ and $0\leq \alpha \leq 1$ to satisfy the NEC, a wormhole can only exist if $\epsilon <0$. On the other hand, comparing the line element (\ref{eq:physmetricShai}) with the canonical form of a wormhole geometry (\ref{eq:WHgeometries}) it follows that $e^{2\Phi}=\psi^2 f$ and $(1-b/r)=f$, which translates the flare-out condition (\ref{eq:flareoutWH}) into $\frac{f'}{2(1-f)^2}>0$. Regarding the regularity of the spacetime one can compute curvature scalars, with the result that they generically diverge, except when the mass is tuned to the value

\begin{equation} \label{eq:massboundSai}
M=-\frac{(\alpha+1)r_0^{2(\alpha+1)}}{3|\epsilon|}I(r_0)=\frac{(\alpha+1)r_0^3}{3(2\alpha-1)|\epsilon|} {}_2F_1\left[\frac{1}{2},\frac{2\alpha-1}{2\alpha+2},\frac{4\alpha+1}{2\alpha+2};1 \right]
\end{equation}
(where the second equality is valid provided that $\alpha \neq 1/2$) for which all curvature scalars are finite. If this constraint on the mass is assumed, then the parameter $x=r_0^2/\vert \epsilon \vert$ separates those states without a horizon, $x<1$, corresponding to traversable wormholes, from black holes with horizons, $x>1$, and for which the curvature divergence is avoided. However, the mass $M$ in Eq.(\ref{eq:massboundSai}) can only be positive if $\alpha > 1/2$. In Fig.\ref{fig:gttgrrShai} the metric components for the case $\alpha=3/4$ are depicted, where this transition between traversable wormholes and regular black holes with a horizon is observed.

Following the analysis of Tamang et al. \cite{Tamang:2015tmd}, Shaikh also discusses the tidal forces upon an observer travelling through the wormhole \cite{Shaikh:2015oha}. Using the tidal acceleration equation (\ref{eq:tidalacceleration}) one can compute the components of such equation in the present case as (the subindex $0$ denotes evaluation of the quantities at the wormhole throat)

\begin{equation}
\Delta a^{\hat{1}}\big|_{r_0}=-\frac{\alpha c^2}{r_0^2}\left(1-\frac{2}{3}x\right)\xi^{\hat{1}}  \hspace{0.05cm};\hspace{0.05cm}
\Delta a^{\hat{2}}\big|_{r_0}=\frac{1}{r_0^2}\left(1-x\right)\gamma_0^2 v_0^2\xi^{\hat{2}}   \hspace{0.05cm};\hspace{0.05cm}
\Delta a^{\hat{3}}\big|_{r_0}=\frac{1}{r_0^2}\left(1-x\right)\gamma_0^2 v_0^2\xi^{\hat{3}}  \label{eq:tidalacc1}
\end{equation}
with the definition $x=r_0^2/\epsilon$ (in terms of this variable, the flare-out condition reads $\frac{f'}{2(1-f)^2}\vert_{r_0}=(1-x)/r_0>0$, which is satisfied if $x<1$, implying that the wormhole throat $r_0<\vert \epsilon \vert^{1/2}$). Restricting the acceleration felt by a traveller of typical size $\xi \sim 2m$ to be below a certain value $g$, i.e, $\Delta a^{\hat{1}'}<g$, one obtains that the minimum wormhole throat radius is

\begin{equation}
r_0^2\geq \frac{2\alpha c^2}{g}\left(1-\frac{2}{3}x\right).
\end{equation}
As $r_0$ is directly related to EiBI constant, this is translated into a maximum bound for $\epsilon$. Now, from solar physics (see section \ref{Sec:ssc}) one has the constraint \cite{Casanellas:2011kf}  $\vert \epsilon\vert/\kappa^2 \lesssim 1.8\times 10^{14}$m$^2$. Take now for instance a model with $\alpha  \gtrsim 1/2$ and $x \lesssim 1$, which implies a bound on the acceleration $|\Delta a^{\hat{1}'}|_{min}\simeq 0.17 \times 10^3$ sec$^{-2}$, or roughly $17$ times Earth's gravity $g_{E}$. However, such a wormhole would have a typical minimum size $r_{0}=\sqrt{c^2/(3g)} =\sqrt{c^2/(51g_E)} \simeq 1.34 \times 10^7$ m, which is roughly $2.1$ Earth's radius. To reduce the wormhole size, one needs to consider smaller values of $\epsilon$, which in turn implies stronger accelerations at the throat. Note that the angular components of the tidal acceleration in Eq.(\ref{eq:tidalacc1}) impose limits upon the radial velocity $v_0$ at the wormhole throat $r_0$. Let us emphasize that the general solution for $\epsilon >0$ gives a singular spacetime, while for $\epsilon <0$ solutions for which the curvature scalars are finite can be found. However, we have not discussed the implications of having solutions with finite or divergence curvature yet. As an exception to this statement, the case $\alpha=1$, for which the structure of the energy-momentum tensor (\ref{eq:anisotropicfluid}) coincides with that of a standard (Maxwell) electromagnetic field, have been derived and studied in detail, which we review thoroughly next in sections \ref{sec:Geons} and \ref{sec:Regular}.

It should be pointed out that there are several difficulties on the consistence and viability of this kind of approaches to construct wormhole geometries supported by exotic fields in the context of EiBI gravity, such as the potential instabilities at the quantum level \cite{Sushkov:2007me}, which would require to perform stability analysis in the context of this theory, something not available in the literature yet.

\subsection{Electromagnetic black holes and geons} \label{sec:Geons}

The solutions we are going to discuss now correspond to EiBI gravity (\ref{Eq:actionEiBI}) coupled to the Maxwell Lagrangian (\ref{eq:actionMaxwell}). However, they will differ from the Ba\~nados-Ferreira solutions \cite{Banados:2010ix} in that i) only the case of $\epsilon<0$ is considered (as comes from the analysis of Shaikh discussed above for the flare-out condition to be satisfied), and ii) the gauge is imposed in such a way that two independent functions are assumed for the auxiliary line element, in contrast with the line elements (\ref{eq:lineBFgEiBI}) and (\ref{eq:lineBFqEiBI}), but runs parallel with the analysis of Harko et al. \cite{Harko:2013aya} reviewed in section \ref{sec:wormholes}. This scenario was first considered in \cite{Olmo:2013gqa}. To start, the matter field equations, $\nabla_{\mu}F^{\mu\nu}=0$, for a generic static, spherically symmetric line element of the form

\begin{equation} \label{eq:metricggeonsEiBI}
\text{d}s^2=g_{tt}dt^2+g_{xx}dx^2+r(x)^2d\Omega^2
\end{equation}
and an electrostatic field, yield the only non-vanishing component of the field strength tensor $F^{tx}=\frac{Q}{r(x)^2 \sqrt{-g_{tt}g_{xx}}}$. Though this component depends explicitly on the metric components $g_{tt}$ and $g_{xx}$, the energy-momentum tensor (\ref{eq:Tmunuem}) does not, which implies

\begin{equation} \label{eq:Tmunuemfield}
{T^\mu}_{\nu}=\frac{X}{8\pi} \begin{pmatrix}
- \hat{I}_{2\times 2} & \hat{0}_{2\times 2}  \\
\hat{0}_{2\times 2} & \hat{I}_{2\times 2}
\end{pmatrix}
 \Rightarrow \vert \hat{\Omega} \vert^{1/2} ({\hat{\Omega}^{-1})^\mu}_{\nu} =
\begin{pmatrix}
 (\lambda + \tilde{X}) \hat{I}_{2\times 2} & \hat{0}_{2\times 2}  \\
\hat{0}_{2\times 2} & (\lambda-\tilde{X}) \hat{I}_{2\times 2}
\end{pmatrix}
\end{equation}
(where we have combined Eqs.(\ref{Eq:metricsimplified})  and (\ref{Eq:defOmega}) for the second equality) and hats denote matrices. Here we have introduced by convenience a new length scale as $\epsilon \rightarrow -2l_{\epsilon}^2$ (to deal with $\epsilon<0$ solutions only) and introduced the object $\tilde{X}=-\frac{l_{\epsilon} \kappa^2}{2\pi}X$. Given the structure of the right-hand-side of (\ref{eq:Tmunuemfield}), one can introduce the ansatz

\begin{equation}
\hat{\Omega}=
\begin{pmatrix}
\Omega_{+} \hat{I} & \hat{0}  \\
\hat{0} & \Omega_{-} \hat{I}
\end{pmatrix}
\Rightarrow \Omega_{-}=(\lambda+X) \hspace{0.1cm};\hspace{0.1cm} \Omega_{+}=(\lambda-X)
\end{equation}
for the $\hat{\Omega}$ matrix, where the explicit expressions of $\Omega_{-}$ and $\Omega_{+}$ follow from solving Eq.(\ref{eq:Tmunuemfield}). This way, the gravitational field equations, with the assumption of vanishing torsion and symmetric Ricci tensor \cite{Olmo:2013lta} become

\begin{equation}\label{eq:Rmn}
{R^{\mu}}_{\nu}(q)=\frac{-1}{2l_{\epsilon}^2}\left(
\begin{array}{cc}
\frac{(\Omega_{-} -1)}{\Omega_{-} }\hat{I}& \hat{0}  \\
\hat{0}& \frac{(\Omega_{+} -1)}{\Omega_{+} }\hat{I} \end{array}
\right)  \ ,
\end{equation}
where ${R^\mu}_{\nu}(q) \equiv q^{\alpha\mu}R_{(\alpha\nu)}$. At this point it should be noted that the length-squared scale $l_{\epsilon}^2$ characterises the high-curvature corrections, as follows from the expansion of the EiBI action in series of $l_{\epsilon}^2 \ll 1$:

\begin{equation}\label{eq:quadraticgravityEiBI}
\Ss{}=\frac{1}{2\kappa^2}\int d^4x \sqrt{-g}\left[\mathcal{R}-2\Lambda +l_{\epsilon}^2\left(-\frac{\mathcal{R}^2}{2}+\mathcal{R}_{(\mu\nu)}\mathcal{R}^{(\mu\nu)}\right)\right] + O(l_{\epsilon}^4) +\Ss{}_M(g_{\mu\nu},\psi_m)
\end{equation}
where $\Lambda=\frac{1-\lambda}{2l_{\epsilon}^2}$ plays the role of the effective cosmological constant. Remarkably, the field equations for the action (\ref{eq:quadraticgravityEiBI}) up to order $l_{\epsilon}^2$, turn out to be exactly the same as those of EiBI gravity in Eq.(\ref{eq:Rmn}), as can be explicitly verified from Ref.\cite{Olmo:2012nx}. The underlying reason for this result lies on the algebraic properties of the EiBI action and goes as follows: given the linear relation between ${T^\mu}_{\nu}$ and $\vert \hat{\Omega} \vert^{1/2} $, the diagonal character of ${T^\mu}_{\nu}$ will make the matrix ${P^\alpha}_{\nu} \equiv g^{\alpha \mu}\mathcal{R}_{(\mu\nu)}(\Gamma)$ to be diagonal as well. Now, if $P$ has two double eigenvalues, like happens in this case, $\hat{P}=\text{diag}(p_1,p_1,p_2,p_2)$, then the fourth-order polynomial $\vert \hat{\Omega} \vert^{1/2} ({\hat{\Omega}^{-1})^\mu}_{\nu} \vert = \vert \hat{I}+\epsilon \hat{P} \vert $ turns into the second-order polynomial appearing in Eq.(\ref{eq:quadraticgravityEiBI}) when the square root is evaluated. Moreover, this quadratic polynomial exactly coincides with the series expansion of the EiBI action. As a result, all the higher-order corrections beyond $l_{\epsilon}^2$ order cancel out, which means that the electrostatic spherically symmetric solutions of EiBI gravity exactly coincides with those obtained for the quadratic Lagrangian at order $l_{\epsilon}^2$ appearing in Eq.(\ref{eq:quadraticgravityEiBI}). Indeed, electrovacuum solutions in the context of such a quadratic gravity models (in Palatini approach) were previously found in Refs.\cite{Olmo:2011np,Olmo:2012hu,Olmo:2012nx}.

Now, to solve the field equations (\ref{eq:Rmn}) we introduce the static, spherically symmetric line element for the geometry $q_{\mu\nu}$ as

\begin{equation} \label{eq:metricgqeonsEiBI}
\text{d}s_q^2=-A(x)e^{2\psi(x)}dt^2+\frac{1}{A(x)}dx^2+\tilde{r}(x)d\Omega^2 \ .
\end{equation}
The three functions in this line element can be reduced to a single one by noting that the combination ${R^t}_t-{R^x}_x=0$ of the field equations, which follows from the symmetry ${T^t}_t={T^x}_x$ of the energy-momentum tensor (\ref{eq:Tmunuemfield}), implies that $\tilde{r}_{xx}=\psi_x\tilde{r}_x$. This allows to redefine the metric function and the radial coordinate as $A \rightarrow A/\tilde{r}_x^2$ and $\tilde{r}_x^2dx^2 \rightarrow dx^2$, respectively, so the line element can be written in the Schwarzschild-like form

\begin{equation}
\text{d}s_q^2=-B(x)dt^2+ \frac{1}{B(x)}dx^2+x^2 d\Omega^2 \ .
\end{equation}
This leaves a single independent function to be determined from the ${R^\theta}_{\theta}$ component of the field equations, which after introducing a standard mass ansatz, $B(x)=1-2M(x)/x$, reads $-4l_{\epsilon}^2 M_x=x^2(\Omega_{-}-1)/\Omega_{-}$. Resolving this equation requires comparison with the spacetime metric  Eq.(\ref{eq:metricggeonsEiBI}) using the transformations (\ref{Eq:defOmega}), which can be split into two $2 \times 2$ blocks as $q_{ab}=g_{ab}\Omega_{+}$ and $q_{mn}=g_{mn}\Omega_{-}$. The latter implies the relation of coordinates in the two line elements

\begin{equation} \label{eq:xrOm}
x^2=r^2 \Omega_{-} \rightarrow \frac{dx}{dr} =\pm \frac{\Omega_{+}}{\Omega_{-}^{1/2}}
\end{equation}
which will play an important role later. This way, the equation to be solved reads $-4l_{\epsilon}^2 M_r=r^2(\Omega_{-}-1)\Omega_{+}/\Omega_{-}^{1/2}$, whose integration can be formally written as $M(z)=M_0(1+\delta_1 G(z))$, where $M_0$ is an integration constant associated to the Schwarzschild mass, $G(z)$ contains the electromagnetic contribution, and $\delta_1$ isolates all the relevant constants out of this integration. A full solution for the spacetime line element (\ref{eq:metricggeonsEiBI}), in the asymptotically flat case, $\lambda=1$, can now be found explicitly as

\begin{equation}
\text{d}s_g^2=-A(x) dt^2  + \frac{dx^2}{A(x)\Omega_{+}^2}+r^2(x) d\Omega^2 \label{eq:solgeonsEiBI}
\end{equation}
with the expressions

\begin{eqnarray}
A(x)&=& \frac{1}{\Omega_+}\left[1-\frac{r_S}{ r(x)  }\frac{(1+\delta_1 G(r(x)))}{\Omega_-^{1/2}}\right] \label{eq:AgeonEiBI} \\
\delta_1&=& \frac{1}{2r_S}\sqrt{\frac{r_Q^3}{l_\epsilon}} \label{eq:deltageonEiBI} \\
\Omega_\pm&=&1\pm \frac{r_c^4}{r^4(x)} \label{eq:omegageonEiBI} \\
r^2(x)&=& \frac{x^2+\sqrt{x^4+4r_c^4}}{2} \label{eq:rxgeonsEiBI}
\end{eqnarray}
where $r_c=\sqrt{r_Ql_{\epsilon}}$, with $r_Q^2=\kappa^2 Q^2/(4\pi)$ a length scale associated to the electric charge which, together with the Schwarzschild radius, $r_S=2M_0$, and the EiBI length scale, $l_{\epsilon}^2$, characterises the solution via the constant $\delta_1$ in Eq.(\ref{eq:deltageonEiBI}). Note that the relation (\ref{eq:rxgeonsEiBI}) follows from explicitly solving Eq.(\ref{eq:xrOm}). The function $G(z)$ in Eq.(\ref{eq:AgeonEiBI}), with the dimensionless variable $z =r/r_c$, follows directly from the field equations as $dG/dz= -\Omega_{+}/(z^2 \Omega_{-}^{1/2})$, and can be explicitly written as

\begin{equation} \label{eq:GzgeonsEiBI}
G(z)=-\frac{1}{\delta_c}+\frac{1}{2}\sqrt{z^4-1}\left({_2}F_1 \left[\frac{1}{2},\frac{3}{4},\frac{3}{2},1-z^4 \right]+{_2}F_1 \left[\frac{1}{2},\frac{3}{7},\frac{3}{2},1-z^4 \right]\right) \ ,
\end{equation}
where ${_2}F_1[a,b,c;t]$ is a hypergeometric function and $\delta_c\approx 0.572069$ is a constant needed to recover the GR solution in the asymptotic regime, $z\gg 1$. In this limit one has $G(z)\simeq -1/z$, $\Omega_{-} \simeq 1$ (so $z^2(x) \simeq x^2$), and the metric function reduces to

\begin{equation} \label{eq:RNsolution}
A(x)\approx 1-\frac{r_S}{ r  }+\frac{r_Q^2}{2r^2} \ ,
\end{equation}
which is the standard Reissner-Nordstr\"om solution of GR. This is confirmed by considering the behaviour of the curvature scalars for $z \gg 1$ as

\begin{eqnarray}\label{eq:Rgr}
R(g)\approx -\frac{48 r_c^8}{r^{10}}+\mathcal{O}\left(\frac{r_c^9}{r^{11}}\right);
Q(g)\approx \frac{r_Q^4}{r^8}\left(1-\frac{16l_{\epsilon}^2}{r^2}+\ldots\right);
K(g)\approx K_{GR}+\frac{144 r_S r_c^4}{r^9}+\ldots \label{eq:Kgr}
\end{eqnarray}
where $R(g)=g^{\mu\nu}R_{(\mu\nu)}$, $Q(g)=R^{(\mu\nu)}R_{(\mu\nu)}$ and $K(g)={R^\alpha}_{\beta\mu\nu}{R_\alpha}^{\beta\mu\nu} $ are the curvature scalar, the Ricci-squared and the Kretchsman, respectively. These expressions smoothly converge to their GR counterparts with higher-order corrections in $l_{\epsilon}^2$.

\subsubsection{Geometry and properties} \label{sec:horizonsgeon}

\begin{figure}[h]
\centering
\includegraphics[width=0.45\textwidth]{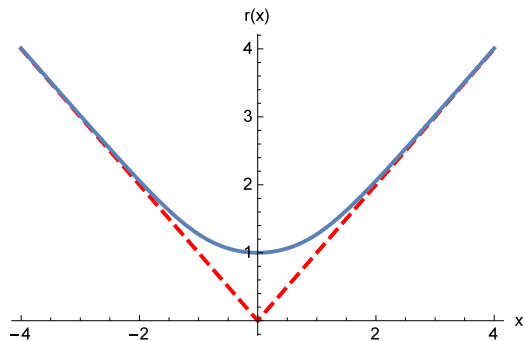}
\caption{Representation of the radial function $r(x)$ in Eq.(\ref{eq:rxgeonsEiBI}) as a function of the coordinate $x$ and measured in units of $r_c$. In this plot the wormhole throat is located at $x=0$ ($r=r_c$). The dotted line represents $\vert x \vert$ (the two asymptotic spaces).  \label{fig:rxEiBI}}
\end{figure}

It should be noted that the line element (\ref{eq:solgeonsEiBI}) can be written in a standard Schwarzschild-like form by absorbing the $\Omega_{+}$ factor via a redefinition of the radial coordinate as $d\tilde{x}^2=dx^2/\Omega_{+}^2$. This must be done with care since the radial coordinate $x\in ]-\infty,+\infty[$, while $r \geq r_c$, as depicted in Fig.\ref{fig:rxEiBI}, where one observes that the area of the two-spheres $S=4\pi r^2(x)$ reaches a minimum of size $S_c=4\pi r_c^2$ at $x=0$, where it bounces off and re-expands again. As already discussed in section \ref{sec:wormholes}, the presence of a minimum value for the radial coordinate allows to infer the presence of a wormhole structure (here the flare-out condition (\ref{eq:flareoutWH}) is automatically satisfied). Indeed, from the relation (\ref{eq:xrOm}) between coordinates, it follows that it is ill-defined at $r=r_c$, because $dr/dx=0$ at this point, and thus the use of $r$ as a radial coordinate is limited to those regions where $r(x)$ is a monotonic function. Thus, in agreement with wormhole physics lore, one needs two charts of $r$ to cover the entire manifold, but a single chart in terms of $x$.

The line element (\ref{eq:metricgqeonsEiBI}) can be alternatively written in Eddington-Filkenstein coordinates using a local redefinition of the time coordinate, $dt=dv-dx/(A\Omega_{+})$, which brings the metric into the form

\begin{equation}\label{eq:ds2_EF}
ds^2=-A(x)dv^2+\frac{2}{\Omega_+}dvdx+r^2(x)d\Omega^2 \ ,
\end{equation}
For null and time-like radial geodesics, $ds^2 \leq 0$, which implies $-A dv^2+\frac{2}{\sigma_+}dvdx\le 0$. Inside the event horizon $A<0$, which means that all such geodesics move in the decreasing direction of $x$ as the advanced time coordinate $v$ moves forward. Now, since the radial function $r(x)$ has a minimum at $x=0$, the relation (\ref{eq:xrOm}) becomes $dx/dr =\Omega_{+}/\Omega_{-}^{1/2}$ in the region $x>0$ and $dx/dr =-\Omega_{+}/\Omega_{-}^{1/2}$ for $x<0$. This way, ingoing geodesics, which always move in the direction of decreasing $x$, propagate in the direction of decreasing area of the radial function $r(x)$ if $x>0$, but in the growing direction if $x<0$, i.e., they approach the wormhole throat if $x>0$, but move away from it if $x<0$ (a similar effect is found for outgoing geodesics). Later in section \ref{sec:Regular} we will discuss the potential troubles of the transit of physical observers across the wormhole throat $x=0$.

As already stated, the geometry described by (\ref{eq:solgeonsEiBI}) reduces to the Reissner-Nordstr\"om of the Einstein-Maxwell field equations, Eq.(\ref{eq:RNsolution}), for $z \gg 1$, but important departures are found as $r \approx r_c$. From the expansion of the $G(z)$ function (\ref{eq:GzgeonsEiBI}) in that region, $G(z)\approx -1/\delta_c + 2(z-1)^{1/2}-(11/6)(z-1)^{3/2} + \ldots$, one finds that the expansion of the metric function $A(z(x))$ there yields the result

\begin{equation}\label{eq:AexpggeonsEiBI}
A(x) \approx  \frac{N_q}{4N_c}\frac{\left(\delta _1-\delta _c\right) }{\delta _1 \delta _c }\sqrt{\frac{r_c}{ r-r_c} }+\frac{N_c-N_q}{2 N_c}+ \mathcal{O}\left(\sqrt{r-r_c}\right) \ ,
\end{equation}
where $N_q=Q/e$ ($e$ is the electron charge) is the number of charges and $N_c =(2/\alpha_{em})^{1/2} \simeq 16.55$ (with $\alpha_{em}$ the fine structure constant). From the expression (\ref{eq:AexpggeonsEiBI}) it is clear that there exists a transition on the behaviour of $A(x)$ for $\delta_1 = \delta_c$, yielding three different situations. This is consistent with the analysis of the horizons of these configurations, as given by the zeroes of $A(x)$. A detailed analysis of such zeroes reveals the following structure for the horizons \cite{Olmo:2013gqa}:

\begin{itemize}
\item If $\delta_1<\delta_c$ a single horizon is located on each side of the wormhole throat $r=r_c$, resembling the structure of the Schwarzschild spacetime.
\item If $\delta_1>\delta_c$ there may be two, one (degenerate) or none horizons, depending on the number of charges $N_q$. These are Reissner-Nordstr\"om-like configurations.
\item If $\delta_1=\delta_c$ a single horizon is found for $N_q>N_c$ and none otherwise. The spacetime metric $g_{\mu\nu}$ is finite at the throat $r=r_c$, and the geometry there is Minkowski-like.
\end{itemize}
It is worth pointing out the resemblance of the three classes (in terms of horizons) of configurations above with those solutions resulting from the coupling of Born-Infeld electrodynamics to GR \cite{Salazar:1987ap,Demianski:1986wx,deOliveira:1994in,Fernando:2003tz,DiazAlonso:2009ak}, described in section \ref{sec:BIBHGRs}. Nonetheless, the presence of a finite-size wormhole throat introduces new features as compared to that case. In this sense, as the region $z \rightarrow 1$ is approached, the resulting expressions strongly deviate from the GR case:

\begin{eqnarray}
r_c^2 R(g)&=& -\frac{1}{2 \delta _2 }\left(1-\frac{\delta_c}{\delta _1}\right)\left[\frac{1}{(z-1)^{3/2}}+\frac{a_1}{\sqrt{z-1}}\right] \nonumber \\
&+& \left(-4+\frac{16 \delta_c}{3 \delta _2}\right)+\mathcal{O}\left({{z-1}}\right) \\
r_c^4 Q(g) &=& \left(1-\frac{\delta_c}{\delta _1}\right)\left[\frac{6 \delta _2-5\delta _1}{3 \delta _2^2 (z-1)^{3/2}}+\frac{a_2}{\sqrt{z-1}}\right] \nonumber \\
&+& \left(10+\frac{86 \delta _1^2}{9 \delta _2^2}-\frac{52 \delta _1}{3 \delta _2}\right)+\mathcal{O}\left({{z-1}}\right) \\
r_c^4K(g)&=& \left(1-\frac{\delta_c}{\delta _1}\right)\left[\frac{2 \left(2\delta _1-3 \delta _2\right)}{3 \delta _2^2 (z-1)^{3/2}}+\frac{a_3}{\sqrt{z-1}}\right] \nonumber \\
&+& \left(16+\frac{88 \delta _1^2}{9 \delta _2^2}-\frac{64 \delta _1}{3 \delta _2}\right)+\mathcal{O}\left({{z-1}}\right)
\end{eqnarray}
where $\{a_1,a_2,a_3\}$ are some constants. These expansion reveal the divergence of all curvature scalars at the wormhole throat, $z=1$, but for the particular choice $\delta_1=\delta_c$ they all become finite. The latter condition sets a particular charge-to-mass ratio, but says nothing on the particular amounts of them. Remember that, from the discussion on the structure of horizons above, the case $\delta_1=\delta_c$ corresponds to the transition between the Reissner-Nordstr\"om-like and Schwarzschild-like case, where the geometry at $z=1$ becomes Minkowskian. As already said, this is somewhat analogous to the case of Born-Infeld electrodynamics coupled to GR, though in such a case curvature divergences at $r=0$ are always present. It should be noted that in the EiBI case, the presence of curvature divergences or not has no influence on the existence of a wormhole structure, so one could wonder about the physical meaning of such divergences (see section \ref{sec:curvaEibIint}).

In the context of GR one could wonder what is the location of the sources generating the mass $M$ and charge $Q$ of the geometry of the Reissner-Nordstr\"om solution. It turns out that it is not possible to have a well defined point-like source generating both mass and charge of the Reissner-Nordstr\"om geometry and, at the same time, being a solution of the Einstein equations everywhere \cite{Ortinbook}. It is equally natural to wonder about the nature of both mass and charge that generate the wormhole geometries discussed here and in section \ref{sec:wormholes}. As first shown by Misner and Wheeler \cite{Misner:1957mt}, the non-trivial topology of the wormhole allows to define by itself a charge without the need of considering sources for the electric field, an effect coined \emph{charge-without-charge} in that paper. Indeed, an electric flux flowing through a $2$-dimensional, spherical $S^2$ surface enclosing one of the sides of the wormhole mouths defines a charge as

\begin{equation} \label{eq:fluxMW}
\Phi=\frac{1}{4\pi} \int_{S^2} *F= \pm  Q
\end{equation}
where $*F$ is the two-form dual to Faradays's tensor and the two signs $\pm$ come from the different orientation of the normal on each side of the wormhole throat. Note that this result holds true regardless of the particular details of the configurations as long as a wormhole throat exists and the topology does not change. In particular, it is not affected by the presence of curvature singularities. For the solutions considered in this section, the density of lines of flux crossing the spherical wormhole throat can be computed as

\begin{equation} \label{eq:fluxMW}
\frac{\Phi}{4\pi r_c^2} =\frac{Q}{r_c^2}=  \sqrt{\frac{2c^7}{(\hbar G)^2}}
\end{equation}
which is independent of the particular amount of charge and mass, i.e, independent on the presence or not of curvature divergences.

In a similar fashion one could wonder about the origin of the mass generating the geometry (\ref{eq:solgeonsEiBI}). Following also the \emph{mass-without-mass} mechanism introduced by Misner and Wheeler \cite{Misner:1957mt}, and in analogy with the energy of an electric field in a Minkowski spacetime, $\Ss{}_M=\int dt \times \mathcal{E}_e$ one can estimate the total mass of the spacetime by evaluating the gravitational + matter action for these configurations, i.e., $\Ss{}=\int dt \times (\mathcal{E}_G + \mathcal{E}_e)$, which can be performed in terms of the variable $dx^2=dz^2/\Omega_{-}$, with the result \cite{Olmo:2013gqa}

\begin{equation}
\Ss{}=2M c^2 \frac{\delta_1}{\delta_c} \int dt
\end{equation}
where the factor $2$ comes from the need of integrating on both sides of the wormhole throat. Like the electric flux above, this result is finite and independent of the existence or not of curvature divergences. The explicit implementation of both charge-without-charge and mass-without-mass mechanisms make these objects be explicit realizations of Wheeler's geon \cite{Wheeler:1955zz}, understood as self-gravitating electromagnetic entities without sources. It remains to be seen whether the case with $\epsilon>0$ (\ref{eq:EBF}), and the wormhole solutions constructed out of anisotropic fluids and described in section \ref{sec:wormholes} admit a similar characterisation, though such an analysis is not available in the literature yet.

We have seen that in the solutions described in this section the presence of curvature divergences at the wormhole throat in the cases $\delta_1 \neq \delta_c$ seems to have no influence on the physical properties of the solutions such as total charge, mass and density of lines of electric field, which are as well defined as in the case $\delta_1=\delta_c$, where no curvature divergences arise. This is somewhat similar to the thin-shell approach to construct wormhole solutions, by which two spacetimes are joined together at a given hypersurface, where the throat is located \cite{Musgrave:1995ka,Dias:2010uh,Garcia:2011aa}. The resulting manifold is geodesically complete by construction, but curvature divergences arise at the wormhole throat \cite{Poissonbook}, which is interpreted as a surface layer with an energy-momentum tensor on it. To get an intuitive idea of the similarities and differences between the smooth, $\delta_1=\delta_c$, and divergent, $\delta_1 \neq \delta_c$ solutions described in this section, one can construct Euclidean embeddings of the spatial equatorial, $\theta=\pi/2$, and $t=$constant section of the line element (\ref{eq:solgeonsEiBI}), expressed in terms of the coordinates $dx^2=\Omega_{+}^2dr^2/\Omega_{-}$, which reads $dl^2=\frac{1}{A\sigma_-}dr^2+r^2d\varphi^2$, to embed it into a three-dimensional space with cylindrical symmetry as

\begin{equation}
dl^2=d\xi^2+dr^2+r^2d\varphi^2 \ ,
\end{equation}
where the function $\xi$ must be chosen so as to match the equatorial $t=$constant line element above. Around the wormhole throat $r_c$ one can make use of the expansions of the metric functions there, together with $\Omega_{-} \simeq 4(r-r_c)/r_c$, which yields

\begin{eqnarray}\label{eq:2Db}
dl^2=\left\{\begin{array}{lr}
\frac{(N_c-N_q)}{8N_c}\frac{r_c}{(r-r_c)}dr^2+r^2d\varphi^2 & \text{ if } \delta_1=\delta_c \\
\text{  } &  \\
\frac{N_c}{N_q}\frac{\delta_1\delta_c}{(\delta_1-\delta_c)}\sqrt{\frac{r_c}{r-r_c}}dr^2+r^2d\varphi^2 & \text{ if } \delta_1\neq \delta_c
\end{array} \right. \\
\end{eqnarray}
so that one has

\begin{eqnarray}
\xi(r)&=&\left\{\begin{array}{lr}
\pm\frac{(N_c-N_q)}{4N_c}\sqrt{r_c}\sqrt{r-r_c} & \text{ if } \delta_1=\delta_c \\
\text{  } &  \\
\pm\frac{4N_c}{3N_q}\frac{\delta_1\delta_c}{(\delta_1-\delta_c)}r_c \left(\frac{r-r_c}{r_c}\right)^{3/4} & \text{ if } \delta_1> \delta_c
\end{array}\right.
\end{eqnarray}
\begin{figure}[h]
\centering
\includegraphics[height=7.5cm,width=8cm]{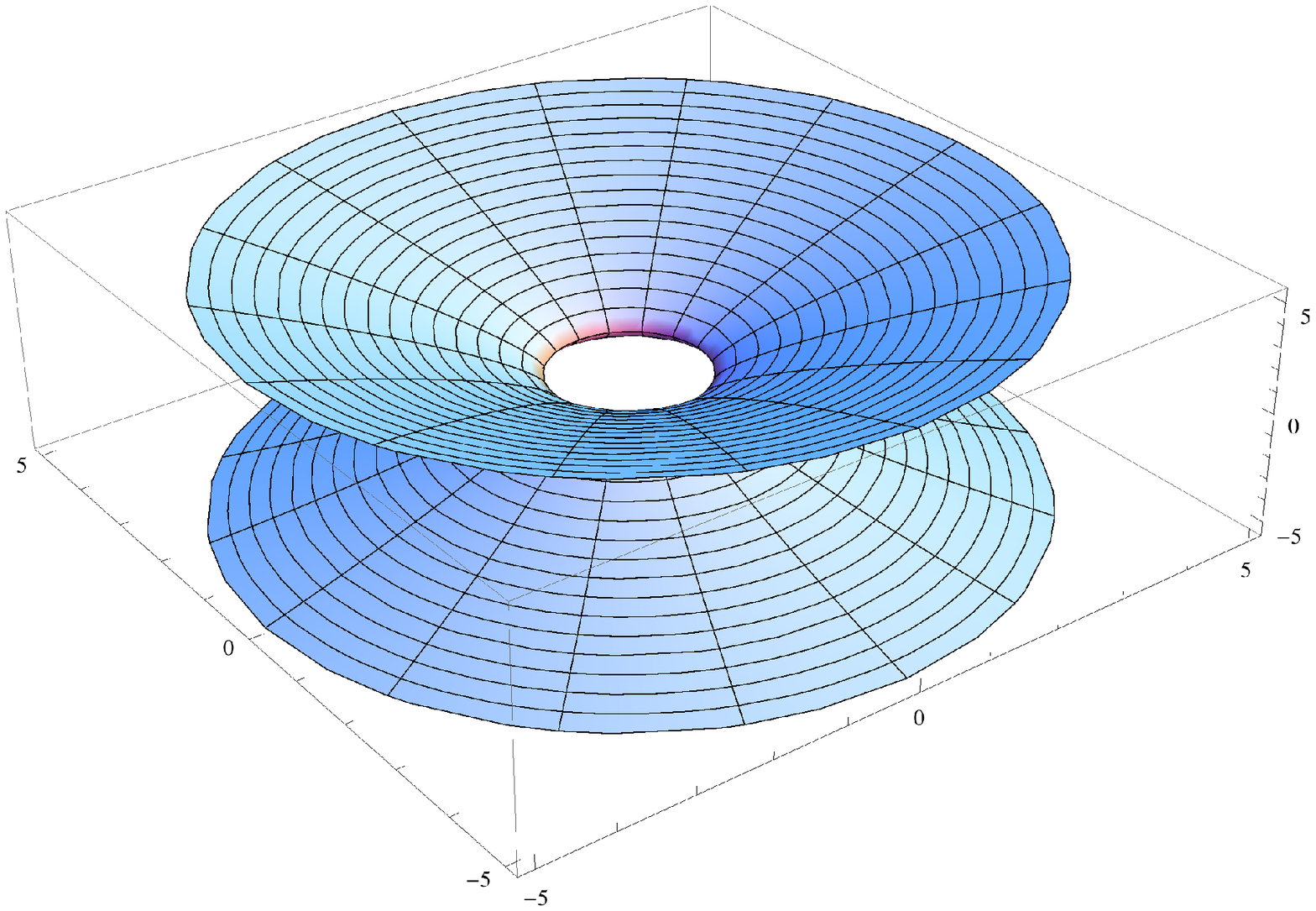}
\includegraphics[width=0.45\textwidth]{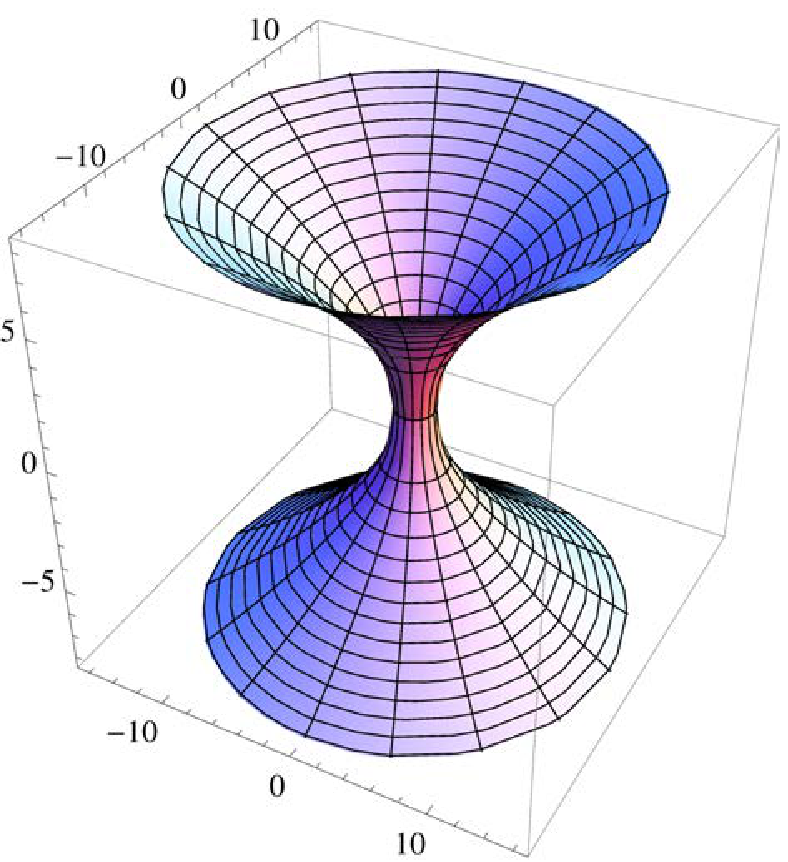}
\caption{Left plot: Euclidean embedding of the equatorial $\theta=\pi/2$ and $t=$constant section of the cases with curvature divergences for electromagnetic geons. The vertical axis represents the function $\xi(r)$. Figure extracted from Ref.\cite{Olmo:2015bya}. Right plot: Euclidean embedding of the wormhole described by Eqs.(\ref{eq:psianishai}) and (\ref{eq:fanishai}) with $\alpha=3/4$, $\epsilon=4$ and $x=1/2$.  Figure extracted from \cite{Shaikh:2015oha}.  \label{fig:eucembgeo}}
\end{figure}
In Fig.\ref{fig:eucembgeo} we have depicted these Euclidean embeddings for $\delta_1=\delta_c$ (top figures) and $\delta_1>\delta_c$ (bottom figures), in those cases where no horizons are present (recall the discussion of section \ref{sec:horizonsgeon}). In both cases, the presence of a wormhole structure is manifest. The two-dimensional curvature, however, as given by the expression of the Kretchsman:

\begin{equation}\label{eq:Kret_2D}
K_{2D}=\left\{\begin{array}{lr}
 \frac{64 \left(N_c-N_q\right)^2}{ N_c^2}\frac{1}{r_c^2 r^2  } & \text{ if } \delta_1=\delta_c \\
\text{  } &  \\
\frac{N_q^2}{N_c^2}\frac{\left(\delta _1-\delta _c\right)^2 }{4 \delta _1^2  \delta _c^2 }\frac{1}{r_c(r-r_c) r^2} & \text{ if } \delta_1> \delta_c
\end{array}\right.
\end{equation}
is finite for the former, but divergent for the latter. This highlights the fact that two similar wormhole structures can show very different properties regarding the behaviour of curvature invariants. This can also be observed in the case of wormholes supported by anisotropic fluids found in \cite{Shaikh:2015oha} and discussed in section \ref{sec:wormholes}, whose Euclidean embedding for the model with $\alpha=3/4$ in Eqs.(\ref{eq:psianishai}) and (\ref{eq:fanishai}) is also depicted in Fig.\ref{fig:eucembgeo}. Remember that in such a case the tidal forces at the throat are finite, regardless of the behaviour of the curvature there. We shall review the issue of non-singular solutions in EiBI gravity in section \ref{sec:Regular}.

\subsubsection{Coupling to Born-Infeld electrodynamics} \label{eq:BIBIelec}

The coupling of EiBI gravity to Born-Infeld electrodynamics (\ref{eq:BINED}) has been considered by Jana and Kar in \cite{Jana:2015cha}. The strategy followed to obtain electrovacuum solutions to the field equations is pretty much the same than the one employed for the Maxwell field above, and therefore we shall omit the details. The resulting line element, with the redefinition $b^2=\alpha/(4\epsilon)$, becomes

\begin{eqnarray}
\text{d}s^2&=&-U(x)e^{2\psi(x)}dt^2+\frac{V(x)}{U(x)}e^{-2\psi(x)} dr^2+r^2\left(d\theta^2 + \sin ^2\theta d\phi^2 \right) \label{eq:linelBI} \\
U(x)&=&\frac{2-\alpha}{2(1-\alpha)}-\frac{\alpha}{2(1-\alpha)}\frac{1}{\sqrt{1+\frac{4\epsilon Q^2(1-\alpha)}{\alpha x^4}}} \nonumber\\
V(x)&=& \frac{2-\alpha}{2(1-\alpha)}-\frac{\alpha}{2(1-\alpha)}\sqrt{1+\frac{4\epsilon Q^2(1-\alpha)}{\alpha x^4}} \ , \nonumber
\label{eq:schw_gauge}
\end{eqnarray}
where the metric function $\psi(x)$ splits into two subcases and takes the form

\begin{eqnarray}
\infty > \alpha >1 : \quad e^{2\psi(x)}&=&1 -\frac{2M}{x} + \frac{\alpha x^2}{6\epsilon (\alpha -1)}\left[\sqrt{1-\frac{4\epsilon Q^2(\alpha -1)}{\alpha x^4}}-1\right] \label{eq:BIalhpa1} \\
&+&
\frac{\alpha^{1/4}(4Q^2)^{3/4}}{3\epsilon^{1/4}(\alpha -1)^{1/4}x}F\left(\arcsin\left(\frac{\left(4\epsilon Q^2(\alpha -1)\right)^{1/4}}{\alpha^{1/4}x}\right), -1\right) \nonumber
\end{eqnarray}
\begin{eqnarray}
-\infty<\alpha<1 : \quad e^{2\psi (x)}&=&1-\frac{2M}{x}-\frac{\alpha x^2}{6\epsilon (1-\alpha)}\left[\sqrt{1+\frac{4\epsilon Q^2 (1-\alpha)}{\alpha x^4}}-1 \right] \label{eq:BIalhpa2} \\
&+& \frac{4Q^2}{3x^2}\, {}_{2}F_1\left(\frac{1}{4},\frac{1}{2};\frac{5}{4};-\frac{4\epsilon Q^2 (1-\alpha)}{\alpha x^4}\right) \ , \nonumber
\end{eqnarray}
and the radial coordinates are related as

\begin{equation} \label{eq:rVx}
r^2=V(x)x^2 \ .
\end{equation}
From the analysis of these solutions it follows that for $\alpha<0$ there is a minimum value of the radial coordinate $r=r_0$ in Eq.(\ref{eq:rVx}), where $r_0=\left(\frac{(\alpha-2)(4\epsilon Q^2 (1-1/\alpha))^{1/2}}{2(\alpha-1)}\right)^{1/2}$, and thus in this case the charge is distributed over a $2$-sphere of radius $r_0$. This is in agreement with the analysis carried out in \cite{Olmo:2013mla} where wormhole solutions with geonic properties were obtained in an extension of GR including quadratic corrections in the curvature and coupled to Born-Infeld electrodynamics. Now, for $\alpha>2$ there is also a minimum, which now occurs at $r_c=(\kappa Q^2 \alpha/(1-\alpha))^{1/4}$, while in the range $0<\alpha \leq 2$, no minimum is found and a point-like charge arises (note that $\alpha=0$ corresponds to the GR case). This implies that, in particular, for $\alpha>0$, depending on the interplay between EiBI gravity and Born-Infeld electrodynamics wormhole solutions might be found, but this is not explicitly investigated by Jana and Kar. Nonetheless, they investigate in detail the case $\alpha=1$. By variation of the EiBI constant $\epsilon$, solutions with one or two horizons may be found in that case (which is thus similar to what is found in Born-Infeld electrodynamics in GR, see section \ref{sec:BIBHGRs}, and to geonic solutions supported by a Maxwell field, see section \ref{sec:horizonsgeon}). Curvature divergences are always present either at the location of the throat $r=r_c$ (when a wormhole is present) or at $r=0$, although the energy density of the electromagnetic field remains finite everywhere.

\begin{figure}[h]
\centering
\includegraphics[width=0.41\textwidth]{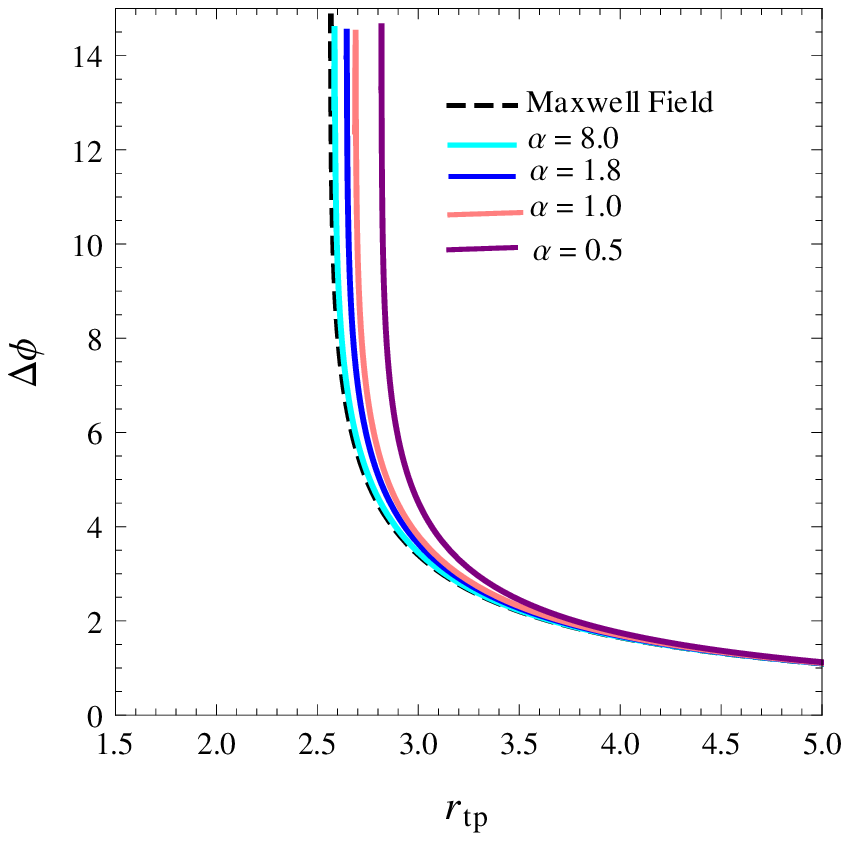}
\includegraphics[width=0.41\textwidth]{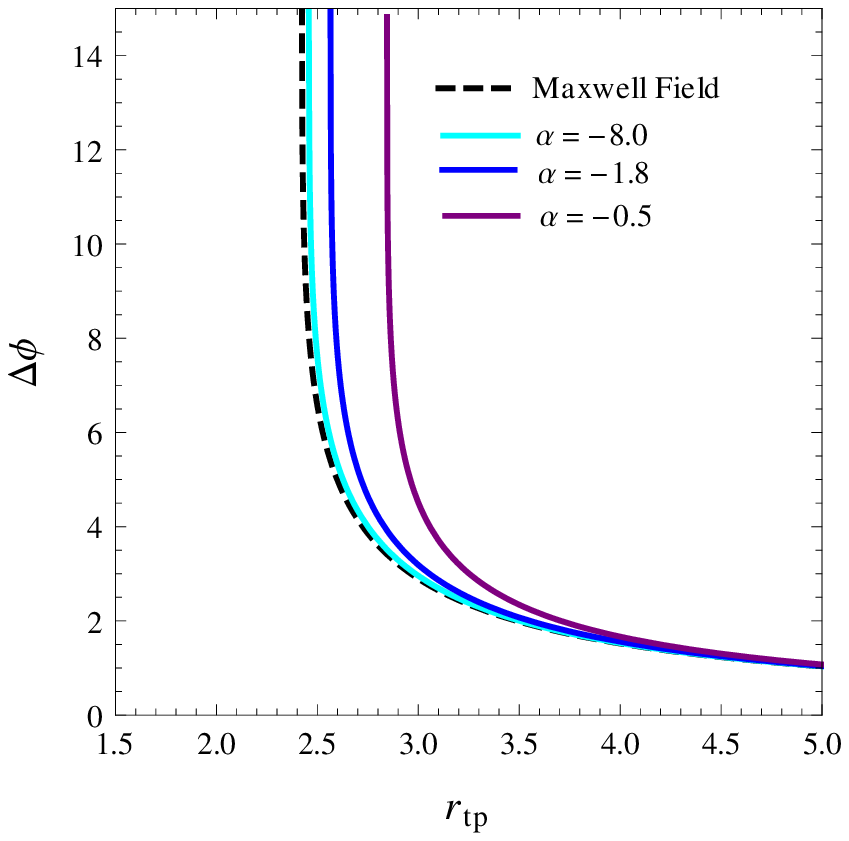}
\caption{Deflection angle $\Delta \Phi$ for EiBI gravity coupled to Born-Infeld electrodynamics as a function of the turning radial point $r_{tp}$ (defined as $dr/d\phi_{r_{th}}=0$),  for different values of $b^2=\alpha/(4\epsilon)$, both positive (left plot) and negative (right plot). The dashed curve represents the coupling of EiBI gravity to Maxwell electrodynamics. Figures taken from \cite{Jana:2015cha}. \label{defangleBI}}
\end{figure}

To investigate the geodesic behaviour in these geometries, one first notices that in non-linear theories of electrodynamics photons do not propagate along null geodesics of the background metric, but instead on null geodesics of an effective geometry \cite{Novello:1999pg} given by $g^{\mu\nu}_{eff}=\left(1+\frac{1}{b^2}F^2\right)g^{\mu\nu}+\frac{1}{b^2}{F^\mu}_{\sigma}F^{\sigma \nu}$, where $F_{\mu\nu}$ is the background electromagnetic field. For any value of $\alpha \neq 1$, the effective metric for the photon propagating in the EiBI background reads \cite{Jana:2015cha}

\begin{equation}
\text{d}s^2_{eff}=-U(x)e^{2\psi(x)}dt^2+U(x)e^{-2\psi(x)}dx^2+\left(\frac{\alpha V^2(x)x^4+4\kappa Q^2}{\alpha V(x)x^2}\right)(d\theta^2 +\sin^2 \theta d\phi^2)
\label{eq:effmetBI}
\end{equation}
from which the expression for the deflection angle (see section \ref{sec:gravlen} for the basic definitions) of the photon moving on this effective metric is obtained as

\begin{equation}
\Delta(\varphi)=2\int^{\infty}_{x_{tp}}\frac{U(x)}{r^2(x)}\left[\frac{U(x_{tp})e^{2\psi(x_{tp})}}{r^2(x_{tp})}-
\frac{U(x)e^{2\psi(x)}}{r^2(x)}\right]^{-1/2}dx -\pi
\label{eq:defangleBI}
\end{equation}
with the expressions appearing in the line element in Eq.(\ref{eq:linelBI}) and with the definitions of Eqs.(\ref{eq:BIalhpa1}) and (\ref{eq:BIalhpa2}). Numerical integration of (\ref{eq:defangleBI}) yields the plots of Figs.\ref{defangleBI}, where the light deflection angle $\Delta \Phi$ is depicted against the turning point radius $r_{tp}$ (for which $dr/d\phi_{r_{th}}=0$) for different values of $\alpha$, both positive and negative, and compared to Maxwell case. This way, like in the case of EiBI gravity coupled to a Maxwell field (see section \ref{sec:gravlen}), gravitational lensing could be used in order to put experimental constraints on the size of $\epsilon$ and on possible nonlinear corrections to Maxwell theory.

\subsection{Non-singular solutions} \label{sec:Regular}

Let us now consider an aspect of utmost importance regarding the internal structure of black holes resulting from gravitational collapse, namely, the presence of a singularity at their center. This is an unavoidable consequence of the singularity theorems provided that i) a trapped surface exists, ii) the null congruence condition holds and iii) global hiperbolicity is fulfilled \cite{Penrose:1964wq,Penrose:1969pc,Hawking:1966vg,Geroch:1968ut} (see \cite{Senovilla:2014gza,Curiel} for more pedagogical discussions of this issue). These theorems are formally based on the notion of geodesic (in)completeness, namely, on the impossibility of extending null and time-like geodesics to arbitrarily large values of their affine parameters. As null geodesics are associated to the transmission of information and time-like geodesics to the free-falling paths of physical observers, geodesic (in)completeness has become the most widely accepted criterium to detect the presence of spacetime singularities\footnote{Note, however, that a given spacetime can be geodesically complete and still be pathological since it can contain finite paths for observers with bounded acceleration, see Geroch \cite{Geroch:1968ut}.}. However, as geodesics are geometrical structures that represent idealized point-like observers without internal structure, it is unclear what a quantum theory of gravity should say about them. Indeed, from an intuitive point of view, since gravity is a matter of curvature, the blow up of curvature scalars could be seen as an indication of the presence of large tidal forces that would potentially rip apart a physical (extended) observer, which has shaped numerous approaches to get rid of spacetime singularities through bounded curvature scalars \cite{AyonBeato:1998ub,Ansoldi:2006vg,Ansoldi:2008jw,Nicolini:2005vd,Modesto:2010uh,Lemos:2011dq,Berej:2006cc}. Indeed, the standard lore of the field states that, as the curvature grows to reach Planckian values, an improved theory of gravity properly incorporating quantum effects should avoid the formation of singularities during the last stages of the gravitational collapse \cite{Hossenfelder:2009fc,Bambi:2013caa,Zhang:2014bea,Rovelli:2014cta,Barcelo:2015noa,Malafarina:2017csn}. In this section we will discuss the regular/singular character of the geonic configurations discussed in section \ref{sec:Geons} making use of these concepts.

\subsubsection{Geodesic completeness} \label{sec:geoEibIint}

\begin{figure}[h]
\centering
\includegraphics[width=0.45\textwidth]{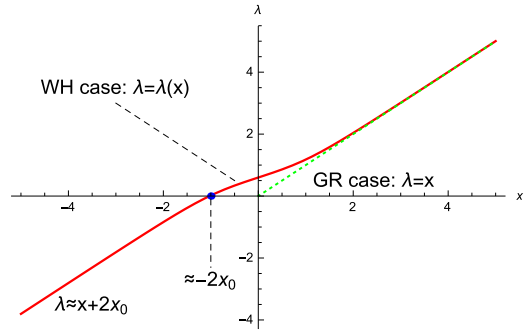}
\caption{Affine parameter $u(x)$ (in this plot $\af \rightarrow \lambda$) as a function of the radial coordinate $x$ for null radial geodesics in Eq.(\ref{eq:nullradial2}), compared to the GR behaviour (dashed green curve). In this plot $E=1$, and the horizon axis is measured in units of $r_c$. Figure taken from \cite{Olmo:2015bya}.  \label{fig:affnulgeons}}
\end{figure}

Our aim in this section is to determine whether the spacetimes considered in section \ref{sec:Geons} are geodesically complete, i.e., whether any time-like or null geodesic can be extended beyond the wormhole throat, since the latter can be reached in finite affine time. We will use the notations and conventions described in section \ref{eq:geodesicbehaviour}. We shall focus on asymptotically flat spacetimes, $\lambda=1$. In terms of the line element (\ref{eq:solgeonsEiBI}) the two conserved quantities of motion read $E=A\frac{dt}{d\af}$ and $L=r^2 \frac{d\varphi}{d\af}$ or, alternatively, in terms of Eddington-Filkenstein coordinates (\ref{eq:ds2_EF}), $E=A\frac{dv}{d\af} - \frac{1}{\Omega_{+}} \frac{dx}{d\af}$. This way, the line element (\ref{eq:solgeonsEiBI}) can be used to write the modulus of the tangent vector $u^{\mu}=dx^{\mu}/d\af$, which satifies $u_{\mu}u^{\mu}=-k$, with $k=0(1)$ for null (time-like) geodesics, as

\begin{equation} \label{eq:geoeqgeons1}
-k =- A \left(\frac{dt}{d\af}\right)^2 + \frac{1}{A \Omega_+^2}\left(\frac{dx}{d\af}\right)^2 +  r^2(x)\left(\frac{d\varphi}{d\af}\right)^2 \ .
\end{equation}
In terms of the conserved quantities above, Eq.(\ref{eq:geoeqgeons1}) reads

\begin{equation}\label{eq:geoeqgeons2}
\frac{1}{\Omega_+^2}\left(\frac{dx}{d\af}\right)^2=E^2-V(x) \hspace{0.1cm};\hspace{0.1cm}  V(x)=A\left(k+\frac{L^2}{r^2(x)}\right) \ ,
\end{equation}
which is just the equation of motion of a particle in a one-dimensional effective potential $V(x)$. For radial ($L=0$) null ($k=0$) geodesics, Eq.(\ref{eq:geoeqgeons2}) simplifies to

\begin{equation}\label{eq:nullradial1}
\frac{1}{\Omega_-}\left(\frac{dr}{d\af}\right)^2=E^2 \ ,
\end{equation}
which admits an analytical integration of the form

\begin{equation}\label{eq:nullradial2}
\pm E \cdot \af(x)=\left\{ \begin{tabular}{lr} ${_{2}{F}}_1[-\frac{1}{4},\frac{1}{2},\frac{3}{4};\frac{r_c^4}{r^4}]  r$ & \text{ if } $x\ge 0$ \\
\text{ }\\
$2x_0- {_{2}{F}}_1[-\frac{1}{4},\frac{1}{2},\frac{3}{4};\frac{r_c^4}{r^4}]  r$ & \text{ if } $x\le 0$
\end{tabular} \right. \ ,
\end{equation}
where $x_0={_{2}{F}}_1[-\frac{1}{4},\frac{1}{2},\frac{3}{4};1] =\frac{\sqrt{\pi}\Gamma[3/4]}{\Gamma[1/4]}\approx 0.59907$ and the sign $\pm$ corresponds to ingoing/outgoing geodesics. For $x \rightarrow \infty$, series expansion of the solution (\ref{eq:nullradial2}) yields $E\af(x)  \approx r \approx x$ and the GR behaviour is naturally recovered. In the GR case one has $(dr/d\af)^2=E^2$ everywhere, whose integration is $r(\af)=\pm E\af$. Since in that case the function $r(\af)$ is strictly positive, then the affine parameter $\af(x)$ is only defined on the positive/negative (ongoing/ingoing) axis and thus geodesics cannot be extended beyond $x=0$, hence such spacetime is geodesically incomplete. In the present case, however, the presence of a wormhole throat introduces significant deviations from the GR solution, and from (\ref{eq:nullradial2}) one finds that at $r=r_c$ ($x=0$) the affine parameter behaves as $E\af(x) \approx \pm x+\sqrt{r-r_c} \approx x \pm x/2$, with the sign $+$ ($-$) corresponding to the region with $x>0$ ($<0$). As depicted in Fig.\ref{fig:affnulgeons}, the affine parameter $\af(x)$ can be smoothly extended beyond $x=0$ and thus radial null geodesics are complete regardless of the value of $\delta_1$. This is a relevant result since in the cases $\delta_1\neq \delta_c$ curvature divergences arise at the wormhole throat, but they do not have any impact on the behaviour of the affine parameter, which is the same in all cases, being free of curvature divergences or not (see section \ref{sec:curvaEibIint} for a discussion on the impact of such divergences).

\begin{figure}[h]
\includegraphics[width=1\textwidth]{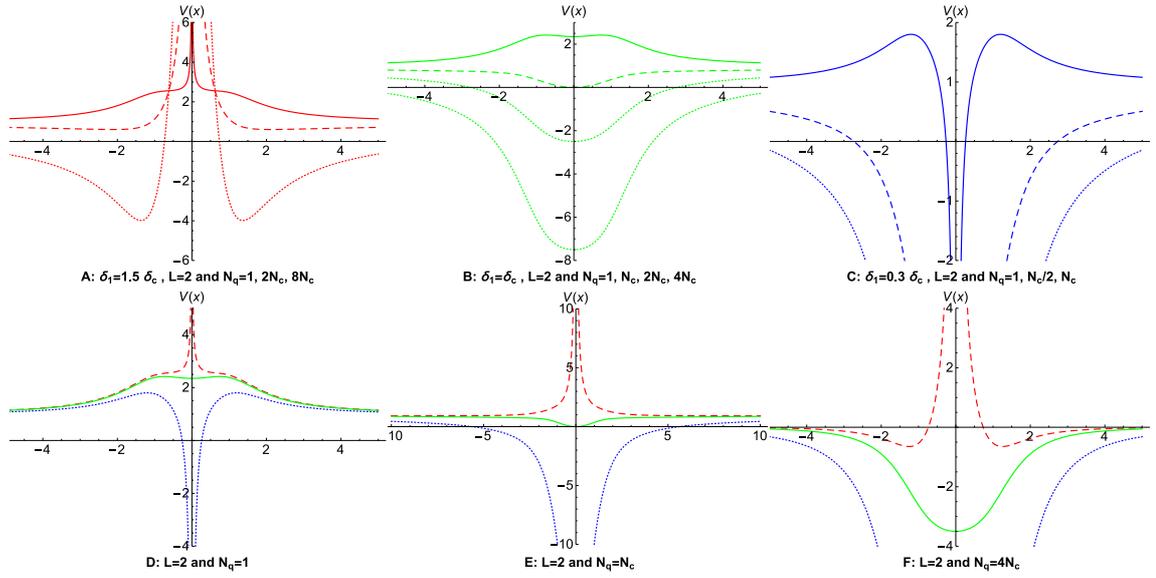}
\caption{Effective potential $V(x)$ for time-like geodesics with $L=2$. Plots A, B and C depict the Reissner-Nordstr\"om-like ($\delta_1>\delta_c$), Minkowski-like ($\delta_1=\delta_c$) and Schwarzschild-like ($\delta_1<\delta_c$) cases, respectively, for three curves corresponding to $N_q=1,N_c,8N_c$ (solid, dashed, dotted, respectively).  Plots D, E and F depict three values of charge $N_q=1,N_c,4N_c$, respectively, for three curves corresponding to $\delta_1=\delta_c, 1.5\delta_c,0.3\delta_c,$ (solid green, dashed red, dotted blue, respectively). Figure taken from \cite{Olmo:2015bya}.  \label{fig:timeLgeons}}
\end{figure}

For null geodesics with $L \neq 0$ and time-like geodesics, the effective potential in (\ref{eq:geoeqgeons2}) can be approximated near the wormhole throat $x=0$ as \cite{Olmo:2015dba}

\begin{equation} \label{eq:effpotapr}
V(x)\approx -\frac{a}{|x|} -a \hspace{0.1cm};\hspace{0.1cm} a=\left(k+\frac{L^2}{r_c^2}\right)\frac{ (\delta_c-\delta_1)}{2\delta_c \delta_2 } \hspace{0.1cm};\hspace{0.1cm}  b=\left(k+\frac{L^2}{r_c^2}\right)\frac{ (\delta_1-\delta_2)}{2\delta_2 } \ .
\end{equation}
From this expression it is clear that if $\delta_1>\delta_c$, corresponding to Reissner-Nordstr\"om-like configurations (see section \ref{sec:horizonsgeon}), an infinite potential barrier prevents any such geodesics to reach the wormhole throat $x=0$, which is the same behaviour found in the GR case. But if $\delta_1<\delta_c$, corresponding to Schwarzschild-like solutions, these geodesics see an infinite attractive potential as $x \rightarrow 0$ and are unavoidably dragged to the wormhole throat. In the GR case, radial time-like geodesics behave near $r=0$ as $\lambda(r)\approx\pm \frac{2}{3}r(r/r_S)^{1/2}$ and, likewise in the case of radial null geodesics above, the fact that $r>0$ makes ingoing/outgoing geodesics to end/start at $x=0$, with no possibility of further extension, and therefore geodesics in this case are incomplete. In the geonic wormhole case, however, the geodesic equation (\ref{eq:geoeqgeons2}) can be integrated as \cite{Olmo:2015bya}

\begin{equation}
\frac{du}{dx}\approx \pm\frac{1}{2}\left|\frac{x}{a}\right|^{\frac{1}{2}}  \ \to u(x)\approx   \pm\frac{x}{3}\left|\frac{x}{a}\right|^{\frac{1}{2}} \ ,
\end{equation}
and again, the fact that $x\in]-\infty,+\infty[$ allows to smoothly extend the affine parameter $u(x)$ across $x=0$ to the whole real axis, which contrasts with the geodesics ending at $x=0$ of the GR case. Finally, if $\delta_1=\delta_c$ (finite curvature cases), the leading order term of the expansion of the effective potential in Eq.(\ref{eq:effpotapr}) vanishes, and the new expression becomes $\frac{V(x)}{L^2}=\frac{1}{2}(1-\frac{N_q}{N_c}) + \mathcal{O}(x)$ (remember that in this case an event horizon is present if $N_q>N_c$) for both null geodesics with $L \neq 0$ and time-like geodesics, which means that the potential is regular at $x=0$. This way, all geodesics with energy $E$ greater than the maximum of the potential $V_{max}$ will be able to go through the wormhole, while bounded orbits can exist for $0<V_{max}<E$. The comparison of the behaviour of the effective potential for the three classes of configurations and different values of the number of charges is depicted in Fig.\ref{fig:timeLgeons}, corresponding to time-like geodesics with $L \neq 0$.

From the description above, it follows that the presence of a spherically symmetric wormhole structure replacing the point-like singularity of GR allows for geodesically complete spacetimes, which is in agreement with the standard lore of wormhole physics \cite{Visserbook}. Nonetheless, the physical meaning of curvature divergences at the wormhole throat requires a separate analysis.

\subsubsection{The physical implications of curvature divergences} \label{sec:curvaEibIint}

We have already discussed in section \ref{sec:wormholes}, following Shaikh \cite{Shaikh:2015oha}, that for wormhole solutions supported by anisotropic fluids, tidal forces at the wormhole throat can be finite. In this section we shall follow a different approach to determine the impact of curvature divergences on physical (extended) observers and review the results of \cite{Olmo:2016fuc}. This approach is based on the concept of \emph{strong} singularities, originally introduced by Ellis and Schmidt \cite{Ellis:1977pj}. Such singularities are identified by the property that all objects approaching them are crushed to zero volume, no matter what their internal constituents or forces holding them might be. This is opposed to \emph{weak} singularities, for which a body could retain its identity while crossing the divergent region. Built on the precise mathematical framework introduced by Tipler \cite{Tipler:1977zzb,Tipler:1977zza}, Clarke and Krolak \cite{ClarkeKrolak} and others \cite{Nolan:1999tw,Ori:2000fi}, the idea is to idealize a physical (extended) observer as a set of points following their own geodesic path (i.e. a congruence), and to determine the relative separation between nearby geodesics as the divergent region is crossed. The congruence is characterised as $x^\mu=x^\mu(u,\xi)$, where $\af$ corresponds to the affine parameter along a given geodesic and $\xi$ labels the different geodesics on such a congruence. The separation between nearby geodesics (for fixed $\af$) is measured by the Jacobi field $Z^\mu\equiv \partial x^\mu/\partial \xi$, which satisfies the geodesic deviation equation

\begin{equation}\label{eq:geoeqdev}
\frac{D^2 Z^\alpha}{du^2}+{R^\alpha}_{\beta\mu\nu}u^\beta Z^\mu u^\nu=0 \ .
\end{equation}
Given the second-order character of this equation, it follows that there are six independent Jacobi fields along a given geodesic, which are obtained as $Z^a(\af)={A^a}_b(\af) Z^b(\af_i)$, where $Z^b(\af_i)$ corresponds to the value of the Jacobi fields at some initial instant $\af_i$ and ${A^a}_b(\af)$ is a $3\times 3$ matrix (the identity if $\af=\af_i$). If all Jacobi fields vanish at $\af=\af_i$, one can instead write $Z^a(\af)={\mathcal{A}^a}_b(\af) \left.\frac{D Z^b}{d\af}\right|_{\af=\af_i}$, where $\mathcal{A}^a$ is a $3\times 3$ matrix that vanishes at $\af=\af_i$. This way, three linearly independent solutions of (\ref{eq:geoeqdev}) allow to define a volume element:

\begin{equation} \label{eq:volele}
V(\af)=\det|A(\af)|V(\af_i)
\end{equation}
(or as $\det|\mathcal{A}(\af)|$ if $Z^a(\af_i)=0$). Thus a strong singularity is met if $\lim_{\af \rightarrow 0} V(\af)=0$ \cite{ClarkeKrolak}, where the singularity is approached if $u \rightarrow 0$.

Following the analysis of Dolan \cite{Nolan:1999tw} for spherically symmetric spacetimes, the Jacobi fields can be written as $ \{Z_{(1)}= B(\af)({u^x}/{A},A u^t,0,0),Z_{(2)}= (0,0,P(\af),0), Z_{(3)}= (0,0,0,Q(\af)/\sin(\theta))\}$, which are orthogonal to the time-like radial geodesic vector $u^\mu=(u^t,u^x,0,0)$, whose components are defined via $u^t\equiv dt/d\af=E/A$ and $u^x\equiv dx/d\af$. The geodesic deviation equation (\ref{eq:geoeqdev}) allows to obtain the functions $B(\af)$, $P(\af)$ and $Q(\af)$ via the equations $P(\af)=P_0+C\int \frac{d\af}{r^2(u)} $, $Q(\af)=Q_0+C'\int \frac{du}{r^2(\af)}$ and $B_{\af\af}+\frac{A_{yy}}{2}B(\af)=0$. Close to the wormhole throat, the behaviour of these functions can be computed and the result compared to their GR counterparts as

\begin{eqnarray}
B_{GR}(u) &\approx& C_1\left(\frac{1}{|\af|^{1/3}}-\frac{|\af|^{4/3}}{|\af_i|^{5/3}} \right) \rightarrow B_{\text{EiBI}}(\af)\approx C_1'\left(\frac{1}{|\af|^{1/3}}-\frac{|\af|^{4/3}}{|\af_i|^{5/3}} \right) \\
P_{GR}(u) &\approx& C_2\left( \frac{1}{|\af_i|^{1/3}} - \frac{1}{|\af|^{1/3}}\right) \rightarrow P_{\text{EiBI}}(\af)\approx C_2' (\af-\af_i) \\
Q_{GR}(u) &\approx& C_3\left( \frac{1}{|\af_i|^{1/3}} - \frac{1}{|\af|^{1/3}}\right) \rightarrow Q_{\text{EiBI}}(\af)\approx C_3' (\af-\af_i) \ ,
\end{eqnarray}
where $\{C_1,C_1',C_2,C_2',C_3,C_3'\}$ are arbitrary constants. Now, from \cite{Nolan:1999tw} the resulting volume from these spacetimes can be written as

\begin{equation}
V(\af)=|B(\af) P(\af) Q(\af)| r^2(\af),
\end{equation}
Now, since in GR one has $r_{GR} \approx (9r_S/4)^{1/3}\af^{2/3}$ and in EiBI geons $r^2(\af) \approx r_c^2 + x^2/2$, then one finds that the volume in the former is $V_{GR} \approx \af^{1/3}$, while in the latter $V_{EiBI} \approx 1/\af^{1/3}$. Thus, in the GR case the volume vanishes as $\af \rightarrow 0$ and, according to Tipler's criterium \cite{Tipler:1977zza,Tipler:1977zzb}, the divergence of curvature scalars is associated to a strong singularity. In the EiBI case, however, the finite radius of the wormhole prevents the convergence of geodesics of the GR case, and the volume element diverges instead as $\af\rightarrow 0$, a scenario that has been independently discussed by Nolan \cite{Nolan:2000rn} and Ori \cite{Ori:2000fi}.

To investigate in more detail the effect of such a divergent volume on physical observers let us rewrite the line element (\ref{eq:solgeonsEiBI}) in free-falling coordinates as \cite{Olmo:2015dba,Olmo:2016fuc}

\begin{equation}\label{eq:frefalgeons}
ds_g^2=-du^2+(u^y)^2d\xi^2  +r^2(\af,\xi)d\Omega^2 \ ,
\end{equation}
where $\xi$ measures the radial separation between nearby geodesics and $u^y\equiv dy/du$, where $dy=dx/(1+r_c^4/r^4(x))$. For the Scharwarzschild-like configurations, $\delta_1<\delta_c$, which is the only case in which time-like observers can go through the wormhole (recall the discussion of section \ref{sec:geoEibIint}), the vector $(u^y)^2$ can be approximated near the wormhole throat as $(u^y)^2 \simeq a/|x| \simeq (\frac{3}{a} |u-E\xi|)^{-\frac{2}{3}}$. This turns (\ref{eq:frefalgeons}) into

\begin{equation}\label{eq:frefalwh}
ds_g^2\approx-d\af^2+\left(\frac{3}{a} |\af-E\xi|\right)^{-2/3}d\xi^2\ .
\end{equation}
This expression states that, as the wormhole throat is approached, the distance between two infinitesimal nearby geodesics diverges, $dl_{Phys}=\left(\frac{3}{a} |u-E\xi|\right)^{-1/3}d\xi$. However, for finite comoving separation between nearby geodesics, $l_\xi\equiv \xi_1-\xi_0$, the physical separation $l_{Phys}\equiv \int |u^y| d\xi$ can be computed as

\begin{equation}\label{eq:lphys}
l_{Phys}\approx \left(\frac{a}{3}\right)^{1/3}\frac{1}{E}\left||u-E \xi_0|^{2/3}-|u-E \xi_1|^{2/3}\right| \ ,
\end{equation}
which is finite. Due to the divergent volume carried by a physical observer, the meaning of this result is that, as the wormhole throat is approached, infinitesimally nearby geodesics are infinitely stretched in the radial direction, followed by an identical contraction as the wormhole is left behind, in a sort of \emph{spaghettisation} process. The danger lies on the possibility that the constituents that make up and keep cohesioned the body could lose causal contact due to the spatial stretching affecting their infinitesimal elements, which would result in the unavoidable destruction of the body. To check this one can consider the propagation of radial null rays, $ds^2=0$, in the background (\ref{eq:frefalwh}), so the photon path satisfies

\begin{equation}
\frac{d\xi}{du}= \pm \left|\frac{3}{a} (u-E\xi)\right|^{1/3}.
\end{equation}
Using a numerical integration, in Fig.(\ref{fig:eucembgeo}) two main results are observed: i) a fiducial observer at $\xi=0$ never loses causal contact with its nearby geodesics (left figure) and ii) the proper time taken in a round trip by a light ray from $\xi=0$ to a nearby geodesic is always finite and casual as the wormhole throat is crossed (right figure), with just an additional delay in the travelling time. Thus, in these geometries, physical observers near the wormhole throat can remain in causal contact despite the spaghettisation process experienced as $u\to E\xi$, and can apparently cross this region with curvature divergences, without experiencing absolutely destructive deformations.

\begin{figure}[h]
\centering
\includegraphics[width=0.35\textwidth]{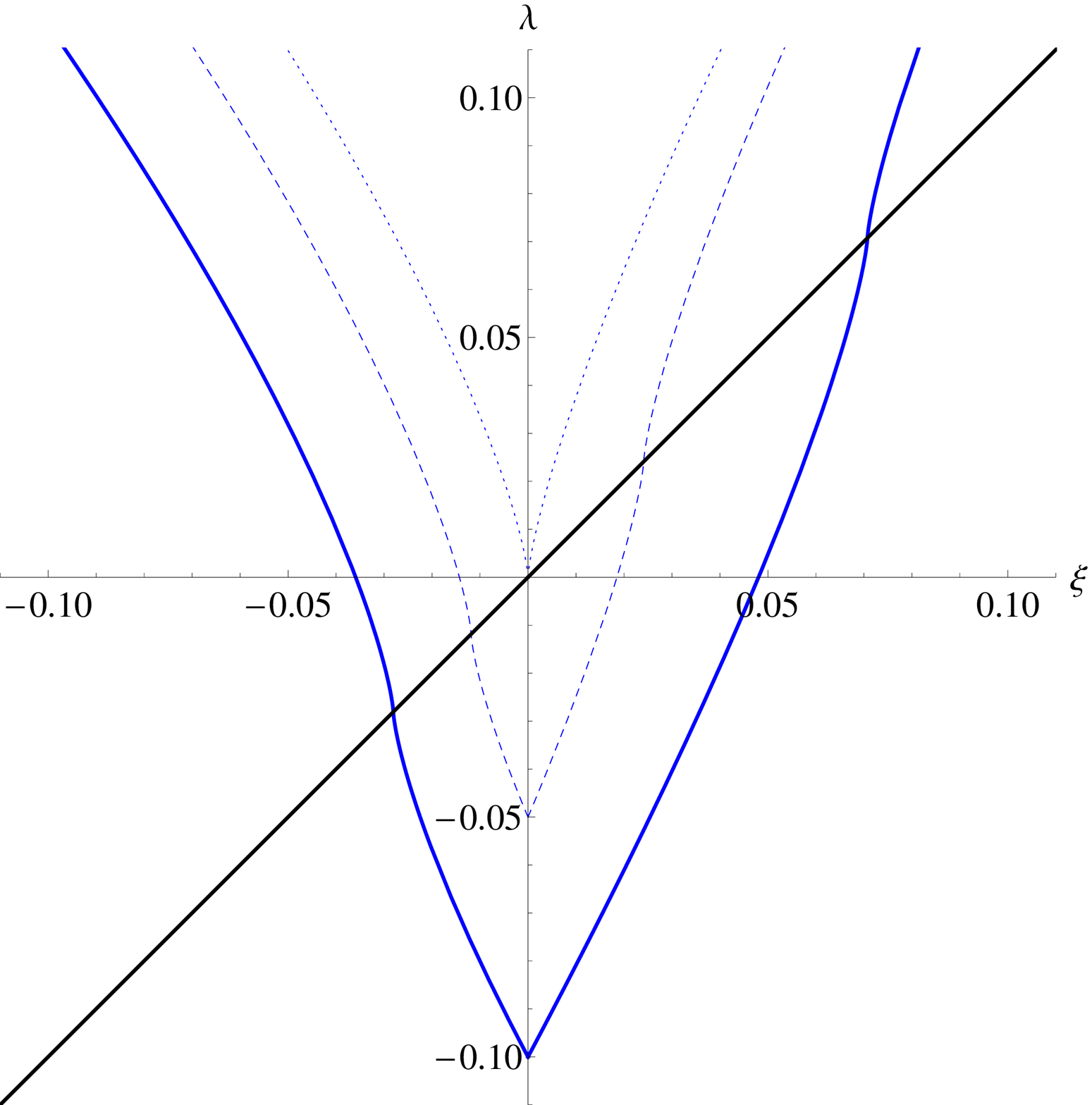}
\includegraphics[width=6cm,height=5.2cm]{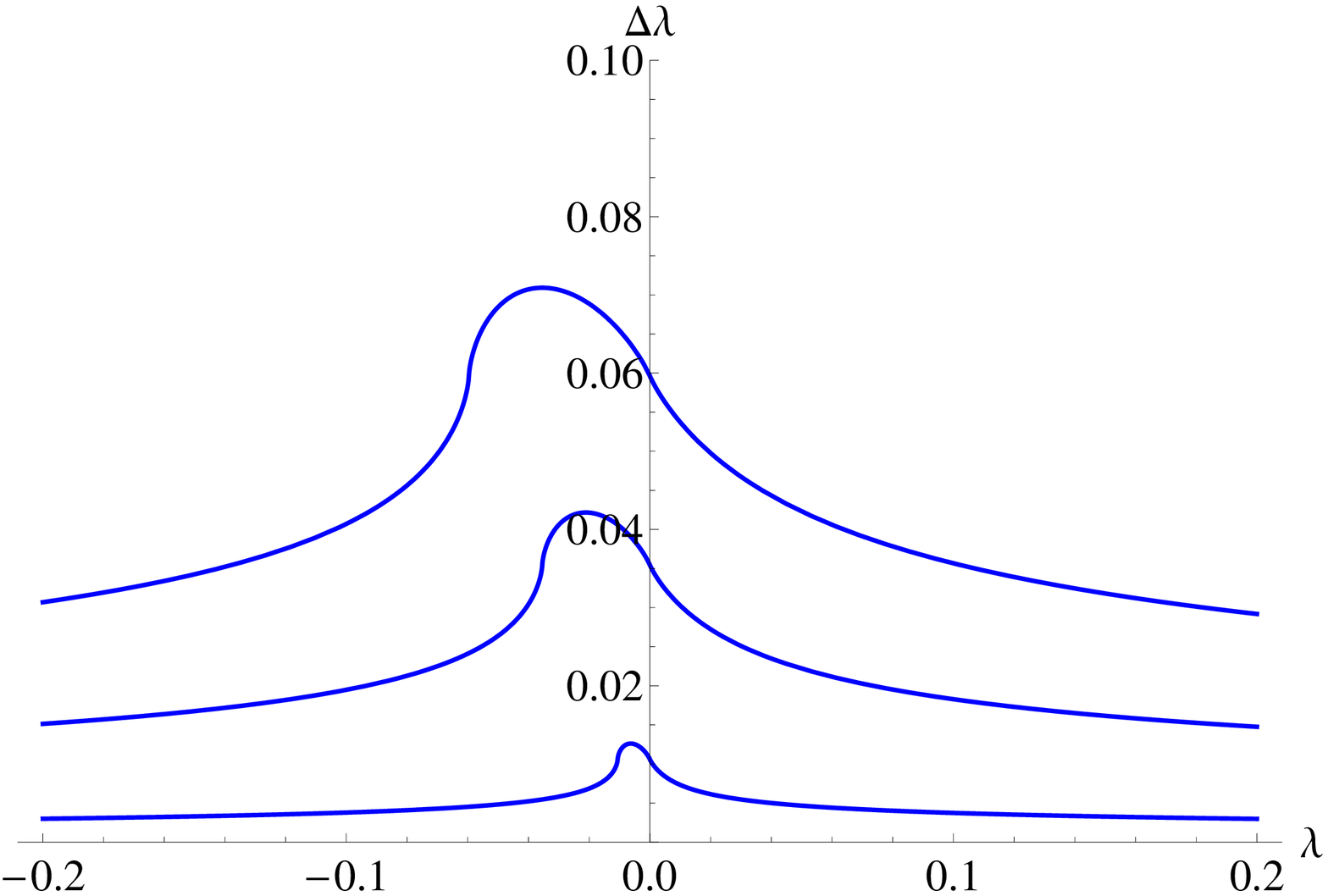}
\caption{Left plot: trajectories of light rays  (in this plot $u \rightarrow \lambda$) emitted by a time-like observer from $\xi=0$ at different times before reaching the wormhole throat (oblique line $u - E\xi= 0$) in a Schwarzschild-like configuration, $\delta_1<\delta_c$ (in this plot, $E=1,a=3$).  Right plot: proper time $\Delta u$ taken in a round trip by a geodesic at $\xi=0$ to another separated by a distance $\xi =\{0.01 r_c,0.005r_c,0.001r_c\}$, as a function of the proper time $u$ at which the light ray was sent. At $u=0$ the divergent region is reached, but the (finite) travelling time tends to zero as the comoving distance tends to zero too. Figures taken from Ref.\cite{Olmo:2016fuc}. \label{fig:scatwave}}
\end{figure}

\subsubsection{Tests with scalar waves}

\begin{figure}[h]
\centering
\includegraphics[width=0.45\textwidth]{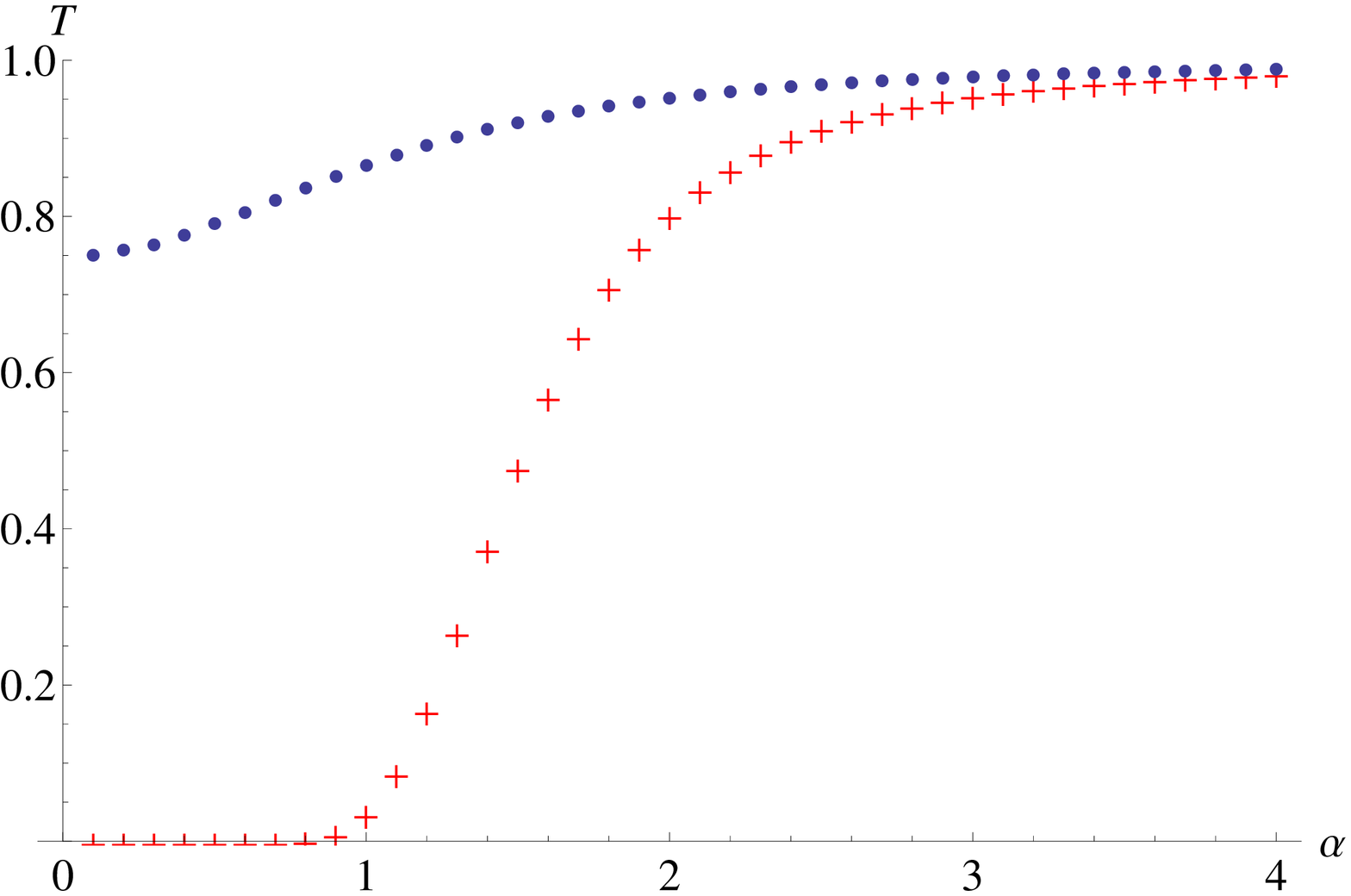}
\includegraphics[width=0.45\textwidth]{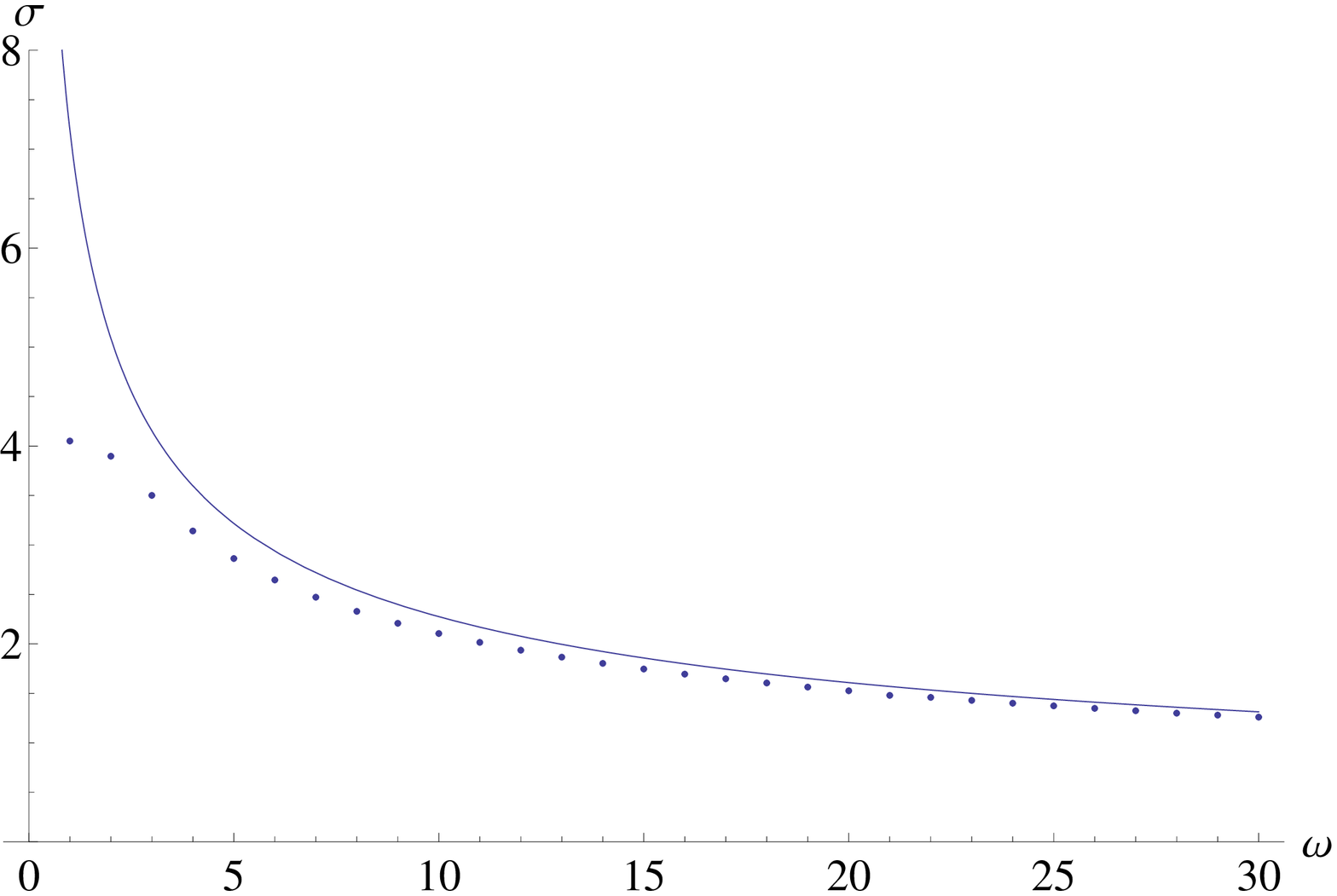}
\caption{Left plot: transmission coefficient for the potential well $-|y'|^{-1/2}$ (blue dots) and the barrier $+|y'|^{-1/2}$ (red crosses).  Right plot: transmission cross section in Eq.(\ref{eq:sigmageons}) as a function of $\omega$ for $\delta_1=1$ and $N_q=10$ (configurations without horizons), from numerical calculation of $V_{eff}$ (dots) in Eq.(\ref{eq:wavegeon}) and compared to the approximation $\sigma \propto \omega^{-1/2}$ (continuous line).  Figures extracted from Ref.\cite{Olmo:2015dba}.  \label{fig:scawavgeon}}
\end{figure}

As a third test to determine whether the presence of curvature divergences endangers the well posedness of the physical laws in these geonic geometries one can study the propagation of scalar waves near the wormhole throat, following the description of \cite{Olmo:2015dba}. This analysis considers the case of Reissner-Nordstr\"om-like configurations, $\delta_1>\delta_c$, where the presence of a time-like Killing vector allows for a separation of variables. The field equation for a massive scalar field, $(\Box-m^2) \phi=0$, can be decomposed in modes of the form $\phi_{\omega, l m}=e^{-i\omega t} Y_{lm}(\theta,\varphi) f_{\omega,l}(x)/r(x)$, where $Y_{lm}(\theta,\varphi)$ are spherical harmonics and the functions $f_{\omega,l}(x)/r(x)$ are governed, in the radial coordinate $dy/dx=1/A(1+r_c^4/r^4)$, by the Schr\"{o}dinger-like equation

\begin{equation}\label{eq:wavegeon}
-f_{yy}+V_{eff}f=\omega^2f \hspace{0.2cm};\hspace{0.2cm} V_{eff}=\frac{r_{yy}}{r}+A(r)\left(m^2+\frac{l(l+1)}{r^2}\right) \ ,
\end{equation}
where the effective potential $V_{eff}$ converges to the GR result for $r \gg r_c$, but behaves near the wormhole throat $r=r_c$ as

\begin{equation}\label{eq:Vefwhtgeon}
V_{eff} \approx \frac{k}{\vert y \vert^{1/2}} \hspace{0.1cm};\hspace{0.1cm}
k\equiv{\sqrt{\frac{(\delta_1-\delta_c)N_q }{\delta_1\delta_c N_c}}}\frac{(N_c[m^2r_c^2+1+l(l+1)]-N_q)}{N_c(8r_c^3)^{1/2}} \ .
\end{equation}
While low-energy modes cannot overcome the potential barrier and are almost entirely reflected, much like in the GR case, high-energy models may overcome such a barrier and end hitting the wormhole throat. Considering an incoming wave packed travelling from null infinity (when no horizon is present, or from the event horizon otherwise), the wave equation (\ref{eq:wavegeon}) reduces to

\begin{equation}
f_{y'y'}+\left(\alpha^2 \pm \frac{1}{\sqrt{y'}} \right)y=0 \ ,
\end{equation}
where the parameter $ \alpha=|k|^{-\frac{2}{3}}\omega$ encodes all the relevant information for this problem. The sign $\pm$ determines an infinite well or potential, the former leading to a transmission coefficient that tends to one as $\alpha$ grows, while the latter has a typical sigmoid profile of barrier experiments, where a threshold around $\alpha=\alpha_{th} \sim 1.5$ from almost complete reflection to almost complete transition is found (see left panel of Fig.\ref{fig:scawavgeon}). For constant $\omega$ there is another threshold, $l=l_{max}$, such that the cross section, $\sigma$, can be roughly estimated by considering that the transmission factor is one for $l>l_{max}$ (almost entire transmission) and zero for $l<l_{max}$ (almost entire reflection) as

\begin{equation} \label{eq:sigmageons}
\sigma=\frac{\pi}{\omega^2}\sum_{l=0}^{l_\text{max}}(2l+1)1=\frac{\pi}{\omega^2}(1+l_\text{max})^2
\end{equation}
which is depicted in the right panel of Fig.\ref{fig:scawavgeon}, where for $\omega \rightarrow \infty$ one has $\sigma \propto \omega^{-1/2}$.

As a summary of this section, the well posedness of the wave scattering problem, together with the geodesic completeness for null and time-like geodesics and all spectrum and mass and charge, and the fact that the constituents making up physical (extended) observers can remain in causal contact as the wormhole throat is crossed, imply the existence of classical non-singular black hole geometries in EiBI gravity. It should be pointed out that similar electromagnetic solutions as those analyzed here and in section \ref{sec:Geons} can be found in functional extensions of the form $f( \mathcal{X} ) = \mathcal{X}^{n}$, where $\mathcal{X}\equiv \det(\hat{g}^{-1}\hat{q})$ and the parameter $n$ labels different models ($n=1/2$ for EiBI gravity). It turns out that for any $1/4<n\leq 1/2$ the corresponding electrically charged solutions, studied in Ref.\cite{Bambi:2016xme}, share a similar wormhole structure as those of EiBI gravity, yielding also geodesically complete structures, while for $n>1/2$ no wormhole solutions were found in that reference.

\subsection{Higher and lower dimensional models and solutions}

\subsubsection{Electromagnetic fields in higher dimensions} \label{sec:hdimBH}

The setup derived and discussed in Sec.(\ref{sec:Geons}) can be extended to their higher-dimensional, $D >4$, counterparts. Most of the corresponding expressions are easily obtained following a similar approach, see Bazeia et al \cite{Bazeia:2015uia}. The field equations in the $q_{\mu\nu}$ geometry
%(ADD EQ)
read now

\begin{equation} \label{eq:Rmnhigher}
{\mathcal{R}^\mu}_{\nu}(q)=\frac{\epsilon^2}{|\hat\Upsilon|^{\frac{1}{D-2}}} \left[\lag_{BI} {\delta^\mu}_{\nu}+ {T^\mu}_{\nu} \right] \hspace{0.1cm};\hspace{0.1cm}
\lag_{BI}=\frac{|\hat\Upsilon|^{\frac{1}{D-2}}-\lambda}{\epsilon \kappa^2} \ ,
\end{equation}
with the definition

\begin{equation} \label{eq:Upshigher}
\hat{\Upsilon} \equiv \vert \hat{\Omega} \vert^{1/2} {(\Omega^{-1})^\mu}_{\nu}=\lambda {\delta^\mu}_\nu -\epsilon \kappa^2 {T^\mu}_\nu
\end{equation}
(remember that $\hat\Omega^{-1}\equiv {\hat q^{-1}}\hat g$, while the definition (\ref{eq:Upshigher}) implies $|\hat\Omega|^{1/2}=|\hat\Upsilon|^{\frac{1}{D-2}}$). The relation between the auxiliary $q_{\mu\nu}$ and the physical metric $g_{\mu\nu}$ is now given by

\begin{equation} \label{eq:gq}
q_{\mu\nu}=   \vert \hat{\Upsilon}\vert^{\frac{1}{D-2}} {(\Upsilon^{-1})_\mu}^{\alpha} g_{\alpha\nu} \hspace{0.1cm};\hspace{0.1cm}
q^{\mu\nu}=  \frac{1}{ \vert \hat{\Upsilon}\vert^{\frac{1}{D-2}}}  g^{\mu \alpha} {\Upsilon_\alpha}^{\nu} \ .
\end{equation}
It is easy to see that the system of equations (\ref{eq:Rmnhigher}), with the definitions (\ref{eq:Upshigher}) and the transformation (\ref{eq:gq}), satisfies the same second-order field equations and ghost-free character of their four-dimensional partners, besides the recovery of the $D$-dimensional Minkowski spacetime in vacuum.

\begin{figure}[h]
\centering
\includegraphics[width=0.45\textwidth]{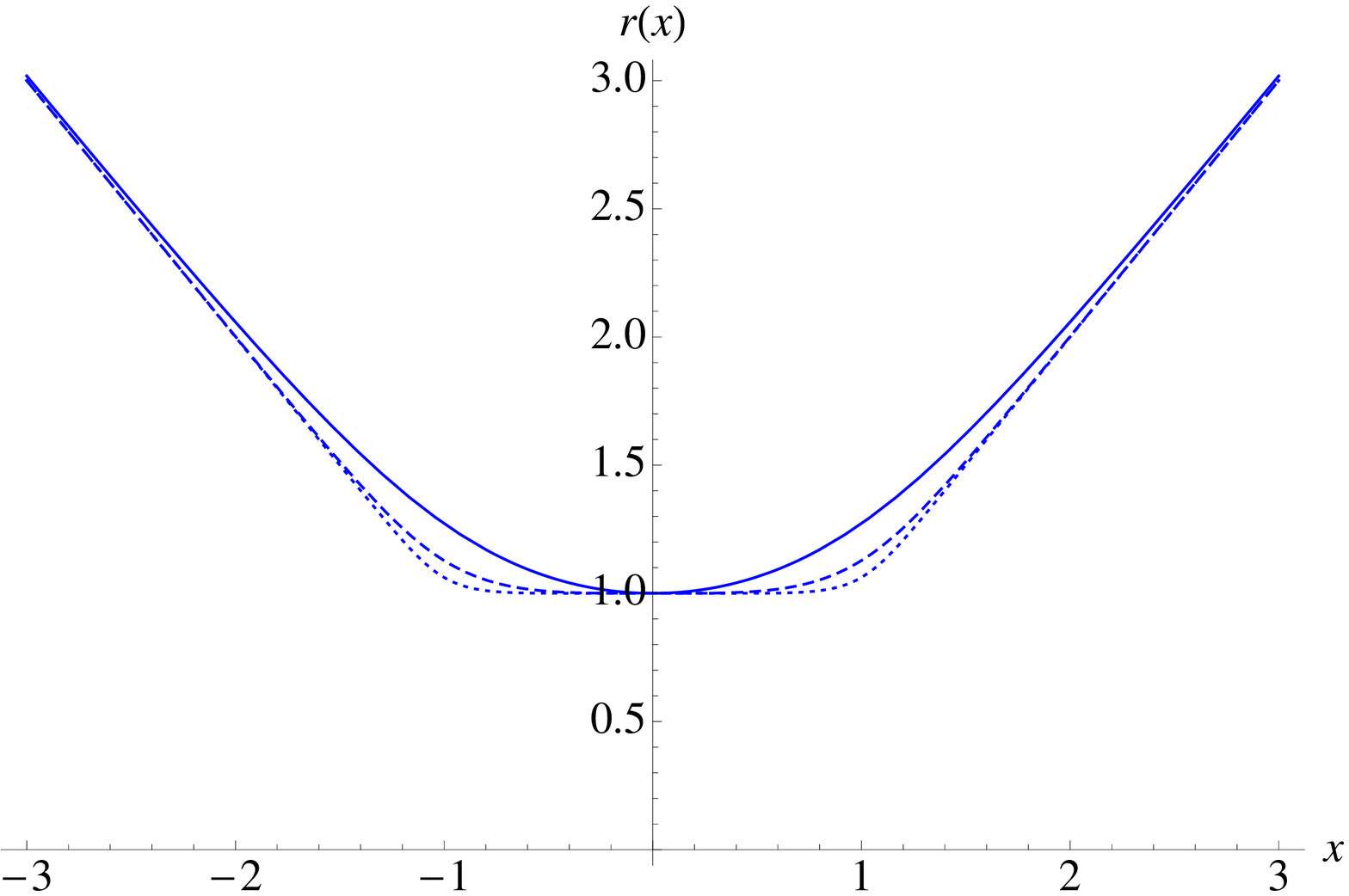}
\includegraphics[width=0.45\textwidth]{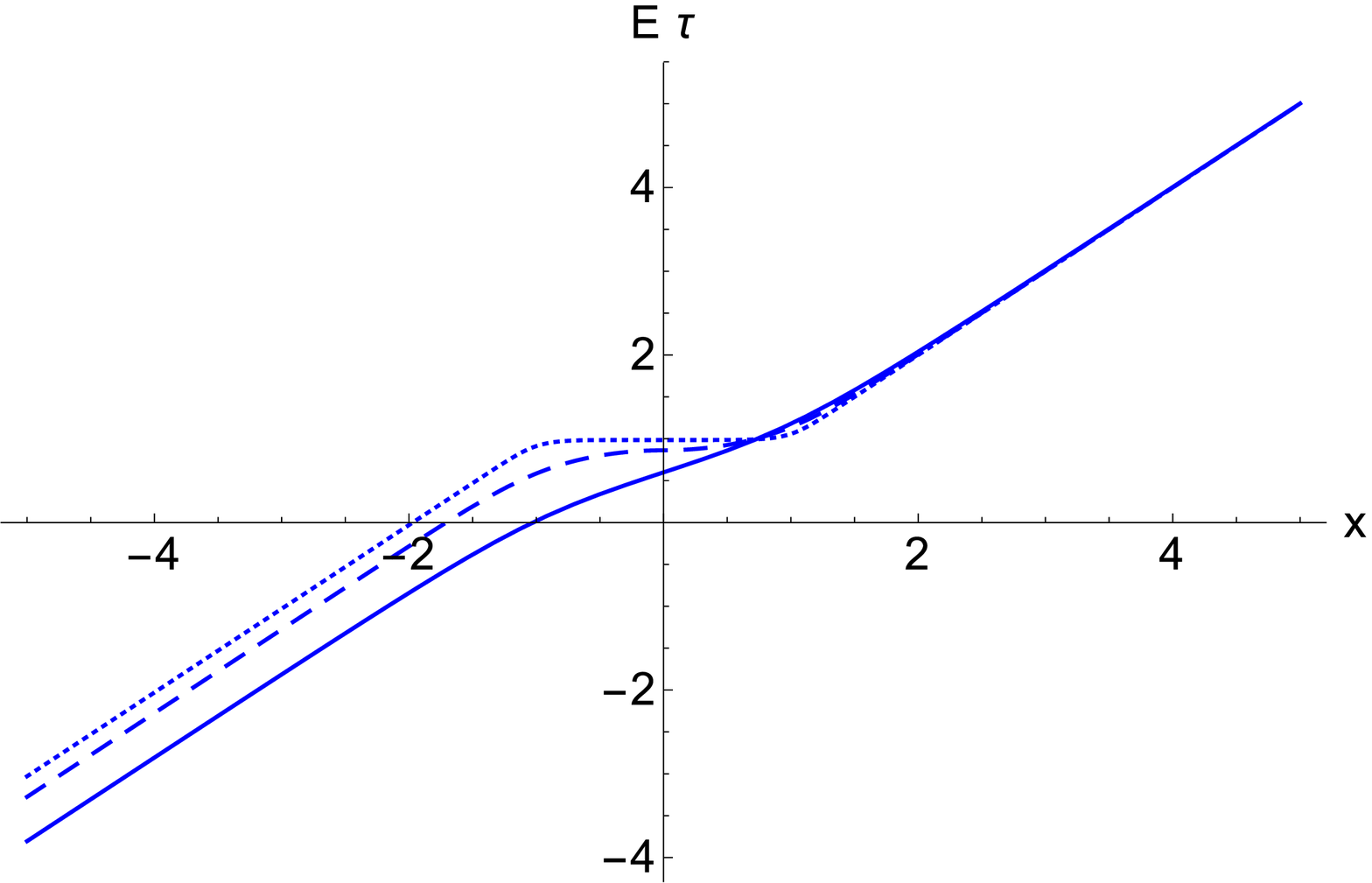}
\caption{Left plot: representation of $r(x)$ in Eq.(\ref{eq:rxhigher}) for $D=4$ (solid), $D=6$ (dashed) and $D=10$ (dotted), with both axes measured in units of $r_c$. The wormhole throat is located at $x=0$. Right plot: representation of the affine (null radial geodesics) parameter $E\af(x)$ (in this plot $\af \rightarrow \tau$) as a function of the radial coordinate $x$ in $D=4$ (solid), $D=5$ (dashed) and $D=10$ (dotted). Figures taken from Ref.\cite{Bazeia:2015uia}.  \label{fig:whafhigher}}
\end{figure}

Electrovacuum solutions of the field equations (\ref{eq:Rmnhigher}) are easily derived following the same steps as in section \ref{sec:Geons}, using now the set of transformations (\ref{eq:Upshigher}). One thus set two static, spherically symmetric line elements of the form

\begin{eqnarray} \label{eq:linhiggeon}
\text{d}s_g^2&=&g_{tt}dt^2+g_{xx}dx^2+r(x)^2 d\Omega^2_{(D-2)} \\
\text{d}s_q^2&=&-A(x)e^{2\psi(x)} dt^2 + \frac{1}{A(x)}dx^2 + x^2 d\Omega^2_{(D-2)},
\end{eqnarray}
(where $d\Omega^2_{(D-2)}$ is the angular sector in the maximally symmetric subspace) for the metric $g_{\mu\nu}$ and $q_{\mu\nu}$, respectively, so that the electromagnetic field satisfies $F^{tx}=\frac{Q}{r(x)^{D-2} \sqrt{-g_{tt}g_{xx}}}$. The ansatz for $\hat{\Omega}$ compatible with the symmetry of the electromagnetic field becomes

\begin{eqnarray} \label{eq:Omegaem}
\hat{\Omega}=
\left(
\begin{array}{cc}
\Omega_{+} \hat{I}_{2\times2} &  \hat{0}_{(D-2) \times 2} \\
\hat{0}_{2 \times (D-2)} & \Omega_{-} \hat{I}_{(D-2) \times (D-2)} \\
\end{array}
\right) \Rightarrow \Omega_{-}=(\lambda + \tilde{X})^{\frac{2}{D-2}}  \hspace{0.1cm};\hspace{0.05cm} \Omega_{+}=\frac{(\lambda - \tilde{X} )}{(\lambda + \tilde{X})^{\frac{D-4}{D-2}}}
\end{eqnarray}
where we have used Eq.(\ref{eq:Upshigher}). Like in the four dimensional case, the combination ${R^t}_t-{R^x}_x=0$ allows to rewrite the line element for $q_{\mu\nu}$ in Eq.(\ref{eq:linhiggeon}) in standard Schwarzschild-like form, while the introduction of a mass ansatz, $A=1-\frac{2M(x)}{(D-3)x^{D-3}}$, allows to solve the field equations for $M(x)$, and transforming that solution back to $g_{\mu\nu}$ using that $\{q_{ab}=g_{ab} \Omega_{+}; q_{mn}=g_{mn}\Omega_{-}\}$, where $(a,b)$ contains the $2\times2$ block and $(m,n)$ the maximally symmetric sector, one obtains the final solution for $g_{\mu\nu}$ as

\begin{eqnarray} \label{eq:metgeonhig}
ds_g^2&=&-\frac{A}{\Omega_{+}} dt^2 +\left( \frac{\Omega_{+}}{A} \right) \left(\frac{dx}{\Omega_{+}} \right)^2 +z^2(x) d\Omega^2 \\
A(z)&=&1-\left(\frac{1+\delta_1 G(z)}{\delta_2  \Omega_{-}^{\frac{D-3}{2}}  z^{D-3}  }\right)  \hspace{0.1cm};\hspace{0.1cm} G_z=-z^{D-2} \left( \frac{\Omega_{-}-1}{\Omega_{-}^{1/2}} \right) \left( \lambda + \frac{1}{z^{2(D-2)}} \right) \\
\delta_1 &\equiv& \frac{(D-3)r_c^{D-1}}{2 M_0 l_{\epsilon}^2}  \hspace{0.1cm};\hspace{0.1cm} \delta_2 \equiv \frac{ (D-3)r_c^{D-3}}{2M_0} \ ,
\end{eqnarray}
with the definition $z\equiv r/r_c$, where $r_c^{2(D-2)} \equiv 2l_{\epsilon}^2 r_Q^{2(D-3)}$ with $\epsilon\equiv-2l_{\epsilon}^2$ and $r_Q^{2(D-3)}\equiv\kappa^2 Q^2/(4\pi)$, while $M_0$ is Schwarzschild mass. Again, to detect the presence of wormhole structures, we just need to inspect the relation between radial coordinates in the two line elements (\ref{eq:linhiggeon}), obtained as

\begin{equation}
x^2= r^2 \Omega_{-} \Rightarrow \left(\frac{|x|}{r_c}\right)^{(D-2)}=\frac{1}{z^{D-2}}\left(z^{2(D-2)}-1\right) \ ,
\end{equation}
which is just a standard quadratic equation for $z^{d-2}$, which can consequently be solved as

\begin{equation}\label{eq:rxhigher}
r^{d-2}=\frac{|x|^{D-2}+\sqrt{|x|^{2(D-2)}+4r_c^{2(D-2)}}}{2}
\end{equation}
where the modulus in $x$ comes from the fact that a square root has been extracted to obtain (\ref{eq:rxhigher}). The behaviour of the radial function $r(x)$ is depicted in Fig.\ref{fig:whafhigher} (left), where we observe the typical bouncing behaviour of a wormhole for any dimension $D$, with the throat located at $x=0$ ($z=1$). As follows from the analysis of Bazeia et al \cite{Bazeia:2015uia}, expansions of the metric functions and the curvature scalars at the throat reveal that, as opposed to the four dimensional case, they always diverge there (in four dimensions, in the case $\delta_1=\delta_c$ they become finite, see section \ref{sec:horizonsgeon}). However, a similar analysis of the geodesic structure near the wormhole throat as in section \ref{sec:geoEibIint} reveals the completeness of null and time-like geodesics for all the spectrum of mass and charge of the solutions. The case of radial null geodesics is depicted in Fig.\ref{fig:whafhigher} (right), where we observe that they can be naturally extended beyond the wormhole throat $x=0$. However, the impact of such curvature divergences on physical observers crossing the wormhole throat has not been analyzed in the literature yet. The geonic properties of such solutions have been also analyzed in \cite{Bazeia:2015uia}, with similar qualitative results as those found in section \ref{sec:horizonsgeon}.

\subsubsection{Kaluza-Klein solutions}

The original idea of extra dimensions was implemented by Kaluza and Klein by assuming that the four dimensional energy-momentum tensor of the electromagnetic field is originated from a part of a five dimensional metric tensor. This idea was reemployed in the EiBI scenario in Ref.\cite{Fernandes:2014bka}, where they assume a five-dimensional metric given by

\begin{eqnarray} \label{eq:KK5D}
\hat{g}_{AB}=
\left(
\begin{array}{cc}
g_{\mu\nu}+\alpha A_{\mu}A_{\nu} &  \alpha  A_{\mu} \\
\alpha  A_{\mu} &  1 \\
\end{array}
\right)
\end{eqnarray}
where $\alpha$ is a parameter, latin indexes run from $A=0, \ldots , 4$ and greek from $\mu=0, \ldots ,3$, while a hat denotes five dimensional objects. Now, tuning the  compactification radius from five to four dimensions, denoted as $\tilde{R}$, to be given by $2\pi \tilde{R}/\hat{G}_5=1/G_4=1$ (where $\hat{G}_5$ and $G_4$ are the five and four dimensional Newton's constant, respectively), and taking by convenience $\alpha=4$, one obtains that the five dimensional EiBI action reduces to \cite{Fernandes:2014bka}

\begin{eqnarray}
\Ss{} &=& \frac{1}{8 \pi \hat{G}_{5} \epsilon} \int d^5x \left[\sqrt{\left|
    (\hat{g}_{AB} + {\epsilon}\hat{\mathcal{R}}_{(AB)}) \right|} -
\lambda \sqrt{\left|
    \hat{g} \right|} \right] \nonumber\\
 &\Rightarrow& \frac{1}{8 \pi G_4 \epsilon} \int d^4x \left[\sqrt{1 +
  \epsilon F^2}\right. \\
&\times & \left.\sqrt{\left| g_{\mu
      \nu} + {\epsilon}(\mathcal{R}_{(\mu \nu)} + 2  F_{\mu
      \beta} F^{\beta}_{\phantom{\beta}\nu}) +
    ({\nabla}_{\delta}F^{\delta}_{\phantom{\delta}\mu}
    {\nabla}_{\beta}F^{\beta}_{\phantom{\beta}\nu})
\sum_{n=0}^{\infty}
    \left(-1 \right)^{n + 1} {\epsilon}^{n+2}
    F^{2n}   \right|} -
\lambda\sqrt{\vert g \vert}\right] \nonumber
\label{KKaction5D4D}
\end{eqnarray}
where it is clearly seen that it transforms into a four dimensional, Class-III gravitational action containing a number of curvature-matter couplings, and where $F_{\mu\nu}$ arises as the field strength tensor associated to the vector potential $A_{\mu}$. The field equations corresponding to the action (\ref{KKaction5D4D}) are highly involved, even to lowest order in $\epsilon$ (see Eqs.(4.4)-(4.7) of Ref.\cite{Fernandes:2014bka}). Nonetheless, in the spherically symmetric case, (electrostatic) solutions to first order in $\epsilon$ can be obtained under the form

\begin{eqnarray}
ds_g^2&=&-f(r)dt^2 +f(r)^{-1}dr^2 + r^2 d\Omega^2 \\
f(r)&=&\left(1-\frac{2M}{r}- \frac{\Lambda r^2}{3}+ \frac{Q^2}{r^2}\right) + \epsilon \left(\frac{3Q^4}{10r^6} - \frac{\Lambda Q^2}{r^2} \right) + \mathcal{O}(\epsilon^2)
\end{eqnarray}
which corresponds to a modification of the Reissner-Nordstr\"om-Anti-de Sitter solution of GR ($\epsilon \rightarrow 0$). Computation of the curvature scalar for these solutions, $R = 4\Lambda + 6\epsilon Q^4/r^8+ \mathcal{O}(\epsilon^2)$, yields a curvature singularity at $r=0$, but no further properties of these solutions (such as horizons, geodesic structure, etc) are investigated in that work.

\subsubsection{Thick branes}

Braneworld scenarios represent an interesting development of the Kaluza-Klein idea, boosted by the proposals introduced by Randall-Sundrum \cite{Randall:1999ee,Randall:1999vf} and Arkani-Hamed-Dimopoulos-Dvali \cite{ArkaniHamed:1998rs,Antoniadis:1998ig} models. They assume that the four-dimensional world to which standard model particles are attached (the brane), is embedded in a higher-dimensional spacetime (the bulk) with a warped geometry, in such a way that gravitons can propagate along the extra dimension (see e.g. \cite{Rattazzi:2003ea} for a review). Though in the original proposals the brane is infinitely thin, in this section we shall consider instead a \emph{thick} brane, namely, a five-dimensional bulk with a scalar field propagating in the extra dimension, and whose energy density is assumed to be localized around the point (say) $y=0$ of the extra dimension. The analysis of this scenario can be carried out to a large degree of generality, by considering a Born-Infeld inspired modification of gravity (in Palatini formalism) defined by an arbitrary function $F$ of the object ${P^\mu}_\nu\equiv g^{\mu\lambda}\mathcal{R}_{(\lambda\nu)}$ and coupled to a scalar field as

\begin{equation} \label{eq:actionbrane}
\Ss{}=\frac{1}{2\kappa^2}\int d^D x\sqrt{-g}F(\hat{P})+\int d^{D}x \sqrt{-g} \lag(X,\phi)
\end{equation}
where $D=d+1$ is the number of spacetime dimensions. The Lagrangian density $\lag(X,\phi)$ contains, in general, a non-canonical contribution from the scalar field kinetic term $X\equiv g^{\alpha\beta}\partial_\alpha\phi \partial_\beta\phi$ (see \cite{ArmendarizPicon:2000dh} for the inception of these theories in Cosmology). The field equations for this system are derived in the usual way, i.e., by performing independent variations of the action (\ref{eq:actionbrane}) with respect to the metric and the connection, which can be handled also by introducing a new metric $q_{\mu\nu}$ as \cite{Bazeia:2015zpa}

\begin{equation} \label{eq:qgbrane}
q^{\mu\nu}=\frac{1}{|\hat{F}_{\hat{P}}|^{\frac{1}{D-2}}}g^{\mu\lambda} {(F_{\hat{P}})_\lambda}^\nu  \hspace{0.1cm};\hspace{0.1cm}
q_{\mu\nu}=|\hat{F}_{\hat{P}}|^{\frac{1}{D-2}}{(F^{-1}_{\hat{P}})_\mu}^\lambda g_{\lambda\nu}
\end{equation}
where ${(F_{\hat{P}})_\lambda}^\nu\equiv \frac{\partial F}{\partial {P^\lambda}_\nu}$ and $\vert \hat{F}_{\hat{P}} \vert$ represents its determinant. The resulting field equations are quite similar to those of EiBI gravity given by (\ref{eq:Rmnhigher}), which are written here by convenience as

\begin{equation}\label{eq:Rmn-q}
{R^\nu}_\alpha(q)=\frac{\kappa^2}{|\hat F_{\hat{P}}|^{\frac{1}{D-2}}}\left(\lag_G{\delta^{\nu}}_\alpha+ {T^{(\phi) \nu}}_{\alpha} \right) \ ,
\end{equation}
where $\lag_G$ corresponds to the particular Lagrangian density considered. To implement the thick brane scenario one sets the line element for the physical metric $g_{\mu\nu}$:

\begin{equation}
\text{d}s_g^2=a^2(y)\eta_{ab} dx^a dx^b + dy^2
\end{equation}
where $a(y)$ is the \emph{warp} factor, which is assumed to depend only on the extra dimension $y$, and $\eta_{ab}$ is the metric on a $d$-dimensional spacetime brane of constant curvature $K$. The corresponding field equations (\ref{eq:Rmn-q}) in this case can be conveniently written with the help of a similar ansatz for the auxiliary metric $q_{\mu\nu}$:

\begin{equation}
\text{d}s_q^2=\tilde{a}^2(\tilde{y})\eta_{ab} dx^a dx^b + d\tilde{y}^2 \ .
\end{equation}
Using (once more) the relation (\ref{Eq:defOmega}) it follows that in this case

\begin{eqnarray} \label{eq:Omunu}
{\Omega_\mu}^\lambda\equiv |\hat F_{\hat{P}}|^{\frac{1}{D-2}}{(F^{-1}_{\hat{P}})_\mu}^\lambda&=&
\left(
\begin{array}{cc}
\Omega_+  I_{d \times d} &  \hat{0} \\
\hat{0}  & \Omega_-  \\
\end{array}
\right) \ ,
\end{eqnarray}
from where one obtains that $\tilde{a}^2(\tilde{y})=\Omega_+ a^2(y) $ and $d\tilde{y}^2=\Omega_-dy^2$. The gravitational field equations follow now immediately as (see Bazeia et al \cite{Bazeia:2015zpa} for details)

\begin{eqnarray}\label{eq:bgd0}
d(d-1)[K-H^2]&=&\frac{\kappa^2}{|\Omega|^{1/2}}\Bigl[(d-1)\lag_G+d \cdot T_+-T_-\Bigr] \label{eq:branegrav1} \\
(d-1)[K+H_{\tilde{y}}]&=& \frac{\kappa^2}{|\Omega|^{1/2}}\left(T_+-T_-\right) \label{eq:branegrav2}
\end{eqnarray}
where $H\equiv {\tilde{a}_{\tilde{y}}}/{\tilde{a}}$, while  $T_+=-\lag(\phi,X)/2$ and $T_-=\lag_X \phi_y^2+T_+$ are the components of the scalar energy-momentum tensor, and $\Omega_{\pm}$ are model-dependent functions of $X$ and $\phi$. For the sake of the search for solutions below, $\lag_G$ in this equation represents EiBI gravity Lagrangian (\ref{Eq:actionEiBI}).

For the case of standard canonical kinetic term with a potential, $\lag=X-V(\phi)$, specific solutions were obtained by Liu et al. \cite{Liu:2012rc}, using a fully equivalent approach to the one depicted above though written directly in terms of the functions $\{a,\tilde{a}\}$, see Eqs.(17a), (17b) and (17c) of that paper. Specifically, they look for a \emph{kink} solution interpolating between different vacua at asymptotic infinity $y=\pm\infty$. This can be achieved by introducing an additional constraint $\phi'(y)=K a^2(y)$, where $K$ is a constant conveniently defined as $K=\pm\big(\frac{7}{3}\big)^{\frac{3}{4}}\frac{1}{2\sqrt{\epsilon \kappa^2}}$. This way, the scalar field equation

\begin{equation}
4\frac{a'}{a}\phi'+\phi''=\frac{\partial V(\phi)}{\partial \phi}
\end{equation}
can be integrated with the result $V(y)=\frac{3}{2}K^2a(y)^4+V_0$, where $V_0$ is an integration constant that can be interpreted as the scalar field vacuum energy, fixed here as $V(\phi)=-\lambda/(\kappa \epsilon)$. Inserting this result into the gravitational field equations (\ref{eq:branegrav1}) and (\ref{eq:branegrav2}) one finds an analytic solution for the warp factor and scalar field profile in closed form as
\begin{eqnarray}
a(y)&=&\text{sech}^{\frac{3}{4}}\left({\frac{2}{\sqrt{21\epsilon}}y}\right),\label{wfsol}\\
\phi(y)&=&\pm\frac{7^{5/4}}{2\times 3^{1/4}\kappa}\left[i \text{E}\left(\frac{iy}{\sqrt{21 \epsilon}},2\right)+\text{sech}^{\frac{1}{2}}\left({\frac{2y}{\sqrt{21\epsilon}}}\right)
\times\sinh\left({\frac{2y}{\sqrt{21\epsilon}}}\right)\right] ,\label{sfsol}
\end{eqnarray}
where $E$ is an elliptic integral of second kind. This way the potential can be expressed as $V(y)=\frac{7\sqrt{21}}{24\epsilon\kappa^2}\text{sech}^{3}({\frac{2y}{\sqrt{21\epsilon}}})-\frac{\lambda}{\epsilon\kappa^2}$, which allows to compute the energy density associated to these solutions as

\begin{equation}
\rho(y)=\frac{7\sqrt{21}}{18\epsilon\kappa^2}\text{sech}^{3}\left({\frac{2y}{\sqrt{21\epsilon}}}\right).
\end{equation}
These scalar field and energy density profiles naturally implement the defining properties of a kink, namely, its interpolating character between different vacua at $y=\pm\infty$, as well as the localized nature of the energy density around the center of the kink, $y=0$. Regarding the curvature of the solutions, a simple calculation yields the result $R=g^{MN}R_{(MN)}=\frac{1}{\epsilon}\left[2-7\tanh^2(\frac{2}{\sqrt{21\epsilon}})y\right]$, which asymptotically approaches the value $R \rightarrow -5/\epsilon<0$, corresponding to an Anti-de Sitter space.

%\begin{figure}[h]
%\centering
%\includegraphics[width=0.45\textwidth]{fig_branescalar.eps}
%\includegraphics[width=0.45\textwidth]{fig_branenergy.eps}
%\caption{Left plot: scalar field profile for two values of the EiBI constant (in this plot $\epsilon \rightarrow b$). Right plot: the corresponding energy density. Figures extracted from Ref.\cite{Liu:2012rc}. \label{fig:kink}}
%\end{figure}

An important aspect of configurations on the brane is to determine its stability against tensorial perturbations there. As shown by Bazeia et al. \cite{Bazeia:2015zpa} this can be also done in full generality for a theory $F(\hat{P})$\footnote{A general treatment of tensorial perturbations in EiBI gravity can be found in \cite{Yang:2013hsa}.}. The idea is to write two perturbed line elements in Gaussian normal coordinates as

\begin{eqnarray}
ds_g^2&=&a^2(y)\left(\eta_{ab}+h_{ab} \right)dx^adx^b+dy^2 \label{eq:tenper} \\
ds_q^2&=&\tilde{a}^2(\tilde{y})\left(\eta_{ab}+h_{ab} \right)dx^adx^b+d\tilde{y}^2 \label{eq:tenpera}
\end{eqnarray}
where the scalar and vector modes are decoupled by imposing the conditions $\delta g_{ab}=a^2(y) h_{ab}$ and $\delta g_{ay}=0=\delta g_{yy}$. From (\ref{eq:tenper}) and (\ref{eq:tenpera}), the perturbation of the field equations (\ref{eq:Rmn-q}) in the $q_{\mu\nu}$ geometry reads simply $\delta {\mathcal{R}^\mu}_\nu (q)=0 \rightarrow \delta \mathcal{R}_{(\mu\nu)}(q)={\mathcal{R}_\mu}^\beta t_{\beta\nu}$, where $t_{ab}=\tilde{a}^2 h_{ab}$ is the only non-vanishing component of $t_{\beta\nu}$. Now, using standard covariant perturbation methods, and after some algebra, the tensorial modes, assumed to be written as ${h_a}^{b}=X(z){\epsilon_a}^b(t,\vec{x})$ (where we have introduced a new coordinate $z$ as $dz^2=d\tilde{y}^2/\tilde{a}^2$), satisfy two sets of equations, namely

\begin{eqnarray}
^{(\eta)}\Box {\epsilon_a}^b-2K{\epsilon_a}^b-p^2{\epsilon_a}^b&=&0 \\  \label{eq:effKG}
-Y_{zz}+V_{eff}(z)Y&=&p^2Y \label{eq:effpot}
\end{eqnarray}
where $p^2$ is a constant. (\ref{eq:effKG}) is a Klein-Gordon-type equation for the massless, $p=0$, and massive, $p \neq 0$, gravitons, while in the Schr\"odinger-like equation (\ref{eq:effpot}) for the Kaluza-Klein modes we have redefined $X=\tilde{a}^{-\frac{(d-1)}{2}}Y$, and the effective potential $V_{eff}(z)$ is given by

\begin{equation} \label{eq:Veffbrane}
V_{eff}=\frac{(d-1)}{2}\mathcal{H}_z+\frac{(d-1)^2}{4}\mathcal{H}^2 \ ,
\end{equation}
where $\mathcal{H}\equiv \tilde{a}_z/\tilde{a}$. The operator on the left-hand side of Eq.(\ref{eq:effpot}) can be factorized as

\begin{equation}
\left(\frac{d}{dz}-\frac{(d-1)}{2}  \mathcal{H} \right) \left( \frac{d}{dz}-\frac{(d-1)}{2}\mathcal{H} \right)
\end{equation}
which is a non-negative operator, guaranteeing in this way that $p^2>0$, which implies the tachyonic-free and stable character of this class of theories of gravity under tensor perturbations. For the particular case of EiBI gravity, Liu et al. \cite{Liu:2012rc} computed the zero mode, $p=0$, as $\Psi_0(z)=N_0~a^{7/2}(z)$, where the normalization condition $\int\Psi^2(z)dz=1$ fixes the constant $N_0^2\approx{0.35}/{\sqrt{\epsilon}}$. This gravity zero mode is localized at the center of the kink, $y=0$, while vanishes at $y=\pm\infty$. The effective potential (\ref{eq:Veffbrane}) has a (asymptotically vanishing) volcano-like profile with a well at the center of the kink, with the result that a continuous set of massive Kaluza-Klein modes (not localized on the brane) arises for $p>0$.

The above study was further generalized by Fu et al. \cite{Fu:2014raa}, where they considered a class of solutions defined by the ansatz $\phi'(y)=Ka(y)^{2n}$, so that Liu et al. case \cite{Liu:2012rc} corresponds just to $n=1$. The corresponding solutions for the warp factor and the scalar field can also be obtained in closed analytical form as

\begin{equation}
a(y)=\text{sech}^{\frac{3}{4n}}\left({ky}\right) \hspace{0.1cm} ;\hspace{0.1cm} \phi(y)=\frac{2K}{k} \Big[i \text{E}(iky/2,2) {\text{sech}^{1/2}(ky)}~\text{sinh}(ky)\Big]
\end{equation}
where the constants $K=\pm \frac{(1+4n/3)^{3/4}}{(n+1)} \sqrt{\frac{n}{\epsilon \kappa^2}}$ and $k=\frac{2n}{\sqrt{3\epsilon(4n+3)}}$, while the energy density can be computed simply as $\rho= \frac{n+1}{n}K^2 \text{sech}^3(ky)$. These configurations show similar features as those of $n=1$ below, namely, interpolation of the kink between two asymptotic vacua at $y=\pm\infty$ and localized character around $y=0$ with a maximum of the energy density there. The impact of increasing the value of $n$ is just to decrease the width of the kink and to lower the maximum of the energy density. One could go on further in the standard strategy in the field, by investigating additional models which allow to modify the physical properties of the kink at will, but we shall stop here. Let us simply emphasize that the zero mode for any $n$ not localized in the brane, while more complex models like $\phi'(y)=K_1a(y)^2(1-K_2a(y)^2)$ allow to find quasi-localized states on the brane for massive KK gravity modes.

\subsubsection{Three dimensions} \label{sec:hdim3d}

\begin{figure}[htb]
\centering
\includegraphics[width=0.45\textwidth]{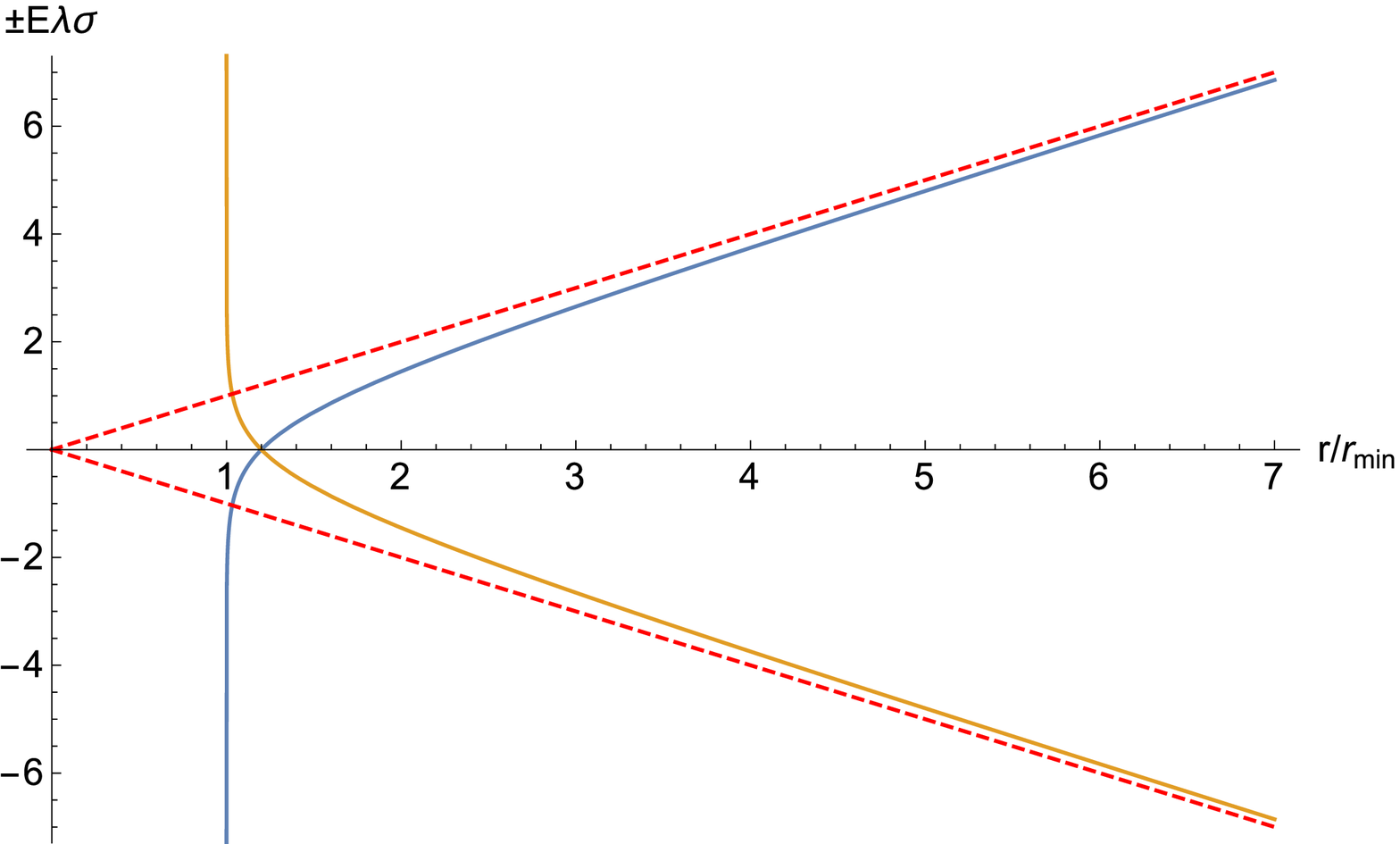}
\includegraphics[width=6.0cm,height=4.5cm]{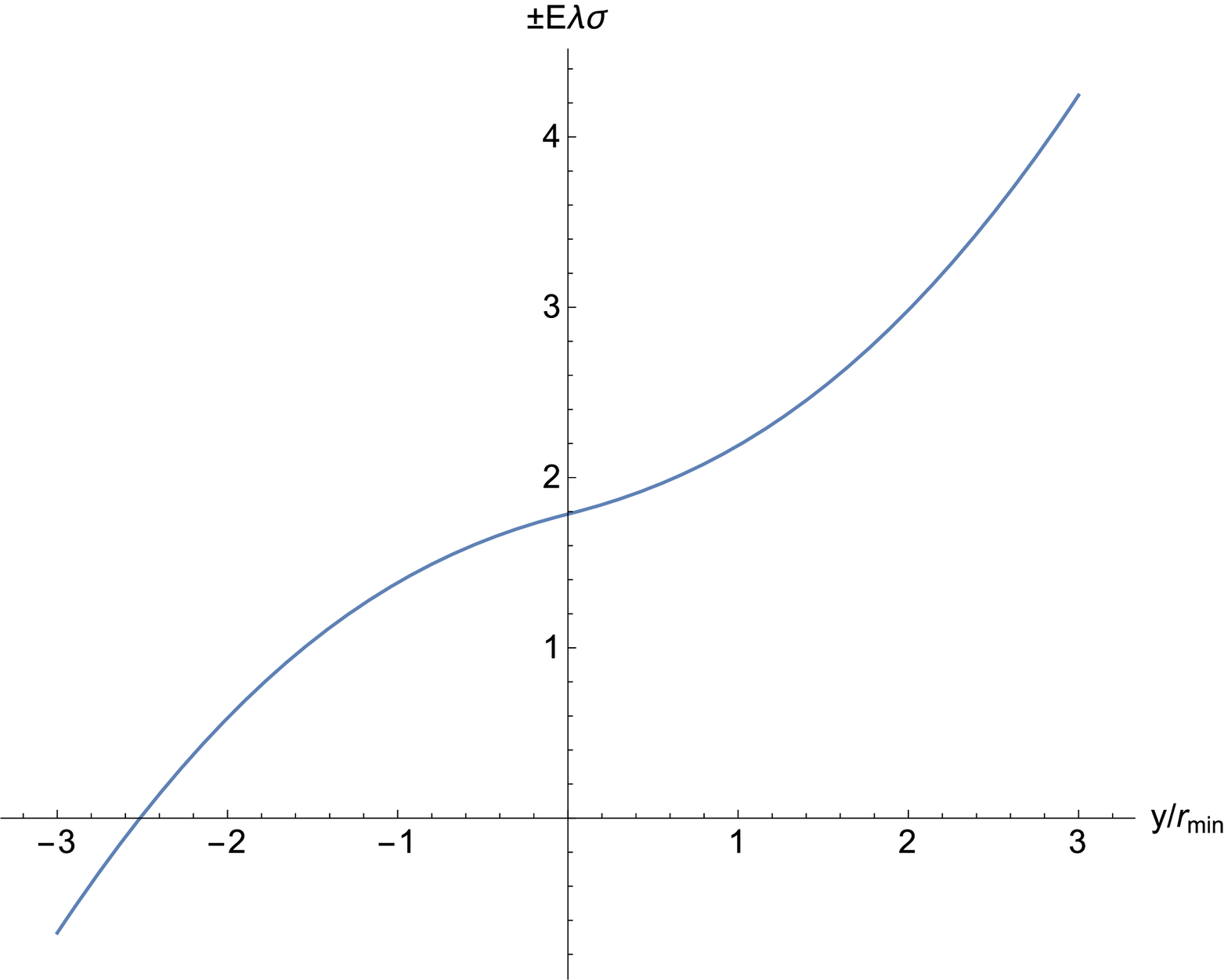}
\caption{Left plot: affine parameter $u(x)$ (in this plot $u\rightarrow \sigma$) for ingoing and outgoing radial null geodesics in the case $s=-1$, as compared to the GR case (dashed lines). In this case, the wormhole lies on the future (or past) boundary of the spacetime. Right plot: Affine parameter $u(r(y))$ (where $y$ is a new suitable radial coordinate) for radial null geodesics in the case $s=+1$, where the wormhole is reached in finite time but can be indefinitely extended. Figures extracted from Ref.\cite{Bazeia:2016rlg}.  \label{fig:affine3dim}}
\end{figure}

Electrovacuum solutions of EiBI gravity in $D=3$ dimensions requires a separate analysis from that of section \ref{sec:hdimBH}, due to the peculiarities of the integration of the metric on such a case. In this sense, the expressions (\ref{eq:Rmnhigher}), (\ref{eq:Upshigher}), (\ref{eq:gq}), (\ref{eq:linhiggeon}), (\ref{eq:Omegaem}) are still valid, but the integration of the metric (with a cosmological constant term, $\lambda \neq 1$) yields now the result
\begin{eqnarray}
ds_g^2&=&-\frac{A(r)}{\Omega_{+}}dt^2 + \frac{1}{A(r)} \left( \frac{dx}{\Omega_{+}^{1/2}}\right)^2 + r^2 (x) d \theta^2  \label{eq:line3dim} \\
A(x) &=&-\lambda^2 M- \frac{\lambda^2-1}{2s|\epsilon|} r^2 -Q^2\left( \frac{sr_c^2}{2r^2}
+\frac{1}{\lambda}  \ln\left[\frac{r^2 + s r_c^2/\lambda}{r_0^2} \right]\right) \ ,
\end{eqnarray}
where $s$ in $\epsilon =s \vert \epsilon \vert$ is the sign of $\epsilon$, and $r_0$ is an integration constant. The line element (\ref{eq:line3dim}) represents a natural generalization of Ba\~nados, Teitelboim and Zanelli (BTZ) solution \cite{Banados:1992wn}, which is recovered both in the limit $\epsilon \rightarrow 0$, and asymptotically, $r \gg 1$. The BTZ solution raised a great deal of interest due to the fact that the states with $M=-1$ does not contain an event horizon but there is no curvature singularity to hide, either. In the EiBI scenario, the function $r(x)$ in Eq.(\ref{eq:line3dim}) can be explicitly written as

\begin{equation} \label{eq:r(x)}
|r(x)|=\frac{|x|\pm \sqrt{|x|^2-4s\lambda r_c^2}}{2\lambda} \ ,
\end{equation}
which attains a minimum at $r=r_c/\lambda^{1/2}$ both for $s= \pm 1$. When $s=-1$ one obtains a wormhole structure similar to that of the higher-dimensional case (compare with Eq.(\ref{eq:rxhigher})), while for $s=+1$ a similar construction as the Einstein-Rosen bridge \cite{Einstein:1935rr} can be obtained. In both cases, null and time-like geodesics can be indefinitely extended despite the presence of curvature divergences at the wormhole throat. However, this is done via two different mechanisms: when $s=-1$ the wormhole lies on the future (or past) boundary of the spacetime, as radial null geodesics take an infinite time to reach the wormhole throat\footnote{A similar result has also been found in other theories of gravity  formulated in the Palatini approach, like $f(\mathcal{R})$ \cite{Olmo:2015axa,Bambi:2015zch,Bejarano:2017fgz}.} (see Fig.\ref{fig:affine3dim}, left), while when $s=+1$, the wormhole is reached on a finite affine time but, like their four and higher dimensional counterparts (see sections \ref{sec:geoEibIint} and \ref{sec:hdimBH}, respectively), it can be extended beyond this point to arbitrarily large values of the affine parameter (see Fig.\ref{fig:affine3dim}, right). This way, all the electrically charged solutions of EiBI gravity with a wormhole structure are geodesically complete in $D \geq 3$ spacetime dimensions.

Three dimensional, asymptotically flat, circularly symmetric charged solutions within the context of Born-Infeld inspired gravity formulated in Weitzenb\"{o}ck spacetime (Class-II) have been found by Ferraro and Fiorini \cite{Ferraro:2009zk}. This is a formulation of classical gravity in terms of a spacetime possessing absolute paralelism (or teleparallel gravity, see Ref.\cite{Hehl:1994ue}). The action considered in this work is defined as

\begin{equation}   \label{acciondet}
\Ss{}_{BIT}=\frac{1}{2\kappa^2 \epsilon (A+B)}\int d^{3}x \left[ \sqrt{|g_{\mu \nu }+2\epsilon F_{\mu \nu }|}-\lambda \sqrt{|g_{\mu \nu }|}\right],
\end{equation}
where $F_{\mu \nu }=AS_{\mu \lambda \rho}T_{\nu }^{\,\,\,\lambda \rho }+BS_{\lambda \mu \rho }T_{\nu }^{\lambda \rho }$ (with $A$ and $B$ some constants) is quadratic in the Weitzenb\"{o}ck torsion  $T_{\ \ \mu \nu }^{\rho}=e_{a}^{\rho }\,(\partial _{\mu }e_{\nu }^{a}-\partial _{\nu}e_{\mu }^{a})$ build out of the set of $3$-forms $\{e^{a}(x)\}$, with the definitions (\ref{Eq:DefSuperS}) and (\ref{Eq:DefTinv}). In the limit $\epsilon \rightarrow 0$ the teleparallel version of GR (TEGR) is obtained, which are equivalent to each other since the curvature scalar of the Levi-Civita connection, $R$, can be written as $R=S_{\rho }^{\ \ \mu \nu } T^{\rho }_{\ \ \mu \nu } $ ($+$ total derivative terms).

Upon resolution of the corresponding field equations for this theory one obtains the line element \cite{Ferraro:2009zk}

\begin{equation} \label{eq:metricweitz}
ds^{2}=\left( \frac{J^{2}}{4r^{2}}+M^{2}\right) \ dt^{2}
-\left( \frac{Y(r)^{2}}{J^{2}/(4\,r^{2})+M^{2}}\right) dr^{2}-r^{2}\left( -%
\frac{J}{2\,r^{2}}\ dt+d\theta \right) ^{2}
\end{equation}
described by a mass $M$ and an angular momentum $J$, while the function $Y$ is determined via the cubic equation $Y^{2}-Y^{3}=\epsilon J^2/(4 r^4)=\Delta$, and out of the three solutions of this system, imposing recovery of the GR limit, $Y=1$ for $\Delta \rightarrow 0$, one gets the result

\begin{equation}
3\ Y=1+\left( 1-\frac{27\Delta}{2}  -\frac{3}{2}\sqrt{3\Delta
\,(27\Delta -4)}\right)^{-1/3}
+\left(1-\frac{27\Delta}{2} -\frac{3}{2}\sqrt{3\Delta
\,(27\Delta -4)} \right)^{1/3}.
\end{equation}
This geometry can be written, using a suitable change of coordinates given by $\{t,r \} \rightarrow \{T=Mt+J\theta/(2M),\rho=M^{-2}(J^2/4+M^2 r^2)^{1/2} \}$, as

\begin{equation} \label{eq:metriconical}
ds^{2}=dT^{2}-Y(\rho)^{2}\ d\rho^{2}-M^{2}\rho^{2}d\theta ^{2}
\end{equation}
so that the TEGR limit, $\epsilon \rightarrow 0$, is naturally recovered. To further understand the geometry (\ref{eq:metriconical}) one can consider the behaviour of the curvature scalars (in the case $\epsilon <0$)

\begin{equation}
R=\frac{2Y(\rho)'}{\rho Y(\rho)^3}=\frac{2Y(r)'}{r Y(r)^3} \hspace{0.1cm};\hspace{0.1cm}
R^{(\mu\nu)}R_{(\mu\nu)}=\frac{1}{2}\, R^2  \hspace{0.1cm};\hspace{0.1cm}
R^\alpha_{\,\,\,\beta\gamma\delta}\,
R_{\alpha}^{\,\,\,\beta\gamma\delta}=R^2 \label{inva1}.
\end{equation}
Ferraro and Fiorini analyse the structure of this spacetime in the two regions of interest. At asymptotic infinity, $\rho \rightarrow \infty$, where $Y \rightarrow 1$, all these scalars vanish, and the geometry (\ref{eq:metriconical}) describes a BTZ-type spacetime with a conical singularity. On the central region, $r \rightarrow 0$, the scalars vanish as well. In particular, the curvature scalar behaves as $R \sim -\frac{16}{3} \left(\frac{\sqrt{2}  r}{|\epsilon|J^2}\right)^{2/3}$ . The physical interpretation of this geometry is that of a spacetime with a deficit angle ranging between $2\pi (1-M)$ at spatial infinity and $2\pi$ at $r=0$, corresponding to the circle of minimum radius $\rho_0=J/(2M^2)$ that can be attained in this geometry. Nonetheless, as radial null geodesics satisfy $dT=Yd\rho$, this means that $T$ diverges as a light ray approaches the minimal circle of radius $\rho_0$, so they take an infinite affine time to reach it and the same applies for time-like geodesics. Therefore, this approach succeeds in removing the conical singularity of GR (and, as the same time, it removes the possibility of existence of closed time-like curves) in much the same way than electrically charged black holes in the $s=+1$ case of EiBI discussed above in this section, i.e., by setting the location of the wormhole throat at the future (or past) boundary of spacetime.

Further analysis in three-dimensional scenarios, involving a Born-Infeld extension of New Massive Gravity \cite{Gullu:2010pc} with a Chern-Simons term (Class-III), and defined by the action

\begin{eqnarray} \label{eq:NMA}
\Ss{}&=&\frac{2m^2}{\kappa^2} \int d^3x \left[\sqrt{-\det(g_{\mu\nu} -m^{-2} G_{\mu\nu} +aF_{\mu\nu}) } -\left(1+\frac{\Lambda}{2m^2} \right) \sqrt{-\det(g_{\mu\nu})} \right] \nonumber \\
&+&\frac{\mu}{2} \int d^3x \varepsilon^{\mu\nu\rho} A_{\mu}\partial_{\nu}A_{\rho}
\end{eqnarray}
where $m$ is a mass scale, $\Lambda$ represents a cosmological constant and $a$, $\mu$ are some constants, has been considered in \cite{Alishahiha:2010iq}. In that work only Anti-de Sitter spacetimes are studied, while black holes were investigated instead in \cite{Ghodsi:2011ua} and subsequently in \cite{Ghodsi:2010ev} where, by expanding the New Massive Gravity action (\ref{eq:NMA}) to four and six derivative terms, the authors develop a method to find evidence of uncharged and charged black holes, but little is said about the deviations of such solutions with respect to the structure of the GR counterparts relevant for this review.

\subsection{Magnetically charged solutions with cylindrical symmetry}

Cylindrically symmetric solutions have only been considered in EiBI theories in the context of magnetically charged configurations (i.e. Melvin-type \cite{Melvin:1963qx}) by Bambi et al \cite{Bambi:2015sla}. The two line elements compatible with such a symmetry can be conveniently written as
\begin{eqnarray}
\text{d}s_g^2&=&f(\rho)(-dt^2+dz^2)+g(\rho)d\rho^2+h(\rho)\rho^2d\varphi^2   \label{eq:metricaxialphys} \\
\text{d}s_q^2&=& \tilde{f}(\rho)(-dt^2+dz^2) + \tilde{g}(\rho)d\rho^2 + \tilde{h}(\rho) \rho^2 d\phi^2 \ . \label{eq:metricaxialaux}
\end{eqnarray}
From the line element (\ref{eq:metricaxialphys}) the only non-vanishing component of Maxwell field equations, $\nabla_{\mu}F^{\mu\nu}=0$, reads $F^{\rho \varphi}=\beta/(\rho f \sqrt{gh})$, where $\beta$ is an integration constants related to the intensity of the magnetic field. The energy-momentum tensor (\ref{eq:Tmunuem}) for these solutions allows to find the matrix $\Omega$ in Eq.(\ref{eq:Upshigher}) as

\begin{equation} \label{eq:Tmnaxialmag}
{T^\mu}_{\nu}=\frac{X}{8\pi} \text{diag}(1,1,-1,-1) \Rightarrow
\hat{\Omega}=
\begin{pmatrix}
\Omega_{+} \hat{I} & \hat{0}  \\
\hat{0} & \Omega_{-} \hat{I}
\end{pmatrix}
\hspace{0.2cm};\hspace{0.2cm} \Omega_{\pm}=1 \pm \frac{f_c^2}{f^2} \ ,
\end{equation}
where $X=-\beta^2/f^2$, $f_c= l_\epsilon/l_{\beta}$ and $l_\beta^2=4\pi/(\kappa^2 \beta^2)$. With this matrix at hand, by the transformation (\ref{Eq:defOmega}) one finds the relations $\{\tilde{f}=\Omega_+ f, \tilde{g}=\Omega_{-}g, \tilde{h}=\Omega_- h\}$ between the metric functions in Eqs.(\ref{eq:metricaxialphys}) and (\ref{eq:metricaxialaux}). The first of these relations can be written as $f=\frac{\tilde{f} + \sqrt{\tilde{f}^2-4f_c^2}}{2}$ and implies that $\tilde{f}\ge 2f_c$, the equality corresponding to $f=f_c$ and $\Omega_{-}=0$. Now, since EiBI Lagrangian density reads now $\lag_G=\frac{\sqrt{ \det \hat{\Omega}}-1}{-2{\kappa}^2 l_\epsilon^2}=\frac{\beta^2f_c^2}{8\pi f^4}$, the field equations for $q_{\mu\nu}$ become

\begin{equation} \label{eq:Rmnqaxial}
{R^\mu}_{\nu}(q)=-\frac{\epsilon^2\beta^2}{2f^2} \left(
\begin{array}{cc}
\frac{1}{\Omega_+} \hat{I} & \hat{0} \\
\hat{0} & -\frac{1}{\Omega_-} \hat{I}  \\
\end{array}
\right) \ .
\end{equation}
Computing the components of the Ricci tensor corresponding to the line element (\ref{eq:metricaxialaux}), and by taking appropriate combinations of the field equations (\ref{eq:Rmnqaxial}) one obtains two independent equations

\begin{equation}
\left(\frac{\tilde{h}_\rho}{\tilde{h}}+\frac{2}{\rho}+\frac{2\tilde{f}_{\rho}}{\tilde{f}}\right)=\frac{2\tilde{f}_{\rho\rho}}{\tilde{f}_\rho} \hspace{0.3cm};\hspace{0.3cm} \label{eq:axial1}
\tilde{f}_{\rho\rho}-\frac{3}{4}\frac{\tilde{f}^2_\rho}{\tilde{f}}=\frac{\kappa^2\beta^2}{8\pi}\frac{\tilde{f}}{f}
\end{equation}
The first equation (\ref{eq:axial1}) can be directly integrated as $\tilde{h}\rho^2=\alpha (\tilde{f}_{\rho}/\tilde{f})^2$, where $\alpha$ is an integration constant. To solve the second one in (\ref{eq:axial1}), in \cite{Bambi:2015sla}  the definitions $\tilde{f}=2f_c \phi(x)$, $\rho^2=\frac{8\pi f_c}{\kappa^2\beta^2}x^2$ are introduced together with the new function $\Omega=\phi_x^2$ (so that $d\Omega/d\phi=2\phi_{xx}$), in terms of which one finds the solution

\begin{equation}
\Omega=C \phi ^{\frac{3}{2}}+\frac{4 \phi ^2}{3} \left(\phi -\sqrt{\phi ^2-1}\right)
- \frac{8 \phi}{3}\  _2F_1\left(\frac{1}{4},\frac{1}{2};\frac{5}{4};\frac{1}{\phi ^2}\right)
\end{equation}
where tuning the integration constant $C=-4/3+2 \sqrt{\pi } \Gamma \left(\frac{5}{4}\right)/\Gamma \left(\frac{7}{4}\right)\approx 2.16274$ guarantees the real character of $\Omega$ when expanded around $\phi \approx 1$. Unfortunately, it is not possible to integrate $\Omega$ to obtain $\phi(x)$ in analytic form, though one can resort to analytical expansions in the relevant regions. For $\phi(x) \gg 1$ one obtains the solution $\phi(x)=4 (1 + (C x)^2/32)^2/C^2$, which is nothing but the Melvin solution of GR \cite{Melvin:1963qx}, such that the line element reads

\begin{equation}\label{eq:linefaraxial}
\frac{ds^2}{f_c}\approx 2\left(\frac{2}{C}\right)^2\left(1+\frac{C^2x^2}{32}\right)^2\left[-dt^2+dz^2+{d\rho^2}\right]
+\left(\frac{C}{2}\right)^2\frac{\rho^2}{\left(1+\frac{C^2x^2}{32}\right)^2} d\varphi^2 \ ,
\end{equation}
which, via a constant rescaling of $(t,z,\rho)\to (\lambda t,\lambda z,\lambda\rho)$ with $\lambda^2=64f_c/C^2$, becomes the GR solution. In the other limit, $\phi(x) \rightarrow 1$, the corresponding field equation $\phi_x^2\approx 2(\phi-1)$, with a rescaling of the form $dx^2/x= dy^2$, yields the line element

\begin{equation}
\frac{ds^2}{f_c}\approx\left[-dt^2+dz^2+\frac{\rho_0^2}{4}{dy^2}\right]+\frac{\alpha}{8f_c\rho_0^2}y^2 d\varphi^2 \ ,
\end{equation}
up to first order in $y^2$. This is just another Minkowski spacetime near the axis as follows from the definitions $r=\rho_0 y/2$ and $\alpha\equiv 2f_c \rho_0^4$ (the constant factor $f_c$ can be reabsorbed via another global rescaling of units). This kind of Melvin-type spacetimes are of great interest in the context of the generation of pairs of entangled black holes in high-intensity magnetic fields via instantons \cite{Garfinkle:1990eq,Garfinkle:1993xk,Dowker:1993bt,Emparan:1995je}. Indeed, very recently $O(4)$ instantons have been studied in the context of the EiBI theory \cite{Arroja:2016ffm}, with the result that both the physical metric and curvature scalars are finite. However, curvature divergences arise on the auxiliary metric, which in turn may induce the formation of singularities, as discussed in detail in section \ref{Sec:Frames}, and be problematic at the quantum level. In view of this, it would be convenient to investigate further and clarify the physical role played by the auxiliary metric.

\subsection{Final remarks}

In this section we have reviewed the developments on black hole physics in Born-Infeld inspired modifications of gravity described in section \ref{Sec:BITheories}. Due to the fact that the Schwarzschild black hole is a vacuum solution of such theories,
the literature on the topic has searched for scenarios going beyond it. In this sense, though astrophysically realistic black holes are not expected to have a significant amount of charge, the investigation of charged black holes is relevant in order to find theoretical insights on the modifications to their innermost structure, as well as observational deviations from the predictions of the Kerr black hole. In the influential paper of Ba\~nados and Ferreira \cite{Banados:2010ix}, where a coupling to Maxwell field was considered, a static, spherically symmetric geometry is obtained (for $\epsilon>0$), whose properties were further extended and complemented by several other authors \cite{Wei:2014dka,Sotani:2014lua}. The case of similar electrically charged black holes for $\epsilon<0$ was also considered  \cite{Olmo:2013gqa}, for which non-singular configurations can be found \cite{Olmo:2015dba}, results partially extended to the coupling to Born-Infeld electrodynamics \cite{Jana:2015cha}. On the other hand, wormhole solutions have been found using anisotropic fluid as the matter source \cite{Harko:2013aya}, though they violate the energy conditions. Finally, higher and lower dimensional models have been the subject of different investigations, but their contributions to fundamental issues has been meager.

There are many open challenges regarding the understanding of black holes in these theories. In particular, rotating solutions in these theories have not been found yet\footnote{In this sense we point out that the applicability of the Janis-Newman method (which allows to obtain a rotating solution from a seed static metric, see Erbin for a review \cite{Erbin:2016lzq}) in the context of Born-Infeld inspired modifications of gravity is still to be understood.}. For $\epsilon>0$, EiBI gravity black holes still require further analysis regarding its innermost structure and the possibility of finding a wormhole core there, and the physics of mass inflation requires further refinement beyond the approximations employed in the analysis of \cite{Avelino:2016kkj,Avelino:2015fve}. On the other hand, though the physics at the photon sphere has been explored and understood to some detail \cite{Wei:2014dka,Sotani:2015ewa}, much research is still needed in order to obtain observational signatures for gravitational waves out of the merging of two such black holes, as well as the potential existence of gravitational echoes in this context \cite{Cardoso:2016oxy,Abedi:2016hgu,Cardoso:2016rao,Barcelo:2017lnx}. For $\epsilon<0$ the existence of non-singular solutions in EiBI gravity has been studied with great detail regarding geodesic completeness \cite{Olmo:2015bya}, but the physical meaning of curvature divergences still calls for an understanding \cite{Olmo:2016fuc}. The seemingly absence of pathologies in such curvature-divergent cases raises questions about what are the geometric degrees of freedom that quantum gravity should quantise, and what infinities should renormalise, if any. Two other interesting issues would be to investigate the existence of hairy black holes and superradiance, found in GR \cite{Herdeiro:2015waa}, in these theories\footnote{Indeed, very recently it was found evidence on the existence of wormhole configurations above a certain mass threshold when a free static and spherically symmetric scalar field is let to gravitate under the Born-Infeld dynamics \cite{Afonso:2017aci}.}, as well as to extend the thermodynamic laws studied in other Palatini theories of gravity such as $f(R)$ \cite{Bamba:2010kf} to the Born-Infeld scenario. To conclude, though many appealing results have been found in the context of Born-Infeld inspired modifications of gravity, there is plenty of room for further research in many different directions.

\section{Cosmology} \label{Sec:Cosmology}

The high precision of the cosmological observations made cosmology an ideal place to test fundamental theories
of gravity \cite{Perlmutter:1998np,Riess:1998cb,Spergel:2003cb,Ade:2013zuv,Eisenstein:2005su}. On the assumption of General Relativity being the underlying theory of gravitational interactions together
with the homogeneity and isotropy, cosmologists were able to construct the standard model of Big Bang
cosmology. Even if this model is simple and stood up to intense scrutiny, it still lacks a fully satisfactory theoretical foundation.
One of the challenges is the cosmological constant problem, posing a naturalness problem due to the giant mismatch between its observed
value and the radiative contributions from known massive particles to the vacuum energy \cite{Weinberg:1988cp,Padmanabhan:2002ji,Martin:2012bt}. On the other hand, the observation
of the accelerated expansion of the universe introduced the necessity of dark energy independently of the cosmological
constant problem \cite{ArmendarizPicon:2000dh,Peebles:2002gy,Copeland:2006wr,Clifton:2011jh,Amendola:2012ys,Bull:2015stt,Joyce:2014kja,Amendola:2016saw}. Furthermore, another problem that one has to face within the realm of General Relativity is the necessity of yet an additional dark component, dubbed dark matter, in order
to correctly account for the formation of large scale structures, the anisotropies of the CMB, weak lensing measurements or observations of rotation curves of galaxies. Albeit great efforts \cite{Bertone:2004pz,Strigari:2013iaa,Bird:2016dcv}, the true nature of dark matter still remains unknown.

The aforementioned challenges concern the late time evolution of the universe and thus, they motivated the consideration of infrared modifications of gravity. Remarkably, the tremendous progress made in observational cosmology also enabled us to probe the underlying physics of the early universe, which in fact shares a similar burden. In order to explain the observations the standard cosmological model is supplemented with the inflationary paradigm requiring an initial phase of accelerated expansion of the universe, that is commonly ascribed to yet another ingredient: the inflaton.
It is believed that the primordial quantum fluctuations during inflation eventually become the seeds in the density field responsible for the cosmic large-scale structure via gravitational instability. Inflation is the most prominent model for a successful
implementation of an extremely rapid exponential expansion, in which the perturbations of the inflationary field successively
translate into the fluctuations of the gravitational potential. Since gravity is coupled to all other fields, these fluctuations are
then imprinted onto all existing cosmic fluids. These density fluctuations leave imprints in the cosmic microwave background
as temperature anisotropies and also in the matter distribution, that then can be probed by gravitational lensing and formation of galaxies.
The inflationary scenario is realised in many different models based on different fields, and observations seem to favour
models with a nearly scale invariant red power spectrum, a small value for the scalar to tensor ratio and a small non-Gaussianity.
While the late time cosmology triggered searches for infrared modifications of gravity, the need for a primordial inflationary phase motivates modifications of gravity in the opposite regime. Furthermore, within the standard picture one is also prone to encounter a primordial classical singularity which calls for new physics beyond
General Relativity at these scales. Moreover, the breakdown of unitarity at the Planck scale requires modifications of gravity
in the ultraviolet regime to describe gravitational effects beyond $\mpl$. These additional challenges motivate to modify gravity at high energies.

It was precisely the cosmological Bing Bang singularities one of the motivations behind the inception of Born-Infeld inspired gravity theories in cosmology.
The original construction by Deser and Gibbons \cite{Deser:1998rj}
formulated in the metric language was an early attempt in this direction. Unfortunately, this approach leads to the presence of  ghostly degrees of freedom due to the presence of higher order field equations (see section \ref{Sec:D&G} for more details), so any regular cosmological solution will not be reliable. In spite of the mentioned ghost instabilities of the
metric formulation, a first quest of the cosmological implications of similar theories was pursued in \cite{Comelli:2004qr}.
There, although different realisations of (quasi) de Sitter solutions were shown to exist for appropriate choices of the parameters,
due to the unavoidable ghost
nature of the higher order derivative interactions, these solutions are unviable.
More promising cosmological solutions
without pathologies were found by considering Born-Infeld inspired gravity theories \`a la Palatini, where the connection is left arbitrary.
In fact, Ba\~nados and Ferreira showed the existence of non-singular solutions in \cite{Banados:2010ix} in the EiBI model, which have since then been extensively studied, and also found in other Born-Infeld theories of gravity. Although the avoidance of the singularities was the initial motivation, they provide very rich cosmological phenomenology, for instance these theories can support quasi de Sitter solutions with more standard forms of matter, like dust or radiation, as a consequence of modifying the high curvature regime of gravity. This behaviour permits to develop inflationary scenarios different from the more traditional models based on some scalar (or more general) degree of freedom. As we will see, in most of the modifications \`a la Born-Infeld, the different cosmological evolution can be traced to a highly non-trivial dependence of the Hubble expansion rate on the density and pressure of the matter fields in a modified Friedman equation. In other words, the effects of the modifications in the gravity
sector translates into a non-linear contribution from the matter fields density to the expansion rate. A remarkable property of these theories is that, while in most modified gravity theories the background expansion is determined by the equation of state parameter, Born-Infeld theories introduce a dependence on the sound speed already at the background level and not only for the perturbations. This is the cosmological analogue of the modified Poisson equation (\ref{eq:Phi-N}) with gradients of the density sourcing the equation for the gravitational potential, with its general case being discussed in section \ref{Sec:SimplifiedEqEiBI}.

The goal of this section will be to review all these cosmological applications and show the novel and interesting phenomenology derived from Born-Infeld inspired gravity theories. However, before starting with that, let us take a moment to fix the notation that we will use throughout this section. Cosmological observations seem to indicate a homogenous and isotropic universe. Compatible with these symmetries,
we will assume the metric tensor to be of the Friedman-Lemaitre-Robertson-Walker (FLRW) form, so the line element will read
\begin{equation}\label{FLRWmetric}
\mathrm{d} s^2=-N(t)^2\mathrm{d} t^2+a^2(t)\mathrm{d} \vec{x}^2\,,
\end{equation}
where $N$ represents the lapse, $a$ the scale factor and $t$ the cosmic time. Sometimes, it will be useful to work in conformal time $\eta$ defined as $a\mathrm{d}\eta=\mathrm{d}t$. We will also extensively refer to the Hubble function $H=\dot{a}/a$ or, in conformal time, $\mathcal{H}=a'/a$, where a dot and a prime denote derivatives with respect to cosmic and conformal time, respectively. It will be sometimes convenient to keep the lapse explicitly because of the presence of two metrics in the Born-Infeld theories, as we extensively discussed in section \ref{Sec:BITheories}.

%%%%%%%%%%%%%%%%%%%%%%%%%%%%%%%%%%%%%%
%%%%%%%%%%%%%%%%%%%%%%%%%%%%%%%%%%%%%%%
\subsection{Eddington-inspired Born-Infeld gravity}\label{sec_cosm_BI}

\begin{figure}
\begin{center}
\includegraphics[width=0.60\textwidth]{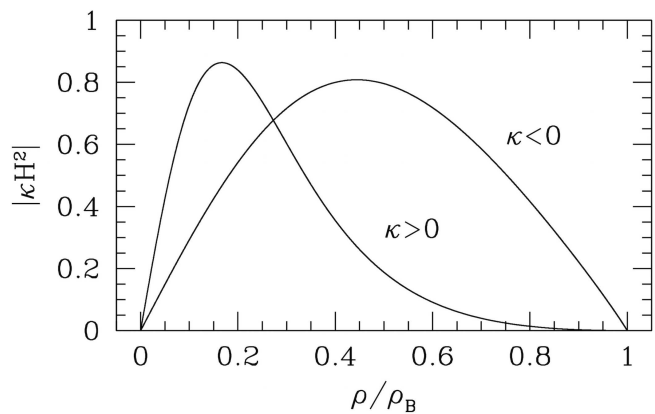}
\end{center}
\caption{\label{cosmfig1} This figure is taken from \cite{Banados:2010ix} and illustrates the dependence of the Hubble rate in terms of the energy density for a radiation dominated universe in the EiBI model. In \cite{Banados:2010ix} the notation $\rho_B$ stands for the maximum energy density where $H^2=0$. Furthermore, $\kappa=\frac{1}{\mbi^2}$ and $\mpl^2=1$ in terms of our notation.
}
\end{figure}

\begin{figure}
\begin{center}
\includegraphics[width=0.60\textwidth]{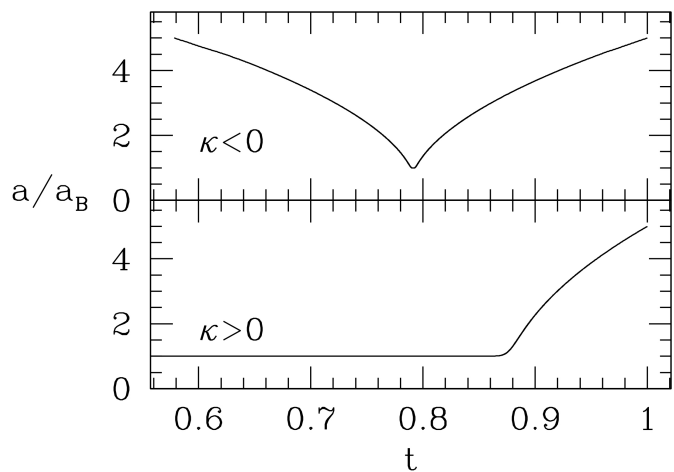}
\end{center}
\caption{\label{cosmfig2} This figure is taken from \cite{Banados:2010ix} and shows the evolution of the scale factor in the EiBI model in the presence of a radiation fluid. The scale factor is normalised by the minimum length scale $a_B$. For $\kappa<0$, the universe undergoes a bounce (in the upper panel), whereas for $\kappa>0$ the universe loiters, where the scale factor approaches the constant value $a_B$ for $t\to-\infty$ (in the lower panel).
}
\end{figure}

We will start our survey on the cosmological applications of Born-Infeld inspired gravity theories by considering the most extensively studied case of EiBI, whose action we rewrite here for convenience as
\begin{equation}
\mathcal{S}_{\rm BI}=\mbi^2\mpl^2\int \mathrm{d}^4x \left\{\sqrt{-\det\left(g_{\mu\nu}+\frac{1}{\mbi^2} \mR_{(\mu\nu)}(\Gamma)\right)}- \lambda\sqrt{-g}\right\}+\mathcal{S}_{\rm matter}\,,
\end{equation}
where $\mathcal{S}_{\rm matter}$ stands for the action of the standard matter fields, that we assume minimally coupled to the metric $g_{\mu\nu}$. As shown in section \ref{Sec:SimplifiedEqEiBI}, varying the above action with respect to the metric
yields the modified field equations
\begin{equation}
\sqrt{\det\Big(g_{\mu\nu}+\frac{1}{\mbi^2} \mR_{(\mu\nu)}(\Gamma)\Big)} \sqrt{-g}\left[\left(\m{g}+\frac{1}{\mbi^2} \m{\mR} \right)^{-1}\right]^{\mu\nu}-\lambda g^{\mu\nu}=-\frac{1}{\mbi^2\mpl^2} T^{\mu\nu}\,.
\end{equation}
Similarly, we can vary the action with respect to the independent connection. Since the connection $\Gamma$ does not carry any dynamics, its algebraic equation can be used to solve it in terms of $g_{\mu\nu}$ and $\mR_{(\mu\nu)}$. The resulting solution is such that the connection can be written as the corresponding Christoffel symbols of the effective metric
\begin{equation}
q_{\mu\nu}=g_{\mu\nu}+\frac{1}{\mbi^2} \mR_{(\mu\nu)}.
\end{equation}
On the other hand, we can use the metric field equations to express $\mR_{(\mu\nu)}$ in terms of $g_{\mu\nu}$ and the matter fields. This amounts to writing the equations as in General Relativity but with a modified non-linear matter coupling. See section \ref{Sec:SimplifiedEqEiBI} for more details on that. This feature becomes more apparent when we write the resulting modified Friedman equation of a homogeneous and isotropic background (\ref{FLRWmetric}). Compatible with the symmetries of the background metric, we assume the following Ansatz for the stress energy tensor $T_\mu{}^\nu={\rm diag}(-\rho(t), p(t), p(t), p(t))$, where $\rho$ and $p$ represents the energy density and pressure of the matter fields, respectively. The Friedman equation modifies into the general form (see section \ref{sec:general_framework_cosmology} for more details)
\begin{equation}\label{modFriedeq_BI}
H^2=f(\rho, p, c_s)\,,
\end{equation}
with a non-trivial function $f$, that depends non-linearly on $\rho$, $p$ and $c_s$.  In the case of the EiBI model, one can compute this function exactly \cite{Banados:2010ix}. In terms of the auxiliary metric we have
\begin{eqnarray}
q_{00}=-\sqrt{\frac{(1-\bar p_T)^3}{(1+\bar \rho_T)}} \qquad \text{and} \qquad q_{ij}=a^2\sqrt{(1+\bar \rho_T)(1-\bar p_T)}\delta_{ij} \,,
\end{eqnarray}
where $\bar{\rho}_T \equiv \frac{\rho_T}{\mbi^2\mpl^2}$ and $\bar{p}_T \equiv \frac{p_T}{\mbi^2\mpl^2}$ with the total energy density $\rho_T=\rho+(\lambda-1)\mbi^2\mpl^2$ and total pressure $p_T=\rho-(\lambda-1)\mbi^2\mpl^2$ and the lapse set to $N=1$. In terms of these quantities, the function $f(\rho, p)$ corresponds to \cite{Banados:2010ix}
\begin{equation}
f(\rho, p)=\frac13 \frac{G}{F^2}\,,
\end{equation}
with the short-cut notations standing for
\begin{eqnarray}
F&=&1-\frac{3(\bar \rho_T+\bar p_T)(1-w-\bar \rho_T-\bar p_T)}{4(1+\bar \rho_T)(1-\bar p_T)} \nonumber\\
G&=&\frac{\mbi^2}{2}\left(1-2q_{00}-3\frac{(1-\bar p_T)}{(1+\bar \rho_T)} \right)\,,
\end{eqnarray}
and the equation of state parameter $w=p/\rho$. Note, that the dependence on $c_s$ drops in $f$ because we have so far $w={\rm const}$. In general, the dependence $\dot{w}$ will appear as well, as we will see in section \ref{varyingEOS} and also in section \ref{sec:general_framework_cosmology} for the more general case. We can study this background equation for two different epochs. At late times for a dust filled universe ($w=0$) together with a cosmological constant, one recovers the standard Friedman equation in General Relativity
\begin{equation}
3H^2 \sim \rho+\Lambda +\left[\frac{\rho^2}{\Lambda}-(\rho+\Lambda)\right]\kappa \Lambda + \mathcal{O}(\kappa \Lambda)^2 \qquad \text{with} \qquad \Lambda=(\lambda-1)/\kappa \,,
\end{equation}
where  $\kappa=\frac{1}{\mbi^2}$ and $\mpl^2=1$ with the notation used in \cite{Banados:2010ix}. On the other hand, at early times, when the universe is dominated by radiation, we have $w=1/3$ and the modified Friedman equation becomes in this case
\begin{equation}
3H^2 =\frac{1}{\kappa}\left( \bar{\rho}-1+\frac{1}{3\sqrt{3}}\sqrt{(1+\bar{\rho})(3-\bar{\rho})^3}\right)\frac{(1+\bar{\rho})(3-\bar{\rho})^2}{(3+\bar{\rho}^2)^2}\,.
\end{equation}
As one can see from the above expression, for $\bar{\rho}=3$ (with $\kappa>0$) one obtains $H^2=0$. The same is true for $\bar{\rho}=-1$ (with $\kappa<0$).
These stationary points correspond to a maximum density. This is shown in figure \ref{cosmfig1}. The evolution of the scale factor in terms of the maximum density is given in figure \ref{cosmfig2}.
%%%%%%%%%%%%%%%%%%%%%%%%%%%%%%
%%%%%%%%%%%%%%%%%%%%%%%%%%%%%%
The maximum energy density would translate into a minimum value of the scale factor of the order $a_B=10^{-32}\kappa^{1/4}a_0$, with $a_0$ representing the scale factor today. Depending on the sign of $\kappa$, the scale factor can evolve in two different ways. If $\kappa<0$ one obtains $H^2\propto a-a_B\propto |t-t_B|^2$, which corresponds to a universe undergoing a bounce. On the other hand, if  $\kappa>0$ one has $H^2\propto (a-a_B)^2$, so that $\ln(a/a_B-1)= \sqrt{8/(3\kappa)}(t-t_B)$. In this scenario there is no bounce, and the universe loiters for a long time. These two behaviours can be visualised nicely by plotting the scale factor normalised by the scale $a_B$ as a function of time, which can be seen in figure \ref{cosmfig2}. A more detailed analysis of these cosmological solutions was further investigated in \cite{Scargill:2012kg,Cho:2012vg,Bouhmadi-Lopez:2013lha,Bouhmadi-Lopez:2014jfa,Bouhmadi-Lopez:2014tna}.
%%%%%%%%%%%%%%%%%%%%%%%%%%%%%%

For positive values of $\kappa$, the primordial nucleosynthesis constraints were used in \cite{Avelino:2012ge} in order to impose stringent restrictions on the allowed region in the parameter space. The agreement between the observed light element abundances and the predictions of the primordial nucleosynthesis is only ensured if the dynamics of the universe deviates from General Relativity only at a few percentage level at the initial epoch of nucleosynthesis. This, on the other hand, imposes the stringent constraint on the energy density at the start of nucleosynthesis to be of the order $\rho_{\rm nuc}\sim 3H^2_{\rm nuc}/(8\pi G)<3/\kappa$, which translates into $\kappa<6\times 10^8 \mathrm{m}^5\mathrm{kg}^{-1}\mathrm{s}^{-2}$.

The Born-Infeld inspired gravity model was also applied to a three dimensional space-time cosmology by S. Jana and S. Kar in \cite{Jana:2013fga}. There the authors compute some explicit analytical and numerical solutions for the scale factor in a curved and flat FLRW background with two different scenarios for the matter fields, namely a pressureless dust field with $p=0$ and a field with $p=\rho/2$. They show that also in three dimensions the branch of solutions with $\mbi^{-2}>0$ is singular, with an exception of specifically conditioned open universes. For the other case with $\mbi^{-2}<0$, they also find non-singular solutions in the same spirit as the four dimensional Born-Infeld gravity model.

\subsubsection{Cosmological tensor instabilities}\label{sec_tensorInst_BI}

In the previous subsection, we have seen that the original EiBI theory yields interesting homogeneous and isotropic solutions, where the cosmological singularities might be avoided by a bounce. We have seen that a bouncing solution with $H^2=0$ at $a_B$ is achievable in the presence of a radiation fluid with $w=1/3$. We have also seen the presence of loitering solutions, where the scale factor approaches $a_B$ for $t\to-\infty$. As next, we shall see whether the perturbations on top of these possible cosmological solutions are stable in order for them to be viable. This was investigated in detail in \cite{EscamillaRivera:2012vz, Lagos:2013aua}. We shall summarise their results here. For this purpose let us start with the tensor perturbations and describe the tensor modes of the spacetime and auxiliary metric in conformal time $\mathrm{d}\eta=\mathrm{d}t/a$ in the following form
\begin{eqnarray}
g_{00}=-a^2, \quad g_{ij}=a^2(\delta_{ij}+h_{ij})  \qquad \text{and}  \qquad q_{00}=-\tilde{N}^2, \quad q_{ij}=\tilde{A}^2(\delta_{ij}+f_{ij}) \,.
\end{eqnarray}
The tensor perturbations $h_{ij}$ and $f_{ij}$ are transverse and traceless, respectively. Since we are interested in the dynamics of the perturbations in the early universe epoch, we will again assume a relativistic perfect fluid for the matter fields and hence the background evolution will be as in section \ref{sec_cosm_BI}. First of all, using the field equations
\begin{equation}
\frac{\tilde{N}\tilde{A}^3}{a^4\tilde{A}^2}f^{ij}+\frac{\lambda}{a^2}h^{ij}=h^{ij}\left( \frac{\tilde{N}\tilde{A}^3}{a^4\tilde{A}^2} +\frac{\lambda}{a^2}\right)
\end{equation}
we immediately observe that the two perturbations are identical even if the background scale factors were different, namely
\begin{equation}
h^{ij}=f^{ij}.
\end{equation}
This is a remarkable property of the EiBI model. In fact, only in the presence of anisotropic stresses, the two tensor perturbations will be different from each other. This proportionality of the tensor perturbations turns out to be a generic feature of Born-Infeld inspired gravity theories beyond the standard formulation. We will see that for a general function of the metric and the Ricci tensor in section \ref{sec:general_framework_cosmology}. See also \cite{Jimenez:2015caa} for more details.
The tensor perturbations of the dynamical metric follow the evolution equation \cite{EscamillaRivera:2012vz}
\begin{equation}\label{tensor_pert_BI}
h_{ij}''+\left( 3\frac{\tilde{A}'}{\tilde{A}}-\frac{\tilde{N}'}{\tilde{N}} \right)h_{ij}'+\left( \frac{\tilde{N}}{\tilde{A}}\right)^2k^2h_{ij}=0\,,
\end{equation}
where we made use of the background equations of motion. In the regime of low energy densities, one recovers the standard evolution equation of the tensor modes as in General Relativity. On the other hand, in the Born-Infeld regime at high energy densities, the modifications in the evolution equation due to the scale factor of the auxiliary metric become appreciable. For the stability of the tensor perturbations, it will be crucial that both scale factors are well-behaved. It is not enough to impose this condition solely on the background variables of the spacetime metric. Similarly, one has to guarantee that the auxiliary metric does not vanish. In fact, as we have seen in the previous section, the evolution of the scale factor for $\kappa>0$ goes as $\ln(a/a_B-1)= \sqrt{8/(3\kappa)}(t-t_B)$, hence the lapse and the scale factor of the auxiliary metric evolve as
\begin{eqnarray}
\tilde{A}&=& 2^{1/4}a\sqrt{\exp\left(\sqrt{8/(3\kappa)}(t-t_B)\right)}\,, \nonumber\\
\tilde{N}&=&\frac{1}{\sqrt{2}} \frac{\tilde{A}^3}{a^2} \,.
\end{eqnarray}
As it becomes clear from these expressions, the scale factor of the auxiliary metric becomes singular for $t\to-\infty$. This non-singular behaviour has
a crucial impact on the tensor perturbations, since their evolution equation scales with the quantities of the auxiliary metric. In the far asymptotic past, the pre-factors of the last two terms in equation (\ref{tensor_pert_BI}) are suppressed and the evolution equation simply becomes $h_{ij}''\sim 0$. The solution for the metric perturbations is hence of the form $h_{ij}\approx A\eta+B$. This represents an unstable growth and therefore, the loitering solution in the case $\kappa>0$ suffers from an instability. This instability is a mild one and can be easily cured by slightly modifying the set-up.

The presence of tensor instabilities in the loitering solution is unfortunately also shared by the bouncing solution and it is even more virulent. For the case $\kappa<0$, we have seen in previous section that a bouncing solution is obtained since $H^2\sim a-a_B\sim |t-t_B|^2$. In terms of the conformal time, the scale factor evolves as
\begin{equation}
a=a_B\left[ 1+\tan^2(\beta \eta) \right]\,,
\end{equation}
with $\beta\equiv a_B\sqrt{2/(3|\kappa|)}$. This, on the other hand, means that the lapse and the scale factor of the auxiliary metric evolve this time as
\begin{eqnarray}
\tilde{N}&=& a^2 \frac{4}{3^{3/2}}\frac{1}{|\tan(\beta\eta)|} \,, \nonumber\\
\tilde{A}&=&a^2 \frac{4}{3^{1/2}}|\tan(\beta\eta)| \,.
\end{eqnarray}
We can Taylor expand these expressions around the bounce $\eta=0$. By doing so, the evolution equation of the tensor perturbations close to the bounce becomes
\begin{equation}\label{tensor_pert_BI_bounce}
h_{ij}''+\frac{2}{\eta}h_{ij}'+\frac{k^2}{3\beta^2\eta}h_{ij}=0\,.
\end{equation}
The solution scales this time as $h_{ij}\approx \eta^n$ with $n=-\frac{1}{2}\pm\frac{1}{2}\sqrt{1-(4k^2/(3\beta^2))}$, representing an unstable growth. Hence, the bouncing solution suffers also from an instability in the same way as the loitering solution, even though in the latter case it was much milder.
Thus, in the EiBI model in the presence of a radiation fluid the interesting loitering and bouncing solutions suffer from tensor instabilities. This unsatisfactory result might change if one considers a more general fluid with varying equation of state parameter or if one extends the EiBI model to a more general Born-Infeld inspired gravity model.

%%%%%%%%%%%%%%%%%%%%%%%%%%%%%%%%%%%
\subsubsection{Varying equation of state parameter}\label{varyingEOS}

\begin{figure}
\begin{center}
\includegraphics[width=0.50\textwidth]{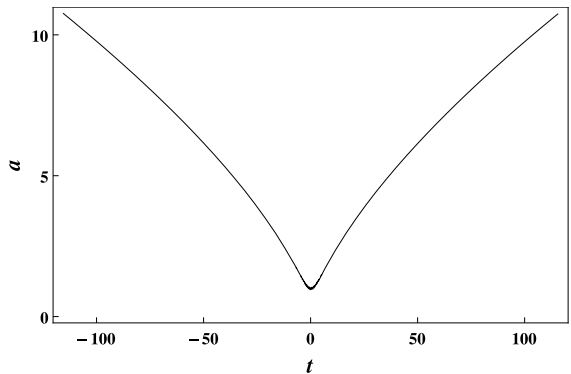}
\end{center}
\caption{\label{cosmfigAvelino} This figure is taken from \cite{Avelino:2012ue} and shows the evolution of the scale factor in the presence of a scalar field with varying equation of state parameter in the EiBI model. The universe undergoes a bounce at $t=0$. The initial values are chosen to be $\rho_i=10^{-4}$ and $w_i=0$ and the mass of the scalar field is assumed to be $m=100$. Note, that they use the unfortunate choice of units $\mpl=1$ and $|\kappa|=1$.}
\end{figure}

In the previous subsections we have seen that the EiBI theory admits interesting bouncing and loitering solutions for early universe cosmology in the presence of a radiation fluid. However, as we have seen, these solutions are plagued by tensor instabilities if the matter field is assumed to be a perfect fluid with the equation of state parameter $w=1/3$. It is possible to find more general solutions if we abandon this restriction and this might alleviate the found tensor instabilities. In fact, this was precisely considered in \cite{Avelino:2012ue}. It could be that additional dynamical fields are present in the early universe, giving rise to matter fields with $\dot{w}\ne0$. In this case, the modified Friedman equation (\ref{modFriedeq_BI}) generalizes to \cite{Avelino:2012ue}
\begin{equation}\label{modFriedeg_BI_wne0}
 H^2=\left( \frac{a\sqrt{g_1}+\dot{w}g_3}{g_2}\right)^2 \,,
\end{equation}
with the functions $g_i$ given by the energy density and pressure of the matter fields
\begin{eqnarray}
g_1&=&2\mbi^{-2}\left(1+\frac{2\rho}{\mbi^{2}}\right)\left(1-\frac{2\rho w}{\mbi^{2}}\right)^2\left[-2+\frac{2}{\mbi^2}(1+3w)+2D\right] \,, \nonumber\\
g_2&=&4+\frac{2}{\mbi^2}\rho\left[1-2w\left(2-\frac{2}{\mbi^2}\rho \right)+3w^2 \left(1+2\frac{2}{\mbi^2}\rho \right) \right] \,, \nonumber\\
g_3&=&-3\rho \left(1+\frac{2}{\mbi^2}\rho\right) \,,
\end{eqnarray}
with $D=\sqrt{(1+ \rho/\mbi^2)(1-p/\mbi^2)^3}$ and the choice of units $\mpl=1$ and $|\kappa|=1$ used in \cite{Avelino:2012ue} (remember that $\kappa=\mbi^{-2}$ in our units). In that work it is shown that the possibility with time varying equation of state parameter can ameliorate the tensor instabilities found for $w={\rm const}$.  As an example, a scalar field with a general kinetic and potential term is considered. In the presence of this scalar field, with the Lagrangian $\mathcal{L}(X, \phi)$ where $X=-\frac12\partial_\mu\phi \partial^\mu\phi$, the equation of state parameter is given by
\begin{equation}\label{modFriedeg_BI_wne0}
w=\frac{\mathcal{L}}{2X \mathcal{L}_{,X}-\mathcal{L} } \,,
\end{equation}
with the pressure $p=\mathcal{L}$ and energy density $\rho=\mathcal{L}_{,X}-\mathcal{L}$ accordingly. It turns out, that for $\kappa=-1$, the instability of the tensor perturbations cannot be avoided. This is the reason why the authors in \cite{Avelino:2012ue} consider the case  $\kappa=1$. Since $\dot{w}\ne0$, one achieves a bouncing solution with $H^2=0$ and $\dot{H}\ne0$, which differs from the case studied in \cite{EscamillaRivera:2012vz}, where $\rho\to w^{-1}$ as $\eta\to-\infty$. Remember that the authors in \cite{Avelino:2012ue} use the units $|\kappa|=1$. For an initial density of $\rho_i=10^{-4}$ and $w_i=0$, this behaviour is illustrated in figure \ref{cosmfigAvelino} for a scalar field with
$\mathcal{L}=X-\frac12m^2\phi^2$.
%%%
%%%
The tensor perturbations on top of this background are given by
\begin{equation}\label{tensor_pert_BI_dwne0}
h_{ij}''+g_4h_{ij}'+g_5k^2h_{ij}=0\,,
\end{equation}
with the two functions
\begin{eqnarray}
g_4&=&2H+ \frac{\kappa\dot\rho}{1+\kappa\rho} \,, \nonumber\\
g_5&=&\frac{1-\kappa\rho w}{1+\kappa\rho} \,.
\end{eqnarray}
For the case $\kappa=1$, the pre-factor in the friction term near the bounce vanishes $g_4\sim0$ and therefore, the tensor instabilities reported in \cite{EscamillaRivera:2012vz} are avoided. This simple example for time varying equation of state parameter was achieved with a standard scalar field with a mass term. As we have seen, this simple set-up already helps with the encountered tensor instabilities. In the next subsection, we will discuss in more detail the presence of a scalar matter field in EiBI model and summarise the works done in this context.

%%%%%%%%%%%%%%%%%%%%%%%%%%%%%%%%%%%

\subsubsection{Born-Infeld with a scalar matter field}

We have seen above that the reported tensor instabilities of the interesting cosmological solutions might be avoided by considering matter fields with varying equation of state parameter. As a specific model, one can consider the presence of a scalar field as matter field. This was for instance done in the works \cite{EscamillaRivera:2012vz,Lagos:2013aua,Cho:2013pea,Yang:2013hsa}. In this way, the underlying physics of the early universe will be determined by both the Born-Infeld modification and the presence of the scalar field. As a simple realisation one can consider a scalar field with a quadratic potential. In standard General Relativity an inflationary scenario with sufficiently long duration based on such a simple scalar field might require very large field values. This is due to the fact that the time derivative of the scalar field increases rapidly as going back in time with the scalar field itself climbing up the potential giving rise to an increasing energy density until the Planck scale is reached quickly. The hope to use this same scalar field in the Born-Infeld inspired gravity theory is to alleviate this requirement. The crucial point with this respect is that the pressure in EiBI gravity is bounded from above due to the square root structure. Hence, there is an upper bound for the value of the field velocity as it was shown in \cite{Cho:2013pea}. This guarantees a real value for the Hubble parameter. Due to this upper limit, one does not run into the same problem as in the standard inflationary model. Let us consider the following action \cite{Lagos:2013aua}

\begin{eqnarray}\label{BI_infl_scalarMatter}
\mathcal{S}_{\rm BI}&=&\mbi^2\mpl^2\int \mathrm{d}^4x \left\{\sqrt{-\det\left(g_{\mu\nu}+\frac{1}{\mbi^2} \mR_{(\mu\nu)}(\Gamma)\right)}- \lambda\sqrt{-g}\right\} \nonumber \\
&+&\int \mathrm{d}^4x \sqrt{-g} \left(-\frac12g^{\mu\nu}\partial_\mu\varphi\partial_\nu\varphi -\frac{m^2}{2}\varphi^2\right) \,.
\end{eqnarray}
In this model, the curvature scale remains finite thanks to the square root structure of EiBI gravity and the early universe undergoes a pre-inflationary accelerated expansion in order then to end in an ordinary chaotic inflationary epoch. Since the scalar Lagrangian is the same as in General Relativity, it follows the same evolution equation. For a homogeneous and isotropic background metric and correspondingly only time dependent scalar field, the equation of the scalar field is simply $\ddot\varphi+3H\dot\varphi+m^2\varphi=0$. The maximum value for the field velocity is achieved when $\dot\varphi^2=m^2\varphi^2+2\lambda\mbi^2$. So we can define this moment of maximum velocity by $\dot\varphi=\sqrt{m^2\varphi^2+2\lambda\mbi^2}$ with the Hubble parameter taking the form $H=-\frac23m^2\varphi/\sqrt{m^2\varphi^2+2\lambda\mbi^2}$ at this point. These equations can be integrated to have the evolution of the scalar field and the scale factor giving rise to solutions that respect the maximal pressure condition. By doing so, the explicit analytic solutions with this bound are given by
\begin{equation}\label{solPhiandScaleFactor_infSM}
\varphi=\frac{\sqrt{2\lambda\mbi^2}}{m}\sinh{[m(t-t_0)]} \qquad \text{and} \qquad a=\frac{a_0}{(2\lambda\mbi^2)^{1/3}}\cosh^{-2/3}[m(t-t_0)]\,.
\end{equation}
These solutions describe a universe that expands until the bouncing stage is achieved at $t=t_0$ and then starts contracting whereas the scalar field tracks the symmetric potential. At early times $t\to -\infty$, there is no singularity and the universe expands exponentially with $a\sim a_0(2/\lambda\mbi^2)^{1/3}e^{\frac23m(t-t_0)}$ and $\varphi\sim-\sqrt{\lambda\mbi^2/(2m^2)}e^{m(t-t_0)}$. During this period, the Hubble parameter is nearly constant and purely determined by the scalar field's mass $H\approx2m/3$ and in this limit $m^2\varphi^2\gg 2\lambda\mbi^2$, i.e. the potential of the scalar field is larger than $\lambda\mbi^2$. Thus, the upper limit in the pressure guarantees that the curvature scale remains finite.
\begin{figure}
\centering
\includegraphics[width=0.45\textwidth]{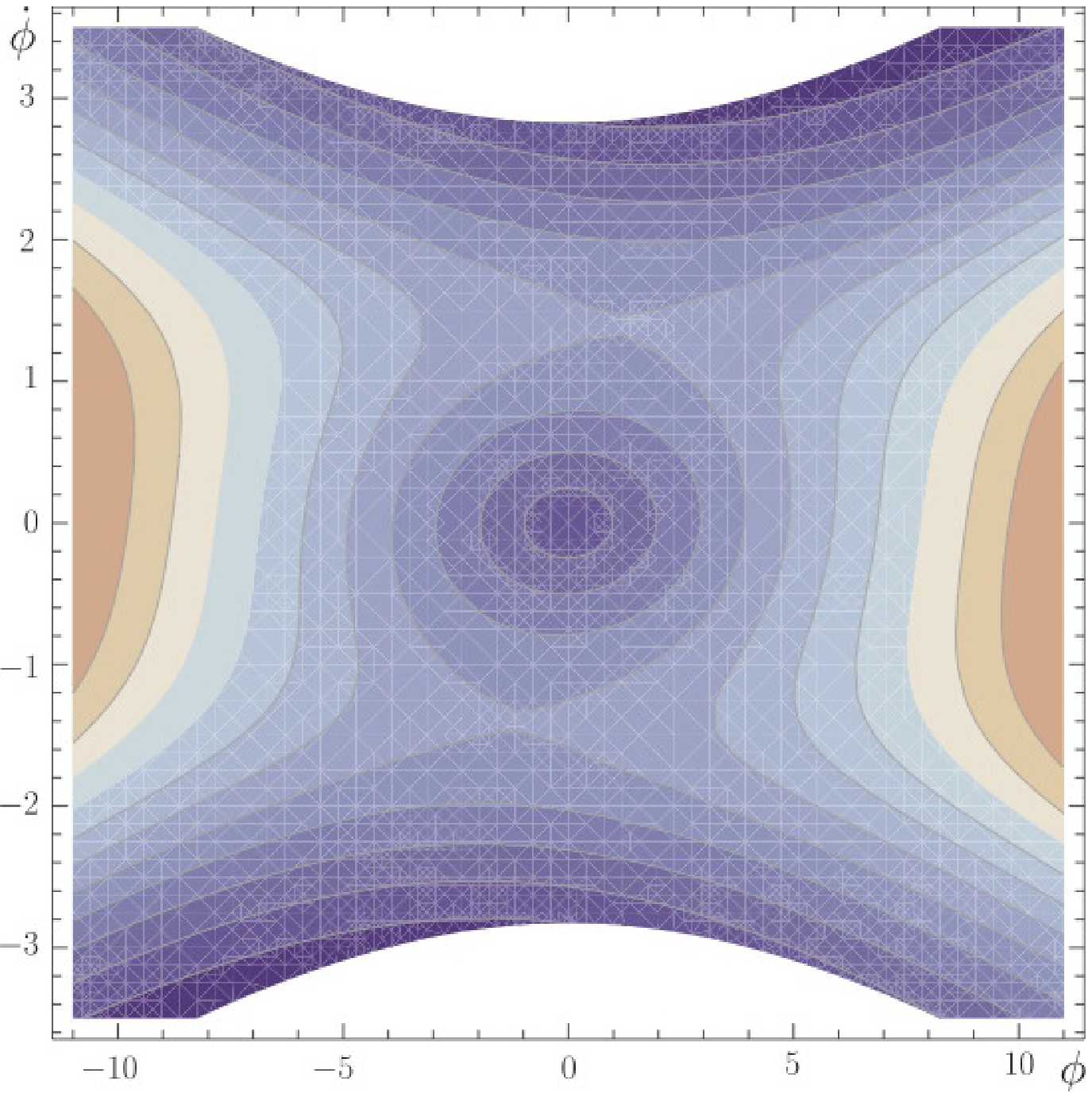}
\includegraphics[width=0.50\textwidth]{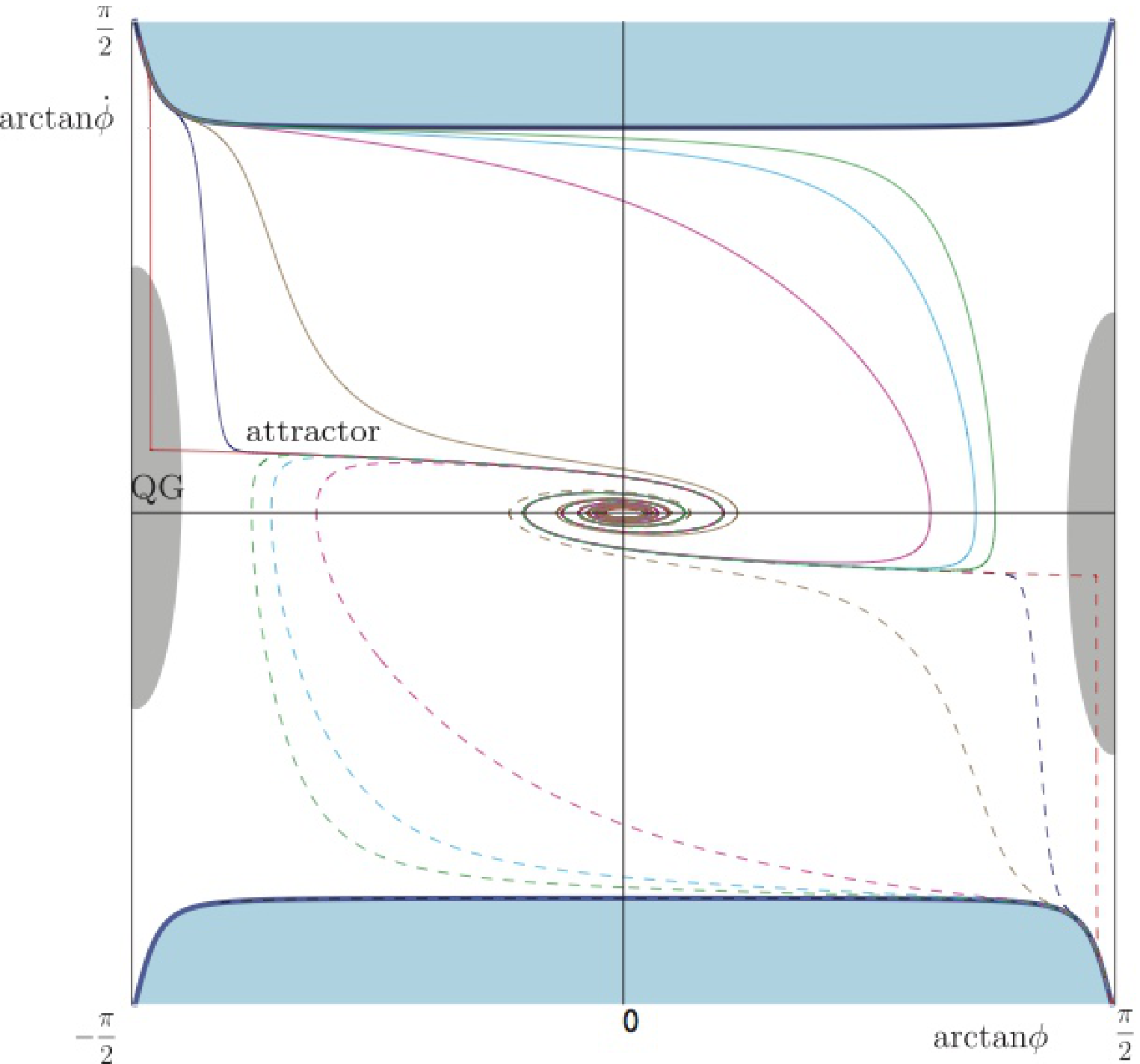}
\caption{This figure illustrates an example of the phase map taken from \cite{Cho:2013pea} with the parameters chosen as $m=1/4$, $\mbi^{-2}=1/4$ and $\lambda=1$, where the authors use the units $\mpl=1$. In the left panel one can see the behaviour of the Hubble parameter denoted by different colours. The red region corresponds to $H>1$, whereas the blue colour shows the regions with small Hubble parameter. The region encoded in white is the physically forbidden region. In the right panel the behaviour of $\dot\varphi$ is represented. The blue region corresponds to the field space where the upper bound limit is violated. The different trajectories correspond to different initial conditions for the scalar field. The grey region is the high-curvature regime and the solid and dashed curves represent trajectories that start from the left top and right bottom, respectively.}
\label{fig:BIwithScalarMatter}
\end{figure}

In figure \ref{fig:BIwithScalarMatter} an example of the phase map is plotted for $\varphi$ and $\dot\varphi$, where the Hubble function is denoted by the colour. We borrowed this figure from \cite{Cho:2013pea}, where one can nicely see the evolution of the Hubble parameter and the scalar field and the realisation of the different phases. One can see that the universe starts off close to the region of the upper bound of the pressure or field velocity respectively and decreases as time passes. Sufficiently away from this region, the universe undergoes the first slow-roll where the field velocity drops rapidly. As time evolves, the universe passes through the second phase of slow roll following the attractor solution represented by the standard chaotic inflationary expansion. This stage ends when the scalar field starts oscillating around the minimum of the potential going over to a possible reheating epoch. As it can be clearly seen in figure \ref{fig:BIwithScalarMatter}, all the trajectories start either from right bottom or from left top, where the forbidden region is avoided and converge to the attractor solution. Note also that, since the scalar field has a non-constant equation of state parameter, the tensor instabilities on top of these cosmological backgrounds can be eluded \cite{Avelino:2012ue}. The tensor and scalar perturbations of this model were investigated in detail in one of the pioneering works \cite{Lagos:2013aua}, where the authors constructed the general algorithm in terms of the bimetric interpretation of the model. They were able to show that the theory admits indeed the expected two tensor modes and one scalar mode, corresponding to the matter field. The authors further found scale-invariant power spectra for the tensor and scalar perturbations. However, they also reported a too large tensor-to-scalar ratio in contradiction with current observations. This is in the case of a scalar matter field that couples minimally to the Born-Infeld gravity. The tensor perturbations within this model were further studied in the work \cite{Cho:2014ija}, where it was shown that the same properties of the standard chaotic inflation are maintained for very short wavelength modes, whereas the model gives rise to a distinctive feature in form of a peculiar rise in the power spectrum for long wavelength modes. This peculiarity could be then tested with the CMB observations.

The preliminary findings of tensor and scalar perturbations of \cite{Lagos:2013aua} were further investigated in great detail in \cite{Cho:2014ija, Cho:2014jta, Cho:2014xaa, Cho:2015yza, Cho:2015yua}, where the authors study the scalar and tensor spectral indices and show that the contributions are second order in the slow roll approximation for the scalar perturbations and first order in the tensor perturbations. In the framework of EiBI gravity the tensor-to-scalar ratio $r$ can be suppressed significantly in difference to the standard chaotic inflation in General Relativity. For the analysis of the scalar perturbations of the model, let us adapt to the useful approach of considering parallel variables for the $g$ metric and the auxiliary metric, as we did above for the tensor perturbations. Let us consider the following scalar perturbations \cite{Lagos:2013aua}

\begin{eqnarray}
\mathrm{d}s_q^2&=&  \tilde{a}^2 \left\{- \frac{(1+2\phi_q)}{\mathcal{Z}} \mathrm{d}\eta^2+2\frac{B_{1,i}}{\sqrt{\mathcal{Z}}} \mathrm{d}\eta\mathrm{d}x^i  +\left[(1-2\psi_1)\delta_{ij}+2E_{1,ij}\right] \mathrm{d}x^i\mathrm{d}x^j\right\} \nonumber\\
\mathrm{d}s_g^2&=& a^2 \left\{- (1+2\phi_g) \mathrm{d}\eta^2+2B_{2,i} \mathrm{d}\eta\mathrm{d}x^i  +\left((1-2\psi_2)\delta_{ij}+2E_{2,ij}\right) \mathrm{d}x^i\mathrm{d}x^j\right\} \,.
\end{eqnarray}
Similarly, we shall perturb the scalar field as $\varphi=\varphi_0+\delta\varphi$. Note that the auxiliary metric carries the additional background quantity $\mathcal{Z}=\frac{1+\rho_0/\mbi^2}{1-p_0/\mbi^2}$ and the overall scale factor $\tilde{a}=(1+\rho_0/\mbi^2)^{1/4}(1-p_0/\mbi^2)^{1/4}a$ with $\rho_0=\varphi'_0/(2a^2)+m^2\varphi^2/2$ and $p_0=\varphi'_0/(2a^2)-m^2\varphi^2/2$. One can use the gauge freedom in order to eliminate some of the perturbations. One could for instance choose $\psi_1=0$ and $E_1=0$. However, not all of the remaining quantities are dynamical. In fact, except for the scalar field, all the remaining perturbations of the metrics can be integrated out using their algebraic equations. This is to be expected, since the $q_{\mu\nu}$ metric is related algebraically with the $g_{\mu\nu}$ metric and the scalar perturbations in the space-time metric are not dynamical (see sections \ref{Sec:SimplifiedEqEiBI} and \ref{Sec:Frames} for more details). After introducing the new perturbation variable $\chi=\mathcal{W}\delta\varphi$ where $\mathcal{W}$ stands for $\mathcal{W}=\left(\frac{3\mathcal{Z}^2-2\mathcal{Z}+3}{\mathcal{Z}(\mathcal{Z}+1)(3\mathcal{Z}-1)}\right)^{1/4}a$ and performing the time transformation $\mathrm{d}\tau=\mathcal{W}^2/f_1\mathrm{d}\eta$, the equation of the dynamical scalar field perturbation takes the simple form at the attractor stage

\begin{equation}\label{solPhiandScaleFactor_infSM}
\ddot\chi+\left(k^2-\frac{2}{(\tau-\tau_0)^2} \right)\chi\approx 0\,,
\end{equation}
where $\tau_0$ denotes the end of inflation and $f_1=\frac{3\mathcal{Z}^2-2\mathcal{Z}+3}{(\mathcal{Z}+1)(3\mathcal{Z}-1)}a^2$ here. In the works \cite{Cho:2014jta, Cho:2014xaa, Cho:2015yua} it was shown that the perturbations start from an initial point where the maximal pressure condition holds and evolve towards an intermediate stage, where the WKB approximation can be applied to then end at the attractor stage. Finally, the solutions of these three stages are matched together. Furthermore, they compute the comoving curvature perturbation $\mathcal{R}\psi_2+H\delta\varphi/\varphi_0$ and from that the scalar power spectrum $\mathcal{P}_\mathcal{R}=k^3|\mathcal{R}|^2/(2\pi^2)$. Last but not least, from this they were able to evaluate the spectral index $n_\mathcal{R}-1=\mathrm{d}\log\mathcal{P}_\mathcal{R}/\mathrm{d}\log k$. They observe that the spectral index is of second order in the slow-roll approximation and a suppression of the tensor-to-scalar ratio. The exact form of the scalar power spectrum and the spectral index can be extracted from \cite{Cho:2015yua,Cho:2015yza} and we refer the reader to these works for more details.

We have seen that in the presence of a scalar field one can realise different epochs in the early universe. One can have a preinflationary scenario followed by a standard chaotic inflationary expansion. Due to the squared root structure of the gravitational interactions, there is an upper limit for the pressure and hence the field velocity. So far we have considered the case where the scalar field is minimally coupled to the gravity and has standard kinetic and mass terms. In the following subsection we will pay attention to the case where the scalar sector obeys the Born-Infeld structure as well.

%%%%%%%%%%%%%%%%%%%%%%%%%%%%%%%%%%%
\subsubsection{Born-Infeld in gravity and matter sector}

In the following we would like to discuss the EiBI gravity theory in the presence of a scalar Born-Infeld matter field. The Born-Infeld structure in both the gravity and matter sector with their corresponding scales might have interesting implications. This idea was pursued by S. Jana and S. Kar in \cite{Jana:2016uvq}, where they provide interesting analytical cosmological solutions for a particular choice of the time derivative of the Born-Infeld scalar. For a positive constant $\mbi^{-2}>0$, they were able to realise solutions with two separate de Sitter expansions with an intermediate sandwiched phase of deceleration. The action of this model is given by \cite{Jana:2016uvq}

\begin{eqnarray}
\mathcal{S}_{\rm BI}&=&\mbi^2\mpl^2\int \mathrm{d}^4x \left\{\sqrt{-\det\left(g_{\mu\nu}+\frac{1}{\mbi^2} \mR_{(\mu\nu)}(\Gamma)\right)}- \lambda\sqrt{-g}\right\} \nonumber \\
&+&\alpha_T^2\int \mathrm{d}^4x \sqrt{-g} V(\phi)\sqrt{1+\alpha_T^{-2}g^{\mu\nu}\partial_\mu\phi\partial_\nu\phi} \,,
\end{eqnarray}
with the scales $\mbi$ and $\alpha_T$ representing the Born-Infeld scales in the gravity and matter sector, respectively, and $V(\phi)$ denoting a potential for the scalar field. The scalar sector is a Dirac-Born-infeld like action. The equation of motion of the scalar field yields

\begin{equation}
\partial_\nu \left( \frac{V\sqrt{-g}g^{\mu\nu}\partial_\mu\phi}{\sqrt{1+\alpha_T^{-2}g^{\mu\nu}\partial_\mu\phi\partial_\nu\phi}}\right)=\alpha_T^2V'\sqrt{-g}\sqrt{1+\alpha_T^{-2}g^{\mu\nu}\partial_\mu\phi\partial_\nu\phi} \,.
\end{equation}
Similarly, the corresponding stress energy tensor reads

\begin{equation}
T^{\mu\nu}=V(\phi) \left[ \frac{(g^{\mu\alpha}g^{\nu\beta}-g^{\mu\nu}g^{\alpha\beta})\partial_\alpha\phi\partial_\beta\phi-g^{\mu\nu}\alpha_T^{-2}}{\sqrt{1+\alpha_T^{-2}g^{\mu\nu}\partial_\mu\phi\partial_\nu\phi}} \right] \,.
\end{equation}
We will be interested in the possible cosmological solutions that one can construct in this particular model. For this purpose, we will again consider a FLRW metric for the background metric $g_{\mu\nu}$ with lapse $N$ and scale factor $a$. For this specific simple background, the scalar field equation becomes

\begin{equation}
\frac{\ddot\phi}{\alpha_T^{2}N-\dot\phi^2}+\frac{3\dot\phi H}{\alpha_T^{2}N}+\frac{V'}{V}-\frac{\dot\phi\dot N}{2N(\alpha_T^{2}N-\dot\phi^2)}=0 \,.
\end{equation}
The equation of motion of the scalar field can also be written as $\dot\rho_\phi/\rho_\phi=-3H\dot{\phi}^2/(N\alpha_T^{2})$, where the corresponding energy density of the field is given as
\begin{equation}
\rho_\phi=\frac{\alpha_T^{2}V}{\sqrt{1-\dot\phi^2N^{-1}\alpha_T^{-2}}} \,.
\end{equation}
Similarly, we can compute the pressure of the scalar field, which for the considered DBI action yields $p_\phi=-\alpha_T^{2}V\sqrt{1-\dot\phi^2N^{-1}\alpha_T^{-2}}$, which we can use to define the corresponding equation of state parameter of the scalar field. In contrast to the standard single field inflation where one assumes a specific form of the potential, here one can choose a specific form for $\dot\phi$. In \cite{Jana:2016uvq} the following solution for $\dot\phi$ is contructed:
\begin{equation}
\dot\phi^2=\frac{N\alpha_T^{2}}{1+C_1a^n} \,,
\end{equation}
with positive constant variables $C_1$ and $n$. For $n=3$ for instance one can obtain a constant negative pressure $p_\phi=-\alpha_T^{2}C_2$ with the integration constant $C_2$. Using the metric field equations
\begin{equation}
\sqrt{-q}q^{\mu\nu}-\lambda\sqrt{-g}g^{\mu\nu}=\frac{\sqrt{-g}T^{\mu\nu}}{\mpl^2\mbi^2} \,,
\end{equation}
we can relate the lapse $\tilde{N}$ and shift $\tilde{a}$ of the auxiliary metric $\hat{q}$ with the scale factor of the $\hat{g}$ metric and the energy density of the scalar field. By doing so, one obtains

\begin{eqnarray}
a&=&\frac{\tilde{a}\sqrt{\tilde{N}/N}}{\tilde{\alpha}_T} \,, \nonumber\\
\rho_\phi&=&\mbi^2\mpl^2\left( \frac{\tilde{\alpha}_T^3N}{\tilde{N}^2}-1 \right)\,,
\end{eqnarray}
with the new constant parameter $\tilde{\alpha}_T=1+\alpha_T^2C_2/(\mbi^2\mpl^2)$. Similarly, the equations of motion for the connection yield the following relations
\begin{eqnarray}\label{eqGamma_sBI}
\frac{\dot{\tilde{a}}^2}{\tilde{a}^2}&=&\frac{\mbi^2}{6}\left( 2\tilde{N}+N-\frac{3\tilde{N}^2}{C_0^2N}\right) \,, \nonumber\\
\frac{ \mathrm{d}}{ \mathrm{d}t}\left( \frac{\dot{\tilde{a}}}{\tilde{a}}\right)-\frac{\dot{\tilde{a}}\dot{\tilde{N}}}{2\tilde{a}\tilde{N}}&=&\frac{\mbi^2}{2}\left( -N+\frac{\tilde{N}^2}{C_0^2N}\right) \,.
\end{eqnarray}
In \cite{Jana:2016uvq} this system of equations is analysed for a particular solution of the scale factor of the auxiliary metric, namely, $\tilde{a}=\tilde{a}_0e^{\bar{H}_bt}$ with two constants $\tilde{a}_0$ and $\bar{H}_b$. Introducing the quantity $Y_\phi=C_0(1-3\bar{H}_b^2\tilde{N}^{-1}\mbi^{-2})$, the last equation of (\ref{eqGamma_sBI}) translates into

\begin{equation}
\frac{\dot{Y}_\phi}{2Y_\phi-\sqrt{Y_\phi^2+3}}=-2\bar{H}_b \,.
\end{equation}
In terms of this new variable, the deceleration parameter $d$ can be expressed as\footnote{Usually, $q$ is used for the deceleration parameter in the literature but we use $d$ here in order to avoid confusion with the auxiliary metric.}
\begin{equation}
d=-\frac{aa''}{a'^2}=-1+\frac{\sqrt{Y_\phi^2+3}-2Y_\phi}{\sqrt{Y_\phi^2+3}-Y_\phi} \left(1+\frac{Y_\phi}{\sqrt{Y_\phi^2+3}}+\frac{\sqrt{Y_\phi^2+3}+Y_\phi C_0}{2(C_0-Y_\phi)} \right) \,,
\end{equation}
\begin{figure}
\centering
\includegraphics[width=0.60\textwidth]{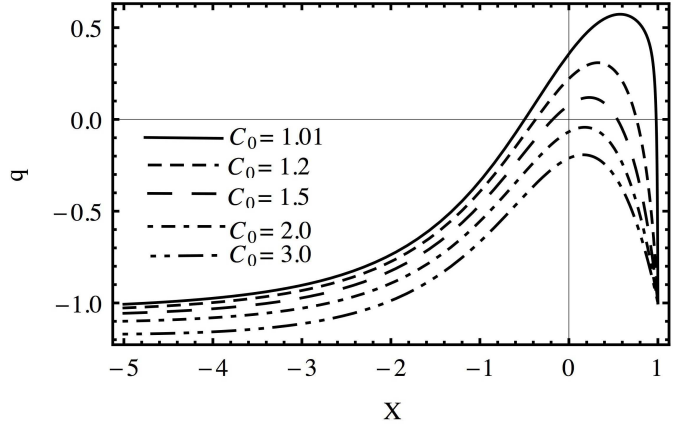}
\caption{\label{fig:decelerationX_sBI} This figure is extracted from \cite{Jana:2016uvq} and illustrates the deceleration parameter (which in their notation is denoted by $q$) as a function of $X$ in the case $\mbi^{-2}>0$. One can see the transition of the deceleration parameter from negative to positive values and then afterwards to become negative again. The outcome for different values of the constant $C_0$ is shown by the different curves.}
\end{figure}
where prime denotes the derivative with respect to cosmological time $\tau=\int \sqrt{N} \mathrm{d}t$. One immediate observation is that for $a\to \infty$ ($Y_\phi\to1$) one has $d=-1$ and similarly for $a\to 0$ ($Y_\phi\to-\infty$) one also has $d=-1$. Hence, one obtains two de Sitter phases, one at early and one at late time universe. In figure \ref{fig:decelerationX_sBI} taken from \cite{Jana:2016uvq} one can see the dependence of the deceleration parameter from $X$ for different values of $C_0$.
Furthermore, one can choose the value of the constant $C_0$ such that one can realise an initial loitering phase with an acceleration and subsequently a decelerated and again an accelerated expansion phase afterwards. This can be seen in figure \ref{fig:scaleFactorDensity_sBI}, where the scale factor is plotted as a function of cosmological time. During the loitering phase, the scale factor grows approximately as $a_{\rm loit}\sim a_0e^{2\sqrt{2}\mbi\tau/\sqrt{3}}$. During the period of inflation the universe grows by $60$ e-folds in $10^{-32}$ seconds. This on the other hand puts the bound $\mbi^{-2}\lesssim 0.67\times 10^{-50}\mathrm{m}^2$. The second phase of accelerated expansion at late times has the scale factor growing as $a_{\rm DE}\sim e^{\sqrt{C_2/3}\alpha_T \tau/\mpl}$. As it becomes clear from the expressions of the scale factor in these two regimes, the evolution at early times depends on the Born-Infeld scale $\mbi$ of the gravity sector whereas at late times on the scalar Born-Infeld parameter $\alpha_T$. The intermediate phase depends on both as $\alpha_T^2\mbi^{-2}$.
%%%
\begin{figure}
\centering
\includegraphics[width=0.95\textwidth]{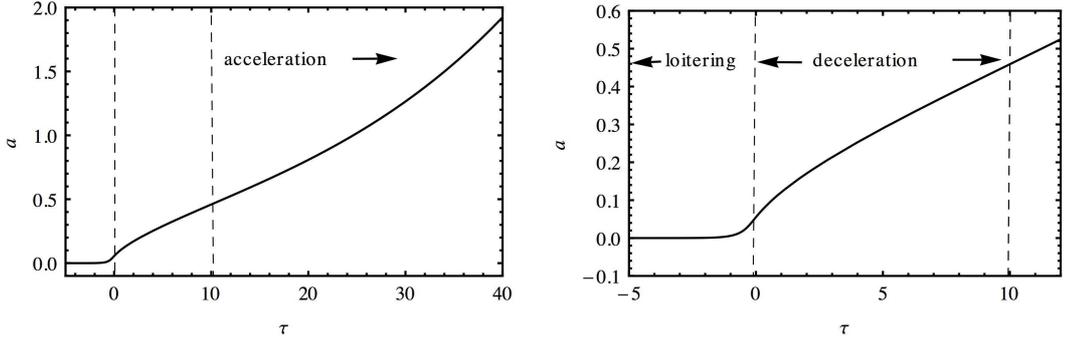}
\caption{\label{fig:scaleFactorDensity_sBI} This figure is borrowed from \cite{Jana:2016uvq}, where one can see the evolution of the scale factor in the different regimes. In \cite{Jana:2016uvq} the following values have been chosen for this plot: $\mpl=1$, $\mbi^{-2}=0.5$, $\alpha_T=5$, $C_2=0.001$ and finally $C_0=1.0025$ with the initial value $a_0=0.06$ at $\tau_0=0.06$. In the left panel, one can see the evolution of the scale factor during the loitering, deceleration and acceleration phases. The right panel is just a zoom of the first two phases.}
\end{figure}
%%%
Similarly, one can study the cosmological solutions of the background equations in the other case when $\mbi^{-2}<0$. Of course, due to the Born-Infeld term in the gravity sector, one can construct bouncing solutions as we have seen before. The novelty of the scalar Born-Infeld term results in an additional accelerating phase at late time universe. Nevertheless, this solution with $\mbi^{-2}<0$ yields a bounce at an unacceptable low redshift as it is shown in \cite{Jana:2016uvq}, therefore we do not report more on this solution here. More details can be taken from \cite{Jana:2016uvq}. There the authors show also the comparison of the obtained solutions with the supernovae Ia Union2.1 data and find that the agreement with the data is as good as in $\Lambda$CDM model. Summarizing, the combined Born-Infeld model in the gravity and scalar sector delivers an interesting framework to study effects both on early and late time universe cosmology. The ongoing physics at early times is dictated by the standard Born-Infeld scale $\mbi$, where one can either realise loitering, accelerating or bouncing solutions, whereas the physics at late times is governed by the scalar Born-Infeld scale $\alpha_T$, which gives rise to an accelerated expansion.

%%%%%%%%%%%%%%%%%%%%%%%%%%%%%%
\subsubsection{Anisotropic cosmological solutions}\label{subsection_anisotropic_cosmology}

So far we have studied the cosmological applications of the EiBI gravity theory for homogeneous and isotropic backgrounds. We have seen the appearance of different interesting cosmological solutions for early universe, including loitering, quasi de Sitter and bouncing solutions and discuss their stability. Another interesting question along this line is the evolution of cosmological backgrounds with anisotropies. Some of the anomalies observed in the cosmic microwave background might be due to the presence of small anisotropies. In this context, Bianchi type models could be a natural and simple extension of the standard FLRW with small anisotropies, which could be for instance at the origin of the power suppression at large scales of the cosmic microwave background. This was exactly pursued in \cite{Kim:2013noa,Harko:2014nya}. Let us consider the following Bianchi type I background for the dynamical metric $g_{\mu\nu}$
\begin{equation}
\mathrm{d}s_g^2=g_{\mu\nu}\mathrm{d}x^\mu\mathrm{d}x^\nu=-\mathrm{d}t^2+g_1(t)\mathrm{d}x^2+g_2(t)\mathrm{d}y^2+g_3(t)\mathrm{d}z^2 \,,
\end{equation}
where $g_i$ denote the scale factors in the $x$, $y$ and $z$ direction, respectively. By defining the quantities $A^2=1+\rho/(\mpl^2\mbi^2)$ and $B_i^2=1-p_i/(\mpl^2\mbi^2)$ we can construct the auxiliary metric with a similar Ansatz with the three different scale factors. It takes the following form
\begin{equation}
\mathrm{d}s_q^2=q_{\mu\nu}\mathrm{d}x^\mu\mathrm{d}x^\nu=-\mathrm{d}t^2+q_1(t)\mathrm{d}x^2+q_2(t)\mathrm{d}y^2+q_3(t)\mathrm{d}z^2 \,,
\end{equation}
with $q_i$ standing for the quantities $q_i=g_i A/B_i$. In terms of these variables, the field equations of the Bianchi Type I geometry of the EiBI theory can be calculated easily. Plugging the two Ansaetze into the covariant field equations yields
\begin{eqnarray}
1-\frac{A}{B_1B_2B_3}&=&\frac{\mbi^{2}\left( \ddot{q}_1q_2q_3+q_1\ddot{q}_2q_3+q_1q_2\ddot{q}_3\right)}{q_1q_2q_3} \,, \\
1-\frac{B_1}{AB_2B_3}&=&\frac{\mbi^{2}\left(\dot{q}_1\dot{q}_2q_3+\dot{q}_1q_2\dot{q}_3+\ddot{q}_1q_2q_3\right)}{q_1q_2q_3} \,, \\
1-\frac{B_2}{AB_1B_3}&=&\frac{\mbi^{2}\left(\dot{q}_1\dot{q}_2q_3+q_1\dot{q}_2\dot{q}_3+q_1\ddot{q}_2q_3\right)}{q_1q_2q_3} \,, \\
1-\frac{B_3}{AB_1B_2}&=&\frac{\mbi^{2}\left(\dot{q}_1q_2\dot{q}_3+q_1\dot{q}_2\dot{q}_3+q_1q_2\ddot{q}_3\right)}{q_1q_2q_3} \,.
\end{eqnarray}
First, we can consider the simple case with isotropic pressure where $p_i=p$ and therefore $B_i=B$. For clarity of the notation, we can further introduce the Hubble functions in the different spatial directions as $H_i=\dot{q}_i/q_i$ and $\Delta H_i=H-H_i$ where $H=1/3\sum_{i=1}^3H_i$ is the mean Hubble expansion rate. We can define the degree of anisotropy as
the shear
\begin{equation}
\sigma=\frac13 \sum_{i=1}^3 \left( \frac{\Delta H_i}{H} \right)\,.
\end{equation}
In the field equations the multiplication of the scale factors appears very often. For this reason, we can define a new variable here as $Q=q_1q_2q_3$ and express the field equations in terms of $H$ and $Q$, which read
\begin{eqnarray}
3\dot{H}+\sum_{i=1}^3H_i^2=\mbi^{-2}\left( 1-\frac{A}{B^3}\right) \,, \\
\frac{1}{Q}\frac{\mathrm{d}}{\mathrm{d}t}(QH_i)=\mbi^{-2}\left( 1-\frac{1}{AB}\right) \,.
\end{eqnarray}
After simple manipulations of the field equations, they can be combined into $\frac{\mathrm{d}}{\mathrm{d}t}\left[Q(H_i-H)\right]=0$, which can be simply integrated to give $H_i=H+C_i/Q$ with integration constants $C_i$. Further integration gives for the scale factors $q_i=q_{i0}Q^{1/3}\exp{\left(C_i\int \left(\frac{1}{Q}\frac{ \mathrm{d}t}{\mathrm{d}Q}\right)\mathrm{d}Q\right)}$. For consistency, the integration constants have to satisfy $C_1+C_2+C_3=0$. Furthermore, the product of the scale factors follows the second order differential equation $\ddot{Q}=3\mbi^{-2}(1-1/(AB))Q$, which can be also integrated easily. From these solutions, we can also determine the quantities of the $\hat{g}$ metric. For instance, the Hubble functions
of the $\hat{g}$ metric in the different directions can be obtained from $H_i^g=H_i+\dot{B}/B-\dot{A}/A$ and similarly the mean Hubble parameter as well. The ordinary matter fields couple to the standard $\hat{g}$ metric, therefore their conservation equation is dictated by the mean Hubble parameter of the $\hat{g}$ metric. Thus, they follow as
\begin{equation}
\dot\rho +3\left( H+\frac{\dot{B}}{B}-\frac{\dot{A}}{A}\right)(\rho+p)=0 \,.
\end{equation}
In terms of the energy density and pressure of the matter fluid, we can also compute the degree of anisotropy in the $\hat{g}$ sector, which takes the following from
\begin{equation}
\sigma^g=\frac{3C^2(\rho+p)^2}{Q^2 \dot{\rho}^2}\,,
\end{equation}
with $C=C_1+C_2+C_3$. In \cite{Harko:2014nya} the quantity $Q$ is used as a parameter in order to obtain the general solution in a parametric form. Furthermore, they provide the general solutions for the Hubble functions and the anisotropy parameter in the $\hat{g}$ sector for three different fluid types: for stiff, radiation and dust fluid. Let us for example consider the dust component with $p=0$ and hence $B=0$.
After making the following change of variables $\theta=t\mbi$ and $\rho=r\mbi^2$, the factor $Q=q_1q_2q_3$ becomes $Q=\mbi^{-2}\rho_0/(r(1+r)^{3/2})$ with the initial density $\rho_0$. From the differential equation $\ddot{Q}=3\mbi^{-2}(1-1/(AB))Q$ one obtains in this case
\begin{equation}
2(r+1)(5r+2)rr''-(7r(5r+4)+9)r'^2-12\sqrt{r+1}r^2(r-(1+r)^{3/2}+1)=0\,
\end{equation}
\begin{figure}[h]
\includegraphics[width=0.48\textwidth]{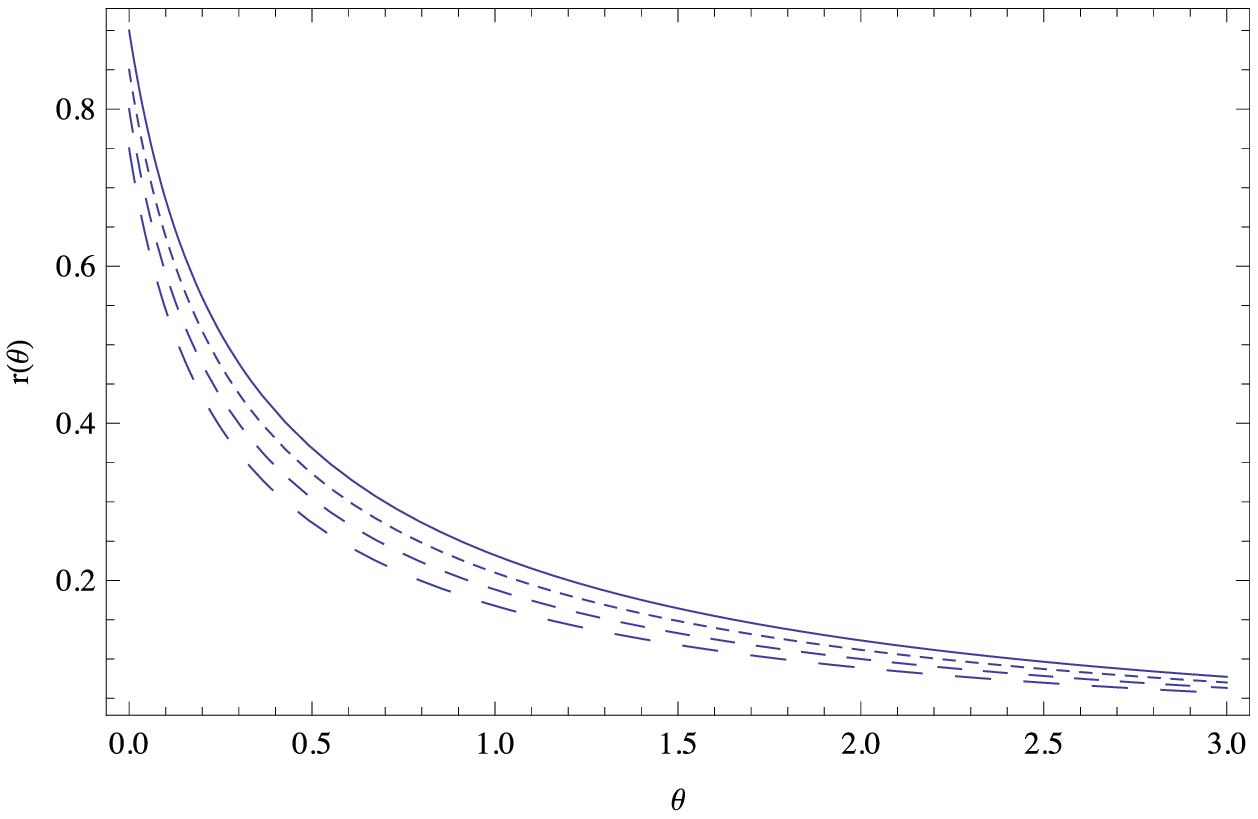}
\includegraphics[width=0.48\textwidth]{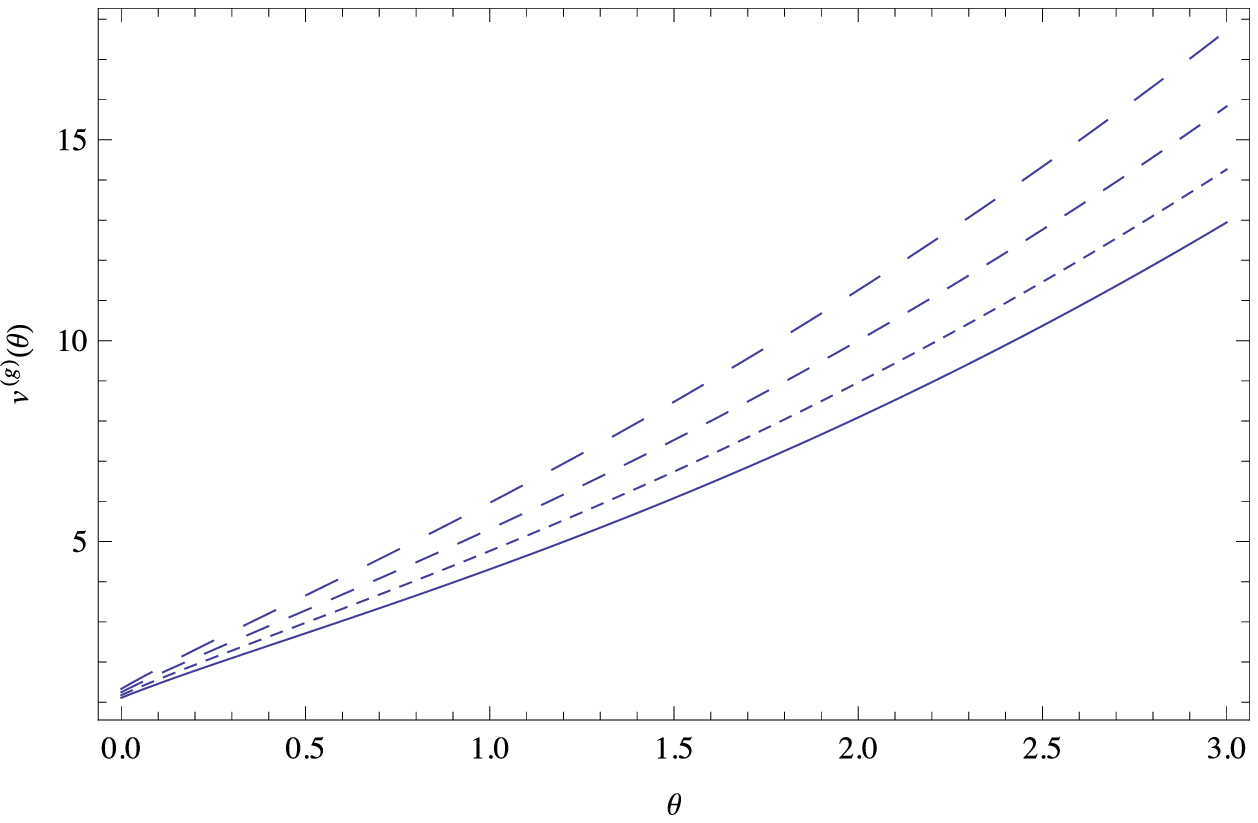}
\caption{This figure represents the evolution of the rescaled energy density $r=\rho\mbi^{-2}$ and the rescaled volume element $\mathcal{G}\mbi^{-2}\rho$ (which is $v^{g}$ in the notation used in \cite{Harko:2014nya} and furthermore $\mpl=1$) as a function of the rescaled time $\theta=t\mbi$ for three different initial conditions: $r(0)=0.9$ (solid line), $r(0)=0.85$ (dotted line), $r_0=0.8$ (short dashed line) and $r(0)=0.75$ (dashed line) in a universe filled with dust. Figures were taken from \cite{Harko:2014nya}.}
\label{fig:AnistropicBI}
\end{figure}
where prime denotes here the derivative with respect to $\theta$. The volume element in the $\hat{g}$ metric sector, so in other words $\mathcal{G}=g_1g_2g_3$, is given by $\mathcal{G}=\mbi^{-2}\rho_0/r$. The evolution of the rescaled energy density $r$ and the volume element is plotted in figure \ref{fig:AnistropicBI} for different choices of the initial energy density, which we took from \cite{Harko:2014nya}. In the presence of a dust fluid, the mean anisotropy parameter is given by $(\mbi^{-2}\rho_0)^2/(3K^2)\sigma^g=r^4(1+r)^3/r'^2$ and is plotted in figure \ref{fig:AnistropicBI_H_sigma}.
\begin{figure}[h]
\includegraphics[width=0.48\textwidth]{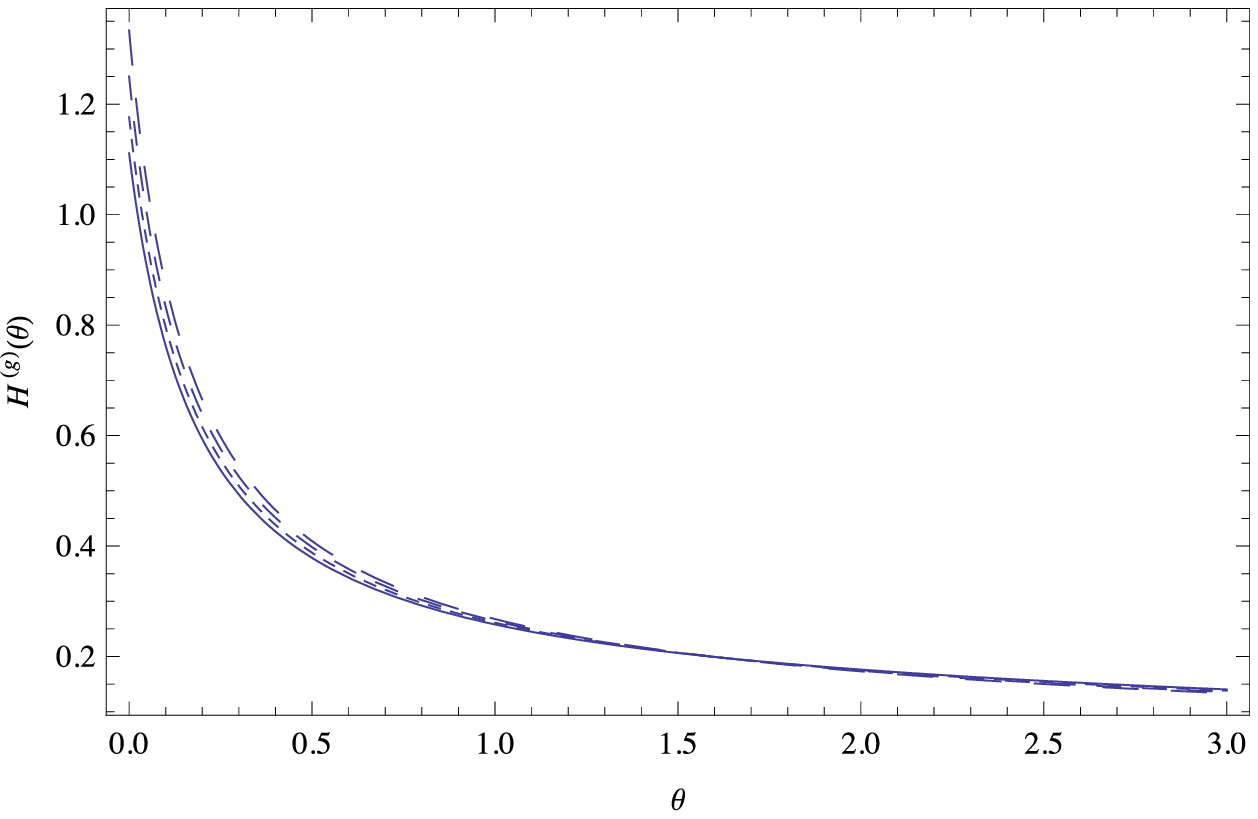}
\includegraphics[width=0.48\textwidth]{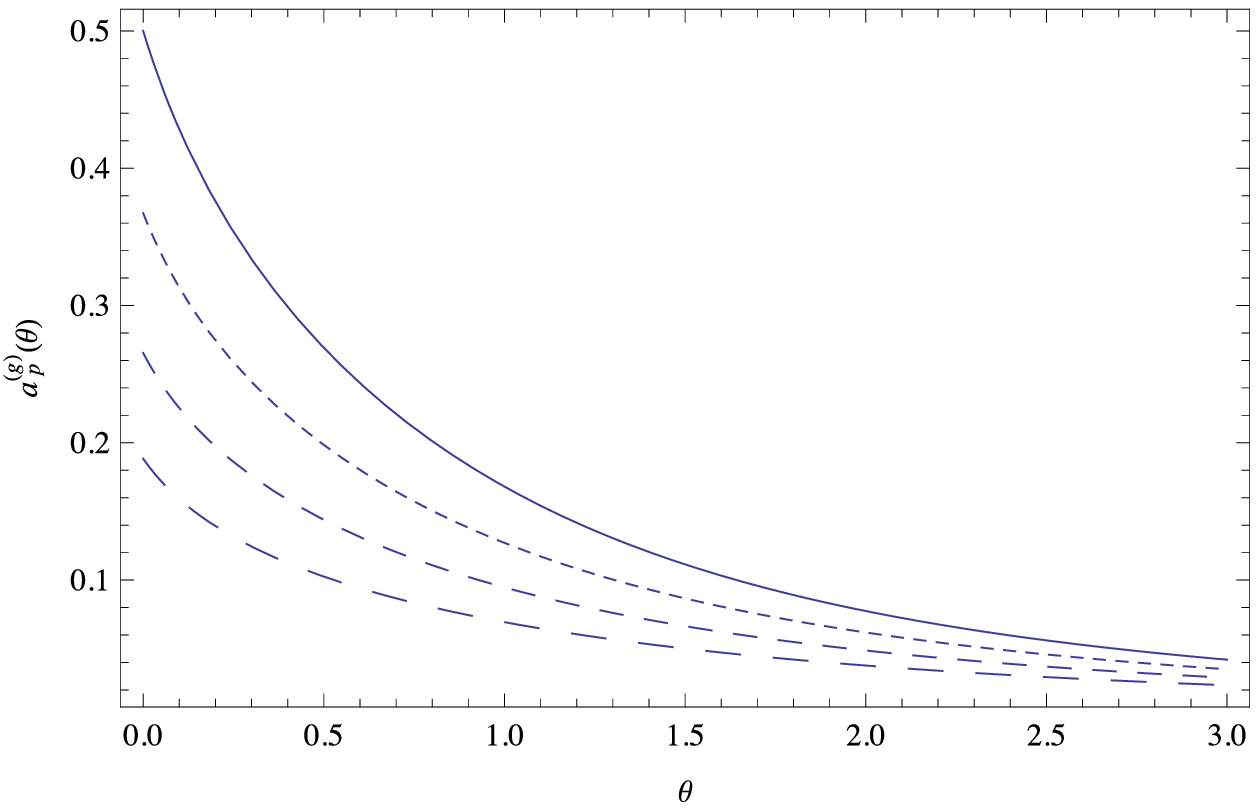}
\caption{In this figure taken from \cite{Harko:2014nya} the evolution of the physical Hubble function in a dust filled universe in a Bianchi type I space-time together with the mean anisotropy parameter $\sigma^g$ (in terms of the notation used in \cite{Harko:2014nya} this corresponds to $a_p^{(g)}$) are shown as a function of the rescaled time $\theta=t\mbi$.}
\label{fig:AnistropicBI_H_sigma}
\end{figure}
As it can be clearly seen in this figure, the mean anisotropy parameter evolves to zero after some time elapses and the universe ends up in an isotropic phase in a Bianchi type I space-time in the EiBI gravity theory. This seems to be complementary to the standard picture in General Relativity where shear decays in the presence of a cosmological constant \cite{PhysRevD.51.3113}. Within the framework of EiBI gravity this property is maintained in the presence of a dust component as well. However, this property does not seem to be general. For instance, instead of a dust fluid, if one considers a radiation fluid with $w=1/3$ or a stiff fluid, then the universe does not isotropise at late times as it was happening for dust. In fact, for a stiff fluid, the mean anisotropy parameter in the $\hat{g}$ metric sector takes rather the form $\sigma^g=r^3(1+r)^3/((1-r)^3r'^2)$. Its evolution together with the evolution of the Hubble function can be extracted from figure \ref{fig:AnistropicBI_H_sigma_Stiff}, which we borrowed from \cite{Harko:2014nya} as well. As one can see in figure \ref{fig:AnistropicBI_H_sigma_Stiff}, the mean anisotropy parameter increases with time. Starting from an initial state with a vanishing anisotropy, the degree of the anisotropy increases until it reaches a maximum constant value.
\begin{figure}[h]
\includegraphics[width=0.48\textwidth]{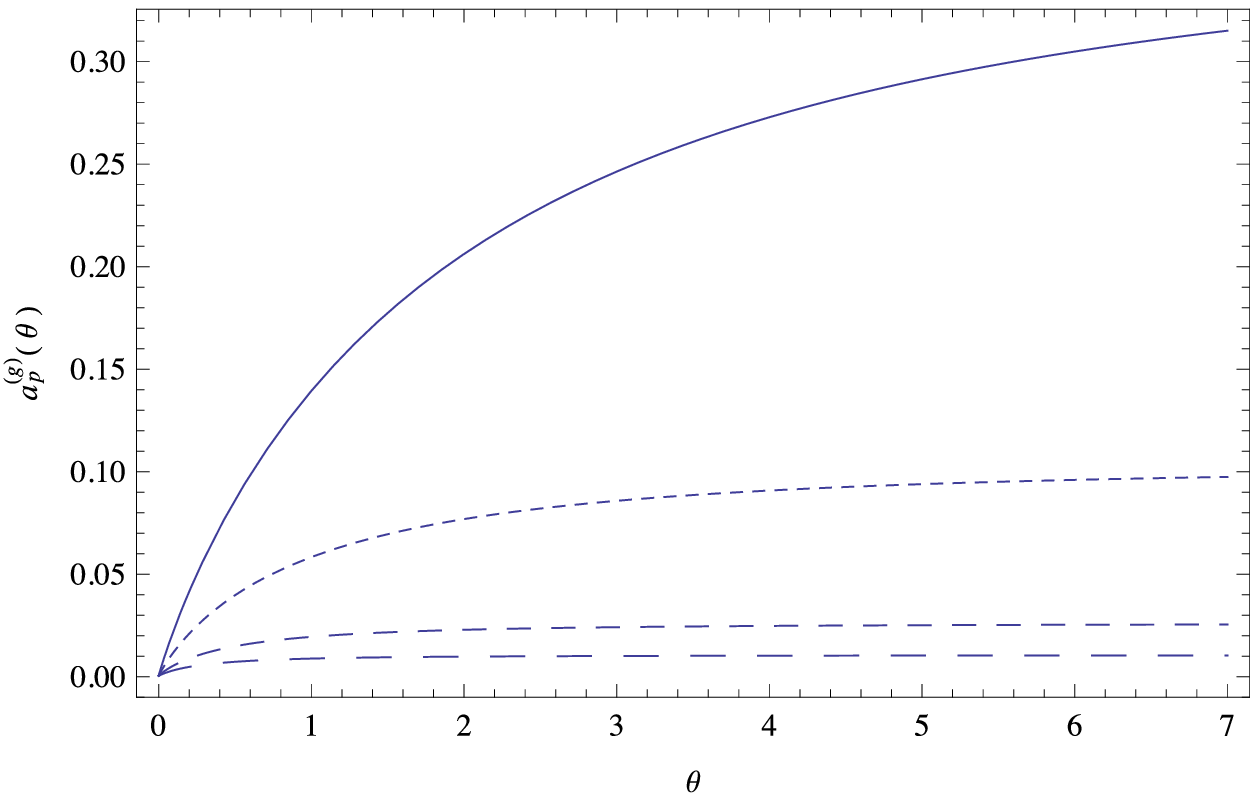}
\includegraphics[width=0.48\textwidth]{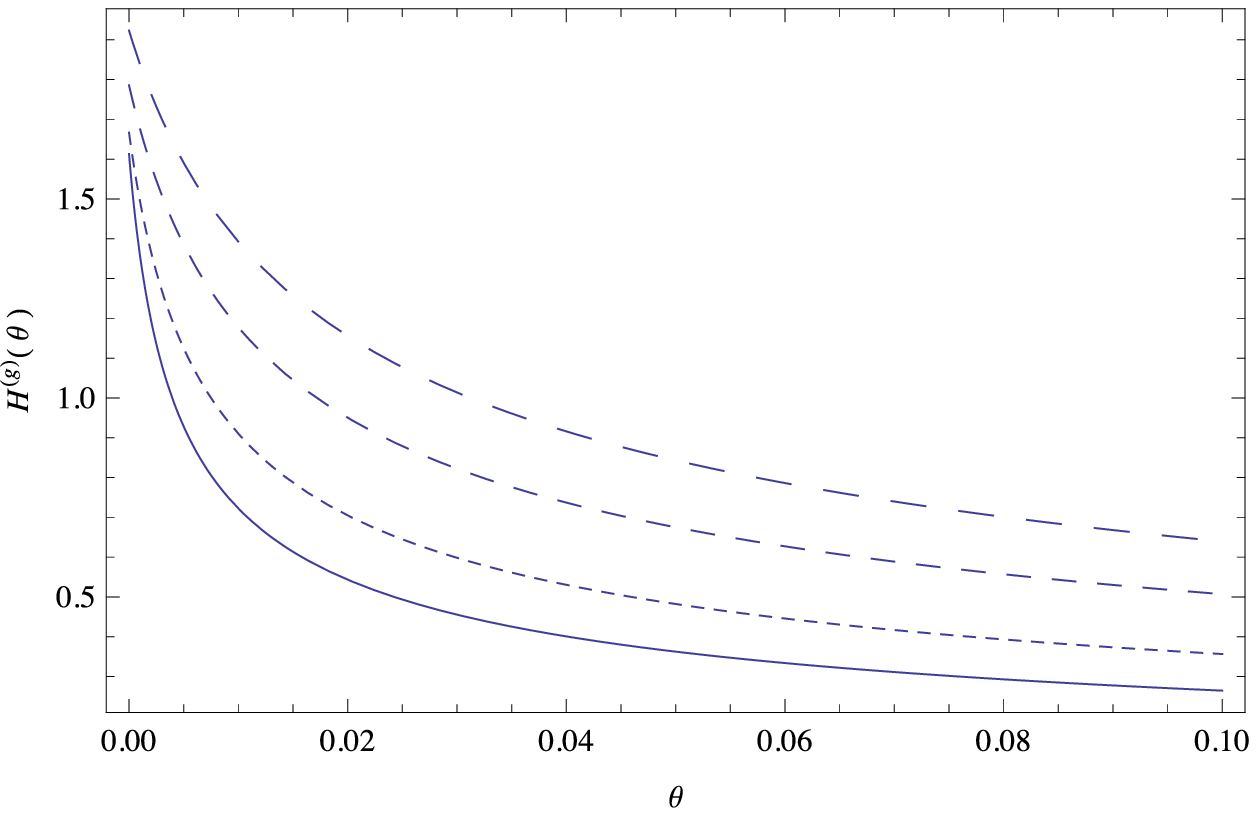}
\caption{This figure borrowed from \cite{Harko:2014nya} illustrates the evolution of the Hubble function and the mean anisotropy parameter $\sigma^g$ (which corresponds to $a_p^{(g)}$ in the notation used in \cite{Harko:2014nya}) as a function of the rescaled time $\theta=t\mbi$ in a universe filled with stiff fluid with different initial values.}
\label{fig:AnistropicBI_H_sigma_Stiff}
\end{figure}

%%%%%%%%%%%%%%%%%%%%%%%%%%%%%

\subsubsection{Late-time cosmology}

As stressed several times throughout this review, the general motivation for Born-Infeld inspired theories of gravity is to modify the gravitational interactions in the high curvatures regime. This means that deviations with respect to GR will typically arise when the curvatures become of the order of the Born-Infeld scale $\mbi^2$. Since the source of gravity is weighted by the Planck scale, an equivalent formulation of this statement is that one only expects deviations from standard gravity when the densities are of the order of\footnote{More precisely, one expects modifications whenever any component of the energy-momentum tensor becomes comparable to $\mbi^2\mpl^2$. In the most standard cosmological backgrounds the energy density is the relevant quantity. but in more general scenarios other components of the energy-momentum tensor could play an important role as well. This is the case for instance of anisotropic cosmological solutions where anisotropic stresses can be present.} $\rho\sim\mbi^2\mpl^2$. For this reason, the natural place where these theories manifest themselves in cosmological scenarios is the early universe, being ideal candidates for inflationary models or bouncing solutions as we have reviewed above. Applications of these theories for models of the late-time universe are instead dissonant as a consequence of their very own defining properties. Since the Born-Infeld effects will become negligible whenever the cosmological energy density drops below $\mbi^2\mpl^2$, from that moment on we will have the usual cosmological evolution with GR governing the gravitational interaction. If we want to have non-negligible effects on cosmological scales today (or somewhere between decoupling and today) that would mean that the whole cosmological evolution would have taken place in the Born-Infeld regime. For this reason, late-time cosmology constitutes an inefficient way to constrain EiBI theories and dark matter and/or dark energy models based on this type of theories are likely to fail in their goal. Models for the dark components of the universe find a better suited arena within the framework of infrared modifications of gravity so that they become relevant in the late-time evolution of the universe. If we want to be on the safe side, we can impose the Born-Infeld corrections be important only before the onset of Big Bang Nucleosynthesis (BBN). Since BBN takes place when the temperature of the universe is roughly $1-10$ MeV, we obtain the conservative constraint $\mbi\gtrsim H_{\rm BBN}\simeq T_{\rm BBN}^2/\mpl\simeq10^{-13}$eV. Notice that this bound is less stringent that the one discussed in section \ref{Sec:Frames} where we obtained $\mbi\gtrsim 10^{-1}$eV from the absence of anomalous interactions in collider experiments.

Despite the general arguments given in the precedent paragraph, there are some works in the literature attempting to explain the dark matter problem in galaxies as an effect of modifying gravity as in EiBI theory \cite{Santos:2015sra}. In \cite{Harko:2013xma}, the authors study spherical dark matter haloes and conclude that the value of the Born-Infeld parameter that allows to realistically reproduce the dark matter haloes is $\mbi^{-2}\simeq 10^{44}$cm$^2$ which translates into $\mbi\simeq 10^{-27}$eV. Even if this value allows to reproduce the haloes, we must remember that the Born-Infeld coupling is universal and this value is in contradiction with the constraints discussed in the previous paragraph so that it is excluded. Analogous studies like e.g. those in \cite{Potapov:2014iva,Izmailov:2015xsa} find similar results and are, thus, subjected to the same limitations. Similarly, the bounds obtained on $\mbi$ from other cosmological and astrophysical probes explain the results found in \cite{Du:2014jka} where the authors compute the matter power spectrum for the EiBI theory. They find that the deviations of the power spectrum with respect to that of $\Lambda$CDM is completely negligible for realistic values of $\mbi$.

Let us also notice that, still within the class of Born-Infeld inspired gravity theories, in order to avoid the aforementioned triviality for late-time cosmological applications, there has been some attempts in the literature to include additional corrections to the EiBI action that could give some effects at late times. We should note however, that this goes against the Born-Infeld spirit and it is very likely that the EiBI sector will not play any role and the whole effect will come from the new terms. As an example of this approach, some works introduced an Einstein-Hilbert term supplementing the Born-Infeld sector, but this class of modifications seriously compromise the stability of the theory. In fact, such variations of the Born-Infeld actions belong to the Class 0 described in the section \ref{Sec:EBIextensionsGeneral} and which are precisely characterised by the presence of pathologies. Thus, even if one can achieve non-negligible effects in the late-time cosmology, this would come at the expense of possibly losing the ghost freedom of the theory. Explicit examples of this type of modifications will be summarised in section \ref{Cos:otherextensions}, but it is worthwhile to stress here that this road of tracing late-time cosmological solutions are doomed to fail due to the mentioned instabilities. As discussed in \ref{Sec:EBIextensionsGeneral}, if one really wants to add an additional Einstein-Hilbert term in the Lagrangian, then the Born-Infeld interactions have to be modified so that the ghost-free massive (bi-) gravity potential interactions in its formulation in terms of the auxiliary metric are recovered.

From the above discussion it is clear that Born-Infeld inspired theories of gravity cannot play a relevant role for the cosmological evolution from roughly BBN (where we need to have standard gravity) until today. There is however a more natural place to study potential effects of EiBI in the late-time cosmology residing within the framework of future cosmological singularities. The properties of dark energy are crucial for the future evolution of the universe and its eventual fate. At this respect, some models predict the existence of future singularities that can be broadly classified according to the divergence of some cosmological quantity (see for instance \cite{Barrow:2004xh,Nojiri:2005sx,Dabrowski:2009kg,Caldwell:2003vq,Fernandez-Jambrina:2014sga,Bouhmadi-Lopez:2014cca,BeltranJimenez:2016dfc,Albarran:2017swy} for some related literature). It is indeed common to perform such a classification attending to which derivative of the scale factor diverges first \cite{FernandezJambrina:2006hj}. In some scenarios with future singularities, the Hubble expansion rate or its derivative show divergences so that the Born-Infeld corrections will eventually be relevant again and one could wonder if such future singularities could be tamed. This was studied in \cite{Bouhmadi-Lopez:2013lha,Bouhmadi-Lopez:2014tna} and it was found that generally future cosmological singularities can remain, although in some cases the divergences can be somewhat smoothed. In \cite{Bouhmadi-Lopez:2016dcf} the authors argued that the classical big rip singularity might be avoided by applying the quantisation based on the Wheeler-DeWitt equation to the EiBI model.

%%%%%%%%%%%%%%%%%%%%%%%%%%%%%%

%%%%%%%%%%%%%%%%%%%%%%%%%%%%%%
\subsection{General cosmological framework for Born-Infeld inspired gravity theories} \label{sec:general_framework_cosmology}

After reviewing the cosmological applications of the EiBI model, we will discuss the cosmological studies performed for other Born-Infeld inspired theories of gravity. Most of them share the same underlying features and mechanisms, although leading to different cosmologies depending on the specific model under consideration. Thus, instead of studying the individual extensions one by one, we will develop here a general framework to study cosmological solutions within these theories. In fact, without increasing the level of difficulty, we can consider the general class of theories already analysed in \ref{Sec:EBIextensionsGeneral} and harvest the results of that section to obtain the relevant equations to study the cosmology of these theories. To avoid the reader to thumb through the review, we will rewrite the main equations here for convenience. The starting action is given by\footnote{We are not assuming any projective symmetry a priori on the Ricci tensor so, in principle, we could consider both the symmetric and the antisymmetric parts of the Ricci tensor. The background cosmological evolution where all the relevant objects are assumed to be diagonal will be, in general, oblivious to the presence of the projective symmetry. However, it is crucial when studying the perturbations. For an analysis of the cosmological scenarios in an extended class of theories see  \cite{Jimenez:2015caa}.}
\begin{equation}\label{actGenF_Pal}
\mS=\frac12\mpl^2\mbi^2\intd\sqrt{-g}F\big(\m{P}\big)\,,
\end{equation}
with $P^\mu{}_\nu=\mbi^{-2}g^{\mu\alpha}\mR_{\alpha\nu}(\Gamma)$.
The analysis of the general field equations, even in the presence of torsion, was discussed in great detail in section \ref{Sec:EBIextensionsGeneral}. For our purposes here, the important equations will be those relating the auxiliary metric $q_{\mu\nu}$, which determines the connection as its Christoffel symbols, with the spacetime metric and the matter content. The two metrics are related by  $\m{q}=\m{g}\m{\Omega}$, where $\m{\Omega}$ is the deformation matrix defined as
\begin{equation}
\m{\Omega}^{-1}=\frac{1}{\sqrt{\det \m{F}_{\m{P}}}}\left(\frac{\partial F}{\partial \m{P}}\right)^T.
\end{equation}
Notice that this definition relates $\m{\Omega}$ and $\m{P}$ so that all the equations below will admit equivalent formulations in terms of $\m{\Omega}$ or $\m{P}$ alone. By using the definition of the deformation matrix, the metric field equations can be expressed as
\begin{equation}\label{metricEOMinOmega}
\m{\Omega}^{-1}\m{P}=\frac{1}{\mbi^2\mpl^2\sqrt{\det\m{\Omega}}}\Big(\lag_G\Id+\m{T}\m{g}\Big)\,
\end{equation}
where $\lag_G=\frac12 \mpl^2\mbi^2 F$ is the Lagrangian. These equations give the deformation matrix $\m{\Omega}$ (or the fundamental object $\m{P}$) in terms of the matter content and the spacetime metric. The resolution of the problem will be completed with the differential equations for the auxiliary metric (see \ref{Sec:EBIextensionsGeneral})
\begin{equation}
R^\mu{}_\nu(q)=\frac{1}{\mpl^2\sqrt{\det\m{\Omega}}}\Big(\lag_G\delta^\mu{}_\nu+T^\mu{}_\nu\Big)\,.
\label{eqR_enWFT}
\end{equation}
After briefly reviewing the relevant equations, we can proceed to the study of cosmological scenarios. As usual, we will consider a homogeneous and isotropic background metric described by the FLRW line element
\be
\d s_g^2=-N^2(t)\d t^2+a^2(t)\d\vec{x}^2
\ee
and a perfect fluid with isotropic pressures as matter sector
\be
T^\mu{}_\nu=
\left(
\begin{array}{cc}
 -\rho & 0  \\
  0&   p\delta^i_j
\end{array}
\right)\ .
\ee
As an additional condition, we will assume that all relevant quantities inherit this form so that we will have
\be
\m{\Omega}=
\left(
\begin{array}{cc}
 \Omega_0 & 0  \\
  0&   \Omega_1\delta^i_j
\end{array}\,
\right)\qquad\text{and}\qquad
\m{P}=
\left(
\begin{array}{cc}
 P_0 & 0  \\
  0&   P_1\delta^i_j
\end{array}
\right).
\ee
As we have explained several times above, the recovery of GR at low curvatures imposes $\m{\Omega}\simeq\Id$ for $P_0, P_1\ll1$. The form of the deformation matrix ensures that the auxiliary metric will also have a FLRW line element
\begin{eqnarray}
\d s_q^2=-\tilde{N}^2(t)\d t^2+\tilde{a}^2(t)\d\vec{x}^2 \, ,
\end{eqnarray}
with $\tilde{N}^2=N^2\Omega_0$ and $\tilde{a}^2=a^2\Omega_1$. We keep the explicit dependence on the lapse function $N(t)$ for later convenience. Once we have specified the assumptions for our homogeneous and isotropic Ans\"atze, we can now proceed to write the background metric field equations (\ref{metricEOMinOmega}), which read
\begin{eqnarray}
\frac{P_0}{\Omega_0}&=&\frac{1}{\mbi^2\mpl^2\sqrt{ \Omega_0\Omega_1^3}} \Big(\lag_G+T^0{}_0 \Big),\nonumber \\
\frac{P_1}{\Omega_1}&=&\frac{1}{\mbi^2\mpl^2\sqrt{ \Omega_0\Omega_1^3}}\Big(\lag_G+\frac13T^i{}_i \Big) \,.
\label{genFMN_EoM_homYani}
\end{eqnarray}
As anticipated above, for a given function $F(\m{P})$, these equations will allow to obtain the components of $\m{\Omega}$ (or those of $\m{P}$) in terms of the energy density $\rho=-T^0{}_0$ and the pressure $p=\frac13 T^i{}_i$ of the matter fields. An important point to keep in mind is that these equations are non-linear so that, in general, we will find several branches. Out of those branches, the condition \refeq{Eq:LimitlowR} on the function $F$ will guarantee the existence of one particular branch that will be continuously connected with GR at low curvatures. This will be the interesting branch for most applications, although other branches might also offer interesting cosmological scenarios.

As we have seen in the previous section for the EiBI model, the crucial step to study the cosmological evolution is to extract the dependence of the Hubble expansion rate in terms of the energy density and pressure of the matter fields. For this purpose, we will make use of the Einstein tensor of the auxiliary metric and express its $00$ component in two different ways. We will start from the definition of the Einstein tensor of $q_{\mu\nu}$ given by

\be
\m{G}(q)=\m{R}-\frac12\m{q}\Tr\Big(\m{q}^{-1}\m{R}\Big).
\ee
First, we will compute its $00$ component in terms of the auxiliary metric, which will simply give the corresponding Hubble expansion rate:
 \begin{equation}
G_{00}(q) =3\tilde{H}^2=3\left(\frac{\d\ln \tilde{a}}{\d t}\right)^2=
3\left[ H^2+\frac12\frac{\d\ln\Omega_1}{\d t}\right]^2.
\label{eq:generalcaseGfl}
\end{equation}
Since $\Omega_1=\Omega_1(\rho,p)$ as obtained from \refeq{genFMN_EoM_homYani}, we can express the time derivative of $\Omega_1$ in terms of derivatives with respect to $\rho$ and $p$. Thus, if we use that matter fields are assumed to be minimally coupled so that they satisfy the usual conservation equation
\be
\dot{\rho}+3H(\rho+p)=0,
\label{eq:conservationmatter}
\ee
we can finally arrive at
\be
G_{00}(q)=3H^2 \left[1 - \frac{3}{2}\big(\rho+p\big)\Big( \partial_\rho\ln \Omega_1+c_s^2\partial_p\ln \Omega_1 \Big) \right]^2\,
\label{eq:generalcaseGfl2}
\ee
where we have introduced the sound speed $c_s^2\equiv\dot{p}/\dot \rho$. If we further assume a barotropic equation of state $p=p(\rho)$ the sound speed can also be written as $c_s^2=\d p/\d \rho$. This completes the first part of our computation of the Hubble expansion rate. The second part consists in writing the Einstein tensor of the auxiliary metric by using the definition $\m{P}=\mbi^{-2}\m{g}^{-1}\m{R}$ and the relation between the two metrics through the deformation matrix $\m{q}=\m{g}\m{\Omega}$ so that we obtain

\begin{equation}
\m{G}(q) =\m{R}-\frac12\m{q}\Tr\Big(\m{q}^{-1}\m{R}\Big)= \mbi^2\,\m{g}\left[ \m{P}-\frac12\m{\Omega}\,{\rm Tr}\Big(\m{\Omega}^{-1}\m{P}\big) \right].
\end{equation}
We can again extract the expression for $G_{00}$, this time in terms of the components of $\m{P}$ and $\m{\Omega}$, as follows:

\begin{equation}
G_{00}(q) = \frac12\mbi^2\,g_{00}\left(P_0 -3\frac{\Omega_0}{\Omega_1}P_1\right)\,.
\label{eq:generalcaseG00inPomega}
\end{equation}
If now we equal the right hand sides of  \refeq{eq:generalcaseGfl2} and \refeq{eq:generalcaseG00inPomega} and solve for $H^2$ we finally obtain

\begin{equation}
\frac{3H^2}{\mbi^2N^2}=\frac{ 3\Omega_0P_1-P_0 \Omega_1 }{2\Omega_1\left[1 - \frac32(\rho+p)\Big( \partial_\rho\ln \Omega_1+c_s^2\partial_p\ln \Omega_1 \Big) \right]^2}\,,
\label{eq:generalcaseH2inPomega}
\end{equation}
where we have used that $g_{00}=-N^2$. This is the master equation providing the modified Friedman equation for the theories under consideration, i.e., it gives the dependence of the Hubble function in terms of the matter field variables. Let us remember that the components of $\m{P}$ and $\m{\Omega}$ are functions of $\rho$ and $p$ as obtained from the resolution of \refeq{genFMN_EoM_homYani} and, hence, the right hand side of the above equation only depends on the matter sector variables. A very distinctive feature of these theories is the appearance of $c_s^2$ in this modified Friedman equation. This means that, unlike the case of GR and many other modified gravity theories, the sound speed not only affects the evolution of the perturbations, but it also affects the background evolution. In particular, this includes one additional {\it parameter} for the homogeneous cosmologies of these theories. While in the most extensively studied modified gravity theories the equation of state fully determines the background evolution, in the theories under consideration here (among which many Born-Infeld inspired theories are included) there is a further dependence encoded in $c_s^2$. Moreover, some matter sources can actually have a non-constant sound speed (like in the case of several interacting fluids) and it could even depend on $H(t)$ so that \refeq{eq:generalcaseH2inPomega} will be an implicit equation for the Hubble expansion rate. We have already encountered a particular case of this result in the EiBI theory rephrased in terms of a time-dependent equation of state parameter and we saw that the background cosmology depends not only on $w(t)$ but also on $\dot{w}$.

There is a number of interesting general features that can be directly inferred from \refeq{eq:generalcaseH2inPomega}. The first thing to notice is that now it is very easy to understand the mechanism by which these theories can give rise to bouncing solutions without violating the NEC. For that, let us rewrite the modified Friedman equation in the more familiar form

\be
H^2=\frac{8\pi G_{\rm eff}(\rho,p,c_s^2)}{3}\rho
\label{eq:modFriedmanG}
\ee
that is closer to its usual form and we have encoded all the modified effects into the effective Newton's constant
\be
8\pi G_{\rm eff}(\rho,p,c_s^2)=\mbi^2\frac{ 3\Omega_0P_1-P_0 \Omega_1 }{2\Omega_1\left[1 - \frac32(\rho+p)\Big( \partial_\rho\ln \Omega_1+c_s^2\partial_p\ln \Omega_1 \Big) \right]^2}
\label{eq:GNeffective}
\ee
where we have momentarily set the lapse to $N=1$. If we take the time derivative of \refeq{eq:modFriedmanG} and use the conservation equation \refeq{eq:conservationmatter} we find

\be
\dot{H}=-4\pi G_{\rm eff}\Big(\rho+p\Big)+\frac{4\pi \dot{G}_{\rm eff}}{3H}\rho.
\ee
In GR with minimally coupled fields, the existence of bouncing solutions (omitting the possible presence of spatial curvature for the sake of simplicity) characterised by an evolution where $\dot{H}$ is initially negative (contracting phase) and becomes positive (expanding phase) is subjected to a regime where the NEC is violated in the initial regime and it holds in the final stage. For the theories under consideration here\footnote{This discussion is not specific of these theories, but it also applies to other modified gravity theories or theories involving non-minimally coupled fields. In general, the argument presented here will be valid for all theories giving rise to a non-constant effective Newton's constant for the cosmological evolution. Once again, the distinctive features of the theories considered here arise from the dependence on $c_{\rm s}^2$ that is not present in other classes of modified gravity theories, and this is what can introduce novel features.}, the presence of the time derivative of the effective Newton's constant makes it possible to have bouncing solutions where the NEC holds throughout the entire evolution. Notice that, at the bounce, the Hubble expansion rate must vanish at a finite but non-vanishing density, so the bouncing will generally occur when
\be
G_{\rm eff}\big(\rho_ b,p_b,c_{s,b}^2\big)=0
\ee
where the subscript $b$ stands for their values at the bounce. By looking at \refeq{eq:GNeffective} we can see that the bounce can generally happen in two ways, namely:
\begin{itemize}
\item$i)$ The numerator vanishes so that $3 \Omega_0P_1-P_0\Omega_1=0$.
\item $ii)$ The denominator diverges so that $\Omega_1\left[1 - \frac32(\rho+p)\Big( \partial_\rho\ln \Omega_1+c_s^2\partial_p\ln \Omega_1 \Big) \right]^2\rightarrow\infty$.
\end{itemize}
Let us stress that these two possibilities are the most straightforward (and perhaps smooth) ways to realise the bouncing solution, but they are not exhaustive. For instance, one could envisage situations where both the numerator and the denominator diverge (or vanish) while the quotient is a well-behaved function with  some roots at $\rho\neq0$. Leaving this possibility aside, the bounce realised by means of $ii)$ will generally rely on the existence of a divergence either in $\Omega_1$ (or one of its derivatives) or in $c_s^2$. Since both of this quantities have a physical relevance, $\Omega_1$ relates the two metrics and $c_s^2$ typically gives the adiabatic sound speed, a divergence in them can potentially give rise to divergent physical effects. On the other hand, the bounce characterised by $i)$ takes place when $3 \Omega_0P_1-P_0\Omega_1=0$. From \refeq{genFMN_EoM_homYani}, we can obtain that
\be
3 \Omega_0P_1-P_0\Omega_1=\frac{1}{\mbi^2\mpl^2}\sqrt{\frac{\Omega_0}{\Omega_1}}\Big(2\lag_G+T\Big)
\ee
with $T=T^\mu{}_\mu$ the trace of the energy momentum tensor. Thus, a bouncing solution where both metrics are regular (i.e., finite and non-vanishing $\Omega_0$ and $\Omega_1$) will be characterised by the equation $2\lag_G+T=0$. Interestingly, for a radiation dominated universe the energy-momentum tensor is traceless and the condition reduces to $F(\m{P})=0$.

In a similar way as we studied the tensor perturbations for the EiBI model in the precedent sections, we can extend the analysis to the general class of theories considered here. We will closely follow the analysis in \cite{Jimenez:2015caa} where tensor perturbations are analyse in detail for an even larger class of theories formulated in the affine formalism. We will recognise that most of the properties we discussed for the EiBI theory are actually generic features for the theories described by \refeq{actGenF_Pal}. Let us then consider tensor perturbations on top of the homogeneous and isotropic background defined as

\begin{equation}
\delta g_{\mu\nu}=
\left(
\begin{array}{cc}
 0 & 0  \\
  0&   a^2h_{ij}
\end{array}\right)\,\quad
\delta q_{\mu\nu}=
\left(
\begin{array}{cc}
 0 & 0  \\
  0&   a^2\Omega_1f_{ij}
\end{array}\,
\right)\quad\text{and}\quad
\delta T^\mu{}_\nu=
\left(
\begin{array}{cc}
 0 & 0  \\
  0&   \Pi^i{}_j
\end{array}\,
\right)
\end{equation}
with $\Pi^i{}_j$ representing the anisotropic stress\footnote{We are dropping here the perfect fluid assumption in the perturbed sector for generality.}. An important property that considerably simplifies the computations with tensor perturbations is that they only live in the spatial 3-dimensional space and all the background quantities are diagonal in that box. This means that, at first order in tensor perturbations, any possible pair of matrices appearing in the equations will commute. Furthermore, the tensor perturbation of any scalar quantity vanishes identically, for instance we will have $\delta\det\m{\Omega}=0$ and so on. The equation \refeq{metricEOMinOmega} at first order in tensor perturbations reads
\be
\delta\m{\Omega}^{-1}\m{P}+\m{\Omega}^{-1}\delta\m{P}=\frac{1}{\mbi^2\mpl^2\sqrt{\Omega_0\Omega^3_1}}\m{\Pi}.
\label{eq:eqdeltaOmega}
\ee
Again, $\m{\Omega}$ and $\m{P}$ are related by means of the definition of $\m{\Omega}$ so the above equation can be seen as an equation for $\delta\m{\Omega}$, whose solution will have the general form
\be
\delta\Omega^i{}_j=\omega(\rho,p)\,\Pi^i{}_j
\ee
with $\omega(\rho,P)$ some function obtained from solving \refeq{eq:eqdeltaOmega} which only depends on background quantities. This expression for the perturbation of $\m{\Omega}$ allows to express the perturbation of the auxiliary metric as
\begin{equation}
\delta\m{q}=\delta\m{g}\,\m{\Omega}+\m{g}\,\delta\m{\Omega}\Rightarrow\delta q_{ij}=\Omega_1\delta g_{ij}+a^2\delta\Omega_{ij}=\Omega_1\delta g_{ij}+a^2\omega\,\Pi_{ij}
\end{equation}
where the spatial indices have been lowered with the Kronecker delta. In terms of $h_{ij}$ and $f_{ij}$, we then have
\be
f_{ij}=h_{ij}+\frac{\omega}{\Omega_1}\Pi_{ij}.
\label{eq:fToh}
\ee
This result generalises the one already found for the EiBI theory in section \ref{sec_tensorInst_BI}. An important property is that, in the absence of any anisotropic stresses, the tensor perturbations of the two metrics are identical and we can simply talk about metric tensor perturbations without referring to any specific metric. In other words, there is only one class of gravitational waves.This roots in the conformal relation for the two background metrics in the spatial 3-hypersurfaces. The field equations for these gravitational waves can be easily computed from \refeq{eqR_enWFT}, whose tensor perturbation yields
\begin{equation}
\delta R^i{}_j(q)= \frac{1}{\mpl^2\sqrt{\Omega_0\Omega_1^3}}\Pi^i{}_j
\label{pertRinFMN}
\end{equation}
As it becomes clear, this evolution equation for gravitational waves is exactly the same as the one found in GR barring the replacement

\begin{equation}
\mpl^2\rightarrow\mpl^2\sqrt{\Omega_0\Omega_1^3}.
\label{GNtoGeffinFMN}
\end{equation}
In terms of the metric perturbations the equation (\ref{pertRinFMN}) can be equivalently written in the familiar form
\begin{equation}
\ddot{f}_{ij}+\left(3\tilde H(t)-\frac{\dot{\tilde N}(t)}{\tilde N(t)}\right)\dot{f}_{ij}-\frac{\tilde N(t)^2}{\tilde a(t)^2}\nabla^2 f_{ij}=16\pi G_{\rm gw}\Pi_{ij},
\label{pert_htequation}
\end{equation}
where $G_{\rm gw}=G_{\rm N}/\sqrt{\Omega_0\Omega_1^3}$. In the regime of small energy densities, $\Omega_0\simeq 1$ and $\Omega_1\simeq1$ so $G_{\rm gw}\simeq G_{\rm N}$. Once the solution for $f_{ij}$ is computed from the above equation, the evolution of the perturbation $h_{ij}$ is determined by the relation \refeq{eq:fToh}. It should not come as a surprise by now that the gravitational waves $f_{ij}$ satisfy the usual equation for cosmological tensor perturbations but with respect to the auxiliary background metric and a modified coupling to the source. This is simply the cosmological application of the discussion presented in \ref{Sec:EBIextensionsGeneral} where it was shown that, in the Einstein frame, the auxiliary metric acquires the standard Einstein-Hilbert kinetic term, but it is coupled in a non-standard way to the matter fields. Since we are working at first order in tensor perturbations, the modified coupling to matter fields was expected to appear as a modified Newton's constant. From \refeq{pert_htequation} we can also understand the rising of tensor perturbations discussed in section \ref{sec_tensorInst_BI} for the bouncing and loitering solutions of EiBI as a consequence of having a non-regular auxiliary metric.

%%%%%%%%%%%%%%%%%%%%%%%%%%%%%%

%%%%%%%%%%%%%%%%%%%%%%%%%%%%%%
\subsection{Elementary symmetric polynomials extension: Minimal model}\label{subsection_Minext_BI}

\begin{figure}[h!]
\begin{center}
\includegraphics[width=14cm]{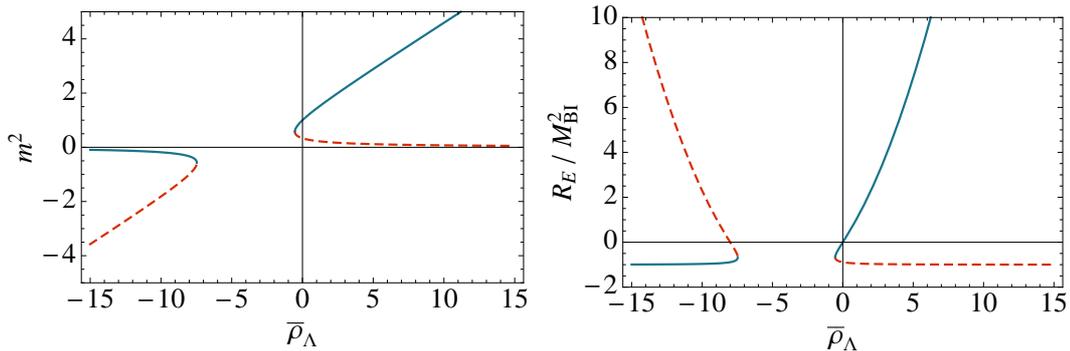}
\caption{In this plot (adapted from \cite{Jimenez:2014fla}) we show the two branches of solutions for $m^2$ as a function of $\bar{\rho}_\Lambda$ given in
\refeq{Eq:solm2Einstein} (left panel) and the characteristic scale $R_E$ normalized to $\mbi^2$ of the corresponding Einstein space (right panel). The blue-solid solution is continuously connected with GR in vacuum, while the red-dashed solution represents the branch giving rise to dS/AdS in vacuum. Interestingly, the dS/AdS branch is almost insensitive to the presence of $\bar{\rho}_\Lambda$. In this branch, the value of $R_E$ quickly saturates to $-\mbi^2$ and remains constant irrespectively of the cosmological constant. Finally, we can see how the positivity of $m^2$ selects the physical solutions and imposes some bounds on the values of the $\bar{\rho}_\Lambda$ that can be accommodated.}
\label{MinModel_m2Einstein}
\end{center}
\end{figure}

We will now consider the Class-I theory introduced in \cite{Jimenez:2014fla} that we already discussed in section \ref{Sec:ClassI}. This family of theories consists in extending the EiBI theory to include all the elementary symmetric polynomials and is described by the actions \refeq{gen_Born_Infeld}. The cosmology of the general case including all the elementary polynomials has not been performed yet in the literature. The fourth polynomial coincides with EiBI so that its cosmology is the one extensively discussed above. The other polynomial whose cosmology has also been investigated is the first one, that was called Minimal model. The corresponding action is given by
\begin{equation}
\mathcal S_{\rm min}=\mbi^2\mpl^2\int \mathrm{d}^4x\sqrt{-g} {\rm Tr}\left[\sqrt{\mathbbm 1+\mbi^{-2}\hat{g}^{-1}\hat{\mR}}-\mathbbm 1\right] \,,
\label{eq:gminimal}
\end{equation}
where the constants have been chosen as to match GR without a cosmological constant in the low curvature regime. A possible cosmological constant term will be considered as part of the matter sector. As shown in the corresponding part of section \ref{Sec:ClassI}, it is convenient to introduce the fundamental matrix of the model given by

\be
\m{M} \equiv \sqrt{\Id + \mbi^{-2}
  \hat{g}^{-1} \hat{\mR}}.
\ee
This fundamental matrix must be positive definite on physically acceptable solutions and is related to the deformation matrix by $\m{\Omega}=\m{M}/\sqrt{\det\m{M}}$, as shown in equation \refeq{Eq:OmegaToMminimal}, thus guaranteeing that both the auxiliary and the spacetime metrics have the same signature. Furthermore, $\m{M}$ satisfies the equation \refeq{Eq:MinimalMatrix}, that we write here again

\be
\m{M}^{-1}-\m{M}-\Big[\Tr\Big(\m{M}-\Id\Big)\Big]\Id=\frac{1}{\mbi^2\mpl^2}\m{T}\m{g}\, .
\label{Eq:MinimalMatrix2}
\ee
Before studying the more general cosmological solutions within this model, we shall first consider Einstein space solutions characterised by $\mR_{(\mu\nu)}(\Gamma)=R_E g_{\mu\nu}$, with $R_E$ some constant curvature. In terms of the fundamental matrix, this means $M^\mu{}_{\nu}=m^2 \delta^\mu{}_{\nu}$ where $m^2=\sqrt{1+R_E\mbi^{-2}}$ and the field equations \refeq{Eq:MinimalMatrix2} simplify to

  \begin{equation}
\left(4-3m^2-\frac{1}{m^2} \right)g_{\mu\nu}=\frac{1}{\mpl^2\mbi^2}T_{\mu\nu}.
\end{equation}
The conservation of the energy-momentum tensor implies that $m^2={\rm const}$. Since we have that $R_E=(m^4-1)\mbi^2$, the curvature $R_E$ must indeed be a constant and cannot be promoted to some arbitrary function. In other words, only a fluid corresponding to a cosmological constant can support Einstein space solutions, as one would have expected. For $T_{\mu\nu}=-\rho_{\Lambda}g_{\mu\nu}$, the above equations
give

\begin{equation}
4-3m^2-\frac{1}{m^2}+\bar{\rho}_\Lambda=0\,,
\end{equation}
where $\bar{\rho}_\Lambda=\rho_\Lambda/(\mpl^2\mbi^2)$. The solution of this equation is

\be
m^2=\frac{4+\bar{\rho}_\Lambda\pm \sqrt{4+\bar{\rho}_\Lambda(8+\bar{\rho}_\Lambda)}}{6}.
\label{Eq:solm2Einstein}
\ee
For these Einstein space solutions the deformation matrix is simply $\m{\Omega}=m^{-2}\Id$ so that both metrics are conformally related as
$q_{\mu\nu}=m^{-2}g_{\mu\nu}$. In the absence of the cosmological constant $\rho_\Lambda=0$, i.e., the vacuum solutions, the two branches give $m^2=1$ and $m^2=1/3$ respectively. The former corresponds to a Ricci-flat space with $\mR_{(\mu\nu)}=0$ and represents the branch continuously connected with GR, whereas the latter gives $\mR_{(\mu\nu)}=(-8\mbi^2/9)g_{\mu\nu}$ and represents a de Sitter (anti-de Sitter) space for negative (positive) $\mbi^2$ without the need for a cosmological constant. The two branches of solutions for $m^2$ in terms of $\rho_\Lambda$ are illustrated in figure \ref{MinModel_m2Einstein}. We can clearly see how the physical condition $m^2>0$ gives the bound $\bar{\rho}_\Lambda\ge2(\sqrt{3}-2)$.
%%%%%%%%%%%%%%%
%%%%%%%%%%%%%%%%%

\begin{figure}[h!]
\begin{center}
\includegraphics[width=14cm]{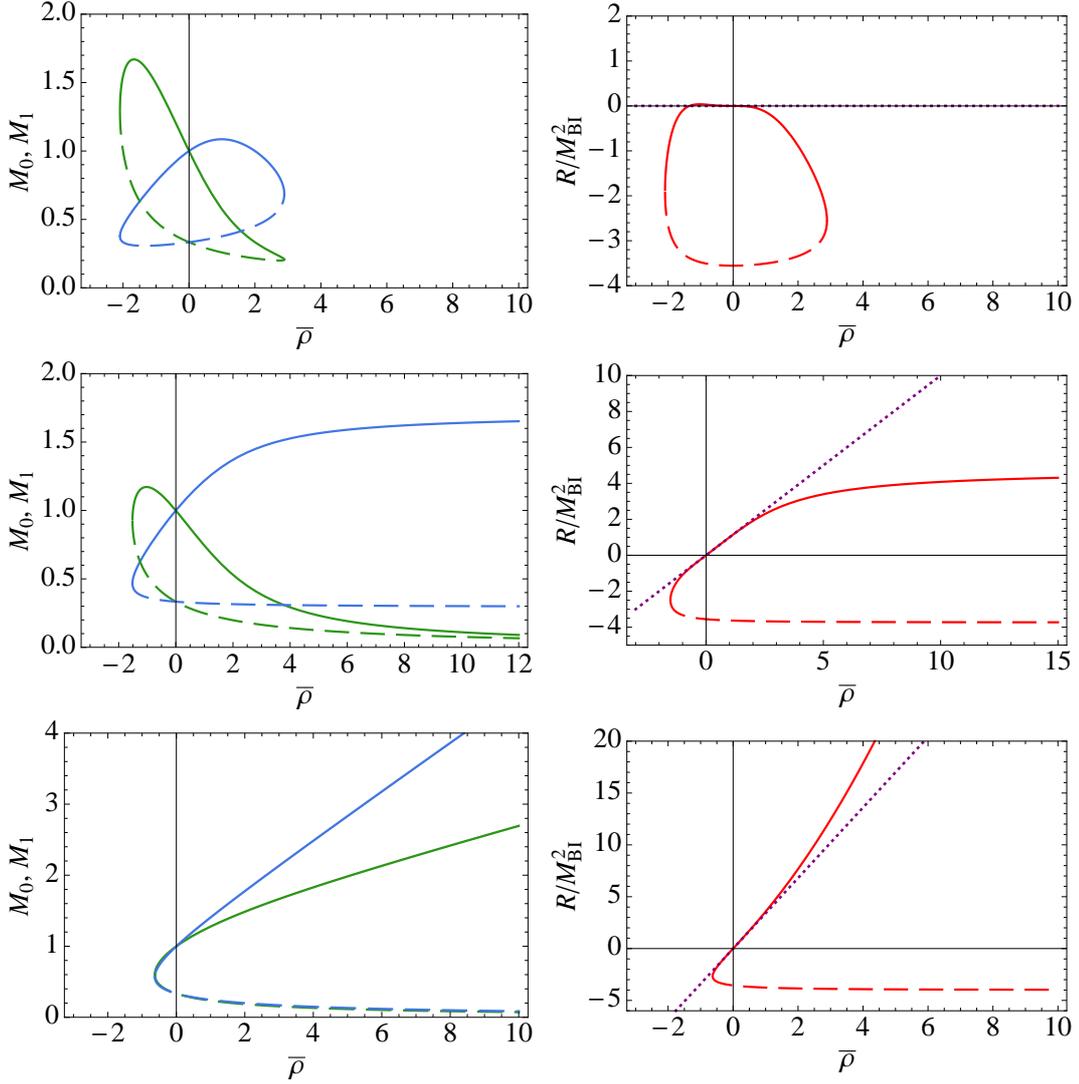}
\caption{Figure adapted from \cite{Jimenez:2014fla}. In the left panels we show the solutions for $M_0$ (green) and $M_1$ (blue) for the Branch I (solid) and Branch II (dashed). In the right panels  the corresponding solutions for the scalar curvature $\mR$ are illustrated. Three types of fluids are considered  from top to bottom: radiation ($p=\rho/3$), dust ($p=0$) and a fluid with $p=-0.8\rho$. The dotted-purple lines represent the corresponding solutions in GR. We can see that the solutions for radiation are bounded for both $\rho$ and $\mR$, for dust the density can grow to infinity but the curvature is bounded by $\sim\mbi^2$ and, finally, for the fluid with $w=-0.8$ neither the density nor the curvature are bounded.}
\label{MinModel_M0yM1}
\end{center}
\end{figure}

\begin{figure}[h!]
\begin{center}
\includegraphics[width=7cm]{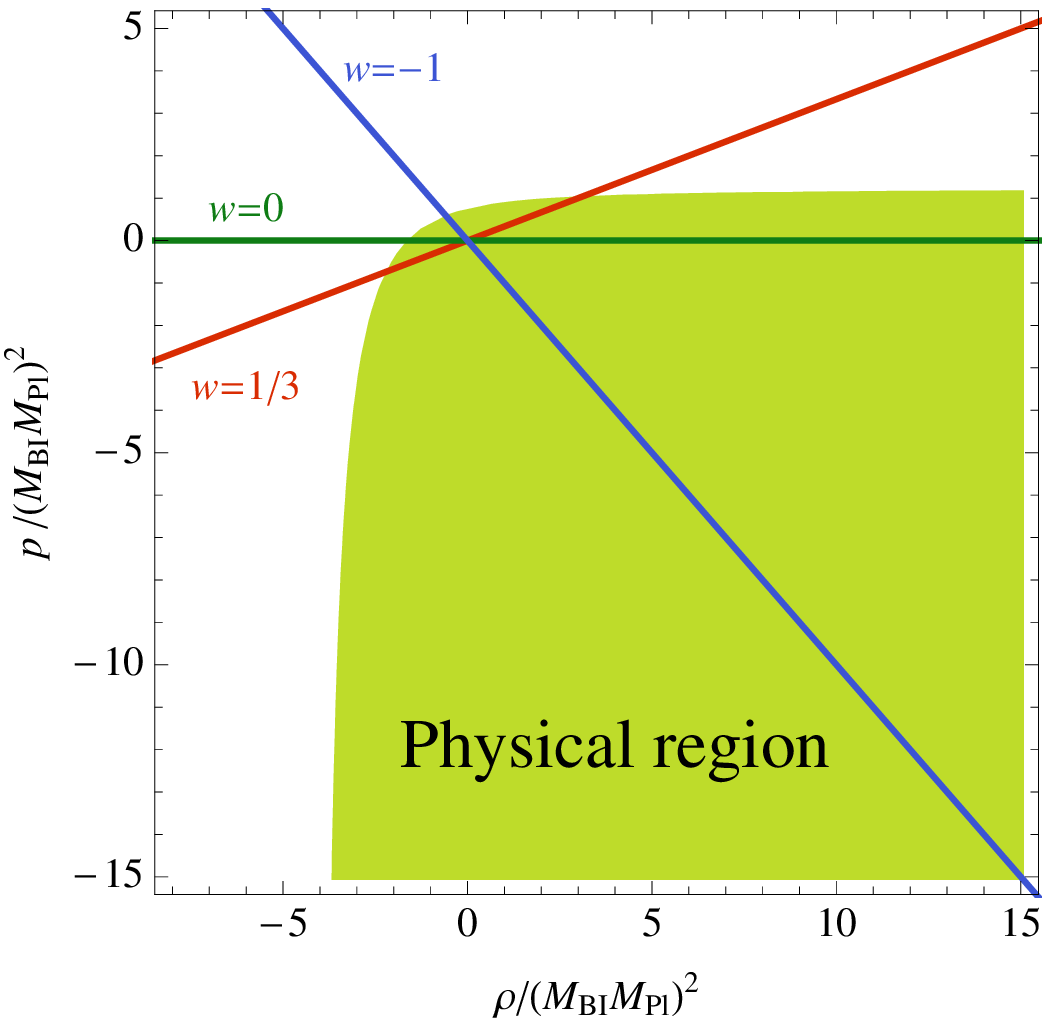}
\includegraphics[width=7cm]{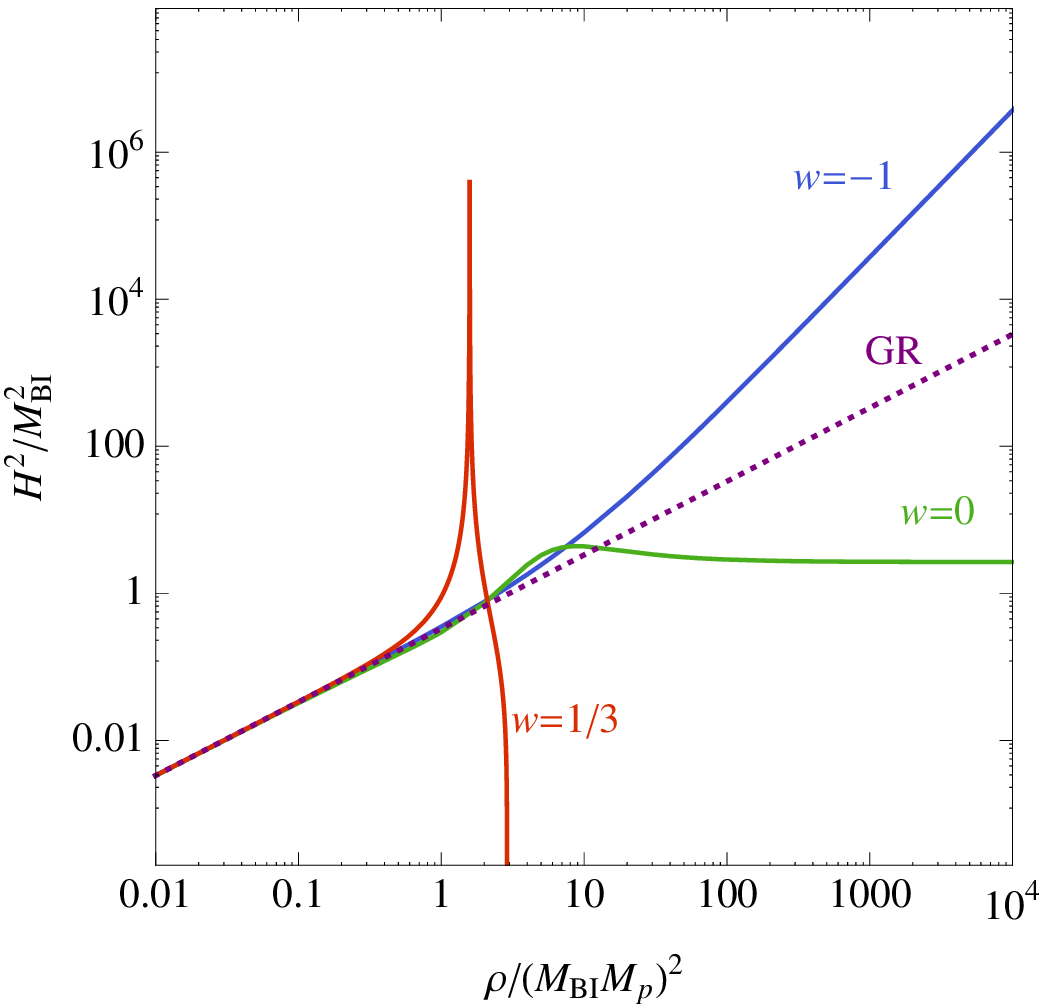}
\caption{Figures adapted from \cite{Jimenez:2015jqa}. Left panel: physical region determined by imposing the positivity of the fundamental matrix $\m{M}$. It is also shown some important equation of state parameters to illustrate the bounds on $\rho$ and $p$ imposed by the Born-Infeld corrections. Right panel:  Evolution of the Hubble function in terms of the energy density for different equation of state parameters. At low energy densities the standard evolution of GR is recovered (depicted by the dotted line), whereas at high energy densities the translated modifications in the matter source from Born-Infeld become dominant. A crucial property of this model is that the Hubble function becomes constant in the Born-Infeld regime for a dust component with $w=0$ giving rise to a de Sitter phase.}
\label{Fig:allowedMinimal}
\end{center}
\end{figure}

%
After briefly going through the simplest case of Einstein and vacuum space solutions, we shall consider the general cosmological solutions. Our homogeneous and isotropic Ansatz make the fundamental matrix take the form $M^\mu{}_{\nu} = {\rm diag}[M_0(t), M_1(t), M_1(t), M_1(t)]$ and the equations \refeq{Eq:MinimalMatrix} read
  \begin{equation}
\begin{aligned}
\frac{1}{M_0}+3M_1&= 4+\bar{\rho}, \\
M_0+2M_1+\frac{1}{M_1}&= 4-\bar{p},
\end{aligned}
\label{eqM0M1}
\end{equation}
with the dimensionless density and pressure

\begin{equation}
\bar{\rho} \equiv \frac{\rho}{\mbi^2\mpl^2}\,,\qquad \bar{p} \equiv \frac{p}{\mbi^2\mpl^2}\,.
\end{equation}
These equations are of course \refeq{metricEOMinOmega} adapted to the present case. We can now obtain from them the quantities  $M_0$ and $M_1$ algebraically
in terms of  $\rhob$ and $\pb$, i.e., we will have $M_0(\rhob,\pb)$ and  $M_1(\rhob,\pb)$. If we solve for $M_1$ from the first equation and plug it in the second one we obtain
\begin{equation}
(4+\bar \rho)M_0^3 +
  \left[\bar P(4+\bar \rho)+\frac{2}{3}(1+\bar\rho)^2-4\right]M_0^2-\left[\bar P+\frac{4}{3}(1+\bar\rho)\right]M_0
  + \frac{2}{3} = 0.
  \label{Eq:M0equation}
\end{equation}
It is possible to solve this equation analytically, but the explicit expression is not very illuminating, so we will omit it here (the interested reader can find it in \cite{Jimenez:2015jqa}) and instead we will plot the solutions in figure \ref{MinModel_M0yM1} for some interesting matter sources. As usual, one finds several branches of solutions, 3 in this case owed to above equation being cubic. Out of those 3, one is always unphysical because either $M_0$ or $M_1$ is negative and we do not consider it. The remaining two branches satisfy the positivity requirement, but only one is continuously connected with GR in the low densities regime (see figure \ref{MinModel_M0yM1}). Even for these physical branches, the positivity of $M_0$ and $M_1$ impose constraints on the allowed values of $\rhob$ and $\pb$ as shown in figure \ref{Fig:allowedMinimal}. From that figure we can see that the Born-Infeld corrections impose the bounds $p\lesssim \mbi^2\mpl^2$ and $\rho\gtrsim-4\mbi^2\mpl^2$. In particular, these constraints make the allowed region for a radiation fluid be compact, i.e., there is a maximum allowed value for its energy density. As can be easily understood from the left panel in figure \ref{Fig:allowedMinimal}, this will be the case for fluids with constant and strictly positive equation of state parameter. However, for dust or a cosmological constant, the energy density can grow arbitrarily large.

The definition of the fundamental matrix $\m{M}$ also allows to obtain the corresponding curvature as
\begin{equation}
\mR(\Gamma)=g^{\mu\nu}\mR_{(\mu\nu)}(\Gamma)=\mbi^2\left(M_0^2+3M_1^2-4 \right).
\end{equation}
Plugging in the physical branches of solutions for $M_0$ and $M_1$ we can then obtain the dependence of the curvature on the density. In the low energy density limit the curvature becomes
\begin{eqnarray}
\mR^{\rm I}&=&\frac{\rho-3p}{\mpl^2}+\mathcal O(\rhob^2)\\
\mR^{\rm II}&=&-\frac{32}{9}\mbi^2-\frac{1}{9\mpl^2}(\rho-3p)+\mathcal O(\rhob^2)
\end{eqnarray}
where the Branch I refers to the solution that connects with GR at low energy densities whereas Branch II stands for the branch that connects with the dS/AdS solutions in vacuum. In figure \ref{MinModel_M0yM1} we show the full solutions. Three different types of fluids are considered: dust with $p=0$, radiation with $p=\rho/3$ and a fluid with the equation of state parameter $w=-0.8$. As we commented above, a radiation fluid naturally shows  an upper bound on its possible energy density, which further translates into an upper bound for the scalar curvature $\sim\mbi^2$. This actually happens for equation of state parameters $0<w<1$. This will have important consequences for early universe cosmology where the relativistic degrees of freedom are supposed to dominate. Concerning dust fluids, the allowed range for $\rho$ is no longer compact and, in fact, $\rhob$ is not bounded from above. However, it is interesting to notice that, even if $\rhob$ can grow arbitrarily large, the scalar curvature saturates beyond $\rhob\simeq 1$ and this makes it be bounded by $\mbi^2$, thus avoiding curvature singularities. Finally, fluids with equation of state close to that of a cosmological constant do not have a compact allowed region and the curvature divergences at high energy densities are even more severe than those in GR in the branch I. While GR gives $R\propto \rho$, in the Minimal model we have $R^{\rm I}\propto \rho^2$. Finally, for the cases with a non-compact allowed region, the branch II shows the same behaviour found for the Einstein space solutions above, i.e., the curvature is insensitive to the value of $\rhob$.

Once we have the fundamental matrix in terms of $\rho$ and $p$ we can easily compute the auxiliary metric by using that $\m{\Omega}=\m{M}/\sqrt{\det\m{M}}$ so that

\begin{equation}
\tilde{N}^2(t)=  N^2(t)\sqrt{M_0 M_1^{-3}}, \qquad \tilde{a}^2(t)=\frac{a^2(t)}{\sqrt{M_0 M_1}}\,.
\label{HtToH}
\end{equation}
Then, we can use the general expression for the Hubble expansion rate \refeq{eq:generalcaseH2inPomega} adapted to the Minimal model to obtain

\begin{equation}
H^2 = \frac{\mbi^2}{3} \frac{M_0^2 + \frac{3}{2}(\bar{p} + \bar\rho) M_0
  -1}{\left \{1 + \frac{3M_0(\rhob+\pb)}{4\left[(4+\bar{\rho})M_0 -1 \right]}
 \left[ 1 + (4+\bar{\rho}) \left(\partial_{\rhob}\ln M_0+c_{\rm s}^2\partial_{\pb}\ln M_0\right) \right] \right\}^2}\,,
\label{eq:Hubble}
\end{equation}
where we have used the equations \refeq{eqM0M1} to express $M_1$ in terms of $M_0$, which can then be solved for from \refeq{Eq:M0equation}. This expression for the Hubble expansion rate shows once again the distinctive property of depending on $\rho$, $p$ and $c_s^2$. Let us notice that the allowed region discussed above arising from the positivity of the fundamental matrix $\m{M}$ constrained the possible values for $\rho$ and $p$. Here we have one additional constraint for the cosmological solutions given by the condition $H^2\geq0$. This condition will in fact be more restrictive since it will depend on $c_s^2$. In other words, if we consider a barotropic fluid with $p=p(\rho)$ (not necessarily linear) so that $c_s^2=\d p/\d\rho$, the constraint $H^2\geq0$ will restrict the parameter space $(\rho,p,\d p/\d\rho )$, while having a positive definite $\m{M}$ only gives constraints on its subspace $(\rho,p)$.

In the right panel of figure \ref{Fig:allowedMinimal} it is shown the Hubble expansion rate as a function of the density for radiation, dust and a cosmological constant. These fluids are important representatives of the following typical behaviour depending on the equation of state parameter $w=p/\rho$:
\begin{itemize}
\item Fluids with the equation of state parameter in the range $0<w<1$ give rise to a maximum value for the energy density $\rho\lesssim \mbi^2M_p^2$. We had already observed this type of behaviour for the standard EiBI theory in the previous section. Thus, it is quite typical to find an upper bound for the allowed energy densities in theories \`a la Born-Infeld.

\item For fluids with  $-2/3<w\leq0$ one does not observe any upper bound for $\rho$. Interestingly, the Hubble function can become constant at high energy densities.

\item Finally, for fluids with $-1\leq w<-2/3$ the Hubble function evolves as $H^2\propto \rho^2$ which is even worse than in GR in terms of singularities at high energy densities. For this type of fluids the realisation of the Born-Infeld mechanism fails.
\end{itemize}

One distinctive and crucial feature of this minimal Born-Infeld extension is the saturation of the Hubble function to a constant value at high energy densities appearing for $-2/3<w\leq0$, which could offer an interesting alternative to realise a de Sitter inflationary epoch in the presence of a dust fluid.
This idea was developed in \cite{Jimenez:2015jqa}, where, in order to achieve an inflationary scenario eventually evolving to a radiation dominated phase, it was considered a cascade of decaying dust fluids at the end of which there is a radiation component. This system is thus described by the following system of equations:
\begin{eqnarray}
&&\dot{\rho}_i+3H\rho_i=\Gamma_{i-1}\rho_{i-1}-\Gamma_i\rho_i\quad\quad i=1,...,n\\
&&\dot{\rho}_r+4H\rho_r=\Gamma_{n}\rho_{n}
\end{eqnarray}
where $\Gamma_0=\rho_0=0$, $\rho_r$ is the energy density of radiation representing the final state and $\Gamma_i$ is the decay rate of the $i$th particle. In order to ensure the stability of the dust components during inflation we need to impose $\Gamma_i<H_{\rm dS}$ with $H_{\rm dS}$ the (nearly constant) Hubble expansion rate during the inflationary phase and that will be $H_{\rm dS}\simeq \mbi$. The idea is then that the quasi de Sitter phase is supported by the dust components as long as $\rho_{\rm dust}\gg \mbi^2\mpl^2$. Since $\rho_{\rm dust}\propto a^{-3}$, the energy density of the dust components will eventually drop below $\mbi^2\mpl^2$ and the Born-Infeld regime will be abandoned. This will determine the end of the inflationary regime and the beginning of the reheating phase. In this phase, the Hubble expansion rate will evolve as $H^2\simeq \rho_{\rm dust}/(3\mpl^2)$ so that the different decay rates will become larger than the expansion rate and, therefore, the dust will start decaying. At the same time, the radiation component will be populated and, after all the dust components have decayed, we will be left with a radiation dominated universe.

\begin{figure}
\includegraphics[width=7cm]{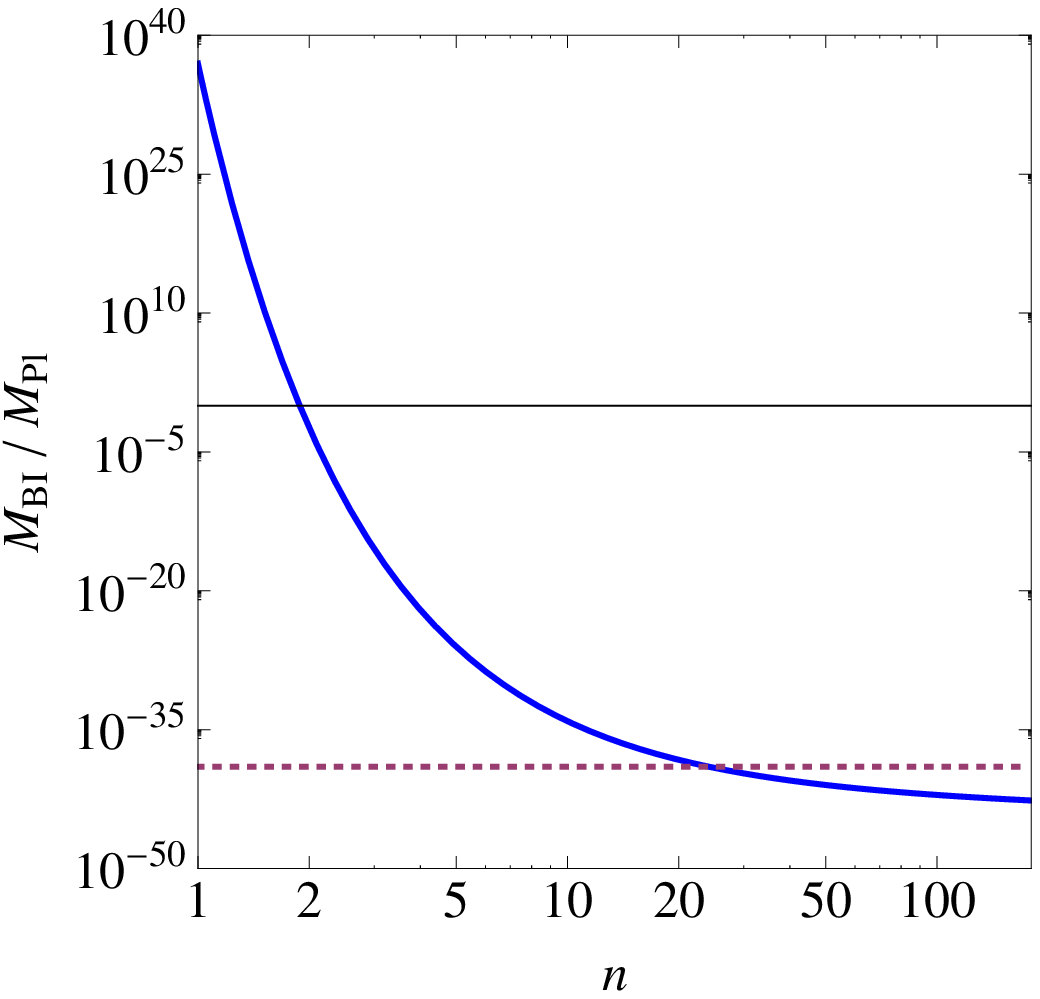}
\includegraphics[width=7cm]{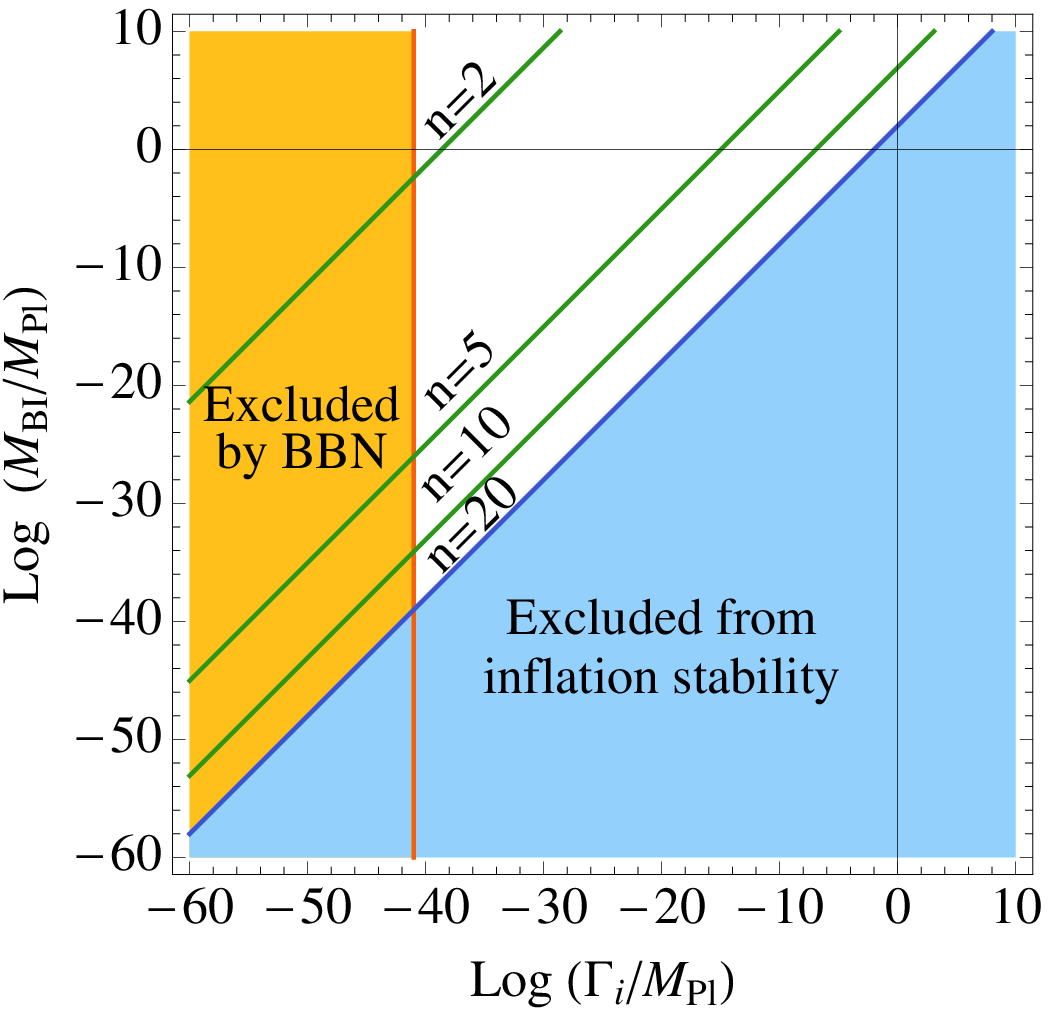}
\caption{Figure adapted from \cite{Jimenez:2015jqa}. The left panel shows the bounds on $\mbi$ to have at least $60$ e-folds of inflation as a function of the number of dust species $n$ and assuming the decay rates so that reheating ends right before BBN. The black line denotes the Planck scale and the dotted line gives the lower bound on $\mbi$ so that the dust components are stable during inflation.  It becomes clear that for $n=1$ the allowed region is above Planck scale and hence no realistic inflationary scenario can be constructed and for more than 20 species BBN constraints and stability during inflation cannot be realized at the same time. In the right panel the bounds on $(\Gamma_i,\mbi)$ are shown together by assuming that all the decay rates are of the same order. The bound on $\Gamma_i$ coming from the condition that reheating should end before BBN is encoded by the orange region. The blue region represents the stable region of dust components during inflation. Furthermore, the green curves indicate the bounds for $\mbi$ in order to have $60$ e-folds of inflation for different number of components.}
\label{fig:bbnbounds_minext}
\end{figure}

This inflationary model has some interesting features that we will summarise here without entering into too many details and refer to \cite{Jimenez:2015jqa} for a more rigorous treatment. The first important property is that there is a maximum value for the allowed energy density in the inflationary regime. This can be traced back to the very presence of the cascade that will lead to a non-trivial and time-dependent sound speed $c_s^2(t)$ and a non-vanishing pressure. As we discussed above, this will give rise to a bounded range of values for the energy density in the physical space. This bound on the energy density will depend on the decay rates and, in turn, it will lead to a maximum number of e-folds for the inflationary phase. Thus, imposing that the inflationary phase lasts for at least $60$ e-folds will give bounds on the parameters of the model, namely $\mbi$ and $\Gamma_i$. Another constraint can be obtained from the fact that the reheating phase should end before the onset of BBN. Since the end of the reheating period is determined by the last decaying dust component, this will give a direct constraint on the smallest decay rate. These constraints are summarised in figure \ref{fig:bbnbounds_minext}. Finally, a remarkable property of this inflationary scenario is that the first slow parameter $\epsilon_1\equiv -\d\log H/\d \log a$ is negative so that we actually have a super-inflationary phase. Nevertheless, one cannot directly infer anything on the perturbations from here since the gravity sector is highly modified with respect to the standard inflationary models. This can be illustrated by looking at the tensor perturbations. Since they propagate on the auxiliary metric we need to compute the effective expansion seen by them. It turns out that they see an effective equation of state $w=1$ and, thus, they are oblivious to the inflationary background and no primordial gravitational waves are generated within this model. Therefore, this inflationary model would come with the distinctive feature of the absence of primordial gravitational waves so that the detection of $B$-modes in the CMB generated by primordial gravitational waves would immediately rule it out.

%%%%%%%%%%%%%%%%%%%%%%%%%%%%%%
\subsection{Functional extensions of Born-Infeld gravity} \label{sec:feoBIg}

\begin{figure}
\centering
\includegraphics[width=0.48\textwidth]{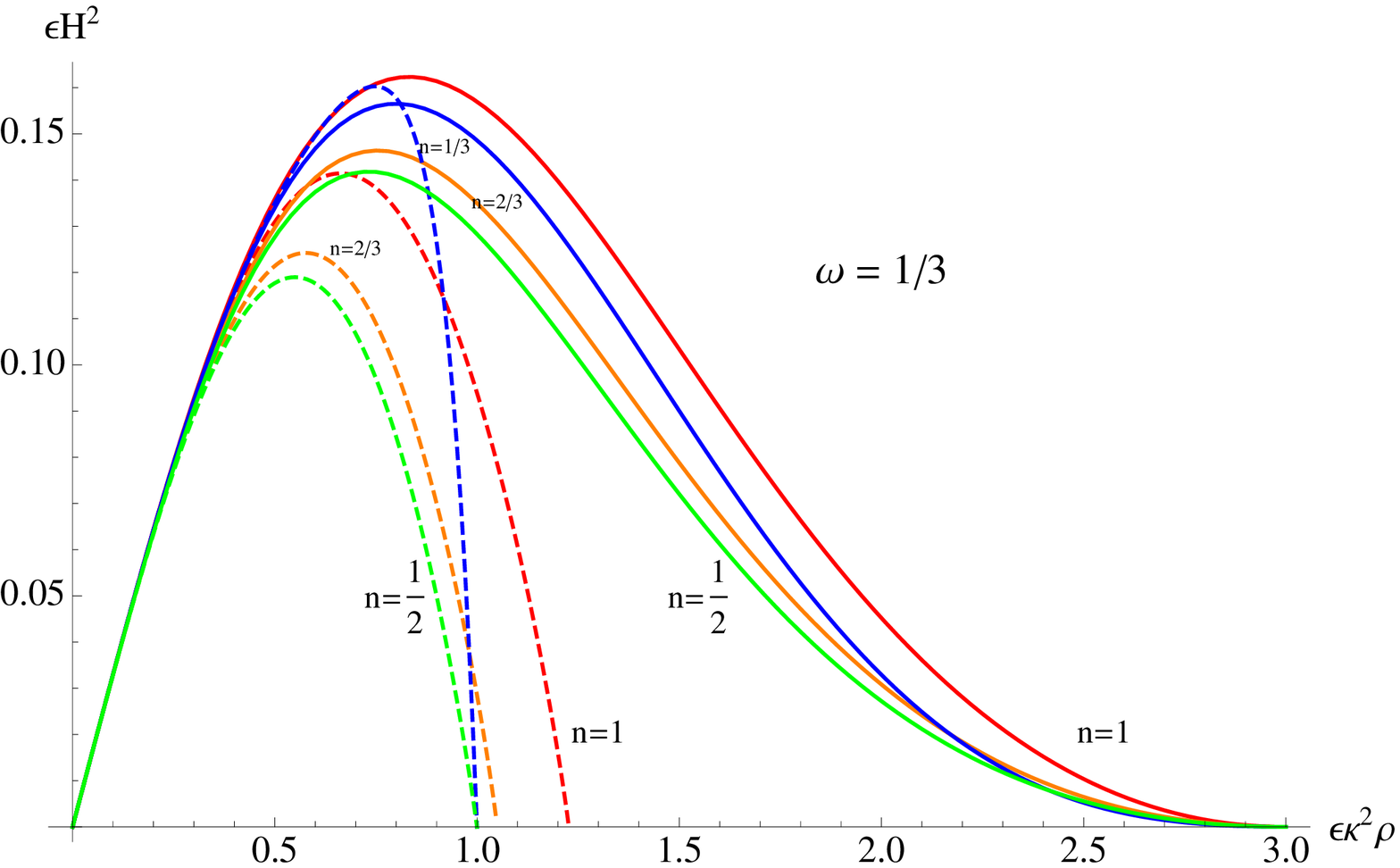}
\includegraphics[width=0.48\textwidth]{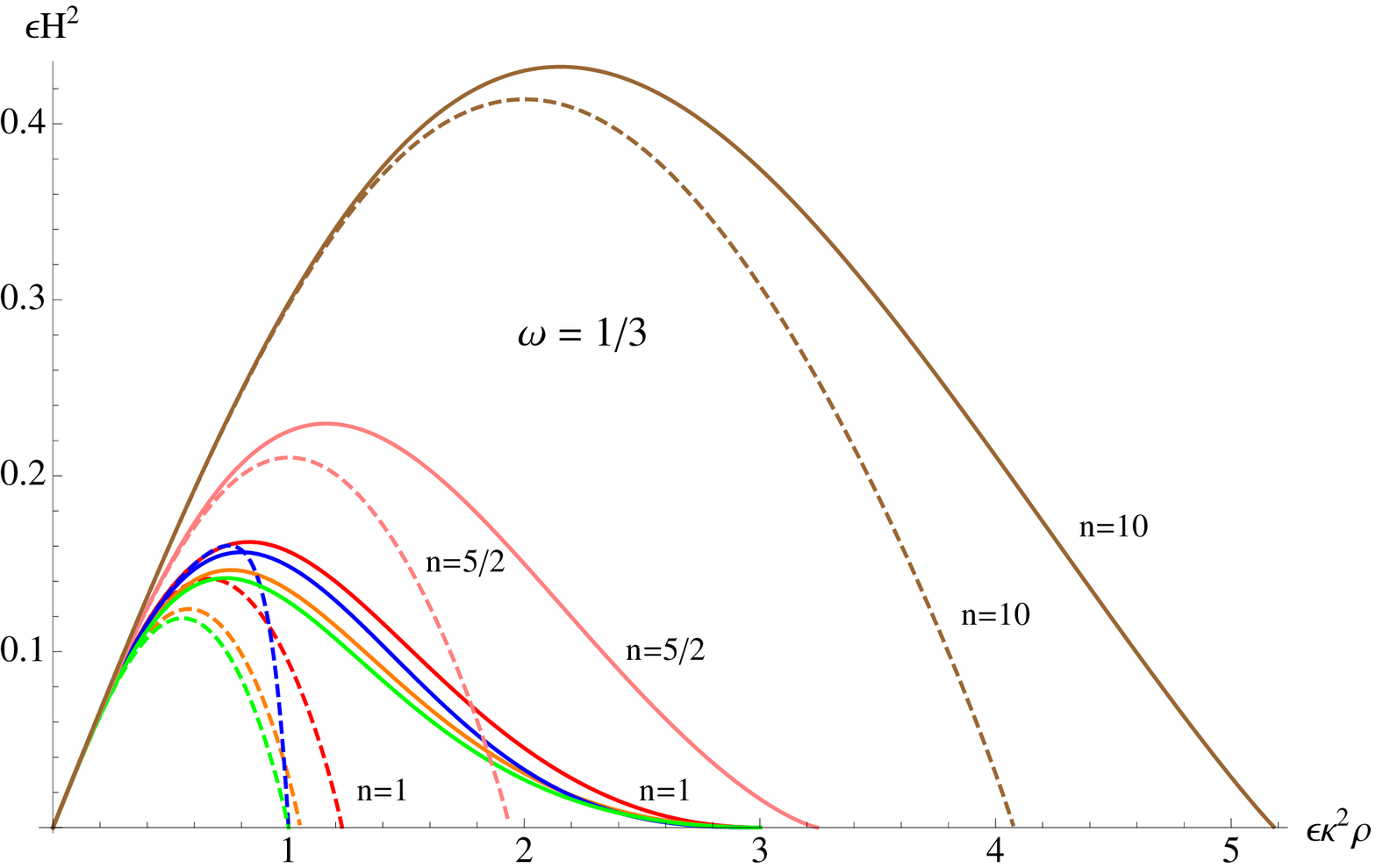}
\caption{These figures are taken from \cite{Odintsov:2014yaa} where the evolution of the Hubble function is plotted in terms of the energy density of the fluid $\rho/(\mpl^2\mbi^2)$ for different values of the index $n$. The notation used there corresponds to $\epsilon \to \mbi^{-2}$ and $\kappa^2\to\mpl^{-2}$ in our notation. In the left panel we see the evolution for a radiation fluid with $w=1/3$. The solutions represented by the dashed lines correspond to $M_{\mbi}^2<0$ (plotted in positive quadrant for graphical convenience) and represent the bouncing solutions whereas the solid lines ($\mbi^2>0$)  represent unstable solutions with $H^2=0$ and $H_{,\rho}=0$ at high energy densities. One can further see that these non-singular solutions have very similar behaviour for small deviations in $n$. In the right panel the evolution is shown for a fluid with equation of state parameter $w=-1/5$. In this case one striking observation is that the solid solutions start resembling bouncing solutions for sufficiently large values for the index $n$.}
\label{fig:functionalExtensionBI_H2rho}
\end{figure}

In the previous section we have studied the cosmological implications of the minimal extension of Born-Infeld inspired gravity theory, which was based on the trace of the square root structure rather than the determinant. We have seen how one can construct interesting quasi de Sitter solutions with a dust component in this model. In this section, we shall draw our attention to another interesting extension of the original EiBI gravity and discuss its potential impact to the early universe cosmology. Instead of a square root one could consider an arbitrary function of the determinant. This modification would still share the same properties as in EiBI gravity in the sense that General Relativity would be recovered in low energy density regimes with the modifications becoming appreciable only at very high energy density regime. Exactly this idea was pursued in detail in \cite{Odintsov:2014yaa} and we shall summarise the main results of this study here. For this purpose, let us adapt to the notation used in \cite{Odintsov:2014yaa} with $\Omega^\alpha{}_\beta=g^{\alpha\mu}q_{\mu\beta}$ where again $q_{\mu\nu}=g_{\mu\nu}+\mbi^{-2}\mR_{(\mu\nu)}(\Gamma)$. In terms of $\Omega$ the Born-Infeld inspired gravity theory can be simply expressed as
\begin{equation}
\mathcal{S}=\mpl^2\mbi^2\int \mathrm{d}^4x\sqrt{-g} \left( \sqrt{|\hat\Omega|}-\lambda\right)+\mathcal{S}_{\rm matter} \,.
\end{equation}
A natural extension arises by promoting the square root to an arbitrary function as it was proposed in \cite{Odintsov:2014yaa}. In this case, the action generalises to
\begin{equation}\label{action_functionalExtension_BI}
\mathcal{S}=\mpl^2\mbi^2\int \mathrm{d}^4x\sqrt{-g} \left( f(|\hat\Omega|)-\lambda\right)+\mathcal{S}_{\rm matter} \,.
\end{equation}
The gravitational Lagrangian density, $\mathcal{L}_G$, in (\ref{action_functionalExtension_BI}), can be conveniently expressed in terms of an auxiliary scalar field $A$ via the general function $f(A)$ with the Lagrange multiplier $(|\hat\Omega|-A)f_A)$ as
\begin{equation}
\mathcal{L}_G=\mpl^2\mbi^2(\Phi |\hat\Omega|-V(\Phi)-\lambda) \,.
\end{equation}
where $\Phi=\mathrm{d}f/\mathrm{d}A$ and $V(\Phi)=Af_A-f(A)$. Written in this language, the connection field equations are simply given by $\nabla_\mu\left( \Phi |\hat\Omega|^{1/2}\sqrt{-q}q^{\mu\nu} \right)=0$. One can use the same trick as in the previous sections to express the Riemann tensor in terms of the quantities of the matter field. By doing so, one obtains \cite{Odintsov:2014yaa}
\begin{equation}
R^\alpha{}_\beta=\frac{\mathcal{L}_G\delta^\alpha{}_\beta+T^\alpha{}_\beta}{2\mpl^2\Phi^2|\hat\Omega|^{3/2}} \,.
\end{equation}
For the cosmological application, we will consider a perfect fluid as a representative of the matter fields. Using exactly the same procedure as above, we can compute the evolution of the Hubble function in terms of the matter field quantities. The resulting Hubble function in this particular extension takes the following form
\begin{equation}
H^2=\frac{\mbi^2}{6(1+\dot{\Delta}/(2H\Delta))^2}\left[ \frac{(\rho+3p)/(\mpl^2\mbi^2)+2(\Phi |\hat\Omega|-V-\lambda)}{\Phi|\hat\Omega|+V+\lambda+\rho/(\mpl^2\mbi^2)} \right]
\end{equation}
where the short-cut notation is introduced $\Delta=2\Phi^2|\hat\Omega|^{3/2}/(\Phi |\hat\Omega|+V+\lambda-p/(\mpl^2\mbi^2))$. In \cite{Odintsov:2014yaa} a family of power law functions $f(|\hat\Omega|)=|\hat\Omega|^n$ was investigated in detail. For values of the parameter close to $n=1/2$, of course the features are very close to the original Born-Infeld gravity. In general, one has again two types of branches of solutions, the branch with $\mbi^{-2}>0$ and the branch with $\mbi^{-2}<0$. It turns out that the first type of solutions are more sensitive to the changes in the index $n$. The second type of solutions representing a bounce are more robust. In figure \ref{fig:functionalExtensionBI_H2rho} extracted from \cite{Odintsov:2014yaa} we can see the evolution of the Hubble function for different values of the index $n$ for a fluid with a positive equation of state parameter in the left panel and with a negative equation of state parameter in the right panel, respectively. The bouncing solutions are depicted by the dashed lines and the solid lines represent the unstable solutions with $H^2=0$ and $H_{,\rho}=0$ for sufficiently high energy densities. The qualitative behaviour of these two branches of solutions remains the same for small deviations in the index parameter. On the other hand, for fluids with negative equation of state parameter the solid line solutions start hitting the horizontal line converting more and more into a rather bouncing solutions for large values of $n$.

%%%%%%%%%%%%%%%%%%%%%%%%%%%%%%
%%%%%%%%%%%%%%%%%%%%%%%%%%%%%%

\subsection{Extensions including a Ricci scalar}

There exist extensions of the original EiBI gravity theory that relies on the presence of an additional Ricci scalar, which we will review in this section. These modifications are constructed either by including a Ricci scalar $\mR(\Gamma)$ directly into the determinantal structure, or by including an additional function as a separate sector into the theory (belonging to the Class II theories in section \ref{subsubsectionClassII}).

\subsubsection{Born-Infeld-$f(\mR)$ gravity}

\begin{figure}[h!]
\begin{center}
\includegraphics[width=9cm,height=9cm]{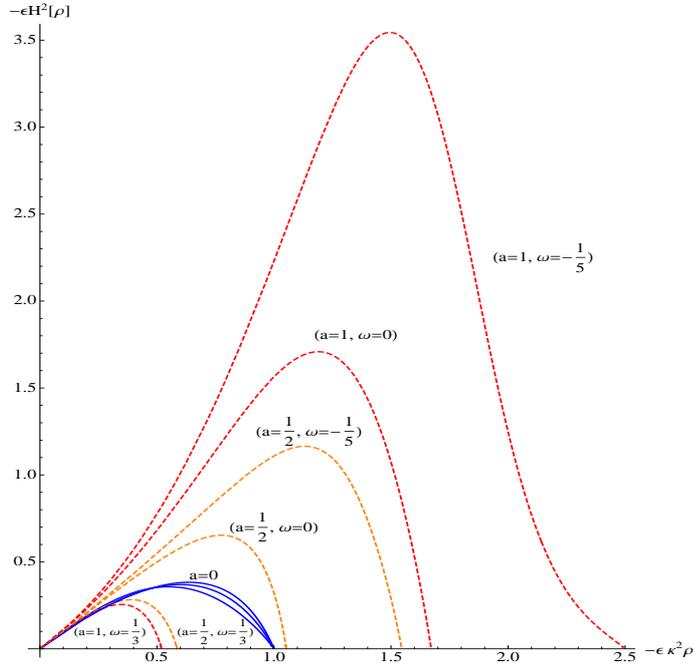}
\caption{This figure from \cite{Makarenko:2014lxa} represents the evolution of the dimensionless Hubble function $- H^2/\mbi^2$ in terms of the dimensionless energy density $-\rho/(\mpl^2\mbi^2)$ for both the original EiBI theory (solid blue) and the modification with the function of the form $f(\mR)=a \mR^2$, with the value $a=1/2$ (dashed orange) and $a=1$ (dashed red), in the presence of a matter fluid with two different equations of state ($w=-1/5,0,$ and $1/3$) respectively. The presence of bouncing solution does not alter with the difference in $a$ and hence is a robust property of the $\mbi^{-2}<0$ branch. The notation in \cite{Makarenko:2014lxa} translated into ours as $1/\kappa \to \mpl^2$ and $1/\epsilon \to \mbi^2$. \label{Fig:BIfR_Negative_Branch}}
\end{center}
\end{figure}

\begin{figure}
\begin{center}
\includegraphics[width=0.90\textwidth]{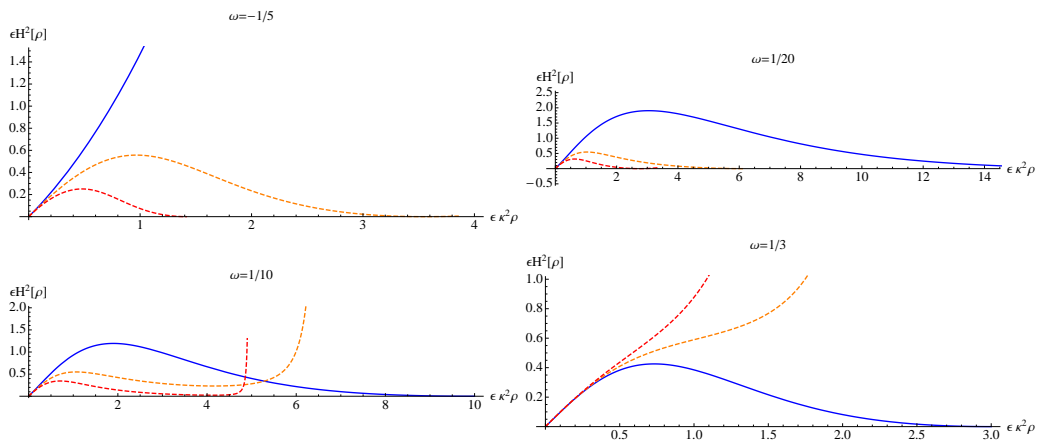}
\caption{In this figure we show the evolution of the dimensionless Hubble function $H^2/\mbi^2$ as a function of  the dimensionless energy density $\rho/(\mpl^2\mbi^2)$ for three different cases: the original EiBI theory $a=0$ (solid blue) and the Born-Infeld-$f(R)$ theory with  $f(R)=a R^2$, with two different values $a=1/2$ (dashed orange) and $a=1$ (dashed red), and different equations of state ($w=-1/5,1/20, 1/10,$ and $1/3$) respectively. One immediate observation is that the zero of  $\epsilon H^2$ in the case $\omega=1/3$ is unstable under the changes of the parameter $a$. Furthermore, when the equation of state saturates to $\omega\to 0$, the Hubble function $ H^2/\mbi^2$ might become again zero for sufficient high densities. However, the corresponding derivative of the function $\dot{H}/\mbi^2$ would vanish, thus representing rather a minimum of $H^2/\mbi^2$. This does not correspond to a bounce but rather signals an instability representing a state of minimum volume.
 \label{Fig:BIfR_Positive_Branch}}
\end{center}
\end{figure}

One of the early predominant extensions of the EiBI gravity theory is the Born-Infeld-$f(\mR)$ gravity, where the original EiBI gravity theory is combined with an additional function, that depends on the Ricci scalar \cite{Makarenko:2014lxa}. In order to avoid any ghost instabilities, the theory is constructed in the Palatini formalism in both sectors. The Riemann tensor and the Ricci scalar both depend only on the connection and not on the metric. This idea of combining Born-Infeld and $f(\mR)$ gravity was further investigated in \cite{Makarenko:2014nca, Makarenko:2014fla, Elizalde:2016vsd}. Due to the presence of the Ricci scalar, the model exhibits more freedom for simultaneous applications to early and late time universe cosmology. In \cite{Makarenko:2014lxa} it was shown, that for $f(\mR)=a\mR^2$, the model does not alter much the physical properties of bouncing solutions found in the original EiBI model, but it does have crucial impact on the loitering solutions. The Lagrangian proposed in \cite{Makarenko:2014lxa} has the following explicit structure:
\begin{eqnarray}
\mathcal{S}_{\rm BI}&=&\mbi^2\mpl^2\int \mathrm{d}^4x \left\{\sqrt{-\det\left(g_{\mu\nu}+\frac{1}{\mbi^2} \mR_{(\mu\nu)}(\Gamma)\right)}- \lambda\sqrt{-g}\right\} \nonumber \\
&+&\frac{\alpha\mbi^2}{2}\int \mathrm{d}^4x\sqrt{-g}f(\mR)+\mathcal{S}_{\rm matter} \,,
\end{eqnarray}
where the first term is the standard EiBI Lagrangian and the second term is the novelty in form on an additional function of $\mR=g^{\mu\nu}\mR_{(\mu\nu)}(\Gamma)$. The matter fields in $\mathcal{S}_{\rm matter}$ couple in a standard manner to the metric.
The variation of this action with respect to the metric yields the modified metric field equations with respect to the standard EiBI model
\begin{equation}
\frac{\sqrt{-q}}{\sqrt{-g}}q^{\mu\nu}-\left[ g^{\mu\nu}\left(\lambda-\frac{\alpha}{2\mbi^2}f\right)+\frac{\alpha f_\mR}{\mbi^2}g^{\mu\beta}g^{\nu\gamma}\mR_{(\beta\gamma)}\right]=0 \,,
\end{equation}
where again we can define the $\hat{q}$ metric as $q_{\mu\nu}=g_{\mu\nu}+\frac{1}{\mbi^2}\mR_{(\mu\nu)}(\Gamma)$ and $f_\mR$ is the derivative with respect to $\mR$. Similarly, the variation with respect to the connection can be written as
\begin{equation}
\nabla_\sigma\left(\sqrt{-q}q^{\mu\nu}+\alpha f_R \sqrt{-g}g^{\mu\nu}\right)=0 \,.
\end{equation}
The connection equation can be written in the for us more useful form $\nabla_\sigma\left( \sqrt{-\tilde{q}}\tilde{q}^{\mu\nu} \right)=0$ where $\tilde{q}$ plays now the role of the auxiliary metric and is defined as $\hat{\tilde{q}}=|\hat{\Sigma}|^{1/2}\hat{\Sigma}^{-1}\hat{g}$ and its inverse as  $\hat{\tilde{q}}^{-1}=|\hat{\Sigma}|^{-1/2}\hat{g}^{-1}\hat{\Sigma}$ with $\Sigma$ representing $\Sigma_\mu{}^{\nu}=|\hat\Omega|^{1/2}(\hat\Omega^{-1})_\mu{}^{\nu}+\alpha f_\mR \delta_\mu{}^{\nu}$ with the standard notation $\hat \Omega=\hat{g}^{-1}\hat{q}$ and $\hat{M}=\sqrt{\hat\Omega}$ in the previous sections.
For the cosmological application of the model, we are interested in homogeneous and isotropic backgrounds. We consider again the metric to be FLRW with $N(t)=1$ and similarly we make an homogeneous and isotropic Ansatz for $\hat{q}$ or $\hat\Sigma$ directly $\hat\Sigma={\rm diag}(\sigma_1, \sigma_2\delta_i{}^j)$. This on the other hand determines the form of the metric $\hat{\tilde{q}}$ to be $\tilde{q}_{00}=-\sqrt{-\sigma_2^3/\sigma_1}$ and $\tilde{q}_{ij}=\sqrt{\sigma_1\sigma_2}a^2\delta_{ij}$. For the matter fields, we again assume a perfect fluid with $\hat{T}=(\rho, p\delta_i^j)$. We can use the same procedure as in standard EiBI gravity model in order to obtain the evolution equation of the Hubble function in terms of the energy density and pressure of the matter fields (see section \ref{sec:general_framework_cosmology} for the general cosmological framework). For that we can use the field equations and the definition of the Einstein tensor and equal them. By doing so, one obtains \cite{Makarenko:2014lxa}

\begin{equation}
H^2=\mbi^2\frac{\sigma_1-3\sigma_2-2|\hat\Omega|^{1/2}(\sigma_1w_1-3\sigma_2w_2)}{2\sigma_1\left( 1-\frac{3(1+w)\rho\Delta_\rho}{2\Delta}\right)^2} \,,
\end{equation}
where $\Delta=\sqrt{\sigma_1\sigma_2}$ and $\Delta_\rho=\partial\Delta/\partial\rho$. In this way, we again have a parametric representation of the Hubble function of the energy density and pressure of the matter fluid. For a particular choice of the function $f(\mR)$ and the equation of state parameter of the matter fluid $w$, one can estimate the evolution of the Hubble function and examine whether different bouncing and loitering solutions exist in this extension of the EiBI model. In \cite{Makarenko:2014lxa} a simple example was studied assuming a quadratic dependence in the form $f(\mR)=a\mR^2$, since this allows to compute $\hat\Omega$ and $H^2$ analytically in terms of the variables of the matter field. For this simple model, the presence of bouncing solutions is assured for $\mbi^{-2}<0$. One obtains $H^2=0$ at $|\rho/(\mpl^2\mbi^2)|=1$ independently of the sign of the equation of state parameter. On the other hand, for $\mbi^{-2}>0$, the Hubble function strongly depends on the sign of $w$ and shows a divergent behaviour for $w\leq0$. The parameter $a$ in the function $f(\mR)$ does not effect significantly the type of bouncing solutions for $\mbi^{-2}<0$ within this model as one can see in figure \ref{Fig:BIfR_Negative_Branch}.
The novelty of this modification coming from $f(\mR)$ becomes apparent in the other branch of solutions when $\mbi^{-2}>0$ and is very sensitive to the sign of the equation of state parameter. For instance, the standard loitering solutions of EiBI gravity theory for $w=1/3$ disappear as one moves away from them due to the presence of $a$ in the function $f(\mR)$. On the contrary, for different equation of state parameters one encounters novel loitering solutions, which were not present in the EiBI theory. These properties are shown in figure \ref{Fig:BIfR_Positive_Branch} taken from \cite{Makarenko:2014lxa}. One additional interesting property is observable for the case $w=1/10$. After reaching a local maximum, $H^2$ evolves towards a non-zero minimum to then diverge at a finite value of large energy densities. The non-zero value of the minimum depends on the parameter $a$. Since $H^2$ does not reach the solution $H^2=0$ in this case, one does not have a bounce. Nevertheless, they could offer an interesting alternative for a quasi de Sitter inflation due to the long plateau between the local minimum and maximum. In this way, one could achieve an inflationary scenario in the presence of radiation. In the minimal extension of the EiBI theory in section \ref{subsection_Minext_BI} we saw that one could realise a quasi de Sitter evolution in the presence of dust with $w=0$. In this modification with $f(\mR)$ this is achievable with radiation.

It is worth mentioning that this same model of Born-Infeld-$f(\mR)$ theories was also used in \cite{Elizalde:2016vsd} in order to construct singular inflationary cosmologies. For this purpose, they borrow ideas from singular $f(\mR)$ inflation \cite{Odintsov:2015zza}. A requirement is that the scale factor evolves in the following form

\begin{equation}
a(t)=e^{-(c_0(-t+t_s)^{1+c_1})/(1+c_1)}\,,
\end{equation}
with the constant variables $c_0$, $c_1$ and $t_s$. The Hubble parameter in this case is $H=c_0(-t+t_s)^{c_1}$. With this Ansatz of the scale factor, one can establish the required relation between the dynamical and auxiliary metric. For large values of the constant $c_1>1$, singular inflation with a graceful exist can be realised \cite{Elizalde:2016vsd}.

\subsubsection{Ricci scalar in the determinant}

\begin{figure}
\begin{center}
\includegraphics[width=0.50\textwidth]{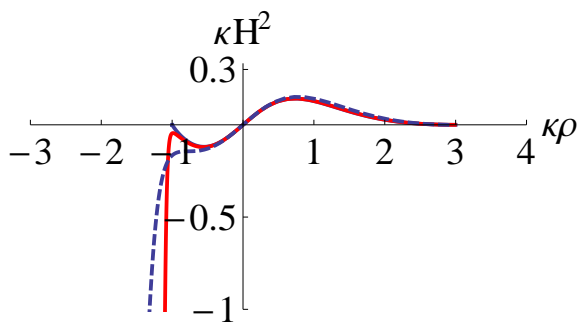}
\caption{This figure is taken from \cite{Chen:2015eha} and shows the dependence of the Hubble expansion function in terms of the energy density for a radiation fluid with $w=1/3$ and for $\beta>0$ but very small values close to zero. The evolution for $\beta=0$ (solid blue), $\beta=10^{-3}$ (solid red) and for $\beta=10^{-2}$ (dashed blue) are plotted respectively. In the notation used in \cite{Chen:2015eha} $\kappa=\mbi^{-2}$. \label{Fig:H2_posBeta_detRRuv}}
\end{center}
\end{figure}

\begin{figure}
\begin{center}
\includegraphics[width=0.45\textwidth]{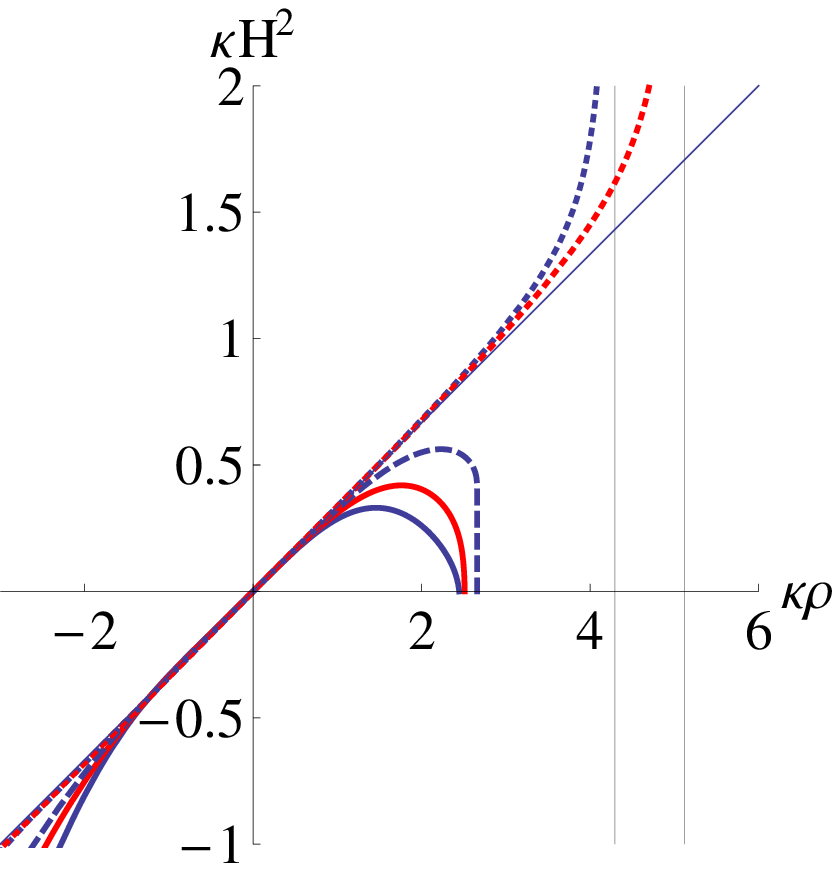}
\includegraphics[width=0.45\textwidth]{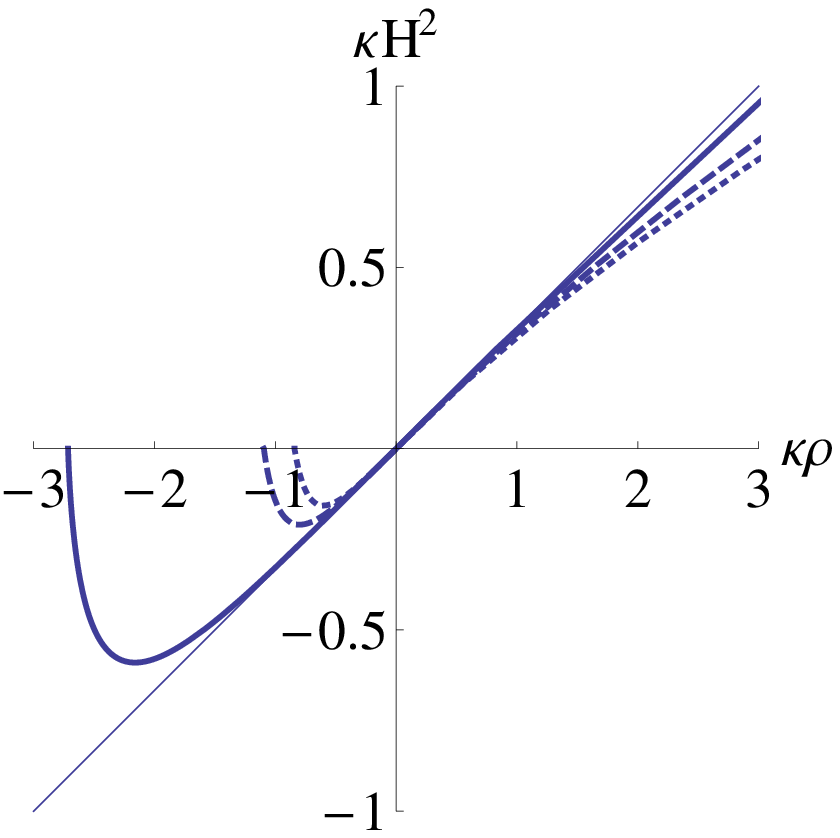}
\caption{This figure is taken from \cite{Chen:2015eha} for the evolution of  the Hubble function in the case $0<\beta\leq1/4$ (left panel) and
$\beta>1/4$ (right panel) with $\kappa=\mbi^{-2}$. \label{Fig:H2_posmedlargeBeta_detRRuv}}
\end{center}
\end{figure}

Other modifications based on the Ricci scalar have been constructed in the literature, where the Ricci scalar enters directly the determinantal structure of Born-Infeld \cite{Chen:2015eha}. The inclusion of this pure trace term in the determinant might offer interesting and promising cosmological implications. The proposed model has the following action \cite{Chen:2015eha}

\begin{equation}\label{action_BIwithRindet}
\mathcal{S}_{\rm BI}=\mbi^2\mpl^2\int \mathrm{d}^4x \left\{\sqrt{-\det\left(g_{\mu\nu}+\frac{1}{\mbi^2} (\alpha \mR_{(\mu\nu)}(\Gamma)+\beta g_{\mu\nu}\mR(\Gamma))\right)}- \lambda\sqrt{-g}\right\}+\mathcal{S}_{\rm matter} \,.
\end{equation}
In order to recover General Relativity in the low energy density limit, the parameters of the theory have to satisfy $\alpha+4\beta=1$. Furthermore, in the corresponding limits, one recovers Palatini $\mR^2$ theories or the original EiBI theory. In the following we will follow the notation of  \cite{Chen:2015eha}, where $\mpl=1$. The variation of the action yields the modified field equations

\begin{equation}\label{EOMg_BIwithRindet}
\frac{\sqrt{-q}}{\sqrt{-g}}\left[\left(1+\frac{\beta \mR}{\mbi^2}\right)q^{\mu\nu} -\frac{\beta}{\mbi^2}q^{\alpha\beta}g_{\alpha\beta}g^{\mu\rho}g^{\nu\sigma}\mR_{(\rho\sigma)} \right]-\lambda g^{\mu\nu}=-\frac{T^{\mu\nu}}{\mbi^2} \,,
\end{equation}
where $q_{\mu\nu}=g_{\mu\nu}+\frac{1}{\mbi^2} (\alpha \mR_{(\mu\nu)}(\Gamma)+\beta g_{\mu\nu}\mR(\Gamma))$ in this particular modification of the EiBI theory. The variation with respect to the connection, on the other hand, results in

\begin{equation}\label{EOMGamma_BIwithRindet}
\nabla_\nu\left[ \sqrt{-q}(\alpha q^{\mu\nu}+\beta q^{\alpha\beta}g_{\alpha\beta}g^{\mu\nu}) \right]=0\,.
\end{equation}
As in the previous sections, we can manipulate the equations on top of a homogeneous and isotropic background such that the Hubble expansion rate can be expressed in terms of the energy density and pressure of the matter fluid. For the homogeneous and isotropic evolution we can make a diagonal Ansatz for $\hat \Omega=\hat{g}^{-1}q$ as $\Omega_0^0=\Omega_1$ and $\Omega_i^j=\Omega_2\delta_i^j$. In terms of a dimensionless parameter $x$ these components can be also written as $\Omega_1=x^3|\hat\Omega|^{1/4}$ and $\Omega_2=|\hat\Omega|^{1/4}/x$. After the adequate manipulations, the dependence of the Hubble expansion rate in terms of the energy density can be expressed as follows \cite{Chen:2015eha}
\begin{equation}\label{H2_BIwithRindet}
H^2=2\mbi^2\left( \frac{\alpha+|\hat\Omega|^{\frac14}(4\beta z-x^3)+3\frac{\sigma_2}{\sigma_1}(|\hat\Omega|^{\frac14}(x^3-4\beta z x^4)-\alpha x^4)}{3\alpha\left(2-\frac{3}{\tilde{q}_2}\frac{\d \tilde{q}_2 }{\d\rho}\rho(1+w)\right)^2}\right)\,,
\end{equation}
with the short-cut notations $\sigma_1=\alpha+\beta(1+3x^4)$, $\sigma_2=\alpha+\beta(x^{-4}+3)$ and $\tilde{q}_2=\sqrt{\sigma_1\sigma_2}\Omega_2$. Furthermore, the variable $z$ satisfies $x^3+3x^{-1}=4z$. After having brought the expression of the Hubble function $H^2$ in the desired form, we can study its evolution for different equation of state parameter of the matter fluid as we did in the previous sections. This will enable us to directly compare the type of bouncing, loitering and quasi de-Sitter solutions within this class of modifications with respect to the standard EiBI gravity theory. In \cite{Chen:2015eha} this analysis was performed for radiation with $w=1/3$ for different values of $\beta$. It was observed, that on the contrary to the previous modification in form of an additional $f(\mR)$, the inclusion of the Ricci scalar into the determinant alters the robustness of the bouncing solutions for $\mbi^{-2}<0$. These solutions seem to be very sensitive to the presence of the parameter $\beta$ for even very small values. This behaviour can be seen in figure \ref{Fig:H2_posBeta_detRRuv}. The Hubble function scales as $H^2\sim \rho$ at large energy densities in this case. Another difference in the model rises for the loitering solutions of the standard EiBI model when $\mbi^{-2}>0$. In this modification the loitering solutions become a bounce with $H^2\sim \rho -\rho_{\rm max}$, where $\rho_{\rm max}$ represents the maximum energy density at the bounce.
The evolution of the Hubble function in the case $0<\beta\leq1/4$ together with $\beta>1/4$ are shown in figure \ref{Fig:H2_posmedlargeBeta_detRRuv}. For the increasing value of $\beta$ getting closer to $1/4$, the cosmological singular solutions resemble more those obtained in $R+R^2$ theories with $H^2\sim \rho/3$ for $\mbi^{-2}<0$. For the opposite case with $\mbi^{-2}>0$, for instance for $\beta=1/10$ and $\beta=3/25$, the loitering solutions of the standard EiBI theory become again a bounce in the past with the Hubble function saturating to $H^2\sim \rho -\rho_{\rm max}$, which has a quasi-sudden singularity in the past. For other values of $\beta$, for example $\beta=1/5$ and $\beta=21/100$, the asymptotic behaviour of the Hubble expansion rate becomes on the other hand $H^2\sim (\rho -\rho_{\rm max})^{-2}$ corresponding to a big freeze singularity in the past.

%%%%%%%%%%%%%%%%%%%%%%%%%%%%%%
\subsection{Other extensions}
\label{Cos:otherextensions}

\subsubsection{Gravity coupled to Born-Infeld}
\begin{figure}
\begin{center}
\includegraphics[width=0.48\textwidth]{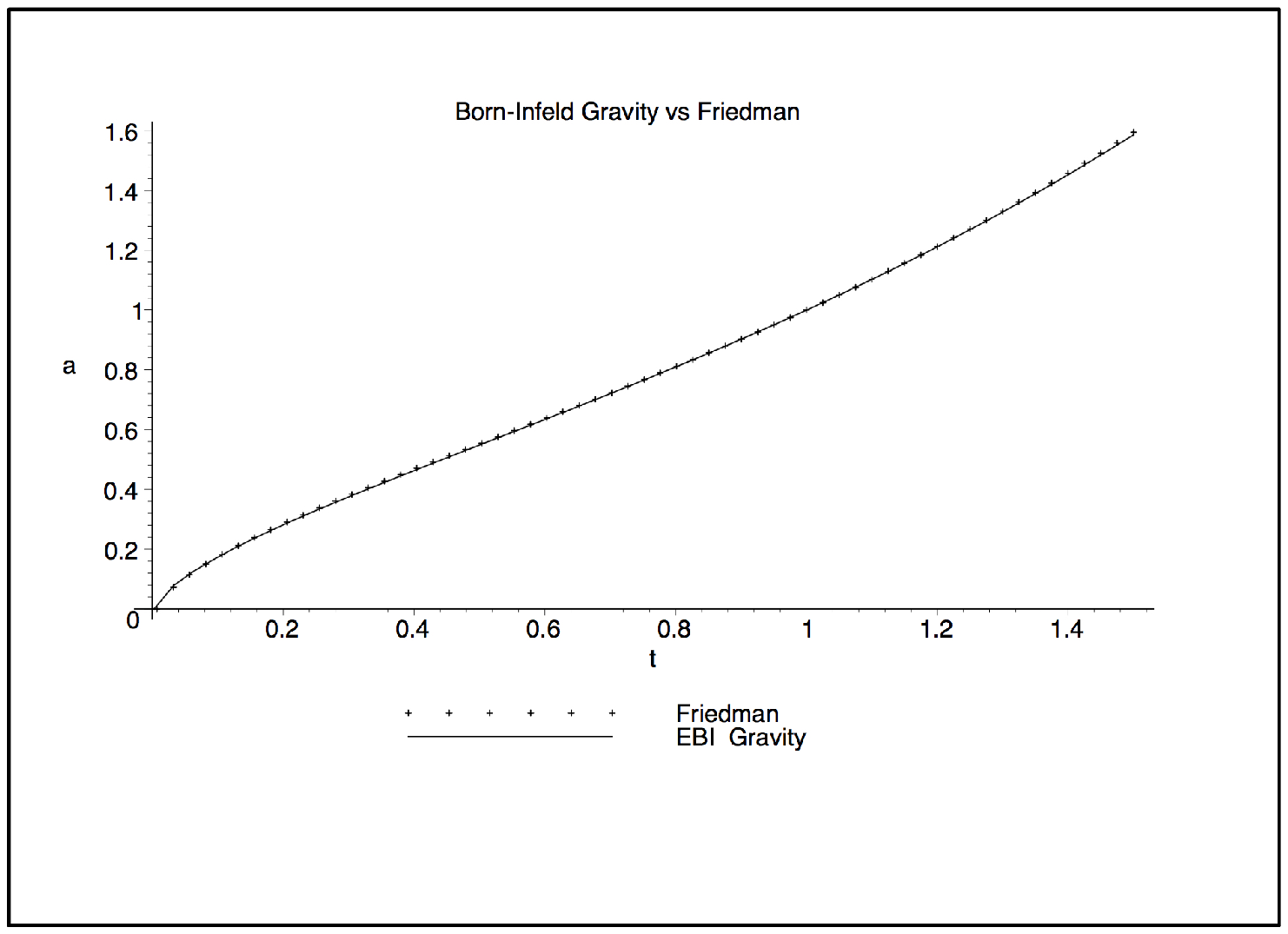}
\includegraphics[width=0.48\textwidth]{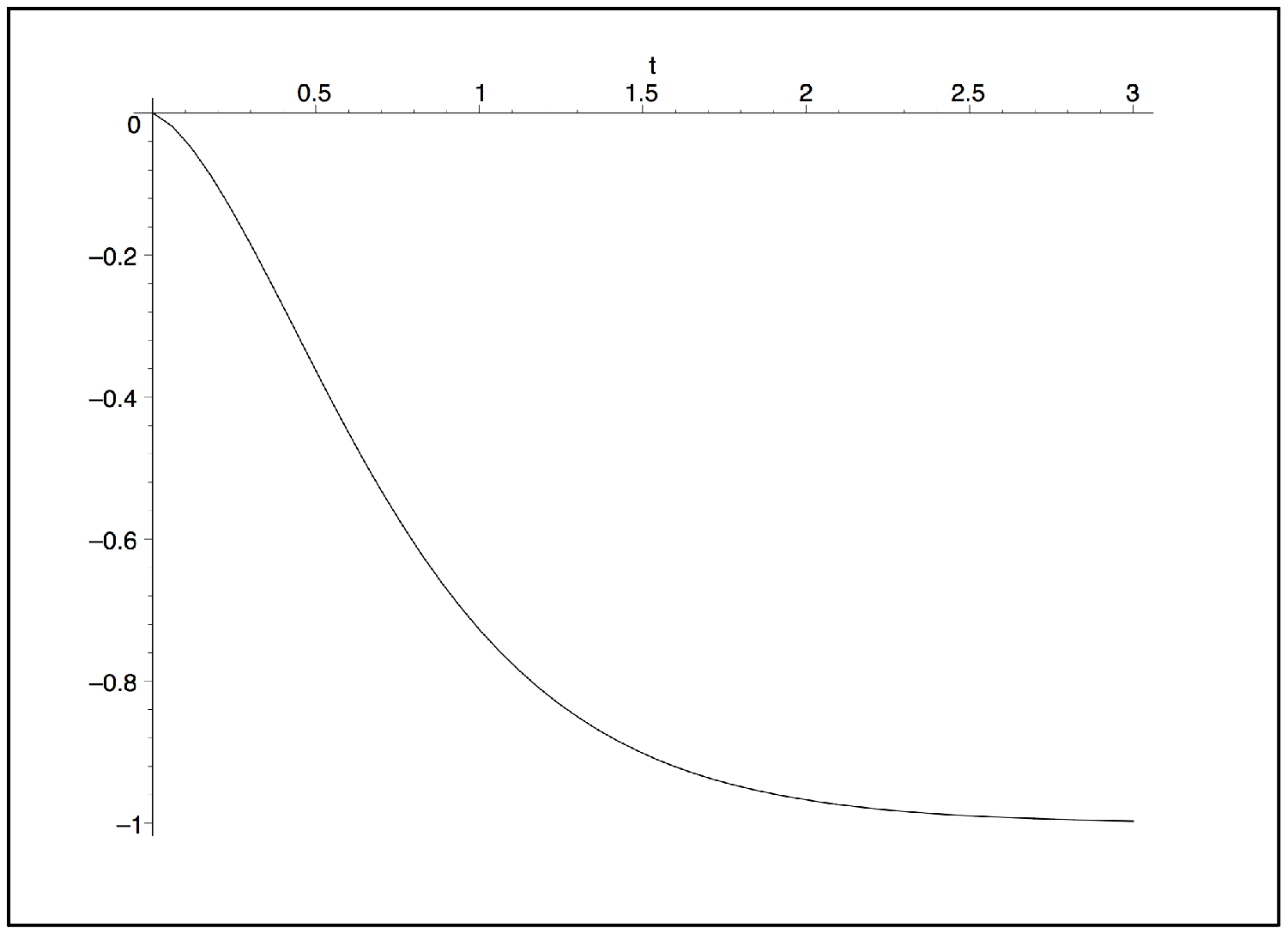}
\end{center}
\caption{\label{fig_gravCouplBI} This figure is borrowed from \cite{Banados:2008rm}, where one can see the evolution of the scale factor in the Born-Infeld gravity model in the presence of the standard Einstein-Hilbert action versus the standard Friedman universe in the left panel. Both evolutions are almost indistinguishable if one chooses $\alpha=0.99$. In the right panel on the other hand the evolution of the effective equation of state parameter $w_{\rm BI}$ is illustrated. }
\end{figure}

Born-Infeld inspired gravity theories were mainly applied to early universe cosmology, since the effects of the modifications become appreciable at high energies. We have seen that interesting alternatives to the standard inflationary paradigm can be constructed within this framework and promising roads to avoid cosmological singularities can be successfully realized. Since the modifications \`a la Born-Infeld are dominant at early times, for a possible application to dark energy and dark matter a change of the framework is needed. This was pursued by Ba\~nados in the work \cite{Banados:2008rm}, where the standard Einstein-Hilbert Lagrangian of GR is coupled to a ``Born-Infeld" field in the hope to reproduce interesting phenomenology for late-time universe. Even if it was proposed as a modification of the original EiBI gravity theory, we would like to emphasise once again that these models do not comply the original Born-Infeld spirit of not modifying the field content. The action considered in \cite{Banados:2008rm} can be expressed as
\begin{equation}
\mathcal{S}=\mpl^2\int \mathrm{d}^4x \left\{\sqrt{-g}R+\frac{2\mbi^2}{\alpha}\sqrt{-\det\left(g_{\mu\nu}-\frac{1}{\mbi^2} \mathcal{R}_{(\mu\nu)}(\Gamma)\right)} \right\}+\mathcal{L}_{\rm matter}\,,
\end{equation}
where $\alpha$ is a dimensionless parameter and $R$ is the Ricci scalar associated to the metric $g$ and $\mathcal{R}_{(\mu\nu)}$ is the Ricci curvature of the independent connection $\Gamma$. This model constitutes General Relativity with the Einstein-Hilbert term coupled to the Born-Infeld connection $\Gamma$. In fact the model can be analogously written as a bimetric theory, where the potential interactions of the two metrics $\hat{g}$ and $\hat{q}$ do not satisfy the potential structure of massive gravity. Therefore, the theory probably might contain dangerous ghostly degrees of freedom. Independently of these ghost issues, the model was studied in \cite{Banados:2008rm}, where it was found that the model admits de Sitter solutions at late times. In fact, it is argued that the parameters can be chosen such that the Born-Infeld field contributes $\sim 73\%$ of the total energy density in form of vacuum energy and $23\%$ in form of dark matter with the equation of state parameter varying between $w=-1$ and $w=0$ respectively. The constructed cosmological solution is such that the scale factor evolves as $a\sim e^{Ht}$ at late times and $a\sim t^{2/3}$ at early times. The field equations of the model are given by
\begin{eqnarray}\label{EoM_grav_coupl_BI}
G_{\mu\nu}&=&-\mbi^2\sqrt{q/g}g_{\mu\rho} q^{\rho\beta} g_{\beta\nu}+\frac{T_{\mu\nu}}{2\mpl^2}\,, \nonumber\\
\mathcal{R}_{\mu\nu}&=& \mbi^2(g_{\mu\nu}+\alpha q_{\mu\nu}) \,,
\end{eqnarray}
where $q_{\mu\nu}$ is the metric associated to the connection $\Gamma$. As it can be seen from the field equations, the structure of the interactions between $\hat{q}$ and $\hat{g}$ in $-\mbi^2\sqrt{q/g}g_{\mu\rho} q^{\rho\beta} g_{\beta\nu}$ does not correspond to the ghost-free massive gravity interactions, signalling the presence of ghostly degrees of freedom. For Einstein space solutions $\mR_{\mu\nu}=\Lambda g_{\mu\nu}$, the two metrics have to be proportional to each other $q_{\mu\nu}=C^2 g_{\mu\nu}$, where the constant $C$ is determined by the field equations to be $C^2=1/(1-\alpha)$.
The modification of the Einstein equations is encoded in the term $-\mbi^2\sqrt{q/g}g_{\mu\rho} q^{\rho\beta} g_{\beta\nu}$ in equation (\ref{EoM_grav_coupl_BI}) and acts as a cosmological constant for the Einstein space Ansatz, where its corresponding value can be expressed as
$\Lambda=C^2\mbi^2=\mbi^2/(1-\alpha)$. As mentioned above, even if these interactions allow for a constant contribution in form of a cosmological constant, they correspond to ghostly interactions, which will render the cosmological solutions unviable.
For general cosmological solutions beyond Einstein space solutions, the following homogeneous and isotropic Ansatz for the two metrics were considered in \cite{Banados:2008rm}:
\begin{eqnarray}
\d s_g^2&=&\left(-N(t)^2\d t^2+a(t)^2\d\vec{x}^2\right) \,, \nonumber\\
\d s_q^2&=&\left(-\tilde{N}(t)^2\d t^2+\tilde{a}(t)^2\d\vec{x}^2\right) \,.
\end{eqnarray}
The background field equations (\ref{EoM_grav_coupl_BI}) for these metrics become
\begin{eqnarray}
H^2&=&\frac{\mbi^2}{3H_0^2}\frac{\tilde{a}^3}{\tilde{N}a^3}+\frac{\rho}{\rho_c} \,, \nonumber\\
\tilde{a}^3&=&3\tilde{N}^2\tilde{a} a^2 H \,,  \nonumber\\
H^2_q&=&\frac{\tilde{N}^2\mbi^2}{3H_0^2}\left(-\frac{1}{2\tilde{N}^2}+\alpha \frac{3}2\frac{a^2}{\tilde{a}^2} \right)\,,
\end{eqnarray}
where $N=1$ and $\rho_c=3H_0^2/{2\mpl^2}$ with $H=\dot{a}/a$, $H_q=\dot{\tilde{a}}/\tilde{a}$ and $H_0$ denoting the Hubble parameter today. The Born-Infeld field contributes to the field equations in form of a fluid with the following effective energy density and pressure:
\begin{equation}
\rho_{\rm BI}=\frac{\mpl^2\mbi^2}{2}\frac{\tilde{a}^3}{\tilde{N}a^3}    \qquad \text{and} \qquad  p_{\rm BI}=-\frac{\mpl^2\mbi^2}{2}\frac{\tilde{N}\tilde{a}}{a}\,.
\end{equation}
We can now study the behaviour of the equations at late and early times. For large values of the scale factor we can neglect the contribution of the ordinary matter fields. In this case, the scale factors evolve as
\begin{equation}
a=a_0 e^{t/c_1}    \qquad \text{and} \qquad \tilde{N}=\frac{1}{\sqrt{1-\alpha}}, \quad \tilde{a}=\frac{a_0}{\sqrt{1-\alpha}}e^{t/c_1} \,,
\end{equation}
with the constant variable $c_1=\sqrt{3(1-\alpha)}\mbi H_0$. This corresponds to the de Sitter solution with $\Omega_\Lambda=1/c_1^2$.
For small values of the scale factor, on the other hand, the solutions can be approximated as
\begin{equation}
a=a_0 t^{2/3}(1+\mathcal{O}(t^{4/3}))    \qquad \text{and} \qquad \tilde{N}=\tilde{N}_0^3(1+\mathcal{O}(t)) , \quad \tilde{a}=\tilde{N}_0(1+\mathcal{O}(t))\,.
\end{equation}
Hence, the scale factor evolves between $a\sim t^{2/3}$ at early times and $a\sim e^{Ht}$ at late times. In \cite{Banados:2008rm} Ba\~nados provides also the numerical solutions to confirm the approximate solutions of these two regimes. In figure \ref{fig_gravCouplBI}, we see the numerical solution for the scale factor. In order to achieve the standard evolution as in $\Lambda$CDM model, the parameter $\alpha$ should be very close to 1, whereas the exact value $\alpha=1$ is singular.

%%%%%%%%%%%%%%%%%%%%%%%%%%%%%%
%%%%%%%%%%%%%%%%%%%%%%%%%%%%%%

As next, we can compute the effective equation of state parameter of the Born-Infeld field. In terms of its energy density and pressure, it can be simply expressed as

\begin{equation}
w_{\rm BI}=\frac{p_{\rm BI}}{\rho_{\rm BI}}=-\left(\frac{a\tilde{N}}{\tilde{a}} \right)^2\,.
\end{equation}
As it can be seen in the right panel of figure \ref{fig_gravCouplBI}, at early times the pressure is $p_{\rm BI}=0$ behaving as matter and at late times $p_{\rm BI}=-\rho_{\rm BI}$ behaving as dark energy.

As mentioned above, even if this model provides an interesting phenomenology for dark energy and dark matter, the ghostly interactions between the $\hat{g}$ and $\hat{q}$ metrics cast serious doubts on the physical viability of these cosmological solutions.
%%%%%%%%%%%%%%%%%%%%%%%%%%%%%%
%%%%%%%%%%%%%%%%%%%%%%%%%%%%%%
\subsubsection{Teleparallel inspired Born-Infeld}\label{Teleparalel_Weizenbock_BI}

\begin{figure}
\begin{center}
\includegraphics[height=2.9in,width=3.1in]{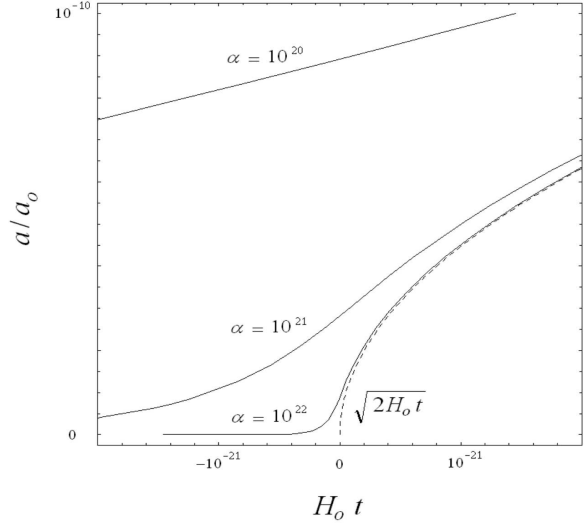}
\end{center}
\caption{\label{fig_scaleFactor_telBI} In this figure we illustrate the evolution of the scale factor of the modified teleparallel model in the presence of a radiation fluid $w=1/3$ for different choices of $\alpha=H_{\rm max}/H_0$, which we extracted from \cite{Ferraro:2006jd}.}
\end{figure}

A Born-Infeld approach to the Teleparallel equivalent of General Relativity was also pursued in the literature in the hope that Born-Infeld teleparallelism might cure the cosmological singularities. For this purpose, Fiorini and Ferraro have considered an extension of a teleparallel model \`a la ``Born-Infeld" in \cite{Ferraro:2006jd, Fiorini:2009ux} with the following action:

\begin{equation}\label{teleparalelBI_actionS}
\mathcal{S}=\mpl^2\mbi^2 \int \mathrm{d}^4x e \left\{\sqrt{1+\frac{2(S\cdot T-2\Lambda)}{\mbi^2}} -1\right\}\,,
\end{equation}
where $e^a_\mu$ represents the four one-forms, $T^{\mu\nu}{}_\rho$ the torsion and $S$ the super-potential

\begin{equation}\label{def_super_potential_telel}
S_\rho{}^{\mu\nu}=-\frac14(T^{\mu\nu}{}_\rho-T^{\nu\mu}{}_\rho-T_\rho{}^{\mu\nu})+\frac12\delta^\mu_\rho T^{\sigma\nu}{}_{\sigma}-\frac12\delta^\nu_\rho T^{\sigma\mu}{}_{\sigma}\,.
\end{equation}
As before, this model can be barely categorised as a Born-Infeld inspired gravity theory according to our criterium, but rather it should be better considered as belonging to the class of $f(T)$ theories (Class-IV).  For a cosmological background $e^a_\mu={\rm diag}(1, a, a, a)$, the field equations read

\begin{eqnarray}\label{teleparalelBI_EOMs}
\frac{1-\frac{4\Lambda}{\mbi^2}}{\sqrt{1-\frac{4\Lambda}{\mbi^2}-\frac{12H^2}{\mbi^2} }}-1&=&\frac{\rho}{\mpl^2\mbi^2} \,, \\
\frac{(1-\frac{4\Lambda}{\mbi^2})\left( \frac{16H^2}{\mbi^2}+\frac{8H^2d}{\mbi^2}-1+\frac{4\Lambda}{\mbi^2}\right)}{\left(1-\frac{4\Lambda}{\mbi^2} -\frac{16H^2}{\mbi^2}\right)^{3/2}}+1&=&\frac{p}{\mpl^2\mbi^2} \,,
\end{eqnarray}
where $d$ is here the deceleration parameter $d=-\frac{a\ddot{a}}{\dot{a}^2}$. A key feature of these modified field equations is that the scale factor approaches

\begin{equation}
a\sim \exp\left(\sqrt{ \frac{\mbi^2(1-\frac{4\Lambda}{\mbi^2})}{12}}t\right)
\end{equation}
at early times $a\to0$. This, on the other hand, means that the Hubble parameter reaches a maximum value $H_{\rm max}=\sqrt{ \frac{\mbi^2-4\Lambda)}{12}}$, regularizing the divergences of standard General Relativity. Let us consider the above action (\ref{teleparalelBI_actionS}) without the cosmological constant $\Lambda=0$. We can combine the two field equations (\ref{teleparalelBI_EOMs}) into a single equation \cite{Ferraro:2006jd}

\begin{equation}
1+d=\frac32 \frac{1+w}{\left(1+\frac{\rho}{\mbi^2\mpl^2}\right)\left(1+\frac{\rho}{2\mbi^2\mpl^2}\right)}\,.
\end{equation}
A negative deceleration parameter can be achieved with a sufficiently large energy density without the need of a fluid with negative pressure as in GR. An accelerated expansion naturally arise if the energy density satisfies the condition

\begin{equation}
\frac{2\rho}{\mpl^2\mbi^2}>-3+\sqrt{13+12w}\,.
\end{equation}
In the case, where the energy density saturates to $\rho\to\infty$, one has $d\to -1$ corresponding to an exponential expansion. For a fluid with $a^{3(1+w)}\rho={\rm const}=a_0^{3(1+w)}\rho_0$, the first field equation can be rewritten as

\begin{equation}\label{reexprFieldeq1_telepar}
\left( \frac{\dot{a}}{a_0}\right)^2+ \frac{\mbi^2a^2}{12a_0^2}\left(\sqrt{1+\beta_0\sum_i\Omega_{0}^i\left( \frac{a}{a_0}\right)^{-3(1+w_i)}}-1 \right)=0\,,
\end{equation}
with the subscript ``$0$" denoting the values today and $\beta_0=(1-12H_0^2/\mbi^2)^{-1/2}-1$. The second term in equation (\ref{reexprFieldeq1_telepar}) is always negative and approaches zero if $w>-2/3$ for $a\to \infty$. In the contrary case, if $w<-2/3$
then this term decreases. In the opposite limit, with $a\to0$ for $w>-1$, one has $a\sim e^{\sqrt{\mbi^2/12}t}$ and hence $H_{\rm max}=\sqrt{ \frac{\mbi^2}{12}}$. The evolution of the scale factor is depicted in figure \ref{fig_scaleFactor_telBI}. Even if one can achieve interesting phenomenology for the early universe cosmology, this model barely represents a modification \`a la Born-Infeld but should be rather considered as a $f(T)$ model. A model more close to the spirit of Born-Infeld will be discussed in the following.
%%%%%%%%%%%%%%%%%%%%%%%%%%%%%%
%%%%%%%%%%%%%%%%%%%%%%%%%%%%%%

%%%%%%%%%%%%%%%%%%%%%%%%%%%%%%
\subsubsection{Born-Infeld in Weitzenb\"ock space-time}

Still within the same framework of the previous section, there has been also the attempt in the literature to consider more general setups in the Weitzenb\"ock space-time and study the consequences for early universe cosmology \cite{Fiorini:2013kba}. The main ingredients are again the super-potential $S$ defined in equation (\ref{def_super_potential_telel}) and torsion $T^{\mu\nu}{}_\rho$. Let us consider the following (Class-III) action

\begin{equation}\label{teleparalelBI_STgeneral}
\mathcal{S}=\mpl^2\mbi^2 \int \mathrm{d}^4x  \left\{\sqrt{\det{\eta_{ab}e^a_\mu e^b_\nu+2\mbi^{-2}\mathcal{F}_{\mu\nu}}} -\lambda\det(e^a_\mu)\right\}\,,
\end{equation}
where the tensor $\mathcal{F}_{\mu\nu}$ has the following general form

\begin{equation}\label{Weitzenboeck_formofF}
\mathcal{F}_{\mu\nu}=\alpha_1S_\mu{}^{\rho\sigma}T_{\nu\rho\sigma}+\alpha_2S_{\rho\mu}{}^{\sigma}T^\rho{}_{\nu\sigma}+\alpha_3 \eta_{ab}e^a_\mu e^b_\nu T \,.
\end{equation}
For the interesting cosmological applications we shall consider a homogeneous and isotropic background for the vielbein $e^a={\rm diag}(1+a(t), a(t), a(t))$ and a barotropic matter field with $p=w\rho$. By varying the action with respect to the vielbein, one obtains the following modified Friedman equation

\begin{equation}
\frac{\sqrt{1+A_2 H^2}}{\sqrt{1-A_1 H^2}}(1+2A_2H^2-3A_1A_2H^4)-1=\frac{\rho}{\mpl^2\mbi^2}\,,
\end{equation}
with the short-cut notations

\begin{equation}
A_1=6(\alpha_2+2\alpha_3)\mbi^{-2} \qquad \text{and} \qquad A_2=2(2\alpha_1+\alpha_2+6\alpha_3)\mbi^{-2} \,.
\end{equation}
Since the matter fluids are assumed to couple minimally to $e$, they follow the standard conservation equation $\dot{\rho}+3(\rho+p)H=0$, which imposes the evolution of the energy density in the form $\rho=\rho_0\left( \frac{a_0}{a}\right)^{3(1+w)}$ with the subscript ``$0$" denoting again the present day value. In order to obtain General Relativity in the low energy density regime, we have to fulfil the condition $\alpha_1+\alpha_2+4\alpha_3=1$. For simplicity, let us first concentrate on the case with $A_2=0$, in other words, $2\alpha_1+\alpha_2+6\alpha_3=0$. This leaves $A_1=12\mbi^{-2}$. In this particular case, the Friedman equation simplifies to

\begin{equation}\label{Friedmaneq_Weizenbock}
\frac{1}{\sqrt{1-12H^2\mbi^{-2}}}-1=\frac{\rho}{\mpl^2\mbi^2} \,.
\end{equation}
This specific case with $A_2=0$ recovers the modified Friedman equation of the previous subsection that we had categorised as $f(T)$ theories. Therefore, in this case one obtains the same non-singular cosmological solutions for radiation and dust matter as the ones reported in the previous subsection \ref{Teleparalel_Weizenbock_BI}. For a more general case of the background with spatial curvature, the Ansatz for the vielbein is a little bit more involved

\begin{equation}
e^0=\mathrm{d}t, \quad e^1=a(t)\tilde{e}^1, \quad e^2=a(t)\tilde{e}^2 \quad \text{and} \quad  e^3=a(t)\tilde{e}^3 \,,
\end{equation}
where the $\tilde{e}^i$ components are given by

\begin{eqnarray}\label{vielbein_e's_weizenbock_FLRW}
\tilde{e}^1&=&K(-K\cos\theta\mathrm{d}\psi+\sin(K\psi)\sin\theta\cos(K\psi)\mathrm{d}\theta-\sin^2(K\psi)\sin^2\theta\mathrm{d}\phi )\,, \nonumber\\
\tilde{e}^2&=&K(K\sin\theta\cos\phi\mathrm{d}\psi-\sin^2(K\psi)(\sin\phi-\cot(K\psi)\cos\theta\cos\phi)\mathrm{d}\theta\nonumber\\
&-&\sin^2(K\psi)\sin\theta(\cot(K\psi)\sin\phi+\cos\theta\cos\phi)\mathrm{d}\phi) \,, \nonumber\\
\tilde{e}^2&=&K(-K\sin\theta\sin\phi\mathrm{d}\psi-\sin^2(K\psi)(\cos\phi+\cot(K\psi)\cos\theta\sin\phi)\mathrm{d}\theta\nonumber\\
&-&\sin^2(K\psi)\sin\theta(\cot(K\psi)\cos\phi-\cos\theta\cos\phi)\mathrm{d}\phi) \,,
\end{eqnarray}
with $K$ denoting the spatial curvature. In this case, the modified Friedman equation (\ref{Friedmaneq_Weizenbock}) becomes

\begin{equation}
\frac{(1\pm \mbi^{-2}a^{-2})^{3/2}}{\sqrt{1-12H^2\mbi^{-2}}}-1=\frac{\rho}{\mpl^2\mbi^2} \,,
\end{equation}
where $\pm$ represents the closed ($K=1$) and open ($K=-1$) universe, respectively. In the high energy density regime, the solution to the modified Friedman equation gives the following evolution for the scale factor in the presence of a radiation fluid

\begin{equation}
a(t)\approx \exp(\sqrt{\mbi^2/12}t) \quad \text{as} \quad a/a_0\to 0 \,.
\end{equation}
This is again the same type of non-singular solution as we found in the previous subsection, which cures the initial singularity and seems to be insensitive to the presence of spatial curvature. Another interesting case arises when one chooses the parameters as $\alpha_2=0$ and $\alpha_1-12\alpha_3=0$. For this particular case, the Friedman equation modifies to

\begin{equation}
\frac{1\pm \mbi^{-2}a^{-2}}{\sqrt{1\pm \mbi^{-2}a^{-2}-12H^2\mbi^{-2}}}-1=\frac{\rho}{\mpl^2\mbi^2} \,.
\end{equation}
We can again abstract the evolution of the scale factor for the high energy density regime. However, now the solution depends highly on the sign of curvature. For the closed universe scenario with the $+$ sign, the scale factor evolves as $a(t)\approx t$ in the high energy regime, which therefore corresponds to a singular solution with the singularity appearing at $t=0$. Maybe a more interesting scenario appears for the case of open universe, where the scale factor evolves as $a(t)\approx a_{\rm min}+\frac{\mbi}{24}t^2$ with $a_{\rm min}=\mbi^{-1}$, constituting a bounce at $t=0$. The accelerated expansion period takes over when the energy density has its maximum value $\rho_{\rm max}\sim a_{\rm min}^{-4}=\mbi^4$ and the volume its minimum value $a_{\rm min}^3=\mbi^{-3}$.

This model with the three components in (\ref{Weitzenboeck_formofF}) was further investigated in \cite{Fiorini:2015hob}, where the realisation of a primordial brusque bounce was studied in detail. The author investigates the unexplored case with $A_1=A_2$ and finds yet other type of interesting cosmological solutions. We shall summarise his findings for this case in the following. In the case with $A_1=A_2$, the modified Friedman equation becomes

\begin{equation}\label{Weitzenboeck_A1=A2caseH2}
6H^2\left(1-\frac{9H^2}{2\mbi^2}\right)=\frac{\rho}{\mpl^2}\,,
\end{equation}
where $\alpha_1=\alpha_2$ and $\alpha_3$ was reabsorbed into $\mbi$. We can solve this equation for $H^2$, which results in $H^2=\mbi^2(1\pm \sqrt{1-3\bar{\rho}})/9$, where $\bar{\rho}$ stands for the dimensionless energy density $\bar{\rho}=\rho/(\mpl^2\mbi^2)$. The conservation equation for the energy density has the standard form $\dot{\bar{\rho}}=-3(1+w)H\bar{\rho}$. Due to the different signs in the expression for $H^2$, we have disconnected two different branches. The branch with the positive sign corresponds to a solution completely disconnected from the GR limit and therefore we can discard this branch. Concerning the negative branch, because of the presence of $\mbi$ one will have different solutions depending on the sign of this parameter. The type of solutions with $\mbi^2<0$ are not interesting either, since they do not admit any regularisation process with the standard diverging behaviour as in GR. On the other hand, the branch with $\mbi^2>0$ gives rise to the wanted regularisation effect with the maximum values $H^2_{\rm max}=\mbi^2/9$ and $\rho_{\rm max}=3\mpl^2\mbi^2$. The conservation equation together with the equation (\ref{Weitzenboeck_A1=A2caseH2}) can be solved exactly. These exact solutions can be found in \cite{Fiorini:2015hob}. We shall only report on the behaviour of these exact solutions in the interesting limits. For a universe filled with radiation ($w=1/3$) and for the branch with $\mbi^2>0$, the scale factor evolves approximately as

\begin{equation}\label{Weitzenboeck_A1=A2caseSF}
\left(\frac{a(t)}{a_0} \right)^4=\frac{\rho_0}{3\mpl^2\mbi^2(1\pm4\mbi t)}+\mathcal{O}(\mbi^2 t^2)\,.
\end{equation}
Accordingly, the dynamics of Hubble function are recast by $H(t)=\mp \mbi/(1\pm4\mbi t)+\mathcal{O}(\mbi^2 t^2)$. From these expressions one immediately observes that there is a minimum value for the scale factor at $t=0$

\begin{equation}
\left(\frac{a_{\rm min}}{a_0} \right)^4=\frac{\rho_0}{3\mpl^2\mbi^2}\,,
\end{equation}
with $H^2(t=0)=\mbi^2$. This represents a brusque bounce. Even if the Ricci scalar suffers from indefiniteness at this point, the solution does not represent a singularity. The author shows explicitly that the geodesics are well behaved at the bounce in the sense that a point particle traveling along causal geodesics does not experience any singular behaviour. Furthermore, the author extends this analysis to a finite size extended object and confirms the same behaviour.

%%%%%%%%%%%%%%%%%%%%%%%%%%%%%%
\subsection{Final remarks}

As we have seen in detail in this section, Eddington-inspired Born-Infeld gravity and its known extensions have received much attention in the literature. Since the modifications \`a la Born-Infeld become appreciable at large energy densities or in high curvature regimes, the direct cosmological applications can be only for early universe physics. The main goal of most of the studies was to construct cosmological solutions curing the standard Big Bang singularities. The inflationary scenario with a standard single field suffers from cosmological singularities. The idea behind using the Eddington-inspired Born-Infeld theory or its extensions was to deliver an alternative scenario for early universe, for instance in form of a bounce or loitering solutions. We have also seen in this section that interesting bouncing and loitering solutions can be constructed within these theories, that are based on a radiation or dust, depending on the explicit model. In the standard inflationary scenario the matter fields couple minimally to the gravity sector. As we have seen in various occasions in this section, Born-Infeld type theories can be seen as nothing else but General Relativity with a non-trivial and non-linear matter coupling. Specially, concerning the cosmological solutions, the essence of the modifications can be encapsulated in the Friedman equation as $H^2=f(\rho,p)$ with a non-linear function. The resulting cosmological evolution correspond to either quasi de-Sitter or bouncing or loitering solutions. We have seen that in the simplest realisation of the bouncing and loitering solutions in the EiBI model, the tensor perturbations were becoming unstable in the presence of matter fields with constant equation of state parameter. This of course renders these simplest realisations unviable. More general scenarios with non-constant equation of state parameters however can alleviate these issues. In this respect, we have seen concrete examples of an additional scalar field as matter field with varying equation of state parameter, which avoids the tensor instabilities in the bouncing and loitering solutions.

After reviewing the works of the standard EiBI model, we then systematically summarised similar cosmological studies of other extensions and modifications of Born-Infeld inspired gravity theories. Since most of these extensions were sharing the same mechanisms and features, before studying the individual cases, we have first developed the general framework of cosmological solutions for a general class of theories constructed out of the Ricci tensor and the inverse metric. Within this general framework, we have derived the master equation that determines the Hubble function in terms of the matter field variables and discussed the general mechanism that provides bouncing solutions. As next, for concrete models we considered the family of Born-Infeld inspired gravity theories based on all of the elementary polynomials and discussed in detail the cosmological solutions of the first polynomial as an example. We considered again fluids with different equation of state parameters and saw that interesting quasi de Sitter solutions can be constructed in a universe with dust component. We summarised briefly the arising inflationary scenario with a cascade of dust components in the early universe. Another interesting extension of the original Born-Infeld gravity is the functional extension in the sense that the square root of the EiBI model is replaced by an arbitrary function of the determinant. The resulting evolution of the Hubble function is such that the bouncing solutions are robust to the functional enlargement, whereas the loitering solutions do undergo notable changes. We have also discussed extensions of the original theory by means of an additional Ricci scalar, appearing either directly in the determinantal structure or as an additional separate function and explored the features of new cosmological solutions beyond the ones present in the EiBI model. Finally, we have also reported on other extensions along the line of teleparallel formulations of Born-Infeld theories and discussed the presence of interesting brusque bouncing solutions.

%%%%%%%%%%%%%%%%%%%%%%%%%%%%%% 
\section{Concluding remarks, open questions and prospects} \label{Sec:Conclusions}

This review has been devoted to provide a comprehensive survey on theories of modified gravity that take inspiration from the Born and Infeld approach to nonlinear electrodynamics. The underlying logic is that a modification of the high curvature/density regime of the gravitational interaction could effectively introduce  upper bounds that cannot be surpassed. As we have seen in this review, the richness of the theory transcends the mere bound of certain invariants, leading to physically sound results even in the presence of curvature divergences in black hole scenarios.

We started our journey on Born-Infeld gravity from the most reasonable place, namely, by reviewing the original construction of Born and Infeld for electromagnetism and the different routes leading to a transcription of its remarkable properties into gravity, specially its determinantal structure. The unsuccessful first attempt of the work of Deser and Gibbons to construct gravity theories \`a la Born-Infeld was rooted in the use of the metric formalism, which inevitably leads to the appearance of ghosts.  So far the only ghostless theories in the metric approach are the so-called $f(R)$ theories, but these can hardly be considered proper Born-Infeld gravity theories. The scrutiny presented in section \ref{Sec:BITheories} suggests that any theory of gravity realizing the Born-Infeld construction and formulated in the metric formalism will either be pathological or reduce to other known theories of gravity. A challenging problem is to find counter-examples to this general no-go result. The story becomes more interesting when resorting to a metric-affine framework, as Vollick did.  When the connection is regarded as an independent field, the aforementioned pathologies arising in the metric formalism are avoided. A further refinement introduced by Ba{\n}ados and Ferreira, making the matter-gravity coupling more standard, resulted in the most extensively explored Born-Infeld inspired gravity theory so far, dubbed EiBI. In this theory, the connection is generated by an auxiliary metric $q_{\mu\nu}$ that is non-trivially related to the metric $g_{\mu\nu}$. Although $q_{\mu\nu}$ made its debut as  an auxiliary object helping to solve the field equations, it soon showed its real significance and allowed to establish the existence of two frames for these theories, similar to what happens in scalar-tensor theories. In the original Born-Infeld frame, matter fields are minimally coupled to the spacetime metric $g_{\mu\nu}$, which satisfies second-order dynamical equations. In the Einstein frame, the metric $q_{\mu\nu}$ behaves as in standard gravity, with an Einstein-Hilbert term governing its dynamics, but it couples in a non-standard way to the matter fields. Elucidating the existence of these two frames allowed to discern that, while matter fields follow geodesics of $g_{\mu\nu}$, the geodesic motion of gravitons is determined by $q_{\mu\nu}$.

Most of the existing developments in the literature make two important assumptions (though not always explicitly said) that we also adopted here. The first one is related to the class of theories considered where only the symmetric part of the Ricci tensor is included. This condition is useful to simplify the field equations and express the solutions for the connection solely in terms of the auxiliary metric. Although it could seem to be rather ad hoc, imposing a projective symmetry naturally results in this type of theories. However, it remains to be explored more general frameworks without the projective invariance and clarify to what extent such a symmetry should be considered as a fundamental ingredient. The second condition that is usually made has to do with the class of solutions that are considered, where the torsion is set to zero. Very little can be found at this respect in the literature of Born-Infeld inspired gravity theories and it is not rare to find works where this issue is simply omitted. Certainly, in most applications assuming vanishing torsion is a consistent Ansatz, but studying the stability of such solutions should also consistently incorporate fluctuations of the torsion. As with the projective invariance, the actual role played by the torsion is to be clarified within the context of these theories. In fact, it would not be too surprising to find links between the projective symmetry and the absence (or irrelevance) of torsion in the solutions. Let us remember that for the Einstein-Hilbert action in the Palatini formalism, the full solution for the connection only contains the trace of the torsion and it precisely enters as a projective mode, thus being pure gauge. %Nos falta hacerlo publico pero hay un paper que aclara completamente el papel de la torsión en la teoría de Born-Infeld: V.I. Afonso, C. Bejarano, Gonzalo J. Olmo, and E. Orazi to appear (2017).

Our first contact with Born-Infeld inspired theories of gravity concluded with a glance at the different approaches adopted in the literature to incorporate the Born-Infeld ideas into gravity and a classification of the existing proposals. We first presented a general formalism showing that most of the features are actually shared by a wide variety of theories. We then decided to perform a classification based on the proximity to the original Born-Infeld construction, and taking the most studied case of EiBI as baseline. We could appreciate the richness of the field where the imagination of the community led to numerous searches along several directions. This was in high contrast with the case of Born-Infeld electrodynamics where the Lagrangian does not admit immediate alterations. This is so because such a Lagrangian was singled out by precise conditions, among which a symmetry guiding principle was paramount. In the case of gravity, an analogous guiding principle permitting to isolate some unique Lagrangian has not been found yet. The projective invariance could be invoked, but we have seen its incapability to sufficiently reduce the number of possible actions. Finding a better suited principle would considerably reduce the different possibilities and would give improved guidance to pursue the exploration of Born-Infeld gravity in closer relation with its electromagnetic ancestor, that turned out to exhibit a number of remarkable features. Until then, a prolific family of different Born-Infeld gravity theories is expected to remain. So far, most of them are based on the EiBI model and extensions along different paths. An interesting alternative was introduced taking TEGR as starting point. This allowed to study a different branch of theories which are formulated in a Weitzenb\"ock space. Since TEGR gives an alternative description of GR as a gauge theory of the translational group, this route could lead to Born-Infeld theories of gravity closer in their structure to the original construction for electromagnetism. This gauge character could be appropriately exploited and it could serve as the desired symmetry principle so it deserves a further exploration.

Once the general landscape of Born-Infeld inspired theories was overviewed, we moved on to the different territories where these theories find applications. The first pertinent place to test the modifications introduced by Born-Infeld inspired theories of gravity was inhabited by the illustrious family of astrophysical objects. Since the Born-Infeld corrections are designed to only appear at high curvatures or densities, compact objects exhibit excellent prospects to put these theories on trial. The first explicit applications, however, already showed some subtleties  in the weak-field limit associated with the energy-density dependence of the modified dynamics proper of metric-affine theories. In diluted systems, the fluid approximation may lead to unphysical effects depending on the weight functions used in the transition from the discrete to the continuum description. Potential pathologies associated to this can be found in Newtonian pressureless fluids and in compact star models based on polytropic equations of state, for instance. As discussed in detail
%in section \ref{sec:astrophysics}
for white dwarfs and neutron stars, polytropic equations of state are very useful to model their structural properties, but the transition to the external (idealized) Schwarzschild solution must be improved in order to construct realistic models able to account for certain observational features (like electromagnetic spectra and radiation fluxes), which at the same time may avoid artificial effects associated with the peculiarities of certain equations of state and/or the fluid approximation. %The explicit construction of such a realistic model is an open question that should be addressed in the future.
After clarifying the importance of correctly modeling the outermost regions of stars, a number of results related with the structural and dynamical properties of compact objects and the Sun were reported. We can highlight the ability of solar observations to constrain the EiBI model via neutrino emission and seismic waves, the possibility of accommodating higher masses with soft equations of state in neutron stars, the stability of these objects against radial perturbations, and the possibility of using the low-mass spectrum of neutron stars to discriminate EiBI from GR. On the other hand, it really came as a surprise the existence of universality relations between quantities constructed using the moment of inertia, the quadrupolar moment, and the Love numbers ($I$-Love-$Q$ relations), which turn out to coincide with those of GR. %Whether this degeneracy is a peculiar property of the EiBI model or a general feature of metric-affine theories is an open question that should be elucidated in the future.
Dipolar magnetic fields also converge to the GR prediction at the crust and surface of neutron stars. These results imply a degeneracy which poses obstacles to the observational discrimination between GR and the EIBI theory.

After spending some time with the best known members of the family of compact objects, we continued our trip and arrived at the place where some of their more exotic acquaintances dwell, namely, black holes and their closest relatives. Obviously, as the black hole terrain has occupied the efforts and imagination of countless theoreticians and astrophysicists alike for decades, is not surprising that a few years of research in the context of Born-Infeld inspired theories of gravity has only allowed to touch a few of the relevant physical aspects regarding these objects. In this sense, our trip quickly went over some dubious proposals for these theories, either because they are formulated in metric approach (and thus plagued by ghostly-like instabilities) or because matter is included in an unconventional form. Nonetheless, this allowed us to naturally introduce the well known black hole solutions for Born-Infeld electromagnetism within GR, of which the familiar Reissner-N\"ordstrom solution is a particular (limiting) case. We thus introduced some of the trademark features of such black holes that bear a close resemblance with those obtained in Born-Infeld inspired theories of gravity, such as the appearance of different number and types of horizons, depending on characteristic parameters of the matter and gravity models. This way we naturally entered into the terrain of black hole solutions within EiBI gravity, where most research in this context has been carried out in the literature. First we reviewed and enlarged the description provided in the paper by Ba\~nados and Ferreira and other works in the field, were we paid special attention to the deviations regarding geodesic motion, strong gravitational lensing and mass inflation. But we also described a different family of solutions, whose study revealed the presence of all types of exotic objects, like geons or wormholes. Geons are self-sustained electromagnetic objects without charges. On the other hand, wormholes represent the promised behaviour of Born-Infeld theories so that the center of black holes in GR is replaced by a regular object of finite size. We saw that the construction of wormhole solutions without any pathologies (i.e. violation of energy conditions) is a hard task, but EiBI gravity managed to surpass our expectations and provided, in analytical form, both wormholes and geons.

Our analysis of the geodesic structure over the innermost region of these objects revealed that, although these places look inhospitable at first, they actually are less perilous than expected and, in fact, geodesics can smoothly pass through, while the impact of curvature divergences on physical observers did not seem to pose any absolutely destructive threat. These results were confirmed by the well posedness of the problem of scattering of scalar waves off the wormhole. Further developments on this field involve higher and lower dimensional models, though with much less impressive results. It should be pointed out that the counterparts of the rotating Kerr solution of GR (and Kerr-Newman when charge is included) in Born-Infeld inspired theories of gravity are not available in the literature and, without such solutions, realistic black hole scenarios for astrophysical purposes cannot be put to a test. This is a very relevant point, since the present (and future) observations from gravitational wave astronomy offer a great opportunity for testing deviations with respect to GR solutions. We cannot but to encourage researchers working in the context of these theories to look for such rotating black hole solutions.

The last stage of our pilgrimage throughout the Born-Infeld land took us to a completely different scenery shaped by cosmological applications. It should not come as a surprise at this point that the natural home for such applications is the early universe because Born-Infeld theories are designed to affect the regime of high densities. In that context, it has been extensively shown that both EiBI and other Born-Inspired theories can provide singular-free solutions of two types, namely: bouncing solutions, where the universe transits from a contracting phase to an expanding one without hitting a singularity, and loitering solutions, where the universe asymptotically approaches a Minkowski universe as the energy density goes to infinity. Both of these solutions were shown to present some tensor instabilities for the original EiBI theory and in the simplest case of one single perfect fluid, although it was later shown that such instabilities could be avoided in more contrived scenarios. These solutions have recurrently been found in other formulations of Born-Infeld inspired gravity. Furthermore, other singular-free solutions have also been found like, e.g. the brusque bounce solution where the Hubble expansion rate is not defined at the bounce, but all relevant geometrical quantities are smooth. In most treatments of these solutions, the analysis is limited to studying the isotropic background evolution and, at most, the tensor perturbations. In some works, homogeneous and anisotropic solutions have also been studied, what is closely related to the analysis of tensor perturbations. However, the full viability of the bouncing solutions can only be claimed once all potential sources of instabilities have been shown to be under control. This is a paramount issue that needs to be properly addressed.

Providing singularity-free cosmological evolutions was precisely the job the Born-Infeld theories were designed for. However, it did not take long to find other jobs for which these theories could serve just as well. In fact, since they are constructed to modify the regime of high densities for gravity, they are also compelling frameworks to have models of inflation. This is achieved for theories that exhibit a nearly constant Hubble expansion rate in the Born-Infeld regime. Such a behavior has been found in several of the proposed models, some even for a radiation dominated universe. In particular, a specific model of inflation was developed in detail where inflation is supported by a dust component that decays into radiation, giving a model of the early universe similar to the usual inflationary scenarios with a re-heating phase, but from a completely different perspective. This in turn led to different predictions, in particular, a super-inflationary phase is achieved where primordial gravitational waves are not produced. One important feature of inflationary models based on these theories is that we have a naturally graceful exit of inflation. This is due to the fact that the density will typically decrease during the inflationary phase so it will eventually become smaller than the transition scale given by $\mbi^2\mpl^2$ and the GR regime will be restored, thus matching the standard cosmological evolutions.  In general, and as with most of the cosmological analysis within Born-Infeld gravity theories, a proper treatment of the scalar and vector perturbations is still to be performed. This is crucial for the viability of these inflationary scenarios since it is of paramount importance to show that a red and nearly scale invariant spectrum of primordial scalar perturbations is generated, in order to be compatible with CMB measurements. However, given the highly non-standard gravitational sector of theses theories, a general and rigorous analysis of the subject will likely be an arduous task. At this respect, simplified models and, perhaps, the use of the Einstein frame could permit advancing in this direction.

Since the expansion of the universe makes the total energy density be diluted during the standard radiation and matter dominated epochs, the Born-Infeld corrections are expected to be negligible for the late-time evolution of the universe. In fact, a safe assumption is to impose that the transition to the GR regime is achieved before BBN. An important consequence of this is that Born-Infeld theories are not fruitful frameworks for dark matter and/or dark energy models and the use of cosmological observables to constrain them by studying their late-time evolution is futile. These theories can only affect the late-time cosmological observables by modifying the initial conditions in the early universe, perhaps set during a Born-Infeld inflationary scenario. There is however a family of cosmologies where Born-Infeld theories can become relevant at late times, namely models with future singularities. If the dark energy component happens to have some exotic features like a phantom behaviour, the asymptotic evolution of the universe in GR will end up with some type of singularity. In the presence of Born-Infeld gravity theories, these models will lead to scenarios where the Born-Infeld regime is reached again when the growing density trespasses the transition density $\mbi^2\mpl^2$. Some works have studied the effects of the Born-Infeld corrections on these future singularities and found that, in general, there is not a universal regularisation of such singularities. This might not be too surprising since the existence of future singularities in GR is tightly linked to exotic properties of dark energy and, thus, the very cosmological model containing those singularities could already present pathologies. At this respect, we find fair to say that Born-Infeld inspired theories are entailed to regularise the Big Bang singularity with standard forms of matter, that is radiation and/or matter. The failure in regularising more general types of singularities should not be regarded as a flaw, but rather their eventual success in this task would be an additional gift granted by these theories.

As we have extensively discussed, the most outstanding feature and the raison d'\^etre of Born-Infeld inspired theories of gravity is the possibility of regularising the singularities of GR without resorting to quantum gravity effects that should appear at the Planck scale $\mpl$. At this respect, we should say that the actual problem in GR is not the existence of singularities per se, but rather that the classical solutions near those divergences go beyond the regime of validity of GR as an effective field theory (presumably near the Planck scale) and, thus, we cannot trust those solutions anymore. The main idea behind Born-Infeld inspired theories is to introduce a new scale $\mbi$ at which the gravitational interaction is modified so that curvature divergences are classically regularised before reaching the quantum regime. However, a proper treatment of the validity of Born-Infeld theories as actual effective field theories is still missing. In particular, an issue that should be clarified is the existence of some regime above $\mbi$, that one could naively identify with the strong coupling scale of the theory, where quantum corrections remain under control and, thus, the resulting classical solutions without singularities can be trusted.  As with other open questions, perhaps the best starting point to address this issue would be the Einstein frame where all the effects are moved to the matter sector. It is not hopeless to expect a nice quantum behaviour, at least for some matter fields. For instance, if we start from a massless scalar field in the Born-Infeld frame, in the Einstein frame we would have a $K-$essence type of theory whose Lagrangian would be of the form $K(X)$, for which the quantum stability of the classical action has already been discussed in detail in \cite{deRham:2014wfa,Brax:2016jjt}. A crucial point to notice here is that for the singular free-solutions provided by Born-Infeld theories, NEC violations are not required and, therefore, the usual arguments for the instability and breaking of unitarity of these solutions do not directly apply.  In general, the question would be as to what extent the Born-Infeld scale determines the strong coupling scale or the cutoff of the theory and the radiative stability for known types of matter.

The voyage undertaken throughout this review has permitted us to encounter an interesting family of gravitational theories that revealed fascinating novel effects in astrophysics, black hole physics and cosmology. They offer excellent opportunities for the exploration of the gravitational interaction and the open questions exposed above should serve, even if not exhaustively, as a guidance for future research within the field. We hope the accompanying traveller profited and enjoyed reading this work as much as we did in its elaboration.





\section*{Acknowledgments}
\addcontentsline{toc}{section}{Acknowledgments}

We are grateful to many colleagues for useful discussions and comments regarding the many topics and results reported in this work. In particular, we thank Pedro Avelino, Alejandro C\'ardenas-Avenda\~no, Joaqu\'in D\'iaz-Alonso, Rafael Ferraro, Franco Fiorini, Soumya Jana, Sayan Kar, Tomi S. Koivisto, Yu-Xiao Liu, Francisco S. N. Lobo, Christophe Ringeval, Diego S\'aez-G\'omez, Bayram Tekin, Ke Yang, and Shao-Wen Wei. J.B.J. acknowledges the financial support
of A*MIDEX project (n ANR-11-IDEX-0001-02) funded by the Investissements d'Avenir French Government program, managed by the French National Research Agency (ANR), MINECO (Spain) projects FIS2014-52837-P, FIS2016-78859-P (AEI/FEDER) and Consolider-Ingenio MULTIDARK CSD2009-00064. L.H. is funded by Dr. Max R\"ossler, the Walter Haefner Foundation and the ETH Zurich Foundation. G. J. O. is supported by a Ramon y Cajal contract, the grant FIS2014-57387-C3-1-P from MINECO and the European Regional Development Fund, and the i-COOPB20105 grant of the Spanish Research Council (CSIC). D. R.-G. is funded by the Funda\c{c}\~ao para a Ci\^encia e a Tecnologia (FCT, Portugal) postdoctoral fellowship No.~SFRH/BPD/102958/2014 and the FCT research grant UID/FIS/04434/2013. This work is also supported by the Consolider Program CPANPHY-1205388, the Severo Ochoa grant SEV-2014-0398 (Spain), and the CNPq (Brazilian agency) project No.~301137/2014-5. J.B.J. thanks the Institute of Astrophysics and Space Sciences (IA) at Lisbon University (Portugal) for financial support under the ``IA Visitor Program" to visit the IA. L.H. thanks the STSM fellowship of CANTATA for financial support to visit the IA. Both J.B.J and L.H. acknowledge the ``Unveiling the dynamics of the Universe" group of IA for their hospitality during the elaboration of this work. This article is based upon work from COST Action CA15117, supported by COST (European Cooperation in Science and Technology).

\section*{References}
\addcontentsline{toc}{section}{References}

\bibliographystyle{elsarticle-num}
\bibliography{BIReview_References}

\begin{thebibliography}{100}

\bibitem{TheLIGOScientific:2016pea}
Abbott, B.~P.  {et~al.} (Virgo, LIGO Scientific), ``{Binary Black Hole Mergers
  in the first Advanced LIGO Observing Run}'', {\em Phys. Rev.}, {\bf X6}(4),
  041015 (2016).
  {\small[\href{http://dx.doi.org/10.1103/PhysRevX.6.041015}{DOI}]},
  {\small[\href{http://arxiv.org/abs/1606.04856}{{arXiv:1606.04856
  {\small[gr-qc]}}}]}.

\bibitem{Abbott:2016blz}
Abbott, B.~P.  {et~al.} (Virgo, LIGO Scientific), ``{Observation of
  Gravitational Waves from a Binary Black Hole Merger}'', {\em Phys. Rev.
  Lett.}, {\bf 116}(6), 061102 (2016).
  {\small[\href{http://dx.doi.org/10.1103/PhysRevLett.116.061102}{DOI}]},
  {\small[\href{http://arxiv.org/abs/1602.03837}{{arXiv:1602.03837
  {\small[gr-qc]}}}]}.

\bibitem{TheLIGOScientific:2016src}
Abbott, B.~P.  {et~al.} (Virgo, LIGO Scientific), ``{Tests of general
  relativity with GW150914}'', {\em Phys. Rev. Lett.}, {\bf 116}(22), 221101
  (2016).
  {\small[\href{http://dx.doi.org/10.1103/PhysRevLett.116.221101}{DOI}]},
  {\small[\href{http://arxiv.org/abs/1602.03841}{{arXiv:1602.03841
  {\small[gr-qc]}}}]}.

\bibitem{Abedi:2016hgu}
Abedi, Jahed, Dykaar, Hannah  and Afshordi, Niayesh, ``{Echoes from the Abyss:
  Evidence for Planck-scale structure at black hole horizons}'' (2016).
  {\small[\href{http://arxiv.org/abs/1612.00266}{{arXiv:1612.00266
  {\small[gr-qc]}}}]}.

\bibitem{Ackermann:2015zua}
Ackermann, M.  {et~al.} (Fermi-LAT), ``{Searching for Dark Matter Annihilation
  from Milky Way Dwarf Spheroidal Galaxies with Six Years of Fermi Large Area
  Telescope Data}'', {\em Phys. Rev. Lett.}, {\bf 115}(23), 231301 (2015).
  {\small[\href{http://dx.doi.org/10.1103/PhysRevLett.115.231301}{DOI}]},
  {\small[\href{http://arxiv.org/abs/1503.02641}{{arXiv:1503.02641
  {\small[astro-ph.HE]}}}]}.

\bibitem{Ade:2013zuv}
Ade, P. A.~R.  {et~al.} (Planck), ``{Planck 2013 results. XVI. Cosmological
  parameters}'', {\em Astron. Astrophys.}, {\bf 571}, A16 (2014).
  {\small[\href{http://dx.doi.org/10.1051/0004-6361/201321591}{DOI}]},
  {\small[\href{http://arxiv.org/abs/1303.5076}{{arXiv:1303.5076
  {\small[astro-ph.CO]}}}]}.

\bibitem{Ade:2015rim}
Ade, P. A.~R.  {et~al.} (Planck), ``{Planck 2015 results. XIV. Dark energy and
  modified gravity}'', {\em Astron. Astrophys.}, {\bf 594}, A14 (2016).
  {\small[\href{http://dx.doi.org/10.1051/0004-6361/201525814}{DOI}]},
  {\small[\href{http://arxiv.org/abs/1502.01590}{{arXiv:1502.01590
  {\small[astro-ph.CO]}}}]}.

\bibitem{Afonso:2017aci}
Afonso, V.~I., Olmo, Gonzalo~J.  and Rubiera-Garcia, D., ``{Scalar geons in
  Born-Infeld gravity}'' (2017).
  {\small[\href{http://arxiv.org/abs/1705.01065}{{arXiv:1705.01065
  {\small[gr-qc]}}}]}.

\bibitem{Akerib:2016vxi}
Akerib, D.~S.  {et~al.} (LUX), ``{Results from a search for dark matter in the
  complete LUX exposure}'', {\em Phys. Rev. Lett.}, {\bf 118}(2), 021303
  (2017).
  {\small[\href{http://dx.doi.org/10.1103/PhysRevLett.118.021303}{DOI}]},
  {\small[\href{http://arxiv.org/abs/1608.07648}{{arXiv:1608.07648
  {\small[astro-ph.CO]}}}]}.

\bibitem{Akmal:1998cf}
Akmal, A., Pandharipande, V.~R.  and Ravenhall, D.~G., ``{The Equation of state
  of nucleon matter and neutron star structure}'', {\em Phys. Rev.}, {\bf C58},
  1804--1828 (1998).
  {\small[\href{http://dx.doi.org/10.1103/PhysRevC.58.1804}{DOI}]},
  {\small[\href{http://arxiv.org/abs/nucl-th/9804027}{{arXiv:nucl-th/9804027
  {\small[nucl-th]}}}]}.

\bibitem{Albarran:2017swy}
Albarran, Imanol, Bouhmadi-L{\'o}pez, Mariam, Chen, Che-Yu  and Chen, Pisin,
  ``{Doomsdays in a modified theory of gravity: A classical and a quantum
  approach}'' (2017).
  {\small[\href{http://arxiv.org/abs/1703.09263}{{arXiv:1703.09263
  {\small[gr-qc]}}}]}.

\bibitem{TelparallelAldrovandi}
Aldrovandi, R.  and Pereira, J.~G., {\em An introduction to Teleparallel
  Gravity}, (Springer Netherlands, 2013).

\bibitem{Alishahiha:2010iq}
Alishahiha, Mohsen, Naseh, Ali  and Soltanpanahi, Hesam, ``{On Born-Infeld
  Gravity in Three Dimensions}'', {\em Phys. Rev.}, {\bf D82}, 024042 (2010).
  {\small[\href{http://dx.doi.org/10.1103/PhysRevD.82.024042}{DOI}]},
  {\small[\href{http://arxiv.org/abs/1006.1757}{{arXiv:1006.1757
  {\small[hep-th]}}}]}.

\bibitem{Almheiri:2012rt}
Almheiri, Ahmed, Marolf, Donald, Polchinski, Joseph  and Sully, James, ``{Black
  Holes: Complementarity or Firewalls?}'', {\em JHEP}, {\bf 02}, 062 (2013).
  {\small[\href{http://dx.doi.org/10.1007/JHEP02(2013)062}{DOI}]},
  {\small[\href{http://arxiv.org/abs/1207.3123}{{arXiv:1207.3123
  {\small[hep-th]}}}]}.

\bibitem{Amendola:2006kh}
Amendola, Luca, Polarski, David  and Tsujikawa, Shinji, ``{Are f(R) dark energy
  models cosmologically viable ?}'', {\em Phys. Rev. Lett.}, {\bf 98}, 131302
  (2007).
  {\small[\href{http://dx.doi.org/10.1103/PhysRevLett.98.131302}{DOI}]},
  {\small[\href{http://arxiv.org/abs/astro-ph/0603703}{{arXiv:astro-ph/0603703
  {\small[astro-ph]}}}]}.

\bibitem{Amendola:2012ys}
Amendola, Luca  {et~al.} (Euclid Theory Working Group), ``{Cosmology and
  fundamental physics with the Euclid satellite}'', {\em Living Rev. Rel.},
  {\bf 16}, 6 (2013).
  {\small[\href{http://dx.doi.org/10.12942/lrr-2013-6}{DOI}]},
  {\small[\href{http://arxiv.org/abs/1206.1225}{{arXiv:1206.1225
  {\small[astro-ph.CO]}}}]}.

\bibitem{Amendola:2016saw}
Amendola, Luca  {et~al.}, ``{Cosmology and Fundamental Physics with the Euclid
  Satellite}'' (2016).
  {\small[\href{http://arxiv.org/abs/1606.00180}{{arXiv:1606.00180
  {\small[astro-ph.CO]}}}]}.

\bibitem{1996PhRvL..77.4134A}
{Andersson}, N.  and {Kokkotas}, K.~D., ``{Gravitational Waves and Pulsating
  Stars: What Can We Learn from Future Observations?}'', {\em Physical Review
  Letters}, {\bf 77}, 4134--4137 (November 1996).
  {\small[\href{http://dx.doi.org/10.1103/PhysRevLett.77.4134}{DOI}]},
  {\small[\href{http://adsabs.harvard.edu/abs/1996PhRvL..77.4134A}{ADS}]},
  {\small[\href{http://arxiv.org/abs/gr-qc/9610035}{{gr-qc/9610035}}]}.

\bibitem{Andersson:1997rn}
Andersson, Nils  and Kokkotas, Kostas~D., ``{Towards gravitational wave
  asteroseismology}'', {\em Mon. Not. Roy. Astron. Soc.}, {\bf 299}, 1059--1068
  (1998).
  {\small[\href{http://dx.doi.org/10.1046/j.1365-8711.1998.01840.x}{DOI}]},
  {\small[\href{http://arxiv.org/abs/gr-qc/9711088}{{arXiv:gr-qc/9711088
  {\small[gr-qc]}}}]}.

\bibitem{Ansoldi:2008jw}
Ansoldi, Stefano, ``{Spherical black holes with regular center: A Review of
  existing models including a recent realization with Gaussian sources}'', in
  {\em {Conference on Black Holes and Naked Singularities Milan, Italy, May
  10-12, 2007}}, (2008).
  {\small[\href{http://arxiv.org/abs/0802.0330}{{arXiv:0802.0330
  {\small[gr-qc]}}}]}URL:
  \newline\url{https://inspirehep.net/record/778724/files/arXiv:0802.0330.pdf}.

\bibitem{Ansoldi:2006vg}
Ansoldi, Stefano, Nicolini, Piero, Smailagic, Anais  and Spallucci, Euro,
  ``{Noncommutative geometry inspired charged black holes}'', {\em Phys.
  Lett.}, {\bf B645}, 261--266 (2007).
  {\small[\href{http://dx.doi.org/10.1016/j.physletb.2006.12.020}{DOI}]},
  {\small[\href{http://arxiv.org/abs/gr-qc/0612035}{{arXiv:gr-qc/0612035
  {\small[gr-qc]}}}]}.

\bibitem{Antoniadis:1998ig}
Antoniadis, Ignatios, Arkani-Hamed, Nima, Dimopoulos, Savas  and Dvali, G.~R.,
  ``{New dimensions at a millimeter to a Fermi and superstrings at a TeV}'',
  {\em Phys. Lett.}, {\bf B436}, 257--263 (1998).
  {\small[\href{http://dx.doi.org/10.1016/S0370-2693(98)00860-0}{DOI}]},
  {\small[\href{http://arxiv.org/abs/hep-ph/9804398}{{arXiv:hep-ph/9804398
  {\small[hep-ph]}}}]}.

\bibitem{Aramaki:2015pii}
Aramaki, T.  {et~al.}, ``{Review of the theoretical and experimental status of
  dark matter identification with cosmic-ray antideuterons}'', {\em Phys.
  Rept.}, {\bf 618}, 1--37 (2016).
  {\small[\href{http://dx.doi.org/10.1016/j.physrep.2016.01.002}{DOI}]},
  {\small[\href{http://arxiv.org/abs/1505.07785}{{arXiv:1505.07785
  {\small[hep-ph]}}}]}.

\bibitem{ArkaniHamed:1998rs}
Arkani-Hamed, Nima, Dimopoulos, Savas  and Dvali, G.~R., ``{The Hierarchy
  problem and new dimensions at a millimeter}'', {\em Phys. Lett.}, {\bf B429},
  263--272 (1998).
  {\small[\href{http://dx.doi.org/10.1016/S0370-2693(98)00466-3}{DOI}]},
  {\small[\href{http://arxiv.org/abs/hep-ph/9803315}{{arXiv:hep-ph/9803315
  {\small[hep-ph]}}}]}.

\bibitem{ArmendarizPicon:2000dh}
Armendariz-Picon, C., Mukhanov, Viatcheslav~F.  and Steinhardt, Paul~J., ``{A
  Dynamical solution to the problem of a small cosmological constant and late
  time cosmic acceleration}'', {\em Phys. Rev. Lett.}, {\bf 85}, 4438--4441
  (2000). {\small[\href{http://dx.doi.org/10.1103/PhysRevLett.85.4438}{DOI}]},
  {\small[\href{http://arxiv.org/abs/astro-ph/0004134}{{arXiv:astro-ph/0004134
  {\small[astro-ph]}}}]}.

\bibitem{Arroja:2016ffm}
Arroja, Frederico, Chen, Che-Yu, Chen, Pisin  and Yeom, Dong-han, ``{Singular
  Instantons in Eddington-inspired-Born-Infeld Gravity}'', {\em JCAP}, {\bf
  1703}(03), 044 (2017).
  {\small[\href{http://dx.doi.org/10.1088/1475-7516/2017/03/044}{DOI}]},
  {\small[\href{http://arxiv.org/abs/1612.00674}{{arXiv:1612.00674
  {\small[gr-qc]}}}]}.

\bibitem{Aschieri:2008ns}
Aschieri, Paolo, Ferrara, Sergio  and Zumino, Bruno, ``{Duality Rotations in
  Nonlinear Electrodynamics and in Extended Supergravity}'', {\em Riv. Nuovo
  Cim.}, {\bf 31}, 625--708 (2008).
  {\small[\href{http://dx.doi.org/10.1393/ncr/i2008-10038-8}{DOI}]},
  {\small[\href{http://arxiv.org/abs/0807.4039}{{arXiv:0807.4039
  {\small[hep-th]}}}]}.

\bibitem{Avelino:2012ge}
Avelino, P.~P., ``{Eddington-inspired Born-Infeld gravity: astrophysical and
  cosmological constraints}'', {\em Phys. Rev.}, {\bf D85}, 104053 (2012).
  {\small[\href{http://dx.doi.org/10.1103/PhysRevD.85.104053}{DOI}]},
  {\small[\href{http://arxiv.org/abs/1201.2544}{{arXiv:1201.2544
  {\small[astro-ph.CO]}}}]}.

\bibitem{Avelino:2012qe}
Avelino, P.~P., ``{Eddington-inspired Born-Infeld gravity: nuclear physics
  constraints and the validity of the continuous fluid approximation}'', {\em
  JCAP}, {\bf 1211}, 022 (2012).
  {\small[\href{http://dx.doi.org/10.1088/1475-7516/2012/11/022}{DOI}]},
  {\small[\href{http://arxiv.org/abs/1207.4730}{{arXiv:1207.4730
  {\small[astro-ph.CO]}}}]}.

\bibitem{Avelino:2015fve}
Avelino, P.~P., ``{Inner Structure of Black Holes in Eddington-inspired
  Born-Infeld gravity: the role of mass inflation}'', {\em Phys. Rev.}, {\bf
  D93}(4), 044067 (2016).
  {\small[\href{http://dx.doi.org/10.1103/PhysRevD.93.044067}{DOI}]},
  {\small[\href{http://arxiv.org/abs/1511.03223}{{arXiv:1511.03223
  {\small[gr-qc]}}}]}.

\bibitem{Avelino:2016kkj}
Avelino, P.~P., ``{Mass inflation in Eddington-inspired Born-Infeld black
  holes: analytical scaling solutions}'', {\em Phys. Rev.}, {\bf D93}(10),
  104054 (2016).
  {\small[\href{http://dx.doi.org/10.1103/PhysRevD.93.104054}{DOI}]},
  {\small[\href{http://arxiv.org/abs/1602.08261}{{arXiv:1602.08261
  {\small[gr-qc]}}}]}.

\bibitem{Avelino:2012ue}
Avelino, P.~P.  and Ferreira, R.~Z., ``{Bouncing Eddington-inspired Born-Infeld
  cosmologies: an alternative to Inflation ?}'', {\em Phys. Rev.}, {\bf D86},
  041501 (2012).
  {\small[\href{http://dx.doi.org/10.1103/PhysRevD.86.041501}{DOI}]},
  {\small[\href{http://arxiv.org/abs/1205.6676}{{arXiv:1205.6676
  {\small[astro-ph.CO]}}}]}.

\bibitem{Avelino:2009vv}
Avelino, P.~P., Hamilton, A. J.~S.  and Herdeiro, C. A.~R., ``{Mass Inflation
  in Brans-Dicke gravity}'', {\em Phys. Rev.}, {\bf D79}, 124045 (2009).
  {\small[\href{http://dx.doi.org/10.1103/PhysRevD.79.124045}{DOI}]},
  {\small[\href{http://arxiv.org/abs/0904.2669}{{arXiv:0904.2669
  {\small[gr-qc]}}}]}.

\bibitem{Avelino:2011ee}
Avelino, P.~P., Hamilton, A. J.~S., Herdeiro, C. A.~R.  and Zilhao, M., ``{Mass
  inflation in a D dimensional Reissner-Nordstrom black hole: a hierarchy of
  particle accelerators ?}'', {\em Phys. Rev.}, {\bf D84}, 024019 (2011).
  {\small[\href{http://dx.doi.org/10.1103/PhysRevD.84.024019}{DOI}]},
  {\small[\href{http://arxiv.org/abs/1105.4434}{{arXiv:1105.4434
  {\small[gr-qc]}}}]}.

\bibitem{AyonBeato:1998ub}
Ayon-Beato, Eloy  and Garcia, Alberto, ``{Regular black hole in general
  relativity coupled to nonlinear electrodynamics}'', {\em Phys. Rev. Lett.},
  {\bf 80}, 5056--5059 (1998).
  {\small[\href{http://dx.doi.org/10.1103/PhysRevLett.80.5056}{DOI}]},
  {\small[\href{http://arxiv.org/abs/gr-qc/9911046}{{arXiv:gr-qc/9911046
  {\small[gr-qc]}}}]}.

\bibitem{Bahcall:1996vj}
Bahcall, John~N.  and Ulmer, Andrew, ``{The Temperature dependence of solar
  neutrino fluxes}'', {\em Phys. Rev.}, {\bf D53}, 4202--4210 (1996).
  {\small[\href{http://dx.doi.org/10.1103/PhysRevD.53.4202}{DOI}]},
  {\small[\href{http://arxiv.org/abs/astro-ph/9602012}{{arXiv:astro-ph/9602012
  {\small[astro-ph]}}}]}.

\bibitem{1997AA328274B}
{Baldo}, M., {Bombaci}, I.  and {Burgio}, G.~F., ``{Microscopic nuclear
  equation of state with three-body forces and neutron star structure}'', {\em
  Astronomy and Astrophysics}, {\bf 328}, 274--282 (December 1997).
  {\small[\href{http://adsabs.harvard.edu/abs/1997A%26A...328..274B}{ADS}]},
  {\small[\href{http://arxiv.org/abs/astro-ph/9707277}{{astro-ph/9707277}}]}.

\bibitem{Bamba:2010kf}
Bamba, Kazuharu  and Geng, Chao-Qiang, ``{Thermodynamics in $f(R)$ gravity in
  the Palatini formalism}'', {\em JCAP}, {\bf 1006}, 014 (2010).
  {\small[\href{http://dx.doi.org/10.1088/1475-7516/2010/06/014}{DOI}]},
  {\small[\href{http://arxiv.org/abs/1005.5234}{{arXiv:1005.5234
  {\small[gr-qc]}}}]}.

\bibitem{Bambi:2012at}
Bambi, Cosimo, ``{Testing the space-time geometry around black hole candidates
  with the analysis of the broad K$\alpha$ iron line}'', {\em Phys. Rev.}, {\bf
  D87}, 023007 (2013).
  {\small[\href{http://dx.doi.org/10.1103/PhysRevD.87.023007}{DOI}]},
  {\small[\href{http://arxiv.org/abs/1211.2513}{{arXiv:1211.2513
  {\small[gr-qc]}}}]}.

\bibitem{Bambi:2015zch}
Bambi, Cosimo, Cardenas-Avendano, Alejandro, Olmo, Gonzalo~J.  and
  Rubiera-Garcia, D., ``{Wormholes and nonsingular spacetimes in Palatini
  $f(R)$ gravity}'', {\em Phys. Rev.}, {\bf D93}(6), 064016 (2016).
  {\small[\href{http://dx.doi.org/10.1103/PhysRevD.93.064016}{DOI}]},
  {\small[\href{http://arxiv.org/abs/1511.03755}{{arXiv:1511.03755
  {\small[gr-qc]}}}]}.

\bibitem{Bambi:2013caa}
Bambi, Cosimo, Malafarina, Daniele  and Modesto, Leonardo, ``{Non-singular
  quantum-inspired gravitational collapse}'', {\em Phys. Rev.}, {\bf D88},
  044009 (2013).
  {\small[\href{http://dx.doi.org/10.1103/PhysRevD.88.044009}{DOI}]},
  {\small[\href{http://arxiv.org/abs/1305.4790}{{arXiv:1305.4790
  {\small[gr-qc]}}}]}.

\bibitem{Bambi:2015sla}
Bambi, Cosimo, Olmo, Gonzalo~J.  and Rubiera-Garcia, D., ``{Melvin Universe in
  Born-Infeld gravity}'', {\em Phys. Rev.}, {\bf D91}(10), 104010 (2015).
  {\small[\href{http://dx.doi.org/10.1103/PhysRevD.91.104010}{DOI}]},
  {\small[\href{http://arxiv.org/abs/1504.01827}{{arXiv:1504.01827
  {\small[gr-qc]}}}]}.

\bibitem{Bambi:2016xme}
Bambi, Cosimo, Rubiera-Garcia, D.  and Wang, Yixu, ``{Black hole solutions in
  functional extensions of Born-Infeld gravity}'', {\em Phys. Rev.}, {\bf
  D94}(6), 064002 (2016).
  {\small[\href{http://dx.doi.org/10.1103/PhysRevD.94.064002}{DOI}]},
  {\small[\href{http://arxiv.org/abs/1608.04873}{{arXiv:1608.04873
  {\small[gr-qc]}}}]}.

\bibitem{Banados:2008rm}
Banados, Maximo, ``{Eddington-Born-Infeld action for dark matter and dark
  energy}'', {\em Phys. Rev.}, {\bf D77}, 123534 (2008).
  {\small[\href{http://dx.doi.org/10.1103/PhysRevD.77.123534}{DOI}]},
  {\small[\href{http://arxiv.org/abs/0801.4103}{{arXiv:0801.4103
  {\small[hep-th]}}}]}.

\bibitem{Banados:2010ix}
Banados, Maximo  and Ferreira, Pedro~G., ``{Eddington's theory of gravity and
  its progeny}'', {\em Phys. Rev. Lett.}, {\bf 105}, 011101 (2010).
  {\small[\href{http://dx.doi.org/10.1103/PhysRevLett.105.011101,
  10.1103/PhysRevLett.113.119901}{DOI}]},
  {\small[\href{http://arxiv.org/abs/1006.1769}{{arXiv:1006.1769
  {\small[astro-ph.CO]}}}]}. [Erratum: Phys. Rev. Lett.113,no.11,119901(2014)].

\bibitem{Banados:2008fj}
Banados, M., Ferreira, P.~G.  and Skordis, C., ``{Eddington-Born-Infeld gravity
  and the large scale structure of the Universe}'', {\em Phys. Rev.}, {\bf
  D79}, 063511 (2009).
  {\small[\href{http://dx.doi.org/10.1103/PhysRevD.79.063511}{DOI}]},
  {\small[\href{http://arxiv.org/abs/0811.1272}{{arXiv:0811.1272
  {\small[astro-ph]}}}]}.

\bibitem{Banados:2008fi}
Banados, Maximo, Gomberoff, Andres, Rodrigues, Davi~C.  and Skordis,
  Constantinos, ``{A Note on bigravity and dark matter}'', {\em Phys. Rev.},
  {\bf D79}, 063515 (2009).
  {\small[\href{http://dx.doi.org/10.1103/PhysRevD.79.063515}{DOI}]},
  {\small[\href{http://arxiv.org/abs/0811.1270}{{arXiv:0811.1270
  {\small[gr-qc]}}}]}.

\bibitem{Banados:1992wn}
Banados, Maximo, Teitelboim, Claudio  and Zanelli, Jorge, ``{The Black hole in
  three-dimensional space-time}'', {\em Phys. Rev. Lett.}, {\bf 69}, 1849--1851
  (1992). {\small[\href{http://dx.doi.org/10.1103/PhysRevLett.69.1849}{DOI}]},
  {\small[\href{http://arxiv.org/abs/hep-th/9204099}{{arXiv:hep-th/9204099
  {\small[hep-th]}}}]}.

\bibitem{Barcelo:2015noa}
Barcel{\'o}, Carlos, Carballo-Rubio, Ra{\'u}l  and Garay, Luis~J., ``{Where
  does the physics of extreme gravitational collapse reside?}'', {\em
  Universe}, {\bf 2}(2), 7 (2016).
  {\small[\href{http://dx.doi.org/10.3390/universe2020007}{DOI}]},
  {\small[\href{http://arxiv.org/abs/1510.04957}{{arXiv:1510.04957
  {\small[gr-qc]}}}]}.

\bibitem{Barcelo:2017lnx}
Barcel{\'o}, Carlos, Carballo-Rubio, Ra{\'u}l  and Garay, Luis~J.,
  ``{Gravitational echoes from macroscopic quantum gravity effects}'' (2017).
  {\small[\href{http://arxiv.org/abs/1701.09156}{{arXiv:1701.09156
  {\small[gr-qc]}}}]}.

\bibitem{PhysRevD.51.3113}
Barrow, John~D., ``Why the Universe is not anisotropic'', {\em Phys. Rev. D},
  {\bf 51}, 3113--3116 (Mar 1995).
  {\small[\href{http://dx.doi.org/10.1103/PhysRevD.51.3113}{DOI}]}URL:
  \newline\url{http://link.aps.org/doi/10.1103/PhysRevD.51.3113}.

\bibitem{Barrow:2004xh}
Barrow, John~D., ``{Sudden future singularities}'', {\em Class. Quant. Grav.},
  {\bf 21}, L79--L82 (2004).
  {\small[\href{http://dx.doi.org/10.1088/0264-9381/21/11/L03}{DOI}]},
  {\small[\href{http://arxiv.org/abs/gr-qc/0403084}{{arXiv:gr-qc/0403084
  {\small[gr-qc]}}}]}.

\bibitem{Bazeia:2015zpa}
Bazeia, D., Losano, L., Menezes, R., Olmo, Gonzalo~J.  and Rubiera-Garcia, D.,
  ``{Robustness of braneworld scenarios against tensorial perturbations}'',
  {\em Class. Quant. Grav.}, {\bf 32}(21), 215011 (2015).
  {\small[\href{http://dx.doi.org/10.1088/0264-9381/32/21/215011}{DOI}]},
  {\small[\href{http://arxiv.org/abs/1509.04895}{{arXiv:1509.04895
  {\small[hep-th]}}}]}.

\bibitem{Bazeia:2016rlg}
Bazeia, D., Losano, L., Olmo, Gonzalo~J.  and Rubiera-Garcia, D.,
  ``{Geodesically complete BTZ-type solutions of $2+1$ Born-Infeld gravity}'',
  {\em Class. Quant. Grav.}, {\bf 34}(4), 045006 (2017).
  {\small[\href{http://dx.doi.org/10.1088/1361-6382/aa56f5}{DOI}]},
  {\small[\href{http://arxiv.org/abs/1609.05827}{{arXiv:1609.05827
  {\small[hep-th]}}}]}.

\bibitem{Bazeia:2015uia}
Bazeia, D., Losano, L., Olmo, Gonzalo~J., Rubiera-Garcia, D.  and
  Sanchez-Puente, A., ``{Classical resolution of black hole singularities in
  arbitrary dimension}'', {\em Phys. Rev.}, {\bf D92}(4), 044018 (2015).
  {\small[\href{http://dx.doi.org/10.1103/PhysRevD.92.044018}{DOI}]},
  {\small[\href{http://arxiv.org/abs/1507.07763}{{arXiv:1507.07763
  {\small[hep-th]}}}]}.

\bibitem{Bejarano:2017fgz}
Bejarano, Cecilia, Olmo, Gonzalo~J.  and Rubiera-Garcia, Diego, ``{What is a
  singular black hole beyond General Relativity?}'', {\em Phys. Rev.}, {\bf
  D95}(6), 064043 (2017).
  {\small[\href{http://dx.doi.org/10.1103/PhysRevD.95.064043}{DOI}]},
  {\small[\href{http://arxiv.org/abs/1702.01292}{{arXiv:1702.01292
  {\small[hep-th]}}}]}.

\bibitem{Bejger}
Bejger, M.  and Haensel, P, ``{Moments of inertia for neutron and strange
  stars: Limits derived for the Crab pulsar}'', {\em Astronomy and
  Astrophysics}, {\bf 396}(3), 917--921 (2002).
  {\small[\href{http://dx.doi.org/10.1051/0004-6361:20021241}{DOI}]}.

\bibitem{BeltranJimenez:2012sz}
Beltran~Jimenez, Jose, Golovnev, Alexey, Karciauskas, Mindaugas  and Koivisto,
  Tomi~S., ``{The Bimetric variational principle for General Relativity}'',
  {\em Phys. Rev.}, {\bf D86}, 084024 (2012).
  {\small[\href{http://dx.doi.org/10.1103/PhysRevD.86.084024}{DOI}]},
  {\small[\href{http://arxiv.org/abs/1201.4018}{{arXiv:1201.4018
  {\small[gr-qc]}}}]}.

\bibitem{Jimenez:2014fla}
Beltran~Jimenez, Jose, Heisenberg, Lavinia  and Olmo, Gonzalo~J., ``{Infrared
  lessons for ultraviolet gravity: the case of massive gravity and
  Born-Infeld}'', {\em JCAP}, {\bf 1411}, 004 (2014).
  {\small[\href{http://dx.doi.org/10.1088/1475-7516/2014/11/004}{DOI}]},
  {\small[\href{http://arxiv.org/abs/1409.0233}{{arXiv:1409.0233
  {\small[hep-th]}}}]}.

\bibitem{Jimenez:2015caa}
Beltran~Jimenez, Jose, Heisenberg, Lavinia  and Olmo, Gonzalo~J., ``{Tensor
  perturbations in a general class of Palatini theories}'', {\em JCAP}, {\bf
  1506}, 026 (2015).
  {\small[\href{http://dx.doi.org/10.1088/1475-7516/2015/06/026}{DOI}]},
  {\small[\href{http://arxiv.org/abs/1504.00295}{{arXiv:1504.00295
  {\small[gr-qc]}}}]}.

\bibitem{Jimenez:2015jqa}
Beltran~Jimenez, Jose, Heisenberg, Lavinia, Olmo, Gonzalo~J.  and Ringeval,
  Christophe, ``{Cascading dust inflation in Born-Infeld gravity}'', {\em
  JCAP}, {\bf 1511}, 046 (2015).
  {\small[\href{http://dx.doi.org/10.1088/1475-7516/2015/11/046}{DOI}]},
  {\small[\href{http://arxiv.org/abs/1509.01188}{{arXiv:1509.01188
  {\small[gr-qc]}}}]}.

\bibitem{BeltranJimenez:2016dfc}
Beltr{\'a}n~Jim{\'e}nez, Jose, Rubiera-Garcia, Diego, S{\'a}ez-G{\'o}mez, Diego
   and Salzano, Vincenzo, ``{Cosmological future singularities in interacting
  dark energy models}'', {\em Phys. Rev.}, {\bf D94}(12), 123520 (2016).
  {\small[\href{http://dx.doi.org/10.1103/PhysRevD.94.123520}{DOI}]},
  {\small[\href{http://arxiv.org/abs/1607.06389}{{arXiv:1607.06389
  {\small[gr-qc]}}}]}.

\bibitem{Benhar:1998au}
Benhar, Omar, Berti, Emanuele  and Ferrari, Valeria, ``{The Imprint of the
  equation of state on the axial w modes of oscillating neutron stars}'', {\em
  Mon. Not. Roy. Astron. Soc.}, {\bf 310}, 797--803 (1999).
  {\small[\href{http://dx.doi.org/10.1046/j.1365-8711.1999.02983.x}{DOI}]},
  {\small[\href{http://arxiv.org/abs/gr-qc/9901037}{{arXiv:gr-qc/9901037
  {\small[gr-qc]}}}]}. [,35(1998)].

\bibitem{BFG2004}
Benhar, Omar, Ferrari, Valeria  and Gualtieri, Leonardo, ``Gravitational wave
  asteroseismology reexamined'', {\em Physical review D: Particles and fields},
  {\bf 70}(12) (12 2004).
  {\small[\href{http://dx.doi.org/10.1103/PhysRevD.70.124015}{DOI}]}.

\bibitem{Berej:2006cc}
Berej, Waldemar, Matyjasek, Jerzy, Tryniecki, Dariusz  and Woronowicz, Mariusz,
  ``{Regular black holes in quadratic gravity}'', {\em Gen. Rel. Grav.}, {\bf
  38}, 885--906 (2006).
  {\small[\href{http://dx.doi.org/10.1007/s10714-006-0270-9}{DOI}]},
  {\small[\href{http://arxiv.org/abs/hep-th/0606185}{{arXiv:hep-th/0606185
  {\small[hep-th]}}}]}.

\bibitem{Bergshoeff:2009hq}
Bergshoeff, Eric~A., Hohm, Olaf  and Townsend, Paul~K., ``{Massive Gravity in
  Three Dimensions}'', {\em Phys. Rev. Lett.}, {\bf 102}, 201301 (2009).
  {\small[\href{http://dx.doi.org/10.1103/PhysRevLett.102.201301}{DOI}]},
  {\small[\href{http://arxiv.org/abs/0901.1766}{{arXiv:0901.1766
  {\small[hep-th]}}}]}.

\bibitem{Berti:2015itd}
Berti, Emanuele  {et~al.}, ``{Testing General Relativity with Present and
  Future Astrophysical Observations}'', {\em Class. Quant. Grav.}, {\bf 32},
  243001 (2015).
  {\small[\href{http://dx.doi.org/10.1088/0264-9381/32/24/243001}{DOI}]},
  {\small[\href{http://arxiv.org/abs/1501.07274}{{arXiv:1501.07274
  {\small[gr-qc]}}}]}.

\bibitem{Bertone:2004pz}
Bertone, Gianfranco, Hooper, Dan  and Silk, Joseph, ``{Particle dark matter:
  Evidence, candidates and constraints}'', {\em Phys. Rept.}, {\bf 405},
  279--390 (2005).
  {\small[\href{http://dx.doi.org/10.1016/j.physrep.2004.08.031}{DOI}]},
  {\small[\href{http://arxiv.org/abs/hep-ph/0404175}{{arXiv:hep-ph/0404175
  {\small[hep-ph]}}}]}.

\bibitem{BialynickiBirula:1984tx}
Bialynicki-Birula, I., ``{Non-linear electrodynamics: Variations on a theme by
  Born and Infeld}'' (1984).

\bibitem{Bird:2016dcv}
Bird, Simeon, Cholis, Ilias, Mu{\~n}oz, Julian~B., Ali-Ha{\"\i}moud, Yacine,
  Kamionkowski, Marc, Kovetz, Ely~D., Raccanelli, Alvise  and Riess, Adam~G.,
  ``{Did LIGO detect dark matter?}'', {\em Phys. Rev. Lett.}, {\bf 116}(20),
  201301 (2016).
  {\small[\href{http://dx.doi.org/10.1103/PhysRevLett.116.201301}{DOI}]},
  {\small[\href{http://arxiv.org/abs/1603.00464}{{arXiv:1603.00464
  {\small[astro-ph.CO]}}}]}.

\bibitem{Biswas:2011ar}
Biswas, Tirthabir, Gerwick, Erik, Koivisto, Tomi  and Mazumdar, Anupam,
  ``{Towards singularity and ghost free theories of gravity}'', {\em Phys. Rev.
  Lett.}, {\bf 108}, 031101 (2012).
  {\small[\href{http://dx.doi.org/10.1103/PhysRevLett.108.031101}{DOI}]},
  {\small[\href{http://arxiv.org/abs/1110.5249}{{arXiv:1110.5249
  {\small[gr-qc]}}}]}.

\bibitem{Born:1933qff}
Born, M., ``{Modified field equations with a finite radius of the electron}'',
  {\em Nature}, {\bf 132}(3329), 282.1 (1933).
  {\small[\href{http://dx.doi.org/10.1038/132282a0}{DOI}]}.

\bibitem{Born410}
Born, Max, ``On the Quantum Theory of the Electromagnetic Field'', {\em
  Proceedings of the Royal Society of London A: Mathematical, Physical and
  Engineering Sciences}, {\bf 143}(849), 410--437 (1934).
  {\small[\href{http://dx.doi.org/10.1098/rspa.1934.0010}{DOI}]},
  {\small[\href{http://arxiv.org/abs/http://rspa.royalsocietypublishing.org/content/143/849/410.full.pdf}{{http://rspa.royalsocietypublishing.org/content/143/849/410.full.pdf}}]}URL:
  \newline\url{http://rspa.royalsocietypublishing.org/content/143/849/410}.

\bibitem{1933Natur.132.1004B}
{Born}, M.  and {Infeld}, L., ``{Foundations of the New Field Theory}'', {\em
  Nature}, {\bf 132}, 1004 (December 1933).
  {\small[\href{http://dx.doi.org/10.1038/1321004b0}{DOI}]},
  {\small[\href{http://adsabs.harvard.edu/abs/1933Natur.132.1004B}{ADS}]}.

\bibitem{Born:1934gh}
Born, M.  and Infeld, L., ``{Foundations of the new field theory}'', {\em Proc.
  Roy. Soc. Lond.}, {\bf A144}, 425--451 (1934).
  {\small[\href{http://dx.doi.org/10.1098/rspa.1934.0059}{DOI}]}.

\bibitem{Borunda:2008kf}
Borunda, Monica, Janssen, Bert  and Bastero-Gil, Mar, ``{Palatini versus metric
  formulation in higher curvature gravity}'', {\em JCAP}, {\bf 0811}, 008
  (2008).
  {\small[\href{http://dx.doi.org/10.1088/1475-7516/2008/11/008}{DOI}]},
  {\small[\href{http://arxiv.org/abs/0804.4440}{{arXiv:0804.4440
  {\small[hep-th]}}}]}.

\bibitem{Bouhmadi-Lopez:2016dcf}
Bouhmadi-L{\'o}pez, Mariam  and Chen, Che-Yu, ``{Towards the Quantization of
  Eddington-inspired-Born-Infeld Theory}'', {\em JCAP}, {\bf 1611}(11), 023
  (2016).
  {\small[\href{http://dx.doi.org/10.1088/1475-7516/2016/11/023}{DOI}]},
  {\small[\href{http://arxiv.org/abs/1609.00700}{{arXiv:1609.00700
  {\small[gr-qc]}}}]}.

\bibitem{Bouhmadi-Lopez:2014tna}
Bouhmadi-Lopez, Mariam, Chen, Che-Yu  and Chen, Pisin, ``{Cosmological
  singularities in Born-Infeld determinantal gravity}'', {\em Phys. Rev.}, {\bf
  D90}, 123518 (2014).
  {\small[\href{http://dx.doi.org/10.1103/PhysRevD.90.123518}{DOI}]},
  {\small[\href{http://arxiv.org/abs/1407.5114}{{arXiv:1407.5114
  {\small[gr-qc]}}}]}.

\bibitem{Bouhmadi-Lopez:2013lha}
Bouhmadi-Lopez, Mariam, Chen, Che-Yu  and Chen, Pisin, ``{Is
  Eddington-Born-Infeld theory really free of cosmological singularities?}'',
  {\em Eur. Phys. J.}, {\bf C74}, 2802 (2014).
  {\small[\href{http://dx.doi.org/10.1140/epjc/s10052-014-2802-x}{DOI}]},
  {\small[\href{http://arxiv.org/abs/1302.5013}{{arXiv:1302.5013
  {\small[gr-qc]}}}]}.

\bibitem{Bouhmadi-Lopez:2014jfa}
Bouhmadi-L{\'o}pez, Mariam, Chen, Che-Yu  and Chen, Pisin,
  ``{Eddington--Born--Infeld cosmology: a cosmographic approach, a tale of
  doomsdays and the fate of bound structures}'', {\em Eur. Phys. J.}, {\bf
  C75}, 90 (2015).
  {\small[\href{http://dx.doi.org/10.1140/epjc/s10052-015-3257-4}{DOI}]},
  {\small[\href{http://arxiv.org/abs/1406.6157}{{arXiv:1406.6157
  {\small[gr-qc]}}}]}.

\bibitem{Bouhmadi-Lopez:2014cca}
Bouhmadi-Lopez, Mariam, Errahmani, Ahmed, Martin-Moruno, Prado, Ouali, Taoufik
  and Tavakoli, Yaser, ``{The little sibling of the big rip singularity}'',
  {\em Int. J. Mod. Phys.}, {\bf D24}(10), 1550078 (2015).
  {\small[\href{http://dx.doi.org/10.1142/S0218271815500789}{DOI}]},
  {\small[\href{http://arxiv.org/abs/1407.2446}{{arXiv:1407.2446
  {\small[gr-qc]}}}]}.

\bibitem{Boulware:1973my}
Boulware, D.~G.  and Deser, Stanley, ``{Can gravitation have a finite
  range?}'', {\em Phys. Rev.}, {\bf D6}, 3368--3382 (1972).
  {\small[\href{http://dx.doi.org/10.1103/PhysRevD.6.3368}{DOI}]}.

\bibitem{Bozza:2002zj}
Bozza, V., ``{Gravitational lensing in the strong field limit}'', {\em Phys.
  Rev.}, {\bf D66}, 103001 (2002).
  {\small[\href{http://dx.doi.org/10.1103/PhysRevD.66.103001}{DOI}]},
  {\small[\href{http://arxiv.org/abs/gr-qc/0208075}{{arXiv:gr-qc/0208075
  {\small[gr-qc]}}}]}.

\bibitem{Bozza:2012by}
Bozza, V.  and Mancini, L., ``{Observing gravitational lensing effects by Sgr
  A* with GRAVITY}'', {\em Astrophys. J.}, {\bf 753}, 56 (2012).
  {\small[\href{http://dx.doi.org/10.1088/0004-637X/753/1/56}{DOI}]},
  {\small[\href{http://arxiv.org/abs/1204.2103}{{arXiv:1204.2103
  {\small[astro-ph.GA]}}}]}.

\bibitem{Brax:2016jjt}
Brax, Philippe  and Valageas, Patrick, ``{Quantum field theory of
  K-mouflage}'', {\em Phys. Rev.}, {\bf D94}(4), 043529 (2016).
  {\small[\href{http://dx.doi.org/10.1103/PhysRevD.94.043529}{DOI}]},
  {\small[\href{http://arxiv.org/abs/1607.01129}{{arXiv:1607.01129
  {\small[astro-ph.CO]}}}]}.

\bibitem{Brecher:1998su}
Brecher, D.  and Perry, M.~J., ``{Bound states of D-branes and the nonAbelian
  Born-Infeld action}'', {\em Nucl. Phys.}, {\bf B527}, 121--141 (1998).
  {\small[\href{http://dx.doi.org/10.1016/S0550-3213(98)00297-1}{DOI}]},
  {\small[\href{http://arxiv.org/abs/hep-th/9801127}{{arXiv:hep-th/9801127
  {\small[hep-th]}}}]}.

\bibitem{Breton:2003tk}
Breton, Nora, ``{Born-Infeld black hole in the isolated horizon framework}'',
  {\em Phys. Rev.}, {\bf D67}, 124004 (2003).
  {\small[\href{http://dx.doi.org/10.1103/PhysRevD.67.124004}{DOI}]},
  {\small[\href{http://arxiv.org/abs/hep-th/0301254}{{arXiv:hep-th/0301254
  {\small[hep-th]}}}]}.

\bibitem{Bronnikov:2004ax}
Bronnikov, K.~A.  and Grinyok, S.~V., ``{Conformal continuations and wormhole
  instability in scalar-tensor gravity}'', {\em Grav. Cosmol.}, {\bf 10}, 237
  (2004).
  {\small[\href{http://arxiv.org/abs/gr-qc/0411063}{{arXiv:gr-qc/0411063
  {\small[gr-qc]}}}]}.

\bibitem{Bull:2015stt}
Bull, Philip  {et~al.}, ``{Beyond $\Lambda$CDM: Problems, solutions, and the
  road ahead}'', {\em Phys. Dark Univ.}, {\bf 12}, 56--99 (2016).
  {\small[\href{http://dx.doi.org/10.1016/j.dark.2016.02.001}{DOI}]},
  {\small[\href{http://arxiv.org/abs/1512.05356}{{arXiv:1512.05356
  {\small[astro-ph.CO]}}}]}.

\bibitem{Burgess:2003jk}
Burgess, C.~P., ``{Quantum gravity in everyday life: General relativity as an
  effective field theory}'', {\em Living Rev. Rel.}, {\bf 7}, 5--56 (2004).
  {\small[\href{http://dx.doi.org/10.12942/lrr-2004-5}{DOI}]},
  {\small[\href{http://arxiv.org/abs/gr-qc/0311082}{{arXiv:gr-qc/0311082
  {\small[gr-qc]}}}]}.

\bibitem{Burrage:2011cr}
Burrage, Clare, de~Rham, Claudia, Heisenberg, Lavinia  and Tolley, Andrew~J.,
  ``{Chronology Protection in Galileon Models and Massive Gravity}'', {\em
  JCAP}, {\bf 1207}, 004 (2012).
  {\small[\href{http://dx.doi.org/10.1088/1475-7516/2012/07/004}{DOI}]},
  {\small[\href{http://arxiv.org/abs/1111.5549}{{arXiv:1111.5549
  {\small[hep-th]}}}]}.

\bibitem{Cai:2015emx}
Cai, Yi-Fu, Capozziello, Salvatore, De~Laurentis, Mariafelicia  and Saridakis,
  Emmanuel~N., ``{f(T) teleparallel gravity and cosmology}'', {\em Rept. Prog.
  Phys.}, {\bf 79}(10), 106901 (2016).
  {\small[\href{http://dx.doi.org/10.1088/0034-4885/79/10/106901}{DOI}]},
  {\small[\href{http://arxiv.org/abs/1511.07586}{{arXiv:1511.07586
  {\small[gr-qc]}}}]}.

\bibitem{Caldwell:2003vq}
Caldwell, Robert~R., Kamionkowski, Marc  and Weinberg, Nevin~N., ``{Phantom
  energy and cosmic doomsday}'', {\em Phys. Rev. Lett.}, {\bf 91}, 071301
  (2003).
  {\small[\href{http://dx.doi.org/10.1103/PhysRevLett.91.071301}{DOI}]},
  {\small[\href{http://arxiv.org/abs/astro-ph/0302506}{{arXiv:astro-ph/0302506
  {\small[astro-ph]}}}]}.

\bibitem{Callan:1997kz}
Callan, Curtis~G.  and Maldacena, Juan~Martin, ``{Brane death and dynamics from
  the Born-Infeld action}'', {\em Nucl. Phys.}, {\bf B513}, 198--212 (1998).
  {\small[\href{http://dx.doi.org/10.1016/S0550-3213(97)00700-1}{DOI}]},
  {\small[\href{http://arxiv.org/abs/hep-th/9708147}{{arXiv:hep-th/9708147
  {\small[hep-th]}}}]}.

\bibitem{Capozziello:2011et}
Capozziello, Salvatore  and De~Laurentis, Mariafelicia, ``{Extended Theories of
  Gravity}'', {\em Phys. Rept.}, {\bf 509}, 167--321 (2011).
  {\small[\href{http://dx.doi.org/10.1016/j.physrep.2011.09.003}{DOI}]},
  {\small[\href{http://arxiv.org/abs/1108.6266}{{arXiv:1108.6266
  {\small[gr-qc]}}}]}.

\bibitem{Capozziello:2015lza}
Capozziello, Salvatore, Harko, Tiberiu, Koivisto, Tomi~S., Lobo, Francisco
  S.~N.  and Olmo, Gonzalo~J., ``{Hybrid metric-Palatini gravity}'', {\em
  Universe}, {\bf 1}(2), 199--238 (2015).
  {\small[\href{http://dx.doi.org/10.3390/universe1020199}{DOI}]},
  {\small[\href{http://arxiv.org/abs/1508.04641}{{arXiv:1508.04641
  {\small[gr-qc]}}}]}.

\bibitem{Cardoso:2016rao}
Cardoso, Vitor, Franzin, Edgardo  and Pani, Paolo, ``{Is the gravitational-wave
  ringdown a probe of the event horizon?}'', {\em Phys. Rev. Lett.}, {\bf
  116}(17), 171101 (2016).
  {\small[\href{http://dx.doi.org/10.1103/PhysRevLett.117.089902,
  10.1103/PhysRevLett.116.171101}{DOI}]},
  {\small[\href{http://arxiv.org/abs/1602.07309}{{arXiv:1602.07309
  {\small[gr-qc]}}}]}. [Erratum: Phys. Rev. Lett.117,no.8,089902(2016)].

\bibitem{Cardoso:2016ryw}
Cardoso, Vitor  and Gualtieri, Leonardo, ``{Testing the black hole `no-hair'
  hypothesis}'', {\em Class. Quant. Grav.}, {\bf 33}(17), 174001 (2016).
  {\small[\href{http://dx.doi.org/10.1088/0264-9381/33/17/174001}{DOI}]},
  {\small[\href{http://arxiv.org/abs/1607.03133}{{arXiv:1607.03133
  {\small[gr-qc]}}}]}.

\bibitem{Cardoso:2016oxy}
Cardoso, Vitor, Hopper, Seth, Macedo, Caio F.~B., Palenzuela, Carlos  and Pani,
  Paolo, ``{Gravitational-wave signatures of exotic compact objects and of
  quantum corrections at the horizon scale}'', {\em Phys. Rev.}, {\bf D94}(8),
  084031 (2016).
  {\small[\href{http://dx.doi.org/10.1103/PhysRevD.94.084031}{DOI}]},
  {\small[\href{http://arxiv.org/abs/1608.08637}{{arXiv:1608.08637
  {\small[gr-qc]}}}]}.

\bibitem{Carpenter:2012rg}
Carpenter, Linda~M., Nelson, Andrew, Shimmin, Chase, Tait, Tim M.~P.  and
  Whiteson, Daniel, ``{Collider searches for dark matter in events with a Z
  boson and missing energy}'', {\em Phys. Rev.}, {\bf D87}(7), 074005 (2013).
  {\small[\href{http://dx.doi.org/10.1103/PhysRevD.87.074005}{DOI}]},
  {\small[\href{http://arxiv.org/abs/1212.3352}{{arXiv:1212.3352
  {\small[hep-ex]}}}]}.

\bibitem{Carriere:2002bx}
Carriere, J., Horowitz, C.~J.  and Piekarewicz, J., ``{Low mass neutron stars
  and the equation of state of dense matter}'', {\em Astrophys. J.}, {\bf 593},
  463--471 (2003). {\small[\href{http://dx.doi.org/10.1086/376515}{DOI}]},
  {\small[\href{http://arxiv.org/abs/nucl-th/0211015}{{arXiv:nucl-th/0211015
  {\small[nucl-th]}}}]}.

\bibitem{Carter:1971zc}
Carter, B., ``{Axisymmetric Black Hole Has Only Two Degrees of Freedom}'', {\em
  Phys. Rev. Lett.}, {\bf 26}, 331--333 (1971).
  {\small[\href{http://dx.doi.org/10.1103/PhysRevLett.26.331}{DOI}]}.

\bibitem{Casanellas:2011kf}
Casanellas, Jordi, Pani, Paolo, Lopes, Ilidio  and Cardoso, Vitor, ``{Testing
  alternative theories of gravity using the Sun}'', {\em Astrophys. J.}, {\bf
  745}, 15 (2012).
  {\small[\href{http://dx.doi.org/10.1088/0004-637X/745/1/15}{DOI}]},
  {\small[\href{http://arxiv.org/abs/1109.0249}{{arXiv:1109.0249
  {\small[astro-ph.SR]}}}]}.

\bibitem{Chandrabook}
Chandrasekhar, S., ``{The Mathematical Theory of Black Holes}'', {\em Oxford
  University Press, New York} (1992).

\bibitem{Chang:2002wy}
Chang, Philip  and Bildsten, Lars, ``{Diffusive nuclear burning on neutron star
  envelopes}'', {\em Astrophys. J.}, {\bf 585}, 464--474 (2003).
  {\small[\href{http://dx.doi.org/10.1086/345551}{DOI}]},
  {\small[\href{http://arxiv.org/abs/astro-ph/0210218}{{arXiv:astro-ph/0210218
  {\small[astro-ph]}}}]}.

\bibitem{Chen:2015eha}
Chen, Che-Yu, Bouhmadi-Lopez, Mariam  and Chen, Pisin, ``{Modified
  Eddington-inspired-Born-Infeld Gravity with a Trace Term}'', {\em Eur. Phys.
  J.}, {\bf C76}, 40 (2016).
  {\small[\href{http://dx.doi.org/10.1140/epjc/s10052-016-3879-1}{DOI}]},
  {\small[\href{http://arxiv.org/abs/1507.00028}{{arXiv:1507.00028
  {\small[gr-qc]}}}]}.

\bibitem{Chiba:2005nz}
Chiba, Takeshi, ``{Generalized gravity and ghost}'', {\em JCAP}, {\bf 0503},
  008 (2005).
  {\small[\href{http://dx.doi.org/10.1088/1475-7516/2005/03/008}{DOI}]},
  {\small[\href{http://arxiv.org/abs/gr-qc/0502070}{{arXiv:gr-qc/0502070
  {\small[gr-qc]}}}]}.

\bibitem{Chiba:2006jp}
Chiba, Takeshi, Smith, Tristan~L.  and Erickcek, Adrienne~L., ``{Solar System
  constraints to general f(R) gravity}'', {\em Phys. Rev.}, {\bf D75}, 124014
  (2007). {\small[\href{http://dx.doi.org/10.1103/PhysRevD.75.124014}{DOI}]},
  {\small[\href{http://arxiv.org/abs/astro-ph/0611867}{{arXiv:astro-ph/0611867
  {\small[astro-ph]}}}]}.

\bibitem{Cho:2015yua}
Cho, Inyong  and Gong, Jinn-Ouk, ``{Spectral indices in Eddington-inspired
  Born-Infeld inflation}'', {\em Phys. Rev.}, {\bf D92}(6), 064046 (2015).
  {\small[\href{http://dx.doi.org/10.1103/PhysRevD.92.064046}{DOI}]},
  {\small[\href{http://arxiv.org/abs/1506.07061}{{arXiv:1506.07061
  {\small[gr-qc]}}}]}.

\bibitem{Cho:2014ija}
Cho, Inyong  and Kim, Hyeong-Chan, ``{Inflationary tensor perturbation in
  Eddington-inspired born-infeld gravity}'', {\em Phys. Rev.}, {\bf D90}(2),
  024063 (2014).
  {\small[\href{http://dx.doi.org/10.1103/PhysRevD.90.024063}{DOI}]},
  {\small[\href{http://arxiv.org/abs/1404.6081}{{arXiv:1404.6081
  {\small[gr-qc]}}}]}.

\bibitem{Cho:2012vg}
Cho, Inyong, Kim, Hyeong-Chan  and Moon, Taeyoon, ``{Universe Driven by Perfect
  Fluid in Eddington-inspired Born-Infeld Gravity}'', {\em Phys. Rev.}, {\bf
  D86}, 084018 (2012).
  {\small[\href{http://dx.doi.org/10.1103/PhysRevD.86.084018}{DOI}]},
  {\small[\href{http://arxiv.org/abs/1208.2146}{{arXiv:1208.2146
  {\small[gr-qc]}}}]}.

\bibitem{Cho:2013pea}
Cho, Inyong, Kim, Hyeong-Chan  and Moon, Taeyoon, ``{Precursor of Inflation}'',
  {\em Phys. Rev. Lett.}, {\bf 111}, 071301 (2013).
  {\small[\href{http://dx.doi.org/10.1103/PhysRevLett.111.071301}{DOI}]},
  {\small[\href{http://arxiv.org/abs/1305.2020}{{arXiv:1305.2020
  {\small[gr-qc]}}}]}.

\bibitem{Cho:2014jta}
Cho, Inyong  and Singh, Naveen~K., ``{Tensor-to-scalar ratio in
  Eddington-inspired Born--Infeld inflation}'', {\em Eur. Phys. J.}, {\bf
  C74}(11), 3155 (2014).
  {\small[\href{http://dx.doi.org/10.1140/epjc/s10052-014-3155-1}{DOI}]},
  {\small[\href{http://arxiv.org/abs/1408.2652}{{arXiv:1408.2652
  {\small[gr-qc]}}}]}.

\bibitem{Cho:2015yza}
Cho, Inyong  and Singh, Naveen~K., ``{Primordial Power Spectra of EiBI
  Inflation in Strong Gravity Limit}'', {\em Phys. Rev.}, {\bf D92}(2), 024038
  (2015). {\small[\href{http://dx.doi.org/10.1103/PhysRevD.92.024038}{DOI}]},
  {\small[\href{http://arxiv.org/abs/1506.02213}{{arXiv:1506.02213
  {\small[gr-qc]}}}]}.

\bibitem{Cho:2014xaa}
Cho, Inyong  and Singh, Naveen~K., ``{Scalar perturbation produced at the
  pre-inflationary stage in Eddington-inspired Born--Infeld gravity}'', {\em
  Eur. Phys. J.}, {\bf C75}(6), 240 (2015).
  {\small[\href{http://dx.doi.org/10.1140/epjc/s10052-015-3458-x}{DOI}]},
  {\small[\href{http://arxiv.org/abs/1412.6344}{{arXiv:1412.6344
  {\small[gr-qc]}}}]}.

\bibitem{ChristensenDalsgaard:2002ur}
Christensen-Dalsgaard, Jorgen, ``{Helioseismology}'', {\em Rev. Mod. Phys.},
  {\bf 74}, 1073--1129 (2003).
  {\small[\href{http://dx.doi.org/10.1103/RevModPhys.74.1073}{DOI}]},
  {\small[\href{http://arxiv.org/abs/astro-ph/0207403}{{arXiv:astro-ph/0207403
  {\small[astro-ph]}}}]}.

\bibitem{ClarkeKrolak}
Clarke, C. J.~S.  and A.., Krolak., {\em J. Geom. Phys.}, {\bf 2}, 127 (1985).

\bibitem{Claudel:2000yi}
Claudel, Clarissa-Marie, Virbhadra, K.~S.  and Ellis, G. F.~R., ``{The Geometry
  of photon surfaces}'', {\em J. Math. Phys.}, {\bf 42}, 818--838 (2001).
  {\small[\href{http://dx.doi.org/10.1063/1.1308507}{DOI}]},
  {\small[\href{http://arxiv.org/abs/gr-qc/0005050}{{arXiv:gr-qc/0005050
  {\small[gr-qc]}}}]}.

\bibitem{Clayton1968}
Clayton, Donald~D., ``{Principles of Stellar Evolution and Nucleosynthesis}'',
  {\em University Of Chicago Press, Chicago} (1968).

\bibitem{Clifton:2011jh}
Clifton, Timothy, Ferreira, Pedro~G., Padilla, Antonio  and Skordis,
  Constantinos, ``{Modified Gravity and Cosmology}'', {\em Phys. Rept.}, {\bf
  513}, 1--189 (2012).
  {\small[\href{http://dx.doi.org/10.1016/j.physrep.2012.01.001}{DOI}]},
  {\small[\href{http://arxiv.org/abs/1106.2476}{{arXiv:1106.2476
  {\small[astro-ph.CO]}}}]}.

\bibitem{Cognola:2007zu}
Cognola, G., Elizalde, E., Nojiri, S., Odintsov, S.~D., Sebastiani, L.  and
  Zerbini, S., ``{A Class of viable modified f(R) gravities describing
  inflation and the onset of accelerated expansion}'', {\em Phys. Rev.}, {\bf
  D77}, 046009 (2008).
  {\small[\href{http://dx.doi.org/10.1103/PhysRevD.77.046009}{DOI}]},
  {\small[\href{http://arxiv.org/abs/0712.4017}{{arXiv:0712.4017
  {\small[hep-th]}}}]}.

\bibitem{Comelli:2005tn}
Comelli, D., ``{Born-Infeld type gravity}'', {\em Phys. Rev.}, {\bf D72},
  064018 (2005).
  {\small[\href{http://dx.doi.org/10.1103/PhysRevD.72.064018}{DOI}]},
  {\small[\href{http://arxiv.org/abs/gr-qc/0505088}{{arXiv:gr-qc/0505088
  {\small[gr-qc]}}}]}.

\bibitem{Comelli:2004qr}
Comelli, D.  and Dolgov, A., ``{Determinant-gravity: Cosmological
  implications}'', {\em JHEP}, {\bf 11}, 062 (2004).
  {\small[\href{http://dx.doi.org/10.1088/1126-6708/2004/11/062}{DOI}]},
  {\small[\href{http://arxiv.org/abs/gr-qc/0404065}{{arXiv:gr-qc/0404065
  {\small[gr-qc]}}}]}.

\bibitem{Copeland:2006wr}
Copeland, Edmund~J., Sami, M.  and Tsujikawa, Shinji, ``{Dynamics of dark
  energy}'', {\em Int. J. Mod. Phys.}, {\bf D15}, 1753--1936 (2006).
  {\small[\href{http://dx.doi.org/10.1142/S021827180600942X}{DOI}]},
  {\small[\href{http://arxiv.org/abs/hep-th/0603057}{{arXiv:hep-th/0603057
  {\small[hep-th]}}}]}.

\bibitem{Curiel}
Curiel, Erik  and Bokulick, Peter, ``{Singularities and Black Holes, The
  Stanford Encyclopedia of Philosophy}'', {\em Edward N. Zalta (ed.).}
  (2012)URL:
  \newline\url{http://plato.stanford.edu/entries/spacetime-singularities/}.

\bibitem{Dabrowski:2009kg}
Dabrowski, Mariusz~P.  and Denkieiwcz, Tomasz, ``{Barotropic index
  w-singularities in cosmology}'', {\em Phys. Rev.}, {\bf D79}, 063521 (2009).
  {\small[\href{http://dx.doi.org/10.1103/PhysRevD.79.063521}{DOI}]},
  {\small[\href{http://arxiv.org/abs/0902.3107}{{arXiv:0902.3107
  {\small[gr-qc]}}}]}.

\bibitem{Dadhich:2010dg}
Dadhich, Naresh  and Pons, Josep~M., ``{Consistent Levi Civita truncation
  uniquely characterizes the Lovelock Lagrangians}'', {\em Phys. Lett.}, {\bf
  B705}, 139--142 (2011).
  {\small[\href{http://dx.doi.org/10.1016/j.physletb.2011.09.108}{DOI}]},
  {\small[\href{http://arxiv.org/abs/1012.1692}{{arXiv:1012.1692
  {\small[gr-qc]}}}]}.

\bibitem{Dadhich:2010xa}
Dadhich, Naresh  and Pons, Josep~M., ``{On the equivalence of the
  Einstein-Hilbert and the Einstein-Palatini formulations of general relativity
  for an arbitrary connection}'', {\em Gen. Rel. Grav.}, {\bf 44}, 2337--2352
  (2012). {\small[\href{http://dx.doi.org/10.1007/s10714-012-1393-9}{DOI}]},
  {\small[\href{http://arxiv.org/abs/1010.0869}{{arXiv:1010.0869
  {\small[gr-qc]}}}]}.

\bibitem{Damour:1992bt}
Damour, T., Deser, Stanley  and McCarthy, James~G., ``{Nonsymmetric gravity
  theories: Inconsistencies and a cure}'', {\em Phys. Rev.}, {\bf D47},
  1541--1556 (1993).
  {\small[\href{http://dx.doi.org/10.1103/PhysRevD.47.1541}{DOI}]},
  {\small[\href{http://arxiv.org/abs/gr-qc/9207003}{{arXiv:gr-qc/9207003
  {\small[gr-qc]}}}]}.

\bibitem{DeFelice:2010aj}
De~Felice, Antonio  and Tsujikawa, Shinji, ``{f(R) theories}'', {\em Living
  Rev. Rel.}, {\bf 13}, 3 (2010).
  {\small[\href{http://dx.doi.org/10.12942/lrr-2010-3}{DOI}]},
  {\small[\href{http://arxiv.org/abs/1002.4928}{{arXiv:1002.4928
  {\small[gr-qc]}}}]}.

\bibitem{delaCruz-Dombriz:2015tye}
de~la Cruz-Dombriz, {\'A}lvaro, Dunsby, Peter K.~S., Kandhai, Sulona  and
  S{\'a}ez-G{\'o}mez, Diego, ``{Theoretical and observational constraints of
  viable f(R) theories of gravity}'', {\em Phys. Rev.}, {\bf D93}(8), 084016
  (2016). {\small[\href{http://dx.doi.org/10.1103/PhysRevD.93.084016}{DOI}]},
  {\small[\href{http://arxiv.org/abs/1511.00102}{{arXiv:1511.00102
  {\small[gr-qc]}}}]}.

\bibitem{deOliveira:1994in}
de~Oliveira, H.~P., ``{Nonlinear charged black holes}'', {\em Class. Quant.
  Grav.}, {\bf 11}, 1469--1482 (1994).
  {\small[\href{http://dx.doi.org/10.1088/0264-9381/11/6/012}{DOI}]}.

\bibitem{deRham:2012ew}
de~Rham, Claudia, Gabadadze, Gregory, Heisenberg, Lavinia  and Pirtskhalava,
  David, ``{Nonrenormalization and naturalness in a class of scalar-tensor
  theories}'', {\em Phys. Rev.}, {\bf D87}(8), 085017 (2013).
  {\small[\href{http://dx.doi.org/10.1103/PhysRevD.87.085017}{DOI}]},
  {\small[\href{http://arxiv.org/abs/1212.4128}{{arXiv:1212.4128
  {\small[hep-th]}}}]}.

\bibitem{deRham:2010kj}
de~Rham, Claudia, Gabadadze, Gregory  and Tolley, Andrew~J., ``{Resummation of
  Massive Gravity}'', {\em Phys. Rev. Lett.}, {\bf 106}, 231101 (2011).
  {\small[\href{http://dx.doi.org/10.1103/PhysRevLett.106.231101}{DOI}]},
  {\small[\href{http://arxiv.org/abs/1011.1232}{{arXiv:1011.1232
  {\small[hep-th]}}}]}.

\bibitem{deRham:2013qqa}
de~Rham, Claudia, Heisenberg, Lavinia  and Ribeiro, Raquel~H., ``{Quantum
  Corrections in Massive Gravity}'', {\em Phys. Rev.}, {\bf D88}, 084058
  (2013). {\small[\href{http://dx.doi.org/10.1103/PhysRevD.88.084058}{DOI}]},
  {\small[\href{http://arxiv.org/abs/1307.7169}{{arXiv:1307.7169
  {\small[hep-th]}}}]}.

\bibitem{deRham:2014naa}
de~Rham, Claudia, Heisenberg, Lavinia  and Ribeiro, Raquel~H., ``{On couplings
  to matter in massive (bi-)gravity}'', {\em Class. Quant. Grav.}, {\bf 32},
  035022 (2015).
  {\small[\href{http://dx.doi.org/10.1088/0264-9381/32/3/035022}{DOI}]},
  {\small[\href{http://arxiv.org/abs/1408.1678}{{arXiv:1408.1678
  {\small[hep-th]}}}]}.

\bibitem{deRham:2014wfa}
de~Rham, Claudia  and Ribeiro, Raquel~H., ``{Riding on irrelevant operators}'',
  {\em JCAP}, {\bf 1411}(11), 016 (2014).
  {\small[\href{http://dx.doi.org/10.1088/1475-7516/2014/11/016}{DOI}]},
  {\small[\href{http://arxiv.org/abs/1405.5213}{{arXiv:1405.5213
  {\small[hep-th]}}}]}.

\bibitem{Delsate:2012ky}
Delsate, Terence  and Steinhoff, Jan, ``{New insights on the matter-gravity
  coupling paradigm}'', {\em Phys. Rev. Lett.}, {\bf 109}, 021101 (2012).
  {\small[\href{http://dx.doi.org/10.1103/PhysRevLett.109.021101}{DOI}]},
  {\small[\href{http://arxiv.org/abs/1201.4989}{{arXiv:1201.4989
  {\small[gr-qc]}}}]}.

\bibitem{Demianski:1986wx}
Demianski, M., ``{Static electromagnetic geon}'', {\em Found. Phys.}, {\bf 16},
  187--190 (1986). {\small[\href{http://dx.doi.org/10.1007/BF01889380}{DOI}]}.

\bibitem{Deser:1998rj}
Deser, Stanley  and Gibbons, G.~W., ``{Born-Infeld-Einstein actions?}'', {\em
  Class. Quant. Grav.}, {\bf 15}, L35--L39 (1998).
  {\small[\href{http://dx.doi.org/10.1088/0264-9381/15/5/001}{DOI}]},
  {\small[\href{http://arxiv.org/abs/hep-th/9803049}{{arXiv:hep-th/9803049
  {\small[hep-th]}}}]}.

\bibitem{Dias:2010uh}
Dias, Goncalo A.~S.  and Lemos, Jose P.~S., ``{Thin-shell wormholes in
  $d$-dimensional general relativity: Solutions, properties, and stability}'',
  {\em Phys. Rev.}, {\bf D82}, 084023 (2010).
  {\small[\href{http://dx.doi.org/10.1103/PhysRevD.82.084023}{DOI}]},
  {\small[\href{http://arxiv.org/abs/1008.3376}{{arXiv:1008.3376
  {\small[gr-qc]}}}]}.

\bibitem{DiazAlonso:2009ak}
Diaz-Alonso, J.  and Rubiera-Garcia, D., ``{Electrostatic spherically symmetric
  configurations in gravitating nonlinear electrodynamics}'', {\em Phys. Rev.},
  {\bf D81}, 064021 (2010).
  {\small[\href{http://dx.doi.org/10.1103/PhysRevD.81.064021}{DOI}]},
  {\small[\href{http://arxiv.org/abs/0908.3303}{{arXiv:0908.3303
  {\small[hep-th]}}}]}.

\bibitem{Douchin:2000kx}
Douchin, F.  and Haensel, P., ``{Inner edge of neutron star crust with SLY
  effective nucleon-nucleon interactions}'', {\em Phys. Lett.}, {\bf B485},
  107--114 (2000).
  {\small[\href{http://dx.doi.org/10.1016/S0370-2693(00)00672-9}{DOI}]},
  {\small[\href{http://arxiv.org/abs/astro-ph/0006135}{{arXiv:astro-ph/0006135
  {\small[astro-ph]}}}]}.

\bibitem{Dowker:1993bt}
Dowker, Fay, Gauntlett, Jerome~P., Kastor, David~A.  and Traschen, Jennie~H.,
  ``{Pair creation of dilaton black holes}'', {\em Phys. Rev.}, {\bf D49},
  2909--2917 (1994).
  {\small[\href{http://dx.doi.org/10.1103/PhysRevD.49.2909}{DOI}]},
  {\small[\href{http://arxiv.org/abs/hep-th/9309075}{{arXiv:hep-th/9309075
  {\small[hep-th]}}}]}.

\bibitem{Du:2014jka}
Du, Xiao-Long, Yang, Ke, Meng, Xin-He  and Liu, Yu-Xiao, ``{Large Scale
  Structure Formation in Eddington-inspired Born-Infeld Gravity}'', {\em Phys.
  Rev.}, {\bf D90}, 044054 (2014).
  {\small[\href{http://dx.doi.org/10.1103/PhysRevD.90.044054}{DOI}]},
  {\small[\href{http://arxiv.org/abs/1403.0083}{{arXiv:1403.0083
  {\small[gr-qc]}}}]}.

\bibitem{Dziembowski:1998nb}
Dziembowski, W.~A., Fiorentini, G., Ricci, B.  and Sienkiewicz, R.,
  ``{Helioseismology and the solar age}'', {\em Astron. Astrophys.}, {\bf 343},
  990 (1999).
  {\small[\href{http://arxiv.org/abs/astro-ph/9809361}{{arXiv:astro-ph/9809361
  {\small[astro-ph]}}}]}.

\bibitem{eddington1924mathematical}
Eddington, Arthur~Stanley, {\em The Mathematical Theory of Relativity},
  (Cambridge University Press, 1924)URL:
  \newline\url{https://books.google.fr/books?id=CqYrAAAAYAAJ}.

\bibitem{EGUCHI1980213}
Eguchi, Tohru, Gilkey, Peter~B.  and Hanson, Andrew~J., ``Gravitation, gauge
  theories and differential geometry'', {\em Physics Reports}, {\bf 66}(6), 213
  -- 393 (1980).
  {\small[\href{http://dx.doi.org/http://dx.doi.org/10.1016/0370-1573(80)90130-1}{DOI}]}URL:
  \newline\url{http://www.sciencedirect.com/science/article/pii/0370157380901301}.

\bibitem{Einstein:1935rr}
Einstein, Albert, Podolsky, Boris  and Rosen, Nathan, ``{Can quantum mechanical
  description of physical reality be considered complete?}'', {\em Phys. Rev.},
  {\bf 47}, 777--780 (1935).
  {\small[\href{http://dx.doi.org/10.1103/PhysRev.47.777}{DOI}]}.

\bibitem{Eisenstein:2005su}
Eisenstein, Daniel~J.  {et~al.} (SDSS), ``{Detection of the Baryon Acoustic
  Peak in the Large-Scale Correlation Function of SDSS Luminous Red
  Galaxies}'', {\em Astrophys. J.}, {\bf 633}, 560--574 (2005).
  {\small[\href{http://dx.doi.org/10.1086/466512}{DOI}]},
  {\small[\href{http://arxiv.org/abs/astro-ph/0501171}{{arXiv:astro-ph/0501171
  {\small[astro-ph]}}}]}.

\bibitem{Elizalde:2016vsd}
Elizalde, Emilio  and Makarenko, Andrey~N., ``{Singular inflation from
  Born--Infeld-f (R) gravity}'', {\em Mod. Phys. Lett.}, {\bf A31}(24), 1650149
  (2016). {\small[\href{http://dx.doi.org/10.1142/S0217732316501492}{DOI}]},
  {\small[\href{http://arxiv.org/abs/1606.05211}{{arXiv:1606.05211
  {\small[gr-qc]}}}]}.

\bibitem{Ellis:1977pj}
Ellis, G. F.~R.  and Schmidt, B.~G., ``{Singular space-times}'', {\em Gen. Rel.
  Grav.}, {\bf 8}, 915--953 (1977).
  {\small[\href{http://dx.doi.org/10.1007/BF00759240}{DOI}]}.

\bibitem{Emparan:1995je}
Emparan, R., ``{Pair creation of black holes joined by cosmic strings}'', {\em
  Phys. Rev. Lett.}, {\bf 75}, 3386--3389 (1995).
  {\small[\href{http://dx.doi.org/10.1103/PhysRevLett.75.3386}{DOI}]},
  {\small[\href{http://arxiv.org/abs/gr-qc/9506025}{{arXiv:gr-qc/9506025
  {\small[gr-qc]}}}]}.

\bibitem{Erbin:2016lzq}
Erbin, Harold, ``{Janis-Newman algorithm: generating rotating and NUT charged
  black holes}'' (2016).
  {\small[\href{http://arxiv.org/abs/1701.00037}{{arXiv:1701.00037
  {\small[gr-qc]}}}]}.

\bibitem{EscamillaRivera:2012vz}
Escamilla-Rivera, Celia, Banados, Maximo  and Ferreira, Pedro~G., ``{A tensor
  instability in the Eddington inspired Born-Infeld Theory of Gravity}'', {\em
  Phys. Rev.}, {\bf D85}, 087302 (2012).
  {\small[\href{http://dx.doi.org/10.1103/PhysRevD.85.087302}{DOI}]},
  {\small[\href{http://arxiv.org/abs/1204.1691}{{arXiv:1204.1691
  {\small[gr-qc]}}}]}.

\bibitem{Exirifard:2007da}
Exirifard, Q.  and Sheikh-Jabbari, M.~M., ``{Lovelock gravity at the crossroads
  of Palatini and metric formulations}'', {\em Phys. Lett.}, {\bf B661},
  158--161 (2008).
  {\small[\href{http://dx.doi.org/10.1016/j.physletb.2008.02.012}{DOI}]},
  {\small[\href{http://arxiv.org/abs/0705.1879}{{arXiv:0705.1879
  {\small[hep-th]}}}]}.

\bibitem{Feigenbaum:1998wy}
Feigenbaum, James~A., ``{Born regulated gravity in four-dimensions}'', {\em
  Phys. Rev.}, {\bf D58}, 124023 (1998).
  {\small[\href{http://dx.doi.org/10.1103/PhysRevD.58.124023}{DOI}]},
  {\small[\href{http://arxiv.org/abs/hep-th/9807114}{{arXiv:hep-th/9807114
  {\small[hep-th]}}}]}.

\bibitem{Feigenbaum:1997pf}
Feigenbaum, James~A., Freund, Peter G.~O.  and Pigli, Mircea, ``{Gravitational
  analogs of nonlinear Born electrodynamics}'', {\em Phys. Rev.}, {\bf D57},
  4738--4744 (1998).
  {\small[\href{http://dx.doi.org/10.1103/PhysRevD.57.4738}{DOI}]},
  {\small[\href{http://arxiv.org/abs/hep-th/9709196}{{arXiv:hep-th/9709196
  {\small[hep-th]}}}]}.

\bibitem{Fernandes:2014bka}
Fernandes, Karan  and Lahiri, Amitabha, ``{Kaluza Ansatz applied to Eddington
  inspired Born-Infeld Gravity}'', {\em Phys. Rev.}, {\bf D91}(4), 044014
  (2015). {\small[\href{http://dx.doi.org/10.1103/PhysRevD.91.044014}{DOI}]},
  {\small[\href{http://arxiv.org/abs/1405.2172}{{arXiv:1405.2172
  {\small[gr-qc]}}}]}.

\bibitem{Fernandez-Jambrina:2014sga}
Fern{\'a}ndez-Jambrina, L., ``{Grand Rip and Grand Bang/Crunch cosmological
  singularities}'', {\em Phys. Rev.}, {\bf D90}, 064014 (2014).
  {\small[\href{http://dx.doi.org/10.1103/PhysRevD.90.064014}{DOI}]},
  {\small[\href{http://arxiv.org/abs/1408.6997}{{arXiv:1408.6997
  {\small[gr-qc]}}}]}.

\bibitem{FernandezJambrina:2006hj}
Fernandez-Jambrina, L.  and Lazkoz, R., ``{Classification of cosmological
  milestones}'', {\em Phys. Rev.}, {\bf D74}, 064030 (2006).
  {\small[\href{http://dx.doi.org/10.1103/PhysRevD.74.064030}{DOI}]},
  {\small[\href{http://arxiv.org/abs/gr-qc/0607073}{{arXiv:gr-qc/0607073
  {\small[gr-qc]}}}]}.

\bibitem{Fernando:2003tz}
Fernando, Sharmanthie  and Krug, Don, ``{Charged black hole solutions in
  Einstein-Born-Infeld gravity with a cosmological constant}'', {\em Gen. Rel.
  Grav.}, {\bf 35}, 129--137 (2003).
  {\small[\href{http://dx.doi.org/10.1023/A:1021315214180}{DOI}]},
  {\small[\href{http://arxiv.org/abs/hep-th/0306120}{{arXiv:hep-th/0306120
  {\small[hep-th]}}}]}.

\bibitem{Ferraris1982}
Ferraris, M., Francaviglia, M.  and Reina, C., ``Variational formulation of
  general relativity from 1915 to 1925 ``Palatini's method'' discovered by
  Einstein in 1925'', {\em General Relativity and Gravitation}, {\bf 14}(3),
  243--254 (1982).
  {\small[\href{http://dx.doi.org/10.1007/BF00756060}{DOI}]}URL:
  \newline\url{http://dx.doi.org/10.1007/BF00756060}.

\bibitem{Ferraris:1992dx}
Ferraris, Marco, Francaviglia, Mauro  and Volovich, Igor, ``{The Universality
  of vacuum Einstein equations with cosmological constant}'', {\em Class.
  Quant. Grav.}, {\bf 11}, 1505--1517 (1994).
  {\small[\href{http://dx.doi.org/10.1088/0264-9381/11/6/015}{DOI}]},
  {\small[\href{http://arxiv.org/abs/gr-qc/9303007}{{arXiv:gr-qc/9303007
  {\small[gr-qc]}}}]}.

\bibitem{Ferraro:2006jd}
Ferraro, Rafael  and Fiorini, Franco, ``{Modified teleparallel gravity:
  Inflation without inflaton}'', {\em Phys. Rev.}, {\bf D75}, 084031 (2007).
  {\small[\href{http://dx.doi.org/10.1103/PhysRevD.75.084031}{DOI}]},
  {\small[\href{http://arxiv.org/abs/gr-qc/0610067}{{arXiv:gr-qc/0610067
  {\small[gr-qc]}}}]}.

\bibitem{Ferraro:2009zk}
Ferraro, Rafael  and Fiorini, Franco, ``{Born-Infeld Determinantal gravity and
  the taming of the conical singularity in 3-dimensional spacetime}'', {\em
  Phys. Lett.}, {\bf B692}, 206--211 (2010).
  {\small[\href{http://dx.doi.org/10.1016/j.physletb.2010.07.040}{DOI}]},
  {\small[\href{http://arxiv.org/abs/0910.4693}{{arXiv:0910.4693
  {\small[gr-qc]}}}]}.

\bibitem{Fiorini:2013kba}
Fiorini, Franco, ``{Nonsingular Promises from Born-Infeld Gravity}'', {\em
  Phys. Rev. Lett.}, {\bf 111}, 041104 (2013).
  {\small[\href{http://dx.doi.org/10.1103/PhysRevLett.111.041104}{DOI}]},
  {\small[\href{http://arxiv.org/abs/1306.4392}{{arXiv:1306.4392
  {\small[gr-qc]}}}]}.

\bibitem{Fiorini:2015hob}
Fiorini, Franco, ``{Primordial brusque bounce in Born-Infeld determinantal
  gravity}'', {\em Phys. Rev.}, {\bf D94}(2), 024030 (2016).
  {\small[\href{http://dx.doi.org/10.1103/PhysRevD.94.024030}{DOI}]},
  {\small[\href{http://arxiv.org/abs/1511.03227}{{arXiv:1511.03227
  {\small[gr-qc]}}}]}.

\bibitem{Fiorini:2009ux}
Fiorini, Franco  and Ferraro, Rafael, ``{A Type of Born-Infeld regular gravity
  and its cosmological consequences}'', {\em Int. J. Mod. Phys.}, {\bf A24},
  1686--1689 (2009).
  {\small[\href{http://dx.doi.org/10.1142/S0217751X09045236}{DOI}]},
  {\small[\href{http://arxiv.org/abs/0904.1767}{{arXiv:0904.1767
  {\small[gr-qc]}}}]}.

\bibitem{Fiorini:2016zrt}
Fiorini, Franco  and Vattuone, Nicolas, ``{An analysis of Born–Infeld
  determinantal gravity in Weitzenböck spacetime}'', {\em Phys. Lett.}, {\bf
  B763}, 45--51 (2016).
  {\small[\href{http://dx.doi.org/10.1016/j.physletb.2016.10.016}{DOI}]},
  {\small[\href{http://arxiv.org/abs/1608.02622}{{arXiv:1608.02622
  {\small[hep-th]}}}]}.

\bibitem{PhysRevD.93.093020}
Fouch\'e, M., Battesti, R.  and Rizzo, C., ``Limits on nonlinear
  electrodynamics'', {\em Phys. Rev. D}, {\bf 93}, 093020 (May 2016).
  {\small[\href{http://dx.doi.org/10.1103/PhysRevD.93.093020}{DOI}]}URL:
  \newline\url{https://link.aps.org/doi/10.1103/PhysRevD.93.093020}.

\bibitem{Fu:2014raa}
Fu, Qi-Ming, Zhao, Li, Yang, Ke, Gu, Bao-Min  and Liu, Yu-Xiao, ``{Stability
  and (quasi)localization of gravitational fluctuations in an
  Eddington-inspired Born-Infeld brane system}'', {\em Phys. Rev.}, {\bf
  D90}(10), 104007 (2014).
  {\small[\href{http://dx.doi.org/10.1103/PhysRevD.90.104007}{DOI}]},
  {\small[\href{http://arxiv.org/abs/1407.6107}{{arXiv:1407.6107
  {\small[hep-th]}}}]}.

\bibitem{Garcia:2011aa}
Garcia, Nadiezhda~Montelongo, Lobo, Francisco S.~N.  and Visser, Matt,
  ``{Generic spherically symmetric dynamic thin-shell traversable wormholes in
  standard general relativity}'', {\em Phys. Rev.}, {\bf D86}, 044026 (2012).
  {\small[\href{http://dx.doi.org/10.1103/PhysRevD.86.044026}{DOI}]},
  {\small[\href{http://arxiv.org/abs/1112.2057}{{arXiv:1112.2057
  {\small[gr-qc]}}}]}.

\bibitem{Garfinkle:1993xk}
Garfinkle, David, Giddings, Steven~B.  and Strominger, Andrew, ``{Entropy in
  black hole pair production}'', {\em Phys. Rev.}, {\bf D49}, 958--965 (1994).
  {\small[\href{http://dx.doi.org/10.1103/PhysRevD.49.958}{DOI}]},
  {\small[\href{http://arxiv.org/abs/gr-qc/9306023}{{arXiv:gr-qc/9306023
  {\small[gr-qc]}}}]}.

\bibitem{Garfinkle:1990eq}
Garfinkle, David  and Strominger, Andrew, ``{Semiclassical Wheeler wormhole
  production}'', {\em Phys. Lett.}, {\bf B256}, 146--149 (1991).
  {\small[\href{http://dx.doi.org/10.1016/0370-2693(91)90665-D}{DOI}]}.

\bibitem{Genzel:2010zy}
Genzel, Reinhard, Eisenhauer, Frank  and Gillessen, Stefan, ``{The Galactic
  Center Massive Black Hole and Nuclear Star Cluster}'', {\em Rev. Mod. Phys.},
  {\bf 82}, 3121--3195 (2010).
  {\small[\href{http://dx.doi.org/10.1103/RevModPhys.82.3121}{DOI}]},
  {\small[\href{http://arxiv.org/abs/1006.0064}{{arXiv:1006.0064
  {\small[astro-ph.GA]}}}]}.

\bibitem{Geroch:1968ut}
Geroch, Robert~P., ``{What is a singularity in general relativity?}'', {\em
  Annals Phys.}, {\bf 48}, 526--540 (1968).
  {\small[\href{http://dx.doi.org/10.1016/0003-4916(68)90144-9}{DOI}]}.

\bibitem{Ghodsi:2010ev}
Ghodsi, Ahmad  and Yekta, Davood~Mahdavian, ``{Black Holes in Born-Infeld
  Extended New Massive Gravity}'', {\em Phys. Rev.}, {\bf D83}, 104004 (2011).
  {\small[\href{http://dx.doi.org/10.1103/PhysRevD.83.104004}{DOI}]},
  {\small[\href{http://arxiv.org/abs/1010.2434}{{arXiv:1010.2434
  {\small[hep-th]}}}]}.

\bibitem{Ghodsi:2011ua}
Ghodsi, Ahmad  and Yekta, Davood~Mahdavian, ``{On Asymptotically AdS-Like
  Solutions of Three Dimensional Massive Gravity}'', {\em JHEP}, {\bf 06}, 131
  (2012). {\small[\href{http://dx.doi.org/10.1007/JHEP06(2012)131}{DOI}]},
  {\small[\href{http://arxiv.org/abs/1112.5402}{{arXiv:1112.5402
  {\small[hep-th]}}}]}.

\bibitem{Gibbons:1997xz}
Gibbons, G.~W., ``{Born-Infeld particles and Dirichlet p-branes}'', {\em Nucl.
  Phys.}, {\bf B514}, 603--639 (1998).
  {\small[\href{http://dx.doi.org/10.1016/S0550-3213(97)00795-5}{DOI}]},
  {\small[\href{http://arxiv.org/abs/hep-th/9709027}{{arXiv:hep-th/9709027
  {\small[hep-th]}}}]}.

\bibitem{Gibbons:2001gy}
Gibbons, G~W, ``{Aspects of Born-Infeld theory and string / M theory}'', {\em
  Rev. Mex. Fis.}, {\bf 49S1}, 19--29 (2003).
  {\small[\href{http://dx.doi.org/10.1063/1.1419338}{DOI}]},
  {\small[\href{http://arxiv.org/abs/hep-th/0106059}{{arXiv:hep-th/0106059
  {\small[hep-th]}}}]}. [AIP Conf. Proc.589,324(2001)].

\bibitem{Gibbons:1995cv}
Gibbons, G.~W.  and Rasheed, D.~A., ``{Electric - magnetic duality rotations in
  nonlinear electrodynamics}'', {\em Nucl. Phys.}, {\bf B454}, 185--206 (1995).
  {\small[\href{http://dx.doi.org/10.1016/0550-3213(95)00409-L}{DOI}]},
  {\small[\href{http://arxiv.org/abs/hep-th/9506035}{{arXiv:hep-th/9506035
  {\small[hep-th]}}}]}.

\bibitem{GlendenningCS}
Glendenning, Norman~K., ``{Compact Stars}'', {\em Astronomy and Astrophysics
  Library} (1997).

\bibitem{Glendenning:2001pe}
Glendenning, N.~K., ``{Phase transitions and crystalline structures in neutron
  star cores}'', {\em Phys. Rept.}, {\bf 342}, 393--447 (2001).
  {\small[\href{http://dx.doi.org/10.1016/S0370-1573(00)00080-6}{DOI}]}.

\bibitem{Gullu:2010pc}
Gullu, Ibrahim, Sisman, Tahsin~Cagri  and Tekin, Bayram, ``{Born-Infeld
  extension of new massive gravity}'', {\em Class. Quant. Grav.}, {\bf 27},
  162001 (2010).
  {\small[\href{http://dx.doi.org/10.1088/0264-9381/27/16/162001}{DOI}]},
  {\small[\href{http://arxiv.org/abs/1003.3935}{{arXiv:1003.3935
  {\small[hep-th]}}}]}.

\bibitem{Gullu:2010wb}
Gullu, Ibrahim, Sisman, Tahsin~Cagri  and Tekin, Bayram, ``{Born-Infeld-Horava
  gravity}'', {\em Phys. Rev.}, {\bf D81}, 104018 (2010).
  {\small[\href{http://dx.doi.org/10.1103/PhysRevD.81.104018}{DOI}]},
  {\small[\href{http://arxiv.org/abs/1004.0611}{{arXiv:1004.0611
  {\small[hep-th]}}}]}.

\bibitem{Gullu:2014gza}
G{\"u}ll{\"u}, Ibrahim, Sisman, Tahsin~Cagri  and Tekin, Bayram, ``{Born-Infeld
  Gravity with a Massless Graviton in Four Dimensions}'', {\em Phys. Rev.},
  {\bf D91}(4), 044007 (2015).
  {\small[\href{http://dx.doi.org/10.1103/PhysRevD.91.044007}{DOI}]},
  {\small[\href{http://arxiv.org/abs/1410.8033}{{arXiv:1410.8033
  {\small[hep-th]}}}]}.

\bibitem{Gullu:2015cha}
G{\"u}ll{\"u}, {\.I}brahim, Sisman, Tahsin~Cagri  and Tekin, Bayram,
  ``{Born-Infeld Gravity with a Unique Vacuum and a Massless Graviton}'', {\em
  Phys. Rev.}, {\bf D92}(10), 104014 (2015).
  {\small[\href{http://dx.doi.org/10.1103/PhysRevD.92.104014}{DOI}]},
  {\small[\href{http://arxiv.org/abs/1510.01184}{{arXiv:1510.01184
  {\small[hep-th]}}}]}.

\bibitem{Hamilton:2008zz}
Hamilton, Andrew J.~S.  and Avelino, Pedro~P., ``{The Physics of the
  relativistic counter-streaming instability that drives mass inflation inside
  black holes}'', {\em Phys. Rept.}, {\bf 495}, 1--32 (2010).
  {\small[\href{http://dx.doi.org/10.1016/j.physrep.2010.06.002}{DOI}]},
  {\small[\href{http://arxiv.org/abs/0811.1926}{{arXiv:0811.1926
  {\small[gr-qc]}}}]}.

\bibitem{Hansen:2005am}
Hansen, Jakob, Khokhlov, Alexei  and Novikov, Igor, ``{Physics of the interior
  of a spherical, charged black hole with a scalar field}'', {\em Phys. Rev.},
  {\bf D71}, 064013 (2005).
  {\small[\href{http://dx.doi.org/10.1103/PhysRevD.71.064013}{DOI}]},
  {\small[\href{http://arxiv.org/abs/gr-qc/0501015}{{arXiv:gr-qc/0501015
  {\small[gr-qc]}}}]}.

\bibitem{Harko:2014nya}
Harko, Tiberiu, Lobo, Francisco S.~N.  and Mak, M.~K., ``{Bianchi type I
  cosmological models in Eddington-inspired Born-Infeld gravity}'', {\em
  Galaxies}, {\bf 2}, 496--519 (2014).
  {\small[\href{http://dx.doi.org/10.3390/galaxies2040496}{DOI}]},
  {\small[\href{http://arxiv.org/abs/1410.5213}{{arXiv:1410.5213
  {\small[gr-qc]}}}]}.

\bibitem{Harko:2013wka}
Harko, Tiberiu, Lobo, Francisco S.~N., Mak, M.~K.  and Sushkov, Sergey~V.,
  ``{Structure of neutron, quark and exotic stars in Eddington-inspired
  Born-Infeld gravity}'', {\em Phys. Rev.}, {\bf D88}, 044032 (2013).
  {\small[\href{http://dx.doi.org/10.1103/PhysRevD.88.044032}{DOI}]},
  {\small[\href{http://arxiv.org/abs/1305.6770}{{arXiv:1305.6770
  {\small[gr-qc]}}}]}.

\bibitem{Harko:2013xma}
Harko, Tiberiu, Lobo, Francisco S.~N., Mak, M.~K.  and Sushkov, Sergey~V.,
  ``{Dark matter density profile and galactic metric in Eddington-inspired
  Born-Infeld gravity}'', {\em Mod. Phys. Lett.}, {\bf A29}(09), 1450049
  (2014). {\small[\href{http://dx.doi.org/10.1142/S0217732314500497}{DOI}]},
  {\small[\href{http://arxiv.org/abs/1305.0820}{{arXiv:1305.0820
  {\small[gr-qc]}}}]}.

\bibitem{Harko:2013aya}
Harko, Tiberiu, Lobo, Francisco S.~N., Mak, M.~K.  and Sushkov, Sergey~V.,
  ``{Wormhole geometries in Eddington-Inspired Born--Infeld gravity}'', {\em
  Mod. Phys. Lett.}, {\bf A30}(35), 1550190 (2015).
  {\small[\href{http://dx.doi.org/10.1142/S0217732315501904}{DOI}]},
  {\small[\href{http://arxiv.org/abs/1307.1883}{{arXiv:1307.1883
  {\small[gr-qc]}}}]}.

\bibitem{Hartle:1967he}
Hartle, James~B., ``{Slowly rotating relativistic stars. 1. Equations of
  structure}'', {\em Astrophys. J.}, {\bf 150}, 1005--1029 (1967).
  {\small[\href{http://dx.doi.org/10.1086/149400}{DOI}]}.

\bibitem{Hassan:2011zd}
Hassan, S.~F.  and Rosen, Rachel~A., ``{Bimetric Gravity from Ghost-free
  Massive Gravity}'', {\em JHEP}, {\bf 02}, 126 (2012).
  {\small[\href{http://dx.doi.org/10.1007/JHEP02(2012)126}{DOI}]},
  {\small[\href{http://arxiv.org/abs/1109.3515}{{arXiv:1109.3515
  {\small[hep-th]}}}]}.

\bibitem{Hawking:1966vg}
Hawking, S.~W., ``{Singularities in the universe}'', {\em Phys. Rev. Lett.},
  {\bf 17}, 444--445 (1966).
  {\small[\href{http://dx.doi.org/10.1103/PhysRevLett.17.444}{DOI}]}.

\bibitem{Hawking:1971vc}
Hawking, S.~W., ``{Black holes in general relativity}'', {\em Commun. Math.
  Phys.}, {\bf 25}, 152--166 (1972).
  {\small[\href{http://dx.doi.org/10.1007/BF01877517}{DOI}]}.

\bibitem{Hawking:1974sw}
Hawking, S.~W., ``{Particle Creation by Black Holes}'', {\em Commun. Math.
  Phys.}, {\bf 43}, 199--220 (1975).
  {\small[\href{http://dx.doi.org/10.1007/BF02345020}{DOI}]}. [,167(1975)].

\bibitem{Hawking:1976ra}
Hawking, S.~W., ``{Breakdown of Predictability in Gravitational Collapse}'',
  {\em Phys. Rev.}, {\bf D14}, 2460--2473 (1976).
  {\small[\href{http://dx.doi.org/10.1103/PhysRevD.14.2460}{DOI}]}.

\bibitem{Hehl1978}
Hehl, Friedrich~W.  and Kerlick, G.~David, ``Metric-affine variational
  principles in general relativity. I. Riemannian space-time'', {\em General
  Relativity and Gravitation}, {\bf 9}(8), 691--710 (1978).
  {\small[\href{http://dx.doi.org/10.1007/BF00760141}{DOI}]}URL:
  \newline\url{http://dx.doi.org/10.1007/BF00760141}.

\bibitem{Hehl:1994ue}
Hehl, Friedrich~W., McCrea, J.~Dermott, Mielke, Eckehard~W.  and Ne'eman,
  Yuval, ``{Metric affine gauge theory of gravity: Field equations, Noether
  identities, world spinors, and breaking of dilation invariance}'', {\em Phys.
  Rept.}, {\bf 258}, 1--171 (1995).
  {\small[\href{http://dx.doi.org/10.1016/0370-1573(94)00111-F}{DOI}]},
  {\small[\href{http://arxiv.org/abs/gr-qc/9402012}{{arXiv:gr-qc/9402012
  {\small[gr-qc]}}}]}.

\bibitem{Heiselberg:1999mq}
Heiselberg, Henning  and Hjorth-Jensen, Morten, ``{Phases of dense matter in
  neutron stars}'', {\em Phys. Rept.}, {\bf 328}, 237--327 (2000).
  {\small[\href{http://dx.doi.org/10.1016/S0370-1573(99)00110-6}{DOI}]},
  {\small[\href{http://arxiv.org/abs/nucl-th/9902033}{{arXiv:nucl-th/9902033
  {\small[nucl-th]}}}]}.

\bibitem{Heisenberg:2014rka}
Heisenberg, Lavinia, ``{Quantum corrections in massive bigravity and new
  effective composite metrics}'', {\em Class. Quant. Grav.}, {\bf 32}(10),
  105011 (2015).
  {\small[\href{http://dx.doi.org/10.1088/0264-9381/32/10/105011}{DOI}]},
  {\small[\href{http://arxiv.org/abs/1410.4239}{{arXiv:1410.4239
  {\small[hep-th]}}}]}.

\bibitem{Herdeiro:2014goa}
Herdeiro, Carlos A.~R.  and Radu, Eugen, ``{Kerr black holes with scalar
  hair}'', {\em Phys. Rev. Lett.}, {\bf 112}, 221101 (2014).
  {\small[\href{http://dx.doi.org/10.1103/PhysRevLett.112.221101}{DOI}]},
  {\small[\href{http://arxiv.org/abs/1403.2757}{{arXiv:1403.2757
  {\small[gr-qc]}}}]}.

\bibitem{Herdeiro:2015waa}
Herdeiro, Carlos A.~R.  and Radu, Eugen, ``{Asymptotically flat black holes
  with scalar hair: a review}'', {\em Int. J. Mod. Phys.}, {\bf D24}(09),
  1542014 (2015).
  {\small[\href{http://dx.doi.org/10.1142/S0218271815420146}{DOI}]},
  {\small[\href{http://arxiv.org/abs/1504.08209}{{arXiv:1504.08209
  {\small[gr-qc]}}}]}.

\bibitem{Hertzberg:2016djj}
Hertzberg, Mark~P., ``{Gravitation, Causality, and Quantum Consistency}''
  (2016). {\small[\href{http://arxiv.org/abs/1610.03065}{{arXiv:1610.03065
  {\small[hep-th]}}}]}.

\bibitem{Hertzberg:2017abn}
Hertzberg, Mark~P.  and Sandora, McCullen, ``{General Relativity from
  Causality}'' (2017).
  {\small[\href{http://arxiv.org/abs/1702.07720}{{arXiv:1702.07720
  {\small[hep-th]}}}]}.

\bibitem{Ho:2006uk}
Ho, Wynn C.~G., Kaplan, David~L., Chang, Philip, van Adelsberg, Matthew  and
  Potekhin, Alexander~Y., ``{Magnetic Hydrogen Atmosphere Models and the
  Neutron Star RX J1856.5-3754}'', {\em Mon. Not. Roy. Astron. Soc.}, {\bf
  375}, 821--830 (2007).
  {\small[\href{http://dx.doi.org/10.1111/j.1365-2966.2006.11376.x}{DOI}]},
  {\small[\href{http://arxiv.org/abs/astro-ph/0612145}{{arXiv:astro-ph/0612145
  {\small[astro-ph]}}}]}.

\bibitem{Horava:2009uw}
Horava, Petr, ``{Quantum Gravity at a Lifshitz Point}'', {\em Phys. Rev.}, {\bf
  D79}, 084008 (2009).
  {\small[\href{http://dx.doi.org/10.1103/PhysRevD.79.084008}{DOI}]},
  {\small[\href{http://arxiv.org/abs/0901.3775}{{arXiv:0901.3775
  {\small[hep-th]}}}]}.

\bibitem{Horava:2009if}
Horava, Petr, ``{Spectral Dimension of the Universe in Quantum Gravity at a
  Lifshitz Point}'', {\em Phys. Rev. Lett.}, {\bf 102}, 161301 (2009).
  {\small[\href{http://dx.doi.org/10.1103/PhysRevLett.102.161301}{DOI}]},
  {\small[\href{http://arxiv.org/abs/0902.3657}{{arXiv:0902.3657
  {\small[hep-th]}}}]}.

\bibitem{Hossenfelder:2009fc}
Hossenfelder, Sabine, Modesto, Leonardo  and Premont-Schwarz, Isabeau, ``{A
  Model for non-singular black hole collapse and evaporation}'', {\em Phys.
  Rev.}, {\bf D81}, 044036 (2010).
  {\small[\href{http://dx.doi.org/10.1103/PhysRevD.81.044036}{DOI}]},
  {\small[\href{http://arxiv.org/abs/0912.1823}{{arXiv:0912.1823
  {\small[gr-qc]}}}]}.

\bibitem{Hulse:1974eb}
Hulse, R.~A.  and Taylor, J.~H., ``{Discovery of a pulsar in a binary
  system}'', {\em Astrophys. J.}, {\bf 195}, L51--L53 (1975).
  {\small[\href{http://dx.doi.org/10.1086/181708}{DOI}]}.

\bibitem{Israel:1967wq}
Israel, Werner, ``{Event horizons in static vacuum space-times}'', {\em Phys.
  Rev.}, {\bf 164}, 1776--1779 (1967).
  {\small[\href{http://dx.doi.org/10.1103/PhysRev.164.1776}{DOI}]}.

\bibitem{Israel:1967za}
Israel, Werner, ``{Event horizons in static electrovac space-times}'', {\em
  Commun. Math. Phys.}, {\bf 8}, 245--260 (1968).
  {\small[\href{http://dx.doi.org/10.1007/BF01645859}{DOI}]}.

\bibitem{Izmailov:2015xsa}
Izmailov, Ramil, Potapov, Alexander~A., Filippov, Alexander~I., Ghosh, Mithun
  and Nandi, Kamal~K., ``{Upper limit on the central density of dark matter in
  the Eddington-inspired Born-Infeld (EiBI) gravity}'', {\em Mod. Phys. Lett.},
  {\bf A30}(11), 1550056 (2015).
  {\small[\href{http://dx.doi.org/10.1142/S021773231550056X}{DOI}]},
  {\small[\href{http://arxiv.org/abs/1506.04023}{{arXiv:1506.04023
  {\small[gr-qc]}}}]}.

\bibitem{Jana:2013fga}
Jana, Soumya  and Kar, Sayan, ``{Three dimensional Eddington-inspired
  Born-Infeld gravity: Solutions}'', {\em Phys. Rev.}, {\bf D88}(2), 024013
  (2013). {\small[\href{http://dx.doi.org/10.1103/PhysRevD.88.024013}{DOI}]},
  {\small[\href{http://arxiv.org/abs/1302.2697}{{arXiv:1302.2697
  {\small[gr-qc]}}}]}.

\bibitem{Jana:2015cha}
Jana, Soumya  and Kar, Sayan, ``{Born-Infeld gravity coupled to Born-Infeld
  electrodynamics}'', {\em Phys. Rev.}, {\bf D92}, 084004 (2015).
  {\small[\href{http://dx.doi.org/10.1103/PhysRevD.92.084004}{DOI}]},
  {\small[\href{http://arxiv.org/abs/1504.05842}{{arXiv:1504.05842
  {\small[gr-qc]}}}]}.

\bibitem{Jana:2016uvq}
Jana, Soumya  and Kar, Sayan, ``{Born-Infeld cosmology with scalar Born-Infeld
  matter}'', {\em Phys. Rev.}, {\bf D94}(6), 064016 (2016).
  {\small[\href{http://dx.doi.org/10.1103/PhysRevD.94.064016}{DOI}]},
  {\small[\href{http://arxiv.org/abs/1605.00820}{{arXiv:1605.00820
  {\small[gr-qc]}}}]}.

\bibitem{Jiang:2015dla}
Jiang, Jiachen, Bambi, Cosimo  and Steiner, James~F., ``{Testing the Kerr
  Nature of Black Hole Candidates using Iron Line Spectra in the CPR
  Framework}'', {\em Astrophys. J.}, {\bf 811}(2), 130 (2015).
  {\small[\href{http://dx.doi.org/10.1088/0004-637X/811/2/130}{DOI}]},
  {\small[\href{http://arxiv.org/abs/1504.01970}{{arXiv:1504.01970
  {\small[gr-qc]}}}]}.

\bibitem{Jiang:2014loa}
Jiang, Jiachen, Bambi, Cosimo  and Steiner, James~F., ``{Using iron line
  reverberation and spectroscopy to distinguish Kerr and non-Kerr black
  holes}'', {\em JCAP}, {\bf 1505}(05), 025 (2015).
  {\small[\href{http://dx.doi.org/10.1088/1475-7516/2015/05/025}{DOI}]},
  {\small[\href{http://arxiv.org/abs/1406.5677}{{arXiv:1406.5677
  {\small[gr-qc]}}}]}.

\bibitem{Joshi}
Joshi, Pankaj, ``{Gravitational Collapse and Space-Time Singularities}'', {\em
  Cambridge University Press, Cambridge, England,} (2007).

\bibitem{Joshibook}
Joshi, Pankaj~S., ``{Gravitational Collapse and Spacetime Singularities}'',
  {\em AIP Press, New York} (2007).
  {\small[\href{http://dx.doi.org/10.1017/CBO9780511536274}{DOI}]}.

\bibitem{Joyce:2014kja}
Joyce, Austin, Jain, Bhuvnesh, Khoury, Justin  and Trodden, Mark, ``{Beyond the
  Cosmological Standard Model}'', {\em Phys. Rept.}, {\bf 568}, 1--98 (2015).
  {\small[\href{http://dx.doi.org/10.1016/j.physrep.2014.12.002}{DOI}]},
  {\small[\href{http://arxiv.org/abs/1407.0059}{{arXiv:1407.0059
  {\small[astro-ph.CO]}}}]}.

\bibitem{Julia:1998ys}
Julia, B.  and Silva, S., ``{Currents and superpotentials in classical gauge
  invariant theories. 1. Local results with applications to perfect fluids and
  general relativity}'', {\em Class. Quant. Grav.}, {\bf 15}, 2173--2215
  (1998). {\small[\href{http://dx.doi.org/10.1088/0264-9381/15/8/006}{DOI}]},
  {\small[\href{http://arxiv.org/abs/gr-qc/9804029}{{arXiv:gr-qc/9804029
  {\small[gr-qc]}}}]}.

\bibitem{2006Msngr12627K}
{Kaper}, L., {van der Meer}, A., {van Kerkwijk}, M.  and {van den Heuvel}, E.,
  ``{Measuring the Masses of Neutron Stars}'', {\em The Messenger}, {\bf 126},
  27--31 (December 2006).
  {\small[\href{http://esoads.eso.org/abs/2006Msngr.126...27K}{ADS}]}.

\bibitem{Kerr:1963ud}
Kerr, Roy~P., ``{Gravitational field of a spinning mass as an example of
  algebraically special metrics}'', {\em Phys. Rev. Lett.}, {\bf 11}, 237--238
  (1963). {\small[\href{http://dx.doi.org/10.1103/PhysRevLett.11.237}{DOI}]}.

\bibitem{Ketov:2001dq}
Ketov, Sergei~V., ``{Many faces of Born-Infeld theory}'', in {\em {7th
  International Wigner Symposium (Wigsym 7) College Park, Maryland, August
  24-29, 2001}}, (2001).
  {\small[\href{http://arxiv.org/abs/hep-th/0108189}{{arXiv:hep-th/0108189
  {\small[hep-th]}}}]}URL:
  \newline\url{http://alice.cern.ch/format/showfull?sysnb=2271075}.

\bibitem{Kim:2013noa}
Kim, Hyeong-Chan, ``{Origin of the universe: A hint from Eddington-inspired
  Born-Infeld gravity}'', {\em J. Korean Phys. Soc.}, {\bf 65}(6), 840--845
  (2014). {\small[\href{http://dx.doi.org/10.3938/jkps.65.840}{DOI}]},
  {\small[\href{http://arxiv.org/abs/1312.0703}{{arXiv:1312.0703
  {\small[gr-qc]}}}]}.

\bibitem{Kim:2013nna}
Kim, Hyeong-Chan, ``{Physics at the surface of a star in Eddington-inspired
  Born-Infeld gravity}'', {\em Phys. Rev.}, {\bf D89}(6), 064001 (2014).
  {\small[\href{http://dx.doi.org/10.1103/PhysRevD.89.064001}{DOI}]},
  {\small[\href{http://arxiv.org/abs/1312.0705}{{arXiv:1312.0705
  {\small[gr-qc]}}}]}.

\bibitem{Koester:2008sh}
Koester, Detlev, ``{White Dwarf Spectra and Atmosphere Models}'', {\em Mem.
  S.A.It.}, {\bf 81}, 921 (2010).
  {\small[\href{http://arxiv.org/abs/0812.0482}{{arXiv:0812.0482
  {\small[astro-ph]}}}]}.

\bibitem{Koivisto:2013kwa}
Koivisto, Tomi~S.  and Tamanini, Nicola, ``{Ghosts in pure and hybrid
  formalisms of gravity theories: A unified analysis}'', {\em Phys. Rev.}, {\bf
  D87}(10), 104030 (2013).
  {\small[\href{http://dx.doi.org/10.1103/PhysRevD.87.104030}{DOI}]},
  {\small[\href{http://arxiv.org/abs/1304.3607}{{arXiv:1304.3607
  {\small[gr-qc]}}}]}.

\bibitem{Kokkotas:1999bd}
Kokkotas, Kostas~D.  and Schmidt, Bernd~G., ``{Quasinormal modes of stars and
  black holes}'', {\em Living Rev. Rel.}, {\bf 2}, 2 (1999).
  {\small[\href{http://dx.doi.org/10.12942/lrr-1999-2}{DOI}]},
  {\small[\href{http://arxiv.org/abs/gr-qc/9909058}{{arXiv:gr-qc/9909058
  {\small[gr-qc]}}}]}.

\bibitem{Kouveliotou:1998ze}
Kouveliotou, C.  {et~al.}, ``{An X-ray pulsar with a superstrong magnetic field
  in the soft gamma-ray repeater SGR 1806-20.}'', {\em Nature}, {\bf 393},
  235--237 (1998). {\small[\href{http://dx.doi.org/10.1038/30410}{DOI}]}.

\bibitem{Kruglov:2012ja}
Kruglov, S.~I., ``{Born-Infeld-like modified gravity}'', {\em Int. J. Theor.
  Phys.}, {\bf 52}, 2477--2484 (2013).
  {\small[\href{http://dx.doi.org/10.1007/s10773-013-1535-1}{DOI}]},
  {\small[\href{http://arxiv.org/abs/1202.4807}{{arXiv:1202.4807
  {\small[gr-qc]}}}]}.

\bibitem{Kruglov:2014gva}
Kruglov, S.~I., ``{Notes on Born–Infeld-like modified gravity}'', {\em
  Astrophys. Space Sci.}, {\bf 361}(2), 73 (2016).
  {\small[\href{http://dx.doi.org/10.1007/s10509-016-2665-8}{DOI}]},
  {\small[\href{http://arxiv.org/abs/1403.0675}{{arXiv:1403.0675
  {\small[gr-qc]}}}]}.

\bibitem{Lagos:2013aua}
Lagos, Macarena, Ba{\~n}ados, M{\'a}ximo, Ferreira, Pedro~G.  and
  Garc{\'\i}a-S{\'a}enz, Sebasti{\'a}n, ``{Noether Identities and Gauge-Fixing
  the Action for Cosmological Perturbations}'', {\em Phys. Rev.}, {\bf D89},
  024034 (2014).
  {\small[\href{http://dx.doi.org/10.1103/PhysRevD.89.024034}{DOI}]},
  {\small[\href{http://arxiv.org/abs/1311.3828}{{arXiv:1311.3828
  {\small[gr-qc]}}}]}.

\bibitem{Lattimer:2006xb}
Lattimer, James~M.  and Prakash, Maddapa, ``{Neutron Star Observations:
  Prognosis for Equation of State Constraints}'', {\em Phys. Rept.}, {\bf 442},
  109--165 (2007).
  {\small[\href{http://dx.doi.org/10.1016/j.physrep.2007.02.003}{DOI}]},
  {\small[\href{http://arxiv.org/abs/astro-ph/0612440}{{arXiv:astro-ph/0612440
  {\small[astro-ph]}}}]}.

\bibitem{Lattimer:2004nj}
Lattimer, James~M.  and Schutz, Bernard~F., ``{Constraining the equation of
  state with moment of inertia measurements}'', {\em Astrophys. J.}, {\bf 629},
  979--984 (2005). {\small[\href{http://dx.doi.org/10.1086/431543}{DOI}]},
  {\small[\href{http://arxiv.org/abs/astro-ph/0411470}{{arXiv:astro-ph/0411470
  {\small[astro-ph]}}}]}.

\bibitem{Lau:2009bu}
Lau, H.~K., Leung, P.~T.  and Lin, L.~M., ``{Inferring physical parameters of
  compact stars from their f-mode gravitational wave signals}'', {\em
  Astrophys. J.}, {\bf 714}, 1234--1238 (2010).
  {\small[\href{http://dx.doi.org/10.1088/0004-637X/714/2/1234}{DOI}]},
  {\small[\href{http://arxiv.org/abs/0911.0131}{{arXiv:0911.0131
  {\small[gr-qc]}}}]}.

\bibitem{Lemos:2011dq}
Lemos, Jose P.~S.  and Zanchin, Vilson~T., ``{Regular black holes: Electrically
  charged solutions, Reissner-Nordstr\'om outside a de Sitter core}'', {\em
  Phys. Rev.}, {\bf D83}, 124005 (2011).
  {\small[\href{http://dx.doi.org/10.1103/PhysRevD.83.124005}{DOI}]},
  {\small[\href{http://arxiv.org/abs/1104.4790}{{arXiv:1104.4790
  {\small[gr-qc]}}}]}.

\bibitem{Liu:2012rc}
Liu, Yu-Xiao, Yang, Ke, Guo, Heng  and Zhong, Yuan, ``{Domain Wall Brane in
  Eddington Inspired Born-Infeld Gravity}'', {\em Phys. Rev.}, {\bf D85},
  124053 (2012).
  {\small[\href{http://dx.doi.org/10.1103/PhysRevD.85.124053}{DOI}]},
  {\small[\href{http://arxiv.org/abs/1203.2349}{{arXiv:1203.2349
  {\small[hep-th]}}}]}.

\bibitem{Lorenz:1992zz}
Lorenz, C.~P., Ravenhall, D.~G.  and Pethick, C.~J., ``{Neutron star crusts}'',
  {\em Phys. Rev. Lett.}, {\bf 70}, 379--382 (1993).
  {\small[\href{http://dx.doi.org/10.1103/PhysRevLett.70.379}{DOI}]}.

\bibitem{Makarenko:2014fla}
Makarenko, Andrey~N., ``{The unification of the inflation with late-time
  acceleration in Born-Infeld-$f(R)$ gravity}'', {\em Astrophys. Space Sci.},
  {\bf 352}, 921--924 (2014).
  {\small[\href{http://dx.doi.org/10.1007/s10509-014-1955-2}{DOI}]},
  {\small[\href{http://arxiv.org/abs/1406.1705}{{arXiv:1406.1705
  {\small[gr-qc]}}}]}.

\bibitem{Makarenko:2014lxa}
Makarenko, Andrey~N., Odintsov, Sergei  and Olmo, Gonzalo~J.,
  ``{Born-Infeld-$f(R)$ gravity}'', {\em Phys. Rev.}, {\bf D90}, 024066 (2014).
  {\small[\href{http://dx.doi.org/10.1103/PhysRevD.90.024066}{DOI}]},
  {\small[\href{http://arxiv.org/abs/1403.7409}{{arXiv:1403.7409
  {\small[hep-th]}}}]}.

\bibitem{Makarenko:2014nca}
Makarenko, Andrey~N., Odintsov, Sergei~D.  and Olmo, Gonzalo~J., ``{Little Rip,
  $\Lambda$CDM and singular dark energy cosmology from Born-Infeld-$f(R)$
  gravity}'', {\em Phys. Lett.}, {\bf B734}, 36--40 (2014).
  {\small[\href{http://dx.doi.org/10.1016/j.physletb.2014.05.024}{DOI}]},
  {\small[\href{http://arxiv.org/abs/1404.2850}{{arXiv:1404.2850
  {\small[gr-qc]}}}]}.

\bibitem{Malafarina:2017csn}
Malafarina, Daniele, ``{Classical collapse to black holes and white hole
  quantum bounces: A review}'' (2017).
  {\small[\href{http://arxiv.org/abs/1703.04138}{{arXiv:1703.04138
  {\small[gr-qc]}}}]}.

\bibitem{Marolf:2017jkr}
Marolf, Donald, ``{The Black Hole information problem: past, present, and
  future}'' (2017).
  {\small[\href{http://arxiv.org/abs/1703.02143}{{arXiv:1703.02143
  {\small[gr-qc]}}}]}.

\bibitem{Martin:2012bt}
Martin, Jerome, ``{Everything You Always Wanted To Know About The Cosmological
  Constant Problem (But Were Afraid To Ask)}'', {\em Comptes Rendus Physique},
  {\bf 13}, 566--665 (2012).
  {\small[\href{http://dx.doi.org/10.1016/j.crhy.2012.04.008}{DOI}]},
  {\small[\href{http://arxiv.org/abs/1205.3365}{{arXiv:1205.3365
  {\small[astro-ph.CO]}}}]}.

\bibitem{Melvin:1963qx}
Melvin, M.~A., ``{Pure magnetic and electric geons}'', {\em Phys. Lett.}, {\bf
  8}, 65--70 (1964).
  {\small[\href{http://dx.doi.org/10.1016/0031-9163(64)90801-7}{DOI}]}.

\bibitem{misner1973gravitation}
Misner, C.W., Thorne, K.S.  and Wheeler, J.A., {\em Gravitation}, number parte
  3 in Gravitation,  (W. H. Freeman, 1973)URL:
  \newline\url{https://books.google.fr/books?id=w4Gigq3tY1kC}.

\bibitem{Misner:1964je}
Misner, Charles~W.  and Sharp, David~H., ``{Relativistic equations for
  adiabatic, spherically symmetric gravitational collapse}'', {\em Phys. Rev.},
  {\bf 136}, B571--B576 (1964).
  {\small[\href{http://dx.doi.org/10.1103/PhysRev.136.B571}{DOI}]}.

\bibitem{Misner:1957mt}
Misner, Charles~W.  and Wheeler, John~A., ``{Classical physics as geometry:
  Gravitation, electromagnetism, unquantized charge, and mass as properties of
  curved empty space}'', {\em Annals Phys.}, {\bf 2}, 525--603 (1957).
  {\small[\href{http://dx.doi.org/10.1016/0003-4916(57)90049-0}{DOI}]}.

\bibitem{Modesto:2010uh}
Modesto, Leonardo, Moffat, John~W.  and Nicolini, Piero, ``{Black holes in an
  ultraviolet complete quantum gravity}'', {\em Phys. Lett.}, {\bf B695},
  397--400 (2011).
  {\small[\href{http://dx.doi.org/10.1016/j.physletb.2010.11.046}{DOI}]},
  {\small[\href{http://arxiv.org/abs/1010.0680}{{arXiv:1010.0680
  {\small[gr-qc]}}}]}.

\bibitem{Moffat79}
Moffat, J.~W., ``New theory of gravitation'', {\em Phys. Rev. D}, {\bf 19},
  3554--3558 (Jun 1979).
  {\small[\href{http://dx.doi.org/10.1103/PhysRevD.19.3554}{DOI}]}URL:
  \newline\url{http://link.aps.org/doi/10.1103/PhysRevD.19.3554}.

\bibitem{Morris:1988cz}
Morris, M.~S.  and Thorne, K.~S., ``{Wormholes in space-time and their use for
  interstellar travel: A tool for teaching general relativity}'', {\em Am. J.
  Phys.}, {\bf 56}, 395--412 (1988).
  {\small[\href{http://dx.doi.org/10.1119/1.15620}{DOI}]}.

\bibitem{Musgrave:1995ka}
Musgrave, Peter  and Lake, Kayll, ``{Junctions and thin shells in general
  relativity using computer algebra. 1: The Darmois-Israel formalism}'', {\em
  Class. Quant. Grav.}, {\bf 13}, 1885--1900 (1996).
  {\small[\href{http://dx.doi.org/10.1088/0264-9381/13/7/018}{DOI}]},
  {\small[\href{http://arxiv.org/abs/gr-qc/9510052}{{arXiv:gr-qc/9510052
  {\small[gr-qc]}}}]}.

\bibitem{Newman:1965my}
Newman, E~T., Couch, R., Chinnapared, K., Exton, A., Prakash, A.  and Torrence,
  R., ``{Metric of a Rotating, Charged Mass}'', {\em J. Math. Phys.}, {\bf 6},
  918--919 (1965). {\small[\href{http://dx.doi.org/10.1063/1.1704351}{DOI}]}.

\bibitem{Nicolini:2005vd}
Nicolini, Piero, Smailagic, Anais  and Spallucci, Euro, ``{Noncommutative
  geometry inspired Schwarzschild black hole}'', {\em Phys. Lett.}, {\bf B632},
  547--551 (2006).
  {\small[\href{http://dx.doi.org/10.1016/j.physletb.2005.11.004}{DOI}]},
  {\small[\href{http://arxiv.org/abs/gr-qc/0510112}{{arXiv:gr-qc/0510112
  {\small[gr-qc]}}}]}.

\bibitem{Nieto:2004qj}
Nieto, J.~A., ``{Born-Infeld gravity in any dimension}'', {\em Phys. Rev.},
  {\bf D70}, 044042 (2004).
  {\small[\href{http://dx.doi.org/10.1103/PhysRevD.70.044042}{DOI}]},
  {\small[\href{http://arxiv.org/abs/hep-th/0402071}{{arXiv:hep-th/0402071
  {\small[hep-th]}}}]}.

\bibitem{Nojiri:2010wj}
Nojiri, Shin'ichi  and Odintsov, Sergei~D., ``{Unified cosmic history in
  modified gravity: from F(R) theory to Lorentz non-invariant models}'', {\em
  Phys. Rept.}, {\bf 505}, 59--144 (2011).
  {\small[\href{http://dx.doi.org/10.1016/j.physrep.2011.04.001}{DOI}]},
  {\small[\href{http://arxiv.org/abs/1011.0544}{{arXiv:1011.0544
  {\small[gr-qc]}}}]}.

\bibitem{Nojiri:2005sx}
Nojiri, Shin'ichi, Odintsov, Sergei~D.  and Tsujikawa, Shinji, ``{Properties of
  singularities in (phantom) dark energy universe}'', {\em Phys. Rev.}, {\bf
  D71}, 063004 (2005).
  {\small[\href{http://dx.doi.org/10.1103/PhysRevD.71.063004}{DOI}]},
  {\small[\href{http://arxiv.org/abs/hep-th/0501025}{{arXiv:hep-th/0501025
  {\small[hep-th]}}}]}.

\bibitem{Nolan:1999tw}
Nolan, Brien~C., ``{Strengths of singularities in spherical symmetry}'', {\em
  Phys. Rev.}, {\bf D60}, 024014 (1999).
  {\small[\href{http://dx.doi.org/10.1103/PhysRevD.60.024014}{DOI}]},
  {\small[\href{http://arxiv.org/abs/gr-qc/9902021}{{arXiv:gr-qc/9902021
  {\small[gr-qc]}}}]}.

\bibitem{Nolan:2000rn}
Nolan, Brien~C., ``{The Central singularity in spherical collapse}'', {\em
  Phys. Rev.}, {\bf D62}, 044015 (2000).
  {\small[\href{http://dx.doi.org/10.1103/PhysRevD.62.044015}{DOI}]},
  {\small[\href{http://arxiv.org/abs/gr-qc/0001026}{{arXiv:gr-qc/0001026
  {\small[gr-qc]}}}]}.

\bibitem{Novello:1999pg}
Novello, M., De~Lorenci, V.~A., Salim, J.~M.  and Klippert, Renato,
  ``{Geometrical aspects of light propagation in nonlinear electrodynamics}'',
  {\em Phys. Rev.}, {\bf D61}, 045001 (2000).
  {\small[\href{http://dx.doi.org/10.1103/PhysRevD.61.045001}{DOI}]},
  {\small[\href{http://arxiv.org/abs/gr-qc/9911085}{{arXiv:gr-qc/9911085
  {\small[gr-qc]}}}]}.

\bibitem{Nunez:2004ts}
Nunez, Alvaro  and Solganik, Slava, ``{Ghost constraints on modified
  gravity}'', {\em Phys. Lett.}, {\bf B608}, 189--193 (2005).
  {\small[\href{http://dx.doi.org/10.1016/j.physletb.2005.01.015}{DOI}]},
  {\small[\href{http://arxiv.org/abs/hep-th/0411102}{{arXiv:hep-th/0411102
  {\small[hep-th]}}}]}.

\bibitem{Odintsov:2015zza}
Odintsov, S.~D.  and Oikonomou, V.~K., ``{Bouncing cosmology with future
  singularity from modified gravity}'', {\em Phys. Rev.}, {\bf D92}(2), 024016
  (2015). {\small[\href{http://dx.doi.org/10.1103/PhysRevD.92.024016}{DOI}]},
  {\small[\href{http://arxiv.org/abs/1504.06866}{{arXiv:1504.06866
  {\small[gr-qc]}}}]}.

\bibitem{Odintsov:2014yaa}
Odintsov, Sergei~D., Olmo, Gonzalo~J.  and Rubiera-Garcia, D., ``{Born-Infeld
  gravity and its functional extensions}'', {\em Phys. Rev.}, {\bf D90}, 044003
  (2014). {\small[\href{http://dx.doi.org/10.1103/PhysRevD.90.044003}{DOI}]},
  {\small[\href{http://arxiv.org/abs/1406.1205}{{arXiv:1406.1205
  {\small[hep-th]}}}]}.

\bibitem{Olmo:2005zr}
Olmo, Gonzalo~J., ``{The Gravity Lagrangian according to solar system
  experiments}'', {\em Phys. Rev. Lett.}, {\bf 95}, 261102 (2005).
  {\small[\href{http://dx.doi.org/10.1103/PhysRevLett.95.261102}{DOI}]},
  {\small[\href{http://arxiv.org/abs/gr-qc/0505101}{{arXiv:gr-qc/0505101
  {\small[gr-qc]}}}]}.

\bibitem{Olmo:2006eh}
Olmo, Gonzalo~J., ``{Limit to general relativity in f(R) theories of
  gravity}'', {\em Phys. Rev.}, {\bf D75}, 023511 (2007).
  {\small[\href{http://dx.doi.org/10.1103/PhysRevD.75.023511}{DOI}]},
  {\small[\href{http://arxiv.org/abs/gr-qc/0612047}{{arXiv:gr-qc/0612047
  {\small[gr-qc]}}}]}.

\bibitem{Olmo:2008ye}
Olmo, Gonzalo~J., ``{Hydrogen atom in Palatini theories of gravity}'', {\em
  Phys. Rev.}, {\bf D77}, 084021 (2008).
  {\small[\href{http://dx.doi.org/10.1103/PhysRevD.77.084021}{DOI}]},
  {\small[\href{http://arxiv.org/abs/0802.4038}{{arXiv:0802.4038
  {\small[gr-qc]}}}]}.

\bibitem{Olmo:2011uz}
Olmo, Gonzalo~J., ``{Palatini Approach to Modified Gravity: f(R) Theories and
  Beyond}'', {\em Int. J. Mod. Phys.}, {\bf D20}, 413--462 (2011).
  {\small[\href{http://dx.doi.org/10.1142/S0218271811018925}{DOI}]},
  {\small[\href{http://arxiv.org/abs/1101.3864}{{arXiv:1101.3864
  {\small[gr-qc]}}}]}.

\bibitem{Olmo:2011np}
Olmo, Gonzalo~J.  and Rubiera-Garcia, D., ``{Nonsingular black holes in
  quadratic Palatini gravity}'', {\em Eur. Phys. J.}, {\bf C72}, 2098 (2012).
  {\small[\href{http://dx.doi.org/10.1140/epjc/s10052-012-2098-7}{DOI}]},
  {\small[\href{http://arxiv.org/abs/1112.0475}{{arXiv:1112.0475
  {\small[gr-qc]}}}]}.

\bibitem{Olmo:2012hu}
Olmo, Gonzalo~J.  and Rubiera-Garcia, Diego, ``{Nonsingular charged black holes
  \`{a} la Palatini}'', {\em Int. J. Mod. Phys.}, {\bf D21}, 1250067 (2012).
  {\small[\href{http://dx.doi.org/10.1142/S0218271812500678}{DOI}]},
  {\small[\href{http://arxiv.org/abs/1207.4303}{{arXiv:1207.4303
  {\small[gr-qc]}}}]}.

\bibitem{Olmo:2012nx}
Olmo, Gonzalo~J.  and Rubiera-Garcia, D., ``{Reissner-Nordstr\'om black holes
  in extended Palatini theories}'', {\em Phys. Rev.}, {\bf D86}, 044014 (2012).
  {\small[\href{http://dx.doi.org/10.1103/PhysRevD.86.044014}{DOI}]},
  {\small[\href{http://arxiv.org/abs/1207.6004}{{arXiv:1207.6004
  {\small[gr-qc]}}}]}.

\bibitem{Olmo:2013lta}
Olmo, Gonzalo~J.  and Rubiera-Garcia, D., ``{Importance of torsion and
  invariant volumes in Palatini theories of gravity}'', {\em Phys. Rev.}, {\bf
  D88}, 084030 (2013).
  {\small[\href{http://dx.doi.org/10.1103/PhysRevD.88.084030}{DOI}]},
  {\small[\href{http://arxiv.org/abs/1306.4210}{{arXiv:1306.4210
  {\small[hep-th]}}}]}.

\bibitem{Olmo:2013mla}
Olmo, Gonzalo~J.  and Rubiera-Garcia, D., ``{Semiclassical geons at particle
  accelerators}'', {\em JCAP}, {\bf 1402}, 010 (2014).
  {\small[\href{http://dx.doi.org/10.1088/1475-7516/2014/02/010}{DOI}]},
  {\small[\href{http://arxiv.org/abs/1306.6537}{{arXiv:1306.6537
  {\small[hep-th]}}}]}.

\bibitem{Olmo:2015axa}
Olmo, Gonzalo~J.  and Rubiera-Garcia, Diego, ``{Nonsingular Black Holes in
  $f(R)$ Theories}'', {\em Universe}, {\bf 1}(2), 173--185 (2015).
  {\small[\href{http://dx.doi.org/10.3390/universe1020173}{DOI}]},
  {\small[\href{http://arxiv.org/abs/1509.02430}{{arXiv:1509.02430
  {\small[hep-th]}}}]}.

\bibitem{Olmo:2015bya}
Olmo, Gonzalo~J., Rubiera-Garcia, D.  and Sanchez-Puente, A., ``{Geodesic
  completeness in a wormhole spacetime with horizons}'', {\em Phys. Rev.}, {\bf
  D92}(4), 044047 (2015).
  {\small[\href{http://dx.doi.org/10.1103/PhysRevD.92.044047}{DOI}]},
  {\small[\href{http://arxiv.org/abs/1508.03272}{{arXiv:1508.03272
  {\small[hep-th]}}}]}.

\bibitem{Olmo:2015dba}
Olmo, Gonzalo~J., Rubiera-Garcia, D.  and Sanchez-Puente, A., ``{Classical
  resolution of black hole singularities via wormholes}'', {\em Eur. Phys. J.},
  {\bf C76}(3), 143 (2016).
  {\small[\href{http://dx.doi.org/10.1140/epjc/s10052-016-3999-7}{DOI}]},
  {\small[\href{http://arxiv.org/abs/1504.07015}{{arXiv:1504.07015
  {\small[hep-th]}}}]}.

\bibitem{Olmo:2016fuc}
Olmo, Gonzalo~J., Rubiera-Garcia, D.  and Sanchez-Puente, A., ``{Impact of
  curvature divergences on physical observers in a wormhole space--time with
  horizons}'', {\em Class. Quant. Grav.}, {\bf 33}(11), 115007 (2016).
  {\small[\href{http://dx.doi.org/10.1088/0264-9381/33/11/115007}{DOI}]},
  {\small[\href{http://arxiv.org/abs/1602.01798}{{arXiv:1602.01798
  {\small[hep-th]}}}]}.

\bibitem{Olmo:2013gqa}
Olmo, Gonzalo~J., Rubiera-Garcia, D.  and Sanchis-Alepuz, Helios, ``{Geonic
  black holes and remnants in Eddington-inspired Born-Infeld gravity}'', {\em
  Eur. Phys. J.}, {\bf C74}, 2804 (2014).
  {\small[\href{http://dx.doi.org/10.1140/epjc/s10052-014-2804-8}{DOI}]},
  {\small[\href{http://arxiv.org/abs/1311.0815}{{arXiv:1311.0815
  {\small[hep-th]}}}]}.

\bibitem{Olmo:2009xy}
Olmo, Gonzalo~J., Sanchis-Alepuz, Helios  and Tripathi, Swapnil, ``{Dynamical
  Aspects of Generalized Palatini Theories of Gravity}'', {\em Phys. Rev.},
  {\bf D80}, 024013 (2009).
  {\small[\href{http://dx.doi.org/10.1103/PhysRevD.80.024013}{DOI}]},
  {\small[\href{http://arxiv.org/abs/0907.2787}{{arXiv:0907.2787
  {\small[gr-qc]}}}]}.

\bibitem{Ori:1991zz}
Ori, Amos, ``{Inner structure of a charged black hole: An exact mass-inflation
  solution}'', {\em Phys. Rev. Lett.}, {\bf 67}, 789--792 (1991).
  {\small[\href{http://dx.doi.org/10.1103/PhysRevLett.67.789}{DOI}]}.

\bibitem{Ori:2000fi}
Ori, Amos, ``{Strength of curvature singularities}'', {\em Phys. Rev.}, {\bf
  D61}, 064016 (2000).
  {\small[\href{http://dx.doi.org/10.1103/PhysRevD.61.064016}{DOI}]}.

\bibitem{Orosz:2011np}
Orosz, Jerome~A., McClintock, Jeffrey~E., Aufdenberg, Jason~P., Remillard,
  Ronald~A., Reid, Mark~J., Narayan, Ramesh  and Gou, Lijun, ``{The Mass of the
  Black Hole in Cygnus X-1}'', {\em Astrophys. J.}, {\bf 742}, 84 (2011).
  {\small[\href{http://dx.doi.org/10.1088/0004-637X/742/2/84}{DOI}]},
  {\small[\href{http://arxiv.org/abs/1106.3689}{{arXiv:1106.3689
  {\small[astro-ph.HE]}}}]}.

\bibitem{Ortinbook}
Ortin, Tomas, ``{Gravity and strings}'', {\em Cambridge Monographs on
  Mathematical Physics, Cambridge University Press} (2004).

\bibitem{ortin2007gravity}
Ort{\'\i}n, T., {\em Gravity and Strings}, Cambridge Monographs on Mathematical
  Physics,  (Cambridge University Press, 2007)URL:
  \newline\url{https://books.google.fr/books?id=HDmucsxABzYC}.

\bibitem{Ozel:2012wu}
Ozel, Feryal, ``{Surface Emission from Neutron Stars and Implications for the
  Physics of their Interiors}'', {\em Rept. Prog. Phys.}, {\bf 76}, 016901
  (2013).
  {\small[\href{http://dx.doi.org/10.1088/0034-4885/76/1/016901}{DOI}]},
  {\small[\href{http://arxiv.org/abs/1210.0916}{{arXiv:1210.0916
  {\small[astro-ph.HE]}}}]}.

\bibitem{Morel1997}
P, Morel, ``{CESAM: A code for stellar evolution calculations}'', {\em Astron.
  Astrophys. Suppl. Ser.}, {\bf 124}(3), 597--614 (1997).
  {\small[\href{http://dx.doi.org/10.1051/aas:1997209}{DOI}]}.

\bibitem{1983ApJ267315P}
{Paczynski}, B., ``{Models of X-ray bursters with radius expansion}'', {\em
  Astrophys. J.}, {\bf 267}, 315--321 (April 1983).
  {\small[\href{http://dx.doi.org/10.1086/160870}{DOI}]},
  {\small[\href{http://adsabs.harvard.edu/abs/1983ApJ...267..315P}{ADS}]}.

\bibitem{Padmanabhan:2002ji}
Padmanabhan, T., ``{Cosmological constant: The Weight of the vacuum}'', {\em
  Phys. Rept.}, {\bf 380}, 235--320 (2003).
  {\small[\href{http://dx.doi.org/10.1016/S0370-1573(03)00120-0}{DOI}]},
  {\small[\href{http://arxiv.org/abs/hep-th/0212290}{{arXiv:hep-th/0212290
  {\small[hep-th]}}}]}.

\bibitem{Palais:1979rca}
Palais, Richard~S., ``{The principle of symmetric criticality}'', {\em Commun.
  Math. Phys.}, {\bf 69}(1), 19--30 (1979).
  {\small[\href{http://dx.doi.org/10.1007/BF01941322}{DOI}]}.

\bibitem{Pani:2011mg}
Pani, Paolo, Cardoso, Vitor  and Delsate, Terence, ``{Compact stars in
  Eddington inspired gravity}'', {\em Phys. Rev. Lett.}, {\bf 107}, 031101
  (2011).
  {\small[\href{http://dx.doi.org/10.1103/PhysRevLett.107.031101}{DOI}]},
  {\small[\href{http://arxiv.org/abs/1106.3569}{{arXiv:1106.3569
  {\small[gr-qc]}}}]}.

\bibitem{Pani:2012qb}
Pani, Paolo, Delsate, Terence  and Cardoso, Vitor, ``{Eddington-inspired
  Born-Infeld gravity. Phenomenology of non-linear gravity-matter coupling}'',
  {\em Phys. Rev.}, {\bf D85}, 084020 (2012).
  {\small[\href{http://dx.doi.org/10.1103/PhysRevD.85.084020}{DOI}]},
  {\small[\href{http://arxiv.org/abs/1201.2814}{{arXiv:1201.2814
  {\small[gr-qc]}}}]}.

\bibitem{Pani:2012qd}
Pani, Paolo  and Sotiriou, Thomas~P., ``{Surface singularities in
  Eddington-inspired Born-Infeld gravity}'', {\em Phys. Rev. Lett.}, {\bf 109},
  251102 (2012).
  {\small[\href{http://dx.doi.org/10.1103/PhysRevLett.109.251102}{DOI}]},
  {\small[\href{http://arxiv.org/abs/1209.2972}{{arXiv:1209.2972
  {\small[gr-qc]}}}]}.

\bibitem{Pani:2013qfa}
Pani, Paolo, Sotiriou, Thomas~P.  and Vernieri, Daniele, ``{Gravity with
  Auxiliary Fields}'', {\em Phys. Rev.}, {\bf D88}(12), 121502 (2013).
  {\small[\href{http://dx.doi.org/10.1103/PhysRevD.88.121502}{DOI}]},
  {\small[\href{http://arxiv.org/abs/1306.1835}{{arXiv:1306.1835
  {\small[gr-qc]}}}]}.

\bibitem{Peebles:2002gy}
Peebles, P. J.~E.  and Ratra, Bharat, ``{The Cosmological constant and dark
  energy}'', {\em Rev. Mod. Phys.}, {\bf 75}, 559--606 (2003).
  {\small[\href{http://dx.doi.org/10.1103/RevModPhys.75.559}{DOI}]},
  {\small[\href{http://arxiv.org/abs/astro-ph/0207347}{{arXiv:astro-ph/0207347
  {\small[astro-ph]}}}]}.

\bibitem{Penrose:1964wq}
Penrose, Roger, ``{Gravitational collapse and space-time singularities}'', {\em
  Phys. Rev. Lett.}, {\bf 14}, 57--59 (1965).
  {\small[\href{http://dx.doi.org/10.1103/PhysRevLett.14.57}{DOI}]}.

\bibitem{Penrose:1969pc}
Penrose, R., ``{Gravitational collapse: The role of general relativity}'', {\em
  Riv. Nuovo Cim.}, {\bf 1}, 252--276 (1969). [Gen. Rel. Grav.34,1141(2002)].

\bibitem{Perlmutter:1998np}
Perlmutter, S.  {et~al.} (Supernova Cosmology Project), ``{Measurements of
  Omega and Lambda from 42 high redshift supernovae}'', {\em Astrophys. J.},
  {\bf 517}, 565--586 (1999).
  {\small[\href{http://dx.doi.org/10.1086/307221}{DOI}]},
  {\small[\href{http://arxiv.org/abs/astro-ph/9812133}{{arXiv:astro-ph/9812133
  {\small[astro-ph]}}}]}.

\bibitem{Piran:2005qu}
Piran, Tsvi, ``{Magnetic fields in gamma-ray bursts: A Short overview}'', {\em
  AIP Conf. Proc.}, {\bf 784}, 164--174 (2005).
  {\small[\href{http://dx.doi.org/10.1063/1.2077181}{DOI}]},
  {\small[\href{http://arxiv.org/abs/astro-ph/0503060}{{arXiv:astro-ph/0503060
  {\small[astro-ph]}}}]}. [,164(2005)].

\bibitem{Plebanski:106680}
Plebanski, J~F, {\em {Lectures on non-linear electrodynamics}},  (NORDITA,
  Copenhagen, 1970)URL: \newline\url{http://cds.cern.ch/record/106680}.
  Lectures given at the Niels Bohr Institute and NORDITA, Copenhagen.

\bibitem{Poissonbook}
Poisson, Erik, ``{A Relativist's Toolkit: The Mathematics of Black-Hole
  Mechanics }'', {\em Cambridge University Press, New York} (2004).

\bibitem{Poisson:1989zz}
Poisson, Eric  and Israel, W., ``{Inner-horizon instability and mass inflation
  in black holes}'', {\em Phys. Rev. Lett.}, {\bf 63}, 1663--1666 (1989).
  {\small[\href{http://dx.doi.org/10.1103/PhysRevLett.63.1663}{DOI}]}.

\bibitem{Poisson:1990eh}
Poisson, Eric  and Israel, W., ``{Internal structure of black holes}'', {\em
  Phys. Rev.}, {\bf D41}, 1796--1809 (1990).
  {\small[\href{http://dx.doi.org/10.1103/PhysRevD.41.1796}{DOI}]}.

\bibitem{polchinski1998stringI}
Polchinski, J., {\em String Theory: Volume 1, An Introduction to the Bosonic
  String}, Cambridge Monographs on Mathematical Physics,  (Cambridge University
  Press, 1998)URL:
  \newline\url{https://books.google.fr/books?id=jbM3t\_usmX0C}.

\bibitem{polchinski1998stringII}
Polchinski, J., {\em String Theory: Volume 2, Superstring Theory and Beyond},
  Cambridge Monographs on Mathematical Physics,  (Cambridge University Press,
  1998)URL: \newline\url{https://books.google.fr/books?id=WKatSc5pjOgC}.

\bibitem{Potapov:2014iva}
Potapov, Alexander~A., Izmailov, Ramil, Mikolaychuk, Olga, Mikolaychuk,
  Nikolay, Ghosh, Mithun  and Nandi, Kamal~K., ``{Constraint on dark matter
  central density in the Eddington inspired Born-Infeld (EiBI) gravity with
  input from Weyl gravity}'', {\em JCAP}, {\bf 1507}(07), 018 (2015).
  {\small[\href{http://dx.doi.org/10.1088/1475-7516/2015/07/018}{DOI}]},
  {\small[\href{http://arxiv.org/abs/1412.7897}{{arXiv:1412.7897
  {\small[gr-qc]}}}]}.

\bibitem{Potekhin:2004jr}
Potekhin, A.~Y., Lai, Dong, Chabrier, G.  and Ho, W. C.~G., ``{Electromagnetic
  polarization in partially ionized plasmas with strong magnetic fields and
  neutron star atmosphere models}'', {\em Astrophys. J.}, {\bf 612}, 1034--1043
  (2004). {\small[\href{http://dx.doi.org/10.1086/422679}{DOI}]},
  {\small[\href{http://arxiv.org/abs/astro-ph/0405383}{{arXiv:astro-ph/0405383
  {\small[astro-ph]}}}]}.

\bibitem{Psaltis:2008bb}
Psaltis, Dimitrios, ``{Probes and Tests of Strong-Field Gravity with
  Observations in the Electromagnetic Spectrum}'', {\em Living Rev. Rel.}, {\bf
  11}, 9 (2008). {\small[\href{http://dx.doi.org/10.12942/lrr-2008-9}{DOI}]},
  {\small[\href{http://arxiv.org/abs/0806.1531}{{arXiv:0806.1531
  {\small[astro-ph]}}}]}.

\bibitem{Qauli:2016vza}
Qauli, A.~I., Iqbal, M., Sulaksono, A.  and Ramadhan, H.~S., ``{Hyperons in
  neutron stars within an Eddington-inspired Born-Infeld theory of gravity}'',
  {\em Phys. Rev.}, {\bf D93}(10), 104056 (2016).
  {\small[\href{http://dx.doi.org/10.1103/PhysRevD.93.104056}{DOI}]},
  {\small[\href{http://arxiv.org/abs/1605.01152}{{arXiv:1605.01152
  {\small[astro-ph.SR]}}}]}.

\bibitem{Randall:1999ee}
Randall, Lisa  and Sundrum, Raman, ``{A Large mass hierarchy from a small extra
  dimension}'', {\em Phys. Rev. Lett.}, {\bf 83}, 3370--3373 (1999).
  {\small[\href{http://dx.doi.org/10.1103/PhysRevLett.83.3370}{DOI}]},
  {\small[\href{http://arxiv.org/abs/hep-ph/9905221}{{arXiv:hep-ph/9905221
  {\small[hep-ph]}}}]}.

\bibitem{Randall:1999vf}
Randall, Lisa  and Sundrum, Raman, ``{An Alternative to compactification}'',
  {\em Phys. Rev. Lett.}, {\bf 83}, 4690--4693 (1999).
  {\small[\href{http://dx.doi.org/10.1103/PhysRevLett.83.4690}{DOI}]},
  {\small[\href{http://arxiv.org/abs/hep-th/9906064}{{arXiv:hep-th/9906064
  {\small[hep-th]}}}]}.

\bibitem{Rattazzi:2003ea}
Rattazzi, R., ``{Cargese lectures on extra-dimensions}'', in {\em {Particle
  physics and cosmology: The interface. Proceedings, NATO Advanced Study
  Institute, School, Cargese, France, August 4-16, 2003}}, pp. 461--517,
  (2003).
  {\small[\href{http://arxiv.org/abs/hep-ph/0607055}{{arXiv:hep-ph/0607055
  {\small[hep-ph]}}}]}URL:
  \newline\url{http://weblib.cern.ch/abstract?CERN-PH-TH-2006-029-JOURNAL-REF:-PARTICLE-PHYSICS}.

\bibitem{Rawls:2011jw}
Rawls, Meredith~L., Orosz, Jerome~A., McClintock, Jeffrey~E., Torres, Manuel
  A.~P., Bailyn, Charles~D.  and Buxton, Michelle~M., ``{Refined Neutron-Star
  Mass Determinations for Six Eclipsing X-Ray Pulsar Binaries}'', {\em
  Astrophys. J.}, {\bf 730}, 25 (2011).
  {\small[\href{http://dx.doi.org/10.1088/0004-637X/730/1/25}{DOI}]},
  {\small[\href{http://arxiv.org/abs/1101.2465}{{arXiv:1101.2465
  {\small[astro-ph.SR]}}}]}.

\bibitem{Riess:1998cb}
Riess, Adam~G.  {et~al.} (Supernova Search Team), ``{Observational evidence
  from supernovae for an accelerating universe and a cosmological constant}'',
  {\em Astron. J.}, {\bf 116}, 1009--1038 (1998).
  {\small[\href{http://dx.doi.org/10.1086/300499}{DOI}]},
  {\small[\href{http://arxiv.org/abs/astro-ph/9805201}{{arXiv:astro-ph/9805201
  {\small[astro-ph]}}}]}.

\bibitem{Robinson:1975bv}
Robinson, D.~C., ``{Uniqueness of the Kerr black hole}'', {\em Phys. Rev.
  Lett.}, {\bf 34}, 905--906 (1975).
  {\small[\href{http://dx.doi.org/10.1103/PhysRevLett.34.905}{DOI}]}.

\bibitem{Rodrigues:2008kv}
Rodrigues, Davi~C., ``{Evolution of Anisotropies in Eddington-Born-Infeld
  Cosmology}'', {\em Phys. Rev.}, {\bf D78}, 063013 (2008).
  {\small[\href{http://dx.doi.org/10.1103/PhysRevD.78.063013}{DOI}]},
  {\small[\href{http://arxiv.org/abs/0806.3613}{{arXiv:0806.3613
  {\small[gr-qc]}}}]}.

\bibitem{Rovelli:2014cta}
Rovelli, Carlo  and Vidotto, Francesca, ``{Planck stars}'', {\em Int. J. Mod.
  Phys.}, {\bf D23}(12), 1442026 (2014).
  {\small[\href{http://dx.doi.org/10.1142/S0218271814420267}{DOI}]},
  {\small[\href{http://arxiv.org/abs/1401.6562}{{arXiv:1401.6562
  {\small[gr-qc]}}}]}.

\bibitem{Salazar:1987ap}
Salazar, I.~H., Garcia, A.  and Plebanski, J., ``{Duality Rotations and Type
  $D$ Solutions to Einstein Equations With Nonlinear Electromagnetic
  Sources}'', {\em J. Math. Phys.}, {\bf 28}, 2171--2181 (1987).
  {\small[\href{http://dx.doi.org/10.1063/1.527430}{DOI}]}.

\bibitem{Santos:2015sra}
Santos, Noelia~S.  and Santos, Janilo, ``{The virial theorem in
  Eddington-Born-Infeld gravity}'', {\em JCAP}, {\bf 1512}(12), 002 (2015).
  {\small[\href{http://dx.doi.org/10.1088/1475-7516/2015/12/002}{DOI}]},
  {\small[\href{http://arxiv.org/abs/1506.04569}{{arXiv:1506.04569
  {\small[gr-qc]}}}]}.

\bibitem{Scargill:2012kg}
Scargill, James H.~C., Banados, Maximo  and Ferreira, Pedro~G., ``{Cosmology
  with Eddington-inspired Gravity}'', {\em Phys. Rev.}, {\bf D86}, 103533
  (2012). {\small[\href{http://dx.doi.org/10.1103/PhysRevD.86.103533}{DOI}]},
  {\small[\href{http://arxiv.org/abs/1210.1521}{{arXiv:1210.1521
  {\small[astro-ph.CO]}}}]}.

\bibitem{Schmidt-May:2014tpa}
Schmidt-May, Angnis  and von Strauss, Mikael, ``{A link between ghost-free
  bimetric and Eddington-inspired Born-Infeld theory}'' (2014).
  {\small[\href{http://arxiv.org/abs/1412.3812}{{arXiv:1412.3812
  {\small[hep-th]}}}]}.

\bibitem{schouten1954ricci}
Schouten, J.A., {\em Ricci-calculus, an introduction to tensor analysis and its
  geometrical applications, by J. A. Schouten,... 2nd edition...},
  (Springer-Verlag, 1954)URL:
  \newline\url{https://books.google.fr/books?id=2\_1iQwAACAAJ}.

\bibitem{Senovilla:2014gza}
Senovilla, Jos{\'e} M.~M.  and Garfinkle, David, ``{The 1965 Penrose
  singularity theorem}'', {\em Class. Quant. Grav.}, {\bf 32}(12), 124008
  (2015).
  {\small[\href{http://dx.doi.org/10.1088/0264-9381/32/12/124008}{DOI}]},
  {\small[\href{http://arxiv.org/abs/1410.5226}{{arXiv:1410.5226
  {\small[gr-qc]}}}]}.

\bibitem{Shaikh:2015oha}
Shaikh, Rajibul, ``{Lorentzian wormholes in Eddington-inspired Born-Infeld
  gravity}'', {\em Phys. Rev.}, {\bf D92}, 024015 (2015).
  {\small[\href{http://dx.doi.org/10.1103/PhysRevD.92.024015}{DOI}]},
  {\small[\href{http://arxiv.org/abs/1505.01314}{{arXiv:1505.01314
  {\small[gr-qc]}}}]}.

\bibitem{Sham:2013sya}
Sham, Y.~H., Leung, P.~T.  and Lin, L.~M., ``{Compact stars in
  Eddington-inspired Born-Infeld gravity: Anomalies associated with phase
  transitions}'', {\em Phys. Rev.}, {\bf D87}(6), 061503 (2013).
  {\small[\href{http://dx.doi.org/10.1103/PhysRevD.87.061503}{DOI}]},
  {\small[\href{http://arxiv.org/abs/1304.0550}{{arXiv:1304.0550
  {\small[gr-qc]}}}]}.

\bibitem{Sham:2012qi}
Sham, Y.~H., Lin, L.~M.  and Leung, P.~T., ``{Radial oscillations and stability
  of compact stars in Eddington inspired Born-Infeld gravity}'', {\em Phys.
  Rev.}, {\bf D86}, 064015 (2012).
  {\small[\href{http://dx.doi.org/10.1103/PhysRevD.86.064015}{DOI}]},
  {\small[\href{http://arxiv.org/abs/1208.1314}{{arXiv:1208.1314
  {\small[gr-qc]}}}]}.

\bibitem{Sham:2013cya}
Sham, Y.~H., Lin, L.~M.  and Leung, P.~T., ``{Testing universal relations of
  neutron stars with a nonlinear matter-gravity coupling theory}'', {\em
  Astrophys. J.}, {\bf 781}, 66 (2014).
  {\small[\href{http://dx.doi.org/10.1088/0004-637X/781/2/66}{DOI}]},
  {\small[\href{http://arxiv.org/abs/1312.1011}{{arXiv:1312.1011
  {\small[gr-qc]}}}]}.

\bibitem{ShapTeuk}
Shapiro, Stuart~L.  and Teukolsky, S.~A., ``{Black holes, white dwarfs and
  neutron stars}'', {\em John Wiley \& Sons, New York} (1983).

\bibitem{Sinha:2010ai}
Sinha, Aninda, ``{On the new massive gravity and AdS/CFT}'', {\em JHEP}, {\bf
  06}, 061 (2010).
  {\small[\href{http://dx.doi.org/10.1007/JHEP06(2010)061}{DOI}]},
  {\small[\href{http://arxiv.org/abs/1003.0683}{{arXiv:1003.0683
  {\small[hep-th]}}}]}.

\bibitem{Sotani:2014goa}
Sotani, Hajime, ``{Observational discrimination of Eddington-inspired
  Born-Infeld gravity from general relativity}'', {\em Phys. Rev.}, {\bf
  D89}(10), 104005 (2014).
  {\small[\href{http://dx.doi.org/10.1103/PhysRevD.89.104005}{DOI}]},
  {\small[\href{http://arxiv.org/abs/1404.5369}{{arXiv:1404.5369
  {\small[astro-ph.HE]}}}]}.

\bibitem{Sotani:2014xoa}
Sotani, Hajime, ``{Stellar oscillations in Eddington-inspired Born-Infeld
  gravity}'', {\em Phys. Rev.}, {\bf D89}(12), 124037 (2014).
  {\small[\href{http://dx.doi.org/10.1103/PhysRevD.89.124037}{DOI}]},
  {\small[\href{http://arxiv.org/abs/1406.3097}{{arXiv:1406.3097
  {\small[astro-ph.HE]}}}]}.

\bibitem{Sotani:2015tya}
Sotani, Hajime, ``{Magnetized relativistic stellar models in Eddington-inspired
  Born-Infeld gravity}'', {\em Phys. Rev.}, {\bf D91}, 084020 (2015).
  {\small[\href{http://dx.doi.org/10.1103/PhysRevD.91.084020}{DOI}]},
  {\small[\href{http://arxiv.org/abs/1503.07942}{{arXiv:1503.07942
  {\small[astro-ph.HE]}}}]}.

\bibitem{Sotani:2014lua}
Sotani, Hajime  and Miyamoto, Umpei, ``{Properties of an electrically charged
  black hole in Eddington-inspired Born-Infeld gravity}'', {\em Phys. Rev.},
  {\bf D90}, 124087 (2014).
  {\small[\href{http://dx.doi.org/10.1103/PhysRevD.90.124087}{DOI}]},
  {\small[\href{http://arxiv.org/abs/1412.4173}{{arXiv:1412.4173
  {\small[gr-qc]}}}]}.

\bibitem{Sotani:2015ewa}
Sotani, Hajime  and Miyamoto, Umpei, ``{Strong gravitational lensing by an
  electrically charged black hole in Eddington-inspired Born-Infeld gravity}'',
  {\em Phys. Rev.}, {\bf D92}(4), 044052 (2015).
  {\small[\href{http://dx.doi.org/10.1103/PhysRevD.92.044052}{DOI}]},
  {\small[\href{http://arxiv.org/abs/1508.03119}{{arXiv:1508.03119
  {\small[gr-qc]}}}]}.

\bibitem{Sotiriou:2008rp}
Sotiriou, Thomas~P.  and Faraoni, Valerio, ``{f(R) Theories Of Gravity}'', {\em
  Rev. Mod. Phys.}, {\bf 82}, 451--497 (2010).
  {\small[\href{http://dx.doi.org/10.1103/RevModPhys.82.451}{DOI}]},
  {\small[\href{http://arxiv.org/abs/0805.1726}{{arXiv:0805.1726
  {\small[gr-qc]}}}]}.

\bibitem{Spergel:2003cb}
Spergel, D.~N.  {et~al.} (WMAP), ``{First year Wilkinson Microwave Anisotropy
  Probe (WMAP) observations: Determination of cosmological parameters}'', {\em
  Astrophys. J. Suppl.}, {\bf 148}, 175--194 (2003).
  {\small[\href{http://dx.doi.org/10.1086/377226}{DOI}]},
  {\small[\href{http://arxiv.org/abs/astro-ph/0302209}{{arXiv:astro-ph/0302209
  {\small[astro-ph]}}}]}.

\bibitem{Stelle:1976gc}
Stelle, K.~S., ``{Renormalization of Higher Derivative Quantum Gravity}'', {\em
  Phys. Rev.}, {\bf D16}, 953--969 (1977).
  {\small[\href{http://dx.doi.org/10.1103/PhysRevD.16.953}{DOI}]}.

\bibitem{Stelle:1977ry}
Stelle, K.~S., ``{Classical Gravity with Higher Derivatives}'', {\em Gen. Rel.
  Grav.}, {\bf 9}, 353--371 (1978).
  {\small[\href{http://dx.doi.org/10.1007/BF00760427}{DOI}]}.

\bibitem{Strigari:2013iaa}
Strigari, Louis~E., ``{Galactic Searches for Dark Matter}'', {\em Phys. Rept.},
  {\bf 531}, 1--88 (2013).
  {\small[\href{http://dx.doi.org/10.1016/j.physrep.2013.05.004}{DOI}]},
  {\small[\href{http://arxiv.org/abs/1211.7090}{{arXiv:1211.7090
  {\small[astro-ph.CO]}}}]}.

\bibitem{Sushkov:2007me}
Sushkov, Sergey~V.  and Zhang, Yuan-Zhong, ``{Scalar wormholes in cosmological
  setting and their instability}'', {\em Phys. Rev.}, {\bf D77}, 024042 (2008).
  {\small[\href{http://dx.doi.org/10.1103/PhysRevD.77.024042}{DOI}]},
  {\small[\href{http://arxiv.org/abs/0712.1727}{{arXiv:0712.1727
  {\small[gr-qc]}}}]}.

\bibitem{Tamang:2015tmd}
Tamang, Amarjit, Potapov, Alexander~A., Lukmanova, Regina, Izmailov, Ramil  and
  Nandi, Kamal~K., ``{On the generalized wormhole in the Eddington-inspired
  Born--Infeld gravity}'', {\em Class. Quant. Grav.}, {\bf 32}(23), 235028
  (2015).
  {\small[\href{http://dx.doi.org/10.1088/0264-9381/32/23/235028}{DOI}]},
  {\small[\href{http://arxiv.org/abs/1512.01451}{{arXiv:1512.01451
  {\small[gr-qc]}}}]}.

\bibitem{Thompson:2001ig}
Thompson, C., Lyutikov, M.  and Kulkarni, S.~R., ``{Electrodynamics of
  magnetars: implications for the persistent x-ray emission and spindown of the
  soft gamma repeaters and anomalous x-ray pulsars}'', {\em Astrophys. J.},
  {\bf 574}, 332--355 (2002).
  {\small[\href{http://dx.doi.org/10.1086/340586}{DOI}]},
  {\small[\href{http://arxiv.org/abs/astro-ph/0110677}{{arXiv:astro-ph/0110677
  {\small[astro-ph]}}}]}.

\bibitem{Tipler:1977zzb}
Tipler, Frank~J., ``{On the nature of singularities in general relativity}'',
  {\em Phys. Rev.}, {\bf D15}, 942--945 (1977).
  {\small[\href{http://dx.doi.org/10.1103/PhysRevD.15.942}{DOI}]}.

\bibitem{Tipler:1977zza}
Tipler, Frank~J., ``{Singularities in conformally flat spacetimes}'', {\em
  Phys. Lett.}, {\bf A64}, 8--10 (1977).
  {\small[\href{http://dx.doi.org/10.1016/0375-9601(77)90508-4}{DOI}]}.

\bibitem{Tsui:2004qd}
Tsui, L.~K.  and Leung, P.~T., ``{Universality in quasi-normal modes of neutron
  stars}'', {\em Mon. Not. Roy. Astron. Soc.}, {\bf 357}, 1029--1037 (2005).
  {\small[\href{http://dx.doi.org/10.1111/j.1365-2966.2005.08710.x}{DOI}]},
  {\small[\href{http://arxiv.org/abs/gr-qc/0412024}{{arXiv:gr-qc/0412024
  {\small[gr-qc]}}}]}.

\bibitem{TurckChieze:2010gc}
Turck-Chieze, Sylvaine  and Couvidat, Sebastien, ``{Solar neutrinos,
  helioseismology and the solar internal dynamics}'', {\em Rept. Prog. Phys.},
  {\bf 74}, 086901 (2011).
  {\small[\href{http://dx.doi.org/10.1088/0034-4885/74/8/086901}{DOI}]},
  {\small[\href{http://arxiv.org/abs/1009.0852}{{arXiv:1009.0852
  {\small[astro-ph.SR]}}}]}.

\bibitem{ValkenburgThesis}
Valkenburg, Wessel, {\em Linearised nonsymmetric metric perturbations in
  cosmology},  (PhD Thesis, 2006)URL:
  \newline\url{https://web.science.uu.nl/ITF/Teaching/2006/Valkenburg.pdf}.

\bibitem{Virbhadra:1999nm}
Virbhadra, K.~S.  and Ellis, George F.~R., ``{Schwarzschild black hole
  lensing}'', {\em Phys. Rev.}, {\bf D62}, 084003 (2000).
  {\small[\href{http://dx.doi.org/10.1103/PhysRevD.62.084003}{DOI}]},
  {\small[\href{http://arxiv.org/abs/astro-ph/9904193}{{arXiv:astro-ph/9904193
  {\small[astro-ph]}}}]}.

\bibitem{Virbhadra:2002ju}
Virbhadra, K.~S.  and Ellis, G. F.~R., ``{Gravitational lensing by naked
  singularities}'', {\em Phys. Rev.}, {\bf D65}, 103004 (2002).
  {\small[\href{http://dx.doi.org/10.1103/PhysRevD.65.103004}{DOI}]}.

\bibitem{Virbhadra:1998dy}
Virbhadra, K.~S., Narasimha, D.  and Chitre, S.~M., ``{Role of the scalar field
  in gravitational lensing}'', {\em Astron. Astrophys.}, {\bf 337}, 1--8
  (1998).
  {\small[\href{http://arxiv.org/abs/astro-ph/9801174}{{arXiv:astro-ph/9801174
  {\small[astro-ph]}}}]}.

\bibitem{Visserbook}
Visser, Matt, ``{Lorentzian Wormholes}'', {\em AIP Press, New York} (1996).

\bibitem{Visser:2003yf}
Visser, Matt, Kar, Sayan  and Dadhich, Naresh, ``{Traversable wormholes with
  arbitrarily small energy condition violations}'', {\em Phys. Rev. Lett.},
  {\bf 90}, 201102 (2003).
  {\small[\href{http://dx.doi.org/10.1103/PhysRevLett.90.201102}{DOI}]},
  {\small[\href{http://arxiv.org/abs/gr-qc/0301003}{{arXiv:gr-qc/0301003
  {\small[gr-qc]}}}]}.

\bibitem{Vollick:2003qp}
Vollick, Dan~N., ``{Palatini approach to Born-Infeld-Einstein theory and a
  geometric description of electrodynamics}'', {\em Phys. Rev.}, {\bf D69},
  064030 (2004).
  {\small[\href{http://dx.doi.org/10.1103/PhysRevD.69.064030}{DOI}]},
  {\small[\href{http://arxiv.org/abs/gr-qc/0309101}{{arXiv:gr-qc/0309101
  {\small[gr-qc]}}}]}.

\bibitem{Vollick:2005gc}
Vollick, Dan~N., ``{Born-Infeld-Einstein theory with matter}'', {\em Phys.
  Rev.}, {\bf D72}, 084026 (2005).
  {\small[\href{http://dx.doi.org/10.1103/PhysRevD.72.084026}{DOI}]},
  {\small[\href{http://arxiv.org/abs/gr-qc/0506091}{{arXiv:gr-qc/0506091
  {\small[gr-qc]}}}]}.

\bibitem{Vollick:2006qd}
Vollick, Dan~N., ``{Black hole and cosmological space-times in
  Born-Infeld-Einstein theory}'' (2006).
  {\small[\href{http://arxiv.org/abs/gr-qc/0601136}{{arXiv:gr-qc/0601136
  {\small[gr-qc]}}}]}.

\bibitem{Waldbook}
Wald, R, M., ``{General Relativity }'', {\em University Press, Chicago} (1984).

\bibitem{Wei:2014dka}
Wei, Shao-Wen, Yang, Ke  and Liu, Yu-Xiao, ``{Black hole solution and strong
  gravitational lensing in Eddington-inspired Born--Infeld gravity}'', {\em
  Eur. Phys. J.}, {\bf C75}, 253 (2015).
  {\small[\href{http://dx.doi.org/10.1140/epjc/s10052-015-3556-9,
  10.1140/epjc/s10052-015-3469-7}{DOI}]},
  {\small[\href{http://arxiv.org/abs/1405.2178}{{arXiv:1405.2178
  {\small[gr-qc]}}}]}. [Erratum: Eur. Phys. J.C75,331(2015)].

\bibitem{Weinberg:1988cp}
Weinberg, Steven, ``{The Cosmological Constant Problem}'', {\em Rev. Mod.
  Phys.}, {\bf 61}, 1--23 (1989).
  {\small[\href{http://dx.doi.org/10.1103/RevModPhys.61.1}{DOI}]}.

\bibitem{2004cgpsbook}
{Weiss}, A., {Hillebrandt}, W., {Thomas}, H.-C.  and {Ritter}, H., {\em {Cox
  and Giuli's Principles of Stellar Structure}}, (2004).
  {\small[\href{http://adsabs.harvard.edu/abs/2004cgps.book.....W}{ADS}]}.

\bibitem{Wheeler:1955zz}
Wheeler, J.~A., ``{Geons}'', {\em Phys. Rev.}, {\bf 97}, 511--536 (1955).
  {\small[\href{http://dx.doi.org/10.1103/PhysRev.97.511}{DOI}]}.

\bibitem{Will:2014kxa}
Will, Clifford~M., ``{The Confrontation between General Relativity and
  Experiment}'', {\em Living Rev. Rel.}, {\bf 17}, 4 (2014).
  {\small[\href{http://dx.doi.org/10.12942/lrr-2014-4}{DOI}]},
  {\small[\href{http://arxiv.org/abs/1403.7377}{{arXiv:1403.7377
  {\small[gr-qc]}}}]}.

\bibitem{Wohlfarth:2003ss}
Wohlfarth, Mattias N.~R., ``{Gravity a la Born-Infeld}'', {\em Class. Quant.
  Grav.}, {\bf 21}, 1927 (2004).
  {\small[\href{http://dx.doi.org/10.1088/0264-9381/21/8/001}{DOI}]},
  {\small[\href{http://arxiv.org/abs/hep-th/0310067}{{arXiv:hep-th/0310067
  {\small[hep-th]}}}]}. [Erratum: Class. Quant. Grav.21,5297(2004)].

\bibitem{Woodard:2006nt}
Woodard, Richard~P., ``{Avoiding dark energy with 1/r modifications of
  gravity}'', {\em Lect. Notes Phys.}, {\bf 720}, 403--433 (2007).
  {\small[\href{http://dx.doi.org/10.1007/978-3-540-71013-4_14}{DOI}]},
  {\small[\href{http://arxiv.org/abs/astro-ph/0601672}{{arXiv:astro-ph/0601672
  {\small[astro-ph]}}}]}.

\bibitem{Woodard:2015zca}
Woodard, Richard~P., ``{Ostrogradsky's theorem on Hamiltonian instability}'',
  {\em Scholarpedia}, {\bf 10}(8), 32243 (2015).
  {\small[\href{http://dx.doi.org/10.4249/scholarpedia.32243}{DOI}]},
  {\small[\href{http://arxiv.org/abs/1506.02210}{{arXiv:1506.02210
  {\small[hep-th]}}}]}.

\bibitem{Woosley:2006ie}
Woosley, Stan  and Janka, Thomas, ``{The physics of core-collapse
  supernovae}'', {\em Nature Phys.}, {\bf 1}, 147 (2005).
  {\small[\href{http://dx.doi.org/10.1038/nphys172}{DOI}]},
  {\small[\href{http://arxiv.org/abs/astro-ph/0601261}{{arXiv:astro-ph/0601261
  {\small[astro-ph]}}}]}.

\bibitem{Yagi:2013bca}
Yagi, Kent  and Yunes, Nicolas, ``{I-Love-Q}'', {\em Science}, {\bf 341},
  365--368 (2013).
  {\small[\href{http://dx.doi.org/10.1126/science.1236462}{DOI}]},
  {\small[\href{http://arxiv.org/abs/1302.4499}{{arXiv:1302.4499
  {\small[gr-qc]}}}]}.

\bibitem{Yagi:2013awa}
Yagi, Kent  and Yunes, Nicolas, ``{I-Love-Q Relations in Neutron Stars and
  their Applications to Astrophysics, Gravitational Waves and Fundamental
  Physics}'', {\em Phys. Rev.}, {\bf D88}(2), 023009 (2013).
  {\small[\href{http://dx.doi.org/10.1103/PhysRevD.88.023009}{DOI}]},
  {\small[\href{http://arxiv.org/abs/1303.1528}{{arXiv:1303.1528
  {\small[gr-qc]}}}]}.

\bibitem{Yang:2013hsa}
Yang, Ke, Du, Xiao-Long  and Liu, Yu-Xiao, ``{Linear perturbations in
  Eddington-inspired Born-Infeld gravity}'', {\em Phys. Rev.}, {\bf D88},
  124037 (2013).
  {\small[\href{http://dx.doi.org/10.1103/PhysRevD.88.124037}{DOI}]},
  {\small[\href{http://arxiv.org/abs/1307.2969}{{arXiv:1307.2969
  {\small[gr-qc]}}}]}.

\bibitem{Zavlin:2002ed}
Zavlin, Vyacheslav~E.  and Pavlov, G.~G., ``{Modeling neutron star
  atmospheres}'', in {\em {Neutron stars, pulsars and supernova remnants.
  Proceedings, 270th WE-Heraeus Seminar, Bad Honnef, Germany, January 21-25,
  2002}}, pp. 262--272, (2002).
  {\small[\href{http://arxiv.org/abs/astro-ph/0206025}{{arXiv:astro-ph/0206025
  {\small[astro-ph]}}}]}.

\bibitem{Zhang:2014bea}
Zhang, Yiyang, Zhu, Yiwei, Modesto, Leonardo  and Bambi, Cosimo, ``{Can static
  regular black holes form from gravitational collapse?}'', {\em Eur. Phys.
  J.}, {\bf C75}(2), 96 (2015).
  {\small[\href{http://dx.doi.org/10.1140/epjc/s10052-015-3311-2}{DOI}]},
  {\small[\href{http://arxiv.org/abs/1404.4770}{{arXiv:1404.4770
  {\small[gr-qc]}}}]}.

\bibitem{zwiebach2009first}
Zwiebach, B., {\em A First Course in String Theory},  (Cambridge University
  Press, 2009)URL: \newline\url{https://books.google.fr/books?id=ih9kI9MEzh0C}.

\end{thebibliography}

\end{document}